%% file: gesamt.tex
\documentclass[11pt,a4paper]{book}

\usepackage[T1]{fontenc}
\usepackage[english]{babel}

\usepackage{epsfig}
\usepackage{latexsym}
\usepackage{ifthen}
\usepackage[small,normal,bf]{caption2}
\usepackage{diss}
\usepackage[linkcolor=blue,citecolor=blue,filecolor=blue,
           menucolor=blue,pagecolor=blue,urlcolor=blue,
           colorlinks=true,
]
           {hyperref}
\usepackage{units}
\usepackage{nicefrac}
\usepackage[intlimits]{amsmath}
\usepackage{amssymb}
\usepackage{amsmath}
\usepackage{slashed}
\usepackage{subfigure}
\usepackage{bm}
\usepackage{feynmp}

\usepackage{float}

\begin{document}
\parindent=0pt

\input{titel_02}

\pagestyle{plain}

{\pagenumbering{roman}
\renewcommand{\baselinestretch}{1.5}
\normalsize \tableofcontents \vfill\eject
\ifthenelse{\isodd{\value{page}}}{}{\rule{0cm}{15cm}\vfill\eject} }

\pagestyle{plain}

\pagenumbering{arabic}
\setcounter{page}{1}

\input{Introduction}

\input{Construction}

\input{Paper}
\input{Doktorarbeit}

\input{Glueball}

\input{Summary}

\input{Zusammenfassung}

\input{Bibliography}
\pagestyle{plain}



\bibliographystyle{alpha}

\input{Danksagung}
\thispagestyle{empty}

\input{cv}

\end{document}

%% file: titel_02.tex
\begin{titlepage}
\linespread{1.3}     
\large
\bigskip \bigskip
\vspace{10cm}
\begin{center}
{\Huge \bf Quarkonium Phenomenology \\ in Vacuum\\
\medskip}
\vspace{12cm}

Dissertation\\
zur\\
Erlangung des Doktorgrades\\
der Naturwissenschaften\\
\bigskip
vorgelegt beim Fachbereich Physik\\
der Johann Wolfgang Goethe-Universit\"at\\
in Frankfurt am Main\\
\bigskip
von\\
Denis Parganlija\\
aus Frankfurt am Main\\
\bigskip
\bigskip
Frankfurt/Main (12.12.2011)\\
\bigskip
(D 30)
\end{center}
\end{titlepage}

\begin{titlepage}
\linespread{1.3}     
\large
\vbox{\vspace{1cm}}
vom Fachbereich Physik 
der Johann Wolfgang Goethe-Universit\"at\\
als Dissertation angenommen

\vspace{11cm}
\begin{center}
\begin{tabular}{llll}
Dekan: & Prof. Dr. Michael Huth\\
       &   \\
Gutachter: & Prof. Dr. Dirk-Hermann Rischke\\
           & Prof. Dr. J\"urgen Schaffner-Bielich\\
   &   \\
Datum der Disputation: & 25.5.2012 \\
\end{tabular}
\end{center}

\vfill\eject
\rule{0cm}{15cm}
\vfill\eject

\end{titlepage}

%% file: Introduction.tex
\begin{fmffile}{introduction}
\chapter{Introduction}

The human race has always been guided by the quest \textit{So that I may perceive
whatever holds // The world together in its inmost folds} -- or in the
original, German version: \textit{Dass ich erkenne, was die Welt // Im
Innersten zusammenh\"{a}lt} \cite{Goethe}. And it has been a long journey: the
first strives of the human spirit to understand the inner principles governing
nature are as old as civilisation itself. It has been remembered, for
example, that more than two and a half millenia ago the ancient Greek
philosopher Leucippus claimed nature to possess inner elements of
structure, the atoms -- or the indivisible ones (as coined by Democritus). An
alternative \textit{theory of everything} foresaw the existence of four basic
elements (ranging from water to fire) upon which nature is built. Although
the idea of the atom is an ancient one, a lot of time passed before the idea of
atomic clusters (that is, molecules) was established by Robert Boyle in 1661.
Approximately a decade later, Isaac Newton developed the corpuscular theory of
light, for which he claimed that it consists of minute particles (corpuscles).

Nowadays the prevalent explanation of the features of light is that of
corpuscle-wave duality: the light may behave more as a set of corpuscles or as
a wave, depending on the experimental surrounding. Indeed the light corpuscles
-- the photons -- act as transmitters of one of the fundamental forces of
nature: the electromagnetic interaction. This interaction is known to affect
all particles carrying electric charge (electrons, protons,...). It is
essential for the attraction of atomic nuclei (consisting of the
positive-charge protons and neutrons that carry no charge) and the electrons, that
build a negative-charge cloud around the atomic nuclei and allow for atoms to
bind into more complex structures, the molecules.

It has been known since the age of Charles Coulomb (18$^{\text{th}}$ century)
that\ the electromagnetic (back then: only electric) interaction possesses an
infinite range -- it is inversely proportional to the squared distance between
two charges:%

\[
F_{C}\sim\frac{1}{r^{2}}\text{.}%
\]

There is, however, another interaction with such a feature: the gravity. This
interaction occurs between objects that possess a mass, be it point-like
objects such as electrons (mass $\sim10^{-31}$ kg or, in units more commonly
used in physics nowadays, $\sim511$ keV) or very large objects such as the sun
($\sim10^{30}$ kg) or even planetary systems, galaxies and galaxy clusters.

According to Newton's law of universal gravity, the magnitude of this force is
also proportional to $1/r^{2}$. Just as the electromagnetic interaction,
gravity is expected to possess its \textit{transmitter particles} (gauge
bosons), denoted as gravitons. Newton himself did not coin the name but was
reportedly dissatisfied with the \textit{action at a distance }implied by his
gravity law (and he expected a \textit{mediator of gravity} to exist). The
problem of action at a distance is indeed solved by the introduction of
gravitons; however, these particles -- if they exist -- have remained elusive
to experimental observation. The basic difference between photons and
gravitons is their spin: photons possess spin one whereas gravitons
are expected to possess spin two rendering them rather difficult to examine
theoretically \cite{Graviton}. However, gravity without gravitons would mark a
special case of an elementary force without gauge particles because not only the
electromagnetic interaction but also the other two elementary forces -- the
strong and the weak interactions -- possess their gauge bosons as well.

The weak force has actually been unified with the electromagnetic force into
the electroweak one in the groundbreaking work of Glashow, Weinberg, Salam and
Ward in the 1960s and 1970s \cite{refsm}. The ensuing Standard Model of
electroweak interactions describes simultaneously the twelve building blocks of
nature known to the modern physics:%

\begin{align*}
\text{6 quarks: }  &  u\text{ (up), }d\text{ (down)}\\
&  s\text{ (strange), }c\text{ (charm)}\\
&  b\text{ (bottom), }t\text{ (top)}%
\end{align*}

and%

\begin{align*}
\text{6 leptons: }  &  e\text{ (electron), }\nu_{e}\text{ (electron
neutrino)}\\
&  \mu\text{ (muon), }\nu_{\mu}\text{ (muon neutrino)}\\
&  \tau\text{ (tau), }\nu_{\tau}\text{ (tau neutrino)}%
\end{align*}

(plus twelve antiparticles). The masses of these particles vary significantly: for
example, as we will see in the next chapter, the mass of the $t$ quark is
approximately 50000 times larger than the masses of the $u$ and $d$ quarks.
Nonetheless, the Standard Model actually starts with the assumption that the
particles possess no mass; this includes the gauge bosons of the electroweak
interaction, labelled as $W^{\pm}$ and $Z^{0}$ as well as the gauge boson of
the electromagnetic interaction, the photon [in the language of modern physics: the local $SU(2)_{L}\times
U(1)_{Y}$ symmetry]. The symmetry is broken by the famous Higgs mechanism that
gives mass to all leptons except neutrinos as well as to $W^{\pm}$ and
$Z^{0}$; the photon remains massless. The predictions of the Standard Model
have been confirmed to a high precision by various experiments
\cite{Pich:2005}. Only the Higgs boson has remained elusive at Tevatron as
well as at the Large Hadron Collider LHC at CERN (but, if it exists, it should
be discovered at the LHC).

There are actually attempts to extend the Standard Model to the physics beyond
(again, if such physics exists -- this is also, in principle, verifyable at
the LHC). One extension is supersymmetry \cite{SUSY}, where one assumes
that to each observed boson and the as-yet unobserved Higgs (integer-spin
particles) there is a supersymmetric fermion counterpart (and analogously for
the observed fermion). This renders the lightest of the supersymmetric
particles -- the LSP -- stable under the so-called $R$-parity; the LSP is
a candidate for a dark-matter particle.

A further possible extension of the Standard Model is represented by
technicolour models \cite{refTC}. These models are based upon the observation that
the Higgs boson -- if it exists -- would be the only elementary
\textit{scalar} particle known to modern physics, i.e., it would possess spin
zero. All the other elementary particles are not scalar: the already mentioned quarks and
leptons are fermions as they possess spin 1/2 whereas the gauge bosons possess spin 1 or spin 2, depending on the
interaction considered. Indeed, until now all scalar
states first assumed to be elementary were eventually determined as composite
(such as the $\sigma$ meson of the strong interaction, discussed below). For this
reason, technicolour models assume the Higgs boson to be composite as
well, consisting of so-called \textit{techniquarks}. Techniquarks are
expected to be several times heavier than the heaviest observed quark (the $t$
quark) but should in principle also be accessible to the LHC. (Note that there
is also a technicolour candidate for the dark matter particle:
technicolour-interacting massive particle or TIMP \cite{Sannino:2008}.)
\newline

The work presented in this thesis will consider a different type of
interaction: the strong one. This interaction is responsible for the stability
of the nucleons, i.e., protons and neutrons (and thus of atoms and molecules);
it is similar to the electromagnetic interaction as it also possesses massless
gauge bosons -- the \textit{gluons}. Given that the gluons are massless just as the
photons, one might expect the range of the strong interaction to be
infinite, just as in the case of the electromagnetic one. However the strong
interaction actually possesses a very short range ($\sim1$ fm = $10^{-15}$ m,
the nucleon radius). Additionally, the gluons, while not charged electrically,
nonetheless carry a different sort of charge: \textit{colour}. They are
transmitters of the strong interaction between quarks; however, their colour
charge allows them to not only interact with quarks but also among themselves.
It is believed that the colour interaction holds quarks and gluons
\textit{confined} within nucleons (which is in turn presumably related to the
nucleon stability) -- but confinement is an experimental observation
without (as yet) a commonly accepted theoretical explanation.

Quarks and gluons were not always confined to nucleons. According to the
theory of the Big Bang, an extremely short-lived phase of the primary matter
($10^{-44}$ s), where no complex matter structures existed, was followed by a
state of the \textit{quark-gluon} plasma, without confinement of quarks and
gluons into nucleons. The expansion of the early universe implied the cooling
of the matter, allowing for the first complex structures to be formed by
quarks. The simplest ones consisted of one quark (${\bar{q}}$) and one
antiquark ($q$). Thus the exploration of the ${\bar{q}q}$ states allows us to
gain insight into the early universe -- and the work presented in this thesis
will have exactly the ${\bar{q}q}$ states as the main topic.

Of course, there can be no true insight into the state of matter in the early
universe from the theoretical standpoint alone; there are various experimental
undertakings attempting to recreate the matter as it was shortly after the Big
Bang ($\sim13$ billion years ago). To this end, heavy ions (Pb, Au) or protons
are collided at velocities comparable to the velocity of light; the collisions
produce very hot (at least $10^{12}$ K) and/or very dense ($\sim10^{15}$
g/cm$^{3}$) matter. Let us mention just three experimental facilities where this is (or will be) accomplished:
proton-proton collisions and heavy-ion collisions are performed at the LHC and
at the Relativistic Heavy-Ion Collider RHIC in Upton, New York/United States; protons and antiprotons will
be collided at the Facility for Antiproton and Ion Research (FAIR), currently
being constructed at the Gesellschaft f\"{u}r Schwerionenforschung (GSI) in
Darmstadt/Germany.\newline

As already indicated, the work presented in this thesis will be concerned with
the strong interaction. Thus in Chapter \ref{sec.QCD} we describe some basic
properties of the theory of strong interactions -- the Quantum Chromodynamics
(QCD). We introduce the concepts of hadrons, quarks, gluons and colour
charge. We observe that the basic equation of QCD -- the \textit{QCD
Lagrangian} -- possesses certain symmetries, most notably the chiral
$U(N_{f})\times U(N_{f})$ symmetry between $N_{f}$ left-handed and
right-handed quark flavours. However, as we discuss in Sec.\ \ref{sec.SSB},
this symmetry is also observed to be broken in vacuum by two mechanisms:
explicitly, by non-vanishing quark masses, and spontaneously, by the quark
condensate. An additional symmetry-breaking mechanism is the so-called chiral
anomaly (a symmetry that is exact classically but broken at the quantum level,
\ discussed in Sec.\ \ref{sec.CM}).

The spontaneous breaking of the chiral symmetry leads to some profound
consequences. Goldstone bosons (for example, the pions) emerge and the
masses are generated for a range of mesons. (The pions obtain their mass from the explicit breaking of the chiral symmetry.)
Most (but not all) mesons can be described as ${\bar{q}q}$ states; considering
the approximate mass degeneration of the non-strange [up ($u$) and down ($d$)]
quarks, there is one scalar isosinglet state that can be constructed:
$\sigma_{N}=(\bar{u}u+\bar{d}d)/\sqrt{2}$. If we consider the strange quark
$s$ as well, then we can construct an additional scalar state: $\sigma
_{S}=\bar{s}s$. However, as we discuss in Chapter \ref{sec.scalarexp},
experimental data demonstrate that the actual number of scalars is
significantly larger: there are six non-strange scalar states [$f_{0}(600)$ or $\sigma$, $f_{0}(980)$,
$f_{0}(1370)$, $f_{0}(1500)$, $f_{0}(1710)$ and $f_{0}(1790)$].
Obviously, at
most two of them can be ${\bar{q}q}$ states -- but the question is
\textit{which two}.

That is the main topic of the work presented in this thesis. In Chapter \ref{chapterC} we develop a generic model of mesons for an arbitrary number of 
flavours, based on the symmetries of QCD. The model can even be studied for various numbers of colours. Then, in Chapters
\ref{chapterQ} -- \ref{ImplicationsFitII}, we apply the model to investigate scalar ${\bar{q}q}$
states in the physical spectrum. It is known as the Linear Sigma Model and it
incorporates not only the global symmetries of QCD (chiral, $CP$) but
also the mechanisms of chiral-symmetry breaking (explicit, spontaneous and the
one induced by the chiral anomaly). However, the Linear Sigma Model contains
not only scalar states; a realistic model of low-energy QCD will inevitably
have to consider other states experimentally established in the region of
interest (in our case: up to $\sim1.8$ GeV). For this reason, our model will
also incorporate vector ($\omega$, $\vec{\rho}$) and axial-vector
[$f_{1}(1285)$, $a_{1}(1260)$] degrees of freedom from the onset. Then, in
Chapter \ref{chapterQ}, we develop a $U(2)_{L}\times U(2)_{R}$ sigma model
with scalars (sigma, $\vec{a}_{0}$), pseudoscalars (pion and the non-strange
component of the physical $\eta$ state), vectors and axial-vectors and
describe their phenomenology. The states present in our model are of ${\bar
{q}q}$ structure, as we discuss in Sec.\ \ref{sec.largen}. Consequently, all
our statements about the physical states depend on the assignment of our
${\bar{q}q}$ model states to the physical ones (conversely, of course,
assigning any of our ${\bar{q}q}$ states to a physical state implies that the
given physical state is of ${\bar{q}q}$ nature).

Given the already mentioned large number of scalar $f_{0}$ states, we work
with two different scenarios in Chapter \ref{chapterQ}: in Scenario I,
Sec.\ \ref{sec.scenarioI}, we assume that the scalar ${\bar{q}q}$ states are
to be looked for in the energy region below 1 GeV. This implies, for example,
that the $f_{0}(600)$ resonance is a ${\bar{q}q}$ state.\ However, this
assumption does not appear to be favoured when its implications are compared
with experimental data. The $f_{0}(600)$ resonance is too narrow.
Therefore, in Sec.\ \ref{sec.scenarioII}, we start with a converse assumption
(Scenario II): that the scalar ${\bar{q}q}$ states are actually above $1$ GeV
[then the $f_{0}(1370)$ resonance is the scalar ${\bar{q}q}$ state]. In this
scenario, the overall description of the data is decisively better:
\textit{the scalar }${\bar{q}q}$ \textit{states appear to be above 1 GeV}
rather than, as one might expect, below.

The discussion of Chapter \ref{chapterQ} is, however, not conclusive. The
reasons are at least twofold: the strange mesons (such as the $K_{0}^{\star}$ states) are missing; additionally, the gauge bosons
of QCD, the gluons, may, just as quarks, form their own bound states --
the glueballs. These states could mix with the scalar ${\bar{q}q}$ states
already present in the model. Thus the question has to be addressed whether
the conclusions of Chapter \ref{chapterQ} hold once the mentioned strange and
glueball states are included into the model. [In principle one could also consider the
admixture of the tetraquark ($\bar{q}\bar{q}qq$) states to the scalar resonances; this can be performed
in succession to the results regarding quarkonium and glueball phenomenology presented in this thesis.]

For this reason, in Chapters \ref{sec.remarks} -- \ref{ImplicationsFitII} we
present the main part of this work: a sigma model containing scalar,
pseudoscalar, vector and axial-vector mesons \textit{both in non-strange and
strange sectors}: an $N_{f}=3$ model. This is {\bf the first time} that all these
states have been considered within a single QCD-based model. Our formalism thus contains, but is not
limited to, $\eta$, $\eta^{\prime}$, $\vec{\pi}$, $K$ (pseudoscalars);
$\omega$, $\varphi(1020)$, $\vec{\rho}$, $K^{\star}$ (vectors) and
$f_{1}(1285)$, $f_{1}(1420)$, $a_{1}(1260)$, $K_{1}$ (axial-vectors). The
model also contains two scalar isosinglet degrees of freedom $\sigma_{N}$
(already present in Chapter \ref{chapterQ}) and additionally $\sigma_{S}%
=\bar{s}s$. We also consider the $\vec{a}_{0}$ triplet\ (already present in
Chapter \ref{chapterQ}) and the scalar-kaon quadruplet $K_{S}$. (Our scalar state $K_{S}$
is to be distinguished from the short-lived pseudoscalar $K_S^0$ state that will not be discussed in this work.)

The model parameters are calculated using all masses except those of the two
$\sigma$ fields. Then, as in Chapter \ref{chapterQ}, we distinguish between
two possibilities (labelled as Fits I and II in Chapters \ref{sec.remarks} --
\ref{ImplicationsFitII}): in Fit I (Chapters \ref{sec.fitI} and
\ref{ImplicationsFitI}) we discuss whether a reasonable meson phenomenology can
be obtained assuming that the scalar ${\bar{q}q}$ states are below 1 GeV. Thus
we work with the assumption that the $f_{0}(600)$ and $K_{0}^{\star}(800)$
resonances are ${\bar{q}q}$ states (analogously to Scenario I presented in
Sec.\ \ref{sec.scenarioI}). This allows us to consider not only scalar-meson
phenomenology, but the broader phenomenology as well -- in particular the
decays of the axial-vector states [e.g., $f_{1}(1285)$, $f_{1}(1420)$ and
$a_{1}(1260)$]. We again obtain a negative result: if the scalars were below 1
GeV, then the axial-vectors would have to have a decay width from 1 GeV up to 20 GeV -- several
orders of magnitude larger than
experimental data. For this reason, we turn to an alternative assignment: that
the scalar ${\bar{q}q}$ states are above 1 GeV. The ensuing fit yields an
extremely improved meson phenomenology: almost all the results are consistent
with experimental data.

As already indicated, our study is motivated by the phenomenology of the
scalar mesons. A realistic description of the scalar states requires the
inclusion of vector and axial-vector states as well. Thus our study will also
include the phenomenology of these states: indeed, in the more general $N_{f}=3$
version of our model in Chapters \ref{sec.fitI} -- \ref{ImplicationsFitII}, we
will calculate widths of all experimentally observed two-body decays of mesons
for which there exist vertices in the model. This will be performed in both
(pseudo)scalar as well as (axial-)vector channels. In addition, three-body and
four-body decay widths will also be calculated utilising sequential decays;
$\pi\pi$ scattering lengths will be calculated as well.
This will in turn provide us with an extremely powerful agent of
discrimination between the two assignments where, respectively, the scalar
states are below and above 1 GeV.

Before the summary and outlook of the work are presented in Chapter
\ref{sec.summary}, we present another extension of the $N_{f}=2$ model of Chapter
\ref{chapterQ} to $N_{f}=2$ + scalar glueball in Chapter \ref{chapterglueball}.
Although Chapter \ref{chapterglueball} does not present results with strange
degrees of freedom, it is still another valuable test of the assertion
obtained in Chapter \ref{chapterQ}: that the scalar ${\bar{q}q}$ states are
above, rather than below, 1 GeV.

\chapter{QCD and Its Symmetries}

\label{sec.QCD}

\section{Introduction}

A large multitude of new particles was discovered in the 1950s and 1960s.
There were usually referred to as \textit{elementary}, implying that they
possessed no inner structure; however, their decay patterns and large numbers
imposed two questions:

\begin{itemize}
\item Why do we observe that the newly discovered particles do not decay into
all other particles into which their decays would be kinematically alowed?

\item Is there a classfication scheme for the new particles, but also for the
already known ones, such as protons and neutrons?
\end{itemize}

In other words: Is there a force binding more elementary blocks into the
observed particles?

A classification scheme was proposed by M.\ Gell-Mann \cite{GM} and G.\ Zweig
\cite{Zweig} in 1964 using the $SU(3)$ flavour symmetry. Zweig proposed the
particle substructure elements to be denoted as \textit{aces} whereas,
according to Gell-Mann's classification, if one considers a unitary triplet
$t$ consisting of an isotopic singlet $s$ of electric charge $z$ and an
isotopic doublet $(u,d)$ with charges $z+1$ and $z$ respectively, then%

\begin{align*}
&  \hspace*{0.65cm}\text{\textit{We can dispense entirely with the basic
baryon} }b\text{ \textit{if we assign to the triplett} }\\
&  \hspace*{0.65cm}\text{\textit{the following properties}}\text{:
\textit{spin} }\frac{1}{2}\text{, }z\text{ }\text{= }-\frac{1}{3}\text{
\textit{and baryon number }}\frac{1}{3}\text{\textit{.}}\\
&  \hspace*{0.65cm}\text{\textit{We then refer to members} }u^{\frac{2}{3}%
}\text{, }d^{-\frac{1}{3}}\text{\textit{, and} }s^{-\frac{1}{3}}\text{\textit{
of the triplet as "quarks" }}\\
&  \hspace*{0.65cm}\text{\textit{and the members of the anti-triplet as
anti-quarks. } }%
\end{align*}
\\
Therefore the particles originally denoted as elementary (protons, neutrons,
hyperons,...) were suggested to possess an inner structure. Strictly speaking,
they are then no longer \textit{elementary} as this role is thereafter played
by their substructure partons, the quarks, but nonetheless they are still
sometimes referred to as elementary. All the particles containing quarks are
subject to the so-called strong interaction, described by

\[
\text{\textbf{Quantum Chromodynamics (QCD).} }%
\]

We will discuss the Lagrangian of QCD later in this chapter. At this
point we note that, due to the electric charge of the quarks, the
electromagnetic interaction also plays a certain, though subdominant, role in
quark interactions. The reason is that the fine-structure constant of the
electromagnetic interaction (that encodes the strength of the electromagnetic
coupling) $\alpha=1/137.035999679(94)$ \cite{PDG} is two orders of magnitude
smaller than the fine-structure constant of the strong interaction $\alpha
_{s}\sim1$ in vacuum. Additionally, the quarks can also interact weakly, by
exchanging weak bosons \cite{Pich:2005}; this mechanism is responsible for the
$\beta$ decay of nucleons.

The particles containing quarks are known as \textit{hadrons} (Greek --
$\alpha\delta\rho\acute{o}\varsigma$: strong). Hadrons are
classified into two groups according to their spin:

\begin{itemize}
\item \textit{Fermionic} hadrons are known as \textit{baryons} (Greek --
$\beta\alpha\rho\acute{\upsilon}\varsigma$, heavy: the lightest baryon, the
proton, is approximately 1836 heavier than the electron).

\item \textit{Bosonic} hadrons are known as \textit{mesons} from the Greek
word -- $\mu\acute{\varepsilon}\sigma o\varsigma$, the middle one: the first
discovered meson was the pion \cite{pion}, approximately 280 times heavier
than the electron but still lighter than the proton; the name has remained although it
is an experimental fact nowadays that baryons and mesons typically accommodate
the same mass region. Note that the mesons are sometimes defined in terms of
their quark structure as antiquark-quark states. This definition is improper
because not all mesons are ${\bar{q}}q$ states (some of them may be of
${\bar{q}\bar{q}}qq$ structure, or even represent bound states of other
mesons). Consequently, this work will utilise the definition of mesons
based on their spin.
\end{itemize}

Current high-energy experimental data suggest that (as already indicated)
there are six building blocks of hadrons -- i.e., six quark \textit{flavours}
with the following masses according to the Particle Data Group (PDG)
\cite{PDG}:
\begin{align*}
m_{u} &  =(1.7-3.1)\text{ MeV};\,m_{d}=(4.1-5.7)\text{ MeV};\\
m_{s} &  =(80-130)\text{ MeV};\\
 m_{c} & =1.29_{-0.11}^{+0.05}\text{ GeV};\,m_{b}=4.19_{-0.06}^{+0.18}\text{
GeV};\\
m_{t} & =(172.9\pm0.6\pm0.9)\text{ GeV}.
\end{align*}

These values are the estimates of the so-called current quark masses. The
values of $m_{u,d,s}$ are not a product of direct experimental observations
but obtained either in lattice calculations \cite{Luds} or in first-principle
calculations \cite{FPuds} at the scale $\mu\approx2$ GeV. Indeed, to our
knowledge, there has recently been only one article by an experimental
collaboration regarding the light-quark masses: the results of the ALEPH
Collaboration suggest $m_{s}=176_{-57}^{+46}$ MeV (at $\mu\approx2$ GeV) from
a $\tau$-decay analysis \cite{Barate:1999hj}. Similarly, the value of $m_{c}$
is also predominantly determined in theoretical calculations \cite{Thc}
although the BABAR Collaboration has recently claimed $m_{c}=(1.196\pm
0.059\pm0.050)$ GeV from $B$ decays \cite{BABAR2010}. The value of
$m_{t}$ stems from direct top-event observations published by the Tevatron
Electroweak Working Group (see Ref.\ \cite{Tevatron} for the latest data and
references therein for the older ones). Similar is true for the $b$ quark \cite{PDG}. Note that the current $u$, $d$ masses
need to be distinguished from their constituent masses $\sim300$ MeV $\simeq
m_{p}/3$ where $m_{p}$ denotes the mass of the proton.\\

Quarks carry electric charges as follows%

\begin{align}
u,c,t  &  \leftrightarrow\frac{2}{3}e\text{,}\\
d,s,b  &  \leftrightarrow-\frac{1}{3}e\text{,}
\end{align}

where $e$ denotes the elementary electric charge. Following the
Gell-Mann--Zweig classification, a proton is a state containing two $u$ quarks
and one $d$ quark (with the charge $2\cdot2e/3-e/3=e$). Given that the total
spin of the proton reads 1/2, then the spin-flavour wave function of this
particle can be written as%

\begin{equation}
|p\rangle=\frac{1}{\sqrt{3}}(|u_{\uparrow}u_{\uparrow}d_{\downarrow}%
\rangle+|u_{\uparrow}u_{\downarrow}d_{\uparrow}\rangle+|u_{\downarrow
}u_{\uparrow}d_{\uparrow}\rangle)\text{.}%
\end{equation}

An analogous relation holds for the neutron upon substituting
$u\longleftrightarrow d$. These relations comply with W.\ Pauli's
Spin-Statistics Theorem \cite{Pauli:1940}. However, in 1965 a baryon with
charge $q=2e$ and spin 3/2 was discovered \cite{Roper:1964}; the particle
could readily be described in terms of $u$ and $d$ quark flavours with spin
$1/2$, but only if the Pauli Principle were violated. The particle was
labelled as $\Delta^{++}$ [or, nowadays, $\Delta(1232)$] and, given the
charge, its spin-flavour wave function had to be composed as%

\begin{equation}
|\Delta^{++}\rangle=|u_{\uparrow}u_{\uparrow}u_{\uparrow}\rangle\text{.}
\label{Dfs}%
\end{equation}

The solution to this paradox was found by introducing an additional degree of
freedom for quarks: colour. If we assume that each quark comes in three
colours, red ($r$), green ($g$) and blue ($b$), then the three quarks
contained in $\Delta^{++}$ can be combined in the following antisymmetric way
in the colour space:%

\begin{equation}
|\Delta^{++}\rangle_{\text{colour}}=\frac{1}{\sqrt{6}}|u_{r}u_{g}u_{b}%
+u_{g}u_{b}u_{r}+u_{b}u_{r}u_{g}-u_{g}u_{r}u_{b}-u_{b}u_{g}u_{r}-u_{r}%
u_{b}u_{g}\rangle\text{.} \label{Dc}%
\end{equation}

Indeed, assuming that any baryon $B$ contains three quarks $q_{1,2,3}$, then
the colour wave function of such a composite object can be
antisymmetrised as%

\begin{align}
|B\rangle_{\text{colour}}  &  =\frac{1}{\sqrt{6}}|q_{1r}q_{2g}q_{3b}%
+q_{1g}q_{2b}q_{3r}+q_{1b}q_{2r}q_{3g}\nonumber\\
&  -q_{1g}q_{2r}q_{3b}-q_{1b}q_{2g}q_{3r}-q_{1r}q_{2b}q_{3g}\rangle
\end{align}

or simply%

\begin{equation}
|B\rangle_{\text{colour}}=\frac{1}{\sqrt{6}}\varepsilon^{\alpha\beta\gamma
}|q_{1\alpha}q_{2\beta}q_{3\gamma}\rangle\label{Bn}\text{,}
\end{equation}

where $\varepsilon^{\alpha\beta\gamma}$ denotes the totally antisymmetric
tensor and $\alpha,\beta,\gamma\in\{r,g,b\}$.

Then the direct product of the $\Delta^{++}$ flavour-spin wave function
(\ref{Dfs}) with the corresponding colour wave function (\ref{Dc}) yields a
total wave function that is antisymmetric under exchange of two quarks -- in
accordance with the Pauli Principle.\\

This is of course valid under the assumption that there are three quark
colours in nature. This statement cannot be validated in vacuum -- it is an
experimental fact that quarks do not appear as free particles in vacuum but
that they are \textit{confined} within hadrons. There is (at least for now) no
analytic proof of confinement from QCD. However, there are indirect
methods from hadron decays allowing us to determine the number of quark colours.\\

\textit{Experiment 1.} The neutral pion decays into $2\gamma$ via a
triangular quark loop; the branching ratio is $\sim100\%$ \cite{PDG}. The
Standard Model determines the corresponding decay width as
\cite{Pich:2005}%

\begin{equation}
\Gamma_{\pi^{0}\rightarrow2\gamma}=\frac{\alpha^{2}m_{\pi}^{3}}{64\pi
^{3}f_{\pi}^{2}}\left(  \frac{N_{c}}{3}\right)  ^{2}\equiv7.73\text{ eV}%
\cdot\left(  \frac{N_{c}}{3}\right)  ^{2}\text{,}
\end{equation}

where $N_{c}$ denotes the number of colours and $f_{\pi}=92.4$ MeV is the pion
decay constant. The experimental result reads $\Gamma_{\pi^{0}\rightarrow
2\gamma}^{\exp}=(7.83\pm0.37)$ eV \cite{PDG} and it can only be described by
the Standard Model if $N_{c}=3$. \\

\textit{Experiment 2. }Consider the ratio of the cross-sections for the
processes $e^{+}e^{-}\rightarrow\gamma$ (or $Z$) $\rightarrow{\bar{q}}q$
$\rightarrow$ hadrons and $e^{+}e^{-}\rightarrow\mu^{+}\mu^{-}$. The ratio
reads \cite{Pich:2005}
\[
\frac{\,\sigma(e^{+}e^{-}\rightarrow\gamma,Z\rightarrow{\bar{q}}q\rightarrow
\text{hadrons})}{\sigma(e^{+}e^{-}\rightarrow\mu^{+}\mu^{-})}=\left\{
\begin{tabular}
[c]{l}%
$\frac{\,2}{\,3}N_{c}\,\ \,(N_{f}=3)$\\
$\frac{10}{\,9}N_{c}\,\,(N_{f}=4)$\\
$\frac{11}{\,9}N_{c}\,\ (N_{f}=5)$.
\end{tabular}
\ \ \right.
\]
The best correspondence with experimental data is obtained if $N_{c}=3$
\cite{Pich:2005}. We thus conclude that the physical world contains three
quarks colours. Note, however, that QCD with two colours can be explored
nonetheless, at least from the theoretical standpoint, see, e.g.,
Ref.\ \cite{Hands}. Additionally, the limit of a large number of colours
(large-$N_{c}$ limit) has also been subject of many studies
\cite{largenc,referenceslargeNc} and represents a valuable tool of model
building (see Sec.\ \ref{sec.largen} for the application to the model
presented in this work). \\

Now that we know the number of colours, it is possible to build colour-neutral
meson states:%

\begin{equation}
|M\rangle_{\text{colour}}=\frac{1}{\sqrt{3}}|{\bar{q}}_{\bar{r}}q_{r}%
+{\bar{q}}_{\bar{g}}q_{g}+{\bar{q}}_{\bar{b}}q_{b}\rangle\text{.}
\label{M}%
\end{equation}

\section{The QCD Lagrangian}

\label{sec.QCDL}

In the previous section we have seen that the necessity to introduce a colour
degree of freedom for quarks arises from the requirement of an antisymmetric
baryon wave function (that adheres to the Pauli Principle). It has allowed us
to construct putative colour wave functions for baryons (\ref{Bn}) as well as
meson ${\bar{q}}q$ states (\ref{M}). In this section we construct a Lagrangian
containing quarks and considering their flavour and colour degrees of freedom. \\

The Lagrangian is constructed utilising the local (gauge) $SU(N_{c}=3)$
symmetry \cite{Han-Nambu}. A quark field $q_{f}$ in the fundamental
representation transforms under the local $SU(3)$ symmetry as%

\begin{equation}
q_{f}\rightarrow q_{f}^{\prime}=\exp\left\{  -i\sum_{a=1}^{N_{c}^{2}-1}%
\alpha^{a}(x)t_{a}\right\}  q_{f}\equiv Uq_{f}\text{,}
\label{qftransformation}%
\end{equation}

where $t_{a}=\lambda_{a}/2$ denotes the generators of the $SU(3)$ group,
$\lambda_{a}$ are the Gell-Mann matrices and $\alpha^{a}(x)$ are the
parameters of the group. Let us remember that the Dirac Lagrangian for a
free fermion $\psi$ possesses this form:%

\begin{equation}%
\mathcal{L}%
_{Dirac}=\bar{\psi}(i\gamma^{\mu}\partial_{\mu}-m_{\psi})\psi\text{.}
\label{DE}%
\end{equation}

Then, in analogy to the Dirac Lagrangian, we can construct the following
Lagrangian involving the quark flavours considering the requirement that the
Lagrangian is \textit{locally} $SU(3)_{c}$ \textit{symmetric} (sum over
flavour index $f$ is implied):%

\begin{equation}%
\mathcal{L}%
_{q}=\bar{q}_{f}(i\gamma^{\mu}D_{\mu}-m_{f})q_{f} \label{lq}%
\end{equation}

where $m_{f}$ denotes the mass of the quark flavour $q_{f}$,
\begin{equation}
D_{\mu}=\partial_{\mu}-ig\mathcal{A_{\mu}} \label{Dmu}%
\end{equation}

represents the $SU(3)$ covariant derivative with the eight gauge fields $A_{\mu}^{a}$

\begin{equation}
\mathcal{A_{\mu}}=\sum_{a=1}^{N_{c}^{2}-1}A_{\mu}^{a}t_{a} \label{Amu}\text{,} 
\end{equation}

referred to as \textit{gluons}. The (adjoint) gluon fields transform as
follows under the local $SU(3)$ group:%

\begin{equation}
\mathcal{A_{\mu}}\rightarrow\mathcal{A}_{\mu}^{\prime}=U\mathcal{A_{\mu}%
}U^{\dagger}-\frac{i}{g}\left(  \partial_{\mu}U\right)  U^{\dagger}\text{.}
\label{Amutransformation}%
\end{equation}

The Lagrangian (\ref{lq}) is invariant under transformations
(\ref{qftransformation}) and (\ref{Amutransformation}). It is possible to
construct an additional gauge invariant term involving only gluons
\cite{RefYM}:%

\begin{equation}%
\mathcal{L}%
_{g}=-\frac{1}{4}\,G_{\mu\nu}^{a}G_{a}^{\mu\nu} \label{lg}%
\end{equation}

(sum over gluon-field index $a$ is implied) where the field strength tensor
$G_{\mu\nu}^{a}$ is defined as%

\begin{equation}
G_{\mu\nu}^{a}=\partial_{\mu}A_{\nu}^{a}-\partial_{\nu}A_{\mu}^{a}%
+gf^{abc}A_{\mu}^{b}A_{\nu}^{c}%
\end{equation}

and $f^{abc}$ denote the antisymmetric structure constants of the $SU(3)$ group.

The sum of the two Lagrangians (\ref{lq}) and (\ref{lg}) yields the
\textit{QCD Lagrangian}:%

\begin{equation}%
\mathcal{L}%
_{QCD}=\bar{q}_{f}(i\gamma^{\mu}D_{\mu}-m_{f})q_{f}-\frac{1}{4}G_{\mu\nu}%
^{a}G_{a}^{\mu\nu}\text{.} \label{lqcd}%
\end{equation}

\section{The Chiral Symmetry}

\label{sec.CM}

In addition to the local $SU(3)_{c}$ colour symmetry, the QCD Lagrangian also
exhibits a global symmetry if quarks are massless -- the chiral symmetry.
To ascertain this symmetry in the Lagrangian (\ref{lq}) in the limit $m_{f}%
=0$, let us first define the following left-handed and right-handed operators
$\mathcal{P}_{R,\;L}$:%
\begin{equation}
\mathcal{P}_{R,\;L}=\frac{1\pm\gamma_{5}}{2} \label{PRLd}\text{,}
\end{equation}

where $\mathcal{P}_{R}$ has the plus sign in the denominator and $\gamma_{5}$
is a matrix defined in terms of the other Dirac matrices as%

\begin{equation}
\gamma_{5}=i\gamma_{0}\gamma^{1}\gamma^{2}\gamma^{3} \label{Gamma5}\text{,}
\end{equation}

with (in the chiral representation)%

\begin{equation}
\gamma_{0}=\left(
\begin{array}
[c]{cc}%
0 & \,\,\,\,1_{2}\\
1_{2} & \,\,0
\end{array}
\right)  ,\;\;\vec{\gamma}=\left(
\begin{array}
[c]{cc}%
0 & \,\,\,\,\vec{\sigma}\\
-\vec{\sigma} & \,\,0
\end{array}
\right)
\end{equation}

and $\vec{\sigma}$ denotes the triplet of the Pauli matrices. Thus we obtain%

\begin{equation}
\gamma_{5}=\left(
\begin{array}
[c]{cc}%
-\,1_{2} & 0\\
0 & \,\,\,1_{2}%
\end{array}
\right)  \text{.} \label{Gamma5-1}%
\end{equation}

Consequently, the two operators $\mathcal{P}_{R,\;L}$ possess the following
form, justifing their labels as right-handed and left-handed%

\begin{equation}
\mathcal{P}_{R}=\left(
\begin{array}
[c]{cc}%
0 & \,\,\,\,0\\
0 & \,\,\,1_{2}%
\end{array}
\right)  ,\;\;\mathcal{P}_{L}=\left(
\begin{array}
[c]{cc}%
\,1_{2} & \,\,\,\,0\\
0 & \,\,0
\end{array}
\right)  \text{.} \label{PRL}%
\end{equation}

By definition (\ref{Gamma5}), the $\gamma_{5}$ matrix has the feature that
$\gamma_{5}^{2}=1$ (i.e., unit matrix). This is demonstrated using the
well-known anticommutation formula of the Dirac matrices%

\begin{equation}
\{\gamma_{\mu},\gamma_{\nu}\}=2g_{\mu\nu} \label{AKD}\text{,}
\end{equation}

where $g_{\mu\nu}=$ diag$(1,-1,-1,-1)$ denotes the metric tensor%

\begin{align}
\gamma_{5}^{2}  &  =-\gamma_{0}\gamma^{1}\gamma^{2}\gamma^{3}\gamma_{0}%
\gamma^{1}\gamma^{2}\gamma^{3}=\gamma_{0}^{2}\gamma^{1}\gamma^{2}\gamma
^{3}\gamma^{1}\gamma^{2}\gamma^{3}\nonumber\\
&  =\gamma^{1}\gamma^{2}\gamma^{3}\gamma^{1}\gamma^{2}\gamma^{3}=(\gamma
^{1})^{2}\gamma^{2}\gamma^{3}\gamma^{2}\gamma^{3}=-\gamma^{2}\gamma^{3}%
\gamma^{2}\gamma^{3}\nonumber\\
&  =(\gamma^{2})^{2} (\gamma^{3})^{2}=1\text{.}%
\end{align}

Additionally,%

\begin{equation}
\{\gamma_{\mu},\gamma_{5}\}=0\text{.} \label{Gamma5-2}%
\end{equation}

Consequently,%

\begin{equation}
\mathcal{P}_{R,\;L}^{2}=\frac{(1\pm\gamma_{5})^{2}}{4}=\frac{1\pm\gamma_{5}%
}{2}=\mathcal{P}_{R,\;L} \label{PRL2}%
\end{equation}

and%

\begin{equation}
\mathcal{P}_{R}\mathcal{P}_{L}=0\text{.} \label{PRPL}%
\end{equation}

$\mathcal{P}_{R,\;L}$ are therefore projection operators -- we refer to them
as \textit{chirality projection operators}. Utilising these operators allows
us to decompose a quark flavour $q_{f}$ into two components, a left-handed and
a right-handed one:%

\begin{equation}
q_{f}=\left(  \mathcal{P}_{R}+\mathcal{P}_{L}\right)  q_{f}=q_{f \, R}+q_{f \,
L}\text{.} \label{qRL}%
\end{equation}

Analogously for the antiquarks:%

\begin{equation}
\bar{q}_{f}=\bar{q}_{f}\left(  \mathcal{P}_{R}+\mathcal{P}_{L}\right)
=\bar{q}_{f \, R}+\bar{q}_{f \, L}\text{.} \label{aqRL}%
\end{equation}

Using Eq.\ (\ref{PRLd}) we obtain from the Lagrangian (\ref{lq})%

\begin{align}%
\mathcal{L}%
_{q}  &  =\bar{q}_{f}(i\gamma^{\mu}D_{\mu}-m_{f})q_{f}\overset{\mathcal{P}%
_{R}+\mathcal{P}_{L}=1}{=}\bar{q}_{f}(\mathcal{P}_{R}+\mathcal{P}_{L}%
)(i\gamma^{\mu}D_{\mu}-m_{f})(\mathcal{P}_{R}+\mathcal{P}_{L})q_{f}\nonumber\\
&  \overset{(\mathcal{P}_{R}+\mathcal{P}_{L})^{2}=1}{=}\bar{q}_{f}%
(\mathcal{P}_{R}\mathcal{P}_{R}+\mathcal{P}_{L}\mathcal{P}_{L})(i\gamma^{\mu
}D_{\mu}-m_{f})(\mathcal{P}_{R}+\mathcal{P}_{L})q_{f}\nonumber\\
&  \overset{\text{Eqs.\ (\ref{Gamma5-2}), (\ref{PRL2}), (\ref{PRPL}) }}{=}%
\bar{q}_{f}\mathcal{P}_{R}i\gamma^{\mu}D_{\mu}\mathcal{P}_{L}q_{f}+\bar{q}%
_{f}\mathcal{P}_{L}i\gamma^{\mu}D_{\mu}\mathcal{P}_{R}q_{f} -\bar{q}_{f}\mathcal{P}_{R}m_{f}\mathcal{P}_{R}q_{f}-\bar{q}_{f}%
\mathcal{P}_{L}m_{f}\mathcal{P}_{L}q_{f}\nonumber\\
&  =q_{f}^{\dagger}\gamma_{0}\mathcal{P}_{R}i\gamma^{\mu}D_{\mu}%
\mathcal{P}_{L}q_{f}+q_{f}^{\dagger}\gamma_{0}\mathcal{P}_{L}i\gamma^{\mu
}D_{\mu}\mathcal{P}_{R}q_{f} -q_{f}^{\dagger}\gamma_{0}\mathcal{P}_{R}m_{f}\mathcal{P}_{R}q_{f}%
-q_{f}^{\dagger}\gamma_{0}\mathcal{P}_{L}m_{f}\mathcal{P}_{L}q_{f} \nonumber \\
&  \overset{\text{Eq.\ (\ref{Gamma5-2}) }}{=}q_{f}^{\dagger}\mathcal{P}%
_{L}\gamma_{0}i\gamma^{\mu}D_{\mu}\mathcal{P}_{L}q_{f}+q_{f}^{\dagger
}\mathcal{P}_{R}\gamma_{0}i\gamma^{\mu}D_{\mu}\mathcal{P}_{R}q_{f} -q_{f}^{\dagger}\mathcal{P}_{L}\gamma_{0}m_{f}\mathcal{P}_{R}q_{f}%
-q_{f}^{\dagger}\mathcal{P}_{R}\gamma_{0}m_{f}\mathcal{P}_{L}q_{f}\nonumber\\
&  \equiv\bar{q}_{f\,L}i\gamma^{\mu}D_{\mu}q_{f\,L}+\bar{q}_{f\,R}i\gamma
^{\mu}D_{\mu}q_{f\,R}-\bar{q}_{f\,L}m_{f}q_{f\,R}-\bar{q}_{f\,R}m_{f}%
q_{f\,L}\text{.} \label{lq1}%
\end{align}

Then the Lagrangian (\ref{lq1}) less the terms $\sim m_{f}$ is symmetric under
the following, global $U(N_{f})\times U(N_{f})$ transformations of the quark
fields in the flavour space ($t_{i}$: group generators)%

\begin{align}
q_{f\,L}  &  \longrightarrow q_{f\,L}^{\prime}=U_{L}q_{f\,L}=\exp\left\{
-i\sum_{i=0}^{N_{f}^{2}-1}\alpha_{L}^{a}t^{a}\right\}  q_{f\,L}\text{,}%
\label{qfLt}\\
q_{f\,R}  &  \longrightarrow q_{f\,R}^{\prime}=U_{R}q_{f\,R}=\exp\left\{
-i\sum_{i=0}^{N_{f}^{2}-1}\alpha_{R}^{a}t^{a}\right\}  q_{f\,R}\text{.}
\label{qfRt}%
\end{align}

This symmetry is referred to as the \textit{chiral symmetry}. As evident from
Eq.\ (\ref{lq1}), the terms proportional to $m_{f}$ break the chiral symmetry
explicitly, i.e., the symmetry is exact only in the case of vanishing quark
masses: $m_{f}=0$.

According to the Noether Theorem \cite{Noether}, a conserved current $J^{\mu}$
is implied by a global symmetry (and vice versa) in a Lagrangian $%
\mathcal{L}%
(\varphi(x^{\mu}))$ that is invariant under transformations of the form
$\varphi(x)\rightarrow\varphi^{\prime}(x)=\varphi(x)+\delta\varphi(x)$ and
$x\rightarrow x^{\prime}(x)=x+\delta x$ with%

\begin{equation}
J^{\mu}=\frac{\partial%
\mathcal{L}%
}{\partial(\partial_{\mu}\varphi)}\delta\varphi+\delta x^{\mu}%
\mathcal{L}%
\text{.} \label{Jmu}%
\end{equation}

Thus the mentioned $U(N_{f})_{L}\times U(N_{f})_{R}$ implies the existence of
the conserved left-handed and right-handed currents $L^{\mu}$ and $R^{\mu}%
$.\ It is usual, however, to work instead with currents of definitive parity
$P$: the vector current $V^{\mu}=(L^{\mu}+R^{\mu})/2$%

\begin{align}
P  &  :V^{0}(t,\vec{x})\longrightarrow V^{0}(t,-\vec{x}),\\
P  &  :V^{i}(t,\vec{x})\longrightarrow-V^{i}(t,-\vec{x})
\end{align}

($i$ denotes the spatial index) and the axial-vector current $A^{\mu}=(L^{\mu
}-R^{\mu})/2$%

\begin{align}
P  &  :A^{i}(t,\vec{x})\longrightarrow A^{i}(t,-\vec{x}),\\
P  &  :A^{0}(t,\vec{x})\longrightarrow-A^{0}(t,-\vec{x})\text{.}%
\end{align}

Indeed the chiral group $U(N_{f})_{L}\times U(N_{f})_{R}$ is isomorphic to the
group $U(N_{f})_{V}\times U(N_{f})_{A}$ of the unitary vector and axial-vector
transformations. From the features of the unitary groups we know that
$U(N_{f})_{V}\times U(N_{f})_{A}\equiv U(1)_{V}\times SU(N_{f})_{V}\times
U(1)_{A}\times SU(N_{f})_{A}$. Let us now discuss the currents obtained from
the Lagrangian (\ref{lq}) without gluon fields ($\mathcal{A_{\mu}}=0$) under
the stated four transformations. 

\begin{itemize}
\item $U(1)_{V}$ implies $\alpha_{L}^{0}=\alpha_{R}^{0}=\alpha_{V}^{0}/2$ in
Eqs.\ (\ref{qfLt}) and (\ref{qfRt}):
\begin{equation}
U_{1V}=\exp(-i\alpha_{V}^{0}t^{0})\text{,} \label{UV0}%
\end{equation}

i.e., $U_{1V}=U_{1L}=U_{1R}$. We define $t^{0}=1_{N_{f}}/\sqrt{2N_{f}}$ [we
denote the other generators of the unitary group $U(N_{f})$ as $t^{i}$ with
$i=1$,... , $N_{f}$]. It is clear that the Lagrangian (\ref{lq}) is symmetric
under the transformation (\ref{UV0}). The corresponding conserved current
obtained by inserting the Lagrangian (\ref{lq}) into Eq.\ (\ref{Jmu}) and
considering infinitesimal transformation $U_{V}\approx1-i\alpha_{V}^{0}t^{0}$
reads%
\begin{equation}
V_{0}^{\mu}=\frac{\partial%
\mathcal{L}%
}{\partial(\partial_{\mu}q_{f})}\delta q_{f}=\bar{q}_{f}\gamma^{\mu}\alpha
_{V}^{0}t^{0}q_{f}\text{.}%
\end{equation}

The parameter $\alpha_{V}^{0}$ can be discarded because the current is
conserved:%
\begin{equation}
\partial_{\mu}V_{0}^{\mu}=0\text{.}%
\end{equation}

Similarly, the generator $t^{0}$ is proportional to the unit matrix and the
corresponding proportionality constant can be absorbed into $\alpha_{V}^{0}$.
Thus we obtain%
\begin{equation}
V_{0}^{\mu}=\bar{q}_{f}\gamma^{\mu}q_{f}\text{.}%
\end{equation}

The zero component of the current reads%
\begin{equation}
V_{0}^{0}=\bar{q}_{f}\gamma^{0}q_{f} \label{Vsc}%
\end{equation}

and it corresponds exactly to the one that we could have obtained also from
the Dirac equation (\ref{DE}). Then we know, however, that integration over
$V_{0}^{0}$ yields a conserved charge $Q$%
\begin{equation}
Q=\int\text{d}^{3}x\bar{q}_{f}\gamma^{0}q_{f}%
\end{equation}

corresponding to the baryon-number conservation.

\item Group parameters for $SU(N_{f})_{V}$ are obtained for $\alpha_{L}%
^{i}=\alpha_{R}^{i}=\alpha_{V}^{i}/2$ with $i=1,...,N_{f}^{2}-1$ or in other
words%
\begin{equation}
U_{V}=\exp(-i\alpha_{V}^{i}t^{i})\text{,}%
\end{equation}

i.e., $U_{V}=U_{L}=U_{R}$. Infinitesimally,%
\begin{equation}
U_{V}\approx1-i\alpha_{V}^{i}t^{i}\text{.} \label{UVi}%
\end{equation}

Varying the quark fields $q_{f}$ in the Lagrangian (\ref{lq})\ under the
transformation (\ref{UVi}) yields (we consider only terms up to order
$\alpha_{V}^{i}$, not higher-order ones):%
\begin{align}%
\mathcal{L}%
_{q}  &  =iq_{f}^{\dagger}(1+i\alpha_{V}^{i}t^{i})\gamma^{0}\gamma^{\mu
}\partial_{\mu}(1-i\alpha_{V}^{i}t^{i})q_{f}\nonumber\\
&  -q_{f}^{\dagger}(1+i\alpha_{V}^{i}t^{i})\gamma^{0}m_{f}(1-i\alpha_{V}%
^{i}t^{i})q_{f}\nonumber\\
&  =i\bar{q}_{f}\gamma^{\mu}\partial_{\mu}q_{f}-i\alpha_{V}^{i}(\bar{q}%
_{f}\,i\,t^{i}\,\gamma^{\mu}\partial_{\mu}\,q_{f}-\bar{q}_{f}\,i\,t^{i}%
\,\gamma^{\mu}\,\partial_{\mu}\,q_{f})\nonumber\\
&  -\bar{q}_{f}m_{f}q_{f}-i\alpha_{V}^{i}(\bar{q}_{f}\,\,[t^{i},m_{f}%
]\,\,q_{f})\nonumber\\
&  =i\bar{q}_{f}\gamma^{\mu}\partial_{\mu}q_{f}-\bar{q}_{f}m_{f}q_{f}%
-i\alpha_{V}^{i}(\bar{q}_{f}\,\,[t^{i},m_{f}]\,\,q_{f})\text{.} \label{LUVi}%
\end{align}

The Lagrangian is only invariant under the vector transformations if the quark
masses are degenerate. The conserved vector current from Eq.\ (\ref{LUVi}) and
the Noether Theorem is%
\begin{equation}
V^{\mu \, i}=\bar{q}_{f}\gamma^{\mu}t^{i}q_{f} \label{Vc}%
\end{equation}

but, as already indicated, the divergence of the current%
\begin{equation}
\partial^{\mu}V_{\mu}^{i}=i\bar{q}_{f}\,[t^{i},m_{f}]\,q_{f}%
\end{equation}

is zero only in the case of degenerate quark masses.

\item An element of $SU(N_{f})_{A}$ is given by%
\begin{equation}
U_{A}=\exp(-i\alpha_{A}^{i}\gamma_{5}t^{i})\text{,}%
\end{equation}

where $U_{A}=U_{L}=U_{R}^{\dagger}$. [$SU(N_{f})_{A}$ is actually not a group
because it is not closed with regard to the product of two elements but this is
not a problem here because the $SU(N_{f})_{A}$ symmetry is spontaneously
broken, see below.] The infinitesimal transformation reads%

\begin{equation}
U_{A}\approx1-i\alpha_{A}^{i}\gamma_{5}t^{i}\text{.} \label{UAi}%
\end{equation}

Varying the quark fields $q_{f}$ in the Lagrangian (\ref{lq})\ under the
transformation (\ref{UVi}) yields%
\begin{align}%
\mathcal{L}%
_{q}  &  =iq_{f}^{\dagger}(1+i\alpha_{A}^{i}\gamma_{5}t^{i})\gamma^{0}%
\gamma^{\mu}\partial_{\mu}(1-i\alpha_{A}^{i}\gamma_{5}t^{i})q_{f}\nonumber\\
&  -q_{f}^{\dagger}(1+i\alpha_{A}^{i}\gamma_{5}t^{i})\gamma^{0}m_{f}%
(1-i\alpha_{A}^{i}\gamma_{5}t^{i})q_{f}\nonumber\\
&  =i\bar{q}_{f}\gamma^{\mu}\partial_{\mu}q_{f}-i\alpha_{A}^{i}(\bar{q}%
_{f}\,i\,t^{i}\,\gamma^{\mu}\partial_{\mu}\,\gamma_{5}\,q_{f}-q_{f}^{\dagger
}\,i\,t^{i}\,\gamma_{5}\,\gamma_{0}\,\gamma^{\mu}\,\partial_{\mu}%
\,q_{f})\nonumber\\
&  -\bar{q}_{f}m_{f}q_{f}+i\alpha_{A}^{i}(\bar{q}_{f}\,\,\{t^{i},m_{f}%
\}\gamma_{5}\,q_{f})\nonumber\\
&  \overset{\text{Eq.\ (\ref{Gamma5-2}) }}{=}i\bar{q}_{f}\gamma^{\mu}%
\partial_{\mu}q_{f}-\bar{q}_{f}m_{f}q_{f}+i\alpha_{A}^{i}(\bar{q}%
_{f}\,\,\{t^{i},m_{f}\}\gamma_{5}\,q_{f})\text{.}%
\end{align}

Thus the axial-vector current%
\begin{equation}
A^{\mu \, i}=\bar{q}_{f}\gamma^{\mu}\gamma_{5}t^{i}q_{f} \label{Ac}%
\end{equation}

is only conserved if all quark masses are zero:%
\begin{equation}
\partial^{\mu}A_{\mu}^{i}=i\bar{q}_{f}\,\{t^{i},m_{f}\}\,q_{f} \text{.} \label{Acd}%
\end{equation}

\item Let us now turn to the axial-vector singlet transformation $U(1)_{A}$.
It has the following form:%
\begin{equation}
U_{1A}=\exp(-i\alpha_{A}^{0}\gamma_{5}t^{0})
\end{equation}

or infinitesimally%
\begin{equation}
U_{1A}\approx1-i\alpha_{A}^{0}\gamma_{5}t^{0}\text{.}%
\end{equation}

Then the Lagrangian (\ref{lq}) transforms under $U(1)_{A}$ as follows:%
\begin{align}%
\mathcal{L}%
_{q}  &  =iq_{f}^{\dagger}(1+i\alpha_{A}^{0}\gamma_{5}t^{0})\gamma^{0}%
\gamma^{\mu}\partial_{\mu}(1-i\alpha_{A}^{0}\gamma_{5}t^0)q_{f}\nonumber\\
&  -q_{f}^{\dagger}(1+i\alpha_{A}^{0}\gamma_{5}t^{0})\gamma^{0}m_{f}%
(1-i\alpha_{A}^{0}\gamma_{5}t^{0})q_{f}\nonumber\\
&  =i\bar{q}_{f}\gamma^{\mu}\partial_{\mu}q_{f}-i\alpha_{A}^{0}(\bar{q}%
_{f}\,i\,t^{0}\,\gamma^{\mu}\partial_{\mu}\,\gamma_{5}\,q_{f}-q_{f}^{\dagger
}\,i\,t^{0}\,\gamma_{5}\,\gamma_{0}\,\gamma^{\mu}\,\partial_{\mu}%
\,q_{f})\nonumber\\
&  -\bar{q}_{f}m_{f}q_{f}+i\alpha_{A}^{0}(\bar{q}_{f}\,\,\{t^{0},m_{f}%
\}\gamma_{5}\,q_{f})\nonumber\\
&  \overset{\text{Eq.\ (\ref{Gamma5-2}) }}{=}i\bar{q}_{f}\gamma^{\mu}%
\partial_{\mu}q_{f}-\bar{q}_{f}m_{f}q_{f}+i\alpha_{A}^{0}(\bar{q}_{f}%
\,m_{f}\,\gamma_{5}\,q_{f})\text{.}%
\end{align}

Thus the axial-vector singlet current%
\begin{equation}
A^{\mu 0}=\bar{q}_{f}\,\gamma^{\mu}\gamma_{5}\,q_{f} \label{Asc}%
\end{equation}

appears to be conserved in the limit $m_{f}=0$:%
\begin{equation}
\partial^{\mu}A_{\mu}^{0}=i\bar{q}_{f}\,m_{f}\,\gamma_{5}q_{f}\text{.}
\label{Asc2}%
\end{equation}

It is, however, only conserved classically. Considering quantum fluctuations
one sees that it is actually not conserved \cite{refanomaly}:%
\begin{equation}
\partial^{\mu}A_{\mu}^{0}|_{m_{f}=0}=-\frac{g^{2}N_{f}}{32\pi^{2}}G_{\mu\nu
}^{a}{\tilde{G}}_{a}^{\mu\nu} \label{Asc1}\text{,}
\end{equation}

where ${\tilde{G}}_{a}^{\mu\nu}$ denotes the dual field-strength tensor
${\tilde{G}}_{a}^{\mu\nu}=\varepsilon^{\mu\nu\rho\sigma}G_{\rho\sigma}^{a}/2$.
Symmetries valid on the classical level but broken on the quantum level are
referred to as anomalies -- Eq.\ (\ref{Asc1}) indicates the \textit{chiral
anomaly}, a very important feature of QCD that has to be considered also
when a model is built (see Sec.\ \ref{sec.anomaly} for the discussion of the
chiral-anomaly term in our model). Nonetheless, given that even the $u$
and $d$ quarks possess non-vanishing mass values, we observe from Eq.\ (\ref{Asc2}) that, even classically,
the axial-vector singlet current is conserved only partially (PCAC).
\end{itemize}

Let us then summarise the results of this section as follows: we observe a
$U(N_{f})_{L}\times U(N_{f})_{R}$ chiral symmetry in the QCD Lagrangian with
$N_{f}$ flavours (\ref{lqcd}); the symmetry is isomorphic to the $U(N_{f}%
)_{V}\times U(N_{f})_{A}\equiv U(1)_{V}\times SU(N_{f})_{V}\times
U(1)_{A}\times SU(N_{f})_{A}$ symmetry and, in the limit of vanishing quark
masses, it appears to be exact (apart from the chiral anomaly). For
non-vanishing but degenerate quark masses, the symmetry is broken explicitly
to $U(N_{f})_{V}\times U(N_{f})_{A}\rightarrow U(1)_{V}\times SU(N_{f})_{V}$
and for non-degenerate quark masses it is broken to $U(N_{f})_{V}\times
U(N_{f})_{A}\rightarrow U(1)_{V}$. Thus, by discussing the properties of the
QCD Lagrangian only, we conclude that the magnitude of the symmetry breaking
in nature should not be large if we consider $u$ and $d$ quarks only because
their masses are small. However, we will see later on that there is another
mechanism of chiral symmetry breaking, the spontaneous one, that leads to a
variety of new conclusions.\newline

Before we turn to the spontaneous breaking of the chiral symmetry, let us
first briefly discuss other symmetries of the QCD Lagrangian.

\section{Other QCD Symmetries}

\label{sec.othersymmetries}

\subsection{\boldmath $CP$ Symmetry}

The parity transformation for fermions (and thus also for quarks) reads%

\begin{equation}
q(t,\vec{x})\overset{P}{\rightarrow}\gamma^{0}q(t,-\vec{x}) \label{qP}%
\end{equation}

and thus%

\begin{equation}
q^{\dagger}(t,\vec{x})\overset{P}{\rightarrow}q^{\dagger}(t,-\vec{x}%
)\gamma^{0}\text{.} \label{qaP}%
\end{equation}

If we transform the quark part of the QCD Lagrangian (\ref{lq}), then we
obtain for $\mu=i\in\{1,2,3\}$
\begin{align}%
\mathcal{L}%
_{q}  &  =\bar{q}_{f}(t,\vec{x})i\gamma^{i}D_{i}q_{f}(t,\vec{x})-\bar
{q}_{f}(t,\vec{x})m_{f}q_{f}(t,\vec{x})\nonumber\\
&  \overset{P}{\rightarrow}q_{f}^{\dagger}(t,-\vec{x})\gamma^{0}\gamma
^{0}i\gamma^{i}(-D_{i})\gamma^{0}q_{f}(t,-\vec{x})-q_{f}^{\dagger}(t,-\vec
{x})\gamma^{0}\gamma^{0}m_{f}\gamma^{0}q_{f}(t,-\vec{x})\nonumber\\
&  \overset{\text{Eq.\ (\ref{AKD}) }}{=}q_{f}^{\dagger}(t,-\vec{x})\gamma
^{0}i\gamma^{i}D_{i}q_{f}(t,-\vec{x})-q_{f}^{\dagger}(t,-\vec{x})m_{f}%
q_{f}(t,-\vec{x})\nonumber\\
&  =\bar{q}_{f}(t,-\vec{x})i\gamma^{i}D_{i}q_{f}(t,-\vec{x})-\bar{q}%
_{f}(t,-\vec{x})m_{f}q_{f}(t,-\vec{x})\text{.} \label{lgP}%
\end{align}

The parity conservation is trivially fulfilled in the case $\mu=0$ due to
Eq.\ (\ref{AKD}). The gauge part of the QCD Lagrangian (\ref{lg}) is obviously parity-conserving.

The charge conjugation of quarks $S_{C}$ is such that%

\begin{equation}
q\overset{C}{\rightarrow}S_{C}\bar{q}^{t}=S_{C}(\gamma^{0})^{t}q^{\ast}
\label{qC}%
\end{equation}

where the superscript $t$ denotes the transposed function and $S_{C}^{-1}%
\gamma^{\mu}S_{C}=(-\gamma^{\mu})^{t}$. (In the case of the Dirac notation,
$S_{C}=-i\gamma^{2}\gamma^{0}$.) We note that, due to $S_{C}^{-1}%
=S_{C}^{\dagger}$ (unitary transformation),%

\begin{equation}
S_{C}^{\dagger}\gamma^{\mu}S_{C}=(-\gamma^{\mu})^{t}\Rightarrow S_{C}%
^{\dagger}\gamma^{\mu}=(-\gamma^{\mu})^{t}S_{C}^{-1}\text{.} \label{qC1}%
\end{equation}

Additionally,%
\begin{equation}
q^{\dagger}\overset{C}{\rightarrow}[S_{C}(\gamma^{0})^{t}q^{\ast}]^{\dagger
}=q^{t}(\gamma^{0})^{\ast}S_{C}^{\dagger}\text{.} \label{qaC}%
\end{equation}

Let us consider how the quark part of the QCD Lagrangian (\ref{lq}) transforms
under charge conjugation. Note that $D_{\mu}=\partial_{\mu}-ig\mathcal{A_{\mu
}}$ contains gluon fields transforming as odd under $C$. Remember that the
quarks are fermions and therefore any commutation of the quark fields in the
following lines yields an additional minus sign.%

\begin{align}%
\mathcal{L}%
_{q} &  =\bar{q}_{f}i\gamma^{\mu}D_{\mu}q_{f}-\bar{q}_{f}m_{f}q_{f}\nonumber\\
&  \overset{C}{\rightarrow}iq_{f}^{t}(\gamma^{0})^{\ast}S_{C}^{\dagger}%
\gamma^{0}\gamma^{\mu}(\partial_{\mu}+ig\mathcal{A_{\mu}})S_{C}(\gamma
^{0})^{t}q_{f}^{\ast}-q_{f}^{t}(\gamma^{0})^{\ast}S_{C}^{\dagger}\gamma
^{0}m_{f}S_{C}(\gamma^{0})^{t}q_{f}^{\ast}\nonumber\\
&  \overset{\text{Eq.\ (\ref{qC1}) }}{=}iq_{f}^{t}(\gamma^{0})^{\ast}%
(-\gamma^{0})^{t}S_{C}^{-1}\gamma^{\mu}S_{C}(\partial_{\mu}+ig\mathcal{A_{\mu
}})(\gamma^{0})^{t}q_{f}^{\ast} -q_{f}^{t}(\gamma^{0})^{\ast}(-\gamma^{0})^{t}S_{C}^{-1}m_{f}S_{C}%
(\gamma^{0})^{t}q_{f}^{\ast}\nonumber\\
&  \overset{S_{C}^{-1}\gamma^{\mu}S_{C}=(-\gamma^{\mu})^{t}\text{ }}{=}%
iq_{f}^{t}(\gamma^{0})^{\ast}(\gamma^{0})^{t}(\gamma^{\mu})^{t}(\partial_{\mu
}+ig\mathcal{A_{\mu}})(\gamma^{0})^{t}q_{f}^{\ast} -q_{f}^{t}(\gamma^{0})^{\ast}(-\gamma^{0})^{t}m_{f}(\gamma^{0})^{t}%
q_{f}^{\ast}\nonumber\\
&  \overset{\gamma^{0}(\gamma^{0})^{\dagger}=1}{=}iq_{f}^{t}(\gamma^{\mu}%
)^{t}(\gamma^{0})^{t}\partial_{\mu}q_{f}^{\ast}+iq_{f}^{t}(\gamma^{\mu}%
)^{t}(ig)\mathcal{A_{\mu}}(\gamma^{0})^{t}q_{f}^{\ast} -q_{f}^{t}(\gamma^{0})^{\ast}(-\gamma^{0})^{t}m_{f}(\gamma^{0})^{t}%
q_{f}^{\ast}\nonumber\\
&  = \left [iq_{f}^{t}(\gamma^{\mu})^{t}(\gamma^{0})^{t}\partial_{\mu
}q_{f}^{\ast} \right ]^{t}+ \left [iq_{f}^{t}(\gamma^{\mu})^{t}(ig)\mathcal{A_{\mu}}%
(\gamma^{0})^{t}q_{f}^{\ast} \right ]^{t} + \left [q_{f}^{t}m_{f}(\gamma^{0})^{t}q_{f}^{\ast} \right ]^{t}\nonumber\\
&  =-i(\partial_{\mu}\bar{q}_{f})\gamma^{\mu}q_{f}-i\bar{q}_{f}\gamma^{\mu
}(ig\mathcal{A_{\mu}})q_{f}-\bar{q}_{f}m_{f}q_{f}\nonumber\\
&  \equiv i\bar{q}_{f}\gamma^{\mu}\partial_{\mu}q_{f}-i\bar{q}_{f}\gamma^{\mu
}(ig\mathcal{A_{\mu}})q_{f}-\bar{q}_{f}m_{f}q_{f}\nonumber\\
&  =i\bar{q}_{f}\gamma^{\mu}D_{\mu}q_{f}-\bar{q}_{f}m_{f}q_{f}\text{,} \label{QCDC}%
\end{align}

where we have used $(\gamma
^{0})^{\ast}(\gamma^{0})^{t}=[\gamma^{0}(\gamma^{0})^{\dagger}]^{t}%
=1^t = 1$ and also the well-known feature that the following equality holds
for an $(N\times N)$\ matrix $M$, $( 1 \times N)$ vector $\vec{v}$ and $(N\times1)$ vector $\vec{u}$ under transposition:%
\begin{equation}
\vec{v}M\vec{u}=(\vec{v}M\vec{u})^{t}\label{vMu}%
\end{equation}

because the result of the multiplication is a number.\\

Then the QCD Lagrangian (\ref{lqcd}) is unchanged under $P$ and $C$
transformations (\ref{qP}), (\ref{qaP}), (\ref{qC}) and (\ref{qaC}) -- the
strong interaction is $CP$ \textit{invariant}. [A review of a possible,
although small, \textit{CP} violation in strong interactions may be found,
e.g.,\ in Ref.\ \cite{Peccei}.]

\subsection{\boldmath $Z$ Symmetry}

This symmetry is a discrete one. The general form of a special unitary
$N_{f}\times N_{f}$ matrix $U$ contains also the centre elements $Z_{n}$
(sometimes referred to simply as $Z$):%

\begin{equation}
U=Z_{n}\exp(-i\alpha^{i}t^{i})\text{, }i=1,...,N_{f}^{2}-1
\end{equation}

where%

\begin{equation}
Z_{n}=\exp\left(  -i\frac{2\pi n}{N_{c}}  t^{0} \right) ,\text{ }n=0,1,...N_{c}-1\text{.}%
\end{equation}

The determinant of $Z_{n}$ reads%

\begin{equation}
\det Z_{n}=\left[  \exp\left(  -i\frac{2\pi n}{N_{c}}\right)  \right]
^{N_{c}}=1\text{.}%
\end{equation}

For this reason, $Z_{n}$ is (as already indicated) a member of the $SU(N_{f})$
group. The quarks and gluons transform under the $Z$ group as%

\begin{align}
q_{f}  &  \rightarrow q_{f}^{^{\prime}}=Z_{n}q_{f}\text{,}\\
A_{\mu}  &  \rightarrow A_{\mu}^{\prime}=Z_{n}A_{\mu}Z_{n}^{\dagger}%
\end{align}

and the Yang-Mills Lagrangian (\ref{lg}) is invariant under these transformations.
The $Z$ symmetry is not exact in the presence of quarks because it
does not fulfill the necessary antisymmetric boundary conditions. Additionally,
note that the symmetry is spontaneously broken in the gauge (i.e., gluon)
sector of QCD at large temperatures. This is an order parameter for the
deconfinement. The order parameter is usually represented (in models such as
the NJL model \cite{NJL} but also in first-principle calculations) by the so-called Polyakov loop (see
Ref.\ \cite{PNJL} for Polyakov-loop extended NJL model).

\subsection{Dilatation Symmetry}

This is a symmetry of the gauge (or Yang-Mills -- YM) sector of QCD,
Eq.\ (\ref{lg}). The dilatation (or scale) transformation is defined as%

\begin{equation}
x^{\mu}\rightarrow x^{\prime\mu}=\lambda^{-1}x^{\mu}\text{,} \label{Dt1}
\end{equation}

where $\lambda$ denotes a scale parameter. Then, for dimensional reasons, the
gauge fields in Eq.\ (\ref{lg}) have to transform as%

\begin{equation}
A_{\mu}^{a}(x)\rightarrow A_{\mu}^{a\prime}(x^{\prime})=\lambda A_{\mu}%
^{a}(x)\text{.} \label{Dt2}%
\end{equation}

Consequently, the Lagrangian (\ref{lg}) obtains a factor $\lambda^{4}$ under
the transformations (\ref{Dt1}) and (\ref{Dt2})%

\begin{equation}%
\mathcal{L}%
_{g}\rightarrow\lambda^{4}%
\mathcal{L}%
_{g}
\end{equation}

but the action%

\begin{equation}
S_{g}=\int\text{d}^{4}x%
\mathcal{L}%
_{g}%
\end{equation}

is invariant. This symmetry is known as the \textit{dilatation} or
\textit{trace} \textit{symmetry}. (Strictly speaking, the notion of a symmetry
always requires the action $S$ to be invariant under a transformation rather than the Lagrangian $%
\mathcal{L}%
$ but in all the other examples
discussed in this chapter there is no transformation of space-time $x$. For
this reason, in all the other cases, the symmetry of the Lagrangian is
simultaneously the symmetry of the action as well.)

Consequently, the obtained conserved current reads%

\begin{equation}
J^{\mu}=x_{\nu}T^{\mu\nu}\text{,}
\end{equation}

where $T^{\mu\nu}$ is the energy-momentum tensor of the gauge-field Lagrangian
(\ref{lg}):%

\begin{equation}
T^{\mu\nu}=\frac{\partial%
\mathcal{L}%
_{g}}{\partial(\partial_{\mu}A_{\xi})}\partial^{\nu}A^{\xi}-g^{\mu\nu}%
\mathcal{L}%
_{g}\text{.}%
\end{equation}

Therefore,%

\begin{equation}
\partial_{\mu}J^{\mu}=T_{\mu}^{\mu}=0\text{.} \label{Tm}%
\end{equation}

Similarly to the singlet axial current (\ref{Asc2}), the dilatation symmetry
is broken both classically and at the quantum level. On the classical level, the
dilatation-symmetry breaking is induced by the inclusion of quark degrees of
freedom. We observe from Eq.\ (\ref{lq}) that the quark fields have to
transform in the following way so that the dilatation symmetry is fulfilled in
the limit $m_{f}=0$:%

\begin{equation}
q_{f}\rightarrow q_{f}^{\prime}=\lambda^{3/2}q_{f}\text{.}%
\end{equation}

If we consider $m_{f}\neq0$, then we observe that the dilatation symmetry is
broken explicitly by non-vanishing quark masses:%

\begin{equation}
T_{\mu}^{\mu}=\sum_{f=1}^{N_{f}}m_{f}\bar{q}q\text{,}
\end{equation}

unlike Eq.\ (\ref{Tm}). The degree of the dilatation-symmetry breaking is of
course small if one considers only light quarks and the symmetry is
exact if one considers $m_{f}=0$.

However, at the quantum level (calculating gluon loops), the symmetry is never
exact. The strong coupling $g$ is known to change with scale $\mu$ (e.g.,
centre-of-mass energy) upon renormalisation of QCD \cite{AF}:
$g\rightarrow g(\mu)$. Perturbative QCD then yields%

\begin{equation}
\partial_{\mu}J^{\mu}=T_{\mu}^{\mu}=\frac{\beta(g)}{4g}G_{\mu\nu}^{a}%
G_{a}^{\mu\nu}\neq0\text{,}
\end{equation}

where $\beta(g)$ denotes the famous $\beta$-function of QCD%

\begin{equation}
\beta(g)=\mu\frac{\partial g}{\partial\mu}%
\end{equation}

that demonstrates how the strong coupling changes with the scale. At 1-loop level:%

\begin{equation}
\beta(g)=-bg^{3}=-\frac{11N_{c}-2N_{f}}{48\pi^{2}}g^{3}\text{.} \label{beta1}%
\end{equation}

If the strong coupling did not change ($g=$ const. $=g_{0}$), then the
dilatation symmetry would not be broken and we would retrieve the result of
Eq.\ (\ref{Tm}). Solving the differential Eq.\ (\ref{beta1}) yields%

\begin{equation}
g^{2}(\mu)=\frac{g_{\ast}^{2}}{1+2bg_{\ast}^{2}\log\frac{\mu}{\mu_{\ast}}%
}\text{.} \label{gmu}%
\end{equation}

Given that $b>0$ for $N_{f}<11N_{c}/2$\ [see Eq.\ (\ref{beta1})], the coupling
decreases with the increasing scale. Thus, at small scales, the coupling is
strong: this is a sign of confinement. However, we also observe that, at a
certain scale, one can expect the interaction strength between quarks and
gluons to decrease sufficiently as to allow for the partons to no longer be
confined within hadrons -- \textit{asymptotic freedom}. Note also that
Eq.\ (\ref{gmu}) implies that the strong coupling $g$ decreases with the
number of colours, a result of major impact also for deliberations in this
work (see Sec.\ \ref{sec.largen}).

Note that Eq.\ (\ref{gmu}) possesses a pole (the so-called Landau pole) at%

\begin{equation}
\mu_{L}\equiv\Lambda_{\text{Landau}}=\mu_{\ast}\exp\left(  -\frac{1}%
{2bg_{\ast}^{2}}\right)  \text{.}%
\end{equation}

Then we can transform Eq.\ (\ref{gmu}) as%

\begin{equation}
g^{2}(\mu)=\frac{1}{2b\log\frac{\mu}{\Lambda_{\text{Landau}}}}\text{.}%
\end{equation}

Of course, this result does not imply that the strong coupling $g$ diverges at
$\mu=\mu_{L}$ but rather indicates that QCD is a strongly bound theory in
the vicinity of the pole. The value of the pole itself is, unfortunately,
unknown because an initial value of $\mu$ [needed to solve Eq.\ (\ref{beta1})]
is unknown as well. However, this nonetheless implies that a scale is
generated in a dimensionless theory via renormalisation -- a mechanism known
as the \textit{dimensional transmutation}.

The breaking of the dilatation (scale) invariance is labelled as the
\textit{trace anomaly}. It leads to the generation of a gluon condensate due
to the non-vanishing vacuum expectation value of the gluon fields:%

\begin{equation}
\left\langle T_{\mu}^{\mu}\right\rangle =-\left\langle \frac{11N_{c}-2N_{f}%
}{48}\frac{\alpha_{s}}{\pi}G_{\mu\nu}^{a}G_{a}^{\mu\nu}\right\rangle
\sim-\frac{11N_{c}-2N_{f}}{48}C^{4}\text{,}
\end{equation}

where $\alpha_{s}=g^{2}/(4\pi)$ is the strong fine-structure constant and the
values
\begin{equation}
C^{4}\simeq(300-600\text{ MeV})^{4}%
\end{equation}

have been obtained through QCD sum rules (lower range of the interval)
\cite{Sumrules} and lattice simulations (higher range of the interval)
\cite{Lattice}.

This raises the possibility to study glueball fields -- bound states of two (or more)
gluons. We will present a calculation involving, among others, a scalar
glueball and a $\bar{q}q$ state in Chapter \ref{chapterglueball}.

\section{Spontaneous Breaking of the Chiral Symmetry}

\label{sec.SSB}

Until now we have only considered quarks and gluons as degrees of freedom. In
this section we turn to structures that are composed of quarks. Concretely, we
will be working with $\bar{q}q$ mesons (flavour index $f$ suppressed). These
states are colour-neutral therefore trivially fulfilling the confinement.\\

States containing an antiquark and a quark can be classified according to
their quantum numbers: total spin $J=L+S$ (where $L$ denotes the relative
orbital angular momentum of the two quarks and $S$ denotes their relative
spin), parity $P$, Eqs.\ (\ref{qP}) and (\ref{qaP}),\ and charge conjugation
$C$, Eq.\ (\ref{qC}) and (\ref{qaC}). Let us restrict ourselves to the case of
the light quarks $u$ and $d$ only, i.e., $\bar{q}\equiv(\bar{u},\bar{d})$ and
$q$ the corresponding column vector; states with heavier quarks are discussed analogously.\\

First we can define a state \cite{Koch}:%
\begin{equation}
\bar{q}q \label{sigmall}%
\end{equation}

for which we observe that it transforms as follows under parity,
Eqs.\ (\ref{qP}) and (\ref{qaP}):%

\begin{equation}
\bar{q}(t,\vec{x})q(t,\vec{x})\overset{P}{\rightarrow}q^{\dagger}(t,-\vec
{x})\gamma^{0}\gamma^{0}\gamma^{0}q(t,-\vec{x})\overset{\text{Eq.\ (\ref{AKD})
}}{=}\bar{q}(t,-\vec{x})q(t,-\vec{x}) \label{sigmaP}%
\end{equation}

and under charge conjugation, Eqs.\ (\ref{qC}) and (\ref{qaC}), as

\begin{align}
\bar{q}q &  \overset{C}{\rightarrow}q^{t}i(\gamma^{0})^{\ast}\gamma^{2}%
\gamma^{0}\gamma^{0}(-i)\gamma^{2}\gamma^{0}(\gamma^{0})^{t}q^{\ast
}\nonumber\\
&  \overset{\text{Eq.\ (\ref{AKD}) }}{=}q^{t}(\gamma^{0})^{\ast}\gamma
^{2}\gamma^{2}\gamma^{0}(\gamma^{0})^{t}q^{\ast}=-q^{t}(\gamma^{0})^{\ast
}\gamma^{0}(\gamma^{0})^{t}q^{\ast}\nonumber\\
&  \overset{!}{=}[-q^{t}(\gamma^{0})^{\ast}\gamma^{0}(\gamma^{0})^{t}q^{\ast
}]^{t} = q^{\dagger}\gamma^{0}(\gamma^{0})^{t}(\gamma^{0})^{\dagger}q=\bar
{q}q\text{.}\label{sigmaC}%
\end{align}

The $\bar{q}q$ state is therefore unchanged under parity and charge
conjugation, and it obviously carries no (total) spin. It is therefore a
scalar. Note, however, that the mere fact of having $J=0$ does not necessarily
imply that $L$ and $S$ vanish as well. Indeed one can demonstrate
\cite{FrancescoDissertation} that, for a system of an antiquark and a quark,%
\begin{equation}
P=(-1)^{L+1}\label{Pc}%
\end{equation}

and
\begin{equation}
C=(-1)^{L+S}\text{.}\label{Cc}%
\end{equation}

For this reason, $P=1=C$ implies $L=1=S$. In
other words: the scalar $\bar{q}q$ state is a $P$-wave state. We can denote it
as $\sigma_{N}$, alluding to the famous $\sigma$ meson (see
Sec.\ \ref{sec.f0(600)}), or in other words%

\begin{equation}
\sigma_{N} \equiv \bar{q}q\text{.}\label{sigmal}%
\end{equation}

If we define a state
\begin{equation}
i\bar{q}\gamma_{5}q\text{,} \label{etal}%
\end{equation}

then we observe that it transforms as follows under parity%

\begin{equation}
i\bar{q}(t,\vec{x})\gamma_{5}q(t,\vec{x})\overset{P}{\rightarrow}iq^{\dagger
}(t,-\vec{x})\gamma^{0}\gamma^{0}\gamma_{5}\gamma^{0}q(t,-\vec{x}%
)\overset{\text{Eq.\ (\ref{AKD}) }}{=}-i\bar{q}(t,-\vec{x})\gamma_{5}%
q(t,-\vec{x}) \label{pionP}%
\end{equation}

and under charge conjugation (here exemplary for the Dirac notation) as:
\begin{align}
i\bar{q}(t,\vec{x})\gamma_{5}q(t,\vec{x})  &  \overset{C}{\rightarrow}%
iq^{t}i\gamma^{0}\gamma^{2}\gamma^{0}\gamma^{0}\gamma_{5}(-i)\gamma^{2}%
q^{\ast}\nonumber\\
&  \overset{\text{Eq.\ (\ref{AKD}) }}{=}iq^{t}i\gamma^{0}\gamma^{2}\gamma
_{5}(-i)\gamma^{2}q^{\ast}\overset{\text{Eq.\ (\ref{Gamma5-2}) }}{=}%
-iq^{t}\gamma^{0}\gamma^{2}\gamma^{2}\gamma_{5}q^{\ast}\nonumber\\
&  =iq^{t}\gamma^{0}\gamma_{5}q^{\ast}\overset{!}{=}(iq^{t}\gamma^{0}%
\gamma_{5}q^{\ast})^{t}=-iq^{\dagger}\gamma_{5}\gamma^{0}q=i\bar{q}\gamma
_{5}q\text{.} \label{pionC}%
\end{align}

The state $\bar{q}\gamma_{5}q$ is thus $P$-odd and $C$-even: it is a pseudoscalar
and we label the state as $\eta_{N}$, alluding to the physical $\eta$ field:
$\eta_{N} \equiv \bar{q}\gamma_{5}q$. [Note, however, that the field defined in
Eq.\ (\ref{etal}) cannot be exactly the physical $\eta$ field because we have
restricted ourselves to two flavours, i.e., non-strange quarks, see
Sec.\ \ref{sec.eta-eta}.] Similarly, we define a pion-like state%
\begin{equation}
\vec{\pi}\equiv i\bar{q}\vec{t}\gamma_{5}q \label{pionl}%
\end{equation}

considering that the pion is an isospin triplett. The calculation of the behaviour
of the state in Eq.\ (\ref{pionl}) under parity and charge conjugation is
analogous to the one demonstrated in Eqs.\ (\ref{pionP}) and (\ref{pionC}).\\

Let us now define a state
\begin{equation}
\bar{q}\gamma^{\mu}q
\end{equation}

for which we observe that it transforms as follows under parity
\begin{align}
\bar{q}(t,\vec{x})\gamma^{\mu}q(t,\vec{x})  &  \overset{P}{\rightarrow
}q^{\dagger}(t,-\vec{x})\gamma^{0}\gamma^{0}\gamma^{\mu}\gamma^{0}q(t,-\vec
{x})\nonumber\\
&  \overset{\text{Eq.\ (\ref{AKD}) }}{=}\left\{
\begin{tabular}
[c]{l}%
$\bar{q}(t,-\vec{x})\gamma^{0}q(t,-\vec{x})$ for $\mu=0$\\
$- \bar{q}(t,-\vec{x})\gamma^{i}q(t,-\vec{x})$ for $\mu=i\in\{1,2,3\}$%
\end{tabular}
\ \ \ \ \right.  \label{omegaP}%
\end{align}

or, in other words, the temporal component is parity-even whereas the spatial
components are parity-odd. Additionally, we observe that the state is odd
under the $C$-transformation [the calculation is analogous to the one in
Eq.\ (\ref{pionC})]. Given that the field combination $\bar{q}\gamma^{\mu}q$
possesses spin 1, we label it as a vector state that we denote as $\omega
_{N}^{\mu}$. Consequently, the state%

\begin{equation}
\vec{\rho}^{\mu} \equiv \bar{q}\vec{t}\gamma^{\mu}q \label{rhol}%
\end{equation}

is an isospin-triplett vector state [just as the $\rho(770)$ meson].\\

Finally, the states
\begin{equation}
f_{1N}^{\mu} \equiv \bar{q}\gamma_{5} \gamma^{\mu} q
\end{equation}

and
\begin{equation}
\vec{a}_{1}^{\mu} \equiv \bar{q}\vec{t}\gamma_{5} \gamma^{\mu} q \label{a1l}%
\end{equation}

are even under both parity and charge conjugation and additionally have spin
1: we label them as axial-vectors.
\newline

Let us now discuss the behaviour of these states under vector and
axial-vector transformations. We observe, for example, that the vector
transformation (\ref{UVi}) of the pion field $\vec{\pi}$ (\ref{pionl}) yields
\begin{align}
\vec{\pi} \equiv i\bar{q}\vec{t}\gamma_{5}q  &  \overset{V}{\rightarrow}iq^{\dagger
}\,(1+i\,\vec{\alpha}_{V}\cdot\vec{t})\,\gamma_{0}\,\vec{t}\,\gamma
_{5}\,(1-i\,\vec{\alpha}_{V}\cdot\vec{t})\,q\nonumber\\
&  =i\bar{q}\vec{t}\gamma_{5}q+iq^{\dagger}\,(i\,t^{i}\,t^{j}\,\alpha_{V}%
^{i}-i\,t^{j}\,t^{i}\,\alpha_{V}^{i})\,\gamma_{0}\,\gamma_{5}\,q\nonumber\\
&  =i\bar{q}\vec{t}\gamma_{5}q-iq^{\dagger}\alpha_{V}^{i}\,\varepsilon^{ijk}\,t^{k}\,\gamma
_{0}\,\gamma_{5}\,q\nonumber\\
&  =i\bar{q}\vec{t}\gamma_{5}q-i\varepsilon^{ijk}\,\alpha_{V}^{i}\,\bar{q}\,t^{k}\,\gamma
_{5}\,q\nonumber\\
&  \equiv\vec{\pi}+i\varepsilon^{ijk}\,\alpha_{V}^{j}\,\bar{q}\,t^{k}%
\,\gamma_{5}\,q \equiv \vec{\pi}+\vec{\alpha}_{V}\times\vec{\pi}\text{,} \label{pionV}%
\end{align}

where we have used the commutator $[t^{i},t^{j}]=i\varepsilon^{ijk}t^{k}$.

Analogously to Eq.\ (\ref{pionV}), we obtain from Eqs.\ (\ref{sigmal}),
(\ref{rhol}) and (\ref{a1l}):
\begin{align}
&  \sigma\overset{V}{\rightarrow}\sigma\text{,}\\
&  \vec{\rho}^{\mu}\overset{V}{\rightarrow}\vec{\rho}^{\mu}+\vec{\alpha}%
_{V}\times\vec{\rho}^{\mu}\text{,}\\
&  \vec{a}_{1}^{\mu}\overset{V}{\rightarrow}\vec{a}_{1}^{\mu}+\vec{\alpha}%
_{V}\times\vec{a}_{1}^{\mu}\text{.}%
\end{align}

Thus the vector transformation corresponds to a rotation in the isospin
space; as we have seen in Sec.\ \ref{sec.CM}, the QCD Lagrangian is invariant
under this transformation (for degenerate quark masses) -- the conserved
vector current can thus be identified with an isospin current.\newline

However, the behaviour of our composite fields is quite different under axial
transformations (\ref{UAi}). Let us again first study the $\vec{\pi}$ field:
\begin{align}
\vec{\pi}=i\bar{q}\vec{t}\gamma_{5}q  &  \overset{A}{\rightarrow}iq^{\dagger
}\,(1+i\,\gamma_{5}\,\vec{\alpha}_{A}\cdot\vec{t})\,\gamma_{0}\,\vec
{t}\,\gamma_{5}\,(1-i\,\gamma_{5}\,\vec{\alpha}_{A}\cdot\vec{t})\,q\nonumber\\
&  =i\bar{q}\vec{t}\gamma_{5}q+iq^{\dagger}\,(i\,\gamma_{5}\,\gamma
_{0}\,\gamma_{5}\,t^{i}\,t^{j}\,\alpha_{A}^{i}-i\,\gamma_{0}\,\gamma_{5}%
^{2}\,t^{j}\,t^{i}\,\alpha_{A}^{i})\,q\nonumber\\
&  \overset{\text{Eq.\ (\ref{Gamma5-2}) }}{=}\vec{\pi}+q^{\dagger}\,\gamma
_{0}\,\alpha_{A}^{i}\,(t^{i}\,t^{j}+t^{j}\,t^{i})q\nonumber\\
&  =\vec{\pi}+q^{\dagger}\,\gamma_{0}\,\alpha_{A}^{i}\,\{t^{i},t^{j}%
\}\,q\equiv\vec{\pi}+\vec{\alpha}_{A}\sigma\text{,} \label{pionA}%
\end{align}

where we have used the anticommutator $\{t^{i},t^{j}\}=\delta^{ij}t^{0}$.
Analogously to Eq.\ (\ref{pionA}) we obtain from Eqs.\ (\ref{sigmal}),
(\ref{rhol}) and (\ref{a1l}):
\begin{align}
&  \sigma\overset{A}{\rightarrow}\sigma-\vec{\alpha}_{A}\cdot\vec{\pi
}\text{,} \label{sigmaA}\\
&  \vec{\rho}^{\mu}\overset{A}{\rightarrow}\vec{\rho}^{\mu}+\vec{\alpha}%
_{A}\times\vec{a}_{1}^{\mu}\text{,} \label{rhoA}\\
&  \vec{a}_{1}^{\mu}\overset{A}{\rightarrow}\vec{a}_{1}^{\mu}-\vec{\alpha}%
_{A}\times\vec{\rho}^{\mu}\text{.}\label{a1A}%
\end{align}

Thus the scalar state $\sigma$ is connected to the pseudoscalar state
$\vec{\pi}$ via the axial transformation (and vice versa); the vector state
$\vec{\rho}^{\mu}$ is connected to the axial-vector state $\vec{a}_{1}^{\mu}$
(and vice versa) in the same way. As we have discussed in Sec.\ \ref{sec.CM},
the axial symmetry $SU(N_{f})_{A}$ is exact within the QCD Lagrangian, up to
the explicit breaking due to non-vanishing quark masses. This implies that, in
the limit of small $u$, $d$ quark masses, the breaking of the axial symmetry
is virtually negligible. Consequently, given that the scalar and the
pseudoscalar can be rotated into each other (just as the vector and the
axial-vector), one would expect these states to possess the same masses.
\textit{Experimentally, this is not the case.} If we assign our vector state
$\vec{\rho}^{\mu}$ to the lowest observed vector excitation, the $\rho(770)$
meson with a mass of $m_{\rho(770)}=(775.49\pm0.34)$ MeV and our axial-vector
state $\vec{a}_{1}^{\mu}$ to the lowest observed axial-vector excitation, the
$a_{1}(1260)$ meson with a mass of $\sim 1230$ MeV \cite{PDG}, then we observe
that the mass difference of these two states is of the order of the
$\rho(770)$ mass itself. Such a large magnitude of symmetry breaking cannot
originate from the (small) quark masses. The symmetry must have been
broken by a different mechanism -- \textit{spontaneously} -- because,
evidently, the axial symmetry realised in the QCD Lagrangian is not realised
in the vacuum states of QCD.

Let us remind ourselves that the chiral symmetry $U(N_{f})_{V}\times U(N_{f})_{A}\equiv U(1)_{V}\times
SU(N_{f})_{V}\times U(1)_{A}\times SU(N_{f})_{A}$ discussed in
Sec.\ \ref{sec.CM} is broken explicitly to $U(1)_{V}\times SU(N_{f}%
)_{V}$ in the case of non-vanishing but degenerate quark masses [note also the existence of the
$U(1)_{A}$ anomaly (\ref{Asc1}) even if all quark masses vanish]. If the quark masses are non-vanishing
and non-degenerate then the residual $U(1)_{V}\times SU(N_{f})_{V}$ symmetry is 
broken completely to $U(1)_{V}$,
indicating baryon-number conservation.\\

On the other hand, the mechanism of spontaneous chiral-symmetry breaking is
based on the existence of the chiral (quark) condensate \cite{refssb}
\begin{equation}
\langle\bar{q}q\rangle=\langle0|\bar{q}q|0\rangle=-i\text{Tr}\lim
_{y\rightarrow x^{+}}S_{F}(x,y) \label{chiralc}
\end{equation}

where $S_{F}(x,y)$ denotes the full quark propagator. Then utilising
Eqs.\ (\ref{PRPL}), (\ref{qRL}) and (\ref{aqRL}) we obtain
\begin{equation}
\langle\bar{q}q\rangle=\langle(\bar{q}_{L}+\bar{q}_{R})(q_{L}+q_{R}
)\rangle=\langle\bar{q}_{R}q_{L}+\bar{q}_{L}q_{R}\rangle\neq0\text{.}
\label{chiralc1}%
\end{equation}

The existence of the quark condensate is a consequence of vacuum polarisation by means of the strong
interaction. The condensate breaks the chiral symmetry $SU(N_{f})_{V}\times SU(N_{f})_{A}$ to $SU(N_{f})_{V}$;
the magnitude of the condensate is a measure of the magnitude of the
spontaneous chiral-symmetry breaking -- for $\langle\bar{q}q\rangle
\rightarrow0$, the axial symmetry is exact again.\\

The spontaneous breaking of the chiral symmetry has at least two important
consequences. According to the Goldstone Theorem \cite{GT}, one expects
$N_{f}^{2}-1$ massless pseudoscalar bosons to emerge as consequence of the
spontaneous breaking of a global symmetry. This is indeed observed: e.g., for
$N_{f}=2$, three pions were discovered a long time ago \cite{pion} and their
masses of $\sim 140$ MeV are several times smaller than the mass of the first
heavier meson. Their non-vanishing mass arises due to the explicit breaking of
the chiral symmetry, rendering them pseudo-Goldstone bosons. For $N_{f}=3$,
experimental observations yield five additional pseudoscalar Goldstone states:
four kaons and the $\eta$ meson. Note, however, that the latter mixes with a
heavier $\eta^{\prime}$ state that would also represent a Goldstone mode of
QCD if the chiral anomaly (\ref{Asc1}) were not present.\\

Additionally, it is expected that the
quark condensate will diminish at non-zero values of temperature and baryon
density; the restoration of the chiral symmetry (the so-called chiral phase
transition) thus denotes the point where the chiral-symmetry breaking is no
longer present (at a temperature $T_{c}\sim190$ MeV \cite{RS}). The chiral
phase transition may occur simultaneously with the deconfinement; however, it
is as yet not known whether this is actually the case.

\section{Calculating the Decay Widths} \label{sec.calcdw}

As already indicated, mesons are very unstable particles. The typical lifetime
of a meson is $(10^{-24} - 10^{-25})$ s, with some notable exceptions such as
the pion with the mean lifetime of approximately $10^{-8}$ s. This work will
analyse various two-body decays of mesons into final states as well as into
states that themselves decay further (sequential decays). For this reason it
is necessary to develop a formalism that allows us to calculate the
corresponding decay widths, denoted as $\Gamma$; these are related to the time
$\tau$ necessary for a particle to decay with the following relation
\begin{equation}
\tau=\Gamma^{-1}\text{.}%
\end{equation}

(There may be deviations from this law under certain conditions which we do not consider here; see Ref.\ \cite{arXiv:1005.4817}.)\\
Let us in the following derive a formula for the decay width
$\Gamma_{\chi\rightarrow\varphi_{1}\varphi_{2}}$ of a particle $\chi$ decaying
into particles $\varphi_{1}$and $\varphi_{2}$.
\begin{align}%
\mathcal{L}%
_{\chi\varphi_{1}\varphi_{2}}  &  =\frac{1}{2}(\partial_{\mu}\chi)^{2}%
-\frac{1}{2}m_{\chi}^{2}\chi^{2}+\frac{1}{2}(\partial_{\mu}\varphi_{1}%
)^{2}-\frac{1}{2}m_{\varphi_{1}}^{2}\varphi_{1}^{2}\nonumber\\
&  +\frac{1}{2}(\partial_{\mu}\varphi_{2})^{2}-\frac{1}{2}m_{\varphi_{2}}%
^{2}\varphi_{2}^{2}+g\chi\varphi_{1}\varphi_{2} \label{chiphiphi}%
\end{align}

where the last term, $g\chi\varphi_{1}\varphi_{2}$, denotes the interaction of
the field $\chi(X)$ with the fields $\varphi_{1}(X)$ and $\varphi_{2}(X)$; $x$
denotes the Minkowski space-time vector. Let us, for simplicity, assume that
$\varphi_{1}=\varphi_{2}\equiv\varphi$. Let us also assume that $m_{\chi
}>2m_{\varphi}$ rendering the tree-level decay $\chi\rightarrow2\varphi$ possible.

The decay amplitude obtained from the Lagrangian in Eq.\ (\ref{chiphiphi})
then reads:
\begin{equation}
-i\mathcal{M}_{\chi\rightarrow2\varphi}=i2g \label{iMchiphiphi}%
\end{equation}

where the symmetry factor of two appears due to the new form of the
interaction part of the Lagrangian (\ref{chiphiphi}): $g\chi\varphi_{1}%
\varphi_{2}\overset{\varphi_{1}=\varphi_{2}}{\rightarrow}g\chi\varphi^{2}$.

Let us consider the scalar fields $\chi$ and $\varphi$\ as confined in a cube
of length $L$ and volume $V=L^3$. Let us furthermore denote the 4-momenta of
$\chi$ and $\varphi$ as $P$ and $K$, respectively; then it is known from
Quantum\ Mechanics that their 3-momenta are quantised: $\vec{p}=2\pi\vec
{n}_{P}/L$, $\vec{k}=2\pi\vec{n}_{K}/L$. We denote the corresponding energies
as $E_{\vec{p}}=\sqrt{m_{\chi}^{2}+\vec{p}^{2}}$ and $E_{\vec{k}}%
=\sqrt{m_{\varphi}^{2}+\vec{k}^{2}}$. It is also known from Quantum Field
Theory that a free scalar bosonic field can be decomposed in terms of creation
and annihilation operators (respectively $\hat{a}^{\dagger}$, $\hat
{b}^{\dagger}$ and $\hat{a}$, $\hat{b}$) utilising the following Fourier transformation:%
\begin{equation}
\hat{\chi}(X)=%
{\displaystyle\int}
\frac{\text{d}^{3}\vec{p}}{\sqrt{(2\pi)^{3}}}\frac{1}{\sqrt{2E_{\vec{p}}}%
}\left[  \hat{b}\left(  \vec{p}\right)  e^{-iP\cdot X}+\hat{b}^{\dagger}%
(\vec{p})e^{iP\cdot X}\right]  \label{sfchi}%
\end{equation}
and%
\begin{equation}
\hat{\varphi}(X)=\int\frac{\text{d}^{3}\vec{k}}{\sqrt{(2\pi)^{3}}}\frac
{1}{\sqrt{2E_{\vec{k}}}}\left[  \hat{a}(\vec{k})e^{-iK\cdot X}+\hat
{a}^{\dagger}(\vec{k})e^{iK\cdot X}\right]  \text{.} \label{sfphi}%
\end{equation}

The operators obey the following commutation relations:
\begin{align}
\lbrack \hat{a}(\vec{k}_{1}),\hat{a}(\vec{k}_{2})]  &  =[\hat{a}^{\dagger}(\vec{k}%
_{1}),\hat{a}^{\dagger}(\vec{k}_{2})]=0\text{,} \label{kr1}\\
\lbrack \hat{a}(\vec{k}_{1}),\hat{a}^{\dagger}(\vec{k}_{2})]  &  =\delta^{(3)}(\vec{k}%
_{1}-\vec{k}_{2})\text{,} \label{kr2}\\
\lbrack \hat{b}(\vec{p}_{1}),\hat{b}(\vec{p}_{2})]  &  =[\hat{b}^{\dagger}(\vec{p}%
_{1}),\hat{b}^{\dagger}(\vec{p}_{2})]=0\text{,} \label{kr3}\\
\lbrack \hat{b}(\vec{p}_{1}),\hat{b}^{\dagger}(\vec{p}_{2})]  &  =\delta^{(3)}(\vec{p}%
_{1}-\vec{p}_{2})\text{.} \label{kr4}%
\end{align}

The $\chi$ resonance represents our initial state $\left\vert i\right\rangle $:%

\begin{equation}
\left\vert i\right\rangle =\sqrt{\frac{(2\pi)^{3}}{V}}\hat{b}^{\dagger}%
(\vec{p})\left\vert 0\right\rangle \label{initial}%
\end{equation}

whereas the two $\varphi$ resonances are our final states:%

\begin{equation}
\left\vert f\right\rangle =\frac{(2\pi)^{3}}{V}\hat{a}^{\dagger}(\vec{k}%
_{1})\hat{a}^{\dagger}(\vec{k}_{2})\left\vert 0\right\rangle \text{.} \label{final}%
\end{equation}

The volume appears in the definition of the initial and final states to ensure
their correct normalisation:
\begin{align}
\langle i\left\vert i\right\rangle  &  =\frac{(2\pi)^{3}}{V}\langle0 \vert
\hat{b}(\vec{p})\hat{b}^{\dagger}(\vec{p})|0 \rangle \overset
{\text{Eq.\ (\ref{kr4})}}{=}\frac{(2\pi)^{3}}{V}\langle0 \vert
\delta^{(3)}(0)-\hat{b}^{\dagger}(\vec{p})\hat{b}(\vec{p})|0 \rangle
\nonumber\\
&  =\frac{(2\pi)^{3}}{V}\langle0 \vert \delta^{(3)}(0)|0 \rangle
=\frac{(2\pi)^{3}}{V}\delta^{(3)}(0)=1\text{,} \label{iverti}%
\end{align}

under the following normalisation condition for the $\delta$ distribution:%

\begin{equation}
\delta^{(3)}(0)=\lim_{V \rightarrow\infty}\frac{V}{(2\pi)^{3}}\text{,}
\end{equation}

obtained from the well-known Fourier transformation%

\begin{equation}
\delta^{(3)}(\vec{p})=\int\frac{\text{d}^{3} \vec x}{(2\pi)^{3}}e^{i\vec
{p}\cdot\vec x}%
\end{equation}

for $\vec{p}=0$. [As an equivalence but not equality, we can state simply
$
\delta^{(3)}(0)\equiv \frac{V}{(2\pi)^{3}}%
$.] The calculation of Eq.\ (\ref{iverti}) can likewise be
repeated for the final state $\left\vert f\right\rangle $ (\ref{final}).

The corresponding element of the scattering matrix then reads

\begin{equation}
\left\langle f\right\vert \mathcal{S}\left\vert i\right\rangle \text{,}
\end{equation}

with the scattering-matrix operator
\begin{equation}
\mathcal{S}=\hat{T}e^{-i\int d^{4}X\mathcal{H}(X)}\text{,}
\end{equation}

where $\hat{T}$ denotes the time-ordering operator and $\mathcal{H}(X)$ is the
interaction Hamiltonian that depends on the interaction part of the Lagrangian
(\ref{chiphiphi}):%

\begin{equation}
\mathcal{H}=-g\chi(X)\varphi^{2}(X)\text{.}%
\end{equation}

For a small coupling, we can consider the scattering matrix up to the first
order only:%

\begin{equation}
\mathcal{S}^{(1)}=-i\int d^{4}X\hat{T}[\mathcal{H}(X)]\text{.}%
\end{equation}

Let us now calculate the expectation value of $\mathcal{S}^{(1)}$ in terms of
the initial and final states:
\begin{align}
\left\langle f\right\vert \mathcal{S}^{(1)}\left\vert i\right\rangle  &
=\left\langle f\right\vert ig\int\text{d}^{4}X\hat{T}[\varphi^{2}%
(X)\chi(X)]\left\vert i\right\rangle \nonumber\\
&  =i\left[  \frac{(2\pi)^{3}}{V}\right]  ^{3/2}g\left\langle 0\right\vert
\hat{a}(\vec{k}_{1})\hat{a}(\vec{k}_{2})\int\text{d}^{4}X\hat{T}[\varphi^{2}%
(X)\chi(X)]\hat{b}^{\dagger}(\vec{p})\left\vert 0\right\rangle \text{.}
\label{fSi1}%
\end{align}

Inserting Eqs.\ (\ref{sfchi}) and (\ref{sfphi}) into Eq.\ (\ref{fSi1}) and
performing time-ordered product of creation and annihilation operators we obtain
\begin{align}
\left\langle f\right\vert \mathcal{S}^{(1)}\left\vert i\right\rangle  &
=i\left[  \frac{(2\pi)^{3}}{V}\right]  ^{3/2}g\frac{1}{\sqrt{2E_{\vec{p}}}%
}\frac{1}{\sqrt{2E_{\vec{k}_{1}}}}\frac{1}{\sqrt{2E_{\vec{k}_{2}}}}%
\int\text{d}^{4}X\frac{\text{d}^{3}\vec{p}}{\sqrt{(2\pi)^{3}}}\frac
{\text{d}^{3}\vec{k}_{1}}{\sqrt{(2\pi)^{3}}}\frac{\text{d}^{3}\vec{k}_{2}%
}{\sqrt{(2\pi)^{3}}}\nonumber\\
&  \times\left\langle 0\right\vert \hat{a}(\vec{k}_{1})\hat{a}(\vec{k}_{2})\left[  \hat
{b}\left(  \vec{p}\right)  e^{-iP\cdot X}+\hat{b}^{\dagger}(\vec{p})e^{iP\cdot
X}\right] \nonumber\\
&  \times\left[  \hat{a}(\vec{k}_{1})e^{-iK_{1}\cdot X}+\hat{a}^{\dagger}%
(\vec{k}_{1})e^{iK_{1}\cdot X}\right]  \left[  \hat{a}(\vec{k}_{2}%
)e^{-iK_{2}\cdot X}+\hat{a}^{\dagger}(\vec{k}_{2})e^{iK_{2}\cdot X}\right]
\hat{b}^{\dagger}(\vec{p})\left\vert 0\right\rangle \nonumber\\
&  =i\left[  \frac{(2\pi)^{3}}{V}\right]  ^{3/2}g\frac{1}{\sqrt{2E_{\vec{p}}}%
}\frac{1}{\sqrt{2E_{\vec{k}_{1}}}}\frac{1}{\sqrt{2E_{\vec{k}_{2}}}}%
\int\text{d}^{4}X\frac{\text{d}^{3}\vec{p}}{\sqrt{(2\pi)^{3}}}\frac
{\text{d}^{3}\vec{k}_{1}}{\sqrt{(2\pi)^{3}}}\frac{\text{d}^{3}\vec{k}_{2}%
}{\sqrt{(2\pi)^{3}}}\nonumber\\
&  \times\left\langle 0\right\vert \hat{a}(\vec{k}_{2})\hat{a}(\vec{k}_{1})\hat{a}(\vec{k}%
_{1})\hat{a}(\vec{k}_{2})\hat{b}\left(  \vec{p}\right)  \hat{b}^{\dagger}\left(
\vec{p}\right)  e^{-i(K_{1}+K_{2}+P)\cdot X}\nonumber\\
&  +\hat{a}(\vec{k}_{2})\hat{a}(\vec{k}_{1})\hat{a}^{\dagger}(\vec{k}_{1})\hat{a}(\vec{k}_{2})\hat
{b}\left(  \vec{p}\right)  \hat{b}^{\dagger}\left(  \vec{p}\right)
e^{i(K_{1}+K_{2}-P)\cdot X}\nonumber\\
&  +\hat{a}(\vec{k}_{2})\hat{a}(\vec{k}_{1})\hat{a}(\vec{k}_{1})\hat{a}^{\dagger}(\vec{k}_{2})\hat
{b}\left(  \vec{p}\right)  \hat{b}^{\dagger}\left(  \vec{p}\right)
e^{-i(K_{1}-K_{2}+P)\cdot X}\nonumber\\
&  +\hat{a}(\vec{k}_{2})\hat{a}(\vec{k}_{1})\hat{a}(\vec{k}_{1})\hat{a}(\vec{k}_{2})\hat{b}^{\dagger
}\left(  \vec{p}\right)  \hat{b}^{\dagger}\left(  \vec{p}\right)
e^{-i(K_{1}+K_{2}-P)\cdot X}\nonumber\\
&  +\hat{a}(\vec{k}_{2})\hat{a}(\vec{k}_{1})\hat{a}^{\dagger}(\vec{k}_{1})\hat{a}(\vec{k}_{2})\hat
{b}^{\dagger}\left(  \vec{p}\right)  \hat{b}^{\dagger}\left(  \vec{p}\right)
e^{i(K_{1}+K_{2}+P)\cdot X}\nonumber\\
&  +\hat{a}(\vec{k}_{2})\hat{a}(\vec{k}_{1})\hat{a}(\vec{k}_{1})\hat{a}^{\dagger}(\vec{k}_{2})\hat
{b}^{\dagger}\left(  \vec{p}\right)  \hat{b}^{\dagger}\left(  \vec{p}\right)
e^{-i(K_{1}-K_{2}-P)\cdot X}\nonumber\\
&  +\hat{a}(\vec{k}_{2})\hat{a}(\vec{k}_{1})\hat{a}^{\dagger}(\vec{k}_{1})\hat{a}^{\dagger}(\vec
{k}_{2})\hat{b}^{\dagger}\left(  \vec{p}\right)  \hat{b}^{\dagger}\left(
\vec{p}\right)  e^{i(K_{1}+K_{2}+P)\cdot X}\nonumber\\
&  +\hat{a}(\vec{k}_{2})\hat{a}(\vec{k}_{1})\hat{a}^{\dagger}(\vec{k}_{1})\hat{a}^{\dagger}(\vec
{k}_{2})\hat{b}\left(  \vec{p}\right)  \hat{b}^{\dagger}\left(  \vec
{p}\right)  e^{i(K_{1}+K_{2}-P)\cdot X}\left\vert 0\right\rangle \text{.}%
\end{align}

Only the term in the last line remains as all the other terms are proportional
to $\hat{a}(\vec{k}_{1,2})\left\vert 0\right\rangle =0$, to $\left\langle
0\right\vert \hat{b}^{\dagger}\left(  \vec{p}\right)  =0$ or to%

\begin{align}
&  \hat{a}(\vec{k}_{2})\hat{a}(\vec{k}_{1})\hat{a}(\vec{k}_{1})\hat{a}^{\dagger}(\vec{k}_{2})\hat
{b}\left(  \vec{p}\right)  \hat{b}^{\dagger}\left(  \vec{p}\right)  \left\vert
0\right\rangle \nonumber\\
&  \overset{\text{Eq.\ (\ref{kr4}) }}{=}\hat{a}(\vec{k}_{2})\hat{a}(\vec{k}_{1})\hat{a}(\vec
{k}_{1})\hat{a}^{\dagger}(\vec{k}_{2})[\delta^{(3)}(0)+\hat{b}^{\dagger}\left(
\vec{p}\right)  \hat{b}\left(  \vec{p}\right)  ]\left\vert 0\right\rangle
\nonumber\\
&  \overset{\text{Eq.\ (\ref{kr2}) }}{=}\hat{a}(\vec{k}_{2})\hat{a}(\vec{k}_{1}%
)[\delta^{(3)}(\vec{k}_{1}-\vec{k}_{2})+\hat{a}^{\dagger}(\vec{k}_{1})\hat{a}(\vec{k}%
_{2})]\delta^{(3)}(0)\left\vert 0\right\rangle = 0\text{.}%
\end{align}

We therefore obtain
\begin{align}
\left\langle f\right\vert \mathcal{S}^{(1)}\left\vert i\right\rangle  &
=i\left[  \frac{(2\pi)^{3}}{V}\right]  ^{3/2}g\frac{1}{\sqrt{2E_{\vec{p}}}%
}\frac{1}{\sqrt{2E_{\vec{k}_{1}}}}\frac{1}{\sqrt{2E_{\vec{k}_{2}}}}%
\int\text{d}^{4}X\frac{\text{d}^{3}\vec{p}}{\sqrt{(2\pi)^{3}}}\frac
{\text{d}^{3}\vec{k}_{1}}{\sqrt{(2\pi)^{3}}}\frac{\text{d}^{3}\vec{k}_{2}%
}{\sqrt{(2\pi)^{3}}}\nonumber\\
&  \times\left\langle 0\right\vert \hat{a}(\vec{k}_{2})\hat{a}(\vec{k}_{1})\hat{a}^{\dagger
}(\vec{k}_{1})\hat{a}^{\dagger}(\vec{k}_{2})\hat{b}\left(  \vec{p}\right)  \hat
{b}^{\dagger}\left(  \vec{p}\right)  e^{i(K_{1}+K_{2}-P)\cdot X}\left\vert
0\right\rangle \text{.}\quad\label{fSi2}%
\end{align}

Eq.\ (\ref{iverti}) implies that our creation and annihilation operators are
normalised as%

\begin{equation}
\langle0 \vert \hat{b}(\vec{p})\hat{b}^{\dagger}(\vec{p})|0\rangle
=\frac{V}{(2\pi)^{3}} \label{iverti1}%
\end{equation}

and consequently Eq.\ (\ref{fSi2}) gains the following form:%

\begin{align}
\left\langle f\right\vert \mathcal{S}^{(1)}\left\vert i\right\rangle =i\left[
\frac{(2\pi)^{3}}{V}\right]  ^{3/2}g\frac{1}{\sqrt{2E_{\vec{p}}}}\frac
{1}{\sqrt{2E_{\vec{k}_{1}}}}\frac{1}{\sqrt{2E_{\vec{k}_{2}}}}\frac{V^{3}%
}{(2\pi)^{9}} \nonumber \\
\times \int\text{d}^{4}X\frac{\text{d}^{3}\vec{p}}{\sqrt{(2\pi)^{3}}%
}\frac{\text{d}^{3}\vec{k}_{1}}{\sqrt{(2\pi)^{3}}}\frac{\text{d}^{3}\vec
{k}_{2}}{\sqrt{(2\pi)^{3}}}e^{i(K_{1}+K_{2}-P)\cdot X}\text{.}%
\end{align}

If we consider the quantised version of the 3-momenta of the particles
involved, then the usual box normalisation yields $\vec{p}=2\pi\vec{n}_{P}/L$
and $\vec{k}=2\pi\vec{n}_{K}/L$, i.e.,

$\Delta\vec{p}=2\pi\Delta\vec{n}_{P}/L$ and $\Delta\vec{k}=2\pi\Delta\vec
{n}_{K}/L$. Substituting $\Delta\vec{p}\rightarrow$ d$\vec{p}$\ and
$\Delta\vec{k}\rightarrow$ d$\vec{k}$\ yields $\int$d$^{3}\vec{p}=(2\pi
)^{3}/V$ = $\int$d$^{3}\vec{k}$ and thus%

\begin{align}
\left\langle f\right\vert \mathcal{S}^{(1)}\left\vert i\right\rangle  &
=i\left(  \frac{1}{V}\right)  ^{3/2}g\frac{1}{\sqrt{2E_{\vec{p}}}}\frac
{1}{\sqrt{2E_{\vec{k}_{1}}}}\frac{1}{\sqrt{2E_{\vec{k}_{2}}}}\frac{V^{3}%
}{(2\pi)^{9}}\int\text{d}^{4}X\frac{(2\pi)^{9}}{V^{3}}e^{i(K_{1}+K_{2}-P)\cdot
X}\nonumber\\
&  =i\left(  \frac{1}{V}\right)  ^{3/2}g\frac{1}{\sqrt{2E_{\vec{p}}}}\frac
{1}{\sqrt{2E_{\vec{k}_{1}}}}\frac{1}{\sqrt{2E_{\vec{k}_{2}}}}\int\text{d}%
^{4}Xe^{i(K_{1}+K_{2}-P)\cdot X}\nonumber\\
&  \equiv\frac{1}{V^{3/2}}\frac{ig}{\sqrt{2E_{\vec{k}_{1}}2E_{\vec{k}_{2}%
}2E_{\vec{p}}}}\int\text{d}^{4}Xe^{i(P-K_{1}-K_{2})X}\nonumber\\
&  =\frac{\sqrt{s_{f}}}{V^{3/2}}\frac{-i\mathcal{M}_{\chi\rightarrow2\varphi}%
}{\sqrt{2E_{\vec{k}_{1}}2E_{\vec{k}_{2}}2E_{\vec{p}}}}(2\pi)^{4}%
\delta^{\left(  4\right)  }(P-K_{1}-K_{2})
\end{align}

where in the last line we have substituted the coupling $g$ by the decay
amplitude $-i\mathcal{M}_{\chi\rightarrow2\varphi}$ (\ref{iMchiphiphi})
multiplied by a symmetry factor $\sqrt{s_{f}}$\ (in our case $s_{f}=2$ because
the decay $\chi\rightarrow2\varphi$ contains two identical particles). The
delta distribution $\delta^{\left(  4\right)  }(P-K_{1}-K_{2})$ corresponds to
energy-momentum conservation at each vertex.

The probability for the decay $\chi\rightarrow2\varphi$ corresponds to the
squared modulus of the scattering matrix:%

\begin{equation}
\left\vert \left\langle f\right\vert \mathcal{S}^{(1)}\left\vert
i\right\rangle \right\vert ^{2}=\frac{s_{f}}{V^{3}}\frac{1}{2E_{\vec{k}_{1}%
}2E_{\vec{k}_{2}}2E_{\vec{p}}}(2\pi)^{8}[\delta^{\left(  4\right)  }%
(P-K_{1}-K_{2})]^{2}\left\vert -i\mathcal{M}_{\chi\rightarrow2\varphi
}\right\vert ^{2}\text{.}%
\end{equation}

The square of the delta distribution can be calculated as follows:
\begin{align}
&  (2\pi)^{8}[(\delta^{\left(  4\right)  }(P-K_{1}-K_{2})]^{2}\nonumber\\
&  =(2\pi)^{4}\delta^{\left(  4\right)  }(P-K_{1}-K_{2})\int\text{d}%
^{4}Xe^{iX(P-K_{1}-K_{2})}\nonumber\\
&  =(2\pi)^{4}\delta^{\left(  4\right)  }(P-K_{1}-K_{2})\int\text{d}%
^{4}X\nonumber\\
&  =(2\pi)^{4}\delta^{\left(  4\right)  }(P-K_{1}-K_{2})\int\text{d}^{3}%
\vec{x}\int_{0}^{t}dt\nonumber\\
&  =(2\pi)^{4}\delta^{\left(  4\right)  }(P-K_{1}-K_{2})Vt\text{.} \label{ft}%
\end{align}

Integrating over $\vec{k}_{1,2}$ we obtain%

\begin{align}
&  \int\int\left\vert \left\langle f\right\vert \mathcal{S}^{(1)}\left\vert
i\right\rangle \right\vert ^{2}\frac{V}{(2\pi)^{3}}\text{d}^{3}\vec{k}%
_{1}\frac{V}{(2\pi)^{3}}\text{d}^{3}\vec{k}_{2}\nonumber\\
&  =(2\pi)^{4}\frac{s_{f}}{V^{2}}\int\int\frac{\left\vert -i\mathcal{M}%
_{\chi\rightarrow2\varphi}\right\vert ^{2}}{2E_{\vec{k}_{1}}2E_{\vec{k}_{2}%
}2E_{\vec{p}}}\delta^{\left(  4\right)  }(P-K_{1}-K_{2})\frac{V}{(2\pi)^{3}%
}\text{d}^{3}\vec{k}_{1}\frac{V}{(2\pi)^{3}}\text{d}^{3}\vec{k}_{2}t \nonumber \\
&  =\Gamma t
\end{align}

where we have defined the decay width for the process $\chi\rightarrow
2\varphi$ as%

\begin{equation}
\Gamma_{\chi\rightarrow2\varphi}=\frac{s_{f}}{(2\pi)^{2}}\int\int
\frac{\left\vert -i\mathcal{M}_{\chi\rightarrow2\varphi}\right\vert ^{2}%
}{2E_{\vec{k}_{1}}2E_{\vec{k}_{2}}2E_{\vec{p}}}\delta^{\left(  4\right)
}(P-K_{1}-K_{2})\text{d}^{3}\vec{k}_{1}\text{d}^{3}\vec{k}_{2}\text{.}
\label{zb1}%
\end{equation}

Consequently, the probability to find two particles at the time $t$ is
$P_{2\varphi}(t)=\Gamma_{\chi\rightarrow2\varphi}t$. Then the probability to
find the particle $\chi$ at the same time point is%

\begin{equation}
P_{\chi}(t)=1-\Gamma_{\chi\rightarrow2\varphi}t
\end{equation}

or, if $\Gamma_{\chi\rightarrow2\varphi}\ll t$
\begin{equation}
P_{\chi}(t)=e^{-\Gamma_{\chi\rightarrow2\varphi}t}\text{.}%
\end{equation}

Consequently, we define the median life-time of the particle $\chi$ as
\begin{equation}
\tau=\Gamma_{\chi\rightarrow2\varphi}^{-1}\text{.} \label{ld}%
\end{equation}

The latter expression is valid in the rest frame of the decaying particle; the life-time of the
particle in the laboratory frame reads
\begin{equation}
\tau^{\prime}=\gamma\tau\text{,}
\end{equation}

where $\gamma=(1-v^{2})^{-1/2}$. We also know that the 4-vectors of all the
particles involved have this form: $P=(m_{\chi},\vec{0})$, $K_{1}=(E_{\vec
{k}_{1}},\vec{k}_{1})$ and $K_{2}=(E_{\vec{k}_{2}},\vec{k}_{2})$. Given that
$\left\vert \vec{k}_{2}\right\vert $ = $\left\vert \vec{k}_{1}\right\vert $,
we obtain $E_{\vec{k}_{1}}$ = $E_{\vec{k}_{2}}$. Let us then rewrite
$\delta^{\left(  4\right)  }(P-K_{1}-K_{2})$ as follows:%
\begin{align}
\delta^{\left(  4\right)  }(P-K_{1}-K_{2})  &  =\delta^{\left(  3\right)
}(\vec{k}_{1}+\vec{k}_{2})\delta(m_{\chi}-E_{\vec{k}_{1}}-E_{\vec{k}_{2}%
})\nonumber\\
&  =\delta^{\left(  3\right)  }(\vec{k}_{1}+\vec{k}_{2})\delta(m_{\chi
}-2E_{\vec{k}_{1}})\text{.}%
\end{align}

Then integrating over $d^{3}\vec{k}_{2}$ in Eq.\ (\ref{zb1}) we obtain%

\begin{equation}
\Gamma=\frac{1}{2(2\pi)^{2}}\int\frac{\left\vert -i\mathcal{M}\right\vert
^{2}}{(2E_{\vec{k}_{1}})^{2}2m_{\chi}}\delta(m_{\chi}-2E_{\vec{k}_{1}%
})\text{d}^{3}\vec{k}_{1}\text{.} \label{zb2}%
\end{equation}

Energy-momentum conservation implies
\begin{equation}
\left\vert \vec{k}_{1}\right\vert =\sqrt{\frac{m_{\chi}^{2}}{4}-m_{\varphi
}^{2}}\equiv k_{f}%
\end{equation}

and therefore the $\delta$ distribution in Eq.\ (\ref{zb2}) can be expressed
in the following way using the generic identity $\delta(g(x))$ = $\sum
_{i}\delta(x-x_{i})/|g^{\prime}(x_{i})|$ where $g(x_{i})=0$:%

\begin{equation}
\delta(m_{\chi}-2E_{\vec{k}_{1}})=\frac{4m_{\chi}}{k_{f}}\delta\left(
\left\vert \vec{k}_{1}\right\vert -k_{f}\right)  \text{.} \label{zb4}%
\end{equation}

Let us then perform the integral in Eq.\ (\ref{zb2}) using the spherical
coordinates: $d^{3}\vec{k}_{1}\equiv\vec{k}_{1}^{2}$d$\left\vert \vec{k}%
_{1}\right\vert $d$\Omega$. Integrating over d$\left\vert \vec{k}%
_{1}\right\vert $ yields%

\begin{equation}
\Gamma=\frac{k_{f}}{32\pi^{2}m_{\chi}^{2}}\int d\Omega\left\vert
-i\mathcal{M}\right\vert ^{2} \label{zb5}%
\end{equation}

and, if the decay amplitude does not depend on $\Omega$%

\begin{equation}
\Gamma=\frac{k_{f}}{8\pi m_{\chi}^{2}}\left\vert -i\mathcal{M}\right\vert
^{2}\text{.} \label{zb0}%
\end{equation}

We have to ensure that there is no double counting in case of a decay into
identical particles. This is performed by introducing a symmetry factor
$s_{f}$ into the formula for the decay width:%

\begin{equation}
\Gamma=s_{f}\frac{k_{f}}{8\pi m_{\chi}^{2}}\left\vert -i\mathcal{M}\right\vert
^{2}\text{.} \label{zb}%
\end{equation}

In the decay discussed in this section, the assumption was made that the
particle $\chi$ decays into two identical particles $\varphi$. Consequently
$s_{f}=1/2$ and from Eq.\ (\ref{zb}) we obtain:%

\begin{equation}
\Gamma_{\chi\rightarrow2\varphi}=\frac{k_{f}}{4\pi m_{\chi}^{2}}g^{2}.
\label{zb3}%
\end{equation}

\subsection{Parametrising the Scattering Amplitude}

Two scattering particles will in general
form an intermediate state before the scattering products subsequently arise
(see, e.g., Sec.\ \ref{sec.SLQ} where $\pi\pi$ scattering entails
contributions of the form $\pi\pi\rightarrow\sigma_{N}\rightarrow\pi\pi$ and
$\pi\pi\rightarrow\rho\rightarrow\pi\pi$, i.e., with intermediate scalar and
vector particles, respectively). A scattering amplitude $\mathcal{M}$ of the
incoming two particles (or, equivalently, the decay amplitude of the
intermediate particle into the incoming two particles) is in general a
complex-valued quantity; we expect it to depend on the centre-of-mass momentum
$p$. Let us then decompose $\mathcal{M}$ in terms of $p$ and a quantity of
dimension [$E^{-1}$] that we will refer to as scattering length $a$ as follows
\cite{Frazer}:%

\begin{equation}
\mathcal{M}=\mathcal{M}(p)=\frac{N}{-a^{-1}-ip}\text{,} \label{Ma}%
\end{equation}

with $N$ a constant of dimension [$E^{2}$] that assures [$\mathcal{M}$]
$=[E]$; $N$ could in principle also be a function (see for instance
Ref.\ \cite{buggf0}\ and references therein for explicit ways how to
parametrise a scattering amplitude) but the exact nature of $N$ is not
important for the statements in this section. It is obvious that the limit
where $p=0$ leads to $\mathcal{M}(p)\sim-a$. If we restrict ourselves to the
behaviour of $\mathcal{M}$ close to threshold, then we observe in
Eq.\ (\ref{Ma}) that the function is analytic in this energy region (i.e.,
$p\sim0$) except for a pole at $p=ia^{-1}\equiv ip_{0}$ (i.e., $p_{0}\equiv
a^{-1}$). The value of the scattering length can be determined from the
scattering amplitude, with the latter being a measurable quantity (for
examples regarding the $\pi\pi$ scattering see also
Refs.\ \cite{Leutwyler,Pelaez1,Peyaud}). The scattering length can have both
positive or negative signs. If $a<0$, then the pole is found in the lower half
of the complex $p$ plane ($\operatorname{Im}p<0$); the pole corresponds to a
"virtual state" \cite{Frazer} or, in the language of the hadronic
physics, to a resonance. The lower half-plane is usually denoted as the
second (unphysical) sheet. Conversely, $a>0$ implies the existence of a pole
in the upper half-plane (denoted as the first, or physical, sheet); it
corresponds to a bound state. It is known, for example, that the
proton-neutron scattering produces such a bound state (deuteron) in the
$^{3}S$ channel. Conversely, $a<0$ is also possible in the $pn$ scattering,
leading to a $^{1}S$ virtual state.

The $p$ plane can also be mapped onto the complex $s$ plane, where $s$ denotes
the Mandelstam variable $s=4(m^{2}+p^{2})$ with $m$ being the mass of the
incoming particles (taken for simplicity to be identical). Thus the condition
$s\geq4m^{2}$ holds in any given experimental environment, notwithstanding
whether it explores a virtual or a bound state. However, the mentioned two
types of states do not behave in the same way below threshold. Although not
accessible to experiments, the region below threshold nonetheless can be
explored mathematically by means of analytic continuation $p\rightarrow
i\tilde{p}$ (with $\tilde{p}$ a positive, real number) yielding from
Eq.\ (\ref{Ma})%

\begin{equation}
\mathcal{M}\sim\frac{1}{\tilde{p}-p_{0}}\text{.}\label{Ma1}%
\end{equation}

\begin{figure}[h]
  \begin{center}
    \begin{tabular}{cc}
      \resizebox{76mm}{!}{\includegraphics{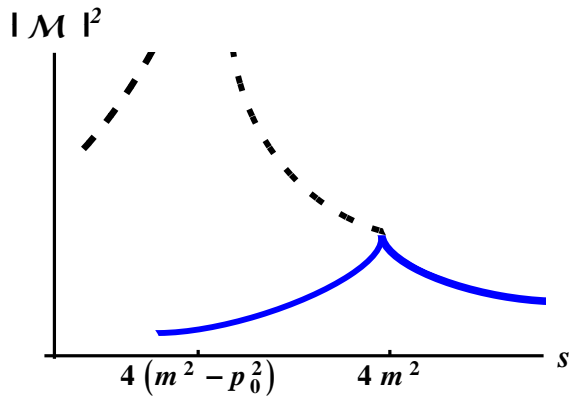}}  
    \end{tabular}
    \caption{Behaviour of $\vert\mathcal{M} \vert^{2}$ for a bound state (dashed
line) and a virtual state (full line).}
    \label{Maf}
  \end{center}
\end{figure}

Then there are two possibilities (see Fig.\ \ref{Maf}): (i) if $p_{0}>0$,
i.e., $a>0$ (bound state), then $|\mathcal{M}|^{2}$ exhibits a pole at
$\tilde{p}=p_{0}$, i.e., $s_0=4(m^{2}-p_{0}^{2})$; (ii) if $p_{0}<0$, i.e.,
$a<0$ (virtual state), then $|\mathcal{M}|^{2}$ exhibits a cusp at the point
$\tilde{p}=0$, i.e., $s=4m^{2}$. Thus the behaviour of the two types of states
is fundamentally different below threshold; the dependence of $|\mathcal{M}%
|^{2}$ on $s$ is a strong indicator whether particle scattering has yielded a
bound state or a virtual-type state.

Let us explore the behaviour of the decay amplitude on a concrete
example. As demonstrated in Eq.\ (\ref{zb}), the square of the decay amplitude
is necessary for the decay width of an unstable state to be calculated.
Similarly to the Lagrangian (\ref{chiphiphi}), let us consider a decay of a
resonance $S\rightarrow2\varphi$ \cite{Francesco2011}:%

\begin{equation}%
\mathcal{L}%
_{S\varphi\varphi}=\frac{1}{2}(\partial_{\mu}S)^{2}-\frac{1}{2}m_{S}^{2}%
S^{2}+\frac{1}{2}(\partial_{\mu}\varphi)^{2}-\frac{1}{2}m_{\varphi}^{2}%
\varphi^{2}+gS\varphi^{2}\text{.}\label{Sphiphi}%
\end{equation}

Let us also introduce the tree-level decay width for the process
$S\rightarrow2\varphi_{1}$ as%

\begin{equation}
\Gamma_{S,0}(x_{S},m_{\varphi},g)=\frac{\sqrt{\frac{x_{S}^{2}}{4}-m_{\varphi
}^{2}}}{8\pi x_{S}^{2}}(\sqrt{2}g)^{2}\theta(x_{S}-2m_{\varphi}%
)\text{,} \label{GSphiphi1}
\end{equation}

where $x_{S}$ denotes the running mass of the state $S$ and the $\theta
$-function implements the tree-level condition that $S$ is above the
$2\varphi$ threshold. The decay width of Eq.\ (\ref{GSphiphi1}) can in
principle be calculated by setting $x_{S}=m_{S}$; however, quantum
fluctuations are known to modify the value of $m_{S}$ (see below) and
consequently we evaluate $\Gamma_{S,0}(x_{S},m_{\varphi},g)$ at the physical
value of the $S$ mass (let us denote it as $m$) rather than at the value of
the Lagrangian parameter $m_{S}$. The total decay width of the state $S$ is then%

\begin{equation}
\Gamma_{S}(m)\equiv\Gamma_{S,0}(m,m_{\varphi},g)\text{.}\label{GSphiphi2}%
\end{equation}

Note that the quantity $\tau_{BW}\equiv1/\Gamma_{S}(m)$ represents the
so-called Breit-Wigner mean life-time of the particle $S$.

\begin{figure}[h]
\begin{center}
\includegraphics[
height=2.0582in,
width=2.9666in
]{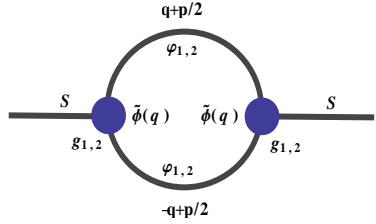}
\end{center}
\caption{Self-energy diagram for the decay process $S \rightarrow\phi_{1,2}
\phi_{1,2}$; $\tilde{\phi} $ is the vertex function, required for
regularisation of the self-energy diagram.}%
\label{Sphiphif}%
\end{figure}

In general, a calculation of the decay width may be performed at tree level
(see Sections \ref{sec.AVP} -- \ref{sec.SVV}). However, the optical theorem
allows us to relate the tree-level decay width with the self-energy of the
decaying particle $S$. To this end, we first have to evaluate the propagator
$G_{S}(p^{2})$ of the state $S$ (with $p$ the centre-of-mass momentum)%

\begin{equation}
G_{S}(p^{2})=\frac{1}{p^{2}-m_{S}^{2}+(\sqrt{2}g)^{2}\Sigma(p^{2},m_{\varphi
}^{2})+i\epsilon}\label{GS}%
\end{equation}

by integrating over the loop diagram presented in Fig.\ \ref{Sphiphif} to
obtain the self-energy%

\begin{equation}
\Sigma(p^{2},m_{\varphi}^{2})=-i\int\frac{\text{d}^{4}q}{(2\pi)^{4}}\frac
{1}{\left[  \left(  \frac{p}{2}+q\right)  ^{2}-m_{\varphi}^{2}+i\varepsilon
\right]  \left[  \left(  \frac{p}{2}-q\right)  ^{2}-m_{\varphi}^{2}%
+i\varepsilon\right]  }\text{.}\label{GS2}%
\end{equation}

The integral stated in Eq.\ (\ref{GS2}) has to be regularised because its real
part is divergent (the imaginary part is convergent). The regularisation is
performed by introducing a function $\tilde{\phi}(q)$ at every vertex in the
loop diagram of Eq.\ (\ref{GS}), see Fig.\ \ref{Sphiphif}. Then we regularise
$\Sigma(p^{2},m_{\varphi}^{2})$ in the following way:%

\begin{equation}
\Sigma(p^{2},m_{\varphi}^{2})=-i\int\frac{\text{d}^{4}q}{(2\pi)^{4}}%
\frac{\left[  \tilde{\phi}(q)\right]  ^{2}}{\left[  \left(  \frac{p}%
{2}+q\right)  ^{2}-m_{\varphi}^{2}+i\varepsilon\right]  \left[  \left(
\frac{p}{2}-q\right)  ^{2}-m_{\varphi}^{2}+i\varepsilon\right]  }%
\text{.}\label{GS3}%
\end{equation}

The function $\tilde{\phi}(q)$ is also referred to as the vertex function. It
does not have a unique form; it can be defined for example as $\tilde{\phi
}(q)=\theta(\Lambda^{2}-q^{2})$ with the cutoff $\Lambda$ [$q$ represents the
off-shell momentum of the state $\varphi$], $\tilde{\phi}(x_{S})=\theta
(\sqrt{\Lambda^{2}+m_{\varphi}^{2}}-x_{S}/2)$ \cite{arXiv:1005.4817},
$\tilde{\phi}(q)=1/[1+(q/\Lambda)^{2}]$ as in Ref.\ \cite{arXiv:1108.2782} or
$\tilde{\phi}(q)=1/(1+\vec{q}^{2}/\Lambda^{2})$ \cite{Giacosa:2007bn}.
Different choices of $\tilde{\phi}(q)$ do not change qualitative statements of
the given model \cite{Francesco2011}.

We can simplify Eq.\ (\ref{GS}) by introducing the loop function%

\begin{equation}
\Pi(p^{2})=(\sqrt{2}g)^{2}\Sigma(p^{2},m_{\varphi}^{2})\label{GS4}%
\end{equation}

obtaining%

\begin{equation}
G_{S}(p^{2})=\frac{1}{p^{2}-m_{S}^{2}+\Pi(p^{2})+i\epsilon}\text{.}
\label{GS1}%
\end{equation}

The optical theorem relates the tree-level decay width in
Eq.\ (\ref{GSphiphi2}), evaluated at any value of the running mass $x_{S}$,
with the imaginary part of the loop function $\Pi(p^{2}=x_{S}^{2})$:%

\begin{equation}
\operatorname{Im}\Pi(x_{S}^{2})=x_{S}\Gamma_{S}(x_{S})\left[  \tilde{\phi
}(q=(0,\vec{q}))\right]  ^{2} \label{OT}%
\end{equation}

with the energy-momentum conservation at vertex yielding $\vec{q}^{2}%
=x_{S}^{2}/4-m_{\varphi}^{2}$. Spatial isotropy implies that $\tilde{\phi
}(q=(0,\vec{q}))$ has to be a function of $\vec{q}^{2}$. As a consequence,

\begin{equation}
\Gamma_{S}(m)\rightarrow\Gamma_{S}(m)\left[  \tilde{\phi}(q=(0,\vec
{q}))\right]  ^{2}\text{,} \label{GSphiphi3}%
\end{equation}

i.e., for the decay widths in Eq.\ (\ref{GSphiphi1})%

\begin{equation}
\Gamma_{S,0}(x_{S},m_{\varphi},g)\rightarrow\Gamma_{S,0}(x_{S},m_{\varphi
},g)\left[  \tilde{\phi}(q=(0,\vec{q}))\right]  ^{2}\text{.}\label{GSphiphi4}%
\end{equation}

We define the {\it spectral function} $d_{S}(x_{S}$) of the resonance $S$ as the
imaginary part of the propagator (\ref{GS1})%

\begin{equation}
d_{S}(x_{S})=\frac{2x_{S}}{\pi}\left\vert \lim_{\varepsilon\rightarrow
0}\operatorname{Im}G_{S}(x_{S}^{2})\right\vert \text{,} \label{SF}%
\end{equation}
i.e.,%

\begin{equation}
d_{S}(x_{S})=\frac{2x_{S}}{\pi}\left\vert \lim_{\varepsilon\rightarrow0}%
\frac{\operatorname{Im}\Pi(x_{S}^{2})+\varepsilon}{\left[  x_{S}^{2}-m_{S}%
^{2}+\operatorname{Re}\Pi(x_{S}^{2})\right]  ^{2}+\left[  \operatorname{Im}%
\Pi(x_{S}^{2})+\varepsilon\right]  ^{2}}\right\vert \text{.} \label{SF1}%
\end{equation}

The differential value $d_{S}(x_{S})$d$x_{S}$ is interpreted as the
probability that the resonance $S$ will have a mass between $x_{S}$ and
$x_{S}+$ d$x_{S}$. For this reason, the spectral function $d_{S}(x_{S})$ has
to be normalised properly:%

\begin{equation}
\int\limits_{0}^{\infty}\text{d}x_{S}d_{S}(x_{S})\overset{!}{=}1\text{.}
\label{SF2}%
\end{equation}

The renormalised (physical) mass $m$ of the resonance $S$ is usually defined
as the zero of the real part of the resonance propagator, i.e., from the
implicit equation%

\begin{equation}
m^{2}-m_{S}^{2}+\operatorname{Re}\Pi(m^{2})=0\text{.} \label{mass}%
\end{equation}

It is common that $\operatorname{Re}\Pi(m^{2})>0$ -- in other words, quantum
fluctuations usually decrease the value of the model mass $m_{S}$ to the
physical mass value $m$. Note, however, that Eq.\ (\ref{mass}) is not the only
way to define the regularised mass: the mass can also be defined as the
position of the minimum of $\operatorname{Im}G_{S}(x_{S}^{2})$ but this
definition leads to qualtiatively the same results (see Ref.\ \cite{UBW} for
an explicit example of the $a_{1}$ mass calculation).\newline

There is an important approximation of Eq.\ (\ref{SF1}). The approximation is
obtained by neglecting the real part of the loop function (justified for
resonances that are not too broad): $\operatorname{Re}\Pi(m^{2})=0$, using
Eq.\ (\ref{OT}) without the vertex function $\tilde{\phi}$ [that is no longer
necessary because the divergent, real part of $\Pi(m^{2})$ is set to zero and
$\operatorname{Im}\Pi(m^{2})$ is convergent] and setting $\varepsilon
\rightarrow0$:
\begin{align}
d_{S}(x_{S})=\bar{N}_{S}\frac{x_{S}^{2}\Gamma_{S}(x_{S})}{(x_{S}^{2}%
-m^{2})^{2}+\left[  x_{S}\Gamma_{S}(x_{S})\right]  ^{2}}\theta(x_{S}%
-2m_{\varphi_{1,2}})\text{,} \label{BW}%
\end{align}

where the constant $2/\pi$ from Eq.\ (\ref{SF1}) has been absorbed into
$\bar{N}_{S}$, the normalisation constant obtained from the condition
(\ref{SF2}). Note that, conversely, the $\theta$ function in Eq.\ (\ref{BW}) is no longer
absorbed into $\Gamma_{S}(x_{S})$ as was the case in Eq.\ (\ref{GSphiphi1}).
Eq.\ (\ref{BW}) can be simplified further by approximating $\Gamma_{S}(x_{S})$\ with the
experimental value $\Gamma_{S}^{\exp}$, i.e., by neglecting the functional
dependence of $\Gamma_{S}(x_{S})$ on $x_{S}$:%

\begin{align}
d_{S}(x_{S})=N_{S}\frac{x_{S}^{2}\Gamma_{S}^{\exp}}{(x_{S}^{2}-m^{2}%
)^{2}+\left (  x_{S}\Gamma_{S}^{\exp}\right )  ^{2}}\theta(x_{S}-2m_{\varphi
}) \text{.} \label{BW1}
\end{align}

Equation (\ref{BW1}) is known as the relativistic Breit-Wigner limit
of the spectral function (or simply the relativistic Breit-Wigner spectral
function). The relativistic Breit-Wigner spectral function will be
used throughout this work to calculate decays via off-shell particles (see,
e.g., Sections \ref{sec.AVP} and \ref{sec.SVV}).\newline

Let us finally note that, in the case of our resonance $S$, the scattering
amplitude discussed in Eq.\ (\ref{Ma}) can also be parametrised in terms of
the Mandelstam variable $s=p^{2}$ with a pole at $s_{0}=(m-i\Gamma_{S}/2)^{2}%
$. The parametrisation can be motivated in the following way: let our starting
point be the propagator (\ref{GS1}) where the loop function $\Pi$ is evaluated
at the physical mass value $m$:%

\begin{align}
\frac{1}{p^{2}-m_{S}^{2}+\Pi(m^{2})}  &  =\frac{1}{p^{2}-m_{S}^{2}%
+\operatorname{Re}\Pi(m^{2})+i\operatorname{Im}\Pi(m^{2})}\nonumber\\
&  \overset{\text{Eq.\ (\ref{OT})}}{=}\frac{1}{p^{2}-m_{S}^{2}%
+\operatorname{Re}\Pi(m^{2})+im\Gamma_{S}(m)}\overset{\text{Eq.\ (\ref{mass}%
)}}{=}\frac{1}{p^{2}-m^{2}+im\Gamma_{S}(m)} \text{.}%
\end{align}

Let us assume that our resonance $S$\ fulfills the condition $\Gamma_{S}%
^{2}(m)\ll m^{2}$, i.e., $\Gamma_{S}^{2}(m)/m^{2}\ll1$. Then we can add the
term $\Gamma_{S}^{2}(m)/4$ to the denominator:%

\begin{align}
\frac{1}{p^{2}-m^{2}+im\Gamma_{S}(m)}  &  \simeq\frac{1}{p^{2}-m^{2}%
+im\Gamma_{S}(m)+\frac{\Gamma_{S}^{2}(m)}{4}}\\
&  =\frac{1}{p^{2}-\left[  m-\frac{i}{2}\Gamma_{S}(m)\right]  ^{2}}\text{.}%
\end{align}

Substituting $p^{2}=s$ and $s_{0}=(m-i\Gamma_{S}/2)^{2}$, we obtain the
following expression%

\begin{equation}
\frac{1}{p^{2}-\left[  m-\frac{i}{2}\Gamma_{S}(m)\right]  ^{2}}=\frac
{1}{s-s_{0}}\text{.}%
\end{equation}

Consequently, \textit{if} an experimentally determined scattering amplitude
can be parametrised as%

\begin{equation}
\mathcal{M}\sim\frac{1}{s-s_{0}}\text{,}
\end{equation}

then it evidently contains a pole at $s=s_{0}$ describing a resonance with
mass $m$ and decay width $\Gamma_{S}(m)$. This is a well-known criterion that
allows one to ascertain whether scattering data entail a resonance signal.

Applications of the discussion in this section are presented in the following,
where some exemplary tree-level decay widths are calculated.


\subsection{Example: Decaying Axial-Vector State I} \label{sec.AVP}

Let us consider a decay process of the form $A\rightarrow V{\tilde{P}}$ where
$A$, $V$ and ${\tilde{P}}$ denote an axial-vector, a vector and a pseudoscalar
state, respectively, with the following interaction Lagrangian describing the
decay of the axial-vector state into neutral modes:

\begin{align}
\mathcal{L}_{AV{\tilde{P}}} &  =A_{AV{\tilde{P}}}A^{\mu0}V_{\mu}^{0}{\tilde
{P}}^{0}\nonumber\\
&  +B_{AV{\tilde{P}}}\left[  A^{\mu0}\left(  \partial_{\nu}V_{\mu}
^{0}-\partial_{\mu}V_{\nu}^{0}\right)  \partial^{\nu}{\tilde{P}}^{0}
+\partial^{\nu}A^{\mu0}\left(  V_{\nu}^{0}\partial_{\mu}{\tilde{P}}^{0}
-V_{\mu}^{0}\partial_{\nu}{\tilde{P}}^{0}\right)  \right] \text{.}\label{AVP}
\end{align}

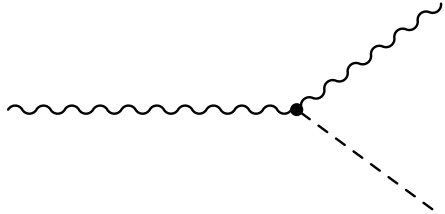
\begin{figure}[h]
\begin{align*}
 \qquad \qquad \qquad  \qquad \qquad \qquad  \quad \; \parbox{180mm}{
  \begin{fmfgraph*}(180,80)
    \fmfleftn{i}{1}\fmfrightn{o}{2}
    \fmfv{label=A,label.dist=45,label.angle=-17}{i1}\fmfv{label=V,label.dist=28,label.angle=-114}{o2}
      \fmf{boson,tension=1,label=\text{\small\(\varepsilon^{(\alpha)}_\mu (P)\)},label.dist=-20}{i1,v1}
    \fmf{boson,label=\text{\small\(\varepsilon^{(\beta)}_\nu (P_1)\)},tension=1,label.dist=-35}{v1,o2}
        \fmf{dashes,tension=1,label=\text{\small\(\tilde{P}(P_2)\)},label.dist=-32}{v1,o1}
        \fmfdot{v1}
      \end{fmfgraph*}} \end{align*}\caption{Decay process $A\rightarrow V{\tilde{P}}$.}\end{figure}

We will consider the possible decay of the axial-vector state into charged
modes at the end of this section. For now, let us consider a generic decay
process of the form $A\rightarrow$ $V^{0}{P}^{0}$.

Let us then denote the momenta of $A$, $V$ and ${\tilde{P}}$\ as $P$, $P_{1}$
and $P_{2}$, respectively. The stated decay process involves two vector
states: $A$\ and$\ {V}$. We therefore have to consider the corresponding
polarisation vectors; let us denote them as $\varepsilon_{\mu}^{(\alpha)}(P)$
for $A$\ and $\varepsilon_{\nu}^{(\beta)}(P_{1})$ for $V$. Then, upon
substitutions $\partial^{\mu}\rightarrow-iP^{\mu}$\ for the decaying particle
and $\partial^{\mu}\rightarrow iP_{1,2}^{\mu}$ for the decay products, we
obtain the following Lorentz-invariant $AV{\tilde{P}}$ scattering amplitude
$-i\mathcal{M}_{A\rightarrow V^{0}{\tilde{P}}^{0}}^{(\alpha,\beta)}$:

\begin{align}
-i\mathcal{M}_{A\rightarrow V^{0}{\tilde{P}}^{0}}^{(\alpha,\beta)} &
=\varepsilon_{\mu}^{(\alpha)}(P)\varepsilon_{\nu}^{(\beta)}(P_{1}
)h_{AV{\tilde{P}}}^{\mu\nu}=i\varepsilon_{\mu}^{(\alpha)}(P)\varepsilon_{\nu
}^{(\beta)}(P_{1})\nonumber\\
&  \times\{A_{AV{\tilde{P}}}g^{\mu\nu}+B_{AV{\tilde{P}}}\left[  P_{1}^{\mu
}P_{2}^{\nu}+P_{2}^{\mu}P^{\nu}-(P_{1}\cdot P_{2})g^{\mu\nu}-(P\cdot
P_{2})g^{\mu\nu}\right]  \}\label{iMAVP}
\end{align}

with

\begin{align}
h_{AV{\tilde{P}}}^{\mu\nu}=i\left\{  A_{AV{\tilde{P}}}g^{\mu\nu}
+B_{AV{\tilde{P}}}[P_{1}^{\mu}P_{2}^{\nu}+P_{2}^{\mu}P^{\nu}-(P_{1}\cdot
P_{2})g^{\mu\nu}-(P\cdot P_{2})g^{\mu\nu}]\right\}\text{,} \label{hAVP}
\end{align}
$\,$\\
where $h_{AV{\tilde{P}}}^{\mu\nu}$ denotes the $AV{\tilde{P}}$ vertex.\newline

It will be necessary to determine the square of the scattering amplitude in
order to calculate the decay width. We note that the scattering amplitude in
Eq.\ (\ref{iMAVP}) depends on the polarisation vectors $\varepsilon_{\mu
}^{(\alpha)}(P)$ and $\varepsilon_{\nu}^{(\beta)}(P_{1})$; therefore, it is
necessary to calculate the average of the squared amplitude for all
polarisation values. Let us denote the masses of the vectors states $A$ and
$V$\ as $m_{A}$ and $m_{V}$, respectively. Then the averaged squared amplitude
$|-i\mathcal{\bar{M}}|^{2}$ is determined as follows:

\begin{align}
-i\mathcal{M}_{A\rightarrow V^{0}{\tilde{P}}^{0}}^{(\alpha,\beta)}  &
=\varepsilon_{\mu}^{(\alpha)}(P)\varepsilon_{\nu}^{(\beta)}(P_{1}%
)h_{AV{\tilde{P}}}^{\mu\nu}\Rightarrow\left\vert -i\mathcal{\bar{M}%
}_{A\rightarrow V^{0}{\tilde{P}}^{0}}\right\vert ^{2}=\frac{1}{3}%
\sum\limits_{\alpha,\beta=1}^{3}\left\vert -i\mathcal{M}_{A\rightarrow
V^{0}{\tilde{P}}^{0}}^{(\alpha,\beta)}\right\vert ^{2}\nonumber\\
&  =\frac{1}{3}\sum\limits_{\alpha,\beta=1}^{3}\varepsilon_{\mu}^{(\alpha
)}(P)\varepsilon_{\nu}^{(\beta)}(P_{1})h_{AV{P}}^{\mu\nu}\varepsilon_{\kappa
}^{(\alpha)}(P)\varepsilon_{\lambda}^{(\beta)}(P_{1})h_{AV{\tilde{P}}}%
^{\ast\kappa\lambda} \text{.} \label{iMK11}%
\end{align}

Given that

\begin{equation}
\sum\limits_{\alpha=1}^{3}\varepsilon_{\mu}^{(\alpha)}(P)\varepsilon_{\kappa
}^{(\alpha)}(P)=\left(  -g_{\mu\kappa}+\frac{P_{\mu}P_{\kappa}}{m_{A}^{2}
}\right) \label{epsilon}
\end{equation}

[an analogous equation holds for $\varepsilon^{(\beta)}$], we then obtain from
Eq.\ (\ref{iMK11}):

\begin{align}
|-i\mathcal{\bar{M}}_{A\rightarrow V^{0}{\tilde{P}}^{0}}|^{2}  &  =\frac{1}%
{3}\left(  -g_{\mu\kappa}+\frac{P_{\mu}P_{\kappa}}{m_{A}^{2}}\right)  \left(
-g_{\nu\lambda}+\frac{P_{1\nu}P_{1\lambda}}{m_{V}^{2}}\right)  h_{AV{\tilde
{P}}}^{\mu\nu}h_{AV{\tilde{P}}}^{\ast\kappa\lambda}\nonumber\\
&  =\frac{1}{3}\left[  h_{\mu\nu AV{\tilde{P}}}h_{AV{\tilde{P}}}^{\ast\mu\nu
}-\frac{\left(  h_{AV{\tilde{P}}}^{\mu\nu}P_{\mu}\right)  \left(  h_{\nu
AV{\tilde{P}}}^{\ast\kappa}P_{\kappa}\right)  }{m_{A}^{2}}-\frac{\left(
h_{AV{\tilde{P}}}^{\mu\nu}P_{1\nu}\right)  \left(  h_{\mu AV{\tilde{P}}}%
^{\ast\lambda}P_{\lambda}\right)  }{m_{V}^{2}}\right. \nonumber\\
&  \left.  +\frac{\left(  h_{AV{\tilde{P}}}^{\mu\nu}P_{\mu}P_{1\nu}\right)
\left(  h_{AV{\tilde{P}}}^{\ast\mu\nu}P_{\mu}P_{1\nu}\right)  }{m_{V}^{2}%
m_{A}^{2}}\right] \nonumber\\
&  =\frac{1}{3}\left[  \left\vert h_{AV{\tilde{P}}}^{\mu\nu}\right\vert
^{2}-\frac{\left\vert h_{AV{\tilde{P}}}^{\mu\nu}P_{\mu}\right\vert ^{2}}%
{m_{A}^{2}}-\frac{\left\vert h_{AV{\tilde{P}}}^{\mu\nu}P_{1\nu}\right\vert
^{2}}{m_{V}^{2}}+\frac{\left\vert h_{AV{\tilde{P}}}^{\mu\nu}P_{\mu}P_{1\nu
}\right\vert ^{2}}{m_{V}^{2}m_{A}^{2}}\right]\text{.} \label{iMAVP2}%
\end{align}

Equation (\ref{iMAVP2}) contains the metric tensor $g_{\mu\nu}=\mathrm{diag}%
(1,-1-1,-1)$. The decay width for the process $A\rightarrow V^{0}{\tilde{P}%
}^{0}$ then reads

\begin{equation}
\Gamma_{A\rightarrow V^{0}{\tilde{P}}^{0}}=\frac{k(m_{A},m_{V},m_{{\tilde{P}}
})}{8\pi m_{A}^{2}}|-i\mathcal{\bar{M}}_{A\rightarrow V^{0}{\tilde{P}}^{0}
}|^{2}\text{.} \label{GAV0P0}
\end{equation}

A non-singlet axial-vector field will in general also posses charged
decay channels. Therefore, in addition to the decay process considered in
Eq.\ (\ref{GAV0P0}),\ we have to consider the contribution of the charged
modes from the process $A\rightarrow V^{\pm}{\tilde{P}}^{\mp}$ to the full
decay width as well. To this end, we multiply the neutral-mode decay width of
Eq.\ (\ref{GAV0P0}) with an isospin factor $I$, i.e., we set $\Gamma
_{A\rightarrow V{\tilde{P}}}=\Gamma_{A\rightarrow V^{0}{\tilde{P}}^{0}}%
+\Gamma_{A\rightarrow V^{\pm}{\tilde{P}}^{\mp}}\equiv I\Gamma_{A\rightarrow
V^{0}{\tilde{P}}^{0}}$, and obtain the following equation for the full decay
width:
\begin{equation}
\Gamma_{A\rightarrow V{\tilde{P}}}=I\frac{k(m_{A},m_{V},m_{{\tilde{P}}})}{8\pi
m_{A}^{2}}|-i\mathcal{\bar{M}}_{A\rightarrow V^{0}{\tilde{P}}^{0}}
|^{2}\text{.}\label{GAVP}
\end{equation}

The exact value of $I$ can be determined from isospin deliberations, or simply
from the interaction Lagrangian of a given decay process (as we will see in
later in this work).

Note that an off-shell vector state can also be considered within our
formalism upon introducing the corresponding spectral function as in
Eq.\ (\ref{BW1})

\begin{equation}
d_{V}(x_{V})=N_{V}\,\frac{x_{V}^{2}\Gamma_{V}^{\exp}}{(x_{V}^{2}-m_{V}%
^{2})^{2}+\left(  x_{V}\Gamma_{V}^{\exp}\right)  ^{2}}\,\theta(m_{A}%
-m_{V}-m_{{\tilde{P}}}) \label{dV}%
\end{equation}

with $x_{V}$ and $\Gamma_{V}^{\exp}$ denoting the off-shell mass and the
tree-level width of the vector state $V$, respectively, and $N_{V}$ determined
such that $\int_{0}^{\infty}\mathrm{d}x_{V}\,d_{V}(x_{V})=1$. This
allows us to calculate the decay width for a sequential decay of the form
$A\rightarrow V{\tilde{P}}\rightarrow{\tilde{P}}_{1}{\tilde{P}}_{2}{\tilde{P}%
}$, i.e., to consider an off-shell decay of the vector particle that possesses
this assumed form: $V\rightarrow{\tilde{P}}_{1}{\tilde{P}}_{2}$. Then
Eq.\ (\ref{GAVP}) is modified as

\begin{equation}
\Gamma_{A\rightarrow V{\tilde{P}}\rightarrow{\tilde{P}}_{1}{\tilde{P}}%
_{2}{\tilde{P}}}=%
{\displaystyle\int\limits_{m_{{\tilde{P}}_{1}}+m_{{\tilde{P}}_{2}}}%
^{m_{A}-m_{{\tilde{P}}}}}
\text{d}x_{V}\, I\frac{k(m_{A},x_{V},m_{{\tilde{P}}})}{8\pi m_{A}^{2}%
} \, d_{V}(x_{V})|-i\mathcal{\bar{M}}_{A\rightarrow V^{0}{\tilde{P}}^{0}}%
|^{2} \text{.} \label{GAVP1}
\end{equation}

For future use we have introduced the momentum function
\begin{equation}
k(m_{a},m_{b},m_{c})=\frac{1}{2m_{a}}\sqrt{m_{a}^{4}-2m_{a}^{2}\,(m_{b}%
^{2}+m_{c}^{2})+(m_{b}^{2}-m_{c}^{2})^{2}}\theta(m_{a}-m_{b}-m_{c})\text{.}
\label{kabc}%
\end{equation}
In the decay process $a\rightarrow b+c$, with masses $m_{a},\,m_{b},\,m_{c}$,
respectively, the quantity $k(m_{a},m_{b},m_{c})$ represents the modulus of
the three-momentum of the outgoing particles $b$ and $c$ in the rest frame of
the decaying particle $a$. The theta function ensures that the decay width
vanishes below threshold.

Note that Eqs.\ (\ref{GAVP}) and (\ref{GAVP1}) will be very useful, e.g., in
Sections \ref{sec.a1rhopion1}, \ref{sec.f1N1}, \ref{sec.f1S1} and
\ref{sec.K11}.

\subsection{Example: Decaying Axial-Vector State II} \label{sec.ASP}

Let us now consider a slightly different example (that will nonetheless also
prove to be useful in the subsequent chapters of this work): a generic decay
of the form $A\rightarrow S{\tilde{P}}$ where $A$, $S$ and ${\tilde{P}}$
denote an axial-vector, a scalar and a pseudoscalar state, respectively, with
the following interaction Lagrangian describing the decay of the neutral
axial-vector component into neutral states:

\begin{equation}
\mathcal{L}_{AS{\tilde{P}}}\equiv A_{AS{\tilde{P}}}A^{\mu0}S^{0}\partial_{\mu
}{\tilde{P}}^{0}+B_{AS{\tilde{P}}}A^{\mu0}{\tilde{P}}^{0}\partial_{\mu}%
S^{0}\text{.}\label{ASP}%
\end{equation}

\begin{figure}[h]
\begin{align*}
\qquad \qquad \qquad  \qquad \qquad \qquad  \quad \; \parbox{180mm}{
  \begin{fmfgraph*}(180,80)
    \fmfleftn{i}{1}\fmfrightn{o}{2}
   \fmfv{label=A,label.dist=45,label.angle=-17}{i1}
      \fmf{boson,tension=1,label=\text{\small\(\varepsilon^{(\alpha)}_\mu (P)\)},label.dist=-20}{i1,v1}
    \fmf{dashes,label=\text{\small\(\tilde{P}(P_1)\)},tension=1,label.dist=-32}{v1,o2}
        \fmf{vanilla,tension=1,label=\text{\small\(S(P_2)\)},label.dist=-32}{v1,o1}\fmfdot{v1}
  \end{fmfgraph*}} 
\end{align*}
\caption{Decay process $A\rightarrow S{\tilde{P}}$.}
\end{figure}
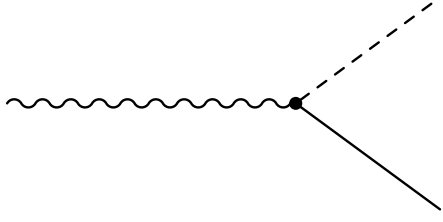

As in the previous section, we denote the momenta of $A$, ${\tilde{P}}$ and
$S$ as $P$, $P_{1}$ and $P_{2}$, respectively. Unlike the previous section,
the decay process now involves only one vector state: $A$. Let us denote the
corresponding polarisation vector as $\varepsilon_{\mu}^{(\alpha)}(P)$. Then,
upon substitution $\partial^{\mu}\rightarrow iP_{1,2}^{\mu}$ for the decay
products, we obtain the following Lorentz-invariant $AS{\tilde{P}}$ scattering
amplitude $-i\mathcal{M}_{A\rightarrow S{\tilde{P}}}^{(\alpha)}$:%

\begin{equation}
-i\mathcal{M}_{A\rightarrow S{\tilde{P}}}^{(\alpha)}=\varepsilon_{\mu
}^{(\alpha)}(P)h_{AS{\tilde{P}}}^{\mu}=-\varepsilon_{\mu}^{(\alpha)}(P)\left(
A_{AS{\tilde{P}}}P_{1}^{\mu}+B_{AS{\tilde{P}}}P_{2}^{\mu}\right)
\label{iMASP}
\end{equation}

with

\begin{equation}
h_{AS{\tilde{P}}}^{\mu}=-\left(  A_{AS{\tilde{P}}}P_{1}^{\mu}+B_{AS{\tilde{P}
}}P_{2}^{\mu}\right)\text{,} \label{hASP}
\end{equation}

where $h_{AS{\tilde{P}}}^{\mu}$ denotes the $AS{\tilde{P}}$ vertex.\newline

As in the previous section, calculation of the decay width will require us to
calculate the square of the average decay amplitude. As apparent from
Eq.\ (\ref{iMASP}), the scattering amplitude $-i\mathcal{M}_{A\rightarrow
S{\tilde{P}}}^{(\alpha)}$ depends on the polarisation vector $\varepsilon
_{\mu}^{(\alpha)}(P)$. Then the averaged squared amplitude $|-i\mathcal{\bar
{M}}|^{2}$ is determined as follows:

\begin{align}
-i\mathcal{M}_{A\rightarrow S{\tilde{P}}}^{(\alpha)}  &  =\varepsilon_{\mu
}^{(\alpha)}(P)h_{AS{\tilde{P}}}^{\mu}\Rightarrow\left\vert -i\mathcal{\bar
{M}}_{A\rightarrow S{\tilde{P}}}\right\vert ^{2}=\frac{1}{3}\sum
\limits_{\alpha=1}^{3}\left\vert -i\mathcal{M}_{A\rightarrow S{\tilde{P}}
}^{(\alpha)}\right\vert ^{2}\nonumber\\
&  =\frac{1}{3}\sum\limits_{\alpha,\beta=1}^{3}\varepsilon_{\mu}^{(\alpha
)}(P)h_{AS{\tilde{P}}}^{\mu}\varepsilon_{\nu}^{(\alpha)}(P)h_{AS{\tilde{P}}
}^{\ast\nu}\text{.}
\end{align}

Let us denote the mass of the state $A$ as $m_{A}$. Then utilising
Eq.\ (\ref{epsilon}) we obtain%

\begin{equation}
|-i\mathcal{\bar{M}}_{A\rightarrow S{\tilde{P}}}|^{2}=\frac{1}{3}\left(
-g_{\mu\nu}+\frac{P_{\mu}P_{\nu}}{m_{A}^{2}}\right)  h_{AS{\tilde{P}}}^{\mu
}h_{AS{\tilde{P}}}^{\ast\nu}=\frac{1}{3}\left[  -\left\vert h_{AS{\tilde{P}}
}^{\mu}\right\vert ^{2}+\frac{\left\vert h_{AS{\tilde{P}}}^{\mu}P_{\mu
}\right\vert ^{2}}{m_{A}^{2}}\right]  \text{.} \label{iMASP1}
\end{equation}

From Eq.\ (\ref{hASP}) we obtain

\begin{equation}
\left\vert h_{AS{\tilde{P}}}^{\mu}\right\vert ^{2}=A_{AS{\tilde{P}}}
^{2}m_{{\tilde{P}}}^{2}+B_{AS{\tilde{P}}}^{2}m_{S}^{2}+2A_{AS{\tilde{P}}
}B_{AS{\tilde{P}}}P_{1}\cdot P_{2}\text{,} \label{hASP2}
\end{equation}

where $m_{{\tilde{P}}}$ and $m_{{S}}$ denote the masses of ${\tilde{P}}$ and
$S$, respectively. Our calculations are performed in the rest frame of the
decaying particle, i.e., $P^{\mu}=(m_{A},\vec{0})$. Consequently,

\begin{equation}
h_{AS{\tilde{P}}}^{\mu}P_{\mu}\equiv h_{AS{\tilde{P}}}^{0}P_{0}=h_{AS{\tilde
{P}}}^{0}m_{A}\overset{\text{Eq.\ (\ref{hASP})}}{=}-\left(  A_{AS{\tilde{P}}
}E_{1}+B_{AS{\tilde{P}}}E_{2}\right)  m_{A} \label{hASP3}
\end{equation}

with $E_{1}=\sqrt{k^{2}(m_{A},m_{S},m_{{\tilde{P}}})+m_{{\tilde{P}}}^{2}}$,
$E_{2}=\sqrt{k^{2}(m_{A},m_{S},m_{{\tilde{P}}})+m_{{S}}^{2}}$\ and
$k(m_{A},m_{S},m_{{\tilde{P}}})$ from Eq.\ (\ref{kabc}).\ Inserting
Eqs.\ (\ref{hASP2}) and (\ref{hASP3}) into Eq.\ (\ref{iMASP1}) then yields

\begin{align}
|-i\mathcal{\bar{M}}_{A\rightarrow S{\tilde{P}}}|^{2}  &  =\frac{1}{3}\left[
\left(  A_{AS{\tilde{P}}}E_{1}+B_{AS{\tilde{P}}}E_{2}\right)  ^{2}
-(A_{AS{\tilde{P}}}^{2}m_{{\tilde{P}}}^{2}+B_{AS{\tilde{P}}}^{2}m_{S}
^{2}+2A_{AS{\tilde{P}}}B_{AS{\tilde{P}}}P_{1}\cdot P_{2})\right] \nonumber\\
&  =\frac{1}{3} \left[ (A_{AS{\tilde{P}}}^{2}+B_{AS{\tilde{P}}}^{2}
)k^{2}(m_{A},m_{S},m_{{\tilde{P}}})+2A_{AS{\tilde{P}}}B_{AS{\tilde{P}}}
(E_{1}E_{2}-P_{1}\cdot P_{2}) \right] \text{.} \label{iMASP2}
\end{align}

Given that $P_{1}^{\mu}=(E_{1},\vec{k}(m_{A},m_{S},m_{{\tilde{P}}}))$ and
$P_{2}^{\mu}=(E_{2},-\vec{k}(m_{A},m_{S},m_{{\tilde{P}}}))$, we obtain
$P_{1}\cdot P_{2}=E_{1}E_{2}+k^{2}(m_{A},m_{S},m_{{\tilde{P}}})$ or, from
Eq.\ (\ref{iMASP2}):

\begin{equation}
|-i\mathcal{\bar{M}}_{A\rightarrow S{\tilde{P}}}|^{2}=\frac{1}{3}
(A_{AS{\tilde{P}}}-B_{AS{\tilde{P}}})^{2}k^{2}(m_{A},m_{S},m_{{\tilde{P}}
})\text{.}
\end{equation}

The formula for the decay width of the process $A\rightarrow S{\tilde{P}}$ may
need to consider not only the decay $A^{0}\rightarrow S^{0}{\tilde{P}}^{0}$
but also $A^{0}\rightarrow S^{\pm}{\tilde{P}}^{\mp}$ (depending on isospin of
the decay products); for this reason, we introduce an isospin factor $I$:

\begin{equation}
\Gamma_{A\rightarrow S{\tilde{P}}}=I\frac{k(m_{A},m_{S},m_{{\tilde{P}}})}{8\pi
m_{A}^{2}}|-i\mathcal{\bar{M}}_{A\rightarrow S{\tilde{P}}}|^{2}=I\frac
{k^{3}(m_{A},m_{S},m_{{\tilde{P}}})}{24\pi m_{A}^{2}}(A_{AS{\tilde{P}}
}-B_{AS{\tilde{P}}})^{2}\text{.}\label{GASP}
\end{equation}

\subsection{Example: Decaying Scalar State} \label{sec.SVV}

Let us consider the decay of a scalar state $S$ into two vector states $V$,
i.e., $S\rightarrow V_{1}V_{2}$. The interaction Lagrangian may be given in
the following simple form:

\begin{equation}
\mathcal{L}_{S{VV}}\equiv A_{S{VV}}SV_{1\mu}^{0}V_{2}^{\mu0}\text{.}
\label{SVV}
\end{equation}

\begin{figure}[h]
\begin{align*} 
\qquad \qquad \qquad  \qquad \qquad \qquad  \quad \; \parbox{180mm}{ \begin{fmfgraph*}(180,80)
    \fmfleftn{i}{1}\fmfrightn{o}{2}
   \fmfv{label=\text{\small\(V_1\)},label.dist=22,label.angle=-112}{o2}\fmfv{label=\text{\small\(\varepsilon^{(\beta)}_\nu (P_2)\)},label.dist=25,label.angle=105}{o1}
      \fmf{dashes,tension=1.2,label=\text{\small\(S(P)\)},label.dist=-20}{i1,v1}
    \fmf{boson,label=\text{\small\(\varepsilon^{(\alpha)}_\mu (P_1)\)},tension=1,label.dist=-34}{v1,o2}
        \fmf{boson,tension=1,label=\text{\small\(V_2\)}}{v1,o1}\fmfdot{v1}
  \end{fmfgraph*}}
\end{align*}\caption{Decay process $S\rightarrow V_{1}V_{2}$.}\end{figure}
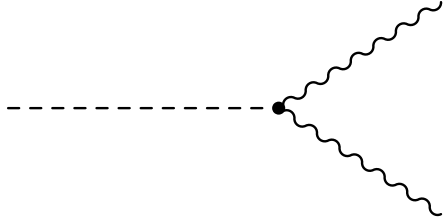

Our calculation of the decay width has to consider polarisations of the two
vector states. We denote the momenta of $S$, ${V}_{1}$ and $V_{2}$ as $P$,
$P_{1}$ and $P_{2}$, respectively, while the polarisation vectors are denoted
as $\varepsilon_{\mu}^{(\alpha)}(P_{1})$ and $\varepsilon_{\nu}^{(\beta
)}(P_{2})$. Then, upon substituting $\partial^{\mu}\rightarrow iP_{1,2}^{\mu}$
for the decay products, we obtain the following Lorentz-invariant
$AS{\tilde{P}}$ scattering amplitude $-i\mathcal{M}_{A\rightarrow S{\tilde{P}%
}}^{(\alpha)}$:

\begin{equation}
-i\mathcal{M}_{S\rightarrow V_{1}V_{2}}^{(\alpha,\beta)}=\varepsilon_{\mu
}^{(\alpha)}(P_{1})\varepsilon_{\nu}^{(\beta)}(P_{2})h_{S{VV}}^{\mu\nu
}=i\varepsilon_{\mu}^{(\alpha)}(P_{1})\varepsilon_{\nu}^{(\beta)}
(P_{2})A_{S{VV}}g^{\mu\nu}\label{iMSVV}
\end{equation}

with

\begin{equation}
h_{S{VV}}^{\mu\nu}=iA_{S{VV}}g^{\mu\nu}\text{,} \label{hSVV}
\end{equation}

where $h_{S{VV}}^{\mu\nu}$ denotes the $S{VV}$ vertex.\newline

The averaged squared amplitude $|-i\mathcal{\bar{M}}|^{2}$ is determined as follows:

\begin{align}
-i\mathcal{M}_{S\rightarrow V_{1}V_{2}}^{(\alpha,\beta)} &  =\varepsilon_{\mu
}^{(\alpha)}(P_{1})\varepsilon_{\nu}^{(\beta)}(P_{2})h_{S{VV}}^{\mu\nu
}\Rightarrow\left\vert -i\mathcal{\bar{M}}_{S\rightarrow V_{1}V_{2}
}\right\vert ^{2}=\frac{1}{3}\sum\limits_{\alpha,\beta=1}^{3}\left\vert
-i\mathcal{M}_{S\rightarrow V_{1}V_{2}}^{(\alpha,\beta)}\right\vert
^{2}\nonumber\\
&  =\frac{1}{3}\sum\limits_{\alpha,\beta=1}^{3}\varepsilon_{\mu}^{(\alpha
)}(P_{1})\varepsilon_{\nu}^{(\beta)}(P_{2})h_{S{VV}}^{\mu\nu}\varepsilon
_{\kappa}^{(\alpha)}(P_{1})\varepsilon_{\lambda}^{(\beta)}(P_{2})h_{S{VV}
}^{\ast\kappa\lambda}\text{.}\label{iMSVV1}
\end{align}

Equation (\ref{epsilon}) then yields the same expression as the one presented in
Eq.\ (\ref{iMAVP2}):

\begin{equation}
|-i\mathcal{\bar{M}}_{S\rightarrow V_{1}V_{2}}|^{2}=\frac{1}{3}\left[
\left\vert h_{S{VV}}^{\mu\nu}\right\vert ^{2}-\frac{\left\vert h_{S{VV}}
^{\mu\nu}P_{1\mu}\right\vert ^{2}}{m_{V_{1}}^{2}}-\frac{\left\vert h_{S{VV}
}^{\mu\nu}P_{2\nu}\right\vert ^{2}}{m_{V_{2}}^{2}}+\frac{\left\vert h_{S{VV}
}^{\mu\nu}P_{1\mu}P_{2\nu}\right\vert ^{2}}{m_{V_{1}}^{2}m_{V_{2}}^{2}
}\right]  \text{.}\label{iMSVV2}
\end{equation}

From Eq.\ (\ref{epsilon}) we obtain $h_{S{VV}}^{\mu\nu}P_{1\mu}=iA_{S{VV}%
}P_{1}^{\nu}$, $h_{S{VV}}^{\mu\nu}P_{2\nu}=iA_{S{VV}}P_{2}^{\mu}$ and
$h_{S{VV}}^{\mu\nu}P_{1\mu}P_{2\nu}$ $=iA_{S{VV}}P_{1}\cdot P_{2}$ and consequently

\begin{equation}
|-i\mathcal{\bar{M}}_{S\rightarrow V_{1}V_{2}}|^{2}=\frac{1}{3}\left[
4-\frac{P_{1}^{2}}{m_{V_{1}}^{2}}-\frac{P_{2}^{2}}{m_{V_{2}}^{2}}+\frac
{(P_{1}\cdot P_{2})^{2}}{m_{V_{1}}^{2}m_{V_{2}}^{2}}\right]  A_{S{VV}}
^{2}\text{.}\label{iMSVV3}
\end{equation}

For on-shell states, $P_{1,2}^{2}=m_{V_{1,2}}^{2}$ and Eq.\ (\ref{iMSVV3})
reduces to

\begin{equation}
|-i\mathcal{\bar{M}}_{S\rightarrow V_{1}V_{2}}|^{2}=\frac{1}{3}\left[
2+\frac{(P_{1}\cdot P_{2})^{2}}{m_{V_{1}}^{2}m_{V_{2}}^{2}}\right]  A_{S{VV}
}^{2}=\frac{1}{3}\left[  2+\frac{(m_{S}^{2}-m_{V_{1}}^{2}-m_{V_{2}}^{2})^{2}
}{4m_{V_{1}}^{2}m_{V_{2}}^{2}}\right]  A_{S{VV}}^{2}\text{.}\label{iMSVV4}
\end{equation}

The decay width is consequently

\begin{equation}
\Gamma_{S\rightarrow V_{1}V_{2}}=I\frac{k(m_{S},m_{V_{1}},m_{{V}_{2}})}{8\pi
m_{S}^{2}}|-i\mathcal{\bar{M}}_{S\rightarrow V_{1}V_{2}}|^{2}\label{GSVV}
\end{equation}

with $k(m_{S},m_{V_{1}},m_{{V}_{2}})$\ from Eq.\ (\ref{kabc}); we have
considered an isospin factor $I$ in case the vector particles are not
isosinglets (then the contribution of the charged modes to the full decay
width would also have to be considered).

Suppose now that the two vector states were unstable themselves and decayed into
pseudoscalars: $V_{1}\rightarrow{\tilde{P}}_{1}{\tilde{P}}_{2}$ and
$V_{2}\rightarrow{\tilde{P}}_{3}{\tilde{P}}_{4}$. Calculation of the decay
width for the process $S\rightarrow V_{1}V_{2}\rightarrow{\tilde{P}}%
_{1}{\tilde{P}}_{2}{\tilde{P}}_{3}{\tilde{P}}_{4}$ requires integration over
the spectral functions of the two vector resonances, Eq.\ (\ref{dV}). The
decay width then reads

\begin{equation}
\Gamma_{S\rightarrow V_{1}V_{2}\rightarrow{\tilde{P}}_{1}{\tilde{P}}%
_{2}{\tilde{P}}_{3}{\tilde{P}}_{4}}=%
{\displaystyle\int\limits_{m_{{\tilde{P}}_{1}}+m_{{\tilde{P}}_{2}}}^{m_{S}}}
\text{d}x_{V_{1}}%
{\displaystyle\int\limits_{m_{{\tilde{P}}_{3}}+m_{{\tilde{P}}_{4}}}%
^{m_{S}-x_{V_{1}}}}
\text{d}x_{V_{2}}\, I\frac{k(m_{S},x_{V_{1}},x_{{V}_{2}})}{8\pi m_{S}%
^{2}}\,d_{V}(x_{V_{1}})d_{V}(x_{V_{2}})|-i\mathcal{\bar{M}}_{S\rightarrow
V_{1}V_{2}}|^{2}\text{.}\label{GSVV1}
\end{equation}

\chapter{Review of Scalar Isosinglets} \label{sec.scalarexp}

The scalar isosinglet mesons have been an extremely interesting topic of investigation from both theoretical
and experimental standpoints for decades. Their features have often been ambiguous due to
large background and various decay channels. This chapter is a brief review of the knowledge about scalar isosinglet
mesons, what we think we know about them and how our understanding of these resonances has developed
in experimental work over the last decades. It will contain a dedicated section regarding
the putative new $f_0(1790)$ resonance which is of particular importance for this work because it
is very close to, and may interfere with, the already established $f_0(1710)$ state -- and the latter is of utmost importance for our 
calculations in the three-flavour version of our model.

\section{The \boldmath $f_{0}(600)$ Resonance} \label{sec.f0(600)}

The $f_{0}(600)$ state (or $\sigma$; in older articles: $\epsilon$ or
$\eta_{0+}$) has a long and troubled history. The existence of this state was
suggested in linear sigma models approximately a decade before it was first
discovered, see, e.g., Ref.\ \cite{gellmanlevy}. The state was introduced
theoretically as the putative chiral partner of the pion; however, it was
shown to be highly non-trivial to ascertain experimentally.\\

The earliest versions of the linear sigma model incorporated only the sigma
and the pion. The pion is a well-established $\bar{q}q$ state and consequently
its chiral partner also had to be a quarkonium. The naive expectation was that
the mass splitting between the pion and its chiral partner would not be large,
or at least that the mass of the $\sigma$ state would be in the interval below
$1$ GeV, with a predominant decay into pions ($\sim100\%$). For this reason,
many experimental collaborations have looked closely into $\pi\pi$ scattering
amplitudes up to 1 GeV (see below) in order to ascertain if an $I(J^{PC})%
=0(0^{++})$ signal could be found. Note, however, that the theoretical $\bar
{q}q$ scalar state possesses the intrinsic angular momentum $L=1$ as well as
the relative spin of the quarks $S=1$. For this reason one could also easily
expect the state to be in the region above $1$ GeV. This is contrary to the
expectation of the first version of the $\sigma$ models. Additionally, four
decades ago Gasiorowicz and Geffen suggested how to introduce vectors ($\rho$,
$\omega$) and axial-vectors ($a_{1}$, $f_{1}$) into linear sigma models
utilising the chiral symmetry \cite{GG}. (The latter article is also important
because it suggested the existence of an axial-vector triplet, known nowadays
as $a_{1}$, almost 10 years before the particle was first established reliably
in the $\rho\pi$ final state produced in $K^{-}p$
reactions\ \cite{Gavillet:1977}.) It was subsequently demonstrated that the
inclusion of (axial-)vectors requires us to assign the scalar-isoscalar state
present in the $\sigma$ models to a resonance above, rather than below, $1$
GeV \cite{Meinereferenzen,Paper1,KR,Lissabon,Zakopane,Krakau,Mainz}%
. Experimental data available nowadays seem to strongly favour this
assignment, particularly in view of the fact that the pure non-strange scalar
state is expected to mix with a pure-strange and a glueball scalar state
leading to experimental observation of three scalar states in the mutual
vicinity: these could possibly be the established resonances $f_{0}(1370)$,
$f_{0}(1500)$ and $f_{0}(1710)$ \cite{PDG}. (See below for a review of these
resonances.) This implies that scalar states below 1 GeV, including the
$f_{0}(600)$ resonance, cannot be of $\bar{q}q$ structure. The mentioned
theoretical ambiguity regarding the structure of this light scalar state is,
however, not the only problem related to $f_{0}(600)$ -- the data on this
resonance suggest that its mass and width are of comparable magnitude,
rendering also the experimental search for this state rather difficult.\\

Before we proceed with a discussion of the experimental evidence for $f_{0}(600)$,
let us summarise the reasons why the hadronic physics requires the existence
of a light scalar meson:

\begin{itemize}
\item It is the putative chiral partner of the pion in the sigma models; the
vacuum expectation value of the $\sigma$ state is used as means of modelling
spontaneous chiral symmetry breaking \cite{gellmanlevy}. Therefore, the
existence of the $\sigma$ meson [expected to possess a mass $<1$ GeV in the
plain sigma models without (axial-)vectors] is a natural consequence of the
existence of the pion and of the chiral symmetry (and the breaking of this symmetry).

\item As already indicated, subsequent calculations suggested that the scalar
$\bar{q}q$ state present in sigma models cannot correspond to a resonance
below 1 GeV but rather to a state above $1$ GeV. However, if we act on the
assumption of the possible existence of tetraquark ($\bar{q}\bar{q}qq$) states
\cite{Jaffeq2q2}, then the search for a light scalar state is nonetheless
justified: it may possess a tetraquark structure.

\item The Nambu--Jona-Lasinio model \cite{NJL} requires the existence of a light
scalar-isoscalar meson with mass $2m_{q}$ where $m_{q}$ denotes the
constituent-quark mass; the vacuum expectation value of the scalar state is
again utilised to model the spontaneous breaking of the chiral symmetry.

\item Nucleon-nucleon scattering is expected to occur with exchange of a light
scalar meson \cite{Susanna}; $f_{0}(980)$ cannot fulfill this role as it
strongly couples to kaons although it was established as a resonance long
before $f_{0}(600)$, see Sec.\ \ref{sec.f0(980)}, and $f_{0}(1370)$ is too
heavy to influence the nucleon scattering at low energies.

\item It is required for a correct description of $\pi\pi$ scattering data,
see below.
\end{itemize}

Experimental evidence for the existence of the $f_{0}(600)$ resonance stems
from analyses of $\pi\pi$ scattering amplitudes. This is true historically as
well as nowadays; however, given the notoriously large decay width of the
state and the limited statistics of the first experiments, the first data on
$\pi\pi$ scattering provided us only with hints rather than definitive proofs
of the existence of this particle.

In 1973, $\pi\pi$ phase shifts were measured at CERN where pions were
scattered off protons: a pion beam was targeted at a 50 cm long liquid-hydrogen
target inducing the reaction ${\pi}^{-}{p\rightarrow}\pi^{+}\pi^{-}n$ at 17.2
GeV, with 300000 events reconstructed \cite{Grayer:1974}. The results of
Ref.\ \cite{Grayer:1974} were later combined with results obtained from the
same ${\pi}^{-}{p}$\ reaction induced by targeting pions on butanol (C$_{4}%
$H$_{9}$OH) \cite{Becker:1978}. A broad enhancement in the $\pi\pi$ $S$-wave
was observed but no definitive conclusions were possible. In 1976, the Particle
Data Group (PDG) removed this state from their listing. The next decade saw
$pp\rightarrow pp\pi^{+}\pi^{-}$ data suggesting a broad $S$-wave enhancement
in the $\pi\pi$ scattering below 1 GeV \cite{Akesson:1986} but still without
definitive conclusions regarding the existence of $f_{0}(600)$. The resonance
was reinstated by the PDG in 1996 after theoretical results amassed suggesting
the existence of the state: the 1993 review of $\pi\pi$\ and $KK$ scattering
data in Ref.\ \cite{Kaminski:1993} found a pole with a mass of $(506\pm10)$
MeV and a width of $(494\pm5)$ MeV; a model of $\pi\pi\rightarrow\pi\pi$ and
$\pi\pi\rightarrow KK$ scattering with crossing symmetry and unitarity found
the data to require a light scalar meson \cite{hep-ph/9511335} and a
model-independent analysis of $\pi\pi$ scattering data below the $KK$
threshold in Ref.\ \cite{Ishida:1995xx} found a pole with a mass of
$(553.3\pm0.5)$ MeV and a width of $(242.6\pm1.2)$ MeV (see also Ref.
\cite{Tornqvist:1995ay}).

Contrarily, high-statistics data from ${\bar{p}p}$ annihilation
\cite{Amsler:1995bf,f0(1500)-CB-1996} did not yield a clear $f_{0}(600)$
signal; additionally, there were no conclusive results from central $pp$
collisions either, although a broad enhancement below $1$ GeV was observed
\cite{Alde:1997}. \newline

However, subsequent experimental results did suggest the existence of
a signal attributed to $f_0(600)$. The E791 Collaboration at Fermilab
\cite{Aitala:2000xu} used a sample of $2\cdot10^{10}$ events from the $\pi
^{-}$-nucleon reaction at 500 GeV to produce charm ($D$) mesons; $1172\pm61$
events $D\rightarrow\pi^{-}\pi^{+}\pi^{+}$ were induced and strong evidence of
a scalar resonance with a mass of $478_{-23}^{+24}\pm17$ MeV and a decay width
of $324_{-40}^{+42}\pm21$ MeV was found in the $\pi\pi$ channel using a Dalitz
plot. Similarly, the CLEO Collaboration \cite{CLEO:2007}\ produced 780000 $DD$
pairs from reaction $e^{+}e^{-}\rightarrow\psi(3770)\rightarrow D^{+}D^{-}$;
various analysis methods were used and a definitive contribution of
$f_{0}(600)\pi^{+}$ in the decay channel $D\rightarrow\pi^{-}\pi^{+}\pi^{+}$
was observed (branching ratio $\sim50\%$). Additionally, the BES II
Collaboration produced 58 million $J/\psi$ events from electron-positron
annihilation and various decay channels involving pions and kaons were analysed
(see the following subsections). We note here that a broad $f_{0}(600)$ peak
was observed in the decay channel $J/\psi\rightarrow$ $\omega\pi^{+}\pi^{-}$
for which the pole mass was determined as $(541\pm39)$ MeV and the decay width
as $(504\pm84)$ MeV \cite{Ablikim:2004qna}. Thus, the analyses found a very
broad resonance with a mass of $\sim500$ MeV and a comparable decay
width.\newline

On the other hand, the theoretical search for this state is confronted with
various problems. As already mentioned, $f_{0}(600)$ is very broad. It is one
of rare meson resonances where the decay width $\Gamma$ is virtually the same
as the mass; this renders an extrapolation of the pole position from the
$\pi\pi$ scattering data highly non-trivial as the pole is very distant from
the real axis. The resonance may easily be distorted by background effects and
by interference with other scalar isosinglets. For this reason, a
parametrisation of $f_{0}(600)$ in terms of a Breit-Wigner distribution [see
Eq.\ (\ref{BW1})] has to be performed with great care (if at all).
Nonetheless, it is possible to determine the pole position of the resonance
utilising Roy equations \cite{Roy} with crossing symmetry, analyticity and
unitarity. This allows one to demonstrate unambiguously that $f_{0}(600)$ is a
genuine resonance -- e.g., in the work of Leutwyler \textit{et al.}
\cite{Leutwyler}, a resonance pole was found at $m_{f_{0}(600)}-i\Gamma
_{f_{0}(600)}/2=(441_{-8}^{+16}-i272_{-12.5}^{+9})$ MeV. Similarly, Pel\'{a}ez
\textit{et al.} \cite{Pelaez1}\ have found $m_{f_{0}(600)}-i\Gamma
_{f_{0}(600)}/2=(461_{-15.5}^{+14.5}-i255\pm16)$ MeV.

These results were obtained from analyses of $\pi\pi$ scattering data with the
pions produced from kaon decays. Let us therefore briefly review kaon decays
in the following. Kaons are produced by targeting protons onto a metal (such
as beryllium). Two types of charged-kaon decays are relevant here -- the
exclusive decays into pions: $K^{\pm}\rightarrow\pi^{\pm}\pi^{0}\pi^{0}$ and
$K^{\pm}\rightarrow\pi^{\pm}\pi^{+}\pi^{-}$ ($K_{3\pi}$ decays; branching
ratio $\sim10^{-2}$ \cite{PDG}) and the semileptonic decays $K^{+}%
\rightarrow\pi^{+}\pi^{-}e^{+}\nu$ (and hermitian conjugate for $K^{-}$). The
latter ones are referred to as $K_{e4}$ decays; they belong to the so-called
rare kaon decays because of the small branching ratio ($\sim10^{-5}$
\cite{PDG}). Note, however, that they also possess a much cleaner environment
than $K_{3\pi}$ decays where pion rescattering may induce an increased error
in the scattering amplitude. First measurements of the $K_{e4}$ decays were
performed in 1977 \cite{Ke4-1977} with a number of events several order of
magnitudes smaller than the latest measurements performed at CERN by the
NA48/2 Collaboration in 2003 and 2004 \cite{Peyaud}. Note that the pion scattering data
allow for determination of $\pi\pi$\ scattering lengths, see Refs.\
\cite{Leutwyler,Pelaez1} and Sections \ref{sec.SLQ} and \ref{SL} in this work.
We will show in the mentioned sections that the pion scattering lengths
require the existence of a light scalar state [i.e., $f_{0}(600)$] as
otherwise their proper description is not possible. However, the broader
phenomenology will disfavour a $\bar{q}q$ structure of this resonance.\newline

We note that the PDG lists $f_{0}(600)$ as having a mass of (400 --
1200) MeV and a width of (600 -- 1000) MeV \cite{PDG}. Let us also note that
$\kappa$ [or $K_{0}^{\star}(800)$], the strange counterpart of $f_{0}(600)$,
has similarly also been subject of a prolonged debate about its existence with
some analyses finding a corresponding pole \cite{KappaY} while others do not
\cite{KappaN}. The fact that $m_\kappa \simeq \Gamma_\kappa$ renders it extremely important not to fit the $\kappa$ meson
with a Breit-Wigner distribution of a constant width; such fits can easily fail to detect this state. 
There is, however, a scalar kaon in the region above 1 GeV as
well, the existence of which appears to be confirmed: this state, denoted as
$K_{0}^{\star}(1430)$, is found in the $K\pi$ channel (see Ref.\ \cite{PDG} and
references therein).

\section{The \boldmath $f_{0}(980)$ Resonance} \label{sec.f0(980)}

This resonance is close to the kaon-kaon threshold rendering an experimental
analysis somewhat difficult, with different collaborations and reviews
obtaining at times very different results. For the same reason, the structure
of the resonance is not clear: it may be interpreted as a quarkonium
\cite{Muenz:1996,Delbourgo:1998,f0(980)asqq}, as a $\bar{q}^{2}q^{2}$ state
\cite{Jaffeq2q2,f0(980)asq2q2,scalars-above1GeVqq-below1GeVq2q2,f0(980)q2q2-f0(1500)glueball,Pelaez-scalars-below1GeVq2q2,f0(980)q2q2-f0(600)qq-f0(1370)q2q2}%
, as a $KK$ bound state
\cite{Barnes:1985,WeinsteinKK,f0(980)asKK,scalars-above1GeVqq-below1GeV-molecules}%
, as a glueball \cite{f0(980)-as-glueball} or even as an $\eta\eta$ bound
state \cite{f0(980)etaeta-f0(1500)glueball-f0(1370)-f0(1710)qq}. One of the
most suitable of ways to ascertain the $f_{0}(980)$ structure is utilising the
decay $f_{0}(980)\rightarrow\gamma\gamma$. The PDG cite world-average value is
$\Gamma_{f_{0}(980)\rightarrow\gamma\gamma}=0.29_{-0.06}^{+0.07}$ keV. Various
approaches have been utilised to calculate this decay width: a relativistic
nonstrange-quark model obtained values between $1.3$ keV and $1.8$ keV
\cite{Muenz:1996}; assuming a $KK$ structure yielded values between $0.2$ keV
and $0.6$ keV \cite{Barnes:1985} and assuming a strange-quarkonium structure of
the resonance resulted in values $\sim0.3$ keV \cite{Delbourgo:1998}. Thus
the only assignment that appears to be excluded is the one where the resonance
is a non-strange quarkonium.

The work presented here contains only quarkonium states; the only possible
interpretation of $f_{0}(980)$ within our model could be as a $\bar{q}q$
state. In the $U(3)\times U(3)$ version of our model, it is not possible to
interpret $f_{0}(980)$\ as a $\bar{q}q$ state\ within our Fit I,
Chapter \ref{ImplicationsFitI} (the decay width would be several times too large), and it is
strongly disfavoured as a $\bar q q$ state within Fit II, Chapter \ref{ImplicationsFitII} [see in particular
the short note on $f_{0}(980)$ at the end of Sec.\ \ref{sec.sigmapionpion2}%
].\newline

The earliest evidence for $f_{0}(980)$ came from Berkeley in 1972
\cite{Berkley}. (There were even earlier data from ${\bar{p}p\rightarrow}%
\pi^{+}\pi^{-}\omega$ at Saclay in 1969 where the analysis required an
$S$-wave isoscalar structure at $\sim940$ MeV \cite{Bizzarri:1970}; however,
no kaon events were considered.) A pion beam of $7.1$ GeV was targeted at
protons (in a hydrogen bubble chamber) and reactions ${\pi}^{+}p\rightarrow{\pi
}^{+}{\pi}^{-}\Delta^{++}$ ($32100$ events) and ${\pi}^{+}p\rightarrow{K}%
^{+}{K}^{-}\Delta^{++}$ ($682$ events) were observed. The ${\pi\pi}$ system
was found to exhibit a rapid drop in the cross-section in the energy region
between $950$ MeV and $980$ MeV (i.e., close to the $KK$ threshold:
$2m_{K^{\pm}}=987.4$ MeV \cite{PDG}). This effect occurred due to strong
coupling of the ${\pi\pi}$ and $KK$ channels upon opening of the kaon
threshold. Subsequent partial-wave analysis of both the pion and kaon
scattering data yielded a pole at $(997\pm6)$ MeV; the pole width was
$(54\pm16)$ MeV. [Incidentally, the same analysis also found a pole at $660$
MeV and the width of $640$ MeV, corresponding to the nowadays $f_{0}(600)$
meson, but the pole was not stable in all parametrisations due to lack of data
below $550$ MeV.]

CERN-Munich data from 1973 confirmed a strong $S$-wave enhancement at
approximately $1$ GeV from the one-pion-exchange (OPE) reaction $\pi
^{-}{p\rightarrow}\pi^{+}\pi^{-}n$ at $17.2$ GeV \cite{CM:1973}. Subsequent
data from $\pi^{-}p\rightarrow\pi^{+}\pi^{-}n$ and $K^{+}K^{-}n$ taken at
Rutherford Laboratory in Chilton, England, also produced a sharp drop in the
$\pi^{+}\pi^{-}$ spectrum close to the $KK$ threshold, assigning this signal
to a $J^{P}=0^{+}$ resonance with pole mass of $(987\pm7)$ MeV and pole width
of $(48\pm14)$ MeV \cite{Binnie:1974}. Further publications regarding
$f_{0}(980)$ are listed in Ref.\ \cite{f0(980)}; let us, however, discuss here
studies of the resonance performed by several collaborations:

\begin{itemize}
\item \textit{WA76 / WA102.} Data were gathered using the CERN Omega
Spectrometer in reactions $pp\rightarrow p_{f}({\pi}^{+}{\pi}^{-})p_{s}$,
$pp\rightarrow p_{f}({\pi}^{0}{\pi}^{0})p_{s}$, $pp\rightarrow p_{f}({K}%
^{+}{K}^{-})p_{s}$ and $pp\rightarrow p_{f}({\bar{K}}_{S}^{0}K_{S}^{0})p_{s}$
at $85$ GeV and at $300$\ GeV and in reaction $\pi^{+}{p\rightarrow}\pi
_{f}^{+}{\pi}^{+}{\pi}^{-}p_{s}$ at $85$ GeV\ (subscripts $f$ and $s$ denote
the fastest and slowest particles in the laboratory frame, respectively). The
$f_{0}(980)$ resonance was identified in both pion and kaon final states;
the coupling of the resonance to kaons ($g_{K}$) was found to be dominant in
comparison to the pion coupling ($g_{\pi}$): $g_{K}/g_{\pi}=2.0\pm0.9$
\cite{Armstrong:1991}.\ A pole mass of $(1001\pm2)$ MeV and a pole width of
$(72\pm8)$ were determined. Note, however, that care is needed when
interpreting these results, as a relativistic form of the Breit-Wigner distribution
was utilised to analyse data, with no dispersive corrections (that are
important due to the effects of the $KK$-threshold opening). In 1999, data
from a higher-resolution reaction $pp\rightarrow p_{f}({K}^{+}{K}^{-})p_{s}$
and $pp\rightarrow p_{f}({\bar{K}}_{S}^{0}K_{S}^{0})p_{s}$ at $450$ GeV
indicated a Breit-Wigner mass of $(985\pm10)$ MeV and a width of $(65\pm20)$ MeV,
with interference effects of $f_{0}(1500)$ and $f_{0}(1710)$ included
\cite{WA102:1999}. A similar analysis was performed for $pp\rightarrow
p_{f}(\pi^{+}\pi^{-})p_{s}$, also at $450$ GeV \cite{WA102:1999_1} allowing
for a combined analysis to be performed in both pion and kaon channels
\cite{Barberis:1999}. Both the T-matrix formalism \cite{Zou-TMatrix} and the K-matrix
formalism \cite{Anisovich-KMatrix} were used. Results were obtained regarding
four scalar resonances: $f_{0}(980)$, $f_{0}(1370)$, $f_{0}(1500)$ and
$f_{0}(1710)$. For $f_{0}(980)$, the obtained mean values were $m_{f_{0}%
(980)}=(987\pm6\pm6)$ MeV and $\Gamma_{f_{0}(1500)}=(96\pm24\pm16)$ MeV. The
$f_0(980)$ coupling to kaons was found to be approximately two times larger
than the coupling to pions.

\item \textit{Crystal Barrel.} The $f_{0}(980)$ resonance appeared in
high-statistics data produced by $16.8$ million ${\bar{p}p}$ collisions at
CERN-LEAR (Low Energy Antiproton Ring) and analysed in 1995. From these
collisions, $712000$ events for ${\bar{p}p\rightarrow3}\pi^{0}$ were selected
\cite{Amsler:1995gf}. The $f_{0}(980)$ resonance was reconstructed with a
K-matrix approach and the values $m_{f_{0}(980)}=(994\pm5)$ MeV and
$\Gamma_{f_{0}(980)}=(26\pm10)$ MeV were obtained. Subsequently, data were
taken from reactions ${\bar{p}p\rightarrow}\pi^{0}\pi^{0}\pi^{0}$ ($712000$
events), ${\bar{p}p\rightarrow}\pi^{0}\pi^{0}\eta$ ($280000$ events) and
${\bar{p}p\rightarrow}\pi^{0}\eta\eta$ ($198000$ events) \cite{Amsler:1995bf}.
Data analysis was not conclusive in that different Riemann sheets in the
T-matrix formalism yielded somewhat different pole masses [$(938$ - $996)$
MeV] and widths [$(70$ - $112)$ MeV]; nonetheless the existence of a pole
(i.e., of a resonance) was ascertained.

\item \textit{GAMS.} Data were taken from the OPE reaction ${\pi}^{-}%
p\rightarrow\pi^{0}\pi^{0}n\rightarrow(4\gamma)n$ at GAMS-IHEP (GAMS: Russian
abbreviation for Hodoscope Automatic Multiphoton Spectrometer). A pion beam at
$38$ GeV was utilised to induce the reaction. A drop in the $\pi\pi$ cross
section just below $1$ GeV was observed allowing for a resonance with a mass of
$(997\pm5)$ MeV and a width of $(48\pm10)$ MeV to be reconstructed
\cite{Alde:1994}. The same experiment was later repeated at the CERN-SPS
accelerator using the electromagnetic hodoscope calorimeter GAMS-4000 but with
a pion beam of $100$ GeV \cite{Alde:1998}. The results were similar to those of
Ref.\ \cite{Alde:1994} although an optimal fit was found for a Breit-Wigner mass
of $(960\pm10)$ MeV, i.e., below the $KK$ threshold.\ The Collaboration also
used two additional production mechanisms. The first involved targeting
antiprotons at $450$ GeV onto liquid hydrogen and inducing the reaction ${\bar
{p}p\rightarrow}p_{f}(\pi^{0}\pi^{0})p_{s}\rightarrow p_{f}(4\gamma)p_{s}$ at
CERN-SPS. A Breit-Wigner fit was optimised at a mass of $(955\pm10)$ MeV and
a width of $(69\pm15)$ MeV in Ref.\ \cite{Alde:1997}. The second involved the
same reaction, however with protons at $450$ GeV: ${pp\rightarrow}p_{f}%
(\pi^{0}\pi^{0})p_{s}\rightarrow p_{f}(4\gamma)p_{s}$ \cite{Bellazzini:1999}.
A fit to data yielded $m_{f_{0}(980)}=(989\pm15)$ MeV and $\Gamma_{f_{0}%
(980)}=(65\pm25)$ MeV. Note that already GAMS data from Ref.\ \cite{Alde:1994}
suggested an interesting feature of the $\pi^{0}\pi^{0}$ invariant mass
spectrum where $f_{0}(980)$ was reconstructed: there was a dip in the spectrum
(for lower momentum transfer of ${\pi}^{-}$ to the two neutral pions) as well
as a peak (if higher momentum transfer was considered). This unusual feature
of $f_{0}(980)$ was analysed in Ref.\ \cite{Achasov:1998pu} where the observed
alteration in the spectrum was suggested to occur due to an $a_{1}$ exchange
contribution in the ${\pi}^{-}p$ amplitude that rises with momentum transfer.

\item \textit{CMD-2.} Data regarding $f_{0}(980)$ obtained by this
Collaboration were result of studies of $\varphi(1020)$ radiative decays. To
this end, $20$ million $\varphi$ events were produced in the annihilation reaction
$e^{+}e^{-}\rightarrow\pi^{+}\pi^{-}\gamma$ and observed by the Cryogenic
Magnetic Detector CMD-2 in Novosibirsk. The $\varphi(1020)$ resonance was
reconstructed in the $\pi^{+}\pi^{-}\gamma$ final state; isolating photons
with energy below $\sim100$ MeV and assuming that the pions in this final
state dominantly coupled to $f_{0}(980)$ yielded $m_{f_{0}(980)}=975$ MeV.
However, the resonance width was not determined from the mentioned
annihilation process but rather held at $40$ MeV \cite{Akhmetshin:1999}. The
same process was subsequently repeated \cite{Akhmetshin:1999di} with $\pi
^{0}\pi^{0}\gamma$ and $\eta\pi^{0}\gamma$ final states. The lack of
bremsstrahlung for the neutral final-state modes allowed for a better
reconstruction of $f_{0}(980)$ in the $\pi^{0}\pi^{0}$ channel. Again,
a dominant coupling of pions to $f_{0}(980)$ was assumed once the photons of
energy below $\sim100$ MeV were isolated. The ensuing results reflected those
of Ref.\ \cite{Akhmetshin:1999} in mass; the width was determined to be
$(56\pm20\pm10)$ MeV. These results are, however, obtained within certain
models -- as discussed in Ref.\ \cite{Akhmetshin:1999di}. Note that the
annihilation process $e^{+}e^{-}\rightarrow\pi\pi\gamma$ (at $1020$ MeV) was
also used by the KLOE Collaboration in the "Frascati $\varphi$ factory", with
results very similar to those of CMD-2 but unfortunately with no determination
of the $f_{0}(980)$ width (see Ref.\ \cite{Ambrosino:2006} and references therein).

\item \textit{Belle.} The Belle Collaboration at KEK (High Energy Accelerator
Research Organization, located in Tsukuba, Japan) have used the annihilation
process $e^{+}e^{-}\rightarrow e^{+}e^{-}\pi^{+}\pi^{-}$\ at $10.58$ GeV
aquiring high-statistics data, see Ref.\ \cite{f0(980)-Belle-2007}. The
$f_{0}(980)$ resonance was reconstructed in the $\pi^{+}\pi^{-}$ final state
with $m_{f_{0}(980)}=985.6_{-1.5-1.6}^{+1.2+1.1}$ MeV and $\Gamma
_{f_{0}(980)\rightarrow\pi\pi}=34.2_{-11.8-2.5}^{+13.9+8.8}$ MeV. The stated
result for the decay width suffers from large errors (particularly at the
upper boundary) and the reason is the possible interference of $e^{+}%
e^{-}\rightarrow e^{+}e^{-}\pi^{+}\pi^{-}$ events with dilepton events
$e^{+}e^{-}\rightarrow\mu^{+}\mu^{-}$ yielding an increased uncertainty in
data evaluation. Subsequent analysis of the same reaction with $\pi^{0}\pi
^{0}$ final states \cite{f0(980)-Belle-2008} yielded $m_{f_{0}(980)}%
=982.2\pm1.0_{-8.0}^{+8.1}$ MeV and $\Gamma_{f_{0}(980)\rightarrow\pi\pi
}=66.9\pm2.2_{-12.5}^{+17.6}$ MeV. The latter is very different from the value
of Ref.\ \cite{f0(980)-Belle-2007} because differential cross-sections (in $S$,
$D$, $G$ waves) were fitted rather than the total one. However, no
consideration was given to the kaon decays of $f_{0}(980)$. A Breit-Wigner
analysis was used in both cases.
\end{itemize}

The PDG estimates $m_{f_{0}(980)}=(980\pm10)$ MeV and $\Gamma_{f_{0}%
(980)}=(40-100)$ MeV \cite{PDG}.

\section{The \boldmath $f_{0}(1370)$ Resonance} \label{sec.f0(1370)}

The $f_{0}(1370)$ resonance decays predominantly into pions and is therefore a
possible candidate for a non-strange quarkonium state. We will discuss this
possibility from the viewpoint of our model in Fit II,
Chapter \ref{ImplicationsFitII}. It is an established experimental fact that
$f_{0}(1370)$ is a broad resonance with $\Gamma_{f_{0}(1370)}\sim(200-500)$
MeV \cite{PDG}. Although the stated value of the decay width is not comparable
to the mass of the resonance, there large width nonetheless needs to be
considered with care when features of the resonance are analysed. One of the
reasons for the large width arises from the fact that $f_{0}(1370)$ is
reconstructed in various decay channels (see below) that may have different
thresholds. For this reason, in this section we will prefer an analysis
combining different sets of data, in various channels and by various
collaborations. The resonance is mostly observed in ${\bar{p}p}$
annihilations, ${\pi}^{-}p$ scattering and $J/\psi$ decays (see below).

There are several reviews offering combined analyses of $f_{0}(1370)$ features
\cite{buggf0,Kaminski:1993,Anisovich-KMatrix,Au:1986,Janssen:1994,Bugg:1994,Tornqvist:1995,Bugg:1996,Anisovich:1996tk,Anisovich:1997zw,Anisovich:2009zza}%
. They are important for at least two reasons. Firstly, they clearly
demonstrate that $f_{0}(600)$ and $f_{0}(1370)$ are distinct resonances
\cite{Kaminski:1993,Au:1986,Janssen:1994,Tornqvist:1995}. Secondly, a broad
resonance with various decay channels -- such as $f_{0}(1370)$ -- is bound to
experience interference among different decay channels due to threshold
openings. These have to be considered within comprehensive reviews combining
different sets of production data (as performed, for example, in references
that have already been stated). In this section, we will in particular
emphasise results from a comprehensive review of $f_{0}(1370)$ by D. Bugg
published in 2007 \cite{buggf0}. Nonetheless, let us first briefly summarise
experimental data where a signal for $f_{0}(1370)$ was seen.

\begin{itemize}
\item \textit{CERN.} A bubble-chamber experiment involving ${\bar
{p}p\rightarrow\pi}^{+}{\pi}^{-}{\pi}^{+}{\pi}^{-}{\pi}^{0}$ (antiprotons at
$1.2$ GeV targeted at hydrogen at rest)\ was analysed in 1969. A possible
$\rho\rho$ enhancement was claimed at $1.4$ GeV \cite{Donald:1969}.\textit{ }

\item \textit{Argonne.} Data were taken from $400000$ events observed at the Argonne National Laboratory
in 1976 from the reaction ${\pi}%
^{-}p$ producing neutrons and neutral pseudoscalar kaons \cite{Cason:1976}. Pions were scattered off a $7.5$ cm-long
liquid-hydrogen target, with ensuing photons (originating from the kaon decays) detected from scintillation
counters in a hodoscope. An enhancement with $\Gamma\sim80$ MeV\ was observed
at approximately $1.25$ GeV but with $I=1$ . Subsequent data from higher
statistics ($110000$ events in ${\pi}^{-}p\rightarrow nK^{-}K^{+}$ and $50000$
events in ${\pi}^{+}n\rightarrow pK^{-}K^{+}$) confirmed the enhancement, but
found it to be rather broad ($\Gamma\sim150$ MeV, at $\sim1.3$ GeV) and with
$I=0$ \cite{Pawlicki:1976af}; see also Ref.\ \cite{Cohen:1980cq}.

\item \textit{BNL.} Data from $15000$ events on $\pi^{-}p\rightarrow{\bar{K}%
}_{S}^{0}K_{S}^{0}n$ taken at Brookhaven National Laboratory suggested a
resonance with a mass of $\sim1463$ MeV and a width of $\Gamma\sim118$ MeV (but with
large errors for the width) \cite{Etkin:1981}.

\item \textit{Crystal Barrel.} The earliest evidence for $f_{0}(1370)$ by the
Crystal Barrel Collaboration was published in 1992 from $\eta\eta$ final
states \cite{f0(1500)-CB-1992}. Data were obtained from the reaction ${\bar
{p}p\rightarrow}\eta\eta\pi^{0}\rightarrow6\gamma$, where the antiprotons were
stopped by a liquid-hydrogen target at the centre of the Crystal Barrel
detector. In this way, ${\bar{p}p}$ annihilation states were limited to $S$
and $P$ waves allowing for a better reconstruction of putative scalar
resonances. The Crystal Barrel itself was in essence comprised of a magnetic
detector with a CsI calorimeter used to detect photons. An optimised fit
suggested the existence of a scalar resonance with a mass of $1430$ MeV and a decay
width of $250$ MeV. The same Collaboration also analysed data from ${\bar
{p}p\rightarrow\pi}^{+}{\pi}^{-}3\pi^{0}\rightarrow\rho^{+}\rho^{-}\pi^{0}$
(antiprotons at $200$ MeV stopped in a 4 cm long liquid-hydrogen target at the
centre of the detector); a strong signal with a mass of $(1374\pm38)$ MeV and
a width of $(375\pm61)$ MeV was reconstructed in both $\rho^{+}\rho^{-}$ and
$\sigma\sigma$ channels \cite{f0(1370)-CB-1994}. Subsequent analysis of data
from ${\bar{p}p\rightarrow3}\pi^{0}$ and ${\bar{p}p\rightarrow\eta\eta}\pi
^{0}$ found both $f_{0}(1370)$ and $f_{0}(1500)$
\cite{Bugg:1994,Anisovich:1994,Anisovich:1994qk}. High-statistics data from
reactions ${\bar{p}p\rightarrow}3\pi^{0}$ ($712000$ events), ${\bar
{p}p\rightarrow}\eta\eta\pi^{0}$ ($280000$ events) and ${\bar{p}p\rightarrow
}\eta\pi^{0}\pi^{0}$ ($198000$ events) were analysed in Refs.\
\cite{Amsler:1995bf,Abele:1996si}. They yielded not only evidence for
$f_{0}(1370)$ and $f_{0}(1500)$ but also for the non-strange isotriplet member
of the scalar nonet above 1 GeV, the $a_{0}(1450)$ resonance, found in the
$\eta\pi^{0}\pi^{0}$ final state \cite{a0(1450)-CB-1994}. Finally, a
simultaneous fit \cite{Bugg:1996} of the Crystal Barrel ${\bar{p}p}$ data with
CERN-Munich data regarding ${\pi}^{-}p{\rightarrow\pi}^{-}{\pi}^{+}n$
\cite{CM:1973} with BNL analyses from Refs.\ \cite{Etkin:1981,Lindenbaum:1991}
and Argonne results from Refs.\ \cite{Cohen:1980cq,Martin:1979} determined that
${\pi\pi}$ scattering data above $1$ GeV require the presence of $f_{0}(1370)$.

\item \textit{Rome-Syracuse. }A review of earlier data in Ref.\
\cite{Gaspero:1992} suggested a resonance with a mass of $(1386\pm10\pm28)$ MeV
and a width of $(310\pm17\pm47)$ MeV.

\item \textit{OBELIX.} The Collaboration utilised reactions induced by
antineutrons; they were produced by the charge-exchange reaction ${\bar
{p}p\rightarrow\bar{n}n}$ in a 15 cm long liquid-hydrogen target inside the
OBELIX spectrometer at CERN-LEAR \cite{OBELIX:1993}. The ensuing beam produced
reactions ${\bar{n}}p\rightarrow{\pi}^{+}{\pi}^{-}{\pi}^{+}$ and ${\bar{n}%
}p\rightarrow{\pi}^{+}{\pi}^{-}{\pi}^{+}{\pi}^{-}{\pi}^{+}$. A scalar state
with a mass of $(1345\pm12)$ MeV and width of $(398\pm26)$ MeV was reconstructed.

\item \textit{Belle.} Recently, the Belle Collaboration\ have claimed
observation of a signal consistent with $f_{0}(1370)$ from $B$ meson decays
produced in $e^{+}e^{-}$ collisions: $e^{+}e^{-}\rightarrow\Upsilon
\rightarrow{\bar{B}}_{S}^{0}B_{S}^{0}${, }${\bar{B}}_{S}^{0}B_{S}^{\ast}$ and
${\bar{B}}_{S}^{\ast}B_{S}^{\ast}$ (with $B_{S}^{\ast}\rightarrow\gamma
B_{S}^{0}$) and{ }$B_{S}^{0}\rightarrow J/\psi f_{0}(1370)\rightarrow
J/\psi\pi^{+}\pi^{-}$. The signal was observed at $1405\pm15_{-7}^{+1}$ MeV;
the width was $54\pm33_{-3}^{+14}$ MeV \cite{Belle:2011}. However, no
interference with the nearby scalar states was considered due to low
statistics, although the observation of a signal for $f_{0}(980)$ was also claimed.
\end{itemize}

However, there are also claims disputing the existence of $f_{0}(1370)$. It
has been claimed that $f_{0}(1370)$ could be merely a broad background in the
energy region up to $1.5$ GeV interfering with $f_{0}(1500)$ and producing a
peak at $1370$ MeV \cite{Ochs:2001}. Additionally, $f_{0}(1370)$ was not
unambiguously identified in the CERN-Munich data of Ref.\ \cite{CM:1973} and
might even, with $f_{0}(600)$, represent a single state - the scalar glueball
\cite{Ochs:1999}; see also Ref.\ \cite{Ochs:2010}. \newline This work will not
follow the assertions of the stated references for several reasons:

\begin{itemize}
\item It is highly unlikely that $f_{0}(600)$ and $f_{0}(1370)$ merely
represent two manifestations of a single state; features of $f_{0}(600)$ have
been discussed in Sec.\ \ref{sec.f0(600)} where we have noted that $\pi\pi$
scattering data unambiguously require a light scalar state corresponding to a
pole in the $\pi\pi$ scattering amplitude. For this reason alone, $f_{0}(600)$
is a state distinct from other resonances, including $f_{0}(1370)$. [We have
already noted at the beginning of this section that Refs.\
\cite{Kaminski:1993,Au:1986,Janssen:1994,Tornqvist:1995} also demonstrate that
$f_{0}(600)$ and $f_{0}(1370)$ are distinct. The same is shown as well in Ref.\
\cite{buggf0} in a simultaneous fit of Crystal Barrel data on ${\bar{p}p}%
$\ and BES II data from $J/\psi$ decays, see below.]

\item The mentioned CERN-Munich data have to be refitted simultaneously with
data that have higher statistics if one would like to make a more elaborate
statement regarding scalar resonances. This has actually been performed in
Ref.\ \cite{Ochs:2010} where, for example, Argonne data on $KK$ scattering from
Ref.\ \cite{Cohen:1980cq} (discussed above) were considered. The Argonne data
possess the largest statistics to date for the $KK$ channel ($\sim10^{5}$
events), as noted in Ref.\ \cite{Ochs:2010}; however, there are even larger
statistics in the $\pi\pi$ channel obtained by Crystal Barrel from ${\bar{p}%
p}$ reactions ($\sim7\cdot10^{5}$ events in $\pi^{0}\pi^{0}$; $\sim
3\cdot10^{5}$ events in $\eta\eta$) \cite{Amsler:1995bf,Abele:1996si}. Reference
\cite{Ochs:2010} does not consider the Crystal Barrel data, although they
appear to be the most definitive ones regarding $f_{0}(1370)$. For this
reason, the ensuing conclusion disputing the existence of $f_{0}(1370)$ is
rather doubtful. The mentioned data from CERN-Munich and Argonne (and also
from BNL \cite{Etkin:1981,Lindenbaum:1991}) were considered in Ref.\
\cite{Bugg:1996}. As discussed above, the combined fit required the presence
of $f_{0}(1370)$ with a mass of $(1300\pm15)$ MeV and a full width of $(230\pm15)$ MeV.

\item The issue whether $f_{0}(1370)$ is actually merely a broad background in
the energy region above $1$ GeV was discussed in some detail in
Ref.\ \cite{buggf0} where it was found to be a genuine resonance. The stated
publication contains various interesting statements regarding $f_{0}(1370)$
and, as a final part of this section, we will discuss the most important ones.
\end{itemize}

Five sets of data have been simultaneously fitted in Ref.\ \cite{buggf0}:
Crystal Barrel data on ${\bar{p}p\rightarrow3\pi}^{0}$ at rest in liquid \cite{Abele:1996si} and
gaseous hydrogen \cite{Anisovich-KMatrix}; Crystal Barrel data on ${\bar
{p}p\rightarrow}\eta\eta\pi^{0}$ at rest in liquid
\cite{f0(1500)-CB-1992,Anisovich:1994,Amsler:1995bz}\ and gaseous hydrogen \cite{Anisovich-KMatrix} and
also BES II data on $J/\psi\rightarrow\varphi\pi^{+}\pi^{-}$
\cite{Ablikim:2004}. The last set of data actually contains a peak at $1.35$
GeV, contributed to interference of $f_{0}(1370)$, $f_{0}(1500)$ and
$f_{2}(1270)$; the corresponding data can be refitted with and without
$f_{0}(1370)$. The main conclusions from the combined fit are:

\begin{itemize}
\item Crystal Barrel data on ${\bar{p}p\rightarrow3\pi}^{0}$ require
$f_{0}(1370)$ as a 32$\sigma$ signal in liquid-hydrogen ${\bar{p}p}$ reactions
($\sigma$: standard deviation) and as a 33$\sigma$ signal in gaseous-hydrogen
${\bar{p}p}$ reactions.

\item Crystal Barrel data on ${\bar{p}p\rightarrow}\eta\eta\pi^{0}$ require
$f_{0}(1370)$ as a 17$\sigma$ signal in liquid-hydrogen ${\bar{p}p}$ reactions
and as an 8$\sigma$ signal in gaseous-hydrogen ${\bar{p}p}$ reactions.

\item BES II data on $J/\psi\rightarrow\varphi\pi^{+}\pi^{-}$ require
$f_{0}(1370)$ as an 8$\sigma$ signal.

\item It is not possible to simulate $f_{0}(1370)$ as a high-tail
representation of $f_{0}(600)$, neither in the ${\pi\pi}$ nor in the $\eta\eta
$\ channels, as the ensuing $\chi^{2}$ fit is noticeably worse than in the case
where $f_{0}(1370)$ is included as a separate resonance.

\item If one fits the $S$-wave $\pi\pi$ scattering amplitude between $1.1$ GeV
and $1.46$ GeV [putative mass range of $f_{0}(1370)$] without assuming a
Breit-Wigner form (i.e., freely in bins of $\pi\pi$ invariant mass), then a
resonance form of the fitted amplitude is still obtained. The resonance is
labelled as $f_{0}(1370)$.

\item CERN-Munich data \cite{CM:1973} can be fitted slightly better with
$f_{0}(1370)$ than without this state but are not definitive in this regard.

\item Due to lack of experimental data on $\pi\pi\rightarrow4\pi$, it is only
possible to constrain $\Gamma_{f_{0}(1370)\rightarrow\pi\pi}/$ $\Gamma
_{f_{0}(1370)\rightarrow4\pi}$ rather than the two decay widths by themselves
($\Gamma$ refers to the Breit-Wigner width). The $2\pi$ line-shape of
$f_{0}(1370)$ can then be fitted with a range of values for both
$\Gamma_{f_{0}(1370)\rightarrow\pi\pi}$ and $\Gamma_{f_{0}(1370)\rightarrow
4\pi}$. The Breit-Wigner width in the $2\pi$ channel optimises the fit at
$\Gamma_{f_{0}(1370)\rightarrow\pi\pi}=325$ MeV and implies $\Gamma
_{f_{0}(1370)\rightarrow4\pi}=(54\pm2\pm5)$ MeV with $m_{f_{0}(1370)}%
=(1309\pm1\pm15)$ MeV. Note, however, that the values of the $2\pi$ and $4\pi$
decay widths are strongly dependent on the value of $m_{f_{0}(1370)}$ at which
they are determined. The reason is that the $4\pi$ phase space increases
rapidly above approximately $1.35$ GeV (see the next point).

\item The quantitative features (mass/width) of $f_{0}(1370)$ are strongly
dependent on the energy range considered. Up to approximately $1.35$ GeV,
the $2\pi$ decay channel of $f_{0}(1370)$ is dominant; thereafter, a rapid rise in
the $4\pi$ phase space and cross-section occurs and thus the $4\pi$ decay
channel becomes dominant and the $2\pi$ contribution decreases rapidly. For
this reason, a proper dispersive analysis of $\pi\pi$ scattering in the energy
region relevant for $f_{0}(1370)$ has to consider contributions from both
$2\pi$ and $4\pi$\ channels but also the $s$-dependence of these channels
($s$: pion invariant mass). This in turn implies that $f_{0}(1370)$ has
\textit{two pion peaks}. The peak in the $2\pi$ channel is at $1282$\ MeV and
possesses a full width at half maximum (FWHM) of $207$ MeV. The resonance mass
at the centre of the FWHM interval is $1269$ MeV. (It does not coincide with
the peak mass because the line shape is not symmetric due to the opening of
the $4\pi$ phase space.) The $4\pi$ peak is shifted by approximately $50$ MeV:
the peak is at $1331$ MeV and possesses a FWHM of $273$ MeV (these results are
in reasonable agreement with $4\pi$ analyses of Refs.\
\cite{f0(1370)-CB-1994,Gaspero:1992,OBELIX:1993}). The resonance mass at the
centre of the FWHM interval is $1377$ MeV -- it is thus shifted by more than
$100$ MeV in comparison with the $2\pi$ channel. We emphasise therefore that
care is needed when one quotes a value of a $f_{0}(1370)$ decay width: the
mass of the resonance always has to be specified as well.\newline Despite
exhibiting the mentioned two peaks, $f_{0}(1370)$ is still a single resonance
for at least two reasons:

\begin{itemize}
\item A combined analysis of both $2\pi$ and $4\pi$ channels yields only one
pole. Depending on the sheet considered, the pole position varies between
$1292$ MeV and $1309$ MeV (close to the $2\pi$ peak because the $s$-dependence
in the $2\pi$ channel is smaller than in the $4\pi$ channel). The pole width
is at average $\sim181$ MeV.

\item Additionally, finding two distinct but near scalar resonances [$\sim(50$
- $100)$ MeV mass difference], one in the $2\pi$ channel and one in the $4\pi$
channel, would appear to violate the well-known level repulsion of states with
the same quantum numbers. Indeed such proximate resonances with the same
quantum numbers are only expected if they possess orthogonal wave functions.
This can obviously not be the case if two hypothetical states were both
reconstructed from pions. Conversely, an example where this may occur is given
by the pair of resonances $f_{0}(1710)$ -- reconstructed predominantly in kaon
final states -- and $f_{0}(1790)$, reconstructed predominantly in pion final
states (see Sections \ref{sec.f0(1710)} and \ref{sec.f0(1790)}). There is
another similar example: the $f_{2}(1565)$ resonance. It possesses a
small-intensity peak in the $\pi\pi$ channel at $1565$ MeV and a
larger-intensity peak in the $\omega\omega$ channel at $1660$ MeV; however, a
dispersive analysis similar to that performed in Ref.\ \cite{buggf0} still
yields a single pole at $1598$ MeV \cite{f2(1565)}.
\end{itemize}
\end{itemize}

The PDG nonetheless accumulate all available data on $f_{0}(1370)$ estimating
$m_{f_{0}(1370)}=(1200$ - $1500)$ MeV and $\Gamma_{f_{0}(1370)}=(200$ - $500)$
MeV \cite{PDG}.

\section{The \boldmath $f_{0}(1500)$ Resonance}

The discovery of the $f_{0}(1500)$ resonance originated in search for the
scalar glueball state. This resonance is found mostly in pion final states
from nucleon-nucleon (or antinucleon-nucleon) and pion-nucleon scattering
processes. If such processes produce four pions, then $f_{0}(1500)$ is
reconstructed from $\rho\rho$ final states in the $2(\pi^{+}\pi^{-})$ channel
and from $\sigma\sigma$ final states in the $2(\pi^{+}\pi^{-})$ or $2(\pi
^{0}\pi^{0})$ channels. The resonance is therefore at least partly
reconstructed in channels containing a double Pomeron exchange rendering the
state a glueball candidate \cite{Close:1987}. Our results from the $U(2)\times
U(2)$ version of the model will confirm this assertion, see Chapter \ref{chapterglueball}. \newline

The $f_{0}(1500)$ resonance was first observed by the Columbia-Syracuse
Collaboration in 1982 \cite{Gray:1983}. A 76-cm bubble chamber at BNL-Columbia
containing deuterium at rest was exposed to antiprotons of various energies
yielding the reactions $\bar{p}+p\rightarrow{\pi}^{0}f_{0}(1500)\rightarrow$
$3{\pi}^{0}$ and $\bar{p}+n\rightarrow{\pi}^{-}f_{0}(1500)\rightarrow{\pi}%
^{-}{\pi}^{+}{\pi}^{-}$. The mass of the resonance was $(1525\pm5)$ MeV; the
decay width was $(101\pm13)$ MeV.\ The new scalar, found in the pion channel,
was determined to be virtually degenerate in mass with the already-known
tensor state $f_{2}^{^{\prime}}(1525)$, produced predominantly in the kaon
channels in the same annihilation process. The state was swiftly confirmed by
the GAMS data (obtained from $\pi^{-}p$ and $\pi^{-}n$ annihilation at IHEP)
in the subsequent few years \cite{f0(1500)GAMS80s,Alde:1985kp}, with mass
values typically $(50-100)$ MeV larger than the value reported by
Columbia-Syracuse. Note that the GAMS Collaboration typically utilised
Breit-Wigner fits, known to shift in different sets of data due to
interference of $f_{0}(1500)$ with the nearby states $f_{0}(1370)$ and
$f_{0}(1710)$. (The same is true for data from $\pi^{-}$Be $\rightarrow
\eta\eta^{\prime}\pi^{-}$Be \cite{VES1} and $\pi^{-}$Be $\rightarrow\eta
\eta\pi^{-}$Be \cite{VES2} from the Vertex Spectrometer VES, also at IHEP.)
Later GAMS publications considered interference effects with $f_{0}(1370)$
\cite{Alde:1998,Bellazzini:1999} and $f_{0}(1710)$ \cite{Binon:2004}. In Ref.\
\cite{Bellazzini:1999}, GAMS data from the reaction $pp\rightarrow p_{f}({\pi}%
^{0}{\pi}^{0})p_{s}\rightarrow$ $p_{f}(4\gamma)p_{s}$ were utilised. The
photons were detected by the Hodoscope Automatic Multiphoton Spectrometer (the
Russian abbreviation for which is, as already indicated, GAMS), momenta of
$p_{f}$ were measured by a magnetic spectrometer with gas chambers while
momenta of $p_{s}$ were measured by a recoil proton detector.

\begin{itemize}
\item \textit{WA76, WA91 and WA102.} A range of data regarding $f_{0}(1500)$
were presented by the WA76, WA91 and WA102 Collaborations using the CERN Omega
Spectrometer. In 1989, data from the reaction $pp\rightarrow p_{f}({\pi}^{+}{\pi
}^{-}{\pi}^{+}{\pi}^{-})p_{s}$ at $300$\ GeV were analysed confirming a scalar
state with a mass of $(1449\pm4)$ MeV and a width of $(78\pm18)$ MeV from a
Breit-Wigner fit \cite{f0(1500)-WA76-1989}. The same reaction and the same
analysis method were used in the subsequent years \cite{f0(1500)-WA91-19945}.
A five-time increase in statistics allowed for a range of resonances between
$1.2$ GeV and $2.0$ GeV to be observed by the WA102 Collaboration. This
included a $J^{PC}=0^{++}$ peak at $1.45$ GeV, found to be a superposition of
two scalar resonances, $f_{0}(1370)$ and $f_{0}(1500)$ \cite{Barberis:1997}.
Subsequently, data from reactions $pp\rightarrow p_{f}(K^{+}K^{-})p_{s}$ and
$pp\rightarrow p_{f}({\bar{K}}_{S}^{0}K_{S}^{0})p_{s}$ \cite{WA102:1999} and
$pp\rightarrow p_{f}(\pi^{+}\pi^{-})p_{s}$ \cite{WA102:1999_1} allowed for a
combined analysis to be performed in both pion and kaon channels
\cite{Barberis:1999}. Both the T-matrix formalism and the K-matrix formalism were used
to obtain results regarding four scalar resonances: $f_{0}(980)$,
$f_{0}(1370)$, $f_{0}(1500)$ and $f_{0}(1710)$. For $f_{0}(1500)$, the mean
obtained values were $m_{f_{0}(1500)}=(1502\pm12\pm10)$ MeV and $\Gamma
_{f_{0}(1500)}=(98\pm18\pm16)$ MeV. The best WA102 data stem from an analysis of
$pp\rightarrow p_{f}({\pi}^{0}{\pi}^{0}{\pi}^{0}{\pi}^{0})p_{s}$,
$pp\rightarrow p_{f}({\pi}^{0}{\pi}^{0}{\pi}^{+}{\pi}^{-})p_{s}$ and
$pp\rightarrow p_{f}({\pi}^{+}{\pi}^{-}{\pi}^{+}{\pi}^{-})p_{s}$ (at $450$
GeV) because they allowed for a study of both $\sigma\sigma$ and $\rho\rho$
contributions to scalars above $1$ GeV. Both decay channels were observed for
$f_{0}(1500)$ with $m_{f_{0}(1500)}=(1511\pm9)$ MeV and $\Gamma_{f_{0}%
(1500)}=(102\pm18)$ MeV \cite{f0(1500)-WA102-2000}. Additionally, data from
$pp\rightarrow p_{f}(\eta\eta)p_{s}$ production (identified via $\eta
\rightarrow\gamma\gamma$ and $\eta\rightarrow\pi^{+}\pi^{-}\pi^{0}$) also
allowed for $f_{0}(1500)$ to be reconstructed \cite{Barberis:2000}, with a pole
position virtually the same as in Ref.\ \cite{f0(1500)-WA102-2000}.

\item \textit{Crystal Barrel.} The earliest evidence for $f_{0}(1500)$ by the
Crystal Barrel Collaboration at CERN-LEAR was published in 1992 from $\eta
\eta$ final states \cite{f0(1500)-CB-1992}. Data were obtained from the reaction
${\bar{p}p\rightarrow}\eta\eta\pi^{0}\rightarrow6\gamma$ involving antiprotons
stopped in liquid hydrogen. Data analysis in Ref.\ \cite{f0(1500)-CB-1992}
suggested $m_{f_{0}(1500)}=(1560\pm25)$ MeV and $\Gamma_{f_{0}(1500)}%
=(245\pm50)$ MeV from a Breit-Wigner fit but no interference with the nearby
scalar states was considered. Further analyses of ${\bar{p}p\rightarrow}%
\pi^{0}\pi^{0}\pi^{0}$ and ${\bar{p}p\rightarrow}\eta\eta\pi^{0}$
\cite{Anisovich:1994}\ as well as ${\bar{p}p\rightarrow}\eta\eta^{\prime}%
\pi^{0}$ \cite{Amsler:1994}\ confirmed the state [with the latter reaction
having a small branching ratio to $f_{0}(1500)$ due to the phase-space
suppression]. Data with highest statistics were analysed in 1995; they
included $16.8$ million ${\bar{p}p}$ collisions. From these collisions,
$712000$ events for ${\bar{p}p\rightarrow3}\pi^{0}$ were selected, see Ref.\
\cite{Amsler:1995gf}. The $f_{0}(1500)$ resonance was reconstructed with
$m_{f_{0}(1500)}=(1500\pm15)$ MeV and $\Gamma_{f_{0}(1500)}=(120\pm25)$ MeV.
Additionally, $198000$ events for ${\bar{p}p\rightarrow\eta\eta}\pi^{0}$ were
selected yielding $m_{f_{0}(1500)}=(1505\pm15)$ MeV and $\Gamma_{f_{0}%
(1500)}=(120\pm30)$ MeV \cite{Amsler:1995bz}. The Collaboration also utilised
the ${\bar{p}p}$ annihilation to study $4\pi^{0}$ decay channel of
$f_{0}(1500)$ via ${\bar{p}p\rightarrow5\pi}^{0}$ in 1996
\cite{f0(1500)-CB-1996}. This channel is nowadays known to represent
approximately $50\%$ of the $f_{0}(1500)$ decays \cite{PDG}. In Ref.\
\cite{f0(1500)-CB-1996}, an enhancement in the $4\pi$ scalar channel was
observed at $1505$ MeV; the width was $169$ MeV. In subsequent work, an
additional production channel was used by the Collaboration: ${\bar
{p}d\rightarrow\pi}^{-}4\pi^{0}p$; in this way, data evaluation was simplified
as only one combination of four pions had to be considered (unlike in
${\bar{p}p\rightarrow5\pi}^{0}$ where four-pion states could be reconstructed
in five different ways from the five pions) \cite{Abele:2001}. Results were
nonetheless consistent with Ref.\ \cite{f0(1500)-CB-1996}. Note that the
Collaboration has also observed subdominant $f_{0}(1500)$ decays into kaons
from ${\bar{p}p\rightarrow}K_{L}^{0}K_{L}^{0}\pi^{0}$ \cite{Abele:1996} and
${\bar{p}p\rightarrow}K^{+}K^{-}\pi^{0}$ \cite{Amsler:2006}, suggesting a
small contribution of $\bar{s}s$ to $f_{0}(1500)$.

\item \textit{OBELIX.} Data taken from the reaction ${\bar{p}p\rightarrow}\pi
^{+}\pi^{-}\pi^{0}$ at CERN-LEAR were analysed in 1997 \cite{OBELIX:1997}.
Three scalar poles were found in a K-matrix formalism: $f_{0}(980)$,
$f_{0}(1370)$ and $f_{0}(1500)$, with $m_{f_{0}(1500)}=(1449\pm20)$ MeV and
$\Gamma_{f_{0}(1500)}=(114\pm30)$ MeV. It was necessary to include
$f_{0}(1500)$ in particular into the fit as the $\chi/$d.o.f. increased from
$1.53$ to $1.71$ if $f_{0}(1500)$ was omitted. The existence of the resonance was
subsequently confirmed by data from $\bar{n}p\rightarrow\pi^{+}\pi^{+}\pi^{-}$
\cite{OBELIX:1998}.
\end{itemize}

The PDG cites a world-average mass $m_{f_{0}(1500)}=(1505\pm6)$ MeV and decay
width $\Gamma_{f_{0}(1710)}=(109\pm7)$ MeV \cite{PDG}.

\section{The \boldmath $f_{0}(1710)$ Resonance}  \label{sec.f0(1710)}

The $f_{0}(1710)$ resonance is of importance for this work because it decays
predominantly into kaons (see below); thus experimental data suggest that it
may be a $\bar{s}s$ state. This is confirmed by our findings in Fit II,
Sec.\ \ref{ImplicationsFitII}. Other approaches suggest that $f_{0}(1710)$ may
possess a large glueball component \cite{f0(1710)asglueball}; however, this
may be in doubt due to the latest ZEUS results that do not exclude coupling of
$f_{0}(1710)$ to photons (see below as well). Experimentally, $f_{0}(1710)$ is
reconstructed in $\pi^{-}p$ and $e^{-}p$ scatterings and $J/\psi$ decays.
\newline

The earliest evidence for the $f_{0}(1710)$ resonance was obtained from the
decay $J/\psi\rightarrow\gamma\eta\eta$ at the SLAC Crystal Ball detector from
$e^{+}e^{-}$ annihilation and published in 1982 \cite{f0(1710)-1981-SLAC}. A
resonance with a mass of $(1640\pm50)$ MeV and a decay width of $220_{-70}^{+100}$ MeV
was found. The resonance was determined to have positive charge-conjugation
quantum number ($C=+1$) because it was produced in a radiative $J/\psi$ decay.
Given that it was reconstructed from two pseudoscalar final states, it
could only have even spin and parity (i.e., $J^{P}=0^{+}$, $2^{+}$,...) and
the initial data analysis in Ref.\ \cite{f0(1710)-1981-SLAC} preferred $J=2$
rather than $J=0$ [but the nearby spin-two state $f_{2}^{\prime}(1525)$ was
omitted from the analysis].

Confirmation of the new resonance was published several months later by the
Brookhaven National Laboratory from data on $\pi^{-}p\rightarrow{\bar{K}}%
_{S}^{0}K_{S}^{0}n$ \cite{f0(1710)-1982-BNL}. Discovery of a resonance with
mass of $(1730\pm10\pm20)$ MeV was claimed; the value of the decay width was
later determined as $200_{-9}^{+156}$ MeV \cite{Etkin:1981}. An isoscalar
resonance of similar mass [$(1650\pm50)$ MeV] and decay width [$(200\pm100)$
MeV] was found by the MARK II Collaboration at SLAC in 1982 from
$J/\psi\rightarrow\gamma\rho^{0}\rho^{0}$ \cite{f0(1710)-1982-MARKII}.
Afterwards, $J/\psi$ decays into etas were used by the Crystal Ball
Collaboration at SLAC to reconstruct the resonance, see Ref.\ \cite{Bloom:1983}%
, confirmed soon thereafter by Ref.\ \cite{Williams:1984} (but with no $J^{P}$ determination).

First DM2 data about this state were published in 1987
\cite{f0(1710)-DM2-1987} citing a resonance at approximately $1.7$ GeV (width:
approximately $140$ MeV). Further analysis yielded a mass $(1707\pm10)$ MeV and
a width $(166.4\pm33.2)$\ MeV\ in 1988 \cite{f0(1710)-DM2-1988}. The two sets of
data were obtained from $J/\psi$ radiative decays into pions and kaons,
respectively. However, no $J^{P}$ determination was possible due to a low
signal-to-background ratio. The same issue prevented a determination of
$J^{P}$ in a further set of data ($J/\psi\rightarrow\varphi\pi^{+}\pi^{-}$,
$\varphi K^{+}K^{-}$, $\varphi{\bar{K}}_{S}^{0}K_{S}^{0}$, $\omega K^{+}K^{-}%
$, $\omega{\bar{K}}_{S}^{0}K_{S}^{0}$, $\varphi\bar{K}K^{\star}$ and
$\varphi{\bar{p}}p$; only decays into kaons were relevant)
\cite{f0(1710)-DM2-1988_1}.

There were several publications claiming this resonance to possess $J=2$
rather than $J=0$ (see Ref.\ \cite{f0(1710)asJ=2,f01370KK2} in addition to
Ref.\ \cite{f0(1710)-1981-SLAC}). However, more recent data suggest that the
resonance is spin-zero:

\begin{itemize}
\item \textit{GAMS}. Experiments regarding $\eta\eta$\ final states, performed
at the GAMS-IHEP proton synchrotron from $\pi^{-}p\rightarrow\eta\eta
n\rightarrow(4\gamma)n$ reactions, suggested a mass of $(1755\pm8)$ MeV and
a width $<50$ MeV in 1986 \cite{f0(1710)-GAMS-1986}. In 1992, an improved
version of the same experiment, considering also reactions $\pi^{-}%
p\rightarrow\eta\eta n^{\ast}\rightarrow(4\gamma)n\pi^{0}$ and $\pi
^{-}p\rightarrow\eta\eta n^{\ast}\rightarrow(4\gamma)n\pi^{0}\pi^{0}$ (where
$n^{\ast}$ denotes an excited neutron), allowed for analysis of new data combined
with the old 1986 data. The Collaboration obtained a mass of $(1744\pm15)$ MeV
and a width $<80$ MeV at 90\% CL \cite{f0(1710)-GAMS-1992}. The resonance was
found to possess $J=0$ already in 1986; the 1992 data suggested that it does
not decay into $\eta\eta^{\prime}$ or $\pi^{0}\pi^{0}$. A Breit-Wigner fit of
GAMS data yielded mass of $(1670\pm20)$ MeV and width $(260\pm50)$ MeV in 2005
\cite{Binon:2004}.

\item \textit{MARK-III.} In 1986 and 1992 the MARK-III Collaboration at SLAC
\cite{MARKIII} published results regarding pion production from decays
$e^{+}e^{-}\rightarrow J/\psi\rightarrow\gamma\pi^{+}\pi^{-}\pi^{+}\pi^{-}$
and $e^{+}e^{-}\rightarrow J/\psi\rightarrow\gamma\pi^{+}\pi^{0}\pi^{-}\pi
^{0}$ claiming the discovery of two pseudoscalar $\rho\rho$ states at $1.55$
and $1.8$ GeV. Similar results were published in 1989 by the DM2 Collaboration
at DCI-Orsay \cite{DM2} where the discovery of three $\eta$-like states in the
region between $1.4$ GeV and $2.2$ GeV\ was claimed (see also results by the
E760 Collaboration at Fermilab published in 1993 \cite{E760}). The MARK-III
and DM2 Collaborations made use only of $^{3}S_{1}$ $\pi\pi$ final states
($\rho$). A re-analysis of MARK-III data was performed in 1995 by the Crystal
Barrel Collaboration \cite{f0(1790)-1995-MARKIII}; here, scalar $^{3}P_{0}$
$\pi\pi$ final states ($\sigma$)\ were considered in addition to the vector
$^{3}S_{1}$ states. A different picture emerged: no pseudoscalar peak was
found (there was a $0^{-}$ signal over the entire energy range between $1.6$ GeV
and $2.4$ GeV but no clear resonance); inclusion of the $\sigma$-like $\pi\pi$
final states yielded a new scalar resonance, denoted\ as $f_{0}(1750)$ with
$m_{f_{0}(1750)}=(1750\pm15)$ MeV and $\Gamma_{f_{0}(1750)}=(160\pm40)$ MeV.
The resonance was found to decay predominantly into $\sigma$ mesons; decay
into $\rho$ states was found to be approximately $4.5$ times less probable.
The mass of this resonance is close to the PDG-preferred value $m_{f_{0}%
(1710)}=(1720\pm6)$ MeV but nonetheless appears to be too large when compared
to the value of $m_{f_{0}(1710)}$ accepted nowadays. Thus the mentioned
MARK-III result may be viewed as evidence that there exists a scalar resonance
between $1.7$ GeV and $1.8$ GeV; however, it may also be viewed as
superposition of two distinct states: $f_{0}(1710)$ and $f_{0}(1790)$, with
evidence for the latter state discussed in the next section.

\item \textit{BES.} Results consistent with the MARK-III reanalysis of Ref.\
\cite{f0(1790)-1995-MARKIII} were obtained by the BES Collaboration at BEPC
(Beijing Electron Positron Collider) in 2000 \cite{f0(1790)-2000-BES} where an
$I(J^{PC})=0(0^{++})$ resonance labelled as $f_{0}(1740)$ with $m_{f_{0}%
(1740)}=1740_{-25}^{+30}$ MeV and $\Gamma_{f_{0}(1740)}=120_{-40}^{+50}$ MeV
was found. This result was again obtained in the decay channel $e^{+}%
e^{-}\rightarrow J/\psi\rightarrow\gamma\pi^{+}\pi^{-}\pi^{+}\pi^{-}$. Note,
however, that these data involved no kaon decays of $J/\psi$. Subsequently,
the same collaboration performed an analysis of a larger number of $J/\psi$ decay
channels: $J/\psi\rightarrow\gamma K^{+}K^{-}$, $\omega K^{+}K^{-}$ and
$\varphi K^{+}K^{-}$ as well as $J/\psi\rightarrow\gamma\pi^{+}\pi^{-}\pi
^{+}\pi^{-}$, $\omega\pi^{+}\pi^{-}\pi^{+}\pi^{-}$ and $\varphi\pi^{+}\pi^{-}$
\cite{f0(1790)-2000-BES_1}. The kaon channels allowed for reconstruction of
the $f_{0}(1710)$ resonance [referred to as $f_{0}(1710\pm20)$] while the pion
channels suggested the existence of a separate $f_{0}(1760\pm20)$ resonance. Thus
a resonance with a mass distinct from $f_{0}(1710)$ appeared to have been found;
it was also produced in different decay channels [i.e., those involving pions
whereas the $f_{0}(1710)$ resonance was reconstructed predominantly in kaon
final states]. This new resonance was later denoted as $f_{0}(1790)$, see next section.

\item \textit{BES II.} An upgrade of BEPC allowed for 58M of $J/\psi$ events to be
collected at BES II. In 2003, $f_{0}(1710)$ was confirmed as a $J^{P}=0^{+}$
resonance reconstructed in kaon final states from $J/\psi\rightarrow\gamma
K^{+}K^{-}$ and $\gamma{\bar{K}}_{S}^{0}K_{S}^{0}$ with $m_{f_{0}%
(1710)}=1740\pm4_{-25}^{+10}$ MeV and $\Gamma_{f_{0}(1710)}=166_{-8-10}%
^{+5+15}$ MeV \cite{f0(1710)-2003-BESII}. These results were confirmed in 2004
from $J/\psi\rightarrow\omega K^{+}K^{-}$ \cite{BESII2004}. Additional
evidence for the existence of $f_{0}(1710)$ was presented in Ref.\
\cite{f0(1790)-2005-BESII} from the $\chi_{c0}\rightarrow\pi^{+}\pi^{-}%
K^{+}K^{-}$ decay; however, this analysis also suggested the existence of a
further scalar state between $1.7$ GeV and $1.8$ GeV, referred to as
$f_{0}(1790)$.

\item \textit{WA102.} Experiments involving $pp$ collisions at $450$ GeV were
performed by the WA102 Collaboration. Final states were reconstructed from
reactions $pp\rightarrow p_{f}(K^{+}K^{-})p_{s}$ and $pp\rightarrow
p_{f}({\bar{K}}_{S}^{0}K_{S}^{0})p_{s}$ (subscripts $f$ and $s$ denote the
fastest and slowest protons in the laboratory frame, respectively). Results
for the $f_{0}(1710)$ resonance suggested $m_{f_{0}(1710)}=(1730\pm15)$ MeV
and $\Gamma_{f_{0}(1710)}=(100\pm25)$ MeV \cite{WA102:1999}. The same
experiment also allowed for resonances in the pion final states to be looked
for \cite{WA102:1999_1}. The corresponding reaction $pp\rightarrow p_{f}%
(\pi^{+}\pi^{-})p_{s}$ allowed for the reconstruction of $f_{0}(980)$,
$f_{0}(1370)$ and $f_{0}(1500)$ but the fit of $f_{0}(1710)$ was conspicuously
worse than in $pp\rightarrow p_{f}(K^{+}K^{-})p_{s}$. The two stated
publications presented results of respective Breit-Wigner fits.
A coupled-channel analysis of both pion and kaon final states yielded a pole at
$m_{f_{0}(1710)}=(1727\pm12\pm11)$ MeV and $\Gamma_{f_{0}(1710)}=(126\pm
16\pm18)$ MeV \cite{Barberis:1999}. Note that a subsequent $T$-matrix analysis
\cite{Barberis:2000} of $pp\rightarrow p_{f}(\eta\eta)p_{s}$ production data
(identified via $\eta\rightarrow\gamma\gamma$ and $\eta\rightarrow\pi^{+}%
\pi^{-}\pi^{0}$) yielded results very close to those of Ref.\
\cite{Barberis:1999}.\ Note also that all the mentioned results implied
$J^{P}=0^{+}$ for $f_{0}(1710)$.

\item \textit{ZEUS.} Electrons at $27.5$ GeV were collided with protons at
$820$ GeV and $920$ GeV at the HERA storage ring in Hamburg (DESY) during the
1996-2000 running period. Reactions were observed using the ZEUS detector and
${\bar{K}}_{S}^{0}K_{S}^{0}$ final states were studied. The $f_{0}(1710)$ was
observed from a $5\sigma$ $J^{P}=0^{+}$ signal yielding $m_{f_{0}%
(1710)}=(1701\pm5_{-2}^{+9})$ MeV and $\Gamma_{f_{0}(1710)}=(100\pm
24_{-22}^{+7})$ MeV \cite{ZEUS:2008}. However, the Collaboration was not able
to exclude the coupling of $f_{0}(1710)$ to photons, implying that this resonance
is not a certain candidate for a predominantly glueball state. Indeed
calculations in our model prefer $f_{0}(1710)$ to be predominantly strange
quarkonium, as we will discuss in Sec.\ \ref{ImplicationsFitI}.
\end{itemize}

The PDG cites a world-average mass $m_{f_{0}(1710)}=(1720\pm6)$ MeV and a decay
width $\Gamma_{f_{0}(1710)}=(135\pm8)$ MeV \cite{PDG}. A more detailed
discussion of the $f_{0}(1710)$ decay channels can be found in
Sec.\ \ref{f0(1710)channels}.

\section{The Peculiar Case of \boldmath $f_{0}(1790)$} \label{sec.f0(1790)}

Our results in Fit II, Chapter \ref{sec.fitII}, will suggest the existence of an
$I(J^{PC})=0(0^{++})$, predominantly $\bar{s}s$ state in the energy region of
approximately $1.6$ - $1.7$ GeV. Assignment of this state to an experimentally
established resonance will depend on decay patterns of our model state;
however, experimental results regarding the $I(J^{PC})=0(0^{++})$ channel in
this energy region are far from clear [although admittedly the issue is less
ambiguous than in the case of $f_{0}(600)$]. The reason is that the existence of
two distinct resonances is claimed within an energy interval of only $100$
MeV: in addition to $f_{0}(1710)$, the BES II Collaboration \cite{Ablikim:2004}
have claimed that a state labelled as $f_{0}(1790)$ also exists. In the following we will
discuss data regarding this resonance; if $f_{0}(1790)$ does
exist, then the disentanglement of data regarding this state from those
regarding the close-by state $f_{0}(1710)$ in experimental observations
becomes imperative, as entangled data are bound to lead to results that could
certainly be described as peculiar (or at least such that they should not be used in models and
theories). Note that data sets regarding $f_{0}(1790)$ -- described in the
following -- provide us with a rather straightforward tool to distinguish this
resonance from $f_{0}(1710)$: the $f_{0}(1790)$ resonance decays predominantly
into pions and only marginally into kaons. For $f_{0}(1710)$, the opposite is
true. Thus a careful analysis of experimental data should be able to
discriminate between these two states. [Additionally, the feature of
predominantly decaying into pions and the mass difference to $f_{0}(1370)$
qualify $f_{0}(1790)$ as a putative radial excitation of $f_{0}(1370)$.]
\newline

Experimental evidence for $f_{0}(1790)$ reads as follows:

\begin{itemize}
\item \textit{MARK-III / Crystal Barrel.} In the previous section we have
already discussed the re-analysis of MARK-III data performed in 1995 by the
Crystal Barrel Collaboration \cite{f0(1790)-1995-MARKIII}. We have indicated
that a careful analysis of the mentioned data yields a scalar resonance
denoted\ as $f_{0}(1750)$ with $m_{f_{0}(1750)}=(1750\pm15)$ MeV and
$\Gamma_{f_{0}(1750)}=(160\pm40)$ MeV. However, the mass of this resonance
appears to be too large to describe $f_{0}(1710)$ alone; it appears more
probable that the mentioned data actually yield a superposition of the
$f_{0}(1710)$ resonance (the existence of which may be regarded as proven)
with a putative new resonance (the existence of which would require more
experimental data, discussed in the following). Therefore, the central value
of the mass of this resonance may indicate a superposition of a state denoted
nowadays as $f_{0}(1790)$ with $f_{0}(1710)$.

\item \textit{Crystal Barrel. }The Crystal Barrel Collaboration also published
analysis of data from the reaction ${\bar{p}}p\rightarrow\eta\eta\pi^{0}$ in
1999 \cite{f0(1790)-1999-CrystalBarrel}. An $8\sigma$ signal was found; the
resonance was referred to as $f_{0}(1770)$ with $m_{f_{0}(1770)}=(1770\pm12)$
MeV and $\Gamma_{f_{0}(1750)}=(220\pm40)$ MeV.

\item \textit{BES. }As mentioned in the previous section, in 2000 the BES
Collaboration claimed the existence of a resonance denoted as $f_{0}%
(1760\pm20)$, found to be distinct from the $f_{0}(1710)$\ state
\cite{f0(1790)-2000-BES_1}. Factors of distinction involved not only the mass but
also production channels: $f_{0}(1710)$ appeared in the $J/\psi\rightarrow
\gamma K^{+}K^{-}$, $\omega K^{+}K^{-}$ and $\varphi K^{+}K^{-}$ channels
while the $f_{0}(1760\pm20)$ was reconstructed in $J/\psi\rightarrow\gamma
\pi^{+}\pi^{-}\pi^{+}\pi^{-}$, $\omega\pi^{+}\pi^{-}\pi^{+}\pi^{-}$ and
$\varphi\pi^{+}\pi^{-}$. More conclusive evidence for a second resonance
between $1.7$ and $1.8$ GeV was obtained by the BES II Collaboration (see
below). Note that already in 1996 the BES Collaboration claimed the existence
of a scalar resonance with a mass of $1781\pm8_{-31}^{+10}$\ MeV and a width of
$85\pm24_{-19}^{+22}$ MeV that appeared to correspond well to a resonance at
$1.79$ GeV [i.e., to the putative $f_{0}(1790)$ resonance] but was, however,
found in the $J/\psi\rightarrow\gamma K^{+}K^{-}$\ channel only
\cite{Bai:1996}. Thus results of Ref.\ \cite{Bai:1996} did not include $J/\psi$
decays in the pion channels and possible interference effects with kaons;
therefore they need to be considered with care.

\item \textit{BES II} -- $J/\psi$\textit{.} An upgrade of BEPC allowed for 58M
of $J/\psi$ events to be collected at BES II. A clear $f_{0}(1790)$ peak
corresponding to a 15$\sigma$ signal was observed in the $J/\psi
\rightarrow\varphi\pi^{+}\pi^{-}$ decay \cite{Ablikim:2004} yielding
$m_{f_{0}(1790)}=1790_{-30}^{+40}$ MeV and $\Gamma_{f_{0}(1790)}%
=270_{-30}^{+60}$ MeV. This is the best available set of data on $f_{0}%
(1790)$. Conversely, the $f_{0}(1710)$ resonance was observed in the
$J/\psi\rightarrow\varphi K^{+}K^{-}$ channel confirming this resonance as
decaying predominantly into kaons [$m_{f_{0}(1710)}$ and $\Gamma_{f_{0}%
(1710)}$ were fixed to the PDG values]. An additional reason for the assertion
that there exist two scalar states in the region between $1.7$ GeV and $1.8$
GeV was presented in Ref.\ \cite{Ablikim:2004}. As already mentioned, a fit of
the $J/\psi\rightarrow\varphi\pi^{+}\pi^{-}$ data allows for determination of
the $f_{0}(1790)$ mass and width. Let us assume that $f_{0}(1710)$ and
$f_{0}(1790)$ actually represent the same resonance and let us denote this
resonance as $\tilde{f}_{0}$ -- i.e., let $\tilde{f}_{0}$ be the only
$I(J^{PC})=0(0^{++})$ resonance between $1.7$ GeV and $1.8$ GeV. We can then
remove (artificially) the $f_{0}(1710)$ resonance from the $J/\psi
\rightarrow\varphi K^{+}K^{-}$ data. This yields the branching ratio
$\Gamma_{\tilde{f}_{0}\rightarrow\pi\pi}/\Gamma_{\tilde{f}_{0}\rightarrow
KK}=1.82\pm0.33$ (in addition to a poorer fit). However, according to Ref.\
\cite{BESII2004} the same ratio for a scalar state between $1.7$ GeV and
$1.8$\ GeV should possess a value $<0.11$, obtained from different production
channels: $J/\psi\rightarrow\omega\pi^{+}\pi^{-}$ and $J/\psi\rightarrow\omega
K^{+}K^{-}$. A single resonance must possess the same value of a branching
ratio in all production channels -- in our case regardless whether it is
produced in $J/\psi\rightarrow\varphi\pi^{+}\pi^{-}$ and $J/\psi
\rightarrow\varphi K^{+}K^{-}$ or in $J/\psi\rightarrow\omega\pi^{+}\pi^{-}$
and $J/\psi\rightarrow\omega K^{+}K^{-}$. For the case of the assumed single
scalar resonance $\tilde{f}_{0}$ between $1.7$ GeV and $1.8$ GeV this is
obviously not true: the branching ratios differ by at least a factor of $17$.
Therefore, there must exist two distinct resonances.

\item \textit{BES II} -- $\chi_{c0}$\textit{. }Additional confirmation of the
$f_{0}(1790)$ state can be found in Ref.\ \cite{f0(1790)-2005-BESII} from a BES
II analysis of the $\chi_{c0}\rightarrow\pi^{+}\pi^{-}K^{+}K^{-}$ decay; see
also Ref.\ \cite{Bugg:2006}.
\end{itemize}

Thus there appears to be sufficient evidence for existence of a sixth
isoscalar resonance below $1.9$ GeV, $f_{0}(1790)$, in addition to
$f_{0}(600)$, $f_{0}(980)$, $f_{0}(1370)$, $f_{0}(1500)$ and $f_{0}(1710)$.
The relative vicinity of $f_{0}(1790)$ to $f_{0}(1710)$ makes it imperative to
consider carefully and, if necessary, to disentangle published results
regarding both resonances. We illustrate this point in the following section
where the partial decay widths of $f_{0}(1710)$ are determined.

\section{Consequences for the \boldmath $f_{0} (1710)$ Decay Channels} \label{f0(1710)channels}

Considering the experimental ambiguities discussed in the previous section, let us
now discuss numerical values regarding the $f_{0}(1710)$ decay channels. The
PDG \cite{PDG} lists five decay channels for this resonance: $f_{0}%
(1710)\rightarrow KK$, $\pi\pi$, $\eta\eta$, $\gamma\gamma$ and $\omega\omega
$. The existence of the decay $f_{0}(1710)\rightarrow\omega\omega$ was determined
only recently by the BES II Collaboration in 2006 \cite{Ablikim:2006}. No
precise determination of the branching ratio was possible because the decay
was reconstructed from the reaction $J/\psi\rightarrow\gamma\omega\omega$, yielding
a strong pseudoscalar contribution and rather weak scalar and tensor
contributions. There is no published value of the corresponding $\Gamma
_{f_{0}(1710)\rightarrow\omega\omega}$ that is expected to be small. The
latter is also true for $\Gamma_{f_{0}(1710)\rightarrow\gamma\gamma}$. We
therefore consider only the first three decays: into kaons, pions and etas:%

\begin{align}
\Gamma_{f_{0}(1710)}  &  \equiv\Gamma_{f_{0}(1710)\rightarrow KK}%
+\Gamma_{f_{0}(1710)\rightarrow\pi\pi}+\Gamma_{f_{0}(1710)\rightarrow\eta\eta
}\nonumber\\
&  =\Gamma_{f_{0}(1710)\rightarrow\pi\pi}\left[  1+\frac{\Gamma_{f_{0}%
(1710)\rightarrow KK}}{\Gamma_{f_{0}(1710)\rightarrow\pi\pi}}+\frac
{\Gamma_{f_{0}(1710)\rightarrow\eta\eta}}{\Gamma_{f_{0}(1710)\rightarrow\pi
\pi}}\right]  \text{.} \label{G_f0(1710)_1}%
\end{align}

In the next three subsections we will calculate decay widths of $f_{0}(1710)$
in various channels using the experimentally known ratios $\Gamma
_{f_{0}(1710)\rightarrow\pi\pi}/\Gamma_{f_{0}(1710)\rightarrow KK}$ and
$\Gamma_{f_{0}(1710)\rightarrow\eta\eta}/\Gamma_{f_{0}(1710)\rightarrow KK}$.
In Section \ref{f0(1710)-PDG-BESII}\ we discuss implications of data on
$f_{0}(1710)$ preferred by the PDG; they include $\Gamma_{f_{0}%
(1710)\rightarrow\pi\pi}/\Gamma_{f_{0}(1710)\rightarrow KK}=0.41_{-0.17}%
^{+0.11}$ from the BES II Collaboration \cite{f0(1710)-2006-BESII} as well as
the ratio $\Gamma_{f_{0}(1710)\rightarrow\eta\eta}/\Gamma_{f_{0}%
(1710)\rightarrow KK}=0.48\pm0.15$ from the WA102 Collaboration
\cite{Barberis:2000} (the latter experiments performed at CERN Omega
Spectrometer). Subsection \ref{aa} contains analogous calculation with the
alternative BES II ratio $\Gamma_{f_{0}(1710)\rightarrow\pi\pi}/\Gamma
_{f_{0}(1710)\rightarrow KK}<0.11$ \cite{BESII2004} (not used by the PDG) but
retaining the WA102 ratio $\Gamma_{f_{0}(1710)\rightarrow\eta\eta}%
/\Gamma_{f_{0}(1710)\rightarrow KK}=0.48\pm0.15$. In Subsection \ref{bb} we
use only the WA102 ratios $\Gamma_{f_{0}(1710)\rightarrow\pi\pi}/\Gamma
_{f_{0}(1710)\rightarrow KK}=0.2\pm0.024\pm0.036$ \cite{Barberis:1999} and
$\Gamma_{f_{0}(1710)\rightarrow\eta\eta}/\Gamma_{f_{0}(1710)\rightarrow
KK}=0.48\pm0.15$.

Note that there are also corresponding results from a combined fit in Ref.\
\cite{Anisovich:2001} that, however, do not constrain the $2\pi/2K$ ratio very
well: $\Gamma_{f_{0}(1710)\rightarrow\pi\pi}/\Gamma_{f_{0}(1710)\rightarrow
KK}=5.8_{-5.5}^{+9.1}$. Additionally, there are older data from the WA76
Collaboration at CERN \cite{Armstrong:1991} reading $\Gamma_{f_{0}%
(1710)\rightarrow\pi\pi}$ $/\,\Gamma_{f_{0}(1710)\rightarrow KK} =0.39\pm0.14$; these are qualitatively consistent with results of Ref.\
\cite{Barberis:1999} and therefore omitted from our discussion.

\subsection{The \boldmath $f_{0}(1710)$ Decay Widths from Data Preferred by
the PDG}

\label{f0(1710)-PDG-BESII}

As already mentioned, the BES II \cite{f0(1710)-2006-BESII}\ ratio cited by
the PDG reads $\Gamma_{f_{0}(1710)\rightarrow\pi\pi}/\Gamma_{f_{0}%
(1710)\rightarrow KK}\equiv\Gamma_{f_{0}(1710)\rightarrow\pi\pi}^{\text{PDG}%
}/\Gamma_{f_{0}(1710)\rightarrow KK}^{\text{PDG}}=0.41_{-0.17}^{+0.11}$. Two
comments are in order for this result. Firstly, data used to extract the
stated ratio ($J/\psi\rightarrow\gamma\pi^{+}\pi^{-}$ and $\gamma\pi^{0}%
\pi^{0}$) suffer from a large background in the $\pi^{+}\pi^{-}$ channel (of
approximately 50\%). This raises doubts about the reliability of the ratio.
Additionally, the ratio was obtained for a scalar resonance with a mass of
$1765_{-3}^{+4}$ MeV and width of $(145\pm8\pm69)$ MeV. Although the resonance
may be assigned to $f_{0}(1710)$ (due to the value of its width; the mass is
too large), the mass of the resonance appears to suggest that it could also be
a superposition of $f_{0}(1710)$ and $f_{0}(1790)$ rather than representing
only a signal for $f_{0}(1710)$. This possibility was also discussed by the
Collaboration itself \cite{f0(1710)-2006-BESII}. Therefore, the stated ratio
for $\Gamma_{f_{0}(1710)\rightarrow\pi\pi}/\Gamma_{f_{0}(1710)\rightarrow KK}$
has to be regarded with care. Indeed we will also consider alternative values
of the $\Gamma_{f_{0}(1710)\rightarrow\pi\pi}/\Gamma_{f_{0}(1710)\rightarrow
KK}$ ratio, such as for example a more reliable result of $\Gamma
_{f_{0}(1710)\rightarrow\pi\pi}/\Gamma_{f_{0}(1710)\rightarrow KK}<0.11$ from
Ref.\ \cite{BESII2004}, also by the BES II Collaboration (see Subsection
\ref{aa}).

Despite the mentioned drawbacks, let us discuss the consequences of
$\Gamma_{f_{0}(1710)\rightarrow\pi\pi}/\Gamma_{f_{0}(1710)\rightarrow
KK}\equiv\Gamma_{f_{0}(1710)\rightarrow\pi\pi}^{\text{PDG}}/\Gamma
_{f_{0}(1710)\rightarrow KK}^{\text{PDG}}=0.41_{-0.17}^{+0.11}$. The inverse
ratio $\Gamma_{f_{0}(1710)\rightarrow KK}/\Gamma_{f_{0}(1710)\rightarrow\pi
\pi}$ has the central value of $2.44$. The error value $\Delta(\Gamma
_{f_{0}(1710)\rightarrow KK}/\Gamma_{f_{0}(1710)\rightarrow\pi\pi})$ is
obtained from%

\begin{align}
\frac{\Gamma_{f_{0}(1710)\rightarrow KK}}{\Gamma_{f_{0}(1710)\rightarrow\pi
\pi}}  &  \equiv\frac{1}{\frac{\Gamma_{f_{0}(1710)\rightarrow\pi\pi}}%
{\Gamma_{f_{0}(1710)\rightarrow KK}}}\Rightarrow\Delta\frac{\Gamma
_{f_{0}(1710)\rightarrow KK}}{\Gamma_{f_{0}(1710)\rightarrow\pi\pi}%
}=\left\vert -\frac{\Delta\frac{\Gamma_{f_{0}(1710)\rightarrow\pi\pi}}%
{\Gamma_{f_{0}(1710)\rightarrow KK}}}{\left[  \frac{\Gamma_{f_{0}%
(1710)\rightarrow\pi\pi}}{\Gamma_{f_{0}(1710)\rightarrow KK}}\right]  ^{2}%
}\right\vert \nonumber\\
&  \Rightarrow\Delta\frac{\Gamma_{f_{0}(1710)\rightarrow KK}}{\Gamma
_{f_{0}(1710)\rightarrow\pi\pi}}=_{-1.01}^{+0.65}\text{.} \label{f0(1710)_2}%
\end{align}

Thus, in total:%

\begin{equation}
\frac{\Gamma_{f_{0}(1710)\rightarrow KK}}{\Gamma_{f_{0}(1710)\rightarrow\pi
\pi}}\equiv\frac{\Gamma_{f_{0}(1710)\rightarrow KK}^{\text{PDG}}}%
{\Gamma_{f_{0}(1710)\rightarrow\pi\pi}^{\text{PDG}}}=2.44_{-1.01}%
^{+0.65}\text{.} \label{f0(1710)_3}%
\end{equation}

The PDG also uses the ratio $\Gamma_{f_{0}(1710)\rightarrow\eta\eta}%
/\Gamma_{f_{0}(1710)\rightarrow KK}=0.48\pm0.15$, published originally by the
WA102 Collaboration in 2000 \cite{Barberis:2000}. Then we obtain for the
central value of the ratio $\Gamma_{f_{0}(1710)\rightarrow\eta\eta}%
/\Gamma_{f_{0}(1710)\rightarrow\pi\pi}$%

\begin{equation}
{\overline{\frac{\Gamma_{f_{0}(1710)\rightarrow\eta\eta}}{\Gamma
_{f_{0}(1710)\rightarrow\pi\pi}}}}={\overline{\frac{\Gamma_{f_{0}%
(1710)\rightarrow\eta\eta}}{\Gamma_{f_{0}(1710)\rightarrow KK}}}}\frac
{1}{{\overline{\frac{\Gamma_{f_{0}(1710)\rightarrow\pi\pi}}{\Gamma
_{f_{0}(1710)\rightarrow KK}}}}}=\frac{0.48}{0.41}=1.17\text{.}
\label{f0(1710)_11}%
\end{equation}

(Note that the line above the observable denotes the central value.) Additionally,%

\begin{align}
\frac{\Gamma_{f_{0}(1710)\rightarrow\eta\eta}}{\Gamma_{f_{0}(1710)\rightarrow
\pi\pi}}  &  =\frac{\Gamma_{f_{0}(1710)\rightarrow\eta\eta}}{\Gamma
_{f_{0}(1710)\rightarrow KK}}\frac{\Gamma_{f_{0}(1710)\rightarrow KK}}%
{\Gamma_{f_{0}(1710)\rightarrow\pi\pi}}\nonumber\\
\Rightarrow\Delta\frac{\Gamma_{f_{0}(1710)\rightarrow\eta\eta}}{\Gamma
_{f_{0}(1710)\rightarrow\pi\pi}}  &  =\sqrt{\left[  \frac{\Gamma
_{f_{0}(1710)\rightarrow\eta\eta}}{\Gamma_{f_{0}(1710)\rightarrow KK}%
}\,\,\Delta\frac{\Gamma_{f_{0}(1710)\rightarrow KK}}{\Gamma_{f_{0}%
(1710)\rightarrow\pi\pi}}\right]  ^{2}+\left[  \frac{\Gamma_{f_{0}%
(1710)\rightarrow KK}}{\Gamma_{f_{0}(1710)\rightarrow\pi\pi}}\,\,\Delta
\frac{\Gamma_{f_{0}(1710)\rightarrow\eta\eta}}{\Gamma_{f_{0}(1710)\rightarrow
KK}}\right]  ^{2}} \label{f0(1710)_4}%
\end{align}

and consequently from Eq.\ (\ref{f0(1710)_2}):%

\begin{equation}
\frac{\Gamma_{f_{0}(1710)\rightarrow\eta\eta}}{\Gamma_{f_{0}(1710)\rightarrow
\pi\pi}}\equiv\frac{\Gamma_{f_{0}(1710)\rightarrow\eta\eta}^{\text{PDG}}%
}{\Gamma_{f_{0}(1710)\rightarrow\pi\pi}^{\text{PDG}}}=1.17_{-0.61}%
^{+0.48}\text{.} \label{f0(1710)_5}%
\end{equation}

From Eq.\ (\ref{G_f0(1710)_1}) we obtain%

\begin{equation}
\Gamma_{f_{0}(1710)\rightarrow\pi\pi}=\frac{\Gamma_{f_{0}(1710)}}%
{1+\frac{\Gamma_{f_{0}(1710)\rightarrow KK}}{\Gamma_{f_{0}(1710)\rightarrow
\pi\pi}}+\frac{\Gamma_{f_{0}(1710)\rightarrow\eta\eta}}{\Gamma_{f_{0}%
(1710)\rightarrow\pi\pi}}} \label{f0(1710)_6}%
\end{equation}

and thus, given that $\Gamma_{f_{0}(1710)}=(135\pm8)$ MeV \cite{PDG},
Eqs.\ (\ref{f0(1710)_3}) and (\ref{f0(1710)_5}) yield the central value
${\overline{\Gamma}}_{f_{0}(1710)\rightarrow\pi\pi}=29.28$ MeV. The
corresponding error value $\Delta\Gamma_{f_{0}(1710)\rightarrow\pi\pi}$ is
obtained from Eq.\ (\ref{f0(1710)_6}) as follows:%

\begin{align}
\Delta\Gamma_{f_{0}(1710)\rightarrow\pi\pi}  &  =\left\{  \left[  \frac
{\Delta\Gamma_{f_{0}(1710)}}{1+\frac{\Gamma_{f_{0}(1710)\rightarrow KK}%
}{\Gamma_{f_{0}(1710)\rightarrow\pi\pi}}+\frac{\Gamma_{f_{0}(1710)\rightarrow
\eta\eta}}{\Gamma_{f_{0}(1710)\rightarrow\pi\pi}}}\right]  ^{2}+\left\{
\frac{\Gamma_{f_{0}(1710)}\,\,\Delta\frac{\Gamma_{f_{0}(1710)\rightarrow
\eta\eta}}{\Gamma_{f_{0}(1710)\rightarrow\pi\pi}}}{\left[  1+\frac
{\Gamma_{f_{0}(1710)\rightarrow KK}}{\Gamma_{f_{0}(1710)\rightarrow\pi\pi}%
}+\frac{\Gamma_{f_{0}(1710)\rightarrow\eta\eta}}{\Gamma_{f_{0}%
(1710)\rightarrow\pi\pi}}\right]  ^{2}}\right\}  ^{2}\right. \nonumber\\
&  \left.  +\left\{  \frac{\Gamma_{f_{0}(1710)}\,\,\Delta\frac{\Gamma
_{f_{0}(1710)\rightarrow KK}}{\Gamma_{f_{0}(1710)\rightarrow\pi\pi}}}{\left[
1+\frac{\Gamma_{f_{0}(1710)\rightarrow KK}}{\Gamma_{f_{0}(1710)\rightarrow
\pi\pi}}+\frac{\Gamma_{f_{0}(1710)\rightarrow\eta\eta}}{\Gamma_{f_{0}%
(1710)\rightarrow\pi\pi}}\right]  ^{2}}\right\}  ^{2}\right\}  ^{\frac{1}{2}}
\label{f0(1710)_7}%
\end{align}

Equations (\ref{f0(1710)_3}), (\ref{f0(1710)_5}) and (\ref{f0(1710)_7}) yield%

\begin{equation}
\Delta\Gamma_{f_{0}(1710)\rightarrow\pi\pi}=_{-7.69}^{+5.42}\text{MeV.}%
\end{equation}

Thus,%

\begin{equation}
\Gamma_{f_{0}(1710)\rightarrow\pi\pi}\equiv\Gamma_{f_{0}(1710)\rightarrow
\pi\pi}^{\text{PDG}}=29.28_{-7.69}^{+5.42}\text{ MeV.} \label{f0(1710)_8}%
\end{equation}

We obtain from Eqs.\ (\ref{f0(1710)_3}) and (\ref{f0(1710)_8}) for
$\Gamma_{f_{0}(1710)\rightarrow KK}$:%

\begin{align}
{\overline{\Gamma}}_{f_{0}(1710)\rightarrow KK}  &  ={\overline{\frac
{\Gamma_{f_{0}(1710)\rightarrow KK}}{\Gamma_{f_{0}(1710)\rightarrow\pi\pi}}}%
}{\overline{\Gamma}}_{f_{0}(1710)\rightarrow\pi\pi}\label{f0(1710)_8a}\\
&  \Rightarrow{\overline{\Gamma}}_{f_{0}(1710)\rightarrow KK}=71.44\text{
MeV.}%
\end{align}

Error values $\Delta\Gamma_{f_{0}(1710)\rightarrow KK}$ are obtained from%

\begin{align}
\Delta\Gamma_{f_{0}(1710)\rightarrow KK}  &  =\left[  \left(  \Gamma
_{f_{0}(1710)\rightarrow\pi\pi}\,\,\Delta\frac{\Gamma_{f_{0}(1710)\rightarrow
KK}}{\Gamma_{f_{0}(1710)\rightarrow\pi\pi}}\right)  ^{2}+\left(  \frac
{\Gamma_{f_{0}(1710)\rightarrow KK}}{\Gamma_{f_{0}(1710)\rightarrow\pi\pi}%
}\,\,\Delta\Gamma_{f_{0}(1710)\rightarrow\pi\pi}\right)  ^{2}\right]
^{\frac{1}{2}}\label{f0(1710)_8b}\\
&  \Rightarrow\Delta\Gamma_{f_{0}(1710)\rightarrow KK}=_{-35.02}%
^{+23.18}\text{MeV.}%
\end{align}

Thus in total:%

\begin{equation}
\Gamma_{f_{0}(1710)\rightarrow KK}\equiv\Gamma_{f_{0}(1710)\rightarrow
KK}^{\text{PDG}}=71.44_{-35.02}^{+23.18}\text{ MeV.} \label{f0(1710)_9}%
\end{equation}

Analogously, in the $f_{0}(1710)\rightarrow\eta\eta$ channel we obtain from
Eqs.\ (\ref{f0(1710)_5}) and (\ref{f0(1710)_8}):%

\begin{align}
{\overline{\Gamma}}_{f_{0}(1710)\rightarrow\eta\eta}  &  ={\overline
{\frac{\Gamma_{f_{0}(1710)\rightarrow\eta\eta}}{\Gamma_{f_{0}(1710)\rightarrow
\pi\pi}}}}{\overline{\Gamma}}_{f_{0}(1710)\rightarrow\pi\pi}%
\label{f0(1710)_9b}\\
&  \Rightarrow{\overline{\Gamma}}_{f_{0}(1710)\rightarrow\eta\eta}=34.26\text{
MeV.}%
\end{align}

while the error values $\Delta\Gamma_{f_{0}(1710)\rightarrow\eta\eta}$ are
obtained from%

\begin{align}
\Delta\Gamma_{f_{0}(1710)\rightarrow\eta\eta}  &  =\left[  \left(
\Gamma_{f_{0}(1710)\rightarrow\pi\pi}\,\,\Delta\frac{\Gamma_{f_{0}%
(1710)\rightarrow\eta\eta}}{\Gamma_{f_{0}(1710)\rightarrow\pi\pi}}\right)
^{2}+\left(  \frac{\Gamma_{f_{0}(1710)\rightarrow\eta\eta}}{\Gamma
_{f_{0}(1710)\rightarrow\pi\pi}}\,\,\Delta\Gamma_{f_{0}(1710)\rightarrow\pi
\pi}\right)  ^{2}\right]  ^{\frac{1}{2}}\label{f0(1710)_9a}\\
&  \Rightarrow\Delta\Gamma_{f_{0}(1710)\rightarrow KK}=_{-20.0}^{+15.42}%
\text{MeV.}%
\end{align}

Therefore,%

\begin{equation}
\Gamma_{f_{0}(1710)\rightarrow\eta\eta}\equiv\Gamma_{f_{0}(1710)\rightarrow
\eta\eta}^{\text{PDG}}=34.26_{-20.0}^{+15.42}\text{ MeV.} \label{f0(1710)_10}%
\end{equation}

\subsection{The \boldmath $f_{0}(1710)$ Decay Widths from Alternative BES II
Data}

\label{aa}

Results for $\Gamma_{f_{0}(1710)\rightarrow\pi\pi}$, $\Gamma_{f_{0}%
(1710)\rightarrow KK}$ and $\Gamma_{f_{0}(1710)\rightarrow\eta\eta}$, stated
respectively in Eqs.\ (\ref{f0(1710)_8}), (\ref{f0(1710)_9}) and
(\ref{f0(1710)_10}), depend among others on the result $\Gamma_{f_{0}%
(1710)\rightarrow\pi\pi}/\Gamma_{f_{0}(1710)\rightarrow KK}=0.41_{-0.17}%
^{+0.11}$\ from Ref.\ \cite{f0(1710)-2006-BESII}, the reliability of which was
discussed at the beginning of Sec.\ \ref{f0(1710)-PDG-BESII}. In this
subsection we discuss implications of an alternative (and more reliable) BES
II result:%
\begin{equation}
\frac{\Gamma_{f_{0}(1710)\rightarrow\pi\pi}}{\Gamma_{f_{0}(1710)\rightarrow
KK}}\equiv\frac{\Gamma_{f_{0}(1710)\rightarrow\pi\pi}^{\text{BES II}}}%
{\Gamma_{f_{0}(1710)\rightarrow KK}^{\text{BES II}}}<0.11\text{.}
\label{f0(1710)_12}%
\end{equation}

This result implies%
\begin{equation}
\frac{\Gamma_{f_{0}(1710)\rightarrow KK}}{\Gamma_{f_{0}(1710)\rightarrow\pi
\pi}}\equiv\frac{\Gamma_{f_{0}(1710)\rightarrow KK}^{\text{BES II}}}%
{\Gamma_{f_{0}(1710)\rightarrow\pi\pi}^{\text{BES II}}}>9.09\text{.}
\label{f0(1710)_13}%
\end{equation}

From $\Gamma_{f_{0}(1710)\rightarrow\eta\eta}/\Gamma_{f_{0}(1710)\rightarrow
KK}=0.48$ \cite{Barberis:2000} we obtain%
\begin{equation}
\frac{\Gamma_{f_{0}(1710)\rightarrow\eta\eta}}{\Gamma_{f_{0}(1710)\rightarrow
\pi\pi}}\equiv\frac{\Gamma_{f_{0}(1710)\rightarrow\eta\eta}^{\text{BES II}}%
}{\Gamma_{f_{0}(1710)\rightarrow\pi\pi}^{\text{BES II}}}=\frac{\Gamma
_{f_{0}(1710)\rightarrow\eta\eta}}{\Gamma_{f_{0}(1710)\rightarrow KK}}\frac
{1}{\frac{\Gamma_{f_{0}(1710)\rightarrow\pi\pi}}{\Gamma_{f_{0}%
(1710)\rightarrow KK}}}>\frac{0.48}{0.11}=4.36\text{.} \label{f0(1710)_14}%
\end{equation}

We will restrain from calculating errors because the ratio $\Gamma
_{f_{0}(1710)\rightarrow\pi\pi}/\Gamma_{f_{0}(1710)\rightarrow KK}<0.11$
provides us only with an upper boundary and no error information. The condition
(\ref{f0(1710)_14}) implies%
\begin{equation}
\frac{\Gamma_{f_{0}(1710)\rightarrow\pi\pi}}{\Gamma_{f_{0}(1710)\rightarrow
\eta\eta}}\equiv\frac{\Gamma_{f_{0}(1710)\rightarrow\pi\pi}^{\text{BES II}}%
}{\Gamma_{f_{0}(1710)\rightarrow\eta\eta}^{\text{BES II}}}<0.23\text{.}
\label{f0(1710)_15}%
\end{equation}

From conditions (\ref{f0(1710)_6}), (\ref{f0(1710)_13}) and (\ref{f0(1710)_14}%
) we obtain%
\begin{equation}
\Gamma_{f_{0}(1710)\rightarrow\pi\pi}\equiv\Gamma_{f_{0}(1710)\rightarrow
\pi\pi}^{\text{BES II}}<9.34\text{ MeV.} \label{f0(1710)_16}%
\end{equation}

Note that\ a similar calculation of an interval for $\Gamma_{f_{0}%
(1710)\rightarrow KK}$\ would yield%
\begin{equation}
\Gamma_{f_{0}(1710)\rightarrow KK}=\frac{\Gamma_{f_{0}(1710)\rightarrow KK}%
}{\Gamma_{f_{0}(1710)\rightarrow\pi\pi}}\Gamma_{f_{0}(1710)\rightarrow\pi\pi
}\text{;}%
\end{equation}

then constraining $\Gamma_{f_{0}(1710)\rightarrow KK}$ is not possible because
condition (\ref{f0(1710)_13}) determines the lower boundary for $\Gamma
_{f_{0}(1710)\rightarrow KK}/\Gamma_{f_{0}(1710)\rightarrow\pi\pi}$ and,
contrarily, condition (\ref{f0(1710)_16}) suggests the upper boundary for
$\Gamma_{f_{0}(1710)\rightarrow\pi\pi}$. For analogous reasons, a calculation of
$\Gamma_{f_{0}(1710)\rightarrow\eta\eta}$ is also not possible. However, given
that $\Gamma_{f_{0}(1710)}\equiv\Gamma_{f_{0}(1710)\rightarrow KK}%
+\Gamma_{f_{0}(1710)\rightarrow\pi\pi}+\Gamma_{f_{0}(1710)\rightarrow\eta\eta
}$, the condition (\ref{f0(1710)_16}) leads to $125.66$ MeV $<\Gamma
_{f_{0}(1710)\rightarrow KK}+\Gamma_{f_{0}(1710)\rightarrow\eta\eta}<135$ MeV
$\equiv\Gamma_{f_{0}(1710)}$.

\subsection{The \boldmath $f_{0}(1710)$ Decay Widths from WA102 Data}

\label{bb}

The WA102 result \cite{Barberis:1999}%
\begin{equation}
\frac{\Gamma_{f_{0}(1710)\rightarrow\pi\pi}}{\Gamma_{f_{0}(1710)\rightarrow
KK}}\equiv\frac{\Gamma_{f_{0}(1710)\rightarrow\pi\pi}^{\text{WA102}}}%
{\Gamma_{f_{0}(1710)\rightarrow KK}^{\text{WA102}}}=0.2\pm0.024\pm
0.036\equiv0.2\pm0.06 \label{f0(1710)_17}%
\end{equation}

implies that the central value of the inverse ratio $\Gamma_{f_{0}%
(1710)\rightarrow KK}/\Gamma_{f_{0}(1710)\rightarrow\pi\pi}$ is $5.0$.
Corresponding error values are calculated using the first line of
Eq.\ (\ref{f0(1710)_2}):%
\begin{equation}
\Delta\frac{\Gamma_{f_{0}(1710)\rightarrow KK}}{\Gamma_{f_{0}(1710)\rightarrow
\pi\pi}}=1.5\text{.}%
\end{equation}

Thus in total we obtain%
\begin{equation}
\frac{\Gamma_{f_{0}(1710)\rightarrow KK}}{\Gamma_{f_{0}(1710)\rightarrow\pi
\pi}}\equiv\frac{\Gamma_{f_{0}(1710)\rightarrow KK}^{\text{WA102}}}%
{\Gamma_{f_{0}(1710)\rightarrow\pi\pi}^{\text{WA102}}}=5.0\pm1.5\text{.}
\label{f0(1710)_18}%
\end{equation}

From $\Gamma_{f_{0}(1710)\rightarrow\eta\eta}/\Gamma_{f_{0}(1710)\rightarrow
KK}=0.48\pm0.15$ \cite{Barberis:2000} we obtain%
\begin{equation}
{\overline{\frac{\Gamma_{f_{0}(1710)\rightarrow\eta\eta}}{\Gamma
_{f_{0}(1710)\rightarrow\pi\pi}}}}={\overline{\frac{\Gamma_{f_{0}%
(1710)\rightarrow\eta\eta}}{\Gamma_{f_{0}(1710)\rightarrow KK}}}}\frac
{1}{{\overline{\frac{\Gamma_{f_{0}(1710)\rightarrow\pi\pi}}{\Gamma
_{f_{0}(1710)\rightarrow KK}}}}}=\frac{0.48}{0.2}=2.4\text{.}%
\end{equation}

Then Eq.\ (\ref{f0(1710)_4}) yields%
\begin{equation}
\frac{\Gamma_{f_{0}(1710)\rightarrow\eta\eta}}{\Gamma_{f_{0}(1710)\rightarrow
\pi\pi}}\equiv\frac{\Gamma_{f_{0}(1710)\rightarrow\eta\eta}^{\text{WA102}}%
}{\Gamma_{f_{0}(1710)\rightarrow\pi\pi}^{\text{WA102}}}=2.4\pm1.04\text{.}
\label{f0(1710)_19}%
\end{equation}

Given that $\Gamma_{f_{0}(1710)}=(135\pm8)$ MeV, we obtain the central value
${\overline{\Gamma}}_{f_{0}(1710)\rightarrow\pi\pi}=16.1$ MeV from
Eq.\ (\ref{f0(1710)_6}). The error is calculated from Eq.\ (\ref{f0(1710)_6});
we obtain $\Delta\Gamma_{f_{0}(1710)\rightarrow\pi\pi}=3.6$ MeV. Thus in total%
\begin{equation}
\Gamma_{f_{0}(1710)\rightarrow\pi\pi}\equiv\Gamma_{f_{0}(1710)\rightarrow
\pi\pi}^{\text{WA102}}=(16.1\pm3.6)\text{ MeV.} \label{f0(1710)_20}%
\end{equation}

Equations (\ref{f0(1710)_8a}), (\ref{f0(1710)_18}) and (\ref{f0(1710)_20}) yield
${\overline{\Gamma}}_{f_{0}(1710)\rightarrow KK}=80.5$ MeV whereas from
Eq.\ (\ref{f0(1710)_8b}) we obtain $\Delta\Gamma_{f_{0}(1710)\rightarrow
KK}=30.1$ MeV. In total:%
\begin{equation}
\Gamma_{f_{0}(1710)\rightarrow KK}\equiv\Gamma_{f_{0}(1710)\rightarrow
KK}^{\text{WA102}}=(80.5\pm30.1)\text{ MeV.} \label{f0(1710)_21}%
\end{equation}

Finally, from Eqs.\ (\ref{f0(1710)_9b}), (\ref{f0(1710)_19}) and
(\ref{f0(1710)_20}) we obtain ${\overline{\Gamma}}_{f_{0}(1710)\rightarrow
\eta\eta}=38.6$ MeV while Eq.\ (\ref{f0(1710)_9a}) yields $\Delta\Gamma
_{f_{0}(1710)\rightarrow\eta\eta}=18.8$ MeV. In total:%
\begin{equation}
\Gamma_{f_{0}(1710)\rightarrow\eta\eta}\equiv\Gamma_{f_{0}(1710)\rightarrow
\eta\eta}^{\text{WA102}}=(38.6\pm18.8)\text{ MeV.} \label{f0(1710)_22}%
\end{equation}

We note from Eq.\ (\ref{f0(1710)_20}) that $\Gamma_{f_{0}(1710)\rightarrow
\pi\pi}$ is approximately by a factor of two smaller in the WA102 data than in
the combined BES II/WA102 data that lead to $\Gamma_{f_{0}(1710)\rightarrow
\pi\pi}=29.28_{-7.69}^{+5.42}$ MeV, Eq.\ (\ref{f0(1710)_8}). This is due to
the difference of the $\Gamma_{f_{0}(1710)\rightarrow\pi\pi}/\Gamma
_{f_{0}(1710)\rightarrow KK}$ ratios\ from Refs.\ \cite{Barberis:1999} 
and \cite{f0(1710)-2006-BESII}.\newline

Therefore, we are now in possession of three distinct sets of data regarding
decay widths of $f_{0}(1710)$: those preferred by the PDG
[Eqs.\ (\ref{f0(1710)_3}), (\ref{f0(1710)_5}), (\ref{f0(1710)_8}),
(\ref{f0(1710)_9}) and (\ref{f0(1710)_10})], those from BES II, not used by
the PDG [Eqs.\ (\ref{f0(1710)_12}) - (\ref{f0(1710)_16})] and those from WA102
[Eqs.\ (\ref{f0(1710)_17}), (\ref{f0(1710)_18}), (\ref{f0(1710)_19}),
(\ref{f0(1710)_20}), (\ref{f0(1710)_21}) and (\ref{f0(1710)_22})]. We will
discuss implications of these results for our model in
Sec.\ \ref{sec.sigmakaonkaon2}.

\end{fmffile}

%% file: Construction.tex
\chapter{Construction of a Meson Model} \label{chapterC}

\section{General Remarks}

We have seen in Chapter \ref{sec.QCD}\ that QCD possesses an exact
$SU(3)_{c}$ local gauge symmetry (the colour symmetry) and an approximate
global $U(N_{f})_{R}\times U(N_{f})_{L}$ symmetry for $N_{f}$ massless quark
flavours (the chiral symmetry). For sufficiently low temperature and density,
quarks and gluons are confined inside colourless hadrons [i.e., $SU(3)_{c}$
invariant configurations]. Thus, it is the chiral symmetry which predominantly
determines hadronic interactions in the low-energy region. However, QCD is
strongly non-perturbative in the low-energy region as is evident from the running
coupling $g^{2}(\mu)$, Eq.\ (\ref{gmu}). Thus, in the non-perturbative regime,
effective theories and models based on the features of the QCD Lagrangian are utilised.

Effective field theories which contain hadrons as degrees of freedom rather
than quarks and gluons have been developed along two lines which differ in the
way in which chiral symmetry is realised: linear \cite{gellmanlevy} and
non-linear \cite{weinberg}. In the non-linear realisation, the so-called
non-linear sigma model, the scalar states are integrated out, leaving the
pseudoscalar states as the only degrees of the freedom. On the other hand, in
the linear representation of the symmetry, the so-called linear sigma model,
both the scalar and pseudoscalar degrees of freedom are present.

In this work, we consider the linear representation of chiral symmetry. Let us
discuss the reasons.

\begin{itemize}
\item \textit{Chiral partners.} The linear sigma model contains not only
pseudoscalar states but also their so-called \textit{chiral partners} from the
onset. The definition of the chiral partners requires us to introduce a
quantum number denoted as $G$-parity [next to the parity $P$ (\ref{qP}) and
charge conjugation $C$ (\ref{qC})]. To this end, consider a special case of
the flavour transformation (here exemplary for two flavours)
\end{itemize}

\[
q_{f}=\left(
\begin{array}
[c]{c}%
u\\
d
\end{array}
\right)  \longrightarrow q_{f}^{\prime}=U_{2}\left(
\begin{array}
[c]{cc}%
u & \\
d &
\end{array}
\right)  \text{,}%
\]

where $U_{2}=\exp(i\pi t_{2})$. Then the $G$-parity operator is defined as
$G=C\cdot U_{2}$ with the value of the corresponding quantum number calculated
from \cite{Halzen}%

\begin{equation}
G=(-1)^{L+S+I}
\end{equation}

where $I$ denotes the isospin. [Remember Eqs.\ (\ref{Pc}) and (\ref{Cc}) for the parity $P$ and the
charge conjugation $C$.] The $G$-parity is defined in such a way that it
possesses true eigenvectors, e.g., for pions%

\begin{align}
G\,|\pi^{0}\rangle\,  &  =-|\pi^{0}\rangle\text{,}\\
G\,|\pi^{+}\rangle &  =-|\pi^{+}\rangle\text{,}\\
G\,|\pi^{-}\rangle &  =-|\pi^{-}\rangle\text{,}
\end{align}

unlike the charge conjugation that, per definition, flips the charge of the
state concerned:%

\begin{align}
C\,|\pi^{+}\rangle &  =-|\pi^{-}\rangle\text{,}\\
C\,|\pi^{-}\rangle &  =-|\pi^{+}\rangle\text{.}%
\end{align}

Note that the $G$-parity is also conserved under strong interactions. [It is
slightly broken, e.g., in the decay $\omega(782)\rightarrow\pi^{+}\pi^{-}$ --
the branching ratio is $\sim1.53\%$ \cite{PDG}.]

Then we define the chiral partners as states that have the same quantum
numbers with the exception of parity and $G$-parity -- for example, the scalar
states sigma and pion are chiral partners, see Eqs.\ (\ref{pionA}) and
(\ref{sigmaA}). The particular version of the model constructed in this work
will in addition also include vector mesons and their chiral partners, the
axial-vectors. For example, the vector state $\rho$ and the axial-vector state
$a_{1}$ are chiral partners [see Eqs.\ (\ref{rhoA}) and (\ref{a1A})]. Thus the
existence of the chiral partners is a consequence of an exactly realised QCD
chiral symmetry (see Sec.\ \ref{sec.SSB}).

\begin{itemize}
\item \textit{Extensions. }The linear sigma model can be extended
straightforwardly to a larger number of flavours. This chapter will see the
construction of a sigma model with two quark flavours (light quarks $u$ and
$d$). The extension of the model to three flavours ($u$, $d$, $s$) will be
presented in Chapters \ref{sec.remarks} -- \ref{ImplicationsFitII}. The model
can also be extended to four flavours ($u$, $d$, $s$, $c$) to account for the
abundant meson spectrum around 2 GeV \cite{Walaa}. The extension of the model
to include a scalar glueball state will be presented in Chapter
\ref{chapterglueball}. The model presented in this work contains mesons up to
spin 1. It can, however, also be extended to include tensor mesons
\cite{Giacosa:2005t}. Additionally, the model can be extended to include a
pseudoscalar glueball \cite{Walaa,StaniD}, tetraquarks, i.e., ${\bar{q}\bar
{q}qq}$ mesons \cite{Achim} and the nucleon and its chiral partner
\cite{Susanna,Susanna-Mainz}.

\item \textit{Non-zero temperatures and densities.} Although this thesis will
be concerned with meson phenomenology in vacuum, the model can be readily
extended to $T\neq0\neq\mu$ to study the chiral phase transition, the critical
point of QCD or matter at finite densities
\cite{RS,Achim,Tneq0,Gallas00,Lenaghan}.
\end{itemize}

A model based on QCD must, of course, implement features of the QCD Lagrangian
demonstrated in Chapter \ref{sec.QCD}. Let us summarise these features now.

\begin{itemize}
\item \textit{Colour symmetry.} The $SU(3)_{c}$\ gauge symmetry is one of the
basic features of QCD. It is an exact symmetry of the QCD Lagrangian (see
Sec.\ \ref{sec.QCDL}). In accordance with the confinement hypothesis, all the
states in our model have to be colour-neutral. As we will be working with
$\bar{q}q$ meson states, the confinement will be trivially fulfilled. Note,
however, that the model will contain no order parameter for
deconfinement.

\item \textit{Chiral symmetry.} As we have discussed in Sec.\ \ref{sec.SSB},
the QCD Lagrangian with $N_{f}$ quark flavours possesses a $U(N_{f})_{L}\times
U(N_{f})_{R}$ chiral symmetry. This symmetry is exact in the limit of
vanishing quark masses and it has to be considered in any field theory or
model based on QCD.

\item \textit{Spontaneous breaking of the chiral symmetry.} Experimental data
in vacuum (and at sufficiently low temperatures and densities of matter)
demonstrate that the chiral $U(N_{f})_{L}\times U(N_{f})_{R}\equiv
U(1)_{V}\times U(1)_{A}\times SU(N_{f})_{V}\times SU(N_{f})_{A}$ symmetry is
broken spontaneously by a non-vanishing expectation value of the quark
condensate (\ref{chiralc}): $\langle\bar{q}q\rangle$ = $\langle\bar{q}%
_{R}q_{L}+\bar{q}_{L}q_{R}\rangle\neq0$. As we have seen in
Sec.\ \ref{sec.SSB}, this symmetry breaking leads to the emergence of
$N_{f}^{2}-1$ pseudoscalar Goldstone bosons. The scalar
states representing the chiral partners of the Goldstone bosons remain massive. For $N_{f}%
=2$, the three lightest meson states, the pions, are identified with these
Goldstone bosons of QCD. They will be present as explicit degrees of freedom
in our model (together with scalar, vector and axial-vector states).

\item \textit{Chiral anomaly.} As we have seen in Sec.\ \ref{sec.CM}, the
$U(N_{f})_{L}\times U(N_{f})_{R}$ symmetry is broken by quantum effects to
$U(1)_{V}\times SU(N_{f})_{V}\times SU(N_{f})_{A}$ [the $U(1)_{A}$ anomaly
(\ref{Asc1})]. The chiral-anomaly term will allow us to generate the splitting
of mass between the pions and the $\eta$ meson (as well as $\eta^{\prime}$ in
Chapter \ref{sec.remarks}).

\item \textit{Explicit breaking of the chiral symmetry.} The explicit breaking
of the axial symmetry $SU(N_{f})_{A}$ is due to non-zero quark masses. The
vector symmetry $SU(N_{f})_{V}$ is broken by non-zero, non-degenerate quark
masses. Our model will therefore contain terms proportional to quark masses to
implement this symmetry-breaking mechanism.

\item $CPT$. QCD also possesses discrete symmetries such as charge
conjugation (\textit{C}), parity (\textit{P}) and time reversal (\textit{T})
symmetry (\textit{CPT}), which are to a very good precision separately
conserved by strong interactions. We have demonstrated in
Sec.\ \ref{sec.othersymmetries} that the $CP$-invariance is a feature of the
QCD Lagrangian; therefore, according to the famous $CPT$ theorem (see
Ref.\ \cite{Greenberg:2002} and references therein), QCD is also
$T$-invariant. This fact offers further constraints in the construction of
effective models of QCD as all the terms in such models have to be $CPT$ invariant.
\end{itemize}

In Chapter \ref{chapterQ} we will study the $N_{f}=2$ version of a linear sigma model which
contains scalar ($\sigma_{N}$, $\vec{a}_{0}$) and pseudoscalar ($\eta_{N}$,
$\vec{\pi}$), and in addition also vector ($\omega_{N}$, $\vec{\rho}$) and
axial-vector ($f_{1N}$, $\vec{a}_{1}$) degrees of freedom. Usually, such
models are constructed under the requirement of local chiral invariance
$U(N_{f})_{R}\times U(N_{f})_{L}$, with the exception of the vector meson mass
term which renders the local symmetry a global one \cite{GG,KR}. In a slight
abuse of terminology, we will refer to these models as locally chirally
invariant models in the following.

As shown in Refs.\ \cite{GG,Paper1,KR,Lissabon,Mainz,Meissner}, the
locally invariant linear sigma model fails to simultaneously describe meson
decay widths and pion-pion scattering lengths in vacuum. As outlined in
Ref.\ \cite{Paper1}, there are at least two ways to solve this issue. One way
is to utilise a model in which the (up to the vector meson mass term) local
invariance of the theory is retained while higher-order terms are added to the
Lagrangian \cite{GG,KR,Meissner}. The second way which is pursued here is the
following: we construct a \textbf{linear sigma model with \emph{global} chiral
invariance} containing all terms up to naive scaling dimension four
\cite{UBW}, see also Ref.\ \cite{Boguta}. (Note that the chiral symmetry of the QCD Lagrangian is also a
global one.) The global invariance allows for additional terms to appear in
our Lagrangian in comparison to the locally invariant case presented,
e.g.,\ in Ref.\ \cite{RS}. (We remark that, introducing a dilaton field, one
can argue \cite{Susanna,dynrec,Stani} that chirally invariant terms of higher
order than scaling dimension four should be absent. The consequences of the
dilaton-field introduction will be discussed in Chapter \ref{chapterglueball}.)

We have to distinguish between two possible assignments for the scalar fields
$\sigma_{N}=(\bar{u}u+\bar{d}d)/\sqrt{2}$ and $a_{0}^{0}=(\bar{u}u-\bar
{d}d)/\sqrt{2}$:

\begin{itemize}
\item They may be identified with $f_{0}(600)$ and $a_{0}(980)$ which are
members of a nonet that in addition consists of $f_{0}(980)$ and $K_{0}%
^{\star}(800)$.

\item They may be identified with $f_{0}(1370)$ and $a_{0}(1450)$ which are
members of a multiplet that also consists of $f_{0}(1500)$,
$f_{0}(1710)$, and $K_{0}^{\star}(1430)$, where the additional
scalar-isoscalar state emerges from the admixture of a glueball field
\cite{AmslerClose,refs2,refs,longglueball}.
\end{itemize}

In the second assignment, scalar mesons below 1 GeV
are not (predominantly) quark-antiquark states. Their spectroscopic wave
functions might contain a dominant tetraquark or mesonic molecular
contribution \cite{Jaffeq2q2,fariborz,refs3,SchechterQ,Giacosa:2006tf}. The correct
assignment of the scalar quark-antiquark fields of the model to physical
resonances is not only important as a contribution to the ongoing debate about
the nature of these resonances, but it is also vital for a study of the
properties of hadrons at nonzero temperature and density, where the chiral
partner of the pion plays a crucial role \cite{Achim}.

It is important to stress that the theoretical $\sigma_{N}$ and $a_{0}$ fields
entering the linear sigma model describe pure quark-antiquark states, just as
all the other fields. This property can be easily proven by using well-known
large-$N_{c}$ results (see Sec.\ \ref{sec.largen} and Ref.\ \cite{largenc}):
the mass and the decay widths of both $\sigma_{N}$ and $a_{0}$ fields scale in
the model as $N_{c}^{0}$ and $N_{c}^{-1}$, respectively.


\section{The Lagrangian with Global Chiral Symmetry}

\label{sec.c}

In this section we conduct the construction of a linear sigma model with
vector and axial-vector mesons in two flavours. The model is constructed based
on the requirements from the QCD Lagrangian discussed in the previous section.
This chapter will discuss the model construction in the meson sector; for a
discussion regarding the model construction, e.g., in the nucleon sector, see
Ref.\ \cite{FrancescoH}. \newline

We first note that all the states in our model will be hadrons, i.e.,
colour-neutral. Thus the confinement hypothesis and the $SU(3)_{c}$ colour
symmetry of the QCD will be fulfilled per construction. Note, however, that
the model parameters will depend on the number of colours ($N_{c}$), as
discussed in Sec.\ \ref{sec.largen}.\newline

The basic step in the construction of our model is the definition of the meson matrix%

\begin{equation}
\Phi_{ij}\equiv\sqrt{2}\bar{q}_{j,R}q_{i,L}. \label{Phiij}%
\end{equation}

The equivalence sign in Eq.\ (\ref{Phiij}) states merely that $\Phi_{ij}$ and
$\bar{q}_{j,R}q_{i,L}$ transform in the same manner under the (left-handed and
right-handed) chiral groups. It is not to be comprehended as the statement
that $\Phi_{ij}$ contains perturbative $\bar{q}q$ pairs as the matrix
$\Phi_{ij}$ is a non-perturbative object. The reason is that the perturbative
(\textit{bare}) quarks are non-perturbatively modified in vacuum due to their
strong interaction and the interaction with gluons. The ensuing
non-perturbative (or constituent) quarks are then, in a good approximation,
elements of the matrix $\Phi_{ij}$. It is actually possible to connect
$\Phi_{ij}$ with the perturbative currents $\bar{q}_{j,R}q_{i,L}$ by rendering
$\Phi_{ij}$ non-local:%

\begin{equation}
\Phi_{ij} \equiv \sqrt{2}\int\text{d}^{4}y \, \bar{q}_{j,R}\left(  x+\frac{y}%
{2}\right)  q_{i,L}\left(  x-\frac{y}{2} \right)  f(y) \label{Phiij1}%
\end{equation}

where $f(y)$ denotes a non-perturbative vertex function and the perturbative
limit is, of course, obtained by setting $f(y)=\delta(y)$. It is clear from
Eq.\ (\ref{Phiij1}) that the global flavour transformations are the same for
the non-perturbative object $\Phi_{ij}$ and the perturbative quarks.
Considering our discussion in the previous section regarding the chiral
symmetry and its breaking mechanisms, it is clear that the transformation
behaviour of the objects in our model will be pivotal for the model
construction. Thus given our interest in the transformation behaviour only, it
is then sufficient to start with the equivalence $\Phi_{ij}\equiv\sqrt{2}%
\bar{q}_{j,R}q_{i,L}$.

Considering transformation properties of the quarks (\ref{qfLt}) and
(\ref{qfRt}), we obtain immediately that the matrix $\Phi$ transforms as%

\begin{equation}
\Phi\rightarrow U_{L}\Phi U_{R}^{\dagger}. \label{Phitr}%
\end{equation}

From Eqs.\ (\ref{PRLd}) and (\ref{Phiij}) we obtain%

\begin{align}
\Phi_{ij}  &  \equiv \sqrt{2}\bar{q}_{j,R}q_{i,L}=\sqrt{2}\bar{q}_{j}\mathcal{P}%
_{L}\mathcal{P}_{L}q_{i}=\sqrt{2}\bar{q}_{j}\mathcal{P}_{L}q_{i}\nonumber\\
&  =\frac{1}{\sqrt{2}}\left(  \bar{q}_{j}q_{i}-\bar{q}_{j}\gamma^{5}%
q_{i}\right)  =\frac{1}{\sqrt{2}}\left(  \bar{q}_{j}q_{i}+i\bar{q}_{j}%
i\gamma^{5}q_{i}\right)  \text{.}%
\end{align}

Then comparing to Eqs.\ (\ref{sigmal}) and (\ref{pionl}) we recognise the
scalar current%

\begin{equation}
S_{ij} \equiv \frac{1}{\sqrt{2}}\bar{q}_{j}q_{i} \label{SQ}%
\end{equation}

and the pseudoscalar current%

\begin{equation}
P_{ij} \equiv \frac{1}{\sqrt{2}}\bar{q}_{j}i\gamma^{5}q_{i}\text{.} \label{PQ}%
\end{equation}

In other words,%

\begin{equation}
\Phi=S+iP\text{.} \label{PhiSP}%
\end{equation}

Thus our matrix $\Phi$ is a combination of scalar and pseudoscalar currents.
Additionally, the matrices $S$ and $P$ are hermitian and therefore they can be
decomposed in terms of generators $t^{a}$ of a unitary group $U(N_{f})$ with
$a=0,...,N_{f}^{2}-1$:%

\begin{align}
S  &  =S^{a}t^{a}\text{,}\\
P  &  =P^{a}t^{a}\text{,}
\end{align}
where
\begin{align}
S^{a}  &  \equiv \sqrt{2}\bar{q}t^{a}q\text{,}\\
P^{a}  &  \equiv \sqrt{2}\bar{q}i\gamma^{5}t^{a}q\text{.}%
\end{align}

As a first step toward the model construction, we consider only terms that
implement the chiral symmetry exactly (note again that the symmetry is exact
in the QCD Lagrangian as well up to the axial anomaly that is of quantum nature):%

\begin{equation}
\mathcal{L}_{\text{sym.}}=\mathrm{Tr}[(\partial^{\mu}\Phi)^{\dagger}%
(\partial_{\mu}\Phi)]-m_{0}^{2}\mathrm{Tr}(\Phi^{\dagger}\Phi)-\lambda
_{1}[\mathrm{Tr}(\Phi^{\dagger}\Phi)]^{2}-\lambda_{2}\mathrm{Tr}(\Phi
^{\dagger}\Phi)^{2}\text{.} \label{Lsim}%
\end{equation}

The Lagrangian in Eq.\ (\ref{Lsim}) is invariant under the transformation
(\ref{Phitr}). This is the original version of the sigma model containing only
scalar and pseudoscalar degrees of freedom. Note that $\mathcal{L}%
_{\text{sym.}}$ contains only terms up to order four in the naive scaling dimension.
Higher-order terms are usually discarded to preserve renormalisability of the
model. However, the model is not valid up to arbitrary large scales
(it is valid only up to the energy of the heaviest resonance
incorporated into the model). For this reason, we consider an alternative
criterion allowing us to constrain the order of terms considered in the
Lagrangian. The criterion is the dilatation invariance rather than
renormalisability. The dilatation invariance of the QCD Lagrangian has already
been discussed in Sec.\ \ref{sec.othersymmetries}. In the language of our
model where only composite states rather than partons are present, once a
dilaton field $G$ has been included then only terms with dimensionless
couplings are allowed in the Lagrangian in order that, in the chiral limit,
the trace anomaly in the model is generated in the same manner as in the QCD
Lagrangian \cite{dynrec,Stani}; see also Chapter \ref{chapterglueball}. Then
terms of the form
\begin{equation}
\alpha\lbrack\mathrm{Tr}(\Phi^{\dagger}\Phi)]^{6} \label{cvc}%
\end{equation}

are disallowed because the coupling $\alpha$ would possess dimension [$E^{-2}%
$]. The coupling $\alpha$ could actually be rendered dimensionless by
modifying the mentioned term as%

\begin{equation}
\frac{\alpha}{G^{2}}[\mathrm{Tr}(\Phi^{\dagger}\Phi)]^{6}%
\end{equation}

that would, however, lead to a singularity for $G\rightarrow0$. Therefore, in
the following, we will only consider terms up to order four in the
fields.\newline

The validity of our model is determined by the energy of
the heaviest state present in the model. In Chapter \ref{sec.scalarexp}, we
will discuss the features of the physical scalar resonances that can in
principle be assigned to the scalar states present in our model. These
resonances possess energies up to $\sim1.8$ GeV -- thus they belong to an
energy region where a multitude of vector and axial-vector states is
experimentally established as well \cite{PDG}. Additionally, (pseudo)scalars
are known to interact with (axial-)vectors \cite{PDG} and thus any realistic
model of QCD should in principle contain as many as possible of all the
mentioned states. For this reason, we need to extend the Lagrangian in
Eq.\ (\ref{Lsim}) to include the (axial-)vector degrees of freedom. Indeed we
will see in Sec.\ \ref{sec.sNppQI} that the inclusion of (axial-)vectors into
our model necessitates the interpretation of scalars above (rather than below)
1 GeV as $\bar{q}q$ states.

We construct the vector and axial-vector matrices analogously to those in
Eqs.\ (\ref{SQ}) and (\ref{PQ}). We first define the right-handed matrix%

\begin{align}
R_{ij}^{\mu}  &  \equiv \sqrt{2}\bar{q}_{j,R}\gamma^{\mu}q_{i,R}\overset
{\text{Eq.\ (\ref{qRL})}}{=}\sqrt{2}q_{j}^{\dagger}\mathcal{P}_{R}\gamma
^{0}\gamma^{\mu}\mathcal{P}_{R}q_{i}\overset{\text{Eq.\ (\ref{PRLd})}}{=}%
\sqrt{2}q_{j}^{\dagger}\frac{1+\gamma_{5}}{2}\gamma^{0}\gamma^{\mu}%
\frac{1+\gamma_{5}}{2}q_{i}\nonumber\\
&  =\frac{\sqrt{2}}{4}(q_{j}^{\dagger}\gamma^{0}\gamma^{\mu}q_{i}%
+q_{j}^{\dagger}\gamma_{5}\gamma^{0}\gamma^{\mu}q_{i}+q_{j}^{\dagger}%
\gamma^{0}\gamma^{\mu}\gamma_{5}q_{i}+q_{j}^{\dagger}\gamma_{5}\gamma
^{0}\gamma^{\mu}\gamma_{5}q_{i})\nonumber\\
&  \overset{\text{Eq.\ (\ref{Gamma5-2})}}{=}\frac{1}{\sqrt{2}}\left(  \bar
{q}_{j}\gamma^{\mu}q_{i}-\bar{q}_{j}\gamma^{5}\gamma^{\mu}q_{i}\right)
\label{RQ1}%
\end{align}

and the left-handed matrix%

\begin{equation}
L_{ij}^{\mu} \equiv \sqrt{2}\bar{q}_{j,L}\gamma^{\mu}q_{i,L}=\frac{1}{\sqrt{2}%
}\left(  \bar{q}_{j}\gamma^{\mu}q_{i}+\bar{q}_{j}\gamma^{5}\gamma^{\mu}%
q_{i}\right)  \text{.} \label{LQ1}%
\end{equation}

As in the case of currents from the QCD Lagrangian, we define%

\begin{align}
R^{\mu} &  =V^{\mu}-A^{\mu}\text{,} \label{RQ}\\
L^{\mu} &  =V^{\mu}+A^{\mu}\label{LQ}%
\end{align}

and thus%

\begin{equation}
V_{ij}^{\mu} \equiv \frac{1}{\sqrt{2}}\bar{q}_{j}\gamma^{\mu}q_{i}\text{, }%
A_{ij}^{\mu} \equiv \frac{1}{\sqrt{2}}\bar{q}_{j}\gamma^{\mu}\gamma^{5}q_{i}%
\end{equation}

or, upon decomposition in terms of the $U(N_{f})$ generators,
\begin{align}
V^{\mu} &  =V^{\mu a}t^{a}\text{,} \label{VQ}\\
A^{\mu} &  =A^{\mu a}t^{a}\text{,} \label{AQ}%
\end{align}

where
\begin{align}
V^{\mu a}  &  \equiv \sqrt{2}\bar{q}\gamma^{\mu}t^{a}q\text{,}\\
A^{\mu a}  &  \equiv \sqrt{2}\bar{q}\gamma^{\mu}\gamma^{5}t^{a}q\text{.}
\end{align}

With Eqs.\ (\ref{RQ1}), (\ref{LQ1}), (\ref{qfLt}) and (\ref{qfRt}), we obtain
immediately that $R^{\mu}$ and $L^{\mu}$ transform as%

\begin{equation}
R^{\mu}\rightarrow U_{R}R^{\mu}U_{R}^{\dagger} \label{URU}%
\end{equation}

and
\begin{equation}
L^{\mu}\rightarrow U_{L}L^{\mu}U_{L}^{\dagger}\text{.} \label{ULU}%
\end{equation}

Let us define the right-handed field-strength tensor $R^{\mu\nu}$ and the
left-handed field strength tensor $L^{\mu\nu}$ as%

\begin{align}
R^{\mu\nu} &  =\partial^{\mu}R^{\nu}-\partial^{\nu}R^{\mu}\text{,} \label{RRmu}\\
L^{\mu\nu} &  =\partial^{\mu}L^{\nu}-\partial^{\nu}L^{\mu}\text{.}\label{LLmu}%
\end{align}

Then considering the transformation properties (\ref{Phitr}), (\ref{URU}) and
(\ref{ULU}) we can construct further chirally invariant terms containing both
(pseudo)scalars and (axial-)vectors, up to order four in the fields:%

\begin{align}
\mathcal{L}_{\text{sym.,1}}  &  =-\frac{1}{4}\mathrm{Tr}(L_{\mu\nu}^{2}%
+R_{\mu\nu}^{2})+\mathrm{Tr}\left[  \frac{m_{1}^{2}}{2}(L_{\mu}^{2}+R_{\mu
}^{2})\right] \nonumber\\
&  +i\frac{g_{2}}{2}(\mathrm{Tr}\{L_{\mu\nu}[L^{\mu},L^{\nu}]\}+\mathrm{Tr}%
\{R_{\mu\nu}[R^{\mu},R^{\nu}]\})\nonumber\\
&  +\frac{h_{1}}{2}\mathrm{Tr}(\Phi^{\dagger}\Phi)\mathrm{Tr}[(L^{\mu}%
)^{2}+(R^{\mu})^{2}]+h_{2}\mathrm{Tr}[|L^{\mu}\Phi|^{2}+|\Phi R^{\mu}%
|^{2}]+2h_{3}\mathrm{Tr}(\Phi R_{\mu}\Phi^{\dagger}L^{\mu})\nonumber\\
&  +g_{3}[\mathrm{Tr}(L_{\mu}L_{\nu}L^{\mu}L^{\nu})+\mathrm{Tr}(R_{\mu}R_{\nu
}R^{\mu}R^{\nu})]+g_{4}[\mathrm{Tr}\left(  L_{\mu}L^{\mu}L_{\nu}L^{\nu
}\right)  +\mathrm{Tr}\left(  R_{\mu}R^{\mu}R_{\nu}R^{\nu}\right)
]\nonumber\\
&  +g_{5}\mathrm{Tr}\left(  L_{\mu}L^{\mu}\right)  \,\mathrm{Tr}\left(
R_{\nu}R^{\nu}\right)  +g_{6}[\mathrm{Tr}(L_{\mu}L^{\mu})\,\mathrm{Tr}(L_{\nu
}L^{\nu})+\mathrm{Tr}(R_{\mu}R^{\mu})\,\mathrm{Tr}(R_{\nu}R^{\nu})]\text{.}
\label{Lsym1}%
\end{align}

The explicit symmetry breaking has to be modelled separately in the
(pseudo)scalar and (axial-)vector channels. In the (pseudo)scalar sector we
introduce the term%

\begin{equation}
\mathcal{L}_{\text{ESB}}=\mathrm{Tr}[H(\Phi+\Phi^{\dagger})]\text{,} \label{LEB}%
\end{equation}

where%

\begin{equation}
H=\mathrm{diag}\left[  h_{0}^{1},h_{0}^{2},...h_{0}^{N_{f}}\right]
\end{equation}

and $h_{0}^{n}$ is proportional to the mass of the $n^{\text{th}}$ quark flavour.
Similarly, in the (axial-)vector channel we introduce the term%

\begin{equation}
\mathcal{L}_{\text{ESB, 1}}=\mathrm{Tr}\left[  \Delta(L_{\mu}^{2}+R_{\mu}%
^{2})\right]\text{,}  \label{LEB1}%
\end{equation}

where%

\begin{equation}
\Delta=\mathrm{diag}\left[  \delta_{u},\delta_{d},\delta_{s}...\right]
\sim\mathrm{diag}\left[  m_{u}^{2},m_{d}^{2},m_{s}^{2}...\right]  \text{.}%
\end{equation}

The chiral anomaly is usually modelled as \cite{Hooft}

\begin{equation}
\mathcal{L}_{\text{anomaly}}=c(\det\Phi+\det\Phi^{\dagger}) \label{Lanomaly}%
\end{equation}

because the determinant is invariant under $SU(N_{f})_{L}\times SU(N_{f})_{R}$
but not under $U(1)_{A}$. Note, however, that the chiral anomaly can also be
modelled as%

\begin{equation}
\mathcal{L}_{\text{anomaly, 1}}=c_{1}(\det\Phi-\det\Phi^{\dagger})^{2}\text{.}%
\end{equation}

We will discuss the implications of the two anomaly terms in
Sec.\ \ref{sec.anomaly}; only the term (\ref{Lanomaly}) will be used in the
two-flavour version of our model (see Chapter \ref{chapterQ}).\newline

Finally, for the modelling of the spontaneous breaking of the chiral symmetry,
let us consider the (pseudo)scalar Lagrangian (\ref{Lsim}) along the axis
$\Phi=\sigma_{N}t^{0}$:%

\begin{equation}
\mathcal{V}_{\text{sym.}}(\sigma_{N})=m_{0}^{2}\sigma_{N}^{2}+(\lambda
_{1}+\lambda_{2})\sigma_{N}^{4}\text{.}%
\end{equation}

The minimum $\sigma_{N}^{(0)}\neq0$ for $m_{0}^{2}<0$. This implies
spontaneous symmetry breaking because the vacuum is no longer symmetric under
the axial transformation. Note that the scalar isosinglet state is the only
one that can condense in the vacuum because that state is the only with the
same quantum numbers as the vacuum ($J$, $P$, $C$ and $I$). \newline

Then utilising the Lagrangians of Eqs.\ (\ref{Lsim}), (\ref{Lsym1}),
(\ref{LEB}), (\ref{LEB1}) and (\ref{Lanomaly}) we construct the following meson
Lagrangian for an arbitrary number of flavours $N_{f}$ and colours $N_{c}$:%

\begin{align}
\mathcal{L}  &  =\mathrm{Tr}[(D^{\mu}\Phi)^{\dagger}(D^{\mu}\Phi)]-m_{0}%
^{2}\mathrm{Tr}(\Phi^{\dagger}\Phi)-\lambda_{1}[\mathrm{Tr}(\Phi^{\dagger}%
\Phi)]^{2}-\lambda_{2}\mathrm{Tr}(\Phi^{\dagger}\Phi)^{2}\nonumber\\
&  -\frac{1}{4}\mathrm{Tr}(L_{\mu\nu}^{2}+R_{\mu\nu}^{2})+\mathrm{Tr}\left[
\left(  \frac{m_{1}^{2}}{2}+\Delta\right)  (L_{\mu}^{2}+R_{\mu}^{2})\right]
+\mathrm{Tr}[H(\Phi+\Phi^{\dagger})]\nonumber\\
&  +c(\det\Phi+\det\Phi^{\dagger})+i\frac{g_{2}}{2}(\mathrm{Tr}\{L_{\mu\nu
}[L^{\mu},L^{\nu}]\}+\mathrm{Tr}\{R_{\mu\nu}[R^{\mu},R^{\nu}]\})\nonumber\\
&  +\frac{h_{1}}{2}\mathrm{Tr}(\Phi^{\dagger}\Phi)\mathrm{Tr}[(L^{\mu}%
)^{2}+(R^{\mu})^{2}]+h_{2}\mathrm{Tr}[|L^{\mu}\Phi|^{2}+|\Phi R^{\mu}%
|^{2}]+2h_{3}\mathrm{Tr}(\Phi R_{\mu}\Phi^{\dagger}L^{\mu})\nonumber\\
&  +g_{3}[\mathrm{Tr}(L_{\mu}L_{\nu}L^{\mu}L^{\nu})+\mathrm{Tr}(R_{\mu}R_{\nu
}R^{\mu}R^{\nu})]+g_{4}[\mathrm{Tr}\left(  L_{\mu}L^{\mu}L_{\nu}L^{\nu
}\right)  +\mathrm{Tr}\left(  R_{\mu}R^{\mu}R_{\nu}R^{\nu}\right)
]\nonumber\\
&  +g_{5}\mathrm{Tr}\left(  L_{\mu}L^{\mu}\right)  \,\mathrm{Tr}\left(
R_{\nu}R^{\nu}\right)  +g_{6}[\mathrm{Tr}(L_{\mu}L^{\mu})\,\mathrm{Tr}(L_{\nu
}L^{\nu})+\mathrm{Tr}(R_{\mu}R^{\mu})\,\mathrm{Tr}(R_{\nu}R^{\nu})]\text{,}
\label{LagrangianGe}%
\end{align}

where

\begin{equation}
D^{\mu}\Phi=\partial^{\mu}\Phi-ig_{1}(L^{\mu}\Phi-\Phi R^{\mu})\text{.}%
\label{derivative}%
\end{equation}

The Lagrangian is invariant under $P$ and $C$ transformations. The
(pseudo)scalar matrix $\Phi$ transforms under parity as%

\begin{equation}
\Phi(t,\vec{x})\overset{P}{\rightarrow}\Phi^{\dagger}(t,-\vec{x})\text{.} \label{PhiP}%
\end{equation}

This is due to Eq.\ (\ref{qP}) and the definition of $\Phi$, Eq.\ (\ref{qP}):%

\begin{align}
\Phi_{ij}(t,\vec{x})  &  \equiv \sqrt{2}\bar{q}_{j,R}(t,\vec{x})q_{i,L}(t,\vec
{x})=\sqrt{2}q_{j,R}^{\dagger}(t,\vec{x})\gamma^{0}q_{i,L}(t,\vec{x}%
)\overset{\text{Eq.\ (\ref{qRL})}}{=}\sqrt{2}q_{j}^{\dagger}(t,\vec
{x})\mathcal{P}_{R}\gamma^{0}\mathcal{P}_{L}q_{i}(t,\vec{x})\nonumber\\
&  \overset{P}{\rightarrow}\sqrt{2}q_{j}^{\dagger}(t,-\vec{x})\gamma
^{0}\mathcal{P}_{R}\gamma^{0}\mathcal{P}_{L}\gamma^{0}q_{i}(t,-\vec
{x})\overset{\text{Eq.\ (\ref{Gamma5-2})}}{=}\sqrt{2}q_{j}^{\dagger}%
(t,-\vec{x})\mathcal{P}_{L}\gamma^{0}\gamma^{0}\gamma^{0}\mathcal{P}_{R}%
q_{i}(t,-\vec{x})\nonumber\\
&  =\sqrt{2}q_{j,L}^{\dagger}(t,-\vec{x})\gamma^{0}q_{i,R}(t,-\vec{x})=\left[
\sqrt{2}q_{j,L}^{\dagger}(t,-\vec{x})\gamma^{0}q_{i,R}(t,-\vec{x})\right]
^{\dagger} \equiv \Phi_{ij}^{\dagger}(t,-\vec{x})\text{.}%
\end{align}

Parity transforms the left-handed matrix $L^{\mu}$ into the right-handed
matrix $R^{\mu}$ and vice versa:%

\begin{align}
&  R^{\mu}(t,\vec{x})\overset{P}{\rightarrow}g^{\mu \nu}L_{\nu}(t,-\vec{x})\text{,} \label{RP}\\
&  L^{\mu}(t,\vec{x})\overset{P}{\rightarrow}g^{\mu\nu}R_{\nu}(t,-\vec{x})\text{,}
\label{LP}%
\end{align}

due to Eq.\ (\ref{qP}) and the definitions (\ref{RQ1}) and (\ref{LQ1}):%

\begin{align}
R_{ij}^{\mu}  &  \equiv \sqrt{2}\bar{q}_{j,R}(t,\vec{x})\gamma^{\mu}q_{i,R}%
(t,\vec{x})=\sqrt{2}q_{j,R}^{\dagger}(t,\vec{x})\gamma^{0}\gamma^{\mu}%
q_{i,R}(t,\vec{x})\overset{\text{Eq.\ (\ref{qRL})}}{=}\sqrt{2}q_{j}^{\dagger
}(t,\vec{x})\mathcal{P}_{R}\gamma^{0}\gamma^{\mu}\mathcal{P}_{R}q_{i}%
(t,\vec{x})\nonumber\\
&  \overset{P}{\rightarrow}\sqrt{2}q_{j}^{\dagger}(t,-\vec{x})\gamma
^{0}\mathcal{P}_{R}\gamma^{0}\gamma^{\mu}\mathcal{P}_{R}\gamma^{0}%
q_{i}(t,-\vec{x})\overset{\text{Eq.\ (\ref{Gamma5-2})}}{=}\sqrt{2}%
q_{j}^{\dagger}(t,-\vec{x})\mathcal{P}_{L}\gamma^{0}\gamma^{0}\gamma^{\mu
}\gamma^{0}\mathcal{P}_{L}q_{i}(t,-\vec{x})\nonumber\\
&  =\sqrt{2}q_{j,L}^{\dagger}(t,-\vec{x})\gamma^{\mu}\gamma^{0}q_{i,L}%
(t,-\vec{x})\overset{\text{Eq.\ (\ref{AKD})}}{=}\left\{
\begin{tabular}
[c]{l}%
$\bar{q}_{j,L}(t,-\vec{x})\gamma^{0}q_{i,L}(t,-\vec{x})$ for $\mu=0$\text{,}\\
$-\bar{q}_{j,L}(t,-\vec{x})\gamma^{k}q_{i,L}(t,-\vec{x})$ for $\mu
=k\in\{1,2,3\}$%
\end{tabular}
\ \ \ \ \ \right.
\end{align}

and analogously for $L^{\mu}$.

The matrix $\Phi$ transforms under charge conjugation as%

\begin{equation}
\Phi\overset{C}{\rightarrow}\Phi^{t}\text{.}%
\end{equation}

The proof is analogous to the calculation in Eq.\ (\ref{sigmaC}). Similarly,
the left-handed matrix $L^{\mu}$ and the right-handed matrix $R^{\mu}$
transform as:%

\begin{align}
&  R_{\mu}\overset{C}{\rightarrow}-L_{\mu}^{t}\text{,} \label{RC}\\
&  L_{\mu}\overset{C}{\rightarrow}- R_{\mu}^{t}\text{.} \label{LC}%
\end{align}

Then it is straightforward to demonstrate that all terms in the Lagrangian
(\ref{LagrangianGe}) fulfill $P$- invariance as well as $C$-invariance; given
that the model is Lorentz-invariant, it is consequently also $T$-invariant
\cite{Greenberg:2002}.

Before we discuss the $N_{f}=2$ and $N_{f}=3$ applications of the Lagrangian
(\ref{LagrangianGe}), let us discuss the large-$N_{c}$ dependence of the model parameters.

\section{Large-\boldmath $N_{c}$ Behaviour of Model Parameters}

\label{sec.largen}

It is important to determine the large-$N_{c}$ dependence of the model
parameters for two reasons:

\begin{itemize}
\item It allows us to estimate the relative magnitudes of parameters
(different parameters will possess different large-$N_{c}$ scaling because
they may be associated to different vertices). In this way, parameters shown
to be suppressed in comparison with other parameters may be set to zero.

\item It enables us to prove that the states present in our model (up to the
dilaton field introduced in Chapter \ref{chapterglueball}) are indeed $\bar
{q}q$ states. This is essential because the goal of this work is to study
whether experimentally ascertained meson states can be interpreted as
quarkonia. Such a study is, of course, only possible if the theoretical
framework presented in this work already contains $\bar{q}q$ states that are
to be assigned to physical states -- and the success of the assignment is
determined by comparison with experimental data.
\end{itemize}

The large-$N_{c}$ dependence of the parameters in Lagrangian
(\ref{LagrangianGe}) is \cite{largenc}:%

\begin{align}
g_{1}\text{, }g_{2}  &  \propto N_{c}^{-1/2}\text{,}\nonumber\\
\lambda_{2}\text{, }h_{2}\text{, }h_{3}\text{, }g_{3}\text{, }g_{4}  &
\propto N_{c}^{-1}\text{,}\nonumber\\
\lambda_{1}\text{, }h_{1}\text{, }g_{5}\text{, }g_{6}  &  \propto N_{c}%
^{-2}\text{,}\nonumber\\
m_{0}^{2}\text{, }m_{1}^{2}\text{, }\delta_{u,d,s...}  &  \propto N_{c}%
^{0}\text{,}\nonumber\\
\text{ }c  &  \propto N_{c}^{-N_{f}/2}\text{,}\nonumber\\
h_{0}^{i}  &  \propto N_{c}^{1/2}\text{.} \label{largen}%
\end{align}
Let us remember that a vertex of $n$ quark-antiquark mesons scales as
$N_{c}^{-(n-2)/2}$. As a consequence, the parameters $g_{1}$, $g_{2}$ scale as
$N_{c}^{-1/2}$, because they are associated with a three-point vertex of
quark-antiquark vector fields (of the kind $\rho^{3}$). This has already been
discussed in Sec.\ \ref{sec.othersymmetries}, see Eq.\ (\ref{gmu}).

Similarly, the parameters $\lambda_{2}$, $h_{2}$, $h_{3}$ scale as $N_{c}%
^{-1}$, because they are associated with quartic terms such as $\pi^{4}$ and
$\pi^{2}\rho^{2}$. The parameters $\lambda_{1}$, $h_{1}$ also describe quartic
interactions, but are further suppressed by a factor $1/N_{c}$ because of the
trace structure of the corresponding terms in the Lagrangian. The quantities
$m_{0}^{2}$, $m_{1}^{2}$ are bare-mass terms and therefore scale as $N_{c}%
^{0}$. Note that our mass terms will be proportional to the square of the pion
and kaon decay constants $f_{\pi}$ and $f_{K}$ [see Eqs.\ (\ref{sigma}) --
(\ref{a1}) in the $N_{f}=2$ case and Eqs.\ (\ref{m_sigma_N}) -- (\ref{m_K_1})
in the $N_{f}=3$ case]. Consequently, $f_{\pi}$ and $f_{K}$ have to\ scale as
$N_{c}^{1/2}$.

The suppression of the parameter $c$ depends on the number of flavours and
colours considered. The axial anomaly is suppressed in the large-$N_{c}$
limit. Note that $c$ possesses a dimension for $N_{f}\neq4$. This is an
exception to the rule illustrated via term (\ref{cvc}) where we have discussed
that only dimensionless couplings should appear in the Lagrangian. This is,
however, not problematic because the chiral anomaly also stems from the gauge
sector of the QCD Lagrangian, see Eq.\ (\ref{Asc1}). [Note that the chiral
anomaly can also be modelled using an alternative term: $c_{1}(\det\Phi
-\det\Phi^{\dagger})^{2}$, see Eq.\ (\ref{Lagrangian}) and
Sec.\ \ref{sec.anomaly}. In this case,$\ c_{1}\propto N_{c}^{-N_{f}}$
holds.]

Note that without any assumptions about the fields, we immediately obtain that
their masses scale as $N_{c}^{0}$ and their decay widths as $N_{c}^{-1}$, as
we shall see in Chapters \ref{chapterQ} -- \ref{ImplicationsFitII}. Therefore,
they must also correspond to quark-antiquark degrees of freedom.

%% file: Paper.tex

%
%
%
\begin{fmffile}{paper}
\chapter{Two-Flavour Linear Sigma Model} \label{chapterQ}

Having constructed a generic Lagrangian containing meson fields for an arbitrary number of flavours and colours, let us now discuss the implications of the Lagrangian for the case of two flavours (and, of course, three colours).

\section{The \boldmath $N_f=2$ Lagrangian} \label{sec.LagrangianQ1}

The globally invariant $U(2)_{L}\times
U(2)_{R}$ Lagrangian possesses the same structure as the one in Eq.\ (\ref{LagrangianGe}):

\begin{align}
\mathcal{L}  &  =\mathrm{Tr}[(D^{\mu}\Phi)^{\dagger}(D^{\mu}\Phi)]-m_{0}%
^{2}\mathrm{Tr}(\Phi^{\dagger}\Phi)-\lambda_{1}[\mathrm{Tr}(\Phi^{\dagger}%
\Phi)]^{2}-\lambda_{2}\mathrm{Tr}(\Phi^{\dagger}\Phi)^{2}\nonumber\\
&  -\frac{1}{4}\mathrm{Tr}(L_{\mu\nu}^{2}+R_{\mu\nu}^{2})+\mathrm{Tr}\left[
\left(  \frac{m_{1}^{2}}{2}+\Delta\right)  (L_{\mu}^{2}+R_{\mu}^{2})\right]
+\mathrm{Tr}[H(\Phi+\Phi^{\dagger})]\nonumber\\
&  +c(\det\Phi+\det\Phi^{\dagger})+i\frac{g_{2}}{2}(\mathrm{Tr}\{L_{\mu\nu
}[L^{\mu},L^{\nu}]\}+\mathrm{Tr}\{R_{\mu\nu}[R^{\mu},R^{\nu}]\})\nonumber\\
&  +\frac{h_{1}}{2}\mathrm{Tr}(\Phi^{\dagger}\Phi)\mathrm{Tr}[(L^{\mu}%
)^{2}+(R^{\mu})^{2}]+h_{2}\mathrm{Tr}[|L^{\mu}\Phi
|^{2} + |\Phi R^{\mu}|^{2}]+2h_{3}\mathrm{Tr}(\Phi R_{\mu}\Phi^{\dagger}L^{\mu})\nonumber\\
&  +g_{3}[\mathrm{Tr}(L_{\mu}L_{\nu}L^{\mu}L^{\nu})+\mathrm{Tr}(R_{\mu}R_{\nu
}R^{\mu}R^{\nu})]+g_{4}[\mathrm{Tr}\left(  L_{\mu}L^{\mu}L_{\nu}L^{\nu
}\right)  +\mathrm{Tr}\left(  R_{\mu}R^{\mu}R_{\nu}R^{\nu}\right)
]\nonumber\\
&  +g_{5}\mathrm{Tr}\left(  L_{\mu}L^{\mu}\right)  \,\mathrm{Tr}\left(
R_{\nu}R^{\nu}\right)  +g_{6}[\mathrm{Tr}(L_{\mu}L^{\mu})\,\mathrm{Tr}(L_{\nu
}L^{\nu})+\mathrm{Tr}(R_{\mu}R^{\mu})\,\mathrm{Tr}(R_{\nu}R^{\nu})]\text{.}
\label{LagrangianQ}%
\end{align}

In Eq.\ (\ref{LagrangianQ}),
\begin{equation}
\Phi=(\sigma_{N}+i\eta_{N})\,t^{0}+(\vec{a}_{0}+i\vec{\pi})\cdot\vec{t}
\label{scalars}%
\end{equation}
contains scalar and pseudoscalar mesons, where $t^{0}$, $\vec{t}$ are the
generators of $U(2)$ in the fundamental representation and $\eta_{N}$ denotes
the non-strange content of the $\eta$ meson (more details will be given in
Sec.\ \ref{sec.eta-eta}). Vector and axial-vector mesons are contained in the
left-handed and right-handed vector fields:

\begin{align}
L^{\mu} &  =(\omega_{N}^{\mu}+f_{1N}^{\mu})\,t^{0}+(\vec{\rho}^{\mu}+\vec
{a}_{1}^{\mu})\cdot\vec{t}\text{,}\label{vectors1}\\
R^{\mu} &  =(\omega_{N}^{\mu}-f_{1N}^{\mu})\,t^{0}+(\vec{\rho}^{\mu}-\vec
{a}_{1}^{\mu})\cdot\vec{t}\text{,}\label{vectors}%
\end{align}
respectively. The covariant derivative

\begin{equation}
D^{\mu}\Phi=\partial^{\mu}\Phi-ig_{1}(L^{\mu}\Phi-\Phi R^{\mu})-ieA^{\mu
}[t^{3},\Phi ] \label{PhiQQ}
\end{equation}
couples scalar and pseudoscalar degrees of freedom to vector and
axial-vector ones as well as to the electromagnetic field $A^{\mu}$. [The
derivative leads to a kinetic term invariant under $U(2)_{L}\times U(2)_{R}$
and will allow us to calculate the decay width $\Gamma_{a_{1}\rightarrow
\pi\gamma}$ in Eq.\ (\ref{a1piongamma}). For this reason it contains $A^{\mu}$, unlike the derivative in Eq.\ (\ref{derivative}).] The left-handed and right-handed
field strength tensors (again with $A^{\mu}$)

\begin{align}
L^{\mu\nu} &  =\partial^{\mu}L^{\nu}-ieA^{\mu}[t^{3},L^{\nu}]-\left\{
\partial^{\nu}L^{\mu}-ieA^{\nu}[t^{3},L^{\mu}]\right\}  \text{,}\label{LQa}\\
R^{\mu\nu} &  =\partial^{\mu}R^{\nu}-ieA^{\mu}[t^{3},R^{\nu}]-\left\{
\partial^{\nu}R^{\mu}-ieA^{\nu}[t^{3},R^{\mu}]\right\}  \text{,}\label{RQa}%
\end{align}
respectively, couple vector and axial-vector mesons to the electromagnetic
field $A^{\mu}$. Explicit breaking of the global symmetry is described by the
term Tr$[H(\Phi+\Phi^{\dagger})]\equiv h_{0N}\sigma$ $(h_{0N}=const.)$ in the
(pseudo)scalar sector and by the term $\mathrm{Tr}\left[  \Delta(L_{\mu
}^{2}+R_{\mu}^{2})\right]  $ in the (axial-)vector channels, with
$\Delta=diag(\delta_{N},\delta_{N})$ and $\delta_{N}\sim m_{u,d}^{2}$. The
chiral anomaly is described by the term $c\,(\det\Phi+\det\Phi^{\dagger})$,
see Sec.\ \ref{sec.c}. (Note that a slightly different form of the
chiral-anomaly term will be utilised in Sec.\ \ref{sec.anomaly}.)

In the pseudoscalar and (axial-)vector sectors the identification of mesons
with particles listed in Ref.\ \cite{PDG} is straightforward, as already
indicated in Eqs.\ (\ref{scalars}) and (\ref{vectors1})-(\ref{vectors}): the
fields $\vec{\pi}$ and $\eta_{N}$ correspond to the pion and the $SU(2)$
counterpart of the $\eta$ meson, $\eta_{N}\equiv(\bar{u}u+\bar{d}d)/\sqrt{2}$,
with a mass of about $700$ MeV. This value can be obtained by "unmixing" the
physical $\eta$ and $\eta^{\prime}$ mesons, which also contain $\bar{s}s$
contributions. The fields $\omega^{\mu}$ and $\vec{\rho}^{\mu}$ represent the
$\omega(782)$ and $\rho(770)$ vector mesons, respectively, while the fields
$f_{1N}^{\mu}$ and $\vec{a_{1}}^{\mu}$ represent the $f_{1}(1285)$ and
$a_{1}(1260)$ axial-vector mesons, respectively. (In principle, the physical
$\omega$ and $f_{1}$ states also contain $\bar{s}s$ contributions, however
their admixture is negligibly small.) Unfortunately, the identification of the
$\sigma_{N}$ and $\vec{a}_{0}$ fields is controversial, the possibilities
being the pairs $\{f_{0}(600),a_{0}(980)\}$ and $\{f_{0}(1370),a_{0}(1450)\}$.
As already mentioned, we will refer to these two assignments as Scenarios I
and II, respectively. We discuss the implications of these two scenarios in
the following.

The inclusion of (axial-)vector mesons in effective models of QCD has been
done also in other ways than the one presented here. Vector and axial-vector
mesons have been included in chiral perturbation theory in Ref.\ \cite{ecker}.
While the mathematical expressions for the interaction terms turn out to be
similar to our results, in our linear approach the number of parameters is
smaller. In Ref.\ \cite{bando} the so-called hidden gauge formalism is used to
introduce vector mesons, and subsequently axial-vector mesons, into a chiral
Lagrangian with a nonlinear realization of chiral symmetry. In this case the
number of parameters is smaller. This approach is closely related to the
locally chirally invariant models \cite{GG,KR} (also called massive Yang-Mills
approaches). We refer also to Ref.\ \cite{birse}, where a comparative analysis
of effective chiral Lagrangians for spin-1 mesons is presented.

One may raise the question whether vector meson dominance (VMD) is still
respected in the globally invariant linear sigma model (\ref{LagrangianQ}). As
outlined in Ref.\ \cite{OCPTW}, there are two ways to realize VMD in a linear
sigma model. The standard version of VMD was introduced by Sakurai
\cite{Sakurai} and considers vector mesons as Yang-Mills gauge fields
\cite{RefYM}; see also Ref.\ \cite{Lissabon}. The gauge symmetry is explicitly broken by the vector meson masses.
Another realization of VMD was first explored by Lurie \cite{Lurie} whose
theory contained a Lagrangian which was globally invariant. It is interesting
to note that Lurie's Lagrangian contained direct couplings of the photon to
pions and $\rho$ mesons, as well as a $\rho$-$\pi$ coupling. It was shown in
Ref.\ \cite{OCPTW} that the two representations of VMD are equivalent if the
$\rho$-$\pi$ coupling $g_{\rho\pi\pi}$ equals the photon-$\rho$ coupling
$g_{\rho}$ (the so-called "universal limit"). It was also shown that, if the
underlying theory is globally invariant, the pion form factor at threshold
$F_{\pi}(q^{2}=0)=1$ for \emph{any\/} value of the above mentioned couplings.
On the other hand, in Sakurai's theory $F_{\pi}(q^{2}=0)\neq1$ unless one
demands $g_{\rho\pi\pi}\overset{!}{=}g_{\rho}$, or other parameters are
adjusted in such a way that $F_{\pi}(q^{2}=0)=1$. In other words, for
\emph{any\/} globally invariant model, and thus also for ours, one has the
liberty of choosing different values for the photon-$\rho$ and $\rho$-$\pi$
couplings, without violating VMD.

\subsection{Tree-Level Masses}

The Lagrangian (\ref{LagrangianQ}) contains 16 parameters. However, the
parameters $g_{k}$ with $k=3$, ..., $6$ are not relevant for the results
presented here; additionally, the explicit symmetry breaking in the
non-strange sector is negligible because the quark masses are small --
therefore, we set $\delta_{N}=0$. Then the number of undetermined parameters
decreases to eleven:

\begin{equation}
m_{0},\text{ }\lambda_{1},\text{ }\lambda_{2},\text{ }m_{1},\text{ }%
g_{1},\text{ }g_{2},\text{ }c,\text{ }h_{0N},\text{ }h_{1},\text{ }h_{2},\text{
}h_{3}\text{.} \label{param}%
\end{equation}
The squared tree-level masses of the mesons in our model contain a
contribution arising from spontaneous symmetry breaking, proportional to
$\phi_{N}^{2}$. The value $\phi_{N}$ is the vacuum expectation value of the
$\sigma_{N}$ field and coincides with the minimum of the potential that
follows from Eq.\ (\ref{LagrangianQ}). The $\sigma_{N}$ field is the only
field with the quantum numbers of the vacuum, $J^{PC}=0^{++}$, i.e., the
condensation of which does not lead to the breaking of parity, charge
conjugation, and Lorentz invariance. The potential for the $\sigma_{N}$ field
reads explicitly
\begin{equation}
V(\sigma_{N})=\frac{1}{2}(m_{0}^{2}-c)\sigma_{N}^{2}+\frac{1}{4}\left(
\lambda_{1}+\frac{\lambda_{2}}{2}\right)  \sigma_{N}^{4}-h_{0N}\sigma
_{N}\text{,}%
\end{equation}
and its minimum is determined by
\begin{equation}
0=\left(  \frac{\mathrm{d}V}{\mathrm{d}\sigma_{N}}\right)  _{\sigma_{N}%
=\phi_{N}}=\left[  m_{0}^{2}-c+\left(  \lambda_{1}+\frac{\lambda_{2}}%
{2}\right)  \phi_{N}^{2}\right]  \phi_{N}-h_{0N}\text{.} \label{minimum}%
\end{equation}
Spontaneous symmetry breaking corresponds to the case when the potential
$V(\phi_{N})$ assumes its minimum for a non-vanishing value $\sigma_{N}%
=\phi_{N}\neq0$. In order to determine the fluctuation of the $\sigma_{N}$
field around the new vacuum, one shifts it by its vacuum expectation value
$\phi_{N}\neq0$, $\sigma_{N}\rightarrow\sigma_{N}+\phi_{N}$. The shift leads
also to $\eta_{N}$-$f_{1}$ and $\vec{\pi}$-$\vec{a}_{1}$ mixing terms and thus
to non-diagonal elements in the scattering matrix:%

\begin{equation}
-g_{1}\phi_{N}(f_{1N}^{\mu}\partial_{\mu}\eta_{N}+{\vec{a}_{1}^{\mu}}%
\cdot\partial_{\mu}{\vec{\pi}})\text{.} \label{mixingtermsQ}%
\end{equation}

These terms are removed from the Lagrangian by shifting the $f_{1}$ and
$\vec{a}_{1}$ fields as follows \cite{DA}:
\begin{align}
&  f_{1N}^{\mu}\rightarrow f_{1N}^{\mu}+Z_{\eta_{N}}w_{f_{1N}}\partial^{\mu
}\eta_{N}\text{,}\;\;\vec{a}_{1}^{\mu}\rightarrow\vec{a}_{1}^{\mu}+Z_{\pi
}w_{a_{1}}\partial^{\mu}\vec{\pi}\text{,}\nonumber\\
&  \eta_{N}\rightarrow Z_{\eta_{N}}\eta_{N}\text{,}\;\;\vec{\pi}\rightarrow
Z_{\pi}\vec{\pi}\text{,} \label{shifts}%
\end{align}
where we defined the quantities
\begin{equation}
w_{f_{1N}}=w_{a_{1}}=\frac{g_{1}\phi_{N}}{m_{a_{1}}^{2}}\text{,}%
\;\;\;Z_{\eta_{N}}=Z_{\pi}=\left(  1-\frac{g_{1}^{2}\phi_{N}^{2}}{m_{a_{1}%
}^{2}}\right)  ^{-1/2}\text{.} \label{Z}%
\end{equation}

More details on these calculations can be found in Ref.\ \cite{DA};
alternatively, see the analogous calculation performed in the $N_{f}=3$ case later
in this work (see Chapters \ref{sec.remarks} - \ref{ImplicationsFitII}). Note
that the field renormalisation of $\eta_{N}$ and $\vec{\pi}$ guarantees the
canonical normalisation of the kinetic terms. This is necessary in order to
interpret the Fourier components of the properly normalized one-meson states
as creation or annihilation operators \cite{GG}. Once the shift $\sigma
_{N}\rightarrow\sigma_{N}+\phi_{N}$ and the transformations (\ref{shifts})
have been performed, the mass terms of the mesons in the Lagrangian
(\ref{LagrangianQ}) read:
\begin{align}
m_{\sigma_{N}}^{2}  &  =m_{0}^{2}-c+3\left(  \lambda_{1}+\frac{\lambda_{2}}%
{2}\right)  \phi_{N}^{2}\text{,}\label{sigma}\\
m_{\eta_{N}}^{2}  &  =Z^{2}\left[  m_{0}^{2}+c+\left(  \lambda_{1}%
+\frac{\lambda_{2}}{2}\right)  \phi_{N}^{2}\right]  =m_{\pi}^{2}+2cZ_{\pi}%
^{2}\text{,}\label{eta1}\\
m_{a_{0}}^{2}  &  =m_{0}^{2}+c+\left(  \lambda_{1}+3\frac{\lambda_{2}}%
{2}\right)  \phi_{N}^{2}\text{,}\label{a0}\\
m_{\pi}^{2}  &  =Z^{2}\left[  m_{0}^{2}-c+\left(  \lambda_{1}+\frac
{\lambda_{2}}{2}\right)  \phi_{N}^{2}\right]  \overset{(\ref{minimum})}%
{=}\frac{Z_{\pi}^{2}h_{0N}}{\phi_{N}}\text{,}\label{pion}\\
m_{\omega_{N}}^{2}  &  =m_{\rho}^{2}=m_{1}^{2}+\frac{\phi_{N}^{2}}{2}%
(h_{1}+h_{2}+h_{3})\text{,}\label{rho}\\
m_{f_{1N}}^{2}  &  =m_{a_{1}}^{2}=m_{1}^{2}+g_{1}^{2}\phi_{N}^{2}+\frac
{\phi_{N}^{2}}{2}(h_{1}+h_{2}-h_{3})\text{.} \label{a1}%
\end{align}
Note that the $\rho$ and $\omega_{N}$ masses as well as the $f_{1N}$ and
$a_{1}$ masses are degenerate. In Sec.\ \ref{sec.LagrangianQ} we show the
Lagrangian in the form when all shifts have been explicitly performed. From
Eqs.\ (\ref{rho}) and (\ref{a1}) we obtain:
\begin{equation}
m_{a_{1}}^{2}=m_{\rho}^{2}+g_{1}^{2}\phi_{N}^{2}-h_{3}\phi_{N}^{2}\text{.}
\label{massdifference}%
\end{equation}

The pion decay constant $f_{\pi}$ is determined from the axial current,
\begin{equation}
J_{A_{\mu}}^{a}=\frac{\phi_{N}}{Z}\partial_{\mu}\pi^{a}+\ldots\equiv f_{\pi
}\partial_{\mu}\pi^{a}+\ldots\;\;\rightarrow\;\;\phi_{N}=Zf_{\pi}\text{.}
\label{jA}%
\end{equation}

Note that the photon coupling entailed in Eqs.\ (\ref{PhiQQ}), (\ref{LQa}) and
(\ref{RQa}) yields the correct coupling of photons to pions as the
corresponding term from the Lagrangian (\ref{LagrangianQ}) reads%

\begin{align}
\mathcal{L}_{\gamma\pi\pi}  &  =eZ_{\pi}^{2}(1-g_{1}w_{a_{1}}\phi_{N})A^{\mu
}\left(  \pi^{1}\partial_{\mu}\pi^{2}-\pi^{2}\partial_{\mu}\pi^{1}\right)
\nonumber\\
&  \overset{w_{a_{1}}=g_{1}\phi_{N}/m_{a_{1}}^{2}}{=}eZ_{\pi}^{2}%
\frac{m_{a_{1}}^{2}-(g_{1}\phi_{N})^{2}}{m_{a_{1}}^{2}}A^{\mu}\left(  \pi
^{1}\partial_{\mu}\pi^{2}-\pi^{2}\partial_{\mu}\pi^{1}\right) \nonumber\\
&  \equiv eZ_{\pi}^{2}Z_{\pi}^{-2}A^{\mu}\left(  \pi^{1}\partial_{\mu}\pi
^{2}-\pi^{2}\partial_{\mu}\pi^{1}\right)  =eA^{\mu}\left(  \pi^{1}%
\partial_{\mu}\pi^{2}-\pi^{2}\partial_{\mu}\pi^{1}\right) \nonumber\\
&  =ieA^{\mu}(\pi^{-}\partial_{\mu}\pi^{+}-\pi^{+}\partial_{\mu}\pi^{-})\text{,}
\end{align}

where in the last line we have substituted $\pi^{1}=(\pi^{+}+\pi
^{-})/\sqrt{2}$ and $\pi^{2}=i(\pi^{+}-\pi^{-})/\sqrt{2}$. The photon-pion
coupling is thus equal to the elementary electric charge $e=\sqrt{4\pi\alpha}$
where $\alpha$ denotes the fine-structure constant $\alpha
=1/137.035999679(94)$ in vacuum \cite{PDG}.\newline

We note that the phenomenology of low-lying axial-vector mesons is also
considered in approaches where the Bethe-Salpeter equation is used to
unitarise the scattering of vector and pseudoscalar mesons -- see, e.g.,
Ref.\ \cite{Oset1}. Here, the Bethe-Salpeter kernel is given by the
lowest-order effective Lagrangian. This leads to the dynamical generation of
resonances, one of which has a pole mass of 1011 MeV and is consequently
assigned to the $a_{1}(1260)$ meson. This unitarised approach is used in
Ref.\ \cite{Oset2} to study the large-$N_{c}$ behaviour of the dynamically
generated resonances, with the conclusion that the $a_{1}(1260)$ resonance is
not a genuine quark-antiquark state.

However, it was shown in Ref.\ \cite{dynrec} that, while unitarising the
chiral Lagrangian by means of a Bethe-Salpeter study allows one to find poles
in the complex plane and identify them with physical resonances, it does not
necessarily allow one to make a conclusion about the structure of those
resonances in the large-$N_{c}$ limit. In order to be able to draw correct
conclusions, a Bethe-Salpeter study requires at least one additional term of
higher order not included in the Lagrangian of Refs.\ \cite{Oset1,Oset2}.
Alternatively, the Inverse Amplitude Method of
Ref.\ \cite{Pelaez-scalars-below1GeVq2q2} can be used.

A very similar approach to the one in Refs.\ \cite{Oset1,Oset2} was also used
in Ref.\ \cite{Leupold} where a very good fit to the $\tau$ decay data from
the ALEPH collaboration \cite{ALEPH} was obtained by fine-tuning the
subtraction point of a loop diagram. Note, however, that detuning the
subtraction point by 5\% will spoil the agreement with experimental data.
Alternately, these data may be described by approaches with the $a_{1}(1260)$
meson as an explicit degree of freedom, such as the one in Ref.\ \cite{UBW},
where $a_{1}(1260)$ is a quark-antiquark state and where the experimental
$a_{1}(1260)$ spectral function is fitted very well. In Ref.\ \cite{UBW},
$m_{a_{1}(1260)}\simeq1150$ MeV and a full width $\Gamma_{a_{1}(1260)}$
$\simeq410$ MeV are obtained. Note that our results, as will be shown later,
give very good results on the $a_{1}(1260)$ phenomenology, for example in the
$a_{1}(1260)\rightarrow\pi\gamma$ and $a_{1}(1260)\rightarrow\rho\pi$ decay
channels, see Sec.\ \ref{sec.a1decaysI}.\\

For the following discussion, it is interesting to note that the $\rho$ meson
mass can be split into two contributions:\textbf{ }%
\begin{equation}
m_{\rho}^{2}=m_{1}^{2}+\frac{\phi_{N}^{2}}{2}(h_{1}+h_{2}+h_{3})\text{.}%
\label{mrhoquad}%
\end{equation}
Without further assumptions, it is not possible to relate the quantity
$m_{1}^{2}$ to microscopic condensates of QCD. However, invoking dilatation
invariance, the term $m_{1}^{2}$Tr$[(L^{\mu})^{2}+(R^{\mu})^{2}]/2$ in
Eq.\ (\ref{LagrangianQ}) arises from a term $aG^{2}$Tr$[(L^{\mu})^{2}+(R^{\mu
})^{2}]/2$ where $G$ is the dilaton field and $a$ a dimensionless constant
[see Eq.\ (\ref{LagrangianG}) and Chapter \ref{chapterglueball}]. Upon
shifting the dilatation field by $G\rightarrow G_{0}+G$, with $G_{0}$ being
the gluon condensate, one obtains the term in our Lagrangian upon identifying
$m_{1}^{2}=aG_{0}^{2}$. Thus, the quantity $m_{\rho}^{2}$ in
Eq.\ (\ref{mrhoquad}) is expressed as a sum of a term which is directly
proportional to the gluon condensate $G_{0}$, and a term which is directly
proportional to the chiral condensate $\phi_{N}^{2}$.

We shall require that none of the two contributions be negative: in fact, a
negative $m_{1}^{2}=aG_{0}^{2}$ would imply that the system is unstable when
$\phi_{N}\rightarrow0$; a negative $\phi_{N}^{2}(h_{1}+h_{2}+h_{3})/2$ would
imply that spontaneous chiral symmetry breaking decreases the $\rho$ mass.
This is clearly unnatural because the breaking of chiral symmetry generates a
sizable effective mass for the light quarks, which is expected to positively
contribute to the meson masses. This positive contribution is a feature of all
known models (such as the Nambu--Jona-Lasinio model and constituent quark
approaches). Indeed, in an important class of hadronic models (see
Ref.\ \cite{harada} and refs.\ therein) the only and obviously positive
contribution to the $\rho$ mass is proportional to $\phi_{N}^{2}$ (i.e.,
$m_{1}=0$).

In the vacuum, the very occurrence of chiral symmetry breaking can be also
traced back to the interaction with the dilaton field: in fact, the quantity
$-m_{0}^{2}$Tr$(\Phi^{\dagger}\Phi)$, where $m_{0}^{2}<0$, arises from a
dilatation-invariant interaction term of the form $bG^{2}$Tr$(\Phi^{\dagger
}\Phi)$ upon the identification $m_{0}^{2}=bG_{0}^{2}$ [see
Eq.\ (\ref{LagrangianG})]. This property also implies that the chiral
condensate $\phi_{N}$ is proportional to the gluon condensate $G_{0}$,
$\phi_{N}\sim G_{0}$. This means that the vacuum expression in
Eq.\ (\ref{mrhoquad}) can be rewritten in the form $m_{\rho}^{2}\sim\phi
_{N}^{2}$, which resembles the KSFR relation \cite{ksfr}. However, the
quantities $G_{0}$ are $\phi_{N}$ may vary independently from each other at
nonzero temperature and density, thus generating a nontrivial behaviour of
$m_{\rho}^{2}$.

\subsection{Equivalent Set of Parameters}

Instead of the eleven parameters in Eq.\ (\ref{param}), it is technically
simpler to use the following, equivalent set of eleven parameters in the
expressions for the physical quantities:
\begin{equation}
m_{\pi}\text{, }m_{\sigma_{N}}\text{, }m_{a_{0}}\text{, }m_{\eta_{N}}\text{,
}m_{\rho}\text{, }m_{a_{1}}\text{, }Z_{\pi}\text{, }\phi_{N}\text{, }%
g_{2}\text{, }h_{1}\text{, }h_{2}\text{.} \label{param2}%
\end{equation}
The quantities $m_{\pi}$, $m_{\rho}$, $m_{a_{1}}$ are taken as the mean values
of the masses of the $\pi$, $\rho$, and $a_{1}$ meson, respectively, as given
by the PDG \cite{PDG}: $m_{\pi}=139.57$ MeV, $m_{\rho}=775.49$ MeV, and
$m_{a_{1}}=1230$ MeV. While $m_{\pi}$ and $m_{\rho}$ are measured to very good
precision, this is not the case for $m_{a_{1}}$. The mass value given above is
referred to as an "educated guess" by the PDG \cite{PDG}. Therefore, we shall
also consider a smaller value, as suggested e.g.\ by the results of
Ref.\ \cite{UBW}. We shall see that, although the overall picture remains
qualitatively unchanged, the description of the decay width of $a_{1}$ into
$\rho\pi$ can be substantially improved.

As outlined in Ref.\ \cite{Mainz},\ the mass of the $\eta_{N}$ meson can be
calculated using the mixing of strange and non-strange contributions in the
physical fields $\eta$ and $\eta^{\prime}(958)$:
\begin{equation}
\eta=\eta_{N}\cos\varphi_{\eta}+\eta_{S}\sin\varphi_{\eta}\text{, }%
\eta^{\prime}=-\eta_{N}\sin\varphi_{\eta}+\eta_{S}\cos\varphi_{\eta}\text{,}
\label{phiGG}%
\end{equation}
where $\eta_{S}$ is a pure $\bar{s}s$ state and $\varphi_{\eta}\simeq
-36^{\circ}$ \cite{Giacosa:2007up}. (A detailed discussion of the $\eta_{N}%
$-$\eta_{S}$ mixing will be presented in Sec.\ \ref{sec.eta-eta}, i.e., in the
$N_{f}=3$ version of our model where the pure-strange field $\eta_{S}$ will be
included as an explicit degree of freedom.) In this way, we obtain the value
$m_{\eta_{N}}=716$ MeV. Given the well-known uncertainty of the value of
$\varphi_{\eta}$, one could also consider other values, e.g., $\varphi_{\eta
}=-41.4^{\circ}$, as published by the KLOE Collaboration \cite{KLOE}. In this
case, $m_{\eta_{N}}=755$ MeV. The variation of the $\eta_{N}$ mass does not
change the results significantly.

The quantities $\phi_{N}$ and $Z_{\pi}$ are linked to the pion decay constant
as $\phi_{N}/Z_{\pi}=f_{\pi}=92.4$ MeV. Therefore, the following six
quantities remain as free parameters:
\begin{equation}
m_{\sigma_{N}}\text{, }m_{a_{0}}\text{, }Z_{\pi}\text{, }g_{2}\text{, }%
h_{1}\text{, }h_{2}\text{.} \label{param3}%
\end{equation}
The masses $m_{\sigma_{N}}$ and $m_{a_{0}}$ depend on the scenario adopted for
the scalar mesons.

At the end of this subsection we report three useful formulas which link the
parameters $g_{1}$, $h_{3}$, and $m_{1}$ of the original set (\ref{param}) to
the second set of parameters (\ref{param2}) [see also Eq.\ (\ref{Z})]:
\begin{align}
g_{1}  &  =g_{1}(Z_{\pi})=\frac{m_{a_{1}}}{Z_{\pi}f_{\pi}}\sqrt{1-\frac
{1}{Z_{\pi}^{2}}}\text{,}\label{g1Q}\\
h_{3}  &  =h_{3}(Z_{\pi})=\frac{m_{a_{1}}^{2}}{Z_{\pi}^{2}f_{\pi}^{2}}\left(
\frac{m_{\rho}^{2}}{m_{a_{1}}^{2}}-\frac{1}{Z_{\pi}^{2}}\right)
\text{,}\label{h3Q}\\
m_{1}^{2}  &  =m_{1}^{2}(Z_{\pi},h_{1},h_{2})=\frac{1}{2}\left[  m_{\rho}%
^{2}+m_{a_{1}}^{2}-Z_{\pi}^{2}f_{\pi}^{2}\left(  g_{1}^{2}+h_{1}+h_{2}\right)
\right]  \text{.} \label{m1eq}%
\end{align}

\section{Decay Widths and \boldmath $\pi\pi$ Scattering Lengths}

In this section, we calculate the formulas for the decay widths and the
$\pi\pi$ scattering lengths and specify their dependence on the parameters
$m_{\sigma}$, $m_{a_{0}}$, $Z_{\pi}$, $g_{2}$, $h_{1}$, and $h_{2}$. Using the
scaling behaviour (\ref{largen}) we obtain that all strong decays and
scattering lengths scale as $N_{c}^{-1}$, as expected. The decay widths are
calculated from the interaction part of the Lagrangian (\ref{LagrangianQ}).

\subsection{Decay Width \boldmath $\rho\rightarrow\pi\pi$} \label{sec.rhopipi}

The $\rho\pi\pi$ interaction Lagrangian obtained from Eq.\ (\ref{LagrangianQ}) reads%

\begin{equation}
\mathcal{L}_{\rho\pi\pi}=A_{\rho\pi\pi}\left(  \partial_{\mu}\vec{\pi}\right)
\cdot\left(  \vec{\rho}^{\mu}\times\vec{\pi}\right)  +B_{\rho\pi\pi}\left(
\partial_{\mu}\vec{\rho}_{\nu}\right)  \cdot\left(  \partial^{\mu}\vec{\pi
}\times\partial^{\nu}\vec{\pi}\right)  \text{.}\label{rhopipiQ}%
\end{equation}

with the following coefficients%

\begin{align}
A_{\rho\pi\pi} &  =Z_{\pi}^{2}\left[  g_{1}(1-g_{1}w_{a_{1}}\phi_{N}%
)+h_{3}w_{a_{1}}\phi_{N}\right]  \equiv Z_{\pi}^{2}g_{1}\frac{m_{\rho}^{2}%
}{m_{a_{1}}^{2}}\text{,} \label{Arhopipi}\\
B_{\rho\pi\pi} &  =-Z_{\pi}^{2}g_{2}w_{a_{1}}^{2}\text{.}\label{Brhopipi}%
\end{align}

Let us consider the decay of $\rho^{0}$ only; the decays of the charged states
are calculated analogously and possess the same values because of the isospin
symmetry that is manifest in our model. The third $\rho$ component possesses
the following interaction Lagrangian

\begin{equation}
\mathcal{L}_{\rho^{0}\pi\pi}=iA_{\rho\pi\pi}\rho_{\mu}^{0}\left(  \pi
^{-}\partial^{\mu}\pi^{+}-\pi^{+}\partial^{\mu}\pi^{-}\right)  +iB_{\rho\pi
\pi}\partial_{\nu}\rho_{\mu}^{0}\left(  \partial^{\nu}\pi^{-}\partial^{\mu}%
\pi^{+}-\partial^{\nu}\pi^{+}\partial^{\mu}\pi^{-}\right)  \text{.}%
\label{rhopipiQ1}
\end{equation}

Let us denote the momenta of $\rho$, $\pi^{+}$ and $\pi^{-}$ as $P$, $P_{1}$
and $P_{2}$, respectively. The $\rho$ meson is a vector state for which we
have to consider the polarisation vector labelled as $\varepsilon_{\mu
}^{(\alpha)}(P)$. 

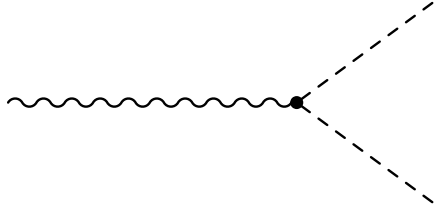
\begin{figure}[h]
\begin{align*} 
\qquad \qquad \qquad  \qquad \qquad \qquad  \quad \; \parbox{180mm}{ \begin{fmfgraph*}(180,80)
    \fmfleftn{i}{1}\fmfrightn{o}{2}
         \fmf{boson,tension=1,label=\text{\small\(\varepsilon^{(\alpha)}_\mu (P)\)},label.dist=-20}{i1,v1}\fmfv{label=\text{\small\(\rho\)},label.dist=45,label.angle=-17}{i1}
    \fmf{dashes,label=\text{\small\(\pi(P_1)\)},tension=1,label.dist=-28}{v1,o2}
        \fmf{dashes,tension=1,label=\text{\small\(\pi (P_2)\)},label.dist=-28}{v1,o1}\fmfdot{v1}
  \end{fmfgraph*}}
\end{align*}\caption{Decay process $\rho \rightarrow \pi\pi$.}\end{figure}

Then, upon substituting $\partial^{\mu}\rightarrow-iP^{\mu
}$\ for the decaying particle and $\partial^{\mu}\rightarrow iP_{1,2}^{\mu}$ for
the decay products, we obtain the following Lorentz-invariant $\rho\pi\pi$
scattering amplitude $-i\mathcal{M}_{\rho^{0}\rightarrow\pi\pi}^{(\alpha)}$
from the Lagrangian (\ref{rhopipiQ1}):

\begin{equation}
-i\mathcal{M}_{\rho^{0}\rightarrow\pi\pi}^{(\alpha)}=\varepsilon_{\mu
}^{(\alpha)}(P)h_{\rho\pi\pi}^{\mu}=\varepsilon_{\mu}^{(\alpha)}(P)[A_{\rho
\pi\pi}(P_{2}^{\mu}-P_{1}^{\mu})+B_{\rho\pi\pi}P_{\nu}(P_{2}^{\mu}P_{1}^{\nu
}-P_{1}^{\mu}P_{2}^{\nu})]\text{,} \label{hrhopipi}%
\end{equation}

where

\begin{equation}
h_{\rho\pi\pi}^{\mu}=A_{\rho\pi\pi}(P_{2}^{\mu}-P_{1}^{\mu})+B_{\rho\pi\pi
}P_{\nu}(P_{2}^{\mu}P_{1}^{\nu}-P_{1}^{\mu}P_{2}^{\nu})\text{.}%
\label{hrhopipi0}
\end{equation}

denotes the $\rho\pi\pi$ vertex.

The vertex can be transformed in the following way:

\begin{align}
h_{\rho\pi\pi}^{\mu} &  =A_{\rho\pi\pi}(P_{2}^{\mu}-P_{1}^{\mu})+B_{\rho\pi
\pi}P_{\nu}(P_{2}^{\mu}P_{1}^{\nu}-P_{1}^{\mu}P_{2}^{\nu})=A_{\rho\pi\pi
}(P_{2}^{\mu}-P_{1}^{\mu})+B_{\rho\pi\pi}\frac{m_{\rho}^{2}}{2}(P_{2}^{\mu
}-P_{1}^{\mu})\nonumber\\
&  =\left(  A_{\rho\pi\pi}+B_{\rho\pi\pi}\frac{m_{\rho}^{2}}{2}\right)
(P_{2}^{\mu}-P_{1}^{\mu})\text{,} \label{hrhopipi1}%
\end{align}

where the equality $P_{\mu}P_{1}^{\mu}=P_{\mu}P_{2}^{\mu}=m_{\rho}^{2}/2$ was
used. The calculation of the decay width will require the determination of the
square of the scattering amplitude. Given that the scattering amplitude in
Eq.\ (\ref{hrhopipi}) depends on the polarisation vector $\varepsilon_{\mu
}^{(\alpha)}(P)$, it is necessary to calculate the average of the amplitude
for all values of $\varepsilon_{\mu}^{(\alpha)}(P)$. For a general scattering
amplitude $-i\mathcal{M}^{(\alpha)}$ $=\varepsilon_{\mu}^{(\alpha)}(P)h^{\mu}$
of a process containing one vector state with mass $m$, the calculation reads
as follows:

\begin{align}
-i\mathcal{M}^{(\alpha)}  &  =\varepsilon_{\mu}^{(\alpha)}(P)h^{\mu
}\Rightarrow|-i\mathcal{\bar{M}}|^{2}=\frac{1}{3}\sum\limits_{\alpha=1}%
^{3}|-i\mathcal{M}^{(\alpha)}|^{2}=\frac{1}{3}\sum\limits_{\alpha=1}%
^{3}\varepsilon_{\mu}^{(\alpha)}(P)h^{\mu}\varepsilon_{\nu}^{(\alpha
)}(P)h^{\nu}\nonumber\\
&  =\frac{1}{3}\sum\limits_{\alpha=1}^{3}\varepsilon_{\mu}^{(\alpha
)}(P)\varepsilon_{\nu}^{(\alpha)}(P)h^{\mu}h^{\nu}=\frac{1}{3}\left(
-g_{\mu\nu}+\frac{P_{\mu}P_{\nu}}{m^{2}}\right)  h^{\mu}h^{\nu}=\frac{1}%
{3}\left[  -(h^{\mu})^{2}+\frac{(P_{\mu}h^{\mu})^{2}}{m^{2}}\right]\text{,}
\label{iM2}
\end{align}

where, in the second line of Eq.\ (\ref{iM2}), we have used%

\begin{equation}
\sum\limits_{\alpha=1}^{3}\varepsilon_{\mu}^{(\alpha)}(P)\varepsilon_{\nu
}^{(\alpha)}(P)=  -g_{\mu\nu}+\frac{P_{\mu}P_{\nu}}{m^{2}}
\text{.} \label{iM21}
\end{equation}

Equation (\ref{iM2}) contains the metric tensor $g_{\mu\nu}=\mathrm{diag}
(1,-1-1,-1)$. Note that, if the vector particle decays, then $P_{\mu}
=(P_{0},\vec{0})$ in the rest frame of the decaying particle and thus

\begin{equation}
\frac{(P_{\mu}h^{\mu})^{2}}{m^{2}}\equiv\frac{(P_{0}h^{0})^{2}}{m^{2}}
=\frac{m^{2}(h^{0})^{2}}{m^{2}}=(h^{0})^{2}\text{.} \label{iM22}
\end{equation}

It is clear that in our case $h_{\rho\pi\pi}^{0}=0$, see
Eq.\ (\ref{hrhopipi1}). Therefore we only have to determine $(h_{\rho\pi\pi
}^{\mu})^{2}$:

\begin{equation}
(h_{\rho\pi\pi}^{\mu})^{2}=\left(  A_{\rho\pi\pi}+B_{\rho\pi\pi}\frac{m_{\rho
}^{2}}{2}\right)  ^{2}\left(  m_{\pi}^{2}+m_{\pi}^{2}-2P_{1}P_{2}\right)
=\left(  A_{\rho\pi\pi}+B_{\rho\pi\pi}\frac{m_{\rho}^{2}}{2}\right)
^{2}\left(  4m_{\pi}^{2}-m_{\rho}^{2}\right)  \text{.}\label{hrhopipi2}%
\end{equation}

Inserting Eq.\ (\ref{hrhopipi2}) into Eq.\ (\ref{iM2}) yields%

\begin{align}
|-i\mathcal{\bar{M}}_{\rho^{0}\rightarrow\pi\pi}|^{2} &  =\frac{1}{3}\left(
A_{\rho\pi\pi}+B_{\rho\pi\pi}\frac{m_{\rho}^{2}}{2}\right)  ^{2}\left(
4m_{\pi}^{2}-m_{\rho}^{2}\right)  \nonumber\\
&  =\frac{4}{3}\left(  A_{\rho\pi\pi}+B_{\rho\pi\pi}\frac{m_{\rho}^{2}}%
{2}\right)  ^{2}k^{2}(m_{\rho},m_{\pi},m_{\pi})\text{,}
\end{align}

where in the second line we have used Eq.\ (\ref{kabc}).

Finally, the full decay width reads

\begin{equation}
\Gamma_{\rho\rightarrow\pi\pi}=\frac{k(m_{\rho},m_{\pi},m_{\pi})}{8\pi
m_{\rho}^{2}}|-i\mathcal{\bar{M}}_{\rho^{0}\rightarrow\pi\pi}|=\frac
{k^{3}(m_{\rho},m_{\pi},m_{\pi})}{6\pi m_{\rho}^{2}}\left(  A_{\rho\pi\pi
}+B_{\rho\pi\pi}\frac{m_{\rho}^{2}}{2}\right)  ^{2}\text{.}\label{Grhopipi}%
\end{equation}

Note that the formula presented in Eq.\ (\ref{Grhopipi}) can be transformed
further using Eqs.\ (\ref{kabc}), (\ref{Arhopipi}) and (\ref{Brhopipi}); we
make explicit the dependence of the decay width on the parameters $Z_{\pi}$ and
$g_{2}$:
\begin{equation}
\Gamma_{\rho\rightarrow\pi\pi}(Z_{\pi},g_{2})=\frac{m_{\rho}^{5}}{48\pi
m_{a_{1}}^{4}}\left[  1-\left(  \frac{2m_{\pi}}{m_{\rho}}\right)  ^{2}\right]
^{3/2}\left[  g_{1}Z_{\pi}^{2}+\left(  1-Z_{\pi}^{2}\right)  \frac{g_{2}}%
{2}\right]  ^{2}\text{.} \label{rhopionpionQ}%
\end{equation}
The experimental value is $\Gamma_{\rho\rightarrow\pi\pi}^{\mathrm{exp}%
}=(149.1\pm0.8)$ MeV \cite{PDG}. The small experimental error can be neglected
and the central value is used as a further constraint allowing us to fix the
parameter $g_{2}$ as function of $Z_{\pi}$:
\begin{equation}
g_{2}=g_{2}(Z_{\pi})=\frac{2}{Z_{\pi}^{2}-1}\left(  g_{1}Z_{\pi}^{2}\pm
\frac{4m_{a_{1}}^{2}}{m_{\rho}}\sqrt{\frac{3\pi\Gamma_{\rho\rightarrow\pi\pi
}^{\mathrm{exp}}}{(m_{\rho}^{2}-4m_{\pi}^{2})^{3/2}}}\,\right)  \text{.}
\label{g2Z}%
\end{equation}
Note that all input values in Eq.\ (\ref{g2Z}) are experimentally known
\cite{PDG}. The parameter $g_{1}=g_{1}(Z_{\pi})$ is fixed via Eq.\ (\ref{g1Q}).

As apparent from Eq.\ (\ref{g2Z}), two solutions for $g_{2}$ are obtained. The
solution with the positive sign in front of the square root may be neglected
because it leads to unphysically large values for the $a_{1}\rightarrow\rho
\pi$ decay width, which is another quantity predicted by our study that also
depends on $g_{2}$ [see Eq.\ (\ref{a1rhopionQ})]. For example, the value
$Z_{\pi}=1.6$ (see below) would lead to $g_{2}\cong40$ which in turn would
give $\Gamma_{a_{1}\rightarrow\rho\pi}\cong14$ GeV -- clearly an unphysically
large value. Therefore, we will take the solution for $g_{2}$ with the
negative sign in front of the square root. In this case, reasonable values for
both $g_{2}$ (see Table~\ref{Table1Q}) and $\Gamma_{a_{1}\rightarrow\rho\pi}$
(see Sec.\ \ref{sec.a1rpQ}) are obtained.

\subsection{Decay Width \boldmath $f_{1}(1285)\rightarrow a_{0}\pi$} \label{sec.f1NQ}

The $f_{1N}a_{0}\pi$ interaction Lagrangian from Eq.\ (\ref{LagrangianQ}) reads%

\begin{equation}
\mathcal{L}_{f_{1N}a_{0}\pi}=A_{f_{1N}a_{0}\pi}f_{1N}^{\mu}\left(
\partial_{\mu}\vec{\pi}\cdot\vec{a}_{0}\right)
+B_{f_{1N}a_{0}\pi}f_{1N}^{\mu}\left(  \partial_{\mu}\vec{a}%
_{0}\cdot\vec{\pi}\right)  \label{f1Na0pi}%
\end{equation}

with the following coefficients:

\begin{align}
A_{f_{1N}a_{0}\pi} &  =Z_{\pi}g_{1}(2g_{1}w_{a_{1}}\phi_{N}-1)+Z_{\pi}%
w_{a_{1}}(h_{2}-h_{3})\phi_{N}\text{,} \label{Af1Na0pi}\\
B_{f_{1N}a_{0}\pi} &  =Z_{\pi}g_{1}\text{.}\label{Bf1Na0pi}%
\end{align}

The decay width of an axial-vector into a scalar and a pseudoscalar has
already been considered in Sec.\ \ref{sec.ASP}; the obtained formula for the
decay width from Eq.\ (\ref{GASP}) can be used here with $I=3$:%

\begin{equation}
\Gamma_{f_{1N}\rightarrow a_{0}\pi}=\frac{k^{3}(m_{f_{1N}},m_{a_{0}},m_{\pi}%
)}{8\pi m_{f_{1N}}^{2}}(A_{f_{1N}a_{0}\pi}-B_{f_{1N}a_{0}\pi})^{2}%
\text{.}\label{Gf1Na0pi}%
\end{equation}

Using Eqs.\ (\ref{kabc}), (\ref{Af1Na0pi}) and (\ref{Bf1Na0pi}),
Eq.\ (\ref{Gf1Na0pi}) can be transformed as follows (we make explicit the
dependence on $Z_{\pi}$ and $h_{2}$):
\begin{equation}
\Gamma_{f_{1N}\rightarrow a_{0}\pi}(m_{a_{0}},Z_{\pi},h_{2})=\frac{g_{1}%
^{2}Z_{\pi}^{2}}{2\pi}\frac{k^{3}(m_{f_{1N}},m_{a_{0}},m_{\pi})}{m_{f_{1N}%
}^{2}m_{a_{1}}^{4}}\left[  m_{\rho}^{2}-\frac{1}{2}(h_{2}+h_{3})\phi_{N}%
^{2}\right]  ^{2} \text{.}\label{f1a0pion}
\end{equation}

There is a subtle point to comment on here. When the
quark-antiquark $a_{0}$ state of our model is identified as the $a_{0}(980)$
meson of the PDG compilation (Scenario I, Sec.\ \ref{sec.scenarioI}), then
this decay width can be used to fix the parameter $h_{2}$ as function of
$Z_{\pi},$ $h_{2}\equiv h_{2}(Z_{\pi})$, by using the corresponding
experimental value $\Gamma_{f_{1N}\rightarrow a_{0}\pi}^{\mathrm{exp}%
}=(8.748\pm2.097)$ MeV \cite{PDG}.
\begin{equation}
h_{2}=h_{2}(Z_{\pi})=\frac{2}{\phi_{N}^{2}}\left(  m_{\rho}^{2}-\frac{h_{3}%
}{2}\phi_{N}^{2}\pm\frac{m_{f_{1N}}m_{a_{0}}^{2}}{g_{1}Z_{\pi}}\sqrt
{\frac{2\pi\Gamma_{f_{1N}\rightarrow a_{0}\pi}^{\mathrm{exp}}}{k^{3}%
(m_{f_{1N}},m_{a_{0}},m_{\pi})}}\right)  \text{.}\label{h2Z}%
\end{equation}
Again, there are two solutions, just as in the case of the parameter $g_{2}$.
How strongly the somewhat uncertain experimental value of $\Gamma
_{f_{1N}\rightarrow a_{0}\pi}$ influences the possible values of $h_{2}$,
depends on the choice of the sign in front of the square root in
Eq.\ (\ref{h2Z}). Varying $\Gamma_{f_{1N}\rightarrow a_{0}\pi}$ within its
experimental range of uncertainty changes the value of $h_{2}$ by an average
of 25\% if the negative sign is chosen, but the same variation of
$\Gamma_{f_{1N}\rightarrow a_{0}\pi}$ changes $h_{2}$ by an average of only
6\% if the positive sign is considered. This is due to the fact that the
solution with the positive square root sign yields larger values of $h_{2}%
\sim80$, while the solution with the negative sign leads to $h_{2}\sim20$. The
absolute change of $h_{2}$ is the same in both cases. Our calculations have
shown that using the negative sign in front of the square root yields a too
small value of the $\eta$-$\eta^{\prime}$ mixing angle $\varphi\cong-9^{\circ
}$. This follows by inserting $h_{2}$ into Eq.\ (\ref{a0etapion}) so that it
is removed as a degree of freedom (i.e., replaced by $Z_{\pi}$) and
calculating the mixing angle $\varphi_{\eta}$ from Eq.\ (\ref{a0etapion0})
using the experimental value of the $a_{0}\rightarrow\eta\pi$ decay amplitude
from Ref.\ \cite{Bugg:1994}. For this reason, we only use the positive sign in
front of the square root in Eq.\ (\ref{h2Z}), i.e., the constraint leading to
higher values of $h_{2}$. Then $\varphi_{\eta}\cong-41.8^{\circ}$ is obtained,
in very good agreement with the central value quoted by the KLOE collaboration
\cite{KLOE}, $\varphi_{\eta}\cong-41.4^{\circ}$ (see also
Sec.\ \ref{sec.fitscenarioI}).

It may be interesting to note that only the (disregarded) lower value of
$h_{2}$ leads to the expected behaviour of the parameter $h_{1}$ which
[according to Eq.\ (\ref{largen})] should be large-$N_{c}$ suppressed: the
lower value of $h_{2}$ yields $h_{1}=1.8$ whereas the higher value of $h_{2}$
yields $h_{1}=-68$ (see Table\textit{ }\ref{Table1Q}).

Note that if the quark-antiquark $a_{0}$ meson of our model is identified as
the $a_{0}(1450)$ meson of the PDG compilation (Scenario II,
Sec.\ \ref{sec.scenarioII}) then the described procedure of replacing $h_{2}$
by $Z_{\pi}$ using Eq.\ (\ref{h2Z}) is no longer applicable because the decay
$f_{1N}\rightarrow a_{0}\pi$ is kinematically not allowed and its counterpart
$a_{0}\rightarrow f_{1N}\pi$ has not been measured.

\subsection{Decay Width \boldmath $\sigma_{N}\rightarrow\pi\pi$} \label{sec.sigmapionpionQ}

The interaction Lagrangian of the scalar state $\sigma_{N}$ with the pions
from Eq.\ (\ref{LagrangianQ}) reads:%

\begin{equation}
\mathcal{L}_{\sigma_{N}\pi\pi}=A_{\sigma_{N}\pi\pi}\sigma_{N}\vec{\pi}%
^{2}+B_{\sigma_{N}\pi\pi}\sigma_{N}\partial_{\mu}\vec{\pi}^{2}+C_{\sigma
_{N}\pi\pi}\sigma_{N}\vec{\pi}\square\vec{\pi} \label{sppQ}%
\end{equation}

with

\begin{align}
A_{\sigma_{N}\pi\pi}  &  =-\left(  \lambda_{1}+\frac{\lambda_{2}}{2}\right)
Z_{\pi}^{2}\phi_{N}\text{,} \label{ANsppQ}\\
B_{\sigma_{N}\pi\pi}  &  =-2g_{1}Z_{\pi}^{2}w_{a_{1}}+\left(  g_{1}^{2}%
+\frac{h_{1}+h_{2}-h_{3}}{2}\right)  Z_{\pi}^{2}w_{a_{1}}^{2}\phi
_{N}\text{,} \label{BNsppQ}\\
C_{\sigma_{N}\pi\pi}  &  =-g_{1}Z_{\pi}^{2}w_{a_{1}}\text{.} \label{CNsppQ}%
\end{align}

The corresponding decay amplitude reads

\begin{equation}
-i\mathcal{M}_{\sigma_{N}\rightarrow\pi\pi}(m_{\sigma_{N}})=i\left (
A_{\sigma_{N}\pi\pi}-B_{\sigma_{N}\pi\pi}\frac{m_{\sigma_{N}}^{2}-2m_{\pi}%
^{2}}{2}-C_{\sigma_{N}\pi\pi}m_{\pi}^{2}\right )  \label{MsNpp}%
\end{equation}

and, consequently, summing over all decay channels $\sigma_{1,2}\rightarrow
\pi^{0}\pi^{0},\pi^{\pm}\pi^{\mp}$ we obtain the following formula for the
decay width $\Gamma_{\sigma_{N}\rightarrow\pi\pi}$:

\begin{equation}
\Gamma_{\sigma_{N}\rightarrow\pi\pi}=\frac{3k(m_{\sigma_{N}},m_{\pi},m_{\pi}%
)}{4\pi m_{\sigma_{N}}^{2}}|-i\mathcal{M}_{\sigma_{N}\rightarrow\pi\pi
}(m_{\sigma_{N}})|^{2}\text{.} \label{GsNpp}
\end{equation}

Note that, using Eqs.\ (\ref{kabc}) and (\ref{ANsppQ}) - (\ref{CNsppQ}), we
can transform Eq.\ (\ref{GsNpp}) as follows (we make explicit the dependence
on free parameters):

\begin{align}
\Gamma_{\sigma_{N}\rightarrow\pi\pi}(m_{\sigma_{N}},Z_{\pi},h_{1}%
,h_{2}) & =\frac{3}{32\pi m_{\sigma_{N}}}\sqrt{1-\left(  \frac{2m_{\pi}%
}{m_{\sigma_{N}}}\right)  ^{2}}\left\{  \frac{m_{\sigma_{N}}^{2}-m_{\pi}^{2}%
}{Z_{\pi}f_{\pi}}\right. \nonumber\\
&  \left.  -\frac{g_{1}^{2}Z_{\pi}^{3}f_{\pi}}{m_{a_{1}}^{4}}\left[  m_{\rho
}^{2}-\frac{\phi_{N}^{2}}{2}(h_{1}+h_{2}+h_{3})\right]  (m_{\sigma_{N}}%
^{2}-2\,m_{\pi}^{2})\right\}  ^{2}\text{.} \label{sigmapionpionQ}%
\end{align}
It is apparent from Eqs.\ (\ref{largen}) that the sigma decay width decreases
as the number of colors $N_{c}$ increases. Thus, the sigma field in our model
is a $\bar{q}q$ state \cite{Pelaez-scalars-below1GeVq2q2}. In Scenario I, Sec.\ \ref{sec.scenarioI}%
, we have assigned the $\sigma_{N}$ field as $f_{0}(600)$, correspondingly we
are working with the assumption that $f_{0}(600)$ [as well as $a_{0}(980)$] is
a $\bar{q}q$ state. In Scenario II, Sec.\ \ref{sec.scenarioII}, the same
assumption is valid for the $f_{0}(1370)$ and $a_{0}(1450)$ states.

Note that in Eq.\ (\ref{sigmapionpionQ}) the first term in braces arises from
the scalar $\sigma_{N}\pi\pi$ vertex, while the second term comes from the
coupling of the $\sigma_{N}$ to the $a_{1}$, which becomes a derivatively
coupled pion after the shift (\ref{shifts}). Because of the different signs,
these two terms interfere destructively. As the decay width of a light
$\sigma_{N}$ meson into two pions can be very well reproduced in the linear
sigma model without vector mesons (corresponding to the case $g_{1}%
\rightarrow0$), this interference prevents obtaining a reasonable value for
this decay width in the present model \textit{with (axial-)vector mesons}, see
Sec.\ \ref{sec.sNppQI}. This problem does not occur for a heavy $\sigma_{N}$
meson, see Sec.\ \ref{sec.sNppQII} and Ref.\ \cite{Zakopane}.

\subsection{Decay Amplitudes \boldmath $a_{0}\rightarrow\eta\pi$ and \boldmath $a_{0}%
\rightarrow\eta^{\prime}\pi$}

Our $N_{f}=2$ Lagrangian (\ref{LagrangianQ}) contains the unphysical field $\eta_{N}$. However, by
making use of Eq.\ (\ref{phi}) and invoking the OZI rule, it is possible to
calculate the decay amplitude for the physical process $a_{0}\rightarrow
\eta\pi$ as%
\begin{equation}
-i\mathcal{M}_{a_{0}\eta\pi}=\cos\varphi(-i\mathcal{M}_{a_{0}\eta_{N}\pi
})\text{.}\label{a0etapion0}%
\end{equation}

The following $a_{0}^{0}\eta_{N}\pi^{0}$ interaction Lagrangian is obtained
from Eq.\ (\ref{LagrangianQ}):%

\begin{equation}
\mathcal{L}_{a_{0}\eta_{N}\pi}=A_{a_{0}\eta_{N}\pi}\vec{a_{0}}\cdot\eta
_{N}\vec{\pi}+B_{a_{0}\eta_{N}\pi}\vec{a_{0}}\cdot\partial_{\mu}\eta
_{N}\partial^{\mu}\vec{\pi}+C_{a_{0}\eta_{N}\pi}\partial_{\mu}\vec{a_{0}}%
\cdot(\vec{\pi}\partial^{\mu}\eta_{N}+\eta_{N}\partial^{\mu}\vec{\pi})
\label{a0etaNpion}%
\end{equation}

with

\begin{align}
A_{a_{0}\eta_{N}\pi}  &  =-\lambda_{2}Z_{\pi}^{2}\phi_{N}\text{,} \label{Aa0epQ}\\
B_{a_{0}\eta_{N}\pi}  &  =2g_{1}Z_{\pi}^{2}w_{a_{1}}(g_{1}w_{a_{1}}\phi
_{N}-1)+(Z_{\pi}w_{a_{1}})^{2}(h_{2}-h_{3})\phi_{N}\nonumber\\
&  \equiv-2\frac{g_{1}^{2}\phi_{N}}{m_{a_{1}}^{2}}\left[  1-\frac{1}{2}%
\frac{Z_{\pi}^{4}f_{\pi}^{2}}{m_{a_{1}}^{2}}(h_{2}-h_{3})\right]\text{,}
\label{Ba0epQ}\\
C_{a_{0}\eta_{N}\pi}  &  =g_{1}w_{a_{1}}Z_{\pi}^{2} \text{.}\label{Ca0epQ}%
\end{align}

As in Sec.\ \ref{sec.sigmapionpionQ}, we obtain

\begin{equation}
-i\mathcal{M}_{a_{0}^{0}\rightarrow\eta_{N}\pi^{0}}=i\left (  A_{a_{0}\eta
_{N}\pi}-B_{a_{0}\eta_{N}\pi}\frac{m_{a_{0}}^{2}-m_{\eta}^{2}-m_{\pi}^{2}}%
{2}+C_{a_{0}\eta_{N}\pi}m_{a_{0}}^{2}\right ) \label{Ma0etaNpionQ}
\end{equation}

Using Eqs.\ (\ref{kabc}) and (\ref{Aa0epQ}) - (\ref{Ca0epQ}),
Eq.\ (\ref{Ma0etaNpionQ}) can be transformed in the following way:%
\begin{align}
-i\mathcal{M}_{a_{0}^{0}\rightarrow\eta_{N}\pi^{0}}(m_{a_{0},}Z,h_{2}%
) & =\frac{i}{Z_{\pi}f_{\pi}}\left\{  m_{\eta_{N}}^{2}-m_{a_{0}}^{2} \right. \nonumber \\
& + \left. \left(
1-\frac{1}{Z_{\pi}^{2}}\right)  \left[  1-\frac{1}{2}\frac{Z_{\pi}^{2}\phi
_{N}^{2}}{m_{a_{1}}^{2}}(h_{2}-h_{3})\right]  (m_{a_{0}}^{2}-m_{\pi}%
^{2}-m_{\eta}^{2})\right\}  \text{.}\label{a0etapion}%
\end{align}

Note that Eqs.\ (\ref{Ma0etaNpionQ}) and (\ref{a0etapion}) contain the unmixed
mass $m_{\eta_{N}}$ which enters when expressing the coupling constants in
terms of the parameters (\ref{param2}), as well as the physical mass $m_{\eta
}=547.8$ MeV. The latter arises because the derivative couplings in the
Lagrangian lead to the appearance of scalar invariants formed from the
four-momenta of the particles emerging from the decay, which can be expressed
in terms of the physical (invariant) masses.

The decay width $\Gamma_{a_{0}\rightarrow\eta\pi}$ follows from
Eq.\ (\ref{a0etapion0}) by including a phase space factor:%
\begin{equation}
\Gamma_{a_{0}\rightarrow\eta\pi}(m_{a_{0}},Z_{\pi},h_{2})=\frac{k(m_{a_{0}%
},m_{\eta},m_{\pi})}{8\pi m_{a_{0}}^{2}}\left\vert -i\mathcal{M}_{a_{0}%
^{0}\rightarrow\eta_{N}\pi^{0}}(m_{a_{0},}Z_{\pi},h_{2})\right\vert
^{2}\text{.}\label{a0etapion2}%
\end{equation}

In the case of Scenario I (Sec.\ \ref{sec.scenarioI}), in which $a_{0}\equiv
a_{0}(980)$, we shall compare the decay amplitude $-i\mathcal{M}_{a_{0}%
^{0}\rightarrow\eta_{N}\pi^{0}}$, Eq.\ (\ref{a0etapion0}), with the
corresponding experimental value deduced from Crystal Barrel data:
$-i\mathcal{M}_{a_{0}^{0}\rightarrow\eta\pi^{0}}^{\mathrm{exp}}=(3330\pm150)$
MeV \cite{Bugg:1994}. This is preferable to the use of the decay width quoted by
the PDG \cite{PDG} for $a_{0}(980)$, which refers to the mean peak width, an
unreliable quantity due to the closeness of the kaon-kaon threshold.

In the case of Scenario II (Sec.\ \ref{sec.scenarioII}), in which $a_{0}\equiv
a_{0}(1450)$, it is also possible to calculate the decay width $a_{0}%
(1450)\rightarrow\eta^{\prime}\pi$, using the OZI rule. The amplitude
$-i\mathcal{M}_{a_{0}\eta^{\prime}\pi}(m_{a_{0}},Z_{\pi},h_{2})$ is obtained
following the same steps as in the previous case, Eq.\ (\ref{a0etapion}):
\begin{align}
-i\mathcal{M}_{a_{0}\eta^{\prime}\pi}(m_{a_{0}},Z_{\pi},h_{2}) &
=-i\frac{\sin\varphi}{Z_{\pi}f_{\pi}}\left\{  m_{\eta_{N}}^{2}-m_{a_{0}}%
^{2}\right.  \nonumber\\
&  +\left.  \left(  1-\frac{1}{Z_{\pi}^{2}}\right)  \left[  1-\frac{1}{2}%
\frac{Z_{\pi}^{2}\phi_{N}^{2}}{m_{a_{1}}^{2}}(h_{2}-h_{3})\right]  (m_{a_{0}%
}^{2}-m_{\pi}^{2}-m_{\eta^{\prime}}^{2})\right\}\text{,} \label{a0etapion00}
\end{align}
where the difference compared to Eqs.\ (\ref{a0etapion0}) and (\ref{a0etapion}%
) is the prefactor $-\sin\varphi$ and the physical $\eta^{\prime}$ mass
$m_{\eta^{\prime}}=958$ MeV. The corresponding decay width reads:
\begin{equation}
\Gamma_{a_{0}(1450)\rightarrow\eta^{\prime}\pi}(m_{a_{0}},Z_{\pi},h_{2}%
)=\frac{k(m_{a_{0}},m_{\eta^{\prime}},m_{\pi})}{8\pi m_{a_{0}}^{2}}\left\vert
-i\mathcal{M}_{a_{0}\eta^{\prime}\pi}(m_{a_{0}},Z_{\pi},h_{2})\right\vert
^{2}\text{.}\label{a0eta'pion2}%
\end{equation}

\subsection{Decay Width \boldmath $a_{1}(1260)\rightarrow\pi\gamma$} \label{sec.a1pg}

The $a_{1}\pi\gamma$ interaction Lagrangian from Eq.\ (\ref{LagrangianQ}) reads

\begin{equation}
\mathcal{L}_{a_{1}\pi\gamma}=eJ_{\mu}A^{\mu}\text{,} \label{a1pg}%
\end{equation}

where $A^{\mu}$ denotes the photon field and $J_{\mu}$ the $a_{1}$-$\pi$
current of the form%

\begin{equation}
J_{\mu}=-iB_{a_{1}\pi\gamma}\left(  a_{1\mu}^{+}\pi^{-}-a_{1\mu}^{-}\pi
^{+}\right)  -iC_{a_{1}\pi\gamma}\left(  a_{1\mu\nu}^{+}\partial^{\nu}\pi
^{-}-a_{1\mu\nu}^{-}\partial^{\nu}\pi^{+}\right)  \label{a1pg1}%
\end{equation}

with%

\begin{align}
B_{a_{1}\pi\gamma}  &  =g_{1}Z_{\pi}^{2}f_{\pi}\text{,} \label{Ba1pg}\\
C_{a_{1}\pi\gamma}  &  =Z_{\pi}w_{a_{1}}=g_{1}\frac{Z_{\pi}^{2}f_{\pi}%
}{m_{a_{1}}^{2}}\equiv\frac{B_{a_{1}\pi\gamma}}{m_{a_{1}}^{2}} \label{Ca1pg}%
\end{align}

and $a_{1\mu\nu}^{\pm}=\partial_{\mu}a_{1\nu}^{\pm}-\partial_{\nu}a_{1\mu
}^{\pm}$. The calculation of $\Gamma_{a_{1}\rightarrow\pi\gamma}$ is performed
analogously to the generic one presented in Sec.\ \ref{sec.AVP}; considering
that the photon has no mass and denoting the momenta of $a_{1}$, $\pi$ and
$\gamma$ respectively as $P$, $P_{1}$ and $P_{2}$, we obtain%

\begin{equation}
\Gamma_{a_1 \rightarrow\pi\gamma}=e^{2}\frac{k(m_{a_{1}},m_{\pi},0)}{24\pi
m_{a_{1}}^{2}} \left\vert 3B_{a_{1}\pi\gamma}^{2}+C_{a_{1}\pi\gamma}^{2}
\left[  m_{a_{1}}^{2}m_{\pi}^{2}+2(P\cdot P_{1})^{2} \right]  -6B_{a_{1}%
\pi\gamma}C_{a_{1}\pi\gamma}(P\cdot P_{1})\right\vert \text{.} \label{Ga1pg}%
\end{equation}

From $P=(m_{a_{1}},\vec{0})$ and $P_{1}=(E_{1},\vec{k}(m_{a_{1}}%
,m_{\pi},0))$ we obtain $P\cdot P_{1}=m_{a_{1}}\sqrt{k^{2}(m_{a_{1}},m_{\pi
},0)+m_{\pi}^{2}}$ $=(m_{a_{1}}^{2}+m_{\pi}^{2})/2$ using Eq.\ (\ref{kabc}).
Additionally, we can substitute $B_{a_{1}\pi\gamma}$ in Eq.\ (\ref{Ga1pg}) by
$C_{a_{1}\pi\gamma}$\ using Eq.\ (\ref{Ca1pg})\ and transform
Eq.\ (\ref{Ga1pg})\ as follows:%

\begin{align}
\Gamma_{a_1 \rightarrow\pi\gamma}  &  =e^{2}\frac{k(m_{a_{1}},m_{\pi},0)}{24\pi
m_{a_{1}}^{2}}\left\vert 3C_{a_{1}\pi\gamma}^{2}m_{a_{1}}^{4}+\frac{1}%
{2}C_{a_{1}\pi\gamma}^{2}(m_{a_{1}}^{4}+m_{\pi}^{4}+4m_{a_{1}}^{2}m_{\pi}%
^{2}) \right. \nonumber \\
& - \left. 3C_{a_{1}\pi\gamma}^{2}m_{a_{1}}^{2}(m_{a_{1}}^{2}+m_{\pi}%
^{2})\right\vert \nonumber\\
&  =e^{2}C_{a_{1}\pi\gamma}^{2}m_{a_{1}}^{2}\frac{k(m_{a_{1}},m_{\pi}%
,0)}{24\pi}\left\vert \frac{1}{2}\left(  \frac{m_{\pi}}{m_{a_{1}}}\right)
^{4}-\left(  \frac{m_{\pi}}{m_{a_{1}}}\right)  ^{2}+\frac{1}{2}\right\vert
\nonumber\\
&  \overset{\text{Eq.\ (\ref{Ca1pg})}}{=}e^{2}g_{1}^{2}\frac{Z_{\pi}^{4}%
f_{\pi}^{2}}{m_{a_{1}}^{2}}\frac{k(m_{a_{1}},m_{\pi},0)}{24\pi}\left\vert
\frac{1}{2}\left(  \frac{m_{\pi}}{m_{a_{1}}}\right)  ^{4}-\left(  \frac
{m_{\pi}}{m_{a_{1}}}\right)  ^{2}+\frac{1}{2}\right\vert \nonumber\\
&  \overset{\text{Eq.\ (\ref{kabc})}}{=}\frac{e^{2}g_{1}^{2}Z_{\pi}^{4}f_{\pi
}^{2}}{48\pi m_{a_{1}}}\left[  1-\left(  \frac{m_{\pi}}{m_{a_{1}}}\right)
^{2}\right]  \left\vert \frac{1}{2}\left(  \frac{m_{\pi}}{m_{a_{1}}}\right)
^{4}-\left(  \frac{m_{\pi}}{m_{a_{1}}}\right)  ^{2}+\frac{1}{2}\right\vert
\nonumber\\
&  \overset{\text{Eq.\ (\ref{g1Q})}}{=}\frac{e^{2}(Z_{\pi}^{2}-1)m_{a_{1}}%
}{48\pi}\left[  1-\left(  \frac{m_{\pi}}{m_{a_{1}}}\right)  ^{2}\right]
\left\vert \frac{1}{2}\left(  \frac{m_{\pi}}{m_{a_{1}}}\right)  ^{4}-\left(
\frac{m_{\pi}}{m_{a_{1}}}\right)  ^{2}+\frac{1}{2}\right\vert
\end{align}

or in other words
\begin{equation}
\Gamma_{a_{1}\rightarrow\pi\gamma}(Z_{\pi})=\frac{e^{2}}{96\pi}\,(Z_{\pi}%
^{2}-1)\,m_{a_{1}}\left[  1-\left(  \frac{m_{\pi}}{m_{a_{1}}}\right)
^{2}\right]  ^{3}\text{.} \label{a1piongamma}
\end{equation}
Note that the $a_{1}\rightarrow\pi\gamma$ decay width depends only on the
renormalisation constant $Z_{\pi}$, explicitly denoted in
Eq.\ (\ref{a1piongamma}). In fact, it is generated via the $a_{1}$-$\pi$
mixing and vanishes in the limit $Z_{\pi}\rightarrow1$. [A similar mechanism
for this decay is described in Ref.\ \cite{ecker}.] The fact that we include
photons following the second realisation of VMD described in
Ref.\ \cite{OCPTW} renders this process possible in our model. Inverting
Eq.\ (\ref{a1piongamma}) we obtain%

\begin{align}
Z_{\pi}^{2}  &  =1+\frac{96\pi\Gamma_{a_{1}\rightarrow\pi\gamma}}%
{e^{2}m_{a_{1}}\left[  1-\left(  \frac{m_{\pi}}{m_{a_{1}}}\right)
^{2}\right]  ^{3}}\nonumber\\
 \Rightarrow Z_{\pi} & =\sqrt{1+\frac{96\pi\Gamma_{a_{1}\rightarrow\pi\gamma}%
}{e^{2}m_{a_{1}}\left[  1-\left(  \frac{m_{\pi}}{m_{a_{1}}}\right)
^{2}\right]  ^{3}}}\text{.} \label{Za1pg}%
\end{align}

The first line of Eq.\ (\ref{Za1pg}) yields in principle two opposite-sign
solutions. We work, however, only with the positive-sign renormalisation.

Using $\Gamma_{a_{1}\rightarrow\pi\gamma}^{\mathrm{exp}}=(0.640\pm0.246)$ MeV
\cite{PDG}, one obtains $Z_{\pi}=1.67\pm0.2$ from Eq.\ (\ref{Za1pg}).
Unfortunately, the experimental error for the quantity $\Gamma_{a_{1}%
\rightarrow\pi\gamma}$ is large. Given that almost all quantities of interest
depend very strongly on $Z_{\pi}$, a better experimental knowledge of this
decay would be useful to constrain $Z_{\pi}$. In the study of Scenario I,
Sec.\ \ref{sec.scenarioI}, this decay width will be part of a $\chi^{2}$
analysis, but still represents the main constraint for $Z_{\pi}$.

\subsection{Decay Width \boldmath $a_{1}(1260)\rightarrow\sigma_{N}\pi$} \label{sec.a1spQ}

The interaction Lagrangian obtained from Eq.\ (\ref{LagrangianQ}) reads

\begin{equation}
\mathcal{L}_{a_{1}\sigma_{N}\pi}=A_{a_{1}\sigma_{N}\pi}\vec{a}_{1\mu}%
\cdot\sigma_{N}\partial^{\mu}\vec{\pi}+B_{a_{1}\sigma_{N}\pi}\vec{a}_{1\mu
}\cdot\vec{\pi}\partial^{\mu}\sigma_{N}\label{a1spQ}
\end{equation}

with the following coefficients:

\begin{align}
A_{a_{1}\sigma_{N}\pi} &  =Z_{\pi}\left[  g_{1}(-1+2g_{1}w_{a_{1}}\phi
_{N})+(h_{1}+h_{2}-h_{3})w_{a_{1}}\phi_{N}\right]\text{,} \label{Aa1sNpQ}\\
B_{a_{1}\sigma_{N}\pi} &  =g_{1}Z_{\pi}\text{.}\label{Ba1sNpQ}%
\end{align}

As in Sec.\ \ref{sec.ASP} we obtain

\begin{equation}
\Gamma_{a_{1}\rightarrow\sigma_{N}\pi}=\frac{k^{3}(m_{a_{1}},m_{\sigma_{N}%
},m_{\pi})}{24\pi m_{a_{1}}^{2}}(A_{a_{1}\sigma_{N}\pi}-B_{a_{1}\sigma_{N}\pi
})^{2}\text{,}\label{Ga1sNp}
\end{equation}

where we have set the isospin factor $I=1$ in Eq.\ (\ref{GASP}).
Equation (\ref{Ga1sNp}) can be further transformed using Eqs.\ (\ref{kabc}),
(\ref{Aa1sNpQ}) and (\ref{Ba1sNpQ}):

\begin{equation}
\Gamma_{a_{1}\rightarrow\sigma_{N}\pi}\equiv\Gamma_{a_{1}\rightarrow\sigma
_{N}\pi}(m_{\sigma_{N}},Z_{\pi},h_{1},h_{2})=\frac{k^{3}(m_{a_{1}}%
,m_{\sigma_{N}},m_{\pi})}{6\pi m_{a_{1}}^{6}}g_{1}^{2}Z_{\pi}^{2}\left[
m_{\rho}^{2}-\frac{\phi_{N}^{2}}{2}(h_{1}+h_{2}+h_{3})\right]  ^{2}%
\text{.}\label{a1sigmapionQ}
\end{equation}

\subsection{Decay Width \boldmath $a_{1}(1260)\rightarrow\rho\pi$} \label{sec.a1rpQ}

We obtain the following $a_{1}\rho\pi$ interaction Lagrangian from
Eq.\ (\ref{LagrangianQ}):%

\begin{align}
\mathcal{L}_{a_{1}\rho\pi}  &  =A_{a_{1}\rho\pi}\vec{a}_{1\mu}\cdot(\vec{\pi
}\times\vec{\rho}^{\mu})+B_{a_{1}\rho\pi}\vec{a}_{1\mu}\cdot\lbrack
(\partial^{\mu}\vec{\rho}^{\nu}-\partial^{\nu}\vec{\rho}^{\mu})\times
\partial_{\nu}\vec{\pi}]\nonumber\\
&  +C_{a_{1}\rho\pi}(\partial_{\nu}\vec{a}_{1\mu}-\partial_{\mu}\vec{a}_{1\nu
})\cdot(\vec{\rho}^{\mu}\times\partial^{\nu}\vec{\pi}) \label{a1rpQ}%
\end{align}

with the following coefficients:

\begin{align}
A_{a_{1}\rho\pi}  &  =Z_{\pi}(g_{1}^{2}-h_{3})\phi_{N}\text{,} \label{Aa1rpQ}\\
B_{a_{1}\rho\pi}  &  =Z_{\pi}g_{2}w_{a_{1}}\text{,} \label{Ba1rpQ}\\
C_{a_{1}\rho\pi}  &  =Z_{\pi}g_{2}w_{a_{1}}\text{.} \label{Ca1rpQ}%
\end{align}

Let us isolate the interaction Lagrangian of the neutral $a_{1}$ component
from Eq.\ (\ref{a1rpQ}):

\begin{align}
\mathcal{L}_{a_{1}^{0}\rho\pi} &  =-iA_{a_{1}\rho\pi}a_{1\mu}^{0}(\rho^{\mu
-}\pi^{+}-\rho^{\mu+}\pi^{-})\nonumber\\
&  -iB_{a_{1}\rho\pi}a_{1\mu}^{0}[(\partial^{\nu}\rho^{\mu-}-\partial^{\mu
}\rho^{\nu-})\partial_{\nu}\pi^{+}-\left(  \partial^{\nu}\rho^{\mu+}%
-\partial^{\mu}\rho^{\nu+}\right)  \partial_{\nu}\pi^{-}]\nonumber\\
&  -iC_{a_{1}\rho\pi}(\partial_{\nu}a_{1\mu}^{0}-\partial_{\mu}a_{1\nu}%
^{0})(\partial^{\nu}\pi^{-}\rho^{\mu+}-\partial^{\nu}\pi^{+}\rho^{\mu
-})\text{.}\label{a1rpQ1}%
\end{align}

Let us for the beginning consider the decay $a_{1}^{0}\rightarrow\rho^{-}%
\pi^{+}$ only:

\begin{align}
\mathcal{L}_{a_{1}^{0}\rho^{-}\pi^{+}} &  =-iA_{a_{1}\rho\pi}a_{1\mu}^{0}%
\rho^{\mu-}\pi^{+}-iB_{a_{1}\rho\pi}a_{1\mu}^{0}(\partial^{\nu}\rho^{\mu
-}-\partial^{\mu}\rho^{\nu-})\partial_{\nu}\pi^{+}\nonumber\\
&  +iC_{a_{1}\rho\pi}(\partial_{\nu}a_{1\mu}^{0}-\partial_{\mu}a_{1\nu}%
^{0})\partial^{\nu}\pi^{+}\rho^{\mu-}\nonumber\\
&  =-iA_{a_{1}\rho\pi}a_{1\mu}^{0}\rho^{\mu-}\pi^{+}-iB_{a_{1}\rho\pi}a_{1\mu
}^{0}(\partial^{\nu}\rho^{\mu-}-\partial^{\mu}\rho^{\nu-})\partial_{\nu}%
\pi^{+}\nonumber\\
&  -iC_{a_{1}\rho\pi}\partial_{\nu}a_{1\mu}^{0}(\partial^{\mu}\pi^{+}\rho
^{\nu-}-\partial^{\nu}\pi^{+}\rho^{\mu-})\text{.}\label{a1rpQ2}%
\end{align}

Let us denote the momenta of $a_{1}$, $\rho$ and $\pi$ as $P$, $P_{1}$ and $P_{2}%
$. Our decay process involves two vector states: $a_{1}$\ and$\ \rho$. For
this reason we have to consider the corresponding polarisation vectors
labelled as $\varepsilon_{\mu}^{(\alpha)}(P)$ for $a_{1}$\ and $\varepsilon
_{\nu}^{(\beta)}(P_{1})$ for $\rho$. Then, upon substituting $\partial^{\mu
}\rightarrow-iP^{\mu}$\ for the decaying particle and $\partial^{\mu}\rightarrow
iP_{1,2}^{\mu}$ for the decay products, we obtain the following Lorentz-invariant
$a_{1}\rho\pi$ scattering amplitude $-i\mathcal{M}_{a_{1}^{0}\rightarrow
\rho^{-}\pi^{+}}^{(\alpha,\beta)}$:

\begin{align}
-i\mathcal{M}_{a_{1}^{0}\rightarrow\rho^{-}\pi^{+}}^{(\alpha,\beta)} &
=\varepsilon_{\mu}^{(\alpha)}(P)\varepsilon_{\nu}^{(\beta)}(P_{1})h_{a_{1}%
\rho\pi}^{\mu\nu}=\varepsilon_{\mu}^{(\alpha)}(P)\varepsilon_{\nu}^{(\beta
)}(P_{1})\nonumber\\
&  \times\left\{  A_{a_{1}\rho\pi}g^{\mu\nu}+B_{a_{1}\rho\pi}\left[
P_{1}^{\mu}P_{2}^{\nu}-(P_{1}\cdot P_{2})g^{\mu\nu}\right]  \right.
\nonumber\\
&  \left.  +C_{a_{1}\rho\pi}\left[  P_{2}^{\mu}P^{\nu}-(P\cdot P_{2})g^{\mu
\nu}\right]  \right\}  \label{Ma1rpQ}%
\end{align}

with

\begin{equation}
h_{a_{1}\rho\pi}^{\mu\nu}=A_{a_{1}\rho\pi}g^{\mu\nu}+B_{a_{1}\rho\pi}\left[
P_{1}^{\mu}P_{2}^{\nu}-(P_{1}\cdot P_{2})g^{\mu\nu}\right]  +C_{a_{1}\rho\pi
}\left[  P_{2}^{\mu}P^{\nu}-(P\cdot P_{2})g^{\mu\nu}\right]\text{,}  \label{ha1rpQ}%
\end{equation}

where $h_{a_{1}\rho\pi}^{\mu\nu}$ denotes the $a_{1}\rho\pi$ vertex [more
precisely, this is only the $a_{1}\rho^{-}\pi^{+}$ vertex but, as evident from
Eq.\ (\ref{a1rpQ1}), it is the same as the $a_{1}\rho^{+}\pi^{-}$ vertex up to
a sign that is of no importance for the calculation of the decay width]. Given
that $B_{a_{1}\rho\pi}=C_{a_{1}\rho\pi}$\ [see Eqs.\ (\ref{Ba1rpQ}) and
(\ref{Ca1rpQ})], we observe that the vertex of Eq.\ (\ref{ha1rpQ}) possesses
exactly the same form as the one presented in Eq.\ (\ref{iMAVP}).
Consequently, we can utilise results from Sec.\ \ref{sec.AVP} where the generic
decay of an axial-vector into a vector and a pseudoscalar was presented. The
squared averaged decay amplitude for the decay $a_{1}\rightarrow\rho\pi$ then reads%

\begin{equation}
|-i\mathcal{\bar{M}}_{a_{1}^{0}\rightarrow\rho^{-}\pi^{+}}|^{2}=\frac{1}%
{3}\left[  \left\vert h_{a_{1}\rho\pi}^{\mu\nu}\right\vert ^{2}-\frac
{\left\vert h_{a_{1}\rho\pi}^{\mu\nu}P_{\mu}\right\vert ^{2}}{m_{a_{1}}^{2}%
}-\frac{\left\vert h_{a_{1}\rho\pi}^{\mu\nu}P_{1\nu}\right\vert ^{2}}{m_{\rho
}^{2}}+\frac{\left\vert h_{a_{1}\rho\pi}^{\mu\nu}P_{\mu}P_{1\nu}\right\vert
^{2}}{m_{\rho}^{2}m_{a_{1}}^{2}}\right]  \text{.}\label{iMa1rpQ}%
\end{equation}

Using the identities $P_{1}\cdot P_{2}=(m_{a_{1}}^{2}-m_{\rho}^{2}-m_{\pi}%
^{2})/2$, $P\cdot P_{1}=m_{a_{1}}E_{1}$ and $P\cdot P_{2}=m_{a_{1}}E_{2}$, we
can now calculate the four contributions to $|-i\mathcal{\bar{M}}_{a_{1}%
^{0}\rightarrow\rho^{-}\pi^{+}}|^{2}$ in Eq.\ (\ref{iMa1rpQ}):%

\begin{align}
\left\vert h_{a_{1}\rho\pi}^{\mu\nu}\right\vert ^{2} &  =4A_{a_{1}\rho\pi}%
^{2}+B_{a_{1}\rho\pi}^{2}[m_{\pi}^{2}m_{\rho}^{2}+2(P_{1}\cdot P_{2}%
)^{2}]+C_{a_{1}\rho\pi}^{2}[m_{\pi}^{2}m_{a_{1}}^{2}+2(P\cdot P_{2}%
)^{2}]\nonumber\\
&  -6A_{a_{1}\rho\pi}B_{a_{1}\rho\pi}(P_{1}\cdot P_{2})-6A_{a_{1}\rho\pi
}C_{a_{1}\rho\pi}(P\cdot P_{2})+6B_{a_{1}\rho\pi}C_{a_{1}\rho\pi}(P_{1}\cdot
P_{2})(P\cdot P_{2})\nonumber\\
&  \overset{\text{Eqs.\ (\ref{Aa1rpQ}) - (\ref{Ca1rpQ})}}{\equiv}Z_{\pi}%
^{4}f_{\pi}^{2}\left\{  4(g_{1}^{2}-h_{3})^{2}+\frac{g_{1}^{2}g_{2}^{2}%
}{m_{a_{1}}^{4}}\left[  m_{a_{1}}^{4}+m_{\pi}^{4}+m_{\rho}^{4}+m_{\pi}%
^{2}m_{\rho}^{2} \right. \right. \nonumber \\
& + \left. \left. m_{a_{1}}^{2}(m_{\pi}^{2}-2m_{\rho}^{2}) + 3(m_{a_{1}}%
^{2}-m_{\rho}^{2}-m_{\pi}^{2})m_{a_{1}}E_{2}\right]  \right.  \nonumber\\
&  -\left.  3\,\frac{g_{1}g_{2}(g_{1}^{2}-h_{3})}{m_{a_{1}}^{2}}\left(
m_{a_{1}}^{2}-m_{\rho}^{2}-m_{\pi}^{2}+2m_{a_{1}}E_{2}\right)  \right\}
\text{,}\label{ha1rpQ1}
\end{align}
\begin{align}
\left\vert h_{a_{1}\rho\pi}^{\mu\nu}P_{\mu}\right\vert ^{2} &  =A_{a_{1}%
\rho\pi}^{2}m_{\rho}^{2}+C_{a_{1}\rho\pi}^{2}\left[  (P\cdot P_{1})^{2}m_{\pi
}^{2}+(P\cdot P_{2})^{2}m_{\rho}^{2}-2(P\cdot P_{1})(P\cdot P_{2})(P_{1}\cdot
P_{2})\right]  \nonumber\\
&  +2A_{a_{1}\rho\pi}C_{a_{1}\rho\pi}\left[  (P\cdot P_{1})(P_{1}\cdot
P_{2})-(P\cdot P_{2})m_{\rho}^{2}\right]  \nonumber\\
&  \overset{\text{Eqs.\ (\ref{Aa1rpQ}) - (\ref{Ca1rpQ})}}{\equiv}Z_{\pi}%
^{4}f_{\pi}^{2}\left\{  (g_{1}^{2}-h_{3})^{2}m_{a_{1}}^{2}+\frac{g_{1}%
^{2}g_{2}^{2}}{4m_{a_{1}}^{4}} \left [(m_{a_{1}}^{2}-m_{\pi}^{2})^{2}(m_{a_{1}}%
^{2}+m_{\pi}^{2}-2m_{\rho}^{2}) \right. \right. \nonumber \\
& + \left. \left. (m_{\pi}^{2}+m_{a_{1}}^{2})m_{\rho}%
^{4} - 4(m_{a_{1}}^{2}-m_{\rho}^{2}-m_{\pi}^{2})m_{a_{1}}^{2}E_{1}%
E_{2} \right ] \right.  \nonumber\\
&  \left.  +\frac{g_{1}g_{2}(g_{1}^{2}-h_{3})}{m_{a_{1}}^{2}}[2m_{a_{1}}^{2}%
E_{1}E_{2}-(m_{a_{1}}^{2}-m_{\rho}^{2}-m_{\pi}^{2})m_{a_{1}}^{2}]\right\}
\text{,}
\end{align}
\begin{align}
\left\vert h_{a_{1}\rho\pi}^{\mu\nu}P_{1\nu}\right\vert ^{2} &  =A_{a_{1}%
\rho\pi}^{2}m_{a_{1}}^{2}+B_{a_{1}\rho\pi}^{2}\left[  (P\cdot P_{1})^{2}%
m_{\pi}^{2}+(P_{1}\cdot P_{2})^{2}m_{a_{1}}^{2}-2(P\cdot P_{1})(P\cdot
P_{2})(P_{1}\cdot P_{2})\right]  \nonumber\\
&  +2A_{a_{1}\rho\pi}B_{a_{1}\rho\pi}\left[  (P\cdot P_{1})(P\cdot
P_{2})-(P_{1}\cdot P_{2})m_{a_{1}}^{2}\right]  \nonumber\\
&  \overset{\text{Eqs.\ (\ref{Aa1rpQ}) - (\ref{Ca1rpQ})}}{\equiv}Z_{\pi}%
^{4}f_{\pi}^{2}\left\{  (g_{1}^{2}-h_{3})^{2}m_{\rho}^{2}+\frac{g_{1}^{2}%
g_{2}^{2}}{4m_{a_{1}}^{4}} \left [(m_{\pi}^{2}-m_{\rho}^{2})^{2}(m_{\pi}^{2}+m_{\rho
}^{2}-2m_{a_{1}}^{2}) \right. \right. \nonumber \\
& + \left. \left. (m_{\pi}^{2}+m_{\rho}^{2})m_{a_{1}}^{4} - 4(m_{a_{1}}^{2}-m_{\rho}^{2}-m_{\pi}^{2})m_{a_{1}}^{2}E_{1}%
E_{2} \right ] \right.
\nonumber\\
&  \left. +\frac{g_{1}g_{2}(g_{1}^{2}-h_{3})}{m_{a_{1}}^{2}}\left[  (m_{a_{1}}%
^{2}-m_{\pi}^{2})m_{a_{1}}E_{1}-2m_{a_{1}}m_{\rho}^{2}\left(  E_{2}%
+\frac{E_{1}}{2}\right)  \right]  \right\}  \text{,}
\end{align}
\begin{align}
\left\vert h_{a_{1}\rho\pi}^{\mu\nu}P_{\mu}P_{1\nu}\right\vert ^{2} &
=[A_{a_{1}\rho\pi}(P\cdot P_{1})]^{2}\overset{\text{Eqs.\ (\ref{Aa1rpQ}) -
(\ref{Ca1rpQ})}}{\equiv}(g_{1}^{2}-h_{3})^{2}Z_{\pi}^{4}f_{\pi}^{2}m_{a_{1}%
}^{2}E_{1}^{2} \label{ha1rpQ4}%
\end{align}
\\
with $E_{1}=\sqrt{k^{2}(m_{a_{1}},m_{\rho},m_{\pi})+m_{\rho}^{2}}$ and
$E_{2}^{2}=\sqrt{k^{2}(m_{a_{1}},m_{\rho},m_{\pi})+m_{\pi}^{2}}$.

The formula for the decay width $\Gamma_{a_{1}\rightarrow\rho\pi}$ is the same
as the one presented in Eq.\ (\ref{GAV0P0}), multiplied by a factor of two in
order to consider the two decay channels $a_{1}^{0}\rightarrow\rho^{-}\pi^{+}$
and $a_{1}^{0}\rightarrow\rho^{+}\pi^{-}$ from Eq.\ (\ref{a1rpQ1}):%

\begin{equation}
\Gamma_{a_{1}\rightarrow\rho\pi}=\frac{k(m_{a_{1}},m_{\rho},m_{\pi})}{4\pi
m_{a_{1}}^{2}}|-i\mathcal{\bar{M}}_{a_{1}^{0}\rightarrow\rho^{-}\pi^{+}}%
|^{2}\label{a1rhopionQ}%
\end{equation}

with $|-i\mathcal{\bar{M}}_{a_{1}^{0}\rightarrow\rho^{-}\pi^{+}}|^{2}$ from
Eq.\ (\ref{iMa1rpQ}), i.e., Eqs.\ (\ref{ha1rpQ1}) - (\ref{ha1rpQ4}).

\subsection{Tree-Level Scattering Lengths} \label{sec.SLQ}

The calculation of the tree-level $\pi\pi$ scattering lengths has been
described in detail in Ref.\ \cite{DA}; in this section we will repeat the
main points.
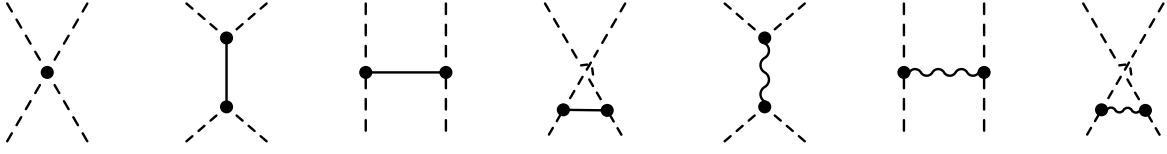
\begin{figure}[h]
\begin{center}
\begin{align*} 
\ \ \parbox{30mm}{ \begin{fmfgraph*}(30,52)
    \fmfstraight\fmftopn{t}{2} \fmfstraight\fmfbottomn{b}{2}
         \fmf{dashes,tension=1}{t1,v1}\fmf{dashes,tension=1}{b2,v1}
         \fmf{dashes,tension=1}{t2,v1}\fmf{dashes,tension=1}{b1,v1}
    \fmfdot{v1}
  \end{fmfgraph*}} \!\!\!\!\!\!\!\!\!\!
\parbox{30mm}{ \begin{fmfgraph*}(30,52)
    \fmfstraight \fmftopn{t}{2}\fmfstraight \fmfbottomn{b}{2}
         \fmf{dashes,tension=1}{t1,v1}\fmf{dashes,tension=1}{t2,v1}
         \fmf{vanilla,tension=1}{v1,v2}
         \fmf{dashes,tension=1}{b1,v2}\fmf{dashes,tension=1}{b2,v2}
    \fmfdot{v1}\fmfdot{v2}
  \end{fmfgraph*}} \!\!\!\!\!\!\!\!\!\!
\parbox{30mm}{ \begin{fmfgraph*}(30,52)
    \fmfstraight \fmftopn{t}{2}\fmfstraight \fmfbottomn{b}{2}
         \fmf{dashes,tension=1}{t1,v1}\fmf{dashes,tension=1}{v1,b1}
         \fmf{dashes,tension=1}{t2,v2}\fmf{dashes,tension=1}{v2,b2}
              \fmfdot{v1}\fmfdot{v2}
              \fmffreeze
              \fmf{vanilla}{v1,v2}
  \end{fmfgraph*}} \!\!\!\!\!\!\!\!\!\!
\parbox{30mm}{ \begin{fmfgraph*}(30,52)
    \fmfstraight \fmftopn{t}{2}\fmfstraight \fmfbottomn{b}{2}
         \fmf{dashes,tension=0.5}{t1,v1}\fmf{dashes,left=1,tension=2.2}{v1,v2} \fmf{dashes}{v2,v3}\fmf{dashes}{v3,b2}
                  \fmf{dashes}{t2,v4}\fmf{phantom,right=1,tension=3}{v4,v5}\fmf{dashes}{v4,v5}\fmf{dashes}{v5,v6}\fmf{dashes,tension=1.5}{v6,b1}
                  \fmffreeze
                    \fmf{vanilla,tension=1}{v3,v6}
                    \fmfdot{v3,v6}
         \end{fmfgraph*}} \!\!\!\!\!\!\!\!\!\!
\parbox{30mm}{ \begin{fmfgraph*}(30,52)
    \fmfstraight \fmftopn{t}{2}\fmfstraight \fmfbottomn{b}{2}
         \fmf{dashes,tension=1}{t1,v1}\fmf{dashes,tension=1}{t2,v1}
         \fmf{boson,tension=1}{v1,v2}
         \fmf{dashes,tension=1}{b1,v2}\fmf{dashes,tension=1}{b2,v2}
    \fmfdot{v1}\fmfdot{v2}
  \end{fmfgraph*}} \!\!\!\!\!\!\!\!\!\!
\parbox{30mm}{ \begin{fmfgraph*}(30,52)
    \fmfstraight \fmftopn{t}{2}\fmfstraight \fmfbottomn{b}{2}
         \fmf{dashes,tension=1}{t1,v1}\fmf{dashes,tension=1}{v1,b1}
         \fmf{dashes,tension=1}{t2,v2}\fmf{dashes,tension=1}{v2,b2}
              \fmfdot{v1}\fmfdot{v2}
              \fmffreeze
              \fmf{boson}{v1,v2}
  \end{fmfgraph*}} \!\!\!\!\!\!\!\!\!\!
\parbox{30mm}{ \begin{fmfgraph*}(30,52)
    \fmfstraight \fmftopn{t}{2}\fmfstraight \fmfbottomn{b}{2}
         \fmf{dashes,tension=0.5}{t1,v1}\fmf{dashes,left=1,tension=2.2}{v1,v2} \fmf{dashes}{v2,v3}\fmf{dashes}{v3,b2}
                  \fmf{dashes}{t2,v4}\fmf{phantom,right=1,tension=3}{v4,v5}\fmf{dashes}{v4,v5}\fmf{dashes}{v5,v6}\fmf{dashes,tension=1.5}{v6,b1}
                  \fmffreeze
                    \fmf{boson,tension=1}{v3,v6}
                    \fmfdot{v3,v6}
         \end{fmfgraph*}}
\end{align*}\end{center}\caption{Diagrams contributing to the $\pi \pi$ scattering lengths. The dashed lines denote the pions, the solid lines denote the intermediate scalar meson whereas
the wavy lines denote the intermediate vector state.}\end{figure}

The scattering lengths are calculated from three contributions: the "pure"
$\pi\pi$ (contact) scattering, $\pi\pi$ scattering via the virtual $\sigma
_{N}$ meson ($s$, $t$, $u$ channels; $s$, $t$, $u$ denote the Mandelstam
variables) and $\pi\pi$ scattering via the virtual $\rho$ meson (also $s$,
$t$, $u$ channels). Consequently, the corresponding scattering Lagrangian
consists of a term containing $4\pi$ vertices ($\mathcal{L}_{4\pi}$) and terms
describing interactions of pions with $\sigma_{N}$ [depicted in $\mathcal{L}%
_{\sigma_{N}\pi\pi}$, Eq.\ (\ref{sppQ})] and interactions of pions with $\rho$
[depicted in $\mathcal{L}_{\rho\pi\pi}$, Eq.\ (\ref{rhopipiQ})]:%

\begin{equation}
\mathcal{L}_{\pi\pi}=\mathcal{L}_{4\pi}+\mathcal{L}_{\sigma_{N}\pi\pi
}+\mathcal{L}_{\rho\pi\pi}\text{,} \label{LSL}%
\end{equation}
where the following form of $\mathcal{L}_{4\pi}$ is obtained from the
Lagrangian (\ref{LagrangianQ}):%

\begin{align}
\mathcal{L}_{4\pi} &  =-\,\frac{1}{4}\,(\lambda_{1}+\frac{\lambda_{2}}%
{2})\,Z_{\pi}^{4}(\vec{\pi}^{2})^{2}+\frac{1}{2}\,(g_{1}^{2}-h_{3})w_{a_{1}%
}^{2}Z_{\pi}^{4}\,(\partial_{\mu}\vec{\pi}\cdot\vec{\pi})^{2}\nonumber\\
&  +\frac{1}{4}(h_{1}+h_{2}+h_{3})w_{a_{1}}^{2}\,Z_{\pi}^{4}\,\vec{\pi}%
^{2}(\partial_{\mu}\vec{\pi})^{2}\text{.}\label{LSL1}%
\end{align}
Note that Eq.\ (\ref{LSL}) may also contain contributions proportional to
$[(\partial^{\mu}\vec{\pi})\times(\partial^{\nu}\vec{\pi})]^{2}$ from the
$g_{3,4}$ terms in the Lagrangian (\ref{LagrangianQ}). However, we do not
consider these terms because all our calculations will be at threshold where
the terms with only pion derivatives do not contribute.

Let us denote the incoming pions with labels $a$ and $b$ and the outgoing
pions with labels $c$ and $d$.\ The $\pi\pi$ scattering amplitude
$\mathcal{M}_{\pi\pi}(s,t,u)$ obtained from the Lagrangian (\ref{LSL}) then
has three contributions, one for the $s$, $t$ and $u$ channels, respectively:%

\begin{equation}
\mathcal{M}_{\pi\pi}(s,t,u)=i\delta^{ab}\delta^{cd}A(s,t,u)+i\delta^{ac}%
\delta^{bd}A(t,u,s)+i\delta^{ad}\delta^{bc}A(u,s,t)\text{,} \label{MSLQ}%
\end{equation}

where

\begin{align}
A(s,t,u)  &  =(g_{1}^{2}-h_{3})Z_{\pi}^{4}w_{a_{1}}^{2}s-2\left(  \lambda
_{1}+\frac{\lambda_{2}}{2}\right)  Z_{\pi}^{4}-(h_{1}+h_{2}+h_{3})Z_{\pi}%
^{4}w_{a_{1}}^{2}(s-2m_{\pi}^{2})\nonumber\\
&  -[-2m_{\pi}^{2}C_{\sigma_{N}\pi\pi}+B_{\sigma_{N}\pi\pi}(2m_{\pi}%
^{2}-s)+2A_{\sigma_{N}\pi\pi}]^{2}\frac{1}{s-m_{\sigma_{N}}^{2}}\nonumber\\
&  +\left(  A_{\rho\pi\pi}+B_{\rho\pi\pi}\frac{t}{2}\right)  ^{2}\frac
{u-s}{t-m_{\rho}^{2}}+\left(  A_{\rho\pi\pi}+B_{\rho\pi\pi}\frac{u}{2}\right)
^{2}\,\frac{t-s}{u-m_{\rho}^{2}}\text{,} \label{ASL1}%
\end{align}
\begin{align}
A(t,u,s)  &  =(g_{1}^{2}-h_{3})Z_{\pi}^{4}w_{a_{1}}^{2}t-2\left(  \lambda
_{1}+\frac{\lambda_{2}}{2}\right)  Z_{\pi}^{4}-(h_{1}+h_{2}+h_{3})Z_{\pi}%
^{4}w_{a_{1}}^{2}(t-2m_{\pi}^{2})\nonumber\\
&  -[-2m_{\pi}^{2}C_{\sigma_{N}\pi\pi}+B_{\sigma_{N}\pi\pi}(2m_{\pi}%
^{2}-t)+2A_{\sigma_{N}\pi\pi}]^{2}\frac{1}{t-m_{\sigma_{N}}^{2}}\nonumber\\
&  +\left(  A_{\rho\pi\pi}+B_{\rho\pi\pi}\frac{s}{2}\right)  ^{2}\frac
{u-t}{s-m_{\rho}^{2}}+\left(  A_{\rho\pi\pi}+B_{\rho\pi\pi}\frac{u}{2}\right)
^{2}\,\frac{s-t}{u-m_{\rho}^{2}}\text{,} \label{ASL2}%
\end{align}
\begin{align}
A(u,s,t)  &  =(g_{1}^{2}-h_{3})Z_{\pi}^{4}w_{a_{1}}^{2}u-2\left(  \lambda
_{1}+\frac{\lambda_{2}}{2}\right)  Z_{\pi}^{4}-(h_{1}+h_{2}+h_{3})Z_{\pi}%
^{4}w_{a_{1}}^{2}(u-2m_{\pi}^{2})\nonumber\\
&  -[-2m_{\pi}^{2}C_{\sigma_{N}\pi\pi}+B_{\sigma_{N}\pi\pi}(2m_{\pi}%
^{2}-u)+2A_{\sigma_{N}\pi\pi}]^{2}\frac{1}{u-m_{\sigma_{N}}^{2}}\nonumber\\
&  +\left(  A_{\rho\pi\pi}+B_{\rho\pi\pi}\frac{s}{2}\right)  ^{2}\,\frac
{t-u}{s-m_{\rho}^{2}}+\left(  A_{\rho\pi\pi}+B_{\rho\pi\pi}\frac{t}{2}\right)
^{2}\,\frac{s-u}{t-m_{\rho}^{2}} \label{ASL3}%
\end{align}

with $A_{\sigma_{N}\pi\pi}$, $B_{\sigma_{N}\pi\pi}$, $C_{\sigma_{N}\pi\pi}$,
$A_{\rho\pi\pi}$ and $B_{\rho\pi\pi}$ respectively from Eqs.\ (\ref{ANsppQ}),
(\ref{BNsppQ}), (\ref{CNsppQ}), (\ref{Arhopipi}) and (\ref{Brhopipi}). Note
that the scattering amplitude $\mathcal{M}_{\pi\pi}$ vanishes at threshold:
$\mathcal{M}_{\pi\pi}(0,0,0)=0$ \cite{DA}.

We can now calculate the three contributions to the scattering amplitude at
threshold ($\vec{p}_{\pi}=0\Rightarrow P_{\pi}^{2}=m_{\pi}^{2}$ and thus
$s\equiv4P_{\pi}^{2}=4m_{\pi}^{2}$, $t=0$, $u=0$). Let us first substitute the
coefficient $A_{\rho\pi\pi}$ in Eqs.\ (\ref{ASL1}) - (\ref{ASL3}) using
Eq.\ (\ref{Arhopipi}); note that, at threshold, there is no contribution from
the terms $\sim B_{\rho\pi\pi}$. We then obtain%

\begin{align}
A(s,t,u)|_{s=4m_{\pi}^{2}}  &  =4g_{1}^{2}Z_{\pi}^{4}w_{a_{1}}^{2}m_{\pi}%
^{2}-2\left(  \lambda_{1}+\frac{\lambda_{2}}{2}\right)  Z_{\pi}^{4}%
-2(h_{1}+h_{2}+3h_{3})Z_{\pi}^{4}w_{a_{1}}^{2}m_{\pi}^{2}\nonumber\\
&  -4[(B_{\sigma_{N}\pi\pi}+C_{\sigma_{N}\pi\pi})m_{\pi}^{2}-A_{\sigma_{N}%
\pi\pi}]^{2}\frac{1}{4m_{\pi}^{2}-m_{\sigma_{N}}^{2}}\nonumber\\
&  +8[g_{1}Z_{\pi}^{2}(1-g_{1}w_{a_{1}}\phi_{N})+h_{3}Z_{\pi}^{2}w_{a_{1}}%
\phi_{N}]^{2}\frac{m_{\pi}^{2}}{m_{\rho}^{2}}\text{,} \label{ASL11}%
\end{align}

\begin{align}
A(t,u,s)|_{s=4m_{\pi}^{2}}  &  =-2\left(  \lambda_{1}+\frac{\lambda_{2}}%
{2}\right)  Z_{\pi}^{4}+2(h_{1}+h_{2}+h_{3})Z_{\pi}^{4}w_{a_{1}}^{2}m_{\pi
}^{2}-4g_{1}^{2}Z_{\pi}^{4}\frac{m_{\pi}^{2}m_{\rho}^{2}}{m_{a_{1}}^{4}%
}\nonumber\\
&  +4[(B_{\sigma_{N}\pi\pi}-C_{\sigma_{N}\pi\pi})m_{\pi}^{2}+A_{\sigma_{N}%
\pi\pi}]^{2}\frac{1}{m_{\sigma_{N}}^{2}} \label{ASL21}%
\end{align}

and

\begin{equation}
A(u,s,t)|_{s=4m_{\pi}^{2}}=A(t,u,s)|_{s=4m_{\pi}^{2}}\text{.} \label{ASL31}%
\end{equation}

The scattering amplitude $T^{0}$ for zero isospin is obtained from \cite{ACGL}%

\begin{equation}
T^{0}|_{s=4m_{\pi}^{2}}=3A(s,t,u)|_{s=4m_{\pi}^{2}}+A(t,u,s)|_{s=4m_{\pi}^{2}%
}+A(u,s,t)|_{s=4m_{\pi}^{2}}\text{.} \label{T0Q}%
\end{equation}

Additionally, the interdependence of $T^{0}$ and the $S$-wave, isospin-zero
$\pi\pi$ scattering length $a_{0}^{0}$ is, at threshold, given by the
following formula [see Ref.\ \cite{DA}, Eq.\ (4.30)]:%

\begin{equation}
a_{0}^{0}|_{s=4m_{\pi}^{2}}=\frac{1}{32\pi}T^{0}|_{s=4m_{\pi}^{2}}\text{.}
\label{a00Q}%
\end{equation}

Inserting Eqs.\ (\ref{ASL11}) - (\ref{ASL31}) into Eq.\ (\ref{T0Q}) and
substituting $T^{0}|_{s=4m_{\pi}^{2}}$ in Eq.\ (\ref{a00Q}) yields (in units
of $m_{\pi}^{-1}$):%

\begin{align}
a_{0}^{0}|_{s=4m_{\pi}^{2}} &  =\frac{1}{32\pi}\left\{  12(g_{1}^{2}%
-h_{3})Z_{\pi}^{4}w_{a_{1}}^{2}m_{\pi}^{2}-10\left(  \lambda_{1}+\frac
{\lambda_{2}}{2}\right)  Z_{\pi}^{4}-2(h_{1}+h_{2}+h_{3})Z_{\pi}^{4}w_{a_{1}%
}^{2}m_{\pi}^{2}\right.  \nonumber\\
&  +\left.  12[(B_{\sigma_{N}\pi\pi}+C_{\sigma_{N}\pi\pi})m_{\pi}%
^{2}-A_{\sigma_{N}\pi\pi}]^{2}\frac{1}{m_{\sigma_{N}}^{2}-4m_{\pi}^{2}%
}\right.  \nonumber\\
&  +\left.  8[(B_{\sigma_{N}\pi\pi}-C_{\sigma_{N}\pi\pi})m_{\pi}^{2}%
+A_{\sigma_{N}\pi\pi}]^{2}\frac{1}{m_{\sigma_{N}}^{2}}+16g_{1}^{2}Z_{\pi}%
^{4}\frac{m_{\pi}^{2}m_{\rho}^{2}}{m_{a_{1}}^{4}}\right\}  \text{.}%
\label{a001Q}%
\end{align}

Upon substitution of $A_{\sigma_{N}\pi\pi}$, $B_{\sigma_{N}\pi\pi}$ and
$C_{\sigma_{N}\pi\pi}$, respectively, from Eqs.\ (\ref{ANsppQ}), (\ref{BNsppQ})
and (\ref{CNsppQ}), we obtain the following formula for the scattering
length:
\begin{align}
a_{0}^{0}|_{s=4m_{\pi}^{2}} &  \equiv a_{0}^{0}|_{s=4m_{\pi}^{2}}(Z_{\pi
},m_{\sigma_{N}},h_{1})=\frac{1}{4\pi}\left(  2g_{1}^{2}Z_{\pi}^{4}%
\frac{m_{\pi}^{2}}{m_{a_{1}}^{4}}\left\{  m_{\rho}^{2}+\frac{\phi_{N}^{2}}%
{16}[12g_{1}^{2}-2(h_{1}+h_{2})-14h_{3}]\right\} \right.  \nonumber\\
& -\frac{3}{2}\left\{  g_{1}^{2}Z_{\pi}^{2}\phi_{N}\frac{m_{\pi
}^{2}}{m_{a_{1}}^{4}}\left[  2m_{a_{1}}^{2}+m_{\rho}^{2}-\frac{\phi_{N}^{2}%
}{2}(h_{1}+h_{2}+h_{3})\right]  -\frac{Z_{\pi}^{2}m_{\sigma_{N}}^{2}-m_{\pi
}^{2}}{2\phi_{N}}\right\}  ^{2}\frac{1}{4m_{\pi}^{2}-m_{\sigma_{N}}^{2}%
}\nonumber\\
& +\left.  \left\{  g_{1}^{2}Z^{2}\phi_{N}\frac{m_{\pi}^{2}%
}{m_{a_{1}}^{4}}\left[  m_{\rho}^{2}-\frac{\phi_{N}^{2}}{2}(h_{1}+h_{2}%
+h_{3})\right]  +\frac{Z_{\pi}^{2}m_{\sigma_{N}}^{2}-m_{\pi}^{2}}{2\phi_{N}%
}\right\}  ^{2}\frac{1}{m_{\sigma_{N}}^{2}}\right) \nonumber\\
& - \frac{5}{8}\frac{Z_{\pi
}^{2}m_{\sigma_{N}}^{2}-m_{\pi}^{2}}{f_{\pi}^{2}}  \text{  .}\label{a00}%
\end{align}

We use the value $a_{0}^{0\,\mathrm{exp}}=0.218\pm0.020$ in accordance with
the data from the NA48/2 collaboration \cite{Peyaud}.

Given that $T^{1}=A(t,u,s)-A(u,s,t)$ \cite{ACGL}, we obtain $T^{1}=0$ at
threshold because of $A(u,s,t)|_{s=4m_{\pi}^{2}}$ $=A(t,u,s)|_{s=4m_{\pi}^{2}}$
[see Eq.\ (\ref{ASL31})]. Therefore,%

\begin{equation}
a_{0}^{1}|_{s=4m_{\pi}^{2}}=0\text{.}\label{a011Q}%
\end{equation}

The $S$-wave, isospin-two $\pi\pi$ scattering length is obtained from the
corresponding $I=2$ scattering amplitude $T^{2}$ given by \cite{ACGL}%

\begin{equation}
T^{2}|_{s=4m_{\pi}^{2}}=A(t,u,s)|_{s=4m_{\pi}^{2}}+A(u,s,t)|_{s=4m_{\pi}^{2}%
}\overset{\text{Eq.\ (\ref{ASL31})}}{\equiv}2A(t,u,s)|_{s=4m_{\pi}^{2}%
}\text{.}\label{T2Q}%
\end{equation}

Analogously to Eq.\ (\ref{a00Q}),%

\begin{equation}
32\pi a_{0}^{2}|_{s=4m_{\pi}^{2}}\equiv T^{2}|_{s=4m_{\pi}^{2}}%
\end{equation}

or

\begin{equation}
16\pi a_{0}^{2}|_{s=4m_{\pi}^{2}} \equiv A(t,u,s)|_{s=4m_{\pi}^{2}}\text{,}%
\end{equation}

implying

\begin{equation}
a_{0}^{2}|_{s=4m_{\pi}^{2}} \equiv\frac{1}{16\pi}A(t,u,s)|_{s=4m_{\pi}^{2}}\text{.}\label{a02Q}%
\end{equation}

Then inserting Eqs.\ (\ref{ANsppQ}), (\ref{BNsppQ}), (\ref{CNsppQ}) and
(\ref{ASL21}) into Eq.\ (\ref{a02Q}) we obtain:
\begin{align}
a_{0}^{2}|_{s=4m_{\pi}^{2}} &  \equiv a_{0}^{2}|_{s=4m_{\pi}^{2}}(Z_{\pi},m_{\sigma_{N}}%
,h_{1})=-\frac{1}{4\pi}\left(g_{1}^{2}Z_{\pi}^{4}\frac{m_{\pi}^{2}}{m_{a_{1}}^{4}%
}\left[  m_{\rho}^{2}-\frac{\phi_{N}^{2}}{2}(h_{1}+h_{2}+h_{3})\right]
\right.  \nonumber\\
& \hspace*{-0.2cm}  -\left.  \left  \{  g_{1}^{2}Z_{\pi}^{2}\phi_{N}\frac{m_{\pi
}^{2}}{m_{a_{1}}^{4}}\left[  m_{\rho}^{2}-\frac{\phi_{N}^{2}}{2}(h_{1}%
+h_{2}+h_{3})\right]  +\frac{Z_{\pi}^{2}m_{\sigma_{N}}^{2}-m_{\pi}^{2}}%
{2\phi_{N}}\right\}  ^{2}\frac{1}{m_{\sigma_{N}}^{2}} + \frac{Z_{\pi}^{2}m_{\sigma_{N}}^{2}-m_{\pi}%
^{2}}{4f_{\pi}^{2}} \right)  \text{.}%
\label{a02}%
\end{align}

The experimental result for $a_{0}^{2}$ from the NA48/2 collaboration is
$a_{0}^{2\,\mathrm{exp}}=-0.0457\pm0.0125$ \cite{Peyaud}. Note that the
$\pi\pi$ scattering lengths were also studied away from threshold in
Ref.\ \cite{Schechter1}, in a model quite similar to ours. We will discuss the scattering lengths also within the extended $U(3)\times U(3)$
version of our model in Sec.\ \ref{SL}.

\section{Scenario I: Light Scalar Quarkonia} \label{sec.scenarioI}

We can now discuss two different interpretations of the scalar mesons.
Sections \ref{sec.fitscenarioI} - \ref{sec.ISL}\ describe the results obtained
when $f_{0}(600)$ and $a_{0}(980)$ are interpreted as scalar quarkonia
(Scenario I). Then, in Sec.\ \ref{sec.scenarioII}, we discuss the results
obtained when $f_{0}(1370)$ and $a_{0}(1450)$ are interpreted as scalar
quarkonia (Scenario II).

\subsection{Fit procedure} \label{sec.fitscenarioI}

As a first step we utilise the central value of the experimental result
$\Gamma_{\rho\rightarrow\pi\pi}^{\mathrm{exp}}=149.1$ MeV \cite{PDG} in order
to express the parameter $g_{2}$ as a function of $Z_{\pi}$ via
Eq.\ (\ref{g2Z}). Moreover, we fix the mass $m_{a_{0}}=980$ MeV \cite{PDG} and
we also use the central value $\Gamma_{f_{1N}\rightarrow a_{0}\pi}(Z_{\pi
},h_{2})=8.748$ MeV to express $h_{2}$ as a function of $Z_{\pi}$. The results
are practically unaffected by the 6\% uncertainty in $h_{2}$ originating from
the uncertainty in $\Gamma_{f_{1N}\rightarrow a_{0}\pi}$, see Eq.\ (\ref{h2Z}).

As a result, the set of free parameters in Eq.\ (\ref{param3}) is further
reduced to three parameters:
\begin{equation}
Z_{\pi}\text{, }m_{\sigma_{N}}\text{, }h_{1}\text{.}%
\end{equation}
Note that in this scenario the field $\sigma_{N}$ is identified with the
resonance $f_{0}(600)$, but the experimental uncertainty on its mass is so
large that it does not allow us to fix $m_{\sigma_{N}}$. We therefore keep
$m_{\sigma_{N}}$ as a free parameter.

We now determine the parameters $Z_{\pi}$, $h_{1}$, and $m_{\sigma_{N}}$ using
known data on the $a_{1}\rightarrow\pi\gamma$ decay width (\ref{a1piongamma})
and on the $\pi\pi$ scattering lengths $a_{0}^{0}$ and $a_{0}^{2}$ reported in
Eqs.\ (\ref{a00}) and (\ref{a02}). This is a system of three equations with
three variables and can be solved uniquely. We make use of the $\chi^{2}$
method in order to determine not only the central values for our parameters
but also their error intervals:%

\begin{equation}
\chi^{2}(Z_{\pi},m_{\sigma_{N}},h_{1})=\left(  \frac{\Gamma_{a_{1}%
\rightarrow\pi\gamma}(Z_{\pi})-\Gamma_{a_{1}\rightarrow\pi\gamma
}^{\mathrm{exp}}}{\triangle\Gamma_{\mathrm{decay}}^{\mathrm{\exp}}}\right)
^{2}+\sum_{i\in\{0,2\}}\left(  \frac{a_{0}^{i}(Z_{\pi},m_{\sigma_{N}}%
,h_{1})-a_{0}^{i\mathrm{,}\text{ }\mathrm{exp}}}{\triangle a_{0}^{i,\text{
}\mathrm{exp}}}\right)  ^{2}\text{.} \label{chi}%
\end{equation}
The errors for the model parameters are calculated as the square roots of the
diagonal elements of the inverted Hessian matrix obtained from $\chi
^{2}(Z_{\pi},m_{\sigma_{N}},h_{1})$. The minimal value is obtained for
$\chi^{2}=0$, as expected given that the parameters are determined from a
uniquely solvable system of equations. The values of the parameters are as
follows:
\begin{equation}
Z_{\pi}=1.67\pm0.2\,\text{,}\;m_{\sigma_{N}}=(332\pm456)\text{ MeV\thinspace
,\ }h_{1}=-68\pm338\text{.} \label{ksquared}%
\end{equation}

\begin{figure}[h]
\begin{center}
\includegraphics[width=8cm]{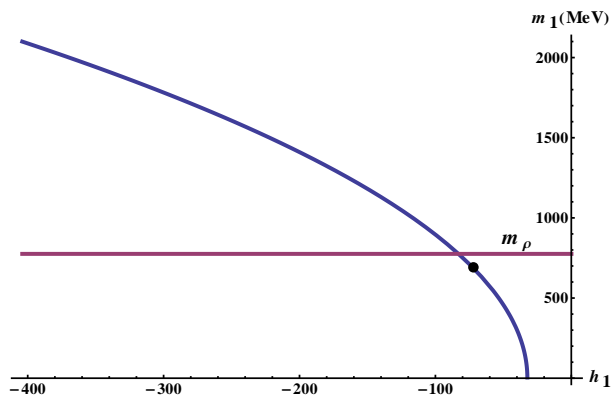}
\end{center}
\caption{$m_{1}$ as function of $h_{1}$, constrained at the central value of
$Z_{\pi}=1.67$. The black dot marks the position of central values $h_{1}=-68$
and $m_{1}=652$ MeV.}%
\label{m1f}%
\end{figure}

Clearly, the error intervals for $m_{\sigma_{N}}$ and $h_{1}$ are very large.
Fortunately, it is possible to constrain the $h_{1}$ error interval as
follows. As evident from Eq.\ (\ref{rho}), $m_\rho^2$
contains two contributions -- \ the bare mass term $m_{1}^{2}$ and the quark
condensate contribution ($\sim\phi_{N}^{2}$). The contribution of the quark
condensate is special for the globally invariant sigma model; in the locally
invariant model $m_{\rho}$ is always equal to $m_{1}$ \cite{RS}. Each of these
contributions should have at most the value of 775.49 MeV ($=m_{\rho}$)
because otherwise either the bare mass or the quark condensate contribution to
the rho mass would be negative, which appears to be unphysical. A plot of the
function $m_{1}=m_{1}(Z_{\pi},h_{1},h_{2}(Z_{\pi})),$ see Eq.\ (\ref{m1eq}),
for the central values of $Z_{\pi}=1.67$ and $\Gamma_{f_{1N}\rightarrow
a_{0}\pi}^{\mathrm{exp}}=8.748$ MeV is shown in Fig.\ \ref{m1f}.

Note that varying the value of $\Gamma_{f_{1N}\rightarrow a_{0}\pi
}^{\mathrm{exp}}$ within its experimental boundaries would only very slightly
change $h_{1}$ by $\pm4$ and this parameter is thus unaffected by the
experimental error for $\Gamma_{f_{1N}\rightarrow a_{0}\pi}^{\mathrm{\exp}}$.
If the value of $m_{1}$ were known exactly, then Eq.\ (\ref{m1eq}) would allow
us to constrain $h_{1}$ via $Z_{\pi}$. However, given that at this point we
can only state that $0\leq m_{1}\leq m_{\rho}$, for each $Z_{\pi}$ one may
consider all values of $h_{1}$ between two boundaries, one obtained from the
condition $m_{1}(Z_{\pi},h_{1},h_{2}(Z_{\pi}))\equiv0$ and another obtained
from the condition $m_{1}(Z_{\pi},h_{1},h_{2}(Z_{\pi}))\equiv m_{\rho}$. For
example, using the central value of $Z_{\pi}=1.67$, we obtain $-83\leq
h_{1}\leq-32$. The lower boundary follows from $m_{1}\equiv m_{\rho}$ and the
upper boundary from $m_{1}\equiv0$, see Fig.\ \ref{m1f}. Note that the central
value $h_{1}=-68$ from Eq.\ (\ref{ksquared}) corresponds to $m_{1}=652$ MeV.
If the minimal value of $Z_{\pi}=1.47$ is used, then $h_{1}=-112$ is obtained
from $m_{1}\equiv m_{\rho}$ and $h_{1}=-46$ from $m_{1}\equiv0$. Thus,
$-112\leq h_{1}\leq-46$ for $Z_{\pi}=1.47$. Analogously, $-64\leq h_{1}%
\leq-24$ is obtained for the maximal value $Z_{\pi}=1.87$.

Clearly, each lower boundary for $h_{1}$ is equivalent to $m_{1}\equiv
m_{\rho}$ and each upper boundary for $h_{1}$ is equivalent to $m_{1}\equiv0$.
Thus, in the following we will only state the values of $Z_{\pi}$ and $m_{1}$;
$h_{1}$ can always be calculated using Eq.\ (\ref{m1eq}). In this way, the
dependence of our results on $m_{1}$ and thus on the origin of the $\rho$ mass
will be exhibited.

\begin{figure}[h]
  \begin{center}
    \begin{tabular}{cc}
      \resizebox{76mm}{!}{\includegraphics{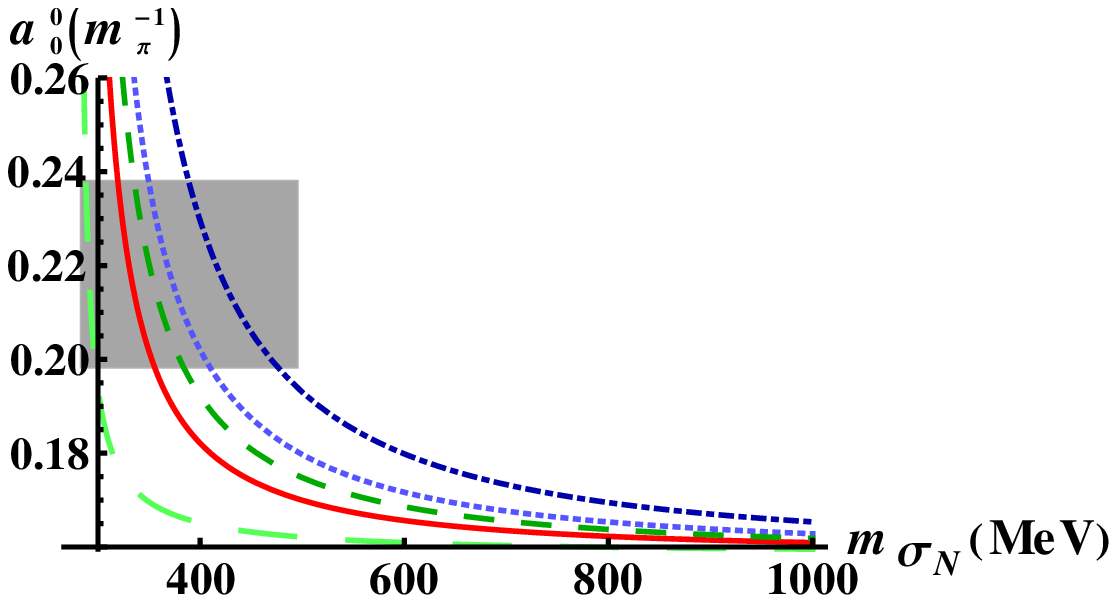}}  &
      \resizebox{80.5mm}{!}{\includegraphics{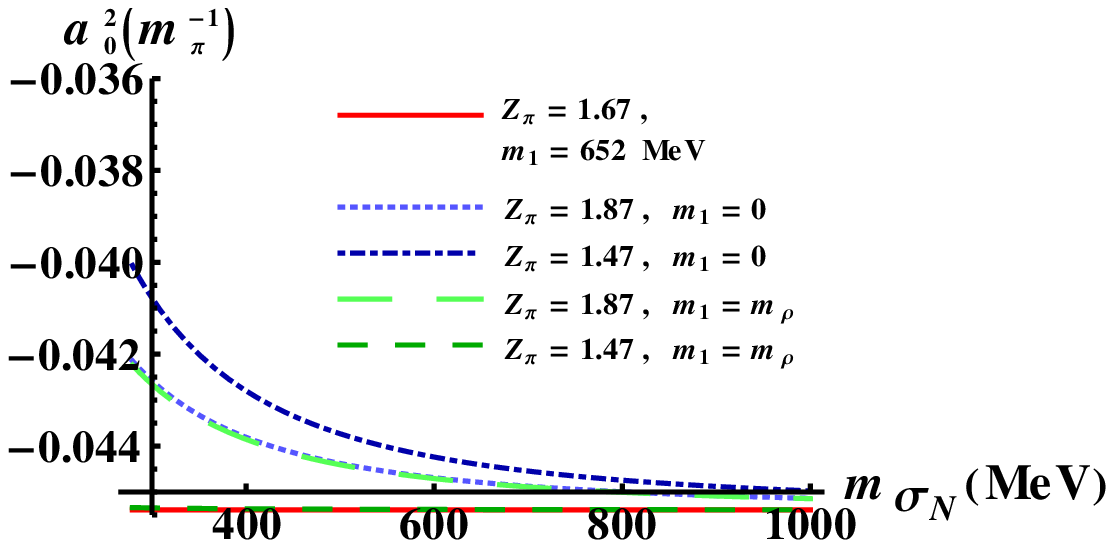}} 
    \end{tabular}
    \caption{Scattering lengths $a_{0}^{0}$ and $a_{0}^{2}$ as function of
$m_{\sigma_N}$ (the shaded band corresponds to the NA48/2 value of $a_{0}^{0}$;
no error interval is shown for $a_{0}^{2}$ due to the large interval size
\cite{Peyaud}).}
    \label{a00a02f}
  \end{center}
\end{figure}

The value of $m_{\sigma_{N}}$ can be constrained in a way similar to $h_{1}$
using the scattering length $a_{0}^{0}$; the scattering length $a_{0}^{2}$
possesses a rather large error interval making it unsuitable to constrain
$m_{\sigma_{N}}$. Figure \ref{a00a02f} shows the different values for
$a_{0}^{0}$ and $a_{0}^{2}$ depending on the choice of $Z_{\pi}$ and $m_{1}$.%

It is obvious that the value of $a_{0}^{0}$ is only consistent with the NA48/2
value \cite{Peyaud} if $m_{\sigma_N}$ is in the interval [288, 477] MeV, i.e.,
$m_{\sigma_N}=332_{-44}^{+145}$ MeV. This value for $m_{\sigma_N}$ follows if the
parameters $Z$ and $m_{1}$ are varied within the allowed boundaries. If we
only consider the $a_{0}^{0}$ curve that is obtained for the central values of
$Z$ and $m_{1}$, a much more constrained value of $m_{\sigma_N}=332_{-13}^{+24}$
MeV follows from Fig.\ \ref{a00a02f}. We will be working with the broader
interval of $m_{\sigma_N}$. Even then, constraining $m_{1}$ to the interval
$[0,m_{\rho}]$, the error bars for $m_{\sigma_N}$ are reduced by at least a
factor of three in comparison to the result (\ref{ksquared}) following from
the $\chi^{2}$ calculation.

We summarise our results for the parameters $Z$ and $m_{\sigma_N}$:
\begin{equation}
Z_\pi =1.67 \pm 0.2\,, \;m_{\sigma_N}=332_{-44}^{+145}\text{ MeV.}%
\end{equation}
The central values of all parameters of the original set (\ref{param}) are
given in Table \ref{Table1Q}. They follow from the $\chi^{2}$\ fit
($m_{\sigma_N}$, $h_{1}$), via decay width constraints ($h_{2}$, $g_{2}$), and
from Eqs.\ (\ref{sigma}) - (\ref{a1}) and (\ref{g1Q}) - (\ref{h3Q}). The
central values of $Z_\pi$, $m_{\sigma_N}$ and $h_{1}$, Eq.\ (\ref{ksquared}), have
been used to calculate all other parameters. We neglect the errors, apart from
those of $m_{1}$, which in this scenario vary in a large range.

\bigskip%
\begin{table}[h] \centering
\begin{tabular}{|c|c|c|c|c|}
\hline
\multicolumn{1}{|c|}{\textit{Parameter}} & $m_{\sigma_N}$ & $h_{1}$ & $h_{2}$
& $h_{3}$ \\ \hline
\multicolumn{1}{|c|}{\textit{Value}} & 332 MeV & -68 & 80 & 2.4 \\ \hline
\multicolumn{1}{|c|}{\textit{Parameter}} & $g_{1}$ & $g_{2}$ & $m_{0}$ & $%
m_{1}$ \\ \hline
\multicolumn{1}{|c|}{\textit{Value}} & 6.4 & 3.1 & 210 MeV & 652$%
_{-652}^{+123}$ MeV \\ \hline
\multicolumn{1}{|c|}{\textit{Parameter}} & $\lambda _{1}$ & $\lambda _{2}$ & 
$c$ & $h_{0N}$ \\ \hline
\multicolumn{1}{|c|}{\textit{Value}} & -14 & 33 & 88744 MeV$^{2}$ & $1\cdot
10^{6}$ MeV$^{3}$ \\ \hline
\end{tabular}
\caption{Central values of parameters for Scenario I.}\label{Table1Q}%
\end{table}%
Note that the values of $a_{0}^{2}$ depend strongly on the choice of the
parameters $Z_{\pi}$ and $m_{1}$. Whereas for the central values of $Z_{\pi}$
and $m_{1}$ this scattering length is constant and has the value $a_{0}%
^{2}=-0.0454$, its value increases if $Z_{\pi}$ and $m_{1}$ are considered at
their respective boundaries, see Fig.\ \ref{a00a02f}.

The value of $Z_{\pi}$ alone allows us to calculate certain decay widths in
the model. For example, as a consistency check we obtain $\Gamma
_{a_{1}\rightarrow\pi\gamma}=0.640_{-0.231}^{+0.261}$ MeV which is in good
agreement with the experimental result. Also, given that the $a_{0}%
\rightarrow\eta_{N}\pi$ decay amplitude only depends on $Z_{\pi}$, it is
possible to calculate the value of this amplitude, Eq.\ (\ref{a0etapion}). For
$Z_{\pi}=1.67$, we obtain the value of $3939$ MeV for the decay
amplitude $a_{0}\rightarrow\eta\pi$ involving the physical $\eta$ field if the
$\eta$-$\eta^{\prime}$ mixing angle of $\varphi_{\eta}=-36%
{{}^\circ}%
$ \cite{Giacosa:2007up} is taken. The Crystal Barrel Collaboration \cite{Bugg:1994} obtained
$3330$ MeV and hence there is an
approximate discrepancy of 20\%. If the KLOE Collaboration \cite{KLOE} value
of $\varphi_{\eta}=-41.4^{\circ}$ is considered, then the value of
$A_{a_{0}\rightarrow\eta\pi}=3373$ MeV follows -- in perfect agreement with
the Crystal Barrel value. From this we conclude that this scenario prefers a
relatively large value of the $\eta$-$\eta^{\prime}$ mixing angle. In fact, if
we use the Crystal Barrel value $A_{a_{0}\rightarrow\eta\pi}^{\mathrm{exp}%
}=3330$ MeV as input, we would predict $\varphi_{\eta}=-41.8^{\circ}$ for the
central value of $Z_{\pi}$ as well as $\varphi_{\eta}=-42.3^{\circ}$ and
$\varphi_{\eta}=-41.6^{\circ}$ for the highest and lowest values of $Z_{\pi}$,
respectively, i.e., $\varphi_{\eta}=-41.8^{\circ}{}_{-0.5^{\circ}%
}^{+0.2^{\circ}}$. This is in excellent agreement with the KLOE collaboration
result $\varphi_{\eta}=-41.4^{\circ}\pm0.5^{\circ}$ but also with the results
from approaches using the Bethe-Salpeter formalism, such as the one in
Ref.\ \cite{Roberts}.

\subsection{Decay Width \boldmath $\sigma_{N}\rightarrow\pi\pi$}  \label{sec.sNppQI}

\begin{figure}[h]
  \begin{center}
    \begin{tabular}{cc}
      \resizebox{94mm}{!}{\includegraphics{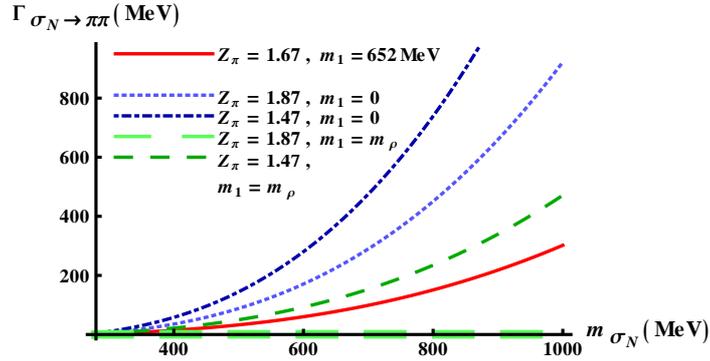}}  
    \end{tabular}
    \caption{$\Gamma_{\sigma_N \rightarrow\pi\pi}$ as function of $m_{\sigma_N}$ for
different values of $Z_\pi$ and $m_{1}$. The PDG \cite{PDG} notes $\Gamma_{\sigma_N
}=(600-1000)$ MeV; the results from the chiral perturbation theory suggest
$\Gamma_{\sigma_N}=544$ MeV \cite{Leutwyler} and $\Gamma_{\sigma_N}=510$ MeV
\cite{Pelaez1}.}
    \label{Sigmaf1}
  \end{center}
\end{figure}

The sigma decay width $\Gamma_{\sigma_{N}\rightarrow\pi\pi}$ depends on all
three parameters $Z_{\pi}$, $m_{1}$ (originally $h_{1}$), and $m_{\sigma_{N}}%
$. In Fig.\ \ref{Sigmaf1} we show the dependence of this decay width on the
sigma mass for fixed values of $Z_{\pi}$ and $m_{1}$, varying the latter
within their respective boundaries.

Generally, the values that we obtain are too small when compared to the PDG
data \cite{PDG} and to other calculations of the sigma meson decay width, such
as the one performed by Leutwyler \textit{et al.} \cite{Leutwyler} who found
$\Gamma_{\sigma_{N}\rightarrow\pi\pi}/2=272_{-12.5}^{+9}$ MeV and Pel\'{a}ez
\textit{et al.} \cite{Pelaez1}\ who found $\Gamma_{\sigma_{N}\rightarrow\pi
\pi}/2=(255\pm16)$ MeV. The largest values for the decay width that we were
able to obtain within our model are for the case when $Z_\pi$ is as small as
possible, $Z_{\pi}=1.47$, and $m_{1}=0$, i.e., when the $\rho$ mass is solely
generated by the quark condensate. As seen above, for this case the scattering
lengths allow a maximum value $m_{\sigma_{N}}=477$ MeV, for which
$\Gamma_{\sigma_{N}\rightarrow\pi\pi}\cong145$ MeV. In all other cases, the
decay width is even smaller. However, as will be discussed in Sec.\ \ref{sec.a1decaysI},
the case $m_{1}=0$ leads to the unphysically small value $\Gamma
_{a_{1}\rightarrow\sigma_{N}\pi}\simeq0$ and should therefore not be taken too
seriously. As apparent from Fig.\ \ref{a00a02f}, excluding small values of
$m_{1}$ would require smaller values for $m_{\sigma_{N}}$ in order to be
consistent with the scattering lengths. According to Fig.\ \ref{Sigmaf1},
however, this in turn leads to even smaller values for the decay width.

Hence, we conclude that the isoscalar meson in our model cannot be
$f_{0}(600)$, thus excluding that this resonance is predominantly a $\bar{q}q$
state and the chiral partner of the pion. Then the interpretation of the
isospin-one state $a_{0}(980)$ as a (predominantly) quarkonium state is also
excluded. The only choice is to consider Scenario II, see Sec.\ \ref{sec.scenarioII}, i.e., to
interpret the scalar states above 1 GeV, $f_{0}(1370)$ and $a_{0}(1450)$, as
being predominantly quarkonia. If the decay width of $f_{0}(1370)$ could be
described by the model, this would be a very strong indication that these
higher-lying states can be indeed interpreted as (predominantly) $\bar{q}q$
states. Note that very similar results about the nature of the light scalar
mesons were also found using different approaches: from an analysis of the
meson behaviour in the large-$N_{c}$ limit in Refs.\ \cite{Pelaez-scalars-below1GeVq2q2} and
\cite{Sannino} as well as from lattice studies, such as those in
Refs.\ \cite{Liu}.

\begin{figure}
[h]
\begin{center}
\includegraphics[
height=2.0582in,
width=4.1666in
]%
{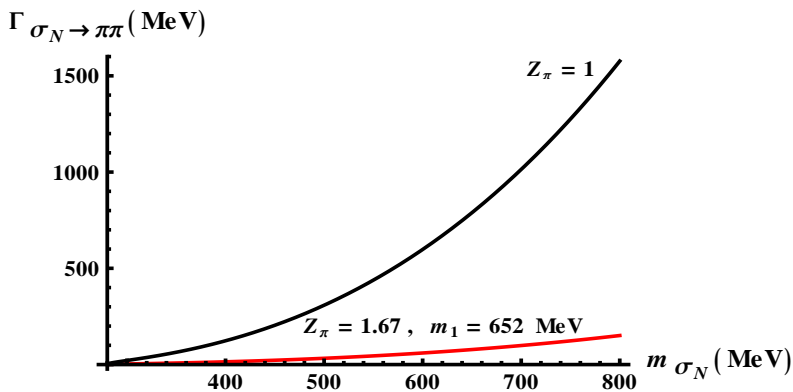}%
\caption{$\Gamma_{\sigma_N \rightarrow\pi\pi}$ as function of $m_{\sigma_N}$ in the case without
(axial-)vectors (upper line, corresponds to $Z_\pi=1$) and in the case with (axial-)vectors (lower line, exemplary for the
central values of $Z_\pi$ and $m_1$). A strong suppression of $\Gamma_{\sigma_N}$ is observed upon inclusion of the
(axial-)vectors into the model.}%
\label{Sigmaf2}%
\end{center}
\end{figure}

We remark that the cause for preventing a reasonable fit of the light sigma
decay width is the interference term arising from the vector mesons in
Eq.\ (\ref{sigmapionpion}). In the unphysical case without vector meson
degrees of freedom, a simultaneous fit of the decay width and the scattering
lengths is possible, see Fig.\ \ref{Sigmaf2} and Ref.\ \cite{Zakopane}.

\subsection{Decays of the \boldmath $a_{1}(1260)$ Meson} \label{sec.a1decaysI}

\begin{figure}
[h]
\begin{center}
\includegraphics[
height=2.0in,
width=3.3884in
]%
{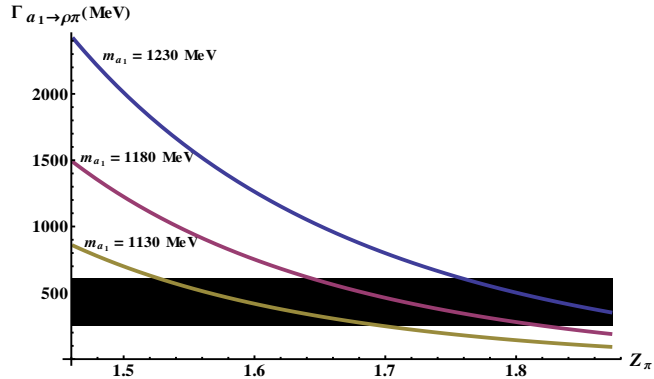}%
\caption{$\Gamma_{a_{1}\rightarrow\rho\pi}$ for different values of $m_{a_{1}%
}$. The shaded area corresponds to the possible values of $\Gamma
_{a_{1}\rightarrow\rho\pi}$ as stated by the PDG. }%
\label{a1rhopif}%
\end{center}
\end{figure}

We first consider the decay width $\Gamma_{a_{1}\rightarrow\rho\pi}$. For a
given $m_{a_{1}}$, this decay width depends only on $Z_{\pi}$. The PDG quotes
a rather large band of values, $\Gamma_{a_{1}\rightarrow\rho\pi}%
^{\mathrm{(exp)}}=(250-600)$ MeV. For $m_{a_{1}}=1230$ MeV, our fit of meson
properties yields $Z_{\pi}=1.67\pm0.2$. The ensuing region is shown as shaded
area in Fig.\ \ref{a1rhopif}. For $m_{a_{1}}=1230$ MeV, $\Gamma_{a_{1}%
\rightarrow\rho\pi}$ decreases from 2.4 GeV to 353 MeV, if $Z_{\pi}$ varies
from 1.47 to 1.87.

We also observe from Fig.\ \ref{a1rhopif} that the range of values for
$Z_{\pi}$, which give values for $\Gamma_{a_{1}\rightarrow\rho\pi}$ consistent
with the experimental error band, becomes larger if one considers smaller
masses for the $a_{1}$ meson. We have taken $m_{a_{1}}=1180$ MeV and
$m_{a_{1}}=1130$ MeV, the latter being similar to the values used in
Refs.\ \cite{UBW} and \cite{Williams:2009rk}. Repeating our calculations, we
obtain a new range of possible values for $Z_{\pi}$, $Z_{\pi}\simeq1.69\pm0.2$
for $m_{a_{1}}=1180$ MeV and $Z_{\pi}\simeq1.71\pm0.2$ for $m_{a_{1}}=1130$
MeV. For the respective central values of $Z_{\pi}$ we then compute
$\Gamma_{a_{1}\rightarrow\rho\pi}^{m_{a_{1}}=1180\mathrm{\ MeV}}=483$ MeV
($Z_{\pi}^{m_{a_{1}}=1180\,\mathrm{MeV}}=1.69$) and $\Gamma_{a_{1}%
\rightarrow\rho\pi}^{m_{a_{1}}=1130\mathrm{\ MeV}}=226$ MeV ($Z_{\pi
}^{m_{a_{1}}=1130\,\mathrm{MeV}}=1.71$), in good agreement with experimental
data. All other results remain valid when $m_{a_{1}}$ is decreased by about
100 MeV. Most notably, the $f_{0}(600)$ decay width remains too small.

We also consider the $a_{1}\rightarrow\sigma_{N}\pi$ decay width. Experimental
data on this decay channel \cite{PDG} are inconclusive. The value
$\Gamma_{a_{1}\rightarrow\sigma_{N}\pi}=56$ MeV is obtained for the central
values of $Z_{\pi}$, $m_{1}$, $m_{\sigma_{N}}$ and $\Gamma_{f_{1N}\rightarrow
a_{0}\pi}$ (which was used to constrain $h_{2}$ via $Z_{\pi}$). Taking the
limit $m_{1}=0$ pulls the value of $\Gamma_{a_{1}\rightarrow\sigma_{N}\pi}$
down to practically zero, regardless of whether $Z_{\pi}=Z_{\pi\min}$ or $Z_{\pi
}=Z_{\pi\max}$. This is an indication that the $m_{1}=0$ limit, where
$m_{\rho}$ is completely generated from the quark condensate, cannot be
physical. Note that the case $Z_{\pi}=Z_{\pi\max}=1.87$ and $m_{1}\equiv
m_{\rho}$, i.e., where the quark condensate contribution to the $\rho$ mass
vanishes, leads to a rather large value of $\Gamma_{a_{1}\rightarrow\sigma
_{N}\pi}$, e.g., for the central value of $m_{\sigma_{N}}=332$ MeV the value
of $\Gamma_{a_{1}\rightarrow\sigma_{N}\pi}=120$ MeV follows. Interestingly,
this picture persists even if lower values of $m_{a_{1}}$ are considered.
Improving experimental data for this decay channel would allow us to further
constrain our parameters.

\subsection{The Case of Isospin-Exact Scattering Lengths} \label{sec.ISL}

So far, the values of the scattering lengths used in our fit, $a_{0}%
^{0}=0.218\pm0.020$ and $a_{0}^{2}=-0.0457\pm0.0125$ \cite{Peyaud}, account
for the small explicit breaking of isospin symmetry due to the difference of
the up and down quark masses. However, in our model the isospin symmetry is
exact. Thus, one should rather use the isospin-exact values $a_{0}%
^{0\,\mathrm{(I)}}=0.244\pm0.020$ and $a_{0}^{2\,\mathrm{(I)}}=-0.0385\pm
0.0125$ \cite{Bloch}. In this section we will briefly show that the
conclusions reached so far remain qualitatively unchanged if the isospin-exact
values for the scattering lengths are considered.

Performing the $\chi^{2}$ fit, Eq.\ (\ref{chi}), with $\Gamma_{a_{1}%
\rightarrow\pi\gamma}$, $a_{0}^{0\,\mathrm{(I)}}$ and $a_{0}^{2\,\mathrm{(I)}%
}$ as experimental input yields $Z_{\pi}=1.67\pm0.2$ -- unchanged in
comparison with the previous case ($Z_{\pi}$ is largely determined by
$\Gamma_{a_{1}\rightarrow\pi\gamma}$ which is the same in both $\chi^{2}$
calculations), $h_{1}=-116\pm70$, and $m_{\sigma_{N}}=(284\pm16)$ MeV. Note
that in this case the errors are much smaller than previously. The reason is
that the mean value of $m_{\sigma_{N}}$ is almost on top of the two-pion decay
threshold and thus leads to an artificially small error band. For such small
values of $m_{\sigma_{N}}$ the decay width $\Gamma_{\sigma_{N}\rightarrow
\pi\pi}$ is at least an order of magnitude smaller than the physical value,
but even for values of $m_{\sigma_{N}}$ up to 500 MeV (not supported by our
error analysis) the decay width never exceeds 150 MeV, see Fig.\ \ref{Sigmaf1}.

\section{Scenario II: Scalar Quarkonia above 1 GeV} \label{sec.scenarioII}

A possible way to resolve the problem of the unphysically small two-pion decay
width of the sigma meson is to identify the fields $\sigma_{N}$ and $a_{0}$ of
the model with the resonances $f_{0}(1370)$ and $a_{0}(1450)$, respectively.
Thus, the scalar quarkonium states are assigned to the energy region above 1
GeV. In the following we investigate the consequences of this assignment.
However, the analysis cannot be conclusive for various reasons:

\begin{itemize}
\item The glueball field is missing. Many studies find that its role in
the mass region at about $1.5$ GeV is crucial, since it mixes with the other
scalar resonances. Indeed, we will extend the $N_{f}=2$ model in Chapter
\ref{chapterglueball} to include the dilaton field representing the scalar
glueball; however, the ensuing result about the structure of $f_{0}(1370)$ as
a $\bar{q}q$ state (see Sec.\ \ref{sec.sNppQII}) will remain unchanged.

\item The light scalar mesons below 1 GeV, such as $f_{0}(600)$ and
$a_{0}(980)$, are not included as elementary fields in our model. The question
is if they can be dynamically generated from the pseudoscalar fields already
present in our model by solving a Bethe-Salpeter equation. If not, they should
be introduced as additional elementary fields from the very beginning [see
also the discussion in Ref.\ \cite{dynrec}].

\item Due to absence of the resonance $f_{0}(600)$, the $\pi\pi$
scattering length $a_{0}^{0}$ cannot be correctly described at tree-level:
whereas $a_{0}^{2}$ stays always within the experimental error band,
$a_{0}^{0}$ clearly requires a light scalar meson for a proper description of experimental
data because a large value of $m_{\sigma_{N}}$ drives this quantity to the
Weinberg limit ($\simeq0.159$ \cite{Weinberg:1966}) which is outside the
experimental error band (see Fig.\ \ref{a00a02f}).
\end{itemize}

Despite these drawbacks, we turn to a quantitative analysis of this scenario.

\subsection{Decays of the \boldmath $a_{0}(1450)$ Meson} \label{sec.a0(1450)Q}

As in Scenario I, the parameter $g_{2}$ can be expressed as a function of
$Z_{\pi}$ by using the $\rho\rightarrow\pi\pi$ decay width (\ref{rhopionpionQ}%
). However, the parameter $h_{2}$ can no longer be fixed by the $f_{1N}%
\rightarrow a_{0}\pi$ decay width: the $a_{0}$ meson is now identified with
the $a_{0}(1450)$ resonance listed in Ref.\ \cite{PDG}, with a central mass of
$m_{a_{0}}=1474$ MeV, and thus $f_{1N}$ is too light to decay into $a_{0}$ and
$\pi$. One would be able to determine $h_{2}$ from the (energetically allowed)
decay $a_{0}(1450)\rightarrow f_{1N}\pi$, but the corresponding decay width is
not experimentally known.

Instead of performing a global fit, it is more convenient to proceed step by
step and calculate the parameters $Z_{\pi}$, $h_{1}$, $h_{2}$ explicitly. We
vary $m_{\sigma_{N}}\equiv m_{f_{0}(1370)}$ within the experimentally known
error band \cite{PDG} and check if our result for $\Gamma_{f_{0}%
(1370)\rightarrow\pi\pi}$ is in agreement with experimental data.

We first determine $Z_{\pi}$ from $a_{1}\rightarrow\pi\gamma$,
Eq.\ (\ref{Za1pg}), and obtain $Z_{\pi}=1.67\pm0.21$. We then
immediately conclude that the $a_{1}\rightarrow\rho\pi$ decay width,
Eq.\ (\ref{a1rhopionQ}), will remain the same as in Scenario I because this
decay width depends on $Z_{\pi}$ (which is virtually the same in both
scenarios) and $g_{2}$ [which is fixed via $\Gamma_{\rho\rightarrow\pi\pi}$,
Eq.\ (\ref{g2Z}), in both scenarios].

The parameter $h_{1}$, being large-$N_{c}$ suppressed, will be set to zero in
the present study. We then only have to determine the parameter $h_{2}$. This
is done by fitting the total decay width of the $a_{0}(1450)$ meson to its
experimental value \cite{PDG},
\begin{align}
\Gamma_{a_{0}(1450)}(Z_{\pi},h_{2})  &  =\Gamma_{a_{0}\rightarrow\pi\eta
}+\Gamma_{a_{0}\rightarrow\pi\eta^{\prime}}+\Gamma_{a_{0}\rightarrow
KK} +\Gamma_{a_{0}\rightarrow\omega_{N}\pi\pi}\equiv\Gamma_{a_{0}%
(1450)}^{\mathrm{exp}}=(265\pm13)\text{ MeV.} \label{a01450}
\end{align}

Although kaons have not yet been included into the calculations, we can easily
evaluate the decay into $KK$ by using flavour symmetry
\begin{align}
\Gamma_{a_{0}(1450)\rightarrow KK}(Z_{\pi},h_{2}) & =2\,\frac{k(m_{a_{0}%
},m_{K},m_{K})}{8\pi m_{a_{0}}^{2}}|-i\mathcal{\bar{M}}_{a_{0}%
(1450)\rightarrow KK}(Z_{\pi},h_{2})|^{2}\text{,} \label{a0KKQ}\\
-i\mathcal{\bar{M}}_{a_{0}(1450)\rightarrow KK}(Z_{\pi},h_{2}) &  =\frac
{i}{2Z_{\pi}f_{\pi}}\left\{  m_{\eta_{N}}^{2}-m_{a_{0}}^{2}+\left(  1-\frac
{1}{Z_{\pi}^{2}}\right)  \right.  \nonumber\\
&  \left.  \times\left[  1-\frac{1}{2}\frac{Z_{\pi}^{2}\phi_{N}^{2}}{m_{a_{1}%
}^{2}}(h_{2}-h_{3})\right]  (m_{a_{0}}^{2}-2m_{K}^{2})\right\}  \text{.}
\end{align}

The remaining, experimentally poorly known decay width $\Gamma
_{a_{0}(1450)\rightarrow\omega_{N}\pi\pi}$ can be calculated from the
sequential decay $a_{0}\rightarrow\omega_{N}\rho\rightarrow\omega_{N}\pi\pi.$
Note that the first decay step requires the $\rho$ to be slightly below its
mass shell, since $m_{a_{0}}<m_{\rho}+m_{\omega_{N}}$. We denote the off-shell
mass of the $\rho$ meson by $x_{\rho}$. From the Lagrangian (\ref{Lagrangian})
we obtain the following $a_{0}\omega_{N}\rho$ interaction Lagrangian:%

\begin{equation}
\mathcal{L}_{a_{0}\omega_{N}\rho}=(h_{2}+h_{3})\phi_{N}\vec{a}_{0}\cdot
\omega_{N\mu}\vec{\rho}^{\mu}\text{.} \label{a0oNr}%
\end{equation}

The generic calculation of the decay width of a scalar state $S$ into two
vector states $V_{1,2}$ has already been presented in Sec.\ \ref{sec.SVV}. We
identify the state $V_{2}$ in the decay amplitude (\ref{iMSVV3})\ with our
off-shell $\rho$ meson; the vertex from the Lagrangian (\ref{a0oNr}) reads
$h_{a_{0}\omega_{N}\rho}^{\mu\nu}=i(h_{2}+h_{3})\phi_{N}g^{\mu\nu}$ and
consequently we obtain from Eq.\ (\ref{GSVV}):
\begin{align}
\Gamma_{a_{0}(1450)\rightarrow\omega_{N}\rho}(x_{\rho}) &  =\frac{k(m_{a_{0}%
},m_{\omega_{N}},x_{\rho})}{8\pi m_{a_{0}}^{2}}(h_{2}+h_{3})^{2}Z_{\pi}%
^{2}f_{\pi}^{2}\nonumber\\
&  \times\left[  3-\frac{x_{\rho}^{2}}{m_{\rho}^{2}}+\frac{(m_{a_{0}}%
^{2}-x_{\rho}^{2}-m_{\omega_{N}}^{2})^{2}}{4m_{\omega_{N}}^{2}m_{\rho}^{2}%
}\right ]
\end{align}
with $I=3$ used in the formula presented in Eq.\ (\ref{GSVV}).

The full decay width $\Gamma_{a_{0}(1450)\rightarrow\omega_{N}\pi\pi}$ is then
obtained from Eq.\ (\ref{GSVV1}):
\begin{equation}
\Gamma_{a_{0}(1450)\rightarrow\omega_{N}\pi\pi}=\int_{0}^{\infty}%
\mathrm{d}x_{\rho}\,\Gamma_{a_{0}\rightarrow\omega_{N}\rho}(x_{\rho}%
)\,d_{\rho}(x_{\rho})\text{,} \label{a0omegapionpion2}%
\end{equation}
where $d_{\rho}(x_{\rho})$ is the mass distribution of the $\rho$ meson,
which is taken to be of relativistic Breit-Wigner form [see Eq.\ (\ref{BW1})]:
\begin{equation}
d_{\rho}(x_{\rho})=N_{\rho}\,\frac{x_{\rho}^{2}\Gamma_{\rho\rightarrow\pi\pi
}^{\exp}}{(x_{\rho}^{2}-m_{\rho}^{2})^{2}+\left(  x_{\rho}\Gamma
_{\rho\rightarrow\pi\pi}^{\exp}\right)  ^{2}}\,\theta(x_{\rho}-2m_{\pi
})\text{,}\label{drho}%
\end{equation}
where $\Gamma_{\rho\rightarrow\pi\pi}^{\exp}=149.1$ MeV and $m_{\rho}=775.49$
MeV \cite{PDG}. [As demonstrated in Eq.\ (\ref{BW}), one should in general use the theoretical quantity
$\Gamma_{\rho\rightarrow\pi\pi} (x_{\rho})$ instead of $\Gamma_{\rho\rightarrow\pi\pi}^{\exp}$, see Refs.\ \cite{Giacosa:2007bn,Elvira}. This is, however, numerically irrelevant
in the following.] The normalisation constant $N_{\rho}$ is chosen
such that
\begin{equation}%
{\displaystyle\int\limits_{0}^{\infty}}
\mathrm{d}x_{\rho}\,d_{\rho}(x_{\rho})=1\text{,}%
\end{equation}
in agreement with the interpretation of $\mathrm{d}x_{\rho}\,d_{\rho}(x_{\rho
})$ as the probability that the off-shell $\rho$ meson has a mass between
$x_{\rho}$ and $x_{\rho}+\mathrm{d}x_{\rho}$.

\begin{figure}
[h]
\begin{center}
\includegraphics[
height=2.0988in,
width=3.1712in
]%
{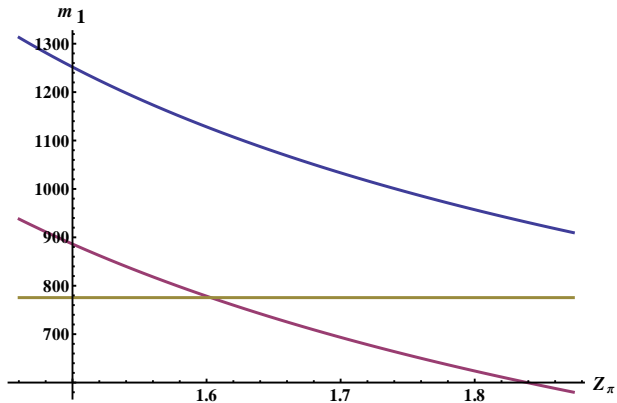}%
\caption{Dependence of $m_{1}$ on $Z_\pi$. The upper curve corresponds to the
lower set of $h_{2}$ values and the lower curve to the higher set of $h_{2}$
values. The horizontal line corresponds to $m_{\rho}$.}%
\label{m1S2f}%
\end{center}
\end{figure}

Inserting Eqs.\ (\ref{a0etapion2}), (\ref{a0eta'pion2}), (\ref{a0KKQ}) and
(\ref{a0omegapionpion2}) into Eq.\ (\ref{a01450}), we can express $h_{2}$ as a
function of $Z_{\pi}$, analogously to Eq.\ (\ref{g2Z}) where $g_{2}$ was
expressed as a function of $Z_{\pi}$. Similarly to that case, we obtain two
bands for $h_{2}$, $-115\leq h_{2}\leq-20$ and $-25\leq h_{2}\leq10$, the
width of the bands corresponding to the uncertainty in determining $Z_{\pi}$,
$Z_{\pi}=1.67\pm0.21$. Both bands for $h_{2}$ remain practically unchanged if
the $5\%$ experimental uncertainty of $\Gamma_{a_{0}(1450)}^{\mathrm{exp}}$ is
taken into account and thus we only use the mean value $265$ MeV in the
following. Since $h_{1}$ is assumed to be zero, Eq.\ (\ref{m1eq}) allows to
express $m_{1}$ as a function of $Z_{\pi}$, $m_{1}=m_{1}(Z_{\pi},h_{1}%
=0,h_{2}(Z_{\pi}))$ (we neglect the experimental uncertainties of $m_{\rho
},\,m_{a_{1}}$, and $f_{\pi}$). The result is shown in Fig.\ \ref{m1S2f}. The
first band of (lower) $h_{2}$ values should be discarded because it leads to
$m_{1}>m_{\rho}$. The second set of (higher) values leads to $m_{1}<m_{\rho}$
only if the lower boundary for $Z_{\pi}$ is 1.60 rather than 1.46. Thus, we
shall use the set of larger $h_{2}$ values and take the constraint
$m_{1}<m_{\rho}$ into account by restricting the values for $Z_{\pi}$ to the
range $Z_{\pi}=1.67_{-0.07}^{+0.21}$. As can be seen from Fig.\ \ref{m1S2f},
this sets a lower boundary for the value of $m_{1}$, $m_{1}\geq580$ MeV. Thus,
in this scenario we obtain $m_{1}=720_{-140}^{+55}$ MeV.

The values for the other parameters can be found in Table \ref{Table2Q} (only
central values are shown with the exception of $m_{1}$ where the corresponding
uncertainties are stated as well).%

\begin{table}[h] \centering
\begin{tabular}
[c]{|c|c|c|c|c}\hline
\multicolumn{1}{|c|}{\textit{Parameter}} & $h_{1}$ & $h_{2}$ & $h_{3}$ &
\multicolumn{1}{c|}{$g_{1}$}\\\hline
\multicolumn{1}{|c|}{\textit{Value}} & 0 & 4.7 & 2.4 &
\multicolumn{1}{c|}{6.4}\\\hline
\multicolumn{1}{|c|}{\textit{Parameter}} & $g_{2}$ & $m_{0}^{2}$ & $m_{1}$ &
\multicolumn{1}{c|}{$\lambda_{1}$}\\\hline
\multicolumn{1}{|c|}{\textit{Value}} & 3.1 & -811987 MeV$^{2}$ &
720$_{-140}^{+55}$ MeV & \multicolumn{1}{c|}{-3.6}\\\hline
\multicolumn{1}{|c|}{\textit{Parameter}} & \multicolumn{1}{c|}{$\lambda_{2} $}
& $c$ & $h_{0N}$ & \multicolumn{1}{l}{}\\\cline{1-4}%
\multicolumn{1}{|c|}{\textit{Value}} & \multicolumn{1}{c|}{84} & 88747
MeV$^{2}$ & $1\cdot10^{6}$ MeV$^{3}$ & \multicolumn{1}{l}{}\\\cline{1-4}%
\end{tabular}
\caption{Central values of the parameters for Scenario II.}\label{Table2Q}%
\end{table}%


Note that $\lambda_{1}\ll\lambda_{2},$ in agreement with the expectations from
the large-$N_{c}$ limit, Eq.\ (\ref{largen}). The value of $m_{1}=720$ MeV is
sizable and constitutes a dominant contribution to the $\rho$ mass. This
implies that non-quark contributions, for instance a gluon condensate, play a
decisive role in the $\rho$ mass generation.

As a final step, we study the ratios $\Gamma_{a_{0}(1450)\rightarrow
\eta^{\prime}\pi}/\Gamma_{a_{0}(1450)\rightarrow\eta\pi}$ and $\Gamma
_{a_{0}(1450)\rightarrow K\overline{K}}/\Gamma_{a_{0}(1450)\rightarrow\eta\pi
}$. Their experimental values read \cite{PDG}
\begin{align}
\frac{\Gamma_{a_{0}(1450)\rightarrow\eta^{\prime}\pi}^{\mathrm{exp}}}%
{\Gamma_{a_{0}(1450)\rightarrow\eta\pi}^{\mathrm{exp}}}  &  =0.35\pm0.16\text{,} \\
\;\;\;\frac{\Gamma_{a_{0}(1450)\rightarrow K\overline{K}}^{\mathrm{exp}}%
}{\Gamma_{a_{0}(1450)\rightarrow\eta\pi}^{\mathrm{exp}}}  &  =0.88\pm
0.23\text{.}%
\end{align}

Using the central value $Z_{\pi}=1.67$ and $\varphi_{\eta}=-36^{\circ}$ for
the $\eta$-$\eta^{\prime}$ mixing angle, we obtain $\Gamma_{a_{0}%
(1450)\rightarrow\eta^{\prime}\pi}$ $/\Gamma_{a_{0}(1450)\rightarrow\eta\pi}=1.0$
and $\Gamma_{a_{0}(1450)\rightarrow K\overline{K}}/\Gamma_{a_{0}%
(1450)\rightarrow\eta\pi}=0.96$. The latter is in very good agreement with the
experiment, the former a factor of two larger. Note, however, that according
to Eqs.\ (\ref{a0etapion0}) and (\ref{a0etapion00}) the value of the ratio
$\Gamma_{a_{0}(1450)\rightarrow\eta^{\prime}\pi}/\Gamma_{a_{0}%
(1450)\rightarrow\eta\pi}$ is proportional to $\sin^{2}\varphi_{\eta}/\cos
^{2}\varphi_{\eta}$. If a lower value of the angle is considered, e.g.,
$\varphi_{\eta}=-30^{\circ}$, then we obtain $\Gamma_{a_{0}(1450)\rightarrow
\eta^{\prime}\pi}/\Gamma_{a_{0}(1450)\rightarrow\eta\pi}=0.58$ for the central
value of $Z_{\pi}$ and the central value of $\Gamma_{a_{0}(1450)}$ in
Eq.\ (\ref{a01450}). Taking $Z_{\pi}=Z_{\pi\max}$ and the upper boundary
$\Gamma_{a_{0}(1450)}^{\mathrm{exp}}=278$ MeV results in $\Gamma
_{a_{0}(1450)\rightarrow\eta^{\prime}\pi}/\Gamma_{a_{0}(1450)\rightarrow
\eta\pi}=0.48$, i.e., in agreement with the experimental value. Therefore, our
results in this scenario favour a smaller value of $\varphi_{\eta}$ than the
one suggested by the KLOE Collaboration {\cite{KLOE}}.

It is possible to calculate the decay width $\Gamma_{a_{0}(1450)\rightarrow
\omega_{N}\pi\pi}$\ using Eq.\ (\ref{a0omegapionpion2}). We have obtained a
very small value $\Gamma_{a_{0}(1450)\rightarrow\omega_{N}\pi\pi}=0.1$ MeV.
From Eq.\ (\ref{a0etapion2}) we obtain $\Gamma_{a_{0}(1450)\rightarrow\eta\pi
}=$ 89.5 MeV, such that the ratio $\Gamma_{a_{0}(1450)\rightarrow\omega_{N}%
\pi\pi}/\Gamma_{a_{0}(1450)\rightarrow\eta\pi}=0.0012$, in contrast to the
results of Ref.\ \cite{Baker}.

\subsection{Decays of the \boldmath $f_{0}(1370)$ Meson} \label{sec.sNppQII}

\begin{figure}
[h]
\begin{center}
\includegraphics[
height=1.714in,
width=3.3419in
]%
{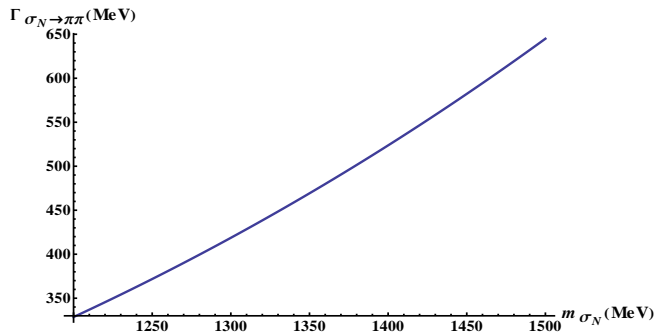}%
\caption{Dependence of the $\sigma_N \equiv f_0 (1370)$ decay width on $m_{\sigma_{N}}$. The
experimental value of the width is expected to be in the range (1200-1500) MeV
\cite{PDG}.}%
\label{f0pionpionf}%
\end{center}
\end{figure}

It is now possible to calculate the width for the $f_{0}(1370)\rightarrow
\pi\pi$ decay using Eq.\ (\ref{sigmapionpion}). The decay width depends on the
$f_{0}(1370)$ mass, $Z_{\pi}$, $h_{1}$, and $h_{2}$ which is expressed via
$Z_{\pi}$ using Eq.\ (\ref{a01450}). The values of the latter three are listed
in Table \ref{Table2Q}. In Fig.\ \ref{f0pionpionf} we show the decay
width as a function of the mass of $f_{0}(1370)$.

Assuming that the two-pion decay dominates the total decay width (true up
to a mass of $1350$ MeV, see Sec.\ \ref{sec.f0(1370)}), we observe a good
agreement with the experimental values. The values of the decay width are
$\sim100$\ MeV larger than those of Ref. \cite{buggf0} but this is not
surprising as the current version of the model contains no strange degrees of
freedom (they are discussed in Chapters \ref{ImplicationsFitI} and
\ref{ImplicationsFitII}) and no glueball (discussed in Chapter
\ref{chapterglueball}).\ We will see in the mentioned chapters that the
currently missing contributions to the decay width will reduce the upper
boundary on $m_{f_{0}(1370)}$. Nevertheless, the correspondence with the
experiment is a lot better in this scenario where we have identified
$f_{0}(1370)$ rather than $f_{0}(600)$ as the (predominantly) isoscalar
$\bar{q}q$ state. Note that this result has been obtained using the decay
width of the $a_{0}(1450)$ meson (in order to express $h_{2}$ via $Z_{\pi}$)
which is also assumed to be a scalar $\bar{q}q$ state in this scenario.

It is remarkable that vector mesons are crucial to obtain realistic values for
the decay width of $f_{0}(1370)$: without vector mesons, the decay width is
$\sim10$ GeV and thus much too large. This is why Scenario II has not been
considered in the standard linear sigma model.

\bigskip%

The four-body decay $f_{0}(1370)\rightarrow4\pi$ can also be studied.
Similarly to the $a_{0}(1450)\rightarrow\omega_{N}\rho$ decay, we view
$f_{0}(1370)\rightarrow4\pi$ as a sequential decay of the form $f_{0}%
(1370)\rightarrow\rho\rho\rightarrow4\pi$. The Lagrangian (\ref{Lagrangian})
leads to
\begin{equation}
\mathcal{L}_{\sigma\rho\rho}=\frac{1}{2}(h_{1}+h_{2}+h_{3})\phi_{N}\sigma
_{N}\vec{\rho}_{\mu}^{2}\label{srrQ}%
\end{equation}
and repeating the calculation of Sec.\ \ref{sec.SVV} we obtain%
\begin{align}
\Gamma_{f_{0}(1370)\rightarrow\rho\rho}(x_{1\rho},x_{2\rho}) &  =\frac
{3}{16\pi}\frac{k(m_{f_{0}},x_{1\rho},x_{2\rho})}{m_{f_{0}}^{2}}\left(
h_{1}+h_{2}+h_{3}\right)  ^{2}\nonumber\\
&  \times Z_{\pi}^{2}f_{\pi}^{2}\left[  4-\frac{x_{1\rho}^{2}+x_{2\rho}^{2}%
}{m_{\rho}^{2}}+\frac{(m_{f_{0}}^{2}-x_{1\rho}^{2}-x_{2\rho}^{2})^{2}%
}{4m_{\rho}^{4}}\right]  \text{,}\label{f0rhorho}%
\end{align}
where $x_{1\rho}$ and $x_{2\rho}$ are the off-shell masses of the $\rho$
mesons. The decay width $\Gamma_{f_{0}\rightarrow4\pi}$ is then given by
\begin{equation}
\Gamma_{f_{0}(1370)\rightarrow4\pi}=%
{\displaystyle\int\limits_{0}^{\infty}}
{\displaystyle\int\limits_{0}^{\infty}}
\mathrm{d}x_{1\rho}\,\mathrm{d}x_{2\rho}\,\Gamma_{f_{0}(1370)\rightarrow
\rho\rho}(x_{1\rho},x_{2\rho})d_{\rho}(x_{1\rho})\,d_{\rho}(x_{2\rho})\text{,}%
\end{equation}
with $\Gamma_{f_{0}(1370)\rightarrow\rho\rho}(x_{1\rho},x_{2\rho})$ from
Eq.\ (\ref{f0rhorho}) and $d_{\rho}(x_{\rho})$ from Eq.\ (\ref{drho}).

Using the previous values for the parameters we obtain that the $\rho\rho$
contribution for the decay is small: $\Gamma_{f_{0}(1370)\rightarrow\rho
\rho\rightarrow4\pi}\simeq10\pm10$ MeV. (The error comes from varying $Z_{\pi
}$ between 1.6 and 1.88.) Reference \cite{buggf0} quotes 54 MeV for the total
$4\pi$ decay width at $\sim 1300$ MeV. Since Ref.\ \cite{Abele:2001pv} ascertains that about 26\% of the
total $4\pi$ decay width originates from the $\rho\rho$ decay channel, our
result is qualitatively consistent with these findings.

\section{Conclusions from the Two-Flavour Version of the Model} \label{sec.conclusionsU2}

In this chapter we have presented the two-flavour version of the generic
Lagrangian with vector mesons and global chiral invariance introduced in
Chapter \ref{chapterC}. This Lagrangian describes mesons as pure quarkonium
states. As shown in Sec.\ \ref{sec.scenarioI}, the resulting low-energy
phenomenology is in general in good agreement with experimental data -- with
one exception: the model fails to correctly describe the $f_{0}%
(600)\rightarrow\pi\pi$ decay width. This led us to conclude that $f_{0}(600)$
and $a_{0}(980)$ cannot be predominantly $\bar{q}q$ states.

Assigning the scalar fields $\sigma_{N}$ and $a_{0}$ of the model to the
$f_{0}(1370)$ and $a_{0}(1450)$ resonances, respectively, improves the results
for the decay widths considerably. We have obtained $\Gamma_{f_{0}%
(1370)\rightarrow\pi\pi}$ $\simeq (300$-$500)$ MeV for $m_{f_{0}(1370)}=(1200$-$1400)$
MeV (see Fig.\ \ref{f0pionpionf}). Thus, the scenario in which the scalar
states above 1 GeV, $f_{0}(1370)$ and $a_{0}(1450)$, are considered to be
(predominantly) $\bar{q}q$ states appears to be favoured over the assignment
in which $f_{0}(600)$ and $a_{0}(980)$ are considered (predominantly) $\bar
{q}q$ states. However, a more detailed study of this scenario is necessary,
because a glueball state with the same quantum numbers mixes with the
quarkonium states. This allows to include the experimentally well-known
resonance $f_{0}(1500)$ into the study.

Of course, interpreting $f_{0}(1370)$ and $a_{0}(1450)$ as $\bar{q}q$ states
leads to question about the nature of $f_{0}(600)$ and $a_{0}(980)$. Their
presence is necessary for the correct description of $\pi\pi$ scattering
lengths that differ from experiment for too large values of the isoscalar mass
(see Sec.\ \ref{sec.fitscenarioI}). We distinguish two possibilities:
(\textit{i}) They can arise as (quasi-)molecular states. This is possible if
the attraction in the $\pi\pi$ and $KK$ channels is large enough. In order to
prove this, one should solve the corresponding Bethe-Salpeter equation in the
framework of Scenario II. In this case $f_{0}(600)$ and $a_{0}(980)$ can be
classified as genuinely dynamically generated states and should not appear in
the Lagrangian, see the discussion in Ref.\ \cite{dynrec}. If, however, the
attraction is not sufficient to generate the two resonances $f_{0}(600)$ and
$a_{0}(980)$ we are led to the alternative possibility that (\textit{ii})
these two scalar states must be incorporated into the model as additional
tetraquark states \cite{Achim}. In this case they shall appear from the very
beginning in the Lagrangian and should not be considered as dynamically
generated states. Of course, the isoscalar tetraquark, quarkonium, and
glueball will mix to produce $f_{0}(600),\,f_{0}(1370),$ and $f_{0}(1500)$,
and the isovector tetraquark and quarkonium will mix to produce $a_{0}(980)$
and $a_{0}(1450)$.

An extension of the model to $N_{f}=3$ will be performed in the next chapters.
One reason is that much more data are available for the strange mesons, which
constitute an important test for the validity of our approach. In addition, an
extremely important question arising from the results presented so far will be
addressed: whether the conclusions reached at $N_{f}=2$ [in particular that
$f_{0}(1370)$ rather than $f_{0}(600)$ is a $\bar{q}q$ state] hold in the more
general three-flavour case.

\section{The Full \boldmath $N_f=2$ Lagrangian} \label{sec.LagrangianQ}

This is the final form of the Lagrangian (\ref{LagrangianQ}) that is obtained
after the shifts (\ref{shifts}) and the renormalisation of the pseudoscalar
wave functions; $\rho^{\mu\nu}\equiv\partial^{\mu}\rho^{\nu}-\partial^{\nu
}\rho^{\mu}$; $a_{1}^{\mu\nu}\equiv\partial^{\mu}a_{1}^{\nu}-\partial^{\nu
}a_{1}^{\mu}$; ($\vec{A}$)$_{3}$ marks the third component of the vector
$\vec{A}$. Note that the term $\mathcal{L}_{4}$ contains the (axial-)vector
four-point vertices [the terms $\sim g_{3,4,5,6}$ in the Lagrangian
(\ref{LagrangianQ})]. We do not give the explicit form of $\mathcal{L}_{4}$
because it is not relevant for the results that are presented here.
\begin{align}
\mathcal{L} & =\frac{1}{2}\,(\partial^{\mu}\sigma_{N}+g_{1}Z_{\pi}%
\,\vec{\pi}\cdot\vec{a}_{1}^{\mu}+g_{1}w_{a_{1}}Z_{\pi}^{2}\,\partial^{\mu
}\vec{\pi}\cdot\vec{\pi}+g_{1}Z_{\pi}\eta_{N}f_{1}^{\mu}+g_{1}w_{a_{1}}Z_{\pi
}^{2}\eta_{N}\,\partial^{\mu}\eta_{N})^{2} \nonumber \\
&  -\,\frac{1}{2}\,\left[  m_{0}^{2}-c+3\left(  \lambda_{1}+\frac{\lambda_{2}%
}{2}\right)  \phi_{N}^{2}\right]  \sigma_{N}^{2} \nonumber \\
&  +\,\frac{1}{2}\,(Z_{\pi}\partial^{\mu}\vec{\pi}+g_{1}Z_{\pi}\vec{\rho}%
^{\mu}\times\vec{\pi}-g_{1}f_{1N}^{\mu}\vec{a}_{0}-g_{1}w_{a_{1}}Z_{\pi
}\partial^{\mu}\eta_{N}\vec{a}_{0}-g_{1}\sigma_{N}\vec{a}_{1}^{\mu}%
-g_{1}w_{a_{1}}Z_{\pi}\sigma_{N}\partial^{\mu}\vec{\pi})^{2} \nonumber\\
&  +\,\frac{1}{2}\,(Z_{\pi}\partial^{\mu}\eta_{N}-g_{1}\sigma_{N}f_{1N}^{\mu
}-g_{1}w_{a_{1}}Z_{\pi}\sigma_{N}\partial^{\mu}\eta_{N}-g_{1}\,\vec{a}%
_{1}^{\mu}\cdot\vec{a}_{0}-g_{1}w_{a_{1}}Z_{\pi}\,\partial^{\mu}\vec{\pi}%
\cdot\vec{a}_{0})^{2} \nonumber \\
&  -\,\frac{1}{2}\,\left[  m_{0}^{2}-c+\left(  \lambda_{1}+\frac{\lambda_{2}%
}{2}\right)  \phi_{N}^{2}\right]  Z_{\pi}^{2}\vec{\pi}^{2}-\frac{1}%
{2}\,\left[  m_{0}^{2}+c+\left(  \lambda_{1}+\frac{\lambda_{2}}{2}\right)
\phi_{N}^{2}\right]  Z_{\pi}^{2}\eta_{N}^{2} \nonumber
\end{align}
\begin{align}
&  +\,\frac{1}{2}\,[\partial^{\mu}\vec{a}_{0}+g_{1}\vec{\rho}^{\mu}\times
\vec{a}_{0}+g_{1}Z_{\pi}f_{1N}^{\mu}\vec{\pi}+g_{1}w_{a_{1}}Z_{\pi}^{2}%
\vec{\pi}\partial^{\mu}\eta_{N}+g_{1}Z_{\pi}\eta_{N}\,\vec{a}_{1}^{\mu}%
+g_{1}w_{a_{1}}Z_{\pi}^{2}\eta\,_{N}\partial^{\mu}\vec{\pi}]^{2} \nonumber \\
&  -\,\frac{1}{2}\,\left[  m_{0}^{2}+c+\left(  \lambda_{1}+\frac{3}{2}%
\lambda_{2}\right)  \phi_{N}^{2}\right]  \vec{a}_{0}^{2}-\frac{\lambda_{2}}%
{2}[(\sigma_{N}\vec{a}_{0}+Z_{\pi}^{2}\eta_{N}\,\vec{\pi})^{2}+Z_{\pi}^{2}%
\vec{a}_{0}^{2}\vec{\pi}^{2}-Z_{\pi}^{2}(\vec{a}_{0}\cdot\vec{\pi})^{2}] \nonumber \\
&  -\,\frac{1}{4}\,\left(  \lambda_{1}+\frac{\lambda_{2}}{2}\right)
\,(\sigma_{N}^{2}+\vec{a}_{0}^{2}+Z_{\pi}^{2}\eta_{N}^{2}+Z_{\pi}^{2}\vec{\pi
}^{2})^{2}-\left(  \lambda_{1}+\frac{\lambda_{2}}{2}\right)  \,\phi_{N}%
\sigma_{N}\,(\sigma_{N}^{2}+\vec{a}_{0}^{2}+Z_{\pi}^{2}\eta_{N}^{2}+Z_{\pi
}^{2}\vec{\pi}^{2}) \nonumber \\
&  -\,\lambda_{2}\phi_{N}\vec{a}_{0}\cdot(\sigma_{N}\vec{a}_{0}+Z_{\pi}%
^{2}\eta_{N}\,\vec{\pi})-\frac{1}{4}\,(\partial^{\mu}\omega_{N}^{\nu}%
-\partial^{\nu}\omega_{N}^{\mu})^{2}+\frac{m_{1}^{2}}{2}\,(\omega_{N}^{\mu
})^{2} \nonumber \\
&  -\,\frac{1}{4}\,[\partial^{\mu}\vec{\rho}^{\nu}-\partial^{\nu}\vec{\rho
}^{\mu}+g_{2}\vec{\rho}^{\mu}\times\vec{\rho}^{\nu}+g_{2}\vec{a}_{1}^{\mu
}\times\vec{a}_{1}^{\nu}+g_{2}w_{a_{1}}Z_{\pi}\partial^{\mu}\vec{\pi}%
\times\vec{a}_{1}^{\nu}+g_{2}w_{a_{1}}Z_{\pi}\vec{a}_{1}^{\mu}\times
\partial^{\nu}\vec{\pi} \nonumber \\
&  +\,g_{2}w_{a_{1}}^{2}Z_{\pi}^{2}(\partial^{\mu}\vec{\pi}\ )\times
(\partial^{\nu}\vec{\pi})]^{2} \nonumber \\
&  +\,\frac{m_{1}^{2}}{2}(\vec{\rho}^{\mu})^{2}-\frac{1}{4}\,(\partial^{\mu
}f_{1N}^{\nu}-\partial^{\nu}f_{1N}^{\mu})^{2}+\frac{m_{1}^{2}+g_{1}^{2}%
\phi_{N}^{2}}{2}\,(f_{1N}^{\mu})^{2}+\frac{m_{1}^{2}+g_{1}^{2}\phi_{N}^{2}}%
{2}\,(\vec{a}_{1}^{\mu})^{2} \nonumber \\
&  -\,\frac{1}{4}\,[\partial^{\mu}\vec{a}_{1}^{\nu}-\partial^{\nu}\vec{a}%
_{1}^{\mu}+g_{2}\vec{\rho}^{\mu}\times\vec{a}_{1}^{\nu}+g_{2}w_{a_{1}}Z_{\pi
}\vec{\rho}^{\mu}\times\partial^{\nu}\vec{\pi}+g_{2}\vec{a}_{1}^{\mu}%
\times\vec{\rho}^{\nu}+g_{2}w_{a_{1}}Z_{\pi}(\partial^{\mu}\vec{\pi}%
)\times\vec{\rho}^{\nu}]^{2} \nonumber\\
&  -\,g_{1}^{2}\,\phi_{N}\,\vec{a}_{1\mu}\cdot\lbrack\vec{\rho}^{\mu}\times
Z_{\pi}\vec{\pi}-f_{1N}^{\mu}\vec{a}_{0}-w_{a_{1}}Z_{\pi}\vec{a}_{0}%
\,\partial^{\mu}\eta_{N}] \nonumber\\
&  -\,g_{1}^{2}w_{a_{1}}Z_{\pi}\,\phi_{N}\,\partial_{\mu}\vec{\pi}%
\cdot\,[Z_{\pi}\vec{\rho}^{\mu}\times\vec{\pi}-f_{1N}^{\mu}\vec{a}%
_{0}-w_{a_{1}}Z_{\pi}\partial^{\mu}\eta_{N}\,\vec{a}_{0}] \nonumber \\
& + \, g_{1}^{2}\,\phi
_{N}f_{1N\mu}\,(\vec{a}_{1}^{\mu}\cdot\vec{a}_{0}+w_{a_{1}}Z_{\pi}%
\partial^{\mu}\vec{\pi}\cdot\vec{a}_{0}) \nonumber \\
&  +\,g_{1}^{2}w_{a_{1}}Z_{\pi}\,\phi_{N}\,\partial_{\mu}\eta_{N}\,(\vec
{a}_{1}^{\mu}\cdot\vec{a}_{0}+w_{a_{1}}Z_{\pi}\partial^{\mu}\vec{\pi}\cdot
\vec{a}_{0}) \nonumber \\
&  +\,g_{1}^{2}\,\phi_{N}\,\sigma_{N}\,[(f_{1N}^{\mu})^{2}+2w_{a_{1}}Z_{\pi
}f_{1N\,\mu}\partial^{\mu}\eta_{N}+w_{a_{1}}^{2}Z_{\pi}^{2}(\partial^{\mu}%
\eta_{N})^{2}] \nonumber \\
&  +\,g_{1}^{2}\,\phi_{N}\,\sigma_{N}\,[(\vec{a}_{1}^{\mu})^{2}+2w_{a_{1}%
}Z_{\pi}\vec{a}_{1\mu}\cdot\partial^{\mu}\vec{\pi}+w_{a_{1}}^{2}Z_{\pi}%
^{2}(\partial^{\mu}\vec{\pi})^{2}] \nonumber \\
&  -\,\frac{1}{2}\,\frac{g_{1}^{2}\phi_{N}^{2}}{\,m_{a_{1}}^{2}}\,Z_{\pi}%
^{2}\,(\partial_{\mu}\eta_{N}\partial^{\mu}\eta_{N}+\partial_{\mu}\vec{\pi
}\cdot\partial^{\mu}\vec{\pi}) \nonumber \\
&  +eA_{\mu}\{(\vec{a}_{0}\times\partial^{\mu}\vec{a}_{0})_{3}+Z_{\pi}%
^{2}(\vec{\pi}\times\partial^{\mu}\vec{\pi})_{3}-4(\vec{\rho}^{\mu\nu}%
\times\vec{\rho}_{\nu})_{3}-4[(\vec{a}_{1\nu}+Z_{\pi}w_{a_{1}}\partial_{\nu
}\vec{\pi})\times\vec{a}_{1}^{\mu\nu}]_{3} \nonumber \\
&  +g_{1}\{2Z_{\pi}(f_{1N}^{\mu}+Z_{\pi}w_{a_{1}}\partial^{\mu}\eta_{N}%
)(\vec{a}_{0}\times\vec{\pi})_{3}+Z_{\pi}(\sigma_{N}+\phi_{N})[(\vec{a}%
_{1}^{\mu}+Z_{\pi}w_{a_{1}}\partial^{\mu}\vec{\pi})\times\vec{\pi}]_{3} \nonumber \\
&  +\eta_{N}[\vec{a}_{0}\times(\vec{a}_{1}^{\mu}+Z_{\pi}w_{a_{1}}\partial
^{\mu}\vec{\pi})]_{3}-a_{0}^{3}(a_{0}^{1}\rho^{\mu1}+a_{0}^{2}\rho^{\mu
2})-Z_{\pi}^{2}\pi^{3}(\pi^{1}\rho^{\mu1}+\pi^{2}\rho^{\mu2}) \nonumber \\
&  +\rho^{\mu3}[(a_{0}^{1})^{2}+(a_{0}^{2})^{2}+Z_{\pi}^{2}(\pi^{1}%
)^{2}+Z_{\pi}^{2}(\pi^{2})^{2}]\} \nonumber \\
&  +4g_{2}\{[\vec{\rho}_{\nu}^{2}+\vec{a}_{1\nu}^{2}+Z_{\pi}^{2}w_{a_{1}}%
^{2}\left(  \partial_{\nu}\vec{\pi}\right)  ^{2}+Z_{\pi}w_{a_{1}}\vec{a}%
_{1\nu}\cdot\partial^{\nu}\vec{\pi}]\rho^{\mu3} \nonumber \\
& +\, 2\vec{\rho}_{\nu}\cdot(\vec
{a}_{1}^{\nu}+Z_{\pi}w_{a_{1}}\partial^{\nu}\vec{\pi})\times(a_{1}^{\mu
3}+Z_{\pi}w_{a_{1}}\partial^{\mu}\pi^{3}) \nonumber\\
&  -(\vec{\rho}_{\nu}\cdot\vec{\rho}^{\mu}+\vec{a}_{1\nu}\cdot\vec{a}_{1}%
^{\mu}+Z_{\pi}w_{a_{1}}\vec{a}_{1\nu}\cdot\partial^{\mu}\vec{\pi}+Z_{\pi
}w_{a_{1}}\vec{a}_{1}^{\mu}\cdot\partial_{\nu}\vec{\pi}+Z_{\pi}^{2}w_{a_{1}%
}^{2}\partial_{\nu}\vec{\pi}\cdot\partial^{\mu}\vec{\pi})\rho^{\nu3} \nonumber \\
&  -(\vec{\rho}^{\mu}\cdot\vec{a}_{1\nu}+\vec{a}_{1}^{\mu}\cdot\vec{\rho}%
_{\nu}+Z_{\pi}w_{a_{1}}\vec{\rho}^{\mu}\cdot\partial_{\nu}\vec{\pi}+Z_{\pi
}w_{a_{1}}\vec{\rho}_{\nu}\cdot\partial^{\mu}\vec{\pi})a_{1}^{\nu3}\}\} \nonumber \\
&  +\frac{e^{2}}{2}A_{\mu}A^{\mu}[(a_{0}^{1})^{2}+(a_{0}^{2})^{2}+Z_{\pi}%
^{2}(\pi^{1})^{2}+Z_{\pi}^{2}(\pi^{2})^{2}+4(\rho^{\nu1})^{2}+4(\rho^{\nu
2})^{2} \nonumber \\
&  +4(a_{1\nu}^{1}+Z_{\pi}w_{a_{1}}\partial_{\nu}\pi^{1})^{2}+4(a_{1\nu}%
^{2}+Z_{\pi}w_{a_{1}}\partial_{\nu}\pi^{2})^{2}] \nonumber \\
&  -2e^{2}A_{\mu}A_{\nu}[\rho^{\mu1}\rho^{\nu1}+\rho^{\mu2}\rho^{\nu2}%
+(a_{1}^{\mu1}+Z_{\pi}w_{a_{1}}\partial^{\mu}\pi^{1})(a_{1}^{\nu1}+Z_{\pi
}w_{a_{1}}\partial^{\nu}\pi^{1}) \nonumber \\
&  +(a_{1}^{\mu2}+Z_{\pi}w_{a_{1}}\partial^{\mu}\pi^{2})(a_{1}^{\nu2}+Z_{\pi
}w_{a_{1}}\partial^{\nu}\pi^{2})]+{\mathcal{L}}_{h_{1,2,3}}+\mathcal{L}_{4}
\end{align}
\begin{align}
%
%
{\mathcal{L}}_{h_{1,2,3}}  &  =\left(  \frac{h_{1}}{4}+\frac{h_{2}}{4}%
+\frac{h_{3}}{4}\right)  \left(  \sigma_{N}^{2}+2\phi_{N}\sigma_{N}+Z_{\pi
}^{2}\eta_{N}^{2}+\vec{a}_{0}^{2}+Z_{\pi}^{2}\vec{\pi}^{2}\right)
(\omega_{N\mu}^{2}+\vec{\rho}_{\mu}^{2})  \nonumber \\
&  +\left(  \frac{h_{1}}{4}+\frac{h_{2}}{4}-\frac{h_{3}}{4}\right)  \left(
\sigma_{N}^{2}+2\phi_{N}\sigma_{N}+Z_{\pi}^{2}\eta_{N}^{2}+\vec{a}_{0}%
^{2}+Z_{\pi}^{2}\vec{\pi}^{2}\right)  \{f_{1N\mu}^{2}+\vec{a}_{1\mu}%
^{2} \nonumber \\
& + \, Z_{\pi}^{2}w_{a_{1}}^{2}[\left(  \partial_{\mu}\eta_{N}\right)
^{2}+\left(  \partial_{\mu}\vec{\pi}\right)  ^{2}] +2Z_{\pi}w_{a_{1}}(f_{1N\mu}\partial^{\mu}\eta_{N}+\vec{a}_{1\mu}%
\cdot\partial^{\mu}\vec{\pi})\} \nonumber \\
&  +\left(  \frac{h_{1}}{4}+\frac{h_{2}}{4}+\frac{h_{3}}{4}\right)  \phi
_{N}^{2}(\omega_{N\mu}^{2}+\vec{\rho}_{\mu}^{2})+\left(  \frac{h_{1}}{4}%
+\frac{h_{2}}{4}-\frac{h_{3}}{4}\right)  \phi_{N}^{2}(f_{1N\mu}^{2}+\vec
{a}_{1\mu}^{2}) \nonumber \\
&  +(h_{2}+h_{3})\omega_{N\mu}[(\sigma_{N}+\phi_{N})\vec{a}_{0}+Z_{\pi}%
^{2}\eta_{N}\vec{\pi}]\cdot\vec{\rho}^{\mu} \nonumber \\
&  +(h_{2}-h_{3})[(\sigma_{N}+\phi_{N})\vec{a}_{0}+Z_{\pi}^{2}\eta_{N}\vec
{\pi}]\cdot\lbrack f_{1N\mu}\vec{a}_{1}^{\mu} \nonumber \\
& + \, Z_{\pi}w_{a_{1}}(\vec{a}_{1\mu
}\partial^{\mu}\eta_{N}+f_{1N\mu}\partial^{\mu}\vec{\pi})+Z_{\pi}^{2}w_{a_{1}%
}^{2}(\partial_{\mu}\eta_{N})(\partial^{\mu}\vec{\pi})] \nonumber \\
&  +(h_{2}+h_{3})Z_{\pi}(\vec{a}_{0}\times\vec{\pi})\cdot(\omega_{N\mu}\vec
{a}_{1}^{\mu}+Z_{\pi}w_{a_{1}}\omega_{N\mu}\partial^{\mu}\vec{\pi}) \nonumber \\
&  +(h_{2}-h_{3})Z_{\pi}(\vec{a}_{0}\times\vec{\pi})\cdot(f_{1N\mu}\vec{\rho
}^{\mu}+Z_{\pi}w_{a_{1}}\vec{\rho}_{\mu}\partial^{\mu}\eta_{N}) \nonumber \\
&  +h_{3}Z_{\pi}[\eta_{N}\vec{a}_{0}-(\sigma_{N}+\phi_{N})\vec{\pi}%
]\cdot\lbrack\vec{\rho}_{\mu}\times(\vec{a}_{1}^{\mu}+Z_{\pi}w_{a_{1}}%
\partial^{\mu}\vec{\pi})] \nonumber \\
&  -\frac{h_{3}}{2}\{(\vec{a}_{0}\times\vec{\rho}^{\mu})^{2}-[\vec{a}%
_{0}\times(\vec{a}_{1}^{\mu}+Z_{\pi}w_{a_{1}}\partial^{\mu}\vec{\pi}%
)]^{2}+Z_{\pi}^{2}(\vec{\pi}\times\vec{\rho}^{\mu})^{2} \nonumber \\
& - \, Z_{\pi}^{2}[\vec{\pi
}\times(\vec{a}_{1}^{\mu}+Z_{\pi}w_{a_{1}}\partial^{\mu}\vec{\pi})]^{2}\} 
\end{align}

\end{fmffile}

%% file: Doktorarbeit.tex
\begin{fmffile}{Doktorarbeit}
\chapter{Three-Flavour Linear Sigma Model}

\label{sec.remarks}

In this chapter we discuss the application of the generic Lagrangian from
Chapter \ref{chapterC} to three flavours: scalar, pseudoscalar, vector and
axial-vector strange mesons are added to the model.

\section{The \boldmath $N_f=3$ Lagrangian}

The globally invariant $U(3)_{L}\times U(3)_{R}$ Lagrangian possesses the same
structure as the one in Eq.\ (\ref{LagrangianGe}) up to the chiral-anomaly
term $c_{1}$ (see discussion in Sec.\ \ref{sec.anomaly}):
\begin{align}
\mathcal{L}  &  =\mathrm{Tr}[(D_{\mu}\Phi)^{\dagger}(D_{\mu}\Phi)]-m_{0}%
^{2}\mathrm{Tr}(\Phi^{\dagger}\Phi)-\lambda_{1}[\mathrm{Tr}(\Phi^{\dagger}%
\Phi)]^{2}-\lambda_{2}\mathrm{Tr}(\Phi^{\dagger}\Phi)^{2}\nonumber\\
&  -\frac{1}{4}\mathrm{Tr}(L_{\mu\nu}^{2}+R_{\mu\nu}^{2})+\mathrm{Tr}\left[
\left(  \frac{m_{1}^{2}}{2}+\Delta\right)  (L_{\mu}^{2}+R_{\mu}^{2})\right]
+\mathrm{Tr}[H(\Phi+\Phi^{\dagger})]\nonumber\\
&  +c_{1}(\det\Phi-\det\Phi^{\dagger})^{2}+i\frac{g_{2}}{2}(\mathrm{Tr}%
\{L_{\mu\nu}[L^{\mu},L^{\nu}]\}+\mathrm{Tr}\{R_{\mu\nu}[R^{\mu},R^{\nu
}]\})\nonumber\\
&  +\frac{h_{1}}{2}\mathrm{Tr}(\Phi^{\dagger}\Phi)\mathrm{Tr}(L_{\mu}%
^{2}+R_{\mu}^{2})+h_{2}\mathrm{Tr}[|L_{\mu}\Phi|^{2}+|\Phi R_{\mu}%
|^{2}]+2h_{3}\mathrm{Tr}(L_{\mu}\Phi R^{\mu}\Phi^{\dagger}).\nonumber\\
&  +g_{3}[\mathrm{Tr}(L_{\mu}L_{\nu}L^{\mu}L^{\nu})+\mathrm{Tr}(R_{\mu}R_{\nu
}R^{\mu}R^{\nu})]+g_{4}[\mathrm{Tr}\left(  L_{\mu}L^{\mu}L_{\nu}L^{\nu
}\right)  +\mathrm{Tr}\left(  R_{\mu}R^{\mu}R_{\nu}R^{\nu}\right)
]\nonumber\\
&  +g_{5}\mathrm{Tr}\left(  L_{\mu}L^{\mu}\right)  \,\mathrm{Tr}\left(
R_{\nu}R^{\nu}\right)  +g_{6}[\mathrm{Tr}(L_{\mu}L^{\mu})\,\mathrm{Tr}(L_{\nu
}L^{\nu})+\mathrm{Tr}(R_{\mu}R^{\mu})\,\mathrm{Tr}(R_{\nu}R^{\nu})]\text{.}
\label{Lagrangian}%
\end{align}

In the three-flavour case, it is more convenient to write the (pseudo)scalar
and (axial-)vector matrices explicitly straightaway rather than implicitly
in terms of the $U(3)$ group generators.

The scalar states present in the $U(3)_{L}\times U(3)_{R}$ version of the
model are [see Eq.\ (\ref{SQ})]:%

\begin{equation}
S=\frac{1}{\sqrt{2}}\left(
\begin{array}
[c]{ccc}%
\frac{\sigma_{N}+a_{0}^{0}}{\sqrt{2}} & a_{0}^{+} & K_{S}^{+}\\
a_{0}^{-} & \frac{\sigma_{N}-a_{0}^{0}}{\sqrt{2}} & K_{S}^{0}\\
K_{S}^{-} & {\bar{K}_{S}^{0}} & \sigma_{S}%
\end{array}
\right)  \label{S}%
\end{equation}

and the pseudoscalar states are [see Eq.\ (\ref{PQ})]:%

\begin{equation}
P=\frac{1}{\sqrt{2}}\left(
\begin{array}
[c]{ccc}%
\frac{\eta_{N}+\pi^{0}}{\sqrt{2}} & \pi^{+} & K^{+}\\
\pi^{-} & \frac{\eta_{N}-\pi^{0}}{\sqrt{2}} & K^{0}\\
K^{-} & {\bar{K}^{0}} & \eta_{S}%
\end{array}
\right)  \text{.} \label{P}%
\end{equation}

Consequently, the $\Phi$ matrix from Eq.\ (\ref{Phiij}) now reads%

\begin{equation}
\Phi=S+iP=\frac{1}{\sqrt{2}}\left(
\begin{array}
[c]{ccc}%
\frac{(\sigma_{N}+a_{0}^{0})+i(\eta_{N}+\pi^{0})}{\sqrt{2}} & a_{0}^{+}%
+i\pi^{+} & K_{S}^{+}+iK^{+}\\
a_{0}^{-}+i\pi^{-} & \frac{(\sigma_{N}-a_{0}^{0})+i(\eta_{N}-\pi^{0})}%
{\sqrt{2}} & K_{S}^{0}+iK^{0}\\
K_{S}^{-}+iK^{-} & {\bar{K}_{S}^{0}}+i{\bar{K}^{0}} & \sigma_{S}+i\eta_{S}%
\end{array}
\right)
\end{equation}

and the adjoint matrix $\Phi^{\dagger}$ is%

\begin{equation}
\Phi^{\dagger}=\frac{1}{\sqrt{2}}\left(
\begin{array}
[c]{ccc}%
\frac{(\sigma_{N}+a_{0}^{0})-i(\eta_{N}+\pi^{0})}{\sqrt{2}} & a_{0}^{+}%
-i\pi^{+} & K_{S}^{+}-iK^{+}\\
a_{0}^{-}-i\pi^{-} & \frac{(\sigma_{N}-a_{0}^{0})-i(\eta_{N}-\pi^{0})}%
{\sqrt{2}} & K_{S}^{0}-iK^{0}\\
K_{S}^{-}-iK^{-} & {\bar{K}_{S}^{0}}-i{\bar{K}^{0}} & \sigma_{S}-i\eta_{S}%
\end{array}
\right)  \text{.}%
\end{equation}

(Note again that, in this work, $K_{S}^{0}$ does not denote the short-lived pseudoscalar kaon but a neutral scalar kaon.)\\

The matrix containing vectors is [see Eq.\ (\ref{VQ})]:%

\begin{equation}
V^{\mu}=\frac{1}{\sqrt{2}}\left(
\begin{array}
[c]{ccc}%
\frac{\omega_{N}^{\mu}+\rho^{\mu0}}{\sqrt{2}} & \rho^{\mu+} & K^{\star\mu+}\\
\rho^{\mu-} & \frac{\omega_{N}^{\mu}-\rho^{\mu0}}{\sqrt{2}} & K^{\star\mu0}\\
K^{\star\mu-} & {\bar{K}}^{\star\mu0} & \omega_{S}^{\mu}%
\end{array}
\right)  \label{Ve}%
\end{equation}

whereas the matrix containing axial-vectors is [see Eq.\ (\ref{AQ})]:%

\begin{equation}
A=\frac{1}{\sqrt{2}}\left(
\begin{array}
[c]{ccc}%
\frac{f_{1N}^{\mu}+a_{1}^{\mu0}}{\sqrt{2}} & a_{1}^{\mu+} & K_{1}^{\mu+}\\
a_{1}^{\mu-} & \frac{f_{1N}^{\mu}-a_{1}^{\mu0}}{\sqrt{2}} & K_{1}^{\mu0}\\
K_{1}^{\mu-} & {\bar{K}}_{1}^{\mu0} & f_{1S}^{\mu}%
\end{array}
\right)  \text{.} \label{A}%
\end{equation}

The matrices from Eqs.\ (\ref{Ve}) and (\ref{A}) are combined into a
right-handed structure just as in Eq.\ (\ref{RQ}):%

\begin{equation}
R^{\mu}=\frac{1}{\sqrt{2}}\left(
\begin{array}
[c]{ccc}%
\frac{\omega_{N}^{\mu}+\rho^{\mu0}}{\sqrt{2}}-\frac{f_{1N}^{\mu}+a_{1}^{\mu0}%
}{\sqrt{2}} & \rho^{\mu+}-a_{1}^{\mu+} & K^{\star\mu+}-K_{1}^{\mu+}\\
\rho^{\mu-}-a_{1}^{\mu-} & \frac{\omega_{N}^{\mu}-\rho^{\mu0}}{\sqrt{2}}%
-\frac{f_{1N}^{\mu}-a_{1}^{\mu0}}{\sqrt{2}} & K^{\star\mu0}-K_{1}^{\mu0}\\
K^{\star\mu-}-K_{1}^{\mu-} & {\bar{K}}^{\star\mu0}-{\bar{K}}_{1}^{\mu0} &
\omega_{S}^{\mu}-f_{1S}^{\mu}%
\end{array}
\right)
\end{equation}

and into a left-handed structure just as in Eq.\ (\ref{LQ}):%

\begin{equation}
L^{\mu}=\frac{1}{\sqrt{2}}\left(
\begin{array}
[c]{ccc}%
\frac{\omega_{N}^{\mu}+\rho^{\mu0}}{\sqrt{2}}+\frac{f_{1N}^{\mu}+a_{1}^{\mu0}%
}{\sqrt{2}} & \rho^{\mu+}+a_{1}^{\mu+} & K^{\star\mu+}+K_{1}^{\mu+}\\
\rho^{\mu-}+a_{1}^{\mu-} & \frac{\omega_{N}^{\mu}-\rho^{\mu0}}{\sqrt{2}}%
+\frac{f_{1N}^{\mu}-a_{1}^{\mu0}}{\sqrt{2}} & K^{\star\mu0}+K_{1}^{\mu0}\\
K^{\star\mu-}+K_{1}^{\mu-} & {\bar{K}}^{\star\mu0}+{\bar{K}}_{1}^{\mu0} &
\omega_{S}^{\mu}+f_{1S}^{\mu}%
\end{array}
\right)  \text{.}%
\end{equation}

The covariant derivative is defined in accordance with Eq.\ (\ref{derivative}%
):
\begin{equation}
D^{\mu}\Phi=\partial^{\mu}\Phi-ig_{1}(L^{\mu}\Phi-\Phi R^{\mu})
\end{equation}

couples scalar and pseudoscalar degrees of freedom to vector and axial-vector
ones and, unlike the one in Eq.\ (\ref{PhiQQ}), it contains no photon field
$A^{\mu}$. The reason is that this and the subsequent chapters will be
dedicated to meson decays into other mesons only; the only exception will be
the decay $a_{1}\rightarrow\pi\gamma$ but the $a_{1}\pi\gamma$ Lagrangian in
the $N_{f}=3$ case would be exactly the same as the one presented in
Eq.\ (\ref{a1pg}) for $N_{f}=2$ and thus we do not need to recalculate it in
this version of the model.

The left-handed and right-handed field strength tensors are defined just as in
Eqs.\ (\ref{RRmu}) and (\ref{LLmu}):
\begin{align}
L^{\mu\nu}  &  =\partial^{\mu}L^{\nu}-\partial^{\nu}L^{\mu}\text{,}\\
R^{\mu\nu}  &  =\partial^{\mu}R^{\nu}-\partial^{\nu}R^{\mu}\text{.}
\end{align}

Explicit breaking of the global symmetry in the (pseudo)scalar channel is
described by the term Tr$[H(\Phi+\Phi^{\dagger})]$, see Eq.\ (\ref{LEB}),\ and
in the (axial-)vector channel by the term $\mathrm{Tr}\left[  \Delta(L_{\mu
}^{2}+R_{\mu}^{2})\right]  $, see Eq.\ (\ref{LEB1}),\ where%

\begin{align}
H  &  =\left(
\begin{array}
[c]{ccc}%
\frac{h_{0N}}{2} & 0 & 0\\
0 & \frac{h_{0N}}{2} & 0\\
0 & {0} & \frac{h_{0S}}{\sqrt{2}}%
\end{array}
\right)  \text{,}\label{H}\\
\Delta &  =\left(
\begin{array}
[c]{ccc}%
\delta_{N} & 0 & 0\\
0 & \delta_{N} & 0\\
0 & {0} & \delta_{S}%
\end{array}
\right)  \text{.} \label{Delta}%
\end{align}

As in Chapter \ref{chapterQ}, the spontaneous breaking of the chiral symmetry
will be implemented by condensing scalar isosinglet states -- in the case of
$N_{f}=3$, there are two such states [see Eq.\ (\ref{S})]: $\sigma_{N}%
\equiv(\bar{u}u+\bar{d}d)/\sqrt{2}$ and $\sigma_{N}\equiv\bar{s}s$. Let us
denote their respective condensates as $\phi_{N}$ and $\phi_{S}$. The
relations between $\phi_{N,S}$ and the pion decay constant $f_{\pi}$ as well
as the kaon decay constant $f_{K}$ are determined analogously to
Eq.\ (\ref{jA})]:%

\begin{align}
\phi_{N}  &  =Z_{\pi}f_{\pi}\text{,}\\
\phi_{S}  &  =\frac{Z_{K}f_{K}}{\sqrt{2}}\text{,}
\end{align}

where $f_{\pi}=92.4$ MeV and $f_{K}=155.5/\sqrt{2}$ MeV \cite{PDG}. Then the
chiral condensates $\phi_{N}$ and $\phi_{S}$ lead to the following mixing
terms in the Lagrangian (\ref{Lagrangian}):

\begin{align}
&  -g_{1}\phi_{N}(f_{1N}^{\mu}\partial_{\mu}\eta_{N}+{\vec{a}_{1}^{\mu}}%
\cdot\partial_{\mu}{\vec{\pi}})-\sqrt{2}g_{1}\phi_{S}f_{1S}^{\mu}\partial
_{\mu}\eta_{S}\nonumber\\
&  -\left(  \frac{g_{1}}{\sqrt{2}}\phi_{S}+\frac{g_{1}}{2}\phi_{N}\right)
\left(  K_{1}^{\mu0}\partial_{\mu}{\bar{K}^{0}+}K_{1}^{\mu+}\partial_{\mu
}K^{-}+\mathrm{h.c.}\right)  \text{ and}\nonumber\\
&  +\left(  i\frac{g_{1}}{\sqrt{2}}\phi_{S}-i\frac{g_{1}}{2}\phi_{N}\right)
\left(  {\bar{K}}^{\star\mu0}\partial_{\mu}{K_{S}^{0}+}K^{\star\mu-}%
\partial_{\mu}{K}_{S}^{+}\right) \nonumber\\
&  +\left(  -i\frac{g_{1}}{\sqrt{2}}\phi_{S}+i\frac{g_{1}}{2}\phi_{N}\right)
\left(  K^{\star\mu0}\partial_{\mu}{\bar{K}_{S}^{0}+}K^{\star\mu+}%
\partial_{\mu}{K}_{S}^{-}\right)  \text{.} \label{mixingterms}%
\end{align}

The first term in the first line of Eq.\ (\ref{mixingterms}) corresponds
exactly to the term (\ref{mixingtermsQ}) obtained in the$\ N_{f}=2$ case; the
other contributions are new. Note that, in the $N_{f}=3$ case, the mixing
between the pseudoscalars and the axial-vectors is accompanied by the
vector-scalar mixing of the fields $K^{\star}$ and $K_{S}$. The corresponding
coupling is imaginary but this is not problematic as the Lagrangian is
nonetheless hermitian, i.e., real.

The mixing terms (\ref{mixingterms}) are removed by the following suitable
shifts of the (axial-)vector states:%
\begin{align}
f_{1N}^{\mu}  &  \rightarrow f_{1N}^{\mu}+w_{f_{1N}}\partial^{\mu}\eta
_{N}\text{,} \label{shift21}\\
{\vec{a}_{1}^{\mu}}  &  \rightarrow{\vec{a}_{1}^{\mu}+}w_{a_{1}}\partial^{\mu
}{\vec{\pi}}\text{,} \label{shift22}\\
f_{1S}^{\mu}  &  \rightarrow f_{1S}^{\mu}+w_{f_{1S}}\partial^{\mu}\eta
_{S}\text{,} \label{shift23}\\
K_{1}^{\mu0}  &  \rightarrow K_{1}^{\mu0}+w_{K_{1}}\partial^{\mu}{K^{0}%
}\mathrm{\,(and\,h.c.)}\text{,} \label{shift24}\\
K^{\star\mu0}  &  \rightarrow K^{\star\mu0}+w_{K^{\star}}\partial^{\mu}%
{K_{S}^{0}}\text{,} \label{shift25}\\
K^{\star\mu+}  &  \rightarrow K^{\star\mu+}+w_{K^{\star}}\partial^{\mu}%
{K_{S}^{+}}\text{,} \label{shift26}\\
{\bar{K}}^{\star\mu0}  &  \rightarrow{\bar{K}}^{\star\mu0}+w_{K^{\star}}%
^{\ast}\partial^{\mu}{\bar{K}_{S}^{0}}\text{,} \label{shift27}\\
K^{\star\mu-}  &  \rightarrow K^{\star\mu-}+w_{K^{\star}}^{\ast}\partial^{\mu
}{K_{S}^{-}}\text{.} \label{shift28}%
\end{align}

The shifts (\ref{shift21}) and (\ref{shift22}) correspond to the one performed
in the $N_{f}=2$ case (\ref{shifts}). The quantities $w_{f_{1N}}$, $w_{a_{1}}%
$, $w_{f_{1S}}$, $w_{K_{1}}$ and $w_{K^{\star}}$ are calculated from the
condition that the mixing terms (\ref{mixingterms}) vanish once the shifts of
the (axial-)vectors have been implemented:
\begin{align}
&  \left[  -g_{1}\phi_{N}+\left(  g_{1}^{2}+\frac{h_{1}}{2}+\frac{h_{2}}%
{2}-\frac{h_{3}}{2}\right)  w_{f_{1N}}\phi_{N}^{2}+(m_{1}^{2}+2\delta
_{N})w_{f_{1N}}+\frac{h_{1}}{2}w_{f_{1N}}\phi_{S}^{2}\right] \nonumber\\
&  \times(f_{1N}^{\mu}\partial_{\mu}\eta_{N}+{\vec{a}_{1}^{\mu}}\cdot
\partial_{\mu}{\vec{\pi}})\overset{!}{=}0\text{,} \label{oldmixing}\\
&  \left[  -\sqrt{2}g_{1}\phi_{S}+\left(  2g_{1}^{2}+\frac{h_{1}}{2}%
+h_{2}-h_{3}\right)  w_{f_{1S}}\phi_{S}^{2}+(m_{1}^{2}+2\delta_{S})w_{f_{1S}%
}+\frac{h_{1}}{2}w_{f_{1S}}\phi_{N}^{2}\right]  f_{1S}^{\mu}\partial_{\mu}%
\eta_{S}\overset{!}{=}0\text{,} \\
&  \left[  -\frac{g_{1}}{\sqrt{2}}\phi_{S}+\frac{g_{1}^{2}-h_{3}}{\sqrt{2}%
}w_{K_{1}}\phi_{N}\phi_{S}+\frac{g_{1}^{2}+h_{1}+h_{2}}{2}w_{K_{1}}\phi
_{S}^{2}+(m_{1}^{2}+\delta_{N}+\delta_{S})w_{K_{1}}-\frac{g_{1}}{2}\phi
_{N}\right. \nonumber\\
&  \left.  +\left(  \frac{g_{1}^{2}}{4}+\frac{h_{1}}{2}+\frac{h_{2}}%
{4}\right)  w_{K_{1}}\phi_{N}^{2}\right]  \left(  K_{1}^{\mu0}\partial_{\mu
}{\bar{K}^{0}+}K_{1}^{\mu+}\partial_{\mu}K^{-}+\mathrm{h.c.}\right)
\overset{!}{=}0\text{,} \\
&  \left[  i\frac{g_{1}}{\sqrt{2}}\phi_{S}+\frac{h_{3}-g_{1}^{2}}{\sqrt{2}%
}w_{K^{\star}}\phi_{N}\phi_{S}+\frac{g_{1}^{2}+h_{1}+h_{2}}{2}w_{K^{\star}%
}\phi_{S}^{2}+(m_{1}^{2}+\delta_{N}+\delta_{S})w_{K^{\star}}-i\frac{g_{1}}%
{2}\phi_{N}\right. \nonumber\\
&  \left.  +\left(  \frac{g_{1}^{2}}{4}+\frac{h_{1}}{2}+\frac{h_{2}}%
{4}\right)  w_{K^{\star}}\phi_{N}^{2}\right]  \left(  {\bar{K}}^{\star\mu
0}\partial_{\mu}{K_{S}^{0}+}K^{\star\mu-}\partial_{\mu}{K}_{S}^{+}\right)
\overset{!}{=}0\text{,} \nonumber\\
+  &  \left[  -i\frac{g_{1}}{\sqrt{2}}\phi_{S}+\frac{h_{3}-g_{1}^{2}}{\sqrt
{2}}w_{K^{\star}}^{\ast}\phi_{N}\phi_{S}+\frac{g_{1}^{2}+h_{1}+h_{2}}%
{2}w_{K^{\star}}^{\ast}\phi_{S}^{2}+(m_{1}^{2}+\delta_{N}+\delta
_{S})w_{K^{\star}}^{\ast}+i\frac{g_{1}}{2}\phi_{N}\right. \nonumber\\
&  \left.  +\left(  \frac{g_{1}^{2}}{4}+\frac{h_{1}}{2}+\frac{h_{2}}%
{4}\right)  w_{K^{\star}}^{\ast}\phi_{N}^{2}\right]  \left(  K^{\star\mu
0}\partial_{\mu}{\bar{K}_{S}^{0}+}K^{\star\mu+}\partial_{\mu}{K}_{S}%
^{-}\right)  \overset{!}{=}0\text{.}  \label{newmixing}%
\end{align}
Equation (\ref{oldmixing}) corresponds exactly to the $f_{1N}^{\mu}$%
-$\partial_{\mu}\eta_{N}$ and ${\vec{a}_{1}^{\mu}}$-$\partial_{\mu}{\vec{\pi}%
}$ mixing terms obtained upon the stated shifts in the $N_{f}=2$ case.
Equations (\ref{oldmixing}) -- (\ref{newmixing}) are fulfilled only if we define:
\begin{align}
w_{f_{1N}}  &  =w_{a_{1}}=\frac{g_{1}\phi_{N}}{m_{a_{1}}^{2}}\text{,} \label{wa1}\\
w_{f_{1S}}  &  =\frac{\sqrt{2}g_{1}\phi_{S}}{m_{f_{1S}}^{2}}\text{,} \label{wf1S}\\
w_{K_{1}}  &  =\frac{g_{1}(\phi_{N}+\sqrt{2}\phi_{S})}{2m_{K_{1}}^{2}%
}\text{,} \label{wK1}\\
w_{K^{\star}}  &  =i\frac{g_{1}(\phi_{N}-\sqrt{2}\phi_{S})}{2m_{K^{\star}}%
^{2}}\text{.} \label{wKstar}%
\end{align}

The definitions (\ref{wa1}) -- (\ref{wKstar}) require the knowledge of the
following mass terms obtained from the Lagrangian (\ref{Lagrangian}):
\begin{align}
m_{\sigma_{N}}^{2}  &  =m_{0}^{2}+3\left(  \lambda_{1}+\frac{\lambda_{2}}%
{2}\right)  \phi_{N}^{2}+\lambda_{1}\phi_{S}^{2}\text{,} \qquad \qquad \qquad \qquad \label{m_sigma_N}\\
m_{\pi}^{2}  &  =Z_{\pi}^{2}\left[  m_{0}^{2}+\left(  \lambda_{1}%
+\frac{\lambda_{2}}{2}\right)  \phi_{N}^{2}+\lambda_{1}\phi_{S}^{2}\right]
\equiv\frac{Z_{\pi}^{2}h_{0N}}{\phi_{N}}\text{,} \qquad \qquad \qquad \qquad \label{m_pi}
\end{align}
\begin{align}
m_{a_{0}}^{2}  &  =m_{0}^{2}+\left(  \lambda_{1}+3\frac{\lambda_{2}}%
{2}\right)  \phi_{N}^{2}+\lambda_{1}\phi_{S}^{2}\text{,}\label{m_a_0}\\
m_{\eta_{N}}^{2}  &  =Z_{\pi}^{2}\left[  m_{0}^{2}+\left(  \lambda_{1}%
+\frac{\lambda_{2}}{2}\right)  \phi_{N}^{2}+\lambda_{1}\phi_{S}^{2}+c_{1}%
\phi_{N}^{2}\phi_{S}^{2}\right]  \equiv m_{\pi}^{2}+c_{1}Z_{\pi}^{2}\phi
_{N}^{2}\phi_{S}^{2}\text{,} \label{m_eta_N} \\
m_{\sigma_{S}}^{2}  &  =m_{0}^{2}+\lambda_{1}\phi_{N}^{2}+3(\lambda
_{1}+\lambda_{2})\phi_{S}^{2}\text{,} \label{m_sigma_S}\\
m_{\eta_{S}}^{2}  &  =Z_{\eta_{S}}^{2}\left[  m_{0}^{2}+\lambda_{1}\phi
_{N}^{2}+(\lambda_{1}+\lambda_{2})\phi_{S}^{2}+c_{1}\frac{\phi_{N}^{4}}%
{4}\right]  =Z_{\eta_{S}}^{2} \left ( \frac{h_{0S}}{\phi_{S}} + \frac{c_1}{4} \phi_N^4 \right )\text{,} \label{m_eta_S}\\
m_{K_{S}}^{2}  &  =Z_{K_{S}}^{2}\left[  m_{0}^{2}+\left(  \lambda_{1}%
+\frac{\lambda_{2}}{2}\right)  \phi_{N}^{2}\ +\frac{\lambda_{2}}{\sqrt{2}}%
\phi_{N}\phi_{S}+(\lambda_{1}+\lambda_{2})\phi_{S}^{2}\right]\text{,} \label{m_K_S}\\
m_{K}^{2}  &  =Z_{K}^{2}\left[  m_{0}^{2}+\left(  \lambda_{1}+\frac
{\lambda_{2}}{2}\right)  \phi_{N}^{2}\ -\frac{\lambda_{2}}{\sqrt{2}}\phi
_{N}\phi_{S}+(\lambda_{1}+\lambda_{2})\phi_{S}^{2}\right]\text{,} \label{m_K}\\
m_{\omega_{N}}^{2}  &  =m_{\rho}^{2}=m_{1}^{2}+2\delta_{N}+\frac{\phi_{N}^{2}%
}{2}(h_{1}+h_{2}+h_{3})+\frac{h_{1}}{2}\phi_{S}^{2}\text{,} \label{m_rho}\\
m_{f_{1N}}^{2}  &  =m_{a_{1}}^{2}=m_{1}^{2}+2\delta_{N}+g_{1}^{2}\phi_{N}%
^{2}+\frac{\phi_{N}^{2}}{2}(h_{1}+h_{2}-h_{3})+\frac{h_{1}}{2}\phi_{S}%
^{2}\text{,} \label{m_a_1}\\
m_{\omega_{S}}^{2}  &  =m_{1}^{2}+2\delta_{S}+\frac{h_{1}}{2}\phi_{N}^{2}%
+\phi_{S}^{2}\left(  \frac{h_{1}}{2}+h_{2}+h_{3}\right)\text{,} \label{m_omega_S}\\
m_{f_{1S}}^{2}  &  =m_{1}^{2}+2\delta_{S}+\frac{h_{1}}{2}\phi_{N}^{2}%
+2g_{1}^{2}\phi_{S}^{2}+\phi_{S}^{2}\left(  \frac{h_{1}}{2}+h_{2}-h_{3}\right)
\text{,} \label{m_f1_S}\\
m_{K^{\star}}^{2}  &  =m_{1}^{2}+\delta_{N}+\delta_{S}+\frac{\phi_{N}^{2}}%
{2}\left(  \frac{g_{1}^{2}}{2}+h_{1}+\frac{h_{2}}{2}\right)  +\frac{1}%
{\sqrt{2}}\phi_{N}\phi_{S}(h_{3}-g_{1}^{2})+\frac{\phi_{S}^{2}}{2}\left(
g_{1}^{2}+h_{1}+h_{2}\right)\text{,} \label{m_K_star}\\
m_{K_{1}}^{2}  &  =m_{1}^{2}+\delta_{N}+\delta_{S}+\frac{\phi_{N}^{2}}%
{2}\left(  \frac{g_{1}^{2}}{2}+h_{1}+\frac{h_{2}}{2}\right)  +\frac{1}%
{\sqrt{2}}\phi_{N}\phi_{S}(g_{1}^{2}-h_{3})+\frac{\phi_{S}^{2}}{2}\left(
g_{1}^{2}+h_{1}+h_{2}\right)\text{,} \label{m_K_1}%
\end{align}

with the renormalisation factors ensuring the canonical normalisation of the
$\eta_{N,S}$, ${\vec{\pi}}$ and $K_{S}$ wave functions
\begin{align}
Z_{\pi}  &  \equiv Z_{\eta_{N}}=\frac{m_{a_{1}}}{\sqrt{m_{a_{1}}^{2}-g_{1}%
^{2}\phi_{N}^{2}}}\mathrm{\,(just\,as\,before)}\text{,} \label{Z_pi}\\
Z_{K}  &  =\frac{2m_{K_{1}}}{\sqrt{4m_{K_{1}}^{2}-g_{1}^{2}(\phi_{N}+\sqrt
{2}\phi_{S})^{2}}}\text{,} \label{Z_K}\\
Z_{\eta_{S}}  &  =\frac{m_{f_{1S}}}{\sqrt{m_{f_{1S}}^{2}-2g_{1}^{2}\phi
_{S}^{2}}}\text{,} \label{Z_eta_S}\\
Z_{K_{S}}  &  =\frac{2m_{K^{_{\star}}}}{\sqrt{4m_{K^{_{\star}}}^{2}-g_{1}%
^{2}(\phi_{N}-\sqrt{2}\phi_{S})^{2}}}\text{.} \label{Z_K_S}%
\end{align}

It is obvious from Eqs.\ (\ref{Z_pi}) - (\ref{Z_K_S}) that all the
renormalisation coefficients will have values larger than 1.

Note that the right-hand sides of Eqs.\ (\ref{m_pi}) and (\ref{m_eta_S}) are
obtained by minimising the potential $\mathcal{V}(\phi_{N},\phi_{S})$ present
in the Lagrangian (\ref{Lagrangian}). The potential reads:%
\begin{equation}
\mathcal{V}(\phi_{N},\phi_{S})=\frac{1}{2}m_{0}^{2}(\phi_{N}^{2}+\phi_{S}%
^{2})+\frac{\lambda_{1}}{4}(\phi_{N}^{4}+2\phi_{N}^{2}\phi_{S}^{2}+\phi
_{S}^{4})+\frac{\lambda_{2}}{4}\left(  \frac{\phi_{N}^{4}}{2}+\phi_{S}%
^{4}\right)  -h_{0N}\phi_{N}-h_{0S}\phi_{S} \label{V}%
\end{equation}

and, consequently, the first derivatives of the potential $\mathcal{V}(\phi
_{N},\phi_{S})$ from the above Eq.\ (\ref{V}) are%

\begin{align}
\frac{\partial\mathcal{V}(\phi_{N},\phi_{S})}{\partial\phi_{N}}  &
=[m_{0}^{2}+\lambda_{1}(\phi_{N}^{2}+\phi_{S}^{2})]\phi_{N}+\frac{\lambda_{2}%
}{2}\phi_{N}^{3}-h_{0N}\text{,} \label{dVN}\\
\frac{\partial\mathcal{V}(\phi_{N},\phi_{S})}{\partial\phi_{S}}  &
=[m_{0}^{2}+\lambda_{1}(\phi_{N}^{2}+\phi_{S}^{2})]\phi_{S}+\lambda_{2}%
\phi_{S}^{3}-h_{0S}\text{.} \label{dVS}%
\end{align}

Then we obtain
\begin{align}
\frac{\partial\mathcal{V}(\phi_{N},\phi_{S})}{\partial\phi_{N}}\overset{!}%
{=}0  &  \Leftrightarrow h_{0N}=m_{0}^{2}+\lambda_{1}(\phi_{N}^{2}+\phi
_{S}^{2})]\phi_{N}+\frac{\lambda_{2}}{2}\phi_{N}^{3}\text{,} \\
\frac{\partial\mathcal{V}(\phi_{N},\phi_{S})}{\partial\phi_{S}}\overset{!}%
{=}0  &  \Leftrightarrow h_{0S}=m_{0}^{2}+\lambda_{1}(\phi_{N}^{2}+\phi
_{S}^{2})]\phi_{S}+\lambda_{2}\phi_{S}^{3}\text{.}%
\end{align}

and the right-hand side of Eqs.\ (\ref{m_pi}) and (\ref{m_eta_S}) ensue.

Note also that from Eq.\ (\ref{Z_pi}) we obtain the $g_{1}=g_{1}(Z_{\pi})$
dependence, just as in Eq.\ (\ref{g1Q}):%

\begin{equation}
g_{1}=g_{1}(Z_{\pi})=\frac{m_{a_{1}}}{Z_{\pi}f_{\pi}}\sqrt{1-\frac{1}{Z_{\pi
}^{2}}} \label{g1}%
\end{equation}

and also from Eqs.\ (\ref{m_rho}) and (\ref{m_a_1}):
\begin{equation}
h_{3}=h_{3}(Z_{\pi})=\frac{m_{a_{1}}^{2}}{Z_{\pi}^{2}f_{\pi}^{2}}\left(
\frac{m_{\rho}^{2}}{m_{a_{1}}^{2}}-\frac{1}{Z_{\pi}^{2}}\right)  . \label{h3}%
\end{equation}

Utilising Eq.\ (\ref{g1}) we can transform Eq.\ (\ref{wa1}) as
follows:\newline%
\begin{equation}
w_{a_{1}}=\frac{\frac{m_{a_{1}}}{Z_{\pi}f_{\pi}}\sqrt{1-\frac{1}{Z_{\pi}^{2}}%
}\phi_{N}}{m_{a_{1}}^{2}}=\frac{\sqrt{1-\frac{1}{Z_{\pi}^{2}}}}{m_{a_{1}}%
}=\frac{\sqrt{Z_{\pi}^{2}-1}}{Z_{\pi}m_{a_{1}}}\text{.} \label{wa11}%
\end{equation}

Additionally, from Eqs.\ (\ref{Z_K}), (\ref{Z_eta_S}) and (\ref{Z_K_S}) we
obtain%
\begin{align}
g_{1}  &  =g_{1}(Z_{\pi},Z_{K})=\frac{2m_{K_{1}}}{Z_{\pi}f_{\pi}+Z_{K}f_{K}%
}\sqrt{1-\frac{1}{Z_{K}^{2}}}\text{,} \label{g11}\\
g_{1}  &  =g_{1}(Z_{K},Z_{\eta_{S}})=\frac{m_{f_{1S}}}{Z_{K}f_{K}}%
\sqrt{1-\frac{1}{Z_{\eta_{S}}^{2}}}\text{,} \label{g12}\\
g_{1}  &  =g_{1}(Z_{\pi},Z_{K_{S}})=\frac{2m_{K^{_{\star}}}}{Z_{\pi}f_{\pi
}-Z_{K}f_{K}}\sqrt{1-\frac{1}{Z_{K_{S}}^{2}}}\text{.} \label{g13}%
\end{align}

Then analogously to Eq.\ (\ref{wa11}) we obtain from Eqs.\ (\ref{wK1}) and
(\ref{g11}):%

\begin{equation}
w_{K_{1}}=\frac{\frac{2m_{K_{1}}}{Z_{\pi}f_{\pi}+Z_{K}f_{K}}\sqrt{1-\frac
{1}{Z_{K}^{2}}}(\phi_{N}+\sqrt{2}\phi_{S})}{2m_{K_{1}}^{2}}=\frac
{\sqrt{1-\frac{1}{Z_{K}^{2}}}}{m_{K_{1}}}=\frac{\sqrt{Z_{K}^{2}-1}}%
{Z_{K}m_{K_{1}}}\text{.} \label{wK11}%
\end{equation}


It is very important to stress that Eq.\ (\ref{Z_K}) is not the only equation
regarding $Z_{K}$. From Eqs.\ (\ref{m_omega_S}) and (\ref{m_f1_S}) we obtain%
\begin{align}
m_{f_{1S}}^{2}-m_{\omega_{S}}^{2} &  =2[g_{1}^{2}(Z_{\pi})-h_{3}(Z_{\pi}%
)]\phi_{S}^{2}\overset{\phi_{S}=\frac{1}{\sqrt{2}}Z_{K}f_{K}}{\equiv}%
[g_{1}^{2}(Z_{\pi})-h_{3}(Z_{\pi})]Z_{K}^{2}f_{K}^{2}\nonumber\\
&  \Rightarrow Z_{K}=\frac{1}{f_{K}}\sqrt{\frac{m_{f_{1S}}^{2}-m_{\omega_{S}%
}^{2}}{g_{1}^{2}(Z_{\pi})-h_{3}(Z_{\pi})}}\label{Z_K1}%
\end{align}
with $g_{1}(Z_{\pi})$ and $h_{3}(Z_{\pi})$ from Eqs.\ (\ref{g1}) and
(\ref{h3}), respectively. Also, from Eqs.\ (\ref{m_K_star}) and (\ref{m_K_1})
we obtain%
\begin{align}
m_{K_{1}}^{2}-m_{K_{\star}}^{2} &  =\sqrt{2}\phi_{N}\phi_{S}[g_{1}^{2}(Z_{\pi
})-h_{3}(Z_{\pi})]\overset{\phi_{S}=\frac{1}{\sqrt{2}}Z_{K}f_{K},\phi
_{N}=Z_{\pi}f_{\pi}}{\equiv}Z_{\pi}f_{\pi}Z_{K}f_{K}[g_{1}^{2}(Z_{\pi}%
)-h_{3}(Z_{\pi})]\nonumber\\
&  \Rightarrow Z_{K}=\frac{m_{K_{1}}^{2}-m_{K_{\star}}^{2}}{Z_{\pi}f_{\pi
}f_{K}[g_{1}^{2}(Z_{\pi})-h_{3}(Z_{\pi})]}\text{.} \label{Z_K2}%
\end{align}
Therefore, in order to be consistent, the values of $Z_{K}$ have to
simultaneously fulfill three equations: (\ref{Z_K}), which is the definition
of $Z_{K}$; (\ref{Z_K1}) and (\ref{Z_K2}). This will represent a very strong
constraint on the values of the parameters in the Lagrangian (\ref{Lagrangian}).\\
We also emphasise that the explicit form of the Lagrangian (\ref{Lagrangian}) is
an extremely complicated function due to the abundance of terms permitted by the global
$U(3)_L \times U(3)_R$ symmetry [increased further by the eight shifts (\ref{shift21}) -- (\ref{shift28})]
and also because of the large number of fields from both
non-strange and strange sectors. For this reason, the evaluation of the Lagrangian
and the extrapolation of vertices necessary for the calculation of decay widths in this chapter as well as 
Chapters \ref{fitstructure} -- \ref{ImplicationsFitII} have been performed using a computer algorithm (available from the author upon request).\\

Most model parameters can be calculated from the meson mass terms (\ref{m_sigma_N}) -- (\ref{m_K_1}). However, using only mass terms as 
means of parameter calculation
would leave some important parameters undetermined,
such as, e.g., $g_2$ [that strongly influences the (axial-)vector phenomenology, see below]. For this reason, our parameter 
calculation will use some decay widths as well, but only
as few as necessary to determine the model parameters. All other decay widths will then be calculated as a consequence,
thus raising the predictive power of the model.\\
Before we determine the parameters we first need to assign the
fields from our model to the physical ones.

\section{Assigning the Fields} \label{sec.assignment}

The model contains four nonets: scalar (\ref{S}), pseudoscalar
(\ref{P}), vector (\ref{Ve}) and axial-vector (\ref{A}) containing both
non-strange and strange states. If we consider isospin multiplets as single degrees of freedom, then there are sixteen resonances that can be described by the model:
$\sigma_{N}$, $\sigma_{S}$, $\vec{a}_{0}$, $K_{S}$ (scalar); $\eta_{N}$,
$\eta_{S}$, $\vec{\pi}$, $K$ (pseudoscalar); $\omega_{N}^{\mu}$, $\omega
_{S}^{\mu}$, $\vec{\rho}^{\mu}$, $K^{\star\mu}$ (vector) and $f_{1N}^{\mu}$,
$f_{1S}^{\mu}$, $\vec{a}_{1}^{\mu}$, $K_{1}$ (axial-vector). All of the states
present in the model possess the ${\bar{q}}q$ structure, for the same reasons
as those presented in Sec.\ \ref{sec.largen}.

As in Ref.\ \cite{Paper1}, in the non-strange sector we assign the fields
$\vec{\pi}$ and $\eta_{N}$ to the pion and the $SU(2)$ counterpart of the
$\eta$ meson, $\eta_{N}\equiv(\bar{u}u+\bar{d}d)/\sqrt{2}$. The
fields $\omega_{N}^{\mu}$ and $\vec{\rho}^{\mu}$ represent the $\omega(782)$
and $\rho(770)$ vector mesons, respectively, and the fields $f_{1N}^{\mu}$ and
$\vec{a}_{1}^{\mu}$ represent the $f_{1}(1285)$ and $a_{1}(1260)$ mesons,
respectively. In the strange sector, we assign the $K$ fields to the kaons;
the $\eta_{S}$ field is the strange contribution to the physical $\eta$ and
$\eta^{\prime}$ fields; the $\omega_{S}$, $f_{1S}$, $K^{\star}$ and $K_{1}$
fields correspond to the $\phi(1020)$, $f_{1}(1420)$, $K^{\star}(892)$, and
$K_{1}(1270)$ mesons, respectively.

Unfortunately, the assignment of the scalar fields is substantially less
clear. Experimental data presented in Chapter \ref{sec.scalarexp} suggest
the existence of six scalar-isoscalar states below $1.8$ GeV: $f_{0}(600)$, $f_{0}(980)$,
$f_{0}(1370)$, $f_{0}(1500)$, $f_{0}(1710)$ and $f_{0}(1790)$. Our model
contains the pure non-strange isoscalar $\sigma_{N}$, the pure strange
isoscalar $\sigma_{S}$ and the scalar kaon $K_{S}$. We will see in the
subsequent chapters that our model yields mixing of $\sigma_{N}$ and
$\sigma_{S}$ producing a predominantly non-strange state, labelled as
$\sigma_{1}$,\ and a predominantly strange state, labelled as $\sigma_{2}$.
Assigning $\sigma_{1}$ and $\sigma_{2}$\ to physical states will be the
primary focus of Chapters \ref{ImplicationsFitI} and \ref{ImplicationsFitII}.
Therefore, a conclusive assignment of the latter states is not possible at this point.

Similarly, the isospin triplet $\vec{a}_{0}$ can be assigned to different
physical resonances, although in this case there are only two candidate
states: $a_{0}(980)$ and $a_{0}(1450)$. An analogous statement holds for the
scalar kaon $K_{S}$ that can be assigned to the resonances $K_{0}^{\star
}(800)$ or $K_{0}^{\star}(1430)$. In the following chapters we will therefore
consider two possibilities for assignments of scalar fields performing two fits:

\begin{itemize}
\item \textit{Fit I}, where $\vec{a}_{0}$ is assigned to $a_{0}(980)$ and
$K_{S}$\ to $K_{0}^{\star}(800)$ (Chapters \ref{sec.fitI} and
\ref{ImplicationsFitI}) -- we assign our scalar states to resonances below $1$ GeV.

\item \textit{Fit II}, where $\vec{a}_{0}$ is assigned to $a_{0}(1450)$ and
$K_{S}$\ to $K_{0}^{\star}(1430)$ (Chapters \ref{sec.fitII} and
\ref{ImplicationsFitII}) -- we assign our scalar states to resonances above $1$ GeV.
\end{itemize}

Note that Fit I implies that $a_{0}(980)$ is a non-strange ${\bar{q}}q$ state
and that $K_{0}^{\star}(800)$ is a strange ${\bar{q}}q$ state. Conversely, Fit II
implies that $a_{0}(1450)$ is a non-strange ${\bar{q}}q$ state and that
$K_{0}^{\star}(1430)$ is a strange ${\bar{q}}q$ state. Given the large number of the
$f_{0}$ resonances, it will not be possible to assign $\sigma_{1}$ and
$\sigma_{2}$ \textit{ab initio }in the two fits; the assignment of these two
states will depend on the fit results.

However, some remarks regarding the model parameters are in order before the
calculations can proceed.

\section{General Discussion of the Model Parameters}

The model contains 18 parameters:
\begin{align}
m_{0}^{2}\text{, } m_{1}^{2}\text{, } c_{1}\text{, }
\delta_{N}\text{, } \delta_{S}\text{, } g_{1}\text{, } g_{2}\text{, } g_{3}\text{, } g_{4}\text{, } g_{5}\text{, }
g_{6}\text{, } h_{0N}\text{, } h_{0S}\text{, } h_{1}\text{, } h_{2}\text{, } h_{3}\text{, } \lambda_{1}\text{, }
\lambda_{2} \text{.} 
\end{align}

Let us make following observations regarding the model
parameters: 

\begin{itemize}
\item The parameters $h_{0N}$ and $h_{0S}$ model the explicit breaking of the
chiral symmetry (ESB) in the (pseudo)scalar sector via the term $\mathrm{Tr}%
[H(\Phi+\Phi^{\dagger})]$ in the Lagrangian (\ref{Lagrangian}); they are
calculated using the extremum equations (\ref{dVN}) and (\ref{dVS}) of the
potential $\mathcal{V}(\phi_{N},\phi_{S})$ or, equivalently, from the mass
terms of the pion, Eq.\ (\ref{m_pi}), and of $\eta_{S}$,\ the strange
counterpart of the $\eta$ meson [see Eq.\ (\ref{m_eta_S})]. From Eqs.\ (\ref{m_pi}), (\ref{m_eta_N}) and (\ref{m_eta_S}) we can then
conclude that the masses of $\vec{\pi}$, $\eta_N$ and $\eta_{S}$ are generated by ESB: the pion mass completely and the $\eta_{N,S}$ masses via interplay
of ESB and the chiral anomaly.
Given that $h_{0N}$ and $h_{0S}$ are determined uniquely from
Eqs.\ (\ref{m_pi}) and (\ref{m_eta_S}), we are then left with 16 free parameters.

\item The parameters $\delta_{N}$ and $\delta_{S}$ model the explicit symmetry
breaking in the vector and axial-vector channels. The ESB stems from the
non-vanishing quark masses and therefore we employ the correspondence $\delta
_{N}\propto m_{u,d}^{2}$ and $\delta_{S}\propto m_{s}^{2}$. Given that
$m_{u,d}\ll m_{s}$, we will set $\delta_{N}=0$ throughout this work;
$\delta_{S}$ will be determined by the fit of (axial-)vector masses.
Consequently, the number of free parameters is decreased to 15.

\item Our fit will make use of all scalar mass terms except $m_{\sigma_{N}}$
and $m_{\sigma_{S}}$ due to the well-known ambiguities regarding the
phenomenology of scalar mesons (see Chapter \ref{sec.scalarexp}). However, the mass terms that
will be used in the fit only contain the linear combination $m_{0}^{2}%
+\lambda_{1}(\phi_{N}^{2}+\phi_{S}^{2})$ rather than the parameters $m_{0}%
^{2}$ and $\lambda_{1}$ (that only appear separately in $m_{\sigma_{N}}$ and
$m_{\sigma_{S}}$).\ Nonetheless, the knowledge of the mentioned linear
combination will allow us to express the parameter $\lambda_{1}$ via the
bare-mass parameter $m_{0}^{2}$. Consequently, the number of free
parameters is decreased to 14.

\item Similarly, the (axial-)vector mass terms [Eqs.\ (\ref{m_rho}) -
(\ref{m_K_1})] allow only for the linear combination $m_{1}^{2}+h_{1}(\phi
_{N}^{2}+\phi_{S}^{2})/2$ rather than the parameters $m_{1}^{2}$ and $h_{1}$
separately to be determined. Given that the parameter $h_{1}$ is suppressed in
the limit of a large number of colours (large-$N_{c}$ limit, see discussion
in Sec.\ \ref{sec.largen}), we will set $h_{1}=0$ throughout this work. Thus the number of
unknown parameters is decreased to 13. [Note that the parameter
$\lambda_{1}$ is also large-$N_{c}$ suppressed and could also in principle be
set to zero; however we do not set $\lambda_{1}=0$ as in that case there would
be no mixing between $\sigma_{N}$ and $\sigma_{S}$, see
Eq.\ (\ref{sigma-sigma}). Nonetheless, the mixing between the two pure $\sigma$
states is shown to be small in Sec.\ \ref{sec.scalarmasses1} (see Fig.\ \ref{phi11}) -- this is in line with expectations
because the mixing is governed by a large-$N_{c}$ suppressed parameter.]

\item The parameter $g_{2}$ is determined by the decay width $\Gamma
_{\rho\rightarrow\pi\pi}$, see Eq.\ (\ref{g2Z}). Additionally, the parameters $g_{3}$,
$g_{4}$, $g_{5}$ and $g_{6}$ do not influence any of the decays to be
discussed in this work and are therefore also left out from the fit. Thus the
number of unknown parameters is decreased to eight.

\item The parameter $c_{1}$ can be substituted by the $\eta$-$\eta^{\prime}$
mixing angle $\varphi_{\eta}$, as we will see in Sec.\ \ref{sec.eta-eta},
Eq.\ (\ref{c1phi}). The value of the mixing angle $\varphi_{\eta}$ can be
determined in a way that the masses of $\eta$ and $\eta^{\prime}$ are as close
to the physical masses as possible. That in turn implies that the value of the
parameter $c_{1}$ is determined uniquely by $m_{\eta}$ and $m_{\eta^{\prime}}$
-- and therefore, $c_{1}$ is not a free parameter in the model.

\item It is obvious from Eqs.\ (\ref{g1}) - (\ref{g13}) that the parameters
$g_{1}$ and $h_{3}$ can be substituted by a pair of renormalisation
coefficients. In principle, any pair of coefficients can be used; however,
given that the coefficients $Z_{\pi}$ and $Z_{K}$ appear respectively in the
chiral condensates $\phi_{N}$ and $\phi_{S}$ (that, in turn, influence all
observables in this work), it is convenient to substitute $g_{1}$ and $h_{3}$
by $Z_{\pi}$ and $Z_{K}$. The coefficients $Z_{\eta_{S}}$ and $Z_{K_{S}}$ can
be calculated from Eqs.\ (\ref{Z_eta_S}) and (\ref{Z_K_S}).
\end{itemize}

We are therefore left with six parameters: $Z_{\pi}$, $Z_{K}$, $m_{1}^{2}$,
$h_{2}$, $\delta_{S}$ and $\lambda_{2}$ and one parameter combination:
$m_{0}^{2}+\lambda_{1}(\phi_{N}^{2}+\phi_{S}^{2})$. Before we turn to the
calculation of the parameters, let us briefly discuss the chiral-anomaly term
in the Lagrangian (\ref{Lagrangian}) and the large-$N_{c}$ behaviour of the
model parameters.

\section{Modelling the Chiral Anomaly} \label{sec.anomaly}

We
turn now to the chiral-anomaly term present in our Lagrangian
(\ref{Lagrangian}). The goal of this subsection is to clarify how different
ways of chiral-anomaly modelling influence the mass terms; the mass terms in
turn represent a crucial part of the fit that will enable us to determine
model parameters.\newline

The term $c_{1}(\det\Phi-\det\Phi^{\dagger})^{2}$ in the Lagrangian
(\ref{Lagrangian}) describes the chiral anomaly of QCD as was first
discussed by Veneziano and Witten in 1979 \cite{Veneziano-Witten}. We will
refer to this chiral-anomaly term as the VW term in the
following. The term has been subject of further calculations in Ref.
\cite{Fariborz-anomaly}.\ However, this is not the only way to model the
chiral anomaly; an alternative had been discussed in 1971 by Kobayashi, Kondo
and Maskawa \cite{KKM}. This alternative term has the form $c(\det\Phi
+\det\Phi^{\dagger})$ as described by 't Hooft \cite{Hooft}; see also Ref.\ \cite{SchaffnerBielich:1999uj}. We will refer
to this term as the Kobayashi-Kondo-Maskawa-'t Hooft (KKMH) term in the
following. Although the two terms model the same (chiral) anomaly of QCD,
there are some notable differences between the two terms.

\begin{itemize}
\item The Veneziano-Witten term is of order $\mathcal{O}(6)$ in the naive
scaling of fields whereas the KKMH term is of order $\mathcal{O}(3)$ in
fields. We have stated in Sec.\ \ref{sec.c} (see also Chapter \ref{chapterglueball}) that, in principle, our Lagrangian only
allows for terms up to order $\mathcal{O}(4)$ in fields so that the dilatation
invariance is satisfied. However, the chirally anomalous terms do not need to
fulfill this invariance and therefore one can choose either of the two terms,
or both, or even additional terms compatible with the anomaly.

\item The Veneziano-Witten term of the chiral anomaly influences only
the phenomenology of the pseudoscalar singlets ($\eta_{N}$ and $\eta_{S}$,\ i.e.,
$\eta$ and $\eta^{\prime}$). The contribution can be inferred from
Eqs.\ (\ref{Lagrangian}), (\ref{m_eta_N}) and (\ref{m_eta_S}):
\begin{align}
(m_{\eta_{N}}^{2})^{\text{VW}}  &  \equiv m_{\eta_{N}}^{2} \sim 
c_{1}Z_{\pi}^{2}\phi_{N}^{2}\phi_{S}^{2}\text{,} \\
(m_{\eta_{S}}^{2})^{\text{VW}}  &  \equiv m_{\eta_{S}}^{2} \sim
c_{1}Z_{\eta_{S}}^{2}\frac{\phi_{N}^{4}}{4}\text{,} \\
\mathcal{L}_{\eta_{N}\eta_{S}}  &  =-c_{1}\frac{Z_{\eta_{S}}Z_{\pi}}{2}%
\phi_{N}^{3}\phi_{S}\eta_{N}\eta_{S} \text{.} \label{u}%
\end{align}
(See also Sec.\ \ref{sec.eta-eta}.)\ Additionally, we can see from Eq.\ (\ref{m_eta_N}) that the chiral anomaly is
responsible for the mass splitting between the pseudoscalar isotriplet
$\vec{\pi}$ and its SU(2) counterpart, the isosinglet $\eta_{N}$. Contrarily, the KKMH
term influences phenomenology of other scalar mesons as well \cite{Klempt}.
The (pseudo)scalar mass terms and other relevant parts of the Lagrangian are
in this case as follows:%
\begin{align}
(m_{\sigma_{N}}^{2})^{\text{KKMH}}  &  =m_{0}^{2}+3\left(  \lambda_{1}%
+\frac{\lambda_{2}}{2}\right)  \phi_{N}^{2}+\lambda_{1}\phi_{S}^{2}-\frac
{c}{\sqrt{2}}\phi_{S}\text{,} \label{q}\\
(m_{\eta_{N}}^{2})^{\text{KKMH}}  &  =Z_{\pi}^{2}\left[  m_{0}^{2}+\left(
\lambda_{1}+\frac{\lambda_{2}}{2}\right)  \phi_{N}^{2}+\lambda_{1}\phi_{S}%
^{2}+\frac{c}{\sqrt{2}}\phi_{S}\right]  \equiv(m_{\pi}^{2})^{\text{KKMH}%
}+\sqrt{2}cZ_{\pi}^{2}\phi_{S}\text{,} \\
(m_{a_{0}}^{2})^{\text{KKMH}}  &  =m_{0}^{2}+\left(  \lambda_{1}%
+3\frac{\lambda_{2}}{2}\right)  \phi_{N}^{2}+\lambda_{1}\phi_{S}^{2}+\frac
{c}{\sqrt{2}}\phi_{S}\text{,} \\
(m_{\pi}^{2})^{\text{KKMH}}  &  =Z_{\pi}^{2}\left[  m_{0}^{2}+\left(
\lambda_{1}+\frac{\lambda_{2}}{2}\right)  \phi_{N}^{2}+\lambda_{1}\phi_{S}%
^{2}-\frac{c}{\sqrt{2}}\phi_{S}\right]  \equiv\frac{Z_{\pi}^{2}h_{0N}}%
{\phi_{N}}\text{,} \\
(m_{\sigma_{S}}^{2})^{\text{KKMH}}  &  =m_{0}^{2}+\lambda_{1}\phi_{N}%
^{2}+3(\lambda_{1}+\lambda_{2})\phi_{S}^{2}\text{,}\\
(m_{\eta_{S}}^{2})^{\text{KKMH}}  &  =Z_{\eta_{S}}^{2}[m_{0}^{2}+\lambda
_{1}\phi_{N}^{2}+(\lambda_{1}+\lambda_{2})\phi_{S}^{2}]=\frac{Z_{\eta_{S}}%
^{2}}{\phi_{S}}(h_{0S}+\frac{c}{2\sqrt{2}}\phi_{N}^{2})\text{,}
\end{align}
\begin{align}
(m_{K_{S}}^{2})^{\text{KKMH}}  &  =Z_{K_{S}}^{2}\left[  m_{0}^{2}+\left(
\lambda_{1}+\frac{\lambda_{2}}{2}\right)  \phi_{N}^{2}\ +\frac{\lambda_{2}%
}{\sqrt{2}}\phi_{N}\phi_{S}+(\lambda_{1}+\lambda_{2})\phi_{S}^{2}+\frac{c}%
{2}\phi_{N}\right]\text{,} \\
(m_{K}^{2})^{\text{KKMH}}  &  =Z_{K}^{2}\left[  m_{0}^{2}+\left(  \lambda
_{1}+\frac{\lambda_{2}}{2}\right)  \phi_{N}^{2}\ -\frac{\lambda_{2}}{\sqrt{2}%
}\phi_{N}\phi_{S}+(\lambda_{1}+\lambda_{2})\phi_{S}^{2}-\frac{c}{2}\phi
_{N}\right]\text{,} \label{w}\\
\mathcal{L}_{\eta_{N}\eta_{S}}^{\text{KKMH}}  &  =-c\frac{Z_{\pi}Z_{\eta_{S}}%
}{\sqrt{2}}\phi_{N}\eta_{N}\eta_{S}\text{,} \label{e}\\
\mathcal{L}_{\sigma_{N}\sigma_{S}}^{\text{KKMH}}  &  \sim \frac{c}{\sqrt{2}}%
\phi_{N}\sigma_{N}\sigma_{S}\text{,} \label{r}\\
\mathcal{V}(\phi_{N},\phi_{S})^{\text{KKMH}}  &  \sim -\frac{c}{2\sqrt{2}}%
\phi_{N}^{2}\phi_{S}\text{.} \label{t}%
\end{align}

From Eqs.\ (\ref{q}) - (\ref{w}) we can see that the KKMH term influences all
scalars and pseudoscalars except for $\bar{s}s$ states (and analogously to the
case of the Veneziano-Witten term, it induces the mass splitting between
$\vec{\pi}$ and $\eta_{N}$); the contributions from the KKMH term have the
same magnitude but opposite sign for pairs of states with the same parity but
different isospin ($\sigma_{N}$-$a_{0}$ and $\eta_{N}$-$\pi$) or with the same
isospin but different parity ($K_{S}$-$K$). Additionally, the KKMH term
influences not only the $\eta_{N}$-$\eta_{S}$ mixing but also the mixing of
$\sigma_{N}$ and $\sigma_{S}$. Note that the contribution to the potential
$\mathcal{V}(\phi_{N},\phi_{S})$, Eq.\ (\ref{t}), does not change the
conditions regarding the spontaneous symmetry breaking, i.e., the potential
with this contribution still yields Eqs.\ (\ref{m02}) and (\ref{l12}).
\end{itemize}

Although one might assume that the results regarding phenomenology to be
presented in this work will differ depending on the choice of the
chiral-anomaly term, this is not the case for the following reasons:
(\textit{i}) for the $\eta_{N}$-$\eta_{S}$ mixing, the constants $c$ and
$c_{1}$ are mutually dependent, as obvious from a comparison of Eqs.\ (\ref{t}%
) and (\ref{u}):%
\begin{equation}
c\frac{\phi_{N}}{\sqrt{2}}\equiv c_{1}\frac{\phi_{N}^{3}\phi_{S}}%
{2}\Leftrightarrow c\equiv c_{1}\frac{\phi_{N}^{2}\phi_{S}}{\sqrt{2}}\text{;}%
\end{equation}

(\textit{ii}) Eqs.\ (\ref{m02}) and (\ref{l12}) regarding the correct
implementation of the spontaneous breaking of the chiral symmetry\ are not
affected by the lack of presence of the chiral-anomaly term in the potential
and (\textit{iii}) for the $\sigma_{N}$-$\sigma_{S}$ mixing, the full mixing
term is the sum of Eqs.\ (\ref{sigma-sigma}) and (\ref{r}): $\mathcal{L}%
_{\sigma_{N}\sigma_{S}}^{\text{KKMH}}=(c/\sqrt{2}-2\lambda_{1}\phi_{N}%
)\phi_{S}\sigma_{N}\sigma_{S}$; the parameter $\lambda_{1}$ cannot be
calculated but only constrained from the linear combination $m_{0}^{2}%
+\lambda_{1}(\phi_{N}^{2}+\phi_{S}^{2})$ and therefore the value of
$\lambda_{1}$ can always be adjusted in such a way that it absorbs the
contribution from the parameter $c$ to the mixing: for this reason, the
results regarding the $\sigma_{N}$-$\sigma_{S}$ mixing are unaffected by our
choice of chiral-anomaly term. Of course, this may not be true for all the
decays that could in principle be calculated from our Lagrangian. The reason
is that the contributions of the full $c$ and $c_{1}$ terms to the Lagrangian
would not be the same (although both describe the same anomaly); we
nonetheless note that the chiral anomaly does not influence any other decays
that will be presented in this work other than those already mentioned. Then
for reason of simplicity we will be using the VW term in the $N_{f}=3$ version
of our model. Note that this form of the chiral-anomaly term allows us to
incorporate a pseudoscalar glueball field $\tilde{G}$ into our model in a very
simple way: the term $i\tilde{c}\tilde{G}(\det\Phi-\det\Phi^{\dagger})$
couples our (pseudo)scalar fields to $\tilde{G}$ with a single constant
$\tilde{c}$.

\chapter{The Fit Structure} \label{fitstructure}

As already mentioned, there are six free parameters: $Z_{\pi}$, $Z_{K}$,
$m_{1}^{2}$, $h_{2}$, $\delta_{S}$ and $\lambda_{2}$ and one parameter
combination: $m_{0}^{2}+\lambda_{1}(\phi_{N}^{2}+\phi_{S}^{2})$ that need to
be determined from our fit. Let us note at the beginning that these parameters
can be combined into three groups: (\textit{i}) parameters influencing only
(pseudo)scalar phenomenology: $m_{0}^{2}+\lambda_{1}(\phi_{N}^{2}+\phi_{S}%
^{2})$ and $\lambda_{2}$; (\textit{ii}) parameters influencing only
(axial-)vector phenomenology: $m_{1}^{2}$, $h_{2}$ and $\delta_{S}$ and
(\textit{iii}) parameters appearing in Lagrangian terms relevant for
the phenomenology of both (pseudo)scalars and (axial-)vectors: $Z_{\pi}$ and
$Z_{K}$. If there were no parameters from the group (\textit{iii}), then the
other parameters would neatly split into two independent groups thus
noticeably simplifying the fit; however, the presence of the renormalisation
coefficients $Z_{\pi}$ and $Z_{K}$ complicates the search for a fit
considerably. The following observables enter the fit:

\begin{itemize}
\item In the (pseudo)scalar sector, we consider all mass terms except for
$m_{\sigma_{N}}$ and $m_{\sigma_{S}}$. We do not consider the $\sigma$ masses
because the data regarding the $I(J^{PC})=0(0^{++})$ states is poor
\cite{PDG}, especially in the region under 1 GeV. Therefore, six mass terms
enter the fit, i.e., Eqs.\ (\ref{m_pi}) - (\ref{m_eta_N}) and (\ref{m_eta_S})
- (\ref{m_K}). Note that we first have to implement the mixing of the pure
non-strange state $\eta_{N}$ and the pure strange state $\eta_{S}$ that will
allow us to calculate $m_{\eta}$ and $m_{\eta^{\prime}}$.\ Note also that the
inclusion of $m_{a_{0}}$ into the fit forces us to consider two different
assignments of the isotriplet scalar state [given that the experimental data
ascertain the existence of two\ $I(J^{PC})=1(0^{++})$ states, $a_{0}(980)$ and
$a_{0}(1450)$]. This in turn means that we will have to consider two different
fits, each containing an assignment of the $\vec{a}_{0}$ state in our model to
the physical state in the region under 1 GeV (referred to as Fit I in the
following) or above 1 GeV (referred to as Fit II in the following). Note that
the scalar kaon state $K_{S}$ present in the model can also be assigned to a
physical state in two different ways: although the PDG confirmes the existence
of only one meson with quantum numbers $I(J^{P})=1/2(0^{+})$, the
$K_{0}^{\star}(1430)$ state, we will nonetheless also work with $K_{0}^{\star
}(800)$ or $\kappa$ in our Fit I. This is necessary as otherwise a scalar
isotriplet from the region under 1 GeV [$a_{0}(980)$] would enter the same fit
as the scalar kaon from the region above 1 GeV [$K_{0}^{\star}(1430)$] which
would be counter-intuitive because it would imply a large mass splitting
($\simeq400$ MeV) between a non-strange meson and a kaon whereas one would
expect the mass splitting to be $\simeq100$ MeV, i.e., close to the
strange-quark mass.

\item In the (axial-)vector channel, we consider all mass terms that follow
from the Lagrangian (\ref{Lagrangian}), i.e., Eqs.\ (\ref{m_rho}) - (\ref{m_K_1}).

\item Our result regarding the decay $a_{1}\rightarrow\pi\gamma$ from the
$N_{f}=2$ version of the model is still valid as the corresponding interaction
Lagrangian remains the same once the $U(3)\times U(3)$ version of the model is
considered. We have seen in Sec.\ \ref{sec.a1pg} that this decay width depends only on
$Z_{\pi}$ but does not constrain $Z_{\pi}$ very well due to lack of precise
experimental data. Still, the fit to be performed in the following sections
should also take this constraint into account (by producing $Z_{\pi}$
within the limits provided by $\Gamma_{a_{1}\rightarrow\pi\gamma}$). Note that
in principle $Z_{\pi}$ can also be determined from $\tau$-lepton decays
\cite{Anja}. However, the model presented in this work neglects weak
interactions and, for this reason, only $\Gamma_{a_{1}\rightarrow
\pi\gamma}$ will be used. 

\item Each of the two fits to be performed will have a decay width that only
depends on $h_{2}$. In the case of Fit I, this will be the decay width
$\Gamma_{f_{1N}\rightarrow a_{0}(980)\pi}$ and in the case of Fit II, this
will be the total decay width of $a_{0}(1450)$. This is analogous to Scenarios
I and II in the $U(2)\times U(2)$ version of the model.

\item We have already noted that there are three formulas for the
renormalisation coefficient $Z_{K}$: Eqs.\ (\ref{Z_K}), (\ref{Z_K1}) and
(\ref{Z_K2}). These equations have to be pairwise the same:%
\begin{align}
&  \frac{2m_{K_{1}}}{\sqrt{4m_{K_{1}}^{2}-g_{1}^{2}(\phi_{N}+\sqrt{2}\phi
_{S})^{2}}}\overset{!}{=}\frac{1}{f_{K}}\sqrt{\frac{m_{f_{1S}}^{2}%
-m_{\omega_{S}}^{2}}{g_{1}^{2}-h_{3}}}\text{,} \label{i}\\
&  \frac{2m_{K_{1}}}{\sqrt{4m_{K_{1}}^{2}-g_{1}^{2}(\phi_{N}+\sqrt{2}\phi
_{S})^{2}}}\overset{!}{=}\frac{m_{K_{1}}^{2}-m_{K^{\star}}^{2}}{Z_{\pi}f_{\pi
}f_{K}(g_{1}^{2}-h_{3})}\text{.} \label{o}%
\end{align}

However, each of the Eqs.\ (\ref{i}), (\ref{o}) contains mass terms
($m_{K^{\star}}$, $m_{\omega_{S}}$, $m_{K_{1}}$, $m_{f_{1S}}$) that are not
mere numbers but also themselves functions of the parameters $Z_{\pi}$, $Z_{K}$,
$m_{1}^{2}$, $h_{2}$ and $\delta_{S}$ [see Eqs.\ (\ref{m_K_star}),
(\ref{m_omega_S}), (\ref{m_K_1}) and (\ref{m_f1_S})]. Additionally, the
equations also contain parameters $g_{1}$ and $h_{3}$ [see Eqs.\ (\ref{g1})
and (\ref{h3})] that depend on $m_{\rho}$ and $m_{a_{1}}$, with the latter two
themselves mass terms depending on the parameters $Z_{\pi}$, $Z_{K}$, $m_{1}^{2}$
and $h_{2}$. Note also that the parameters $g_{1}$ and $h_{3}$ enter all of
the mass terms mentioned.

Therefore, Eqs.\ (\ref{i}) and (\ref{o}) actually represent a system of
implicit equations for the fit parameters:%
\begin{align}
&  \frac{2m_{K_{1}}[Z_{\pi},Z_{K},m_{1}^{2},h_{2},\delta_{S}]}{\sqrt
{4m_{K_{1}}^{2}[Z_{\pi},Z_{K},m_{1}^{2},h_{2},\delta_{S}]-g_{1}^{2}[Z_{\pi
},Z_{K},m_{1}^{2},h_{2}](\phi_{N}[Z_{\pi}]+\sqrt{2}\phi_{S}[Z_{K}])^{2}}%
}\nonumber\\
&  \overset{!}{=}\frac{1}{f_{K}}\sqrt{\frac{m_{f_{1S}}^{2}[Z_{\pi},Z_{K}%
,m_{1}^{2},h_{2},\delta_{S}]-m_{\omega_{S}}^{2}[Z_{\pi},Z_{K},m_{1}^{2}%
,h_{2},\delta_{S}]}{g_{1}^{2}[Z_{\pi},Z_{K},m_{1}^{2},h_{2}]-h_{3}[Z_{\pi
},Z_{K},m_{1}^{2},h_{2}]}}\text{,} \label{Z_K3}\\
&  \frac{2m_{K_{1}}[Z_{\pi},Z_{K},m_{1}^{2},h_{2},\delta_{S}]}{\sqrt
{4m_{K_{1}}^{2}[Z_{\pi},Z_{K},m_{1}^{2},h_{2},\delta_{S}]-g_{1}^{2}[Z_{\pi
},Z_{K},m_{1}^{2},h_{2}](\phi_{N}[Z_{\pi}]+\sqrt{2}\phi_{S}[Z_{K}])^{2}}%
}\nonumber\\
&  \overset{!}{=}\frac{m_{K_{1}}^{2}[Z_{\pi},Z_{K},m_{1}^{2},h_{2},\delta
_{S}]-m_{K^{\star}}^{2}[Z_{\pi},Z_{K},m_{1}^{2},h_{2},\delta_{S}]}{Z_{\pi
}f_{\pi}f_{K}(g_{1}^{2}[Z_{\pi},Z_{K},m_{1}^{2},h_{2}]-h_{3}[Z_{\pi}%
,Z_{K},m_{1}^{2},h_{2}])}\text{.} \label{Z_K4}%
\end{align}

Equations (\ref{Z_K3}) and (\ref{Z_K4}) have to be considered as well, and
represent an additional, strong constraint in the fit.
\end{itemize}

\textit{We therefore have 16 equations for seven unknowns [}$Z_{\pi}$\textit{,
}$Z_{K}$\textit{, }$m_{1}^{2}$\textit{, }$h_{2}$\textit{, }$\delta_{S}%
$\textit{, }$\lambda_{2}$\textit{ and }$m_{0}^{2}+\lambda_{1}(\phi_{N}%
^{2}+\phi_{S}^{2})$\textit{].}

Before we perform the fits, it is necessary to discuss the issue of the $\eta
$-$\eta^{\prime}$ mixing in the model given that the corresponding masses
enter the fit (and also determine the value of the parameter $c_{1}$, as we
have already mentioned).

\section{Mixing of \boldmath $\eta$ and \boldmath $\eta^{\prime}$} \label{sec.eta-eta}

The pure non-strange and strange fields $\eta_{N}$ and $\eta_{S}$ mix in the
Lagrangian (\ref{Lagrangian}):%
\begin{equation}
\mathcal{L}_{\eta_{N}\eta_{S}}=-c_{1}\frac{Z_{\eta_{S}}Z_{\pi}}{2}\phi_{N}%
^{3}\phi_{S}\eta_{N}\eta_{S}\text{.} \label{eta-eta}%
\end{equation}

The full $\eta_{N}$-$\eta_{S}$\ interaction Lagrangian has the form%
\begin{equation}
\mathcal{L}_{\eta_{N}\eta_{S,}\,\mathrm{full}}=\frac{1}{2}(\partial_{\mu}%
\eta_{N})^{2}+\frac{1}{2}(\partial_{\mu}\eta_{S})^{2}-\frac{1}{2}m_{\eta_{N}%
}^{2}\eta_{N}{}^{2}-\frac{1}{2}m_{\eta_{S}}^{2}\eta_{S}{}^{2}+z_{\eta}\eta
_{N}\eta_{S}\text{,} \label{eta-eta-1}%
\end{equation}

where $z_{\eta}$ is the mixing term of pure states $\eta_{N}\equiv(\bar
{u}u-\bar{d}d)/\sqrt{2}$ and $\eta_{S}\equiv\bar{s}s$.

Comparing Eqs.\ (\ref{eta-eta}) and (\ref{eta-eta-1}) we see that in the case
of our model Lagrangian the mixing term $z_{\eta}$ is%
\begin{equation}
z_{\eta}=-c_{1}\frac{Z_{\eta_{S}}Z_{\pi}}{2}\phi_{N}^{3}\phi_{S}\text{.}
\label{z1}%
\end{equation}

However, the mixing between the pure states $\eta_{N}$ and $\eta_{S}$ can be
equivalently expressed as the mixing between the octet state%
\begin{equation}
\eta_{8}=\sqrt{\frac{1}{6}}(\bar{u}u+\bar{d}d-2\bar{s}s)\equiv\sqrt{\frac
{1}{3}}\eta_{N}-\sqrt{\frac{2}{3}}\eta_{S} \label{eta_8}%
\end{equation}

and the singlet state%

\begin{equation}
\eta_{0}=\sqrt{\frac{1}{3}}(\bar{u}u+\bar{d}d+\bar{s}s)\equiv\sqrt{\frac{2}%
{3}}\eta_{N}+\sqrt{\frac{1}{3}}\eta_{S}\text{.} \label{eta_0}%
\end{equation}

We determine the physical states $\eta$ and $\eta^{\prime}$ as mixture
of the octet and singlet states with a mixing angle $\varphi_{P}$:%
\begin{equation}
\left(
\begin{array}
[c]{c}%
\eta\\
\eta^{\prime}%
\end{array}
\right)  =\left(
\begin{array}
[c]{cc}%
\cos\varphi_{P} & -\sin\varphi_{P}\\
\sin\varphi_{P} & \cos\varphi_{P}%
\end{array}
\right)  \left(
\begin{array}
[c]{c}%
\eta_{8}\\
\eta_{0}%
\end{array}
\right) \label{aaa}
\end{equation}

or, using Eqs.\ (\ref{eta_8}) and (\ref{eta_0}),

\begin{equation}
\left(
\begin{array}
[c]{c}%
\eta\\
\eta^{\prime}%
\end{array}
\right)  =\left(
\begin{array}
[c]{cc}%
\cos\varphi_{P} & -\sin\varphi_{P}\\
\sin\varphi_{P} & \cos\varphi_{P}%
\end{array}
\right)  \left(
\begin{array}
[c]{cc}%
\sqrt{\frac{1}{3}} & -\sqrt{\frac{2}{3}}\\
\sqrt{\frac{2}{3}} & \sqrt{\frac{1}{3}}%
\end{array}
\right)  \left(
\begin{array}
[c]{c}%
\eta_{8}\\
\eta_{0}%
\end{array}
\right)  \text{.}%
\end{equation}

If we introduce $\arcsin(\sqrt{2/3})=54.7456%
{{}^\circ}%
\equiv\varphi_{I}$, then the trigonometric addition formulas lead to

\begin{equation}
\left(
\begin{array}
[c]{c}%
\eta\\
\eta^{\prime}%
\end{array}
\right)  =\left(
\begin{array}
[c]{cc}%
\cos(\varphi_{P}+\varphi_{I}) & -\sin(\varphi_{P}+\varphi_{I})\\
\sin(\varphi_{P}+\varphi_{I}) & \cos(\varphi_{P}+\varphi_{I})
\end{array}
\right)  \left(
\begin{array}
[c]{c}%
\eta_{N}\\
\eta_{S}%
\end{array}
\right)  \text{.}%
\end{equation}

Defining the $\eta$-$\eta^{\prime}$ mixing angle $\varphi_{\eta}$
\begin{equation}
\varphi_{\eta}=-(\varphi_{P}+\varphi_{I})\text{,} \label{phi_eta}%
\end{equation}

we obtain

\begin{equation}
\left(
\begin{array}
[c]{c}%
\eta\\
\eta^{\prime}%
\end{array}
\right)  =\left(
\begin{array}
[c]{cc}%
\cos\varphi_{\eta} & \sin\varphi_{\eta}\\
-\sin\varphi_{\eta} & \cos\varphi_{\eta}%
\end{array}
\right)  \left(
\begin{array}
[c]{c}%
\eta_{N}\\
\eta_{S}%
\end{array}
\right)
\end{equation}

or in other words
\begin{align}
\eta &  =\cos\varphi_{\eta}\eta_{N}+\sin\varphi_{\eta}\eta_{S}\text{,} \label{eta}\\
\eta^{\prime}  &  =-\sin\varphi_{\eta}\eta_{N}+\cos\varphi_{\eta}\eta
_{S}\text{.} \label{etap}%
\end{align}

The interaction Lagrangian, Eq.\ (\ref{eta-eta-1}), contains only pure states
$\eta_{N}$ and $\eta_{S}$. Inverting Eqs.\ (\ref{eta}) and (\ref{etap})%

\begin{align}
\eta_{N}  &  =\cos\varphi_{\eta}\eta-\sin\varphi_{\eta}\eta^{\prime
}\text{,} \label{etaN}\\
\eta_{S}  &  =\sin\varphi_{\eta}\eta+\cos\varphi_{\eta}\eta^{\prime}\text{,}
\label{etaS}%
\end{align}

and substituting $\eta_{N}$ and $\eta_{S}$ by $\eta$ and $\eta^{\prime}$ in
Eq.\ (\ref{eta-eta-1}) we obtain%
\begin{align}
\mathcal{L}_{\eta\eta^{\prime}}  &  =\frac{1}{2}[(\partial_{\mu}\eta)^{2}%
(\cos\varphi_{\eta})^{2}+(\partial_{\mu}\eta^{\prime})^{2}(\sin\varphi_{\eta
})^{2}-\sin(2\varphi_{\eta})\partial_{\mu}\eta\partial^{\mu}\eta^{\prime
}]\nonumber\\
&  +\frac{1}{2}[(\partial_{\mu}\eta)^{2}(\sin\varphi_{\eta})^{2}%
+(\partial_{\mu}\eta^{\prime})^{2}(\cos\varphi_{\eta})^{2}+\sin(2\varphi
_{\eta})\partial_{\mu}\eta\partial^{\mu}\eta^{\prime}]\nonumber\\
&  -\frac{1}{2}m_{\eta_{N}}^{2}[\eta^{2}(\cos\varphi_{\eta})^{2}+(\eta
^{\prime})^{2}(\sin\varphi_{\eta})^{2}-\sin(2\varphi_{\eta})\eta\eta^{\prime
}]\nonumber\\
&  -\frac{1}{2}m_{\eta_{S}}^{2}[\eta^{2}(\sin\varphi_{\eta})^{2}+(\eta
^{\prime})^{2}(\cos\varphi_{\eta})^{2}+\sin(2\varphi_{\eta})\eta\eta^{\prime
}]\nonumber\\
&  +z_{\eta}\{[\eta^{2}-(\eta^{\prime})^{2}]\sin\varphi_{\eta}\cos
\varphi_{\eta}+\cos(2\varphi_{\eta})\eta\eta^{\prime}\}\nonumber\\
&  =\frac{1}{2}(\partial_{\mu}\eta)^{2}+\frac{1}{2}(\partial_{\mu}\eta
^{\prime})^{2}-\frac{1}{2}[m_{\eta_{N}}^{2}(\cos\varphi_{\eta})^{2}%
+m_{\eta_{S}}^{2}(\sin\varphi_{\eta})^{2}-z_{\eta}\sin(2\varphi_{\eta}%
)]\eta^{2}\nonumber\\
&  -\frac{1}{2}[m_{\eta_{N}}^{2}(\sin\varphi_{\eta})^{2}+m_{\eta_{S}}^{2}%
(\cos\varphi_{\eta})^{2}+z_{\eta}\sin ( 2\varphi_{\eta})](\eta^{\prime
})^{2}\nonumber\\
&  -\frac{1}{2}[(m_{\eta_{S}}^{2}-m_{\eta_{N}}^{2})\sin(2\varphi_{\eta
})-2z_{\eta}\cos(2\varphi_{\eta})]\eta\eta^{\prime}\text{.} \label{eta-eta-2}%
\end{align}

From Eq.\ (\ref{eta-eta-2}) we obtain the following relations for $m_{\eta}$
and $m_{\eta^{\prime}}$ in terms of the pure non-strange and strange mass
terms $m_{\eta_{N}}$ and $m_{\eta_{S}}$ [with the latter known from
Eqs.\ (\ref{m_eta_N}) and (\ref{m_eta_S}), respectively]:%

\begin{align}
m_{\eta}^{2}  &  =m_{\eta_{N}}^{2}\cos^{2}\varphi_{\eta}+m_{\eta_{S}}^{2}%
\sin^{2}\varphi_{\eta}-z_{\eta}\sin(2\varphi_{\eta})\text{,} \label{m_eta}\\
m_{\eta^{\prime}}^{2}  &  =m_{\eta_{N}}^{2}\sin^{2}\varphi_{\eta}+m_{\eta_{S}%
}^{2}\cos^{2}\varphi_{\eta}+z_{\eta}\sin(2\varphi_{\eta})\text{.}
\label{m_eta'}%
\end{align}

Additionally, assigning our fields $\eta$ and $\eta^{\prime}$ to physical (asymptotic) states requires that
the Lagrangian $\mathcal{L}_{\eta\eta^{\prime}}$ does not contain
any $\eta$-$\eta^{\prime}$ mixing terms and thus from Eq.\ (\ref{eta-eta-2})
we obtain%

\begin{equation}
z_{\eta}\overset{!}{=}(m_{\eta_{S}}^{2}-m_{\eta_{N}}^{2})\tan(2\varphi_{\eta
})/2\text{.} \label{z}%
\end{equation}

consequently, from Eqs.\ (\ref{z1}) and (\ref{z}) we obtain%

\begin{equation}
(m_{\eta_{S}}^{2}-m_{\eta_{N}}^{2})\tan(2\varphi_{\eta})=-c_{1}Z_{\eta_{S}%
}Z_{\pi}\phi_{N}^{3}\phi_{S}\text{.} \label{phi}%
\end{equation}

Given that $m_{\eta_{N}}$ and $m_{\eta_{S}}$ depend on $c_{1}$ [see
Eqs.\ (\ref{m_eta_N}) and (\ref{m_eta_S})], we obtain from Eq.\ (\ref{phi})
that%
\begin{align}
&  \left\{  Z_{\eta_{S}}^{2}\left[  m_{0}^{2}+\lambda_{1}\phi_{N}^{2}%
+(\lambda_{1}+\lambda_{2})\phi_{S}^{2}+c_{1}\frac{\phi_{N}^{4}}{4}\right]
-Z_{\pi}^{2}\left[  m_{0}^{2}+\left(  \lambda_{1}+\frac{\lambda_{2}}%
{2}\right)  \phi_{N}^{2}+\lambda_{1}\phi_{S}^{2}+c_{1}\phi_{N}^{2}\phi_{S}%
^{2}\right]  \right\} \nonumber\\
&  \times\tan(2\varphi_{\eta})=-c_{1}Z_{\eta_{S}}Z_{\pi}\phi_{N}^{3}\phi
_{S}\nonumber\\
&  \Leftrightarrow\left\{  (Z_{\eta_{S}}^{2}-Z_{\pi}^{2})[m_{0}^{2}%
+\lambda_{1}(\phi_{N}^{2}+\phi_{S}^{2})]+\lambda_{2}\left(  Z_{\eta_{S}}%
^{2}\phi_{S}^{2}-\frac{Z_{\pi}^{2}}{2}\phi_{N}^{2}\right)  +c_{1}\left(
\frac{Z_{\eta_{S}}^{2}}{4}\phi_{N}^{2}-Z_{\pi}^{2}\phi_{S}^{2}\right)
\phi_{N}^{2}\right\} \nonumber\\
&  \times\tan(2\varphi_{\eta})=-c_{1}Z_{\eta_{S}}Z_{\pi}\phi_{N}^{3}\phi
_{S}\nonumber\\
&  \Leftrightarrow c_{1}=\frac{4(Z_{\pi}^{2}-Z_{\eta_{S}}^{2})[m_{0}%
^{2}+\lambda_{1}(\phi_{N}^{2}+\phi_{S}^{2})]+2\lambda_{2}\left(  Z_{\pi}%
^{2}\phi_{N}^{2}-2Z_{\eta_{S}}^{2}\phi_{S}^{2}\right)  }{\left(  Z_{\eta_{S}%
}^{2}\phi_{N}^{2}-4Z_{\pi}^{2}\phi_{S}^{2}\right)  \phi_{N}^{2}\tan
(2\varphi_{\eta})+4Z_{\eta_{S}}Z_{\pi}\phi_{N}^{3}\phi_{S}}\tan(2\varphi
_{\eta})\text{.} \label{c1phi}%
\end{align}

Using Eq.\ (\ref{c1phi}) it is possible to calculate $c_{1}$ from the $\eta
$-$\eta^{\prime}$ mixing angle $\varphi_{\eta}$ (and vice versa) if the other
parameters are known. We will choose $\varphi_{\eta}$ such that our results
for $m_{\eta_{N}}$ and $m_{\eta_{S}}$ are as close as possible to their
experimental values; then $c_{1}$ is no longer a free parameter, as already
mentioned in the general discussion of the fit.

\chapter{Fit I: Scalars below 1 GeV} \label{sec.fitI}

Seven unknowns [$Z_{\pi}$\textit{, }$Z_{K}$\textit{, }$m_{1}^{2}$\textit{,
}$h_{2}$\textit{, }$\delta_{S}$\textit{, }$\lambda_{2}$\textit{ }and\textit{
}$m_{0}^{2}+\lambda_{1}(\phi_{N}^{2}+\phi_{S}^{2})$] enter the fit together
with 16 equations: for $m_{\pi}$, $m_{K}$, $m_{K_{S}}\equiv m_{K_{0}^{\star
}(800)\equiv\kappa}$, $m_{a_{0}}\equiv m_{a_{0}(980)}$, $m_{\eta}$,
$m_{\eta^{\prime}}$ [the latter two via Eqs.\ (\ref{m_eta}) and (\ref{m_eta'})
from $m_{\eta_{N}}$ and $m_{\eta_{S}}$], $m_{\rho}$, $m_{K^{\star}}$,
$m_{\omega_{S}}$, $m_{a_{1}}$, $m_{K_{1}}$, $m_{f_{1S}}$, $\Gamma
_{a_{1}\rightarrow\pi\gamma}$, $\Gamma_{f_{1N}\rightarrow a_{0}(980)\pi}$ (Fit
I) or $\Gamma_{a_{0}(1450)}$ (Fit II) as well as Eqs.\ (\ref{Z_K3}) and
(\ref{Z_K4}). Explicitly, the fit results should satisfy the following
equations (experimental central values from the PDG \cite{PDG}; at this point
we disregard the experimental uncertainties):%

\begin{align}
&  Z_{\pi}^{2}\left[  m_{0}^{2}+\lambda_{1}(\phi_{N}^{2}+\phi_{S}^{2}%
)+\frac{\lambda_{2}}{2}\phi_{N}^{2}\right]  =(139.57\text{ MeV})^{2}\equiv
m_{\pi}^{2}\text{,} \label{fit11}\\
&  Z_{K}^{2}\left[  m_{0}^{2}+\lambda_{1}(\phi_{N}^{2}+\phi_{S}^{2}%
)+\lambda_{2}\left(  \frac{\phi_{N}^{2}}{2}-\frac{\phi_{N}\phi_{S}}{\sqrt{2}%
}+\phi_{S}^{2}\right)  \right]  =(493.677\text{ MeV})^{2}\equiv m_{K}%
^{2}\text{,} \label{fit12}\\
&  Z_{K_{S}}^{2}\left[  m_{0}^{2}+\lambda_{1}(\phi_{N}^{2}+\phi_{S}%
^{2})+\lambda_{2}\left(  \frac{\phi_{N}^{2}}{2}+\frac{\phi_{N}\phi_{S}}%
{\sqrt{2}}+\phi_{S}^{2}\right)  \right]  =(676\text{ MeV})^{2}\equiv
m_{\kappa}^{2}\text{,} \label{fit13}\\
&  m_{0}^{2}+\lambda_{1}(\phi_{N}^{2}+\phi_{S}^{2})+\frac{3}{2}\lambda_{2}%
\phi_{N}^{2}=(980\text{ MeV})^{2}\equiv m_{a_{0}(980)}^{2}\text{,} \label{fit14}\\
&  Z_{\pi}^{2}\left[  m_{0}^{2}+\lambda_{1}(\phi_{N}^{2}+\phi_{S}^{2}%
)+\frac{\lambda_{2}}{2}\phi_{N}^{2}+c_{1}\phi_{N}^{2}\phi_{S}^{2}\right]
\cos^{2}\varphi_{\eta} \nonumber \\
& +Z_{\eta_{S}}^{2}\left[  m_{0}^{2}+\lambda_{1}(\phi
_{N}^{2}+\phi_{S}^{2})+\lambda_{2}\phi_{S}^{2}+c_{1}\frac{\phi_{N}^{4}}%
{4}\right]  \sin^{2}\varphi_{\eta}\nonumber\\
&  +c_{1}\frac{Z_{\eta_{S}}Z_{\pi}}{2}\phi_{N}^{3}\phi_{S}\sin(2\varphi_{\eta
})=(547.853\text{ MeV})^{2}\equiv m_{\eta}^{2}\text{,} \label{fit15}\\
&  Z_{\pi}^{2}\left[  m_{0}^{2}+\lambda_{1}(\phi_{N}^{2}+\phi_{S}^{2}%
)+\frac{\lambda_{2}}{2}\phi_{N}^{2}+c_{1}\phi_{N}^{2}\phi_{S}^{2}\right]
\sin^{2}\varphi_{\eta} \nonumber \\
& +Z_{\eta_{S}}^{2}\left[  m_{0}^{2}+\lambda_{1}(\phi
_{N}^{2}+\phi_{S}^{2})+\lambda_{2}\phi_{S}^{2}+c_{1}\frac{\phi_{N}^{4}}%
{4}\right]  \cos^{2}\varphi_{\eta}\nonumber\\
&  -c_{1}\frac{Z_{\eta_{S}}Z_{\pi}}{2}\phi_{N}^{3}\phi_{S}\sin(2\varphi_{\eta
})=(957.78\text{ MeV})^{2}\equiv m_{\eta^{\prime}}^{2}\text{,} \label{fit16}\\
&  m_{1}^{2}+(h_{2}+h_{3})\frac{\phi_{N}^{2}}{2}=(775.49\text{ MeV})^{2}\equiv
m_{\rho}^{2}\text{,} \label{fit17}\\
&  m_{1}^{2}+g_{1}^{2}\phi_{N}^{2}+(h_{2}-h_{3})\frac{\phi_{N}^{2}}%
{2}=(1230\text{ MeV})^{2}\equiv m_{a_{1}}^{2}\text{,} \label{fit18}\\
&  m_{1}^{2}+\delta_{S}+\left(  g_{1}^{2}+h_{2}\right)  \frac{\phi_{N}^{2}}%
{4}+\frac{1}{\sqrt{2}}(h_{3}-g_{1}^{2})\phi_{N}\phi_{S}+\left(  g_{1}%
^{2}+h_{2}\right)  \frac{\phi_{S}^{2}}{2}=(891.66\text{ MeV})^{2}\equiv
m_{K^{\star}}^{2}\text{,} \label{fit19}\\
&  m_{1}^{2}+\delta_{S}+\left(  g_{1}^{2}+h_{2}\right)  \frac{\phi_{N}^{2}}%
{4}+\frac{1}{\sqrt{2}}(g_{1}^{2}-h_{3})\phi_{N}\phi_{S}+\left(  g_{1}%
^{2}+h_{2}\right)  \frac{\phi_{S}^{2}}{2}=(1272\text{ MeV})^{2}\equiv
m_{K_{1}}^{2}\text{,} \label{fit110} \\
&  m_{1}^{2}+2\delta_{S}+\left(  h_{2}+h_{3}\right)  \phi_{S}^{2}%
=(1019.455\text{ MeV})^{2}\equiv m_{\omega_{S}}^{2}\text{,} \label{fit111}\\
&  m_{1}^{2}+2\delta_{S}+2g_{1}^{2}\phi_{S}^{2}+\left(  h_{2}-h_{3}\right)
\phi_{S}^{2}=(1426.4\text{ MeV})^{2}\equiv m_{f_{1S}}^{2}\text{,} \label{fit112}\\
&  \frac{e^{2}}{96\pi}(Z_{\pi}^{2}-1)m_{a_{1}}\left[  1-\left(  \frac{m_{\pi}%
}{m_{a_{1}}}\right)  ^{2}\right]  ^{3}=0.640\text{ MeV}\equiv\Gamma
_{a_{1}\rightarrow\pi\gamma}\text{,} \label{fit113}
\end{align}
\begin{align}
&  \frac{g_{1}^{2}Z_{\pi}^{2}}{16\pi}\frac{[m_{f_{1N}}^{4}-2m_{f_{1N}}%
^{2}\,(m_{a_{0}}^{2}+m_{\pi}^{2})+(m_{a_{0}}^{2}-m_{\pi}^{2})^{2}]^{3/2}%
}{m_{f_{1N}}^{5}m_{a_{1}}^{4}}\nonumber\\
&  \times\left[  m_{\rho}^{2}-\frac{1}{2}(h_{2}+h_{3})\phi_{N}^{2}\right]
^{2}=8.748\text{ MeV}\equiv\Gamma_{f_{1N}\rightarrow a_{0}(980)\pi}\text{,}
\label{fit114}%
\end{align}

and additionally the $Z_{K}$ Eqs.\ (\ref{Z_K3}) and (\ref{Z_K4}). We have set
$h_{1}=0=\delta_{N}$; note that $c_{1}=c_{1}(\varphi_{\eta})$ by
Eq.\ (\ref{c1phi}) and that we also use $\phi_{N}=Z_{\pi}f_{\pi}$ ($f_{\pi
}=92.4$ MeV), $\phi_{S}=Z_{K}f_{K}/\sqrt{2}$ ($f_{K}=155.5/\sqrt{2}$ MeV),
$g_{1}$ from Eq.\ (\ref{g1}), $h_{3}$ from Eq.\ (\ref{h3}), $Z_{K_{S}}$ from
Eq.\ (\ref{Z_K_S}) and $Z_{\eta_{S}}$ from Eq.\ (\ref{Z_eta_S}). Note also
that the mass terms present in Eqs.\ (\ref{g1}), (\ref{h3}), (\ref{Z_K_S}) and
(\ref{Z_eta_S}) are themselves functions of the parameters stated at the
beginning of this section.

Therefore, a comment is necessary before we proceed with the parameter
determination. Equations (\ref{fit11}) - (\ref{fit114}), (\ref{Z_K3}) and
(\ref{Z_K4}) could in principle be subject of a $\chi^{2}$ fit analogous to
the one performed in the $N_{f}=2$ version of our model, see
Sec.\ \ref{sec.scenarioI}. However, the $N_{f}=3$ version of the model
requires us to consider a significantly larger number of equations and
parameters. For this reason, a $\chi^{2}$ fit\ in $N_{f}=3$ is extremely
complicated to perform numerically, not least because the large number of
input equations would imply an extremely large number of local minima that
would have to be considered. In addition, the minima strongly depend on
the initial conditions of the parameters. To circumvent the technical issues
that a $\chi^{2}$ fit would bring about, we will use an iterative procedure
(described below) rather than a fit to determine the parameters of the model. The
procedure does not allow for errors to be determined (this would be possible
in a fit) but nonetheless the parameters determined in this way can in turn be
used as initial conditions in a $\chi^{2}$ fit ascertaining that a global
minimum has been found. Such a $\chi^{2}$ fit is currently under development \cite{Gyuri}
and it appears to assign small errors to our parameters (of the order of several percent).
Thus all results presented in this and the subsequent chapters \ref{ImplicationsFitI} -- \ref{ImplicationsFitII}
should be considered as possessing an error $\lesssim 10 \%$. 
Then we can refer to the
parameter determinations in the stated chapters as
"Fit I" and "Fit II".\\

The parameter values can be found iteratively in the following four steps:

\begin{itemize}
\item \textit{Step 1: }$Z_{\pi}$\textit{, }$Z_{K}$\textit{, (pseudo)scalar
parameters.} Solve the first four equations in the fit [Eqs.\ (\ref{fit11}) -
(\ref{fit14})] and determine $Z_{\pi}$ and $Z_{K}$ (among others).

\item \textit{Step 2: (axial-)vector masses.}\ Constrain the values of
$m_{\rho}$, $m_{a_{1}}$, $m_{K^{\star}}$, $m_{\omega_{S}}$, $m_{K_{1}}$ and
$m_{f_{1S}}$ via the $Z_{K}$ Eqs.\ (\ref{Z_K3}) and (\ref{Z_K4}).

\item \textit{Step 3: (axial-)vector parameters.} Calculate $h_{2}$,
$m_{1}^{2}$ and $\delta_{S}$ from the mass values determined in Step 2.

\item \textit{Step 4: }$\eta$\textit{-}$\eta^{\prime}$\textit{ mixing angle
}$\varphi_{\eta}$\textit{.} Calculate $\varphi_{\eta}$ from $m_{\eta}$ and
$m_{\eta^{\prime}}$; additionally, $c_{1}$ is calculated from
Eq.\ (\ref{c1phi}).
\end{itemize}

\begin{figure}[h]
  \begin{center}
    \begin{tabular}{cc}
      \resizebox{77mm}{!}{\includegraphics{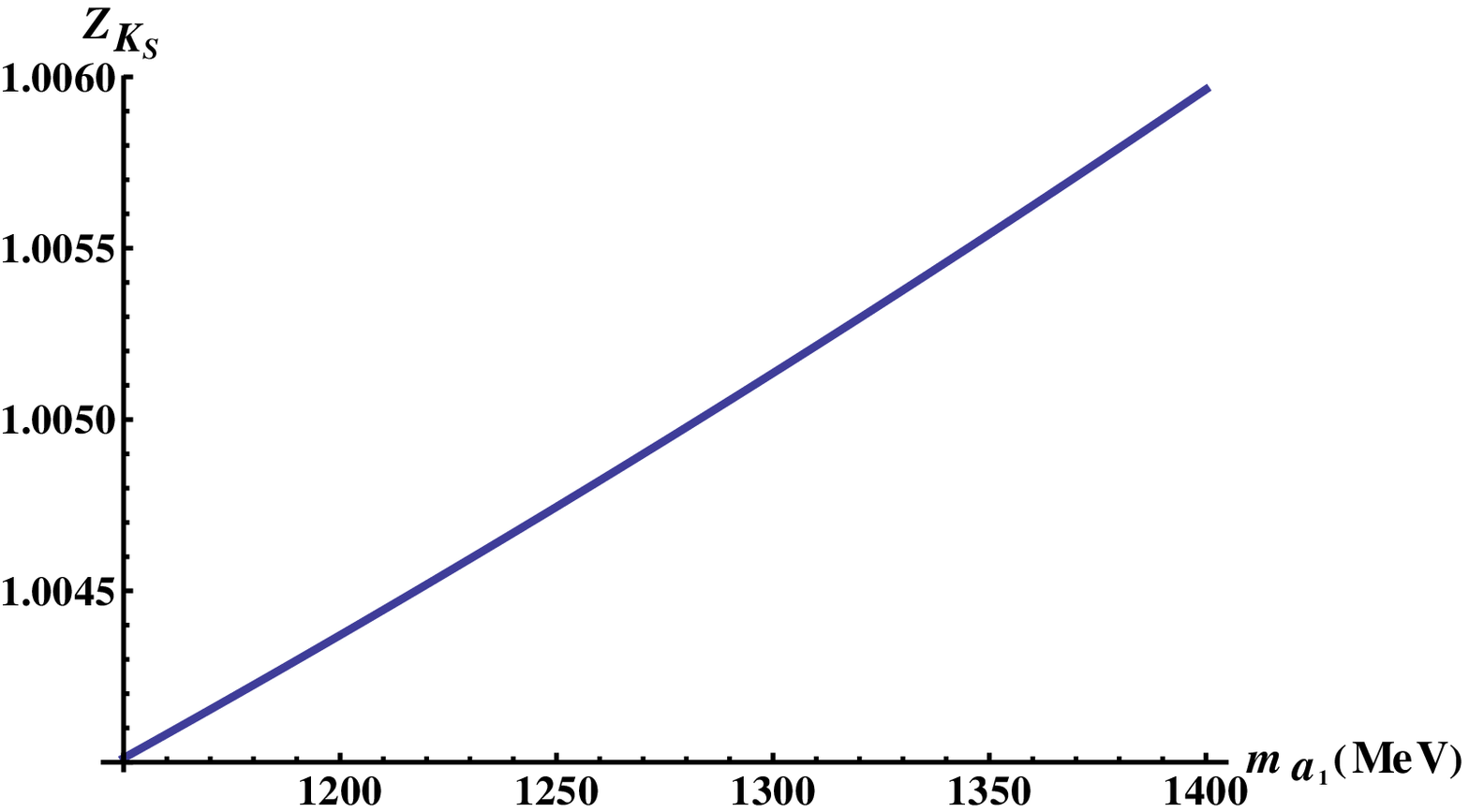}} &
      \resizebox{77mm}{!}{\includegraphics{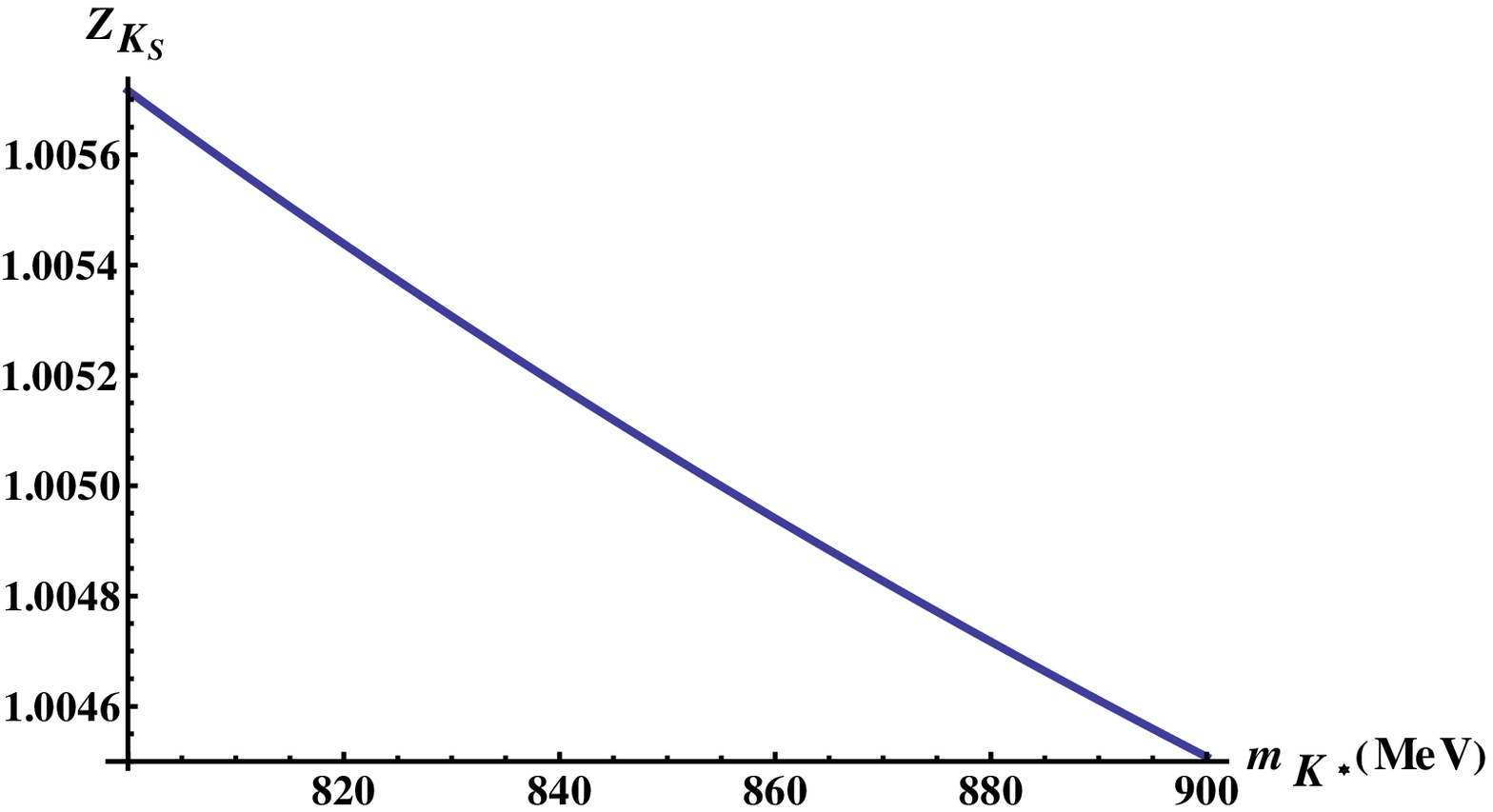}} 
    \end{tabular}
    \caption{Dependence of the renormalisation coefficient $Z_{K_{S}}$ on $m_{a_{1}}$
(left panel) and $m_{K^{\star}}$ (right panel). 
}
    \label{ZKS}
  \end{center}
\end{figure}

\textit{Step 1.} Let us note that the first four equations entering the fit,
i.e., Eqs.\ (\ref{fit11}) - (\ref{fit14}) contain four variables: $Z_{\pi}$,
$Z_{K}$, $\lambda_{2}$,\textit{ }$m_{0}^{2}+\lambda_{1}(\phi_{N}^{2}+\phi
_{S}^{2})$ if we assign certain values to $m_{a_{1}}$ and $m_{K^{\star}}$
(e.g., $m_{a_{1}}=1230$ MeV and $m_{K^{\star}}=891.66$ MeV \cite{PDG}) in
order for $Z_{K_{S}}$ to be calculated [subsequently, it is possible to choose
(axial-)vector parameters in the mass terms for $m_{a_{1}}$ and $m_{K^{\star}%
}$ -- see Eqs.\ (\ref{fit18}) and (\ref{fit19}) -- in such a way that the
values assigned are correct]. Note, however, that the starting values of
$m_{a_{1}}$ and $m_{K^{\star}}$\ are not strongly constrained because
$Z_{K_{S}}$ changes by approximately 0.1\% if $m_{a_{1}}$ and $m_{K^{\star}}$
are varied, see Fig.\ \ref{ZKS}. Indeed, experimental data allow for a rather
large interval in particular of $m_{a_{1}}$ because $a_{1}(1260)$ is a very
broad resonance, with $\Gamma_{a_{1}(1260)}=(250-600)$ MeV \cite{PDG}.%

We thus obtain a system of four equations with four unknowns. This equation
system can be solved exactly with a numerical analysis yielding the following
parameter values:%
\begin{align*}
Z_{\pi}  &  =0.31\text{,} \\
Z_{K}  &  =0.51\text{,} \\
\lambda_{2}  &  =931\text{,} \\
m_{0}^{2}+\lambda_{1}(\phi_{N}^{2}+\phi_{S}^{2})  &  =-172665\text{ MeV}%
^{2}\text{.}%
\end{align*}

Unfortunately, the stated solutions cannot be used because that would imply
$Z_{\pi}<1$ and $Z_{K}<1$ that cannot be true due to the definitions of
$Z_{\pi}$, Eq.\ (\ref{Z_pi}), and $Z_{K}$, Eq.\ (\ref{Z_K}) as otherwise one
would have to allow either for imaginary scalar-vector coupling $g_{1}$ in the
Lagrangian (\ref{Lagrangian}) or for imaginary condensates $\phi_{N,S}$.
Therefore, we have to consider other (approximate) solutions of
Eqs.\ (\ref{fit11}) - (\ref{fit14}). A numerical analysis leads to the
parameter values shown in Table \ref{Fit1-1}.%

\begin{table}[h] \centering
\begin{tabular}
[c]{|c|c|c|c|}\hline
Parameter & Value & Observable & Value [MeV]\\\hline
$Z_{\pi}$ & $1.38$ & $m_{\pi}$ & $138.04$\\\hline
$Z_{K}$ & $1.39$ & $m_{K}$ & $490.84$\\\hline
$\lambda_{2}$ & $58.5$ & $m_{a_{0}(980)}$ & $978$\\\hline
$m_{0}^{2}+\lambda_{1}(\phi_{N}^{2}+\phi_{S}^{2})$ & $-463425$ MeV$^{2}$ &
$m_{\kappa}$ & $1129$\\\hline
\end{tabular}%
\caption{Best solutions of Eqs.\ (\ref{fit11}) - (\ref{fit14}%
) under the conditions $Z_\pi\overset{!}{>}1$, $Z_K \overset{!}{>}1$.\label
{Fit1-1}}%
\end{table}%

The parameters produce an excellent agreement with all input masses except
$m_{\kappa}$ where the value from the fit is almost by a factor of two larger
than the PDG value $m_{\kappa}^{\exp}=(676\pm40)$ MeV. However, we note that
the $\kappa$ resonance is very broad [$\Gamma_{\kappa}^{\exp}=(548\pm24)$ MeV]
and therefore we will, for the moment, disregard the large mass difference
between the fit result and the PDG value. Additionally, $\Gamma_{a_{1}%
\rightarrow\pi\gamma}=0.322$ MeV is obtained from the parameter values in Table
\ref{Fit1-1}, slightly smaller than the lower boundary on this decay width
cited by the PDG to be $0.394$ MeV.\newline

\textit{Step 2.} Let us now turn to the parameters in (axial-)vector mass terms.
The most convenient way to proceed is to first determine the values of vector
and axial-vector masses that lead to the pairwise equality of the three
$Z_{K}$ formulas, Eqs.\ (\ref{Z_K3}) and (\ref{Z_K4}). Note that the
calculation involving Eqs.\ (\ref{Z_K3}) and (\ref{Z_K4}) requires knowledge
of $Z_{\pi}$ and $Z_{K}$ (see Table \ref{Fit1-1}) and also of $m_{\rho}$,
$m_{a_{1}}$, $m_{K^{\star}}$, $m_{\omega_{S}}$, $m_{K_{1}}$ and $m_{f_{1S}}$.
We start with the PDG values of all masses except $m_{a_{1}}$ (as already
mentioned, the variation of $m_{a_{1}}$ is experimentally allowed by the large
decay width of this resonance and it does not lead to an inconsistency with
the determination of scalar parameters in Table \ref{Fit1-1}). We then look for
conditions under which the pairwise equality of the three $Z_{K}$ formulas can
be obtained. Unfortunately, the mentioned equality does not exist for the PDG
values of masses. We therefore alternate all the mass values (holding all
masses except $m_{a_{1}}$ as close as possible to their respective
experimental values) until the pairwise equality of the three $Z_{K}$ formulas
has been reached. In this way we obtain (axial-)vector masses as follows:%

\begin{align*}
m_{a_{1}}  &  =1396\text{ MeV, }m_{\rho}=775.49\text{ MeV, }m_{K^{\star}%
}=832.53\text{ MeV,}\\
m_{\omega_{S}}  &  =870.35\text{ MeV, }m_{K_{1}}=1520\text{ MeV, }m_{f_{1S}%
}=1643.4\text{ MeV.}%
\end{align*}

\textit{Step 3.} Once the values of the (axial-)vector masses are known, then
the (axial-)vector fit parameters are determined in such a way that the mass
values determined by the three $Z_{K}$ formulas are reproduced. Note, however,
that this does not require for many parameters to be calculated: $h_{2}$ is
already known from $\Gamma_{f_{1N}\rightarrow a_{0}(980)\pi}$; $g_{1}$ and
$h_{3}$ are determined from $m_{\rho}$ and $m_{a_{1}}$ [see Eqs.\ (\ref{g1})
and (\ref{h3})] and consequently we need to calculate only the values of
$m_{1}^{2}$ and $\delta_{S}$.

As already mentioned in Sec.\ \ref{fitstructure}, it is possible to calculate
the parameter $h_{2}$\ via $\Gamma_{f_{1N}\rightarrow a_{0}(980)\pi}$,
Eq.\ (\ref{fit114}).\ In Scenario I of the two-flavour version of the model
(Sec.\ \ref{sec.scenarioI}) we have seen that in this way two sets of $h_{2}$ values arise, a set
of relatively lower and a set of relatively higher values, see Eq.\ (\ref{h2Z}). We have
also seen that the set of relatively lower $h_{2}$ values does not yield a
correct value of the $a_{0}(980)\rightarrow\eta\pi$ decay amplitude.
Therefore, we are also in this case naturally inclined to use the set of
higher $h_{2}$ values, i.e., $h_{2}\sim80$. However, the only way to obtain a
fit in this case is to allow for negative values of $m_{1}^{2}$ and
$\delta_{S}$ (see Table \ref{Fit1-2}). However, such a fit could not be
considered physical as it would imply an imaginary vector meson mass in the
chirally restored phase.%

\begin{table}[h] \centering
\begin{tabular}
[c]{|c|c|}\hline
Parameter & Value\\\hline
$m_{1}^{2}$ & $-697^{2}$ MeV$^2$\\\hline
$\delta_{S}$ & $-404^{2}$ MeV$^2$\\\hline
$h_{2}$ & $161$\\\hline
\end{tabular}%
\caption{(Axial-)vector parameters from Eqs.\ (\ref{fit17}) - (\ref
{fit112}) using the higher set of
$h_2$ values from Eq.\ (\ref{fit114}%
). The parameter $h_2$ has a rather large value due to the large value of
$m_{a_1}$, constrained from the $Z_K$ formulas (\ref{Z_K3}) and (\ref
{Z_K4}).  \label{Fit1-2}}%
\end{table}%

For these reasons, we have to use the smaller set of $h_{2}$ values [and later
ascertain whether it is still possible to obtain a correct value of the
$a_{0}(980)\rightarrow\eta\pi$ decay amplitude, see Sec.\ \ref{sec.a0etapion}%
]. In this case, the fit yields positive values of $m_{1}^{2}$ and $\delta
_{S}$.%

\begin{table}[h] \centering
\begin{tabular}
[c]{|c|c|}\hline
Parameter & Value\\\hline
$m_{1}^{2}$ & $697^{2}$ MeV$^2$\\\hline
$\delta_{S}$ & $229^{2}$ MeV$^2$\\\hline
$h_{2}$ & $40.6$\\\hline
\end{tabular}%
\caption{(Axial-)vector parameters from Eqs.\ (\ref{fit17}) - (\ref
{fit112}) using the lower set of
$h_2$ values from Eq.\ (\ref{fit114}).  \label{Fit1-3}}%
\end{table}%

\textit{Step 4.} Using Eqs.\ (\ref{m_eta}) and (\ref{m_eta'}) we can calculate
the $\eta$-$\eta^{\prime}$ mixing angle $\varphi_{\eta}$ under the conditions
that $m_{\eta}$ and $m_{\eta^{\prime}}$ are as close as possible to their
respective experimental values. Additionally, we also require $\varphi_{\eta
}<\mid45%
{{}^\circ}%
\mid$ as otherwise we would have the (counter-intuitive) ordering $m_{\eta
_{S}}<m_{\eta_{N}}$. Under the latter condition it is actually not possible to
exactly obtain the experimental value of $m_{\eta}$ but rather a slightly
lower one: $m_{\eta}=517.13$ MeV. We obtain also $m_{\eta^{\prime}}=957.78$
MeV $\equiv m_{\eta^{\prime}}^{\exp}$ with $\varphi_{\eta}=-42^{\circ}$;
Eq.\ (\ref{c1phi}) yields $c_{1}=0.0015$ MeV$^{-2}$.

Table \ref{Fit1-4} shows results for all parameters from Fit I.%

\begin{table}[h] \centering
\begin{tabular}
[c]{|c|c|c|c|}\hline
Parameter & Value & Parameter & Value\\\hline
$Z_{\pi}$ & $1.38$ & $g_{1}$, Eq.\ (\ref{g1}) & $7.54$\\\hline
$Z_{K}$ & $1.39$ & $g_{2}$, Eq.\ (\ref{g2Z}) & $-11.2$\\\hline
$\lambda_{2}$ & $58.5$ & $h_{3}$, Eq.\ (\ref{h3}) & $-26.3$\\\hline
$m_{0}^{2}+\lambda_{1}(\phi_{N}^{2}+\phi_{S}^{2})$ & $-463425$ MeV$^{2}$ &
$h_{0N}$, Eq.\ (\ref{m_pi}) & $1.279\cdot10^{6}$ MeV$^{3}$\\\hline
$m_{1}$ & $697$ MeV & $h_{0S}$, Eq.\ (\ref{m_eta_S}) & $3.443\cdot
10^{7}$ MeV$^{3}$\\\hline
$\delta_{S}$ & $229^{2}$ MeV$^2$ & $h_{1}$ & $0$\\\hline
$h_{2}$ & $40.6$ & $\delta_{N}$ & $0$\\\hline
$c_{1}$ & $0.0015$ MeV$^{-2}$ & $g_{3,4,5,6}$ & $0$\\\hline
\end{tabular}%
\caption{Best values of parameters from Fit I (experimental uncertainties are omitted).  \label{Fit1-4}}%
\end{table}%

Table \ref{Fit1-5} shows the results for all observables from Fit I.
Note that the implemented iterative calculation of the parameters does not allow
for an error determination and thus we also do not cite experimental errors in
Table \ref{Fit1-5}. Additionally, some mass values (e.g., $m_{\pi}$ and
$m_{K}$) are known very precisely (up to several decimals), i.e., the
corresponding errors are very small. Our model does not aim to reproduce these
mass values to such a high precision -- it suffices to reproduce the
experimental masses sufficiently closely. Then our results for some masses
[such as $m_{a_{0}(980)}$] will be within errors, others will not ($m_{\pi}$
and $m_{K}$) but they will still be sufficiently close to the experimental
result (within several MeV) rendering them acceptable.

Nonetheless, the proximity of our results to the experiment is actually not
accomplished very well at this point (because the underlying assumption of
scalar $\bar{q}q$ states below $1$ GeV is generally disfavoured by our model,
see below) -- we will see that the correspondence of our results with the data
is significantly improved once the scalar $\bar{q}q$ states are assumed to be above
$1$ GeV, see Table \ref{Fit2-5}.

\begin{table}[h] \centering
\begin{tabular}
[c]{|c|c|c|}\hline
Observable & Our Value [MeV] & Experimental Value [MeV]\\\hline
$m_{\pi}$ & $138.04$ & $139.57$\\\hline
$m_{K}$ & $490.84$ & $493.68$\\\hline
$m_{a_{0}(980)}$ & $978.27$ & $980$\\\hline
$m_{\kappa}$ & $1128.7$ & $676$\\\hline
$m_{\eta}$ & $517.13$ & $547.85$\\\hline
$m_{\eta^{\prime}}$ & $957.78$ & $957.78$\\\hline
$m_{\rho}$ & $775.49$ & $775.49$\\\hline
$m_{a_{1}}$ & $1396$ & $1230$\\\hline
$\text{ }m_{K^{\star}}$ & $832.53$ & $891.66$\\\hline
$m_{\omega_{S}}\text{ }$ & $870.35$ & $1019.46$\\\hline
$m_{K_{1}}$ & $1520$ & $1272$\\\hline
$m_{f_{1S}}$ & $1643.4$ & $1426.4$\\\hline
$\Gamma_{a_{1}\rightarrow\pi\gamma}$ & $0.369$ & $0.640$\\\hline
$\Gamma_{f_{1N}\rightarrow a_{0}(980)\pi}$ & $8.748$ & $8.748$\\\hline
\end{tabular}%
\caption{Observables from Fit I. \label{Fit1-5}}%
\end{table}%

We observe from Table \ref{Fit1-5} that, in addition to a rather large value
of $m_{\kappa}$, the fit also yields too large values of $m_{a_{1}}$ and
$m_{f_{1S}}$. The $a_{1}(1260)$ resonance is very broad: $\Gamma_{a_{1}%
(1260)}^{\exp}=(250-600)$ MeV \cite{PDG} and thus the discrepancy between our
and the experimental results is not too serious; however, the $f_{1}%
(1420)\equiv f_{1S}$ resonance is much narrower [$\Gamma_{f_{1}(1420)}^{\exp
}=(54.9\pm2.6)$ MeV] and therefore, in this case, the discrepancy with the
experimental value is rather large. The same holds for the $\omega_{S}%
\equiv\varphi(1020)$ resonance, a sharp peak with a width of $4.26\pm0.04$ MeV
\cite{PDG} and also for $K^{\star}(892)$, although for the latter resonance
the discrepancy with the experimental mass is of the order of the decay width,
i.e., $(50.8\pm0.9)$ MeV. Note that the discrepancy between the fit value and
experimental result is also very large for our $K_{1}$ resonance; however,
this can be amended by assigning the $K_{1}$ state in the model to the
$K_{1}(1400)$ resonance rather than to $K_{1}(1270)$. Data regarding the
former resonance suggest $m_{K_{1}(1400)}=(1403\pm7)$ MeV and $\Gamma
_{K_{1}(1400)}=(174\pm13)$ MeV and then the discrepancy between our value
$m_{K_{1}}=1520$ MeV and the experimental result is smaller than the value of
the $K_{1}(1400)$ decay width. [Note, however, that the stated correspondence
to $K_{1}(1400)$ is actually in itself problematic because axial-vector kaons
are expected to mix, see Sec.\ \ref{2K1}. The mixing of the $K_{1}$\ states is
well-established \cite{PDG,Goldman1998,Godfrey,Lipkin,Cheng:2011,Cheng:2003};
thus the absence of the mixing within Fit I represents another discrepancy
with experiment.]

Note that the results also imply $m_{1}=697$ MeV, i.e., non-quark
contributions are favoured to play a decisive role in the $\rho$ mass generation.

\chapter{Implications of Fit I} \label{ImplicationsFitI}

Despite some discrepancies between results stemming from the fit and
experimental data, we will proceed with calculations of hadronic decay widths
in scalar and axial-vector channels (as these channels possess the most
ambiguities regarding not only the decay widths but also regarding the
structure of resonances).

\section{Phenomenology in the \boldmath $I(J^{PC})=0(0^{++})$ Channel}

As apparent from Eqs.\ (\ref{m_sigma_N}) and (\ref{m_sigma_S}), the masses of the
strange and non-strange sigma fields, $m_{\sigma_{N}}$ and $m_{\sigma_{S}}$,
depend on $m_{0}^{2}+3\lambda_{1}\phi_{N}^{2}+\lambda_{1}\phi_{S}^{2}$ and
$m_{0}^{2}+\lambda_{1}\phi_{N}^{2}+3\lambda_{1}\phi_{S}^{2}$, respectively,
and thus cannot be calculated with the knowledge of the parameter combination
$m_{0}^{2}+\lambda_{1}(\phi_{N}^{2}+\phi_{S}^{2})$ stated in Table
\ref{Fit1-4}. However, if the linear combination $m_{0}^{2}+\lambda_{1}%
(\phi_{N}^{2}+\phi_{S}^{2})$ is known, then the parameter $\lambda_{1}$ can be
expressed in terms of the mass parameter $m_{0}^{2}$\ (given that $Z_{\pi}%
$\ and $Z_{K}$ are also known). Nonetheless, this is not satisfactory because
it does not allow us to constrain the masses and decay widths of the two
$I(J^{PC})=0(0^{++})$ resonances present in the model. In
the next two subsections we will therefore derive a constraint on $m_{0}^{2}$ and $\lambda_{1}$,
using the spontaneous breaking of chiral symmetry. We will discuss
conditions under which the vacuum potential $\mathcal{V}(\phi_{N},\phi_{S})$
arising from the Lagrangian (\ref{Lagrangian}) allows for the Spontaneous
Symmetry Breaking (SSB) to occur while having the correct behaviour in the
limit of large values of condensates $\phi_{N}$ and $\phi_{S}$ ($\lim
_{\phi_{N,S}\rightarrow\infty}\mathcal{V}(\phi_{N},\phi_{S})\rightarrow\infty$).

\subsection{A Necessary Condition for the Spontaneous Symmetry Breaking}

Calculating the elements of the Hesse matrix from the potential $\mathcal{V}%
(\phi_{N},\phi_{S})$ with respect to the condensates $\phi_{N}$ and $\phi
_{S}$ yields:%

\begin{align}
\frac{\partial^{2}\mathcal{V}(\phi_{N},\phi_{S})}{\partial\phi_{N}^{2}}  &
=m_{0}^{2}+\lambda_{1}(3\phi_{N}^{2}+\phi_{S}^{2})+\frac{3}{2}\lambda_{2}%
\phi_{N}^{2}\text{,} \label{d2VN}\\
\frac{\partial^{2}\mathcal{V}(\phi_{N},\phi_{S})}{\partial\phi_{S}^{2}}  &
=m_{0}^{2}+\lambda_{1}\phi_{N}^{2}+3(\lambda_{1}+\lambda_{2})\phi_{S}%
^{2}\text{,} \label{d2VS}\\
\frac{\partial^{2}\mathcal{V}(\phi_{N},\phi_{S})}{\partial\phi_{N}\partial
\phi_{S}}  &  =2\lambda_{1}\phi_{N}\phi_{S}\text{.} \label{d2VNS}%
\end{align}

From Eqs.\ (\ref{d2VN}) - (\ref{d2VS}) we obtain the following form of the
Hesse matrix in the limit $\phi_{N}=\phi_{S}=0$:%

\[
H(m_{0}^{2})=\left(
\begin{array}
[c]{cc}%
m_{0}^{2} & 0\\
0 & m_{0}^{2}%
\end{array}
\right)
\]

and the vacuum is unstable only if the Hesse matrix has
negative eigenvalues or in other words%
\begin{equation}
m_{0}^{2}\overset{!}{<}0\text{.} \label{m02}%
\end{equation}
This is a necessary condition for the Spontaneous Symmetry Breaking to occur.
However, we still need to ascertain whether the potential $\mathcal{V}%
(\phi_{N},\phi_{S})$ from Eq.\ (\ref{V}) has the right behaviour in the limit
$\phi_{N,S}\rightarrow\infty$. This will be verified in the following subsection.

\subsection{A Condition for \boldmath $\lambda_{1,2}$ from SSB}

Let us isolate the quartic terms from the potential $\mathcal{V}(\phi_{N}%
,\phi_{S})$, Eq.\ (\ref{V}), in the following expression $\mathcal{V}_{4}(\phi
_{N},\phi_{S})$:%

\begin{equation}
\mathcal{V}_{4}(\phi_{N},\phi_{S})=\frac{\lambda_{1}}{4}(\phi_{N}^{4}%
+2\phi_{N}^{2}\phi_{S}^{2}+\phi_{S}^{4})+\frac{\lambda_{2}}{4}\left(
\frac{\phi_{N}^{4}}{2}+\phi_{S}^{4}\right)  \text{.} \label{V4}%
\end{equation}

The quadratic terms in the potential $\mathcal{V}(\phi_{N},\phi_{S})$
represent a negative-sign contribution due to the condition $m_{0}^{2}%
\overset{!}{<}0$.\ Thus, a correct implementation of the Spontaneous Symmetry
Breaking requires that the quartic term $\mathcal{V}_{4}(\phi_{N},\phi_{S})$
is a positive-sign contribution to $\mathcal{V}(\phi_{N},\phi_{S})$ because
otherwise the potential $\mathcal{V}(\phi_{N},\phi_{S})$ would not exhibit
minima. Let us now define the variables $x_{\sigma}\equiv\phi_{N}^{2}$ and
$y_{\sigma}\equiv\phi_{S}^{2}$, bringing $\mathcal{V}_{4}(\phi_{N},\phi_{S})$,
Eq.\ (\ref{V4}), to the following form:%
\begin{equation}
\mathcal{V}_{4}(\phi_{N},\phi_{S})=\frac{2\lambda_{1}+\lambda_{2}}{8}%
x_{\sigma}^{2}+\frac{\lambda_{1}+\lambda_{2}}{4}y_{\sigma}^{2}+\frac
{\lambda_{1}}{2}x_{\sigma}y_{\sigma}\text{ \ \ \ \ }(x_{\sigma}\geq
0,y_{\sigma}\geq0)\text{.} \label{V41}%
\end{equation}
Obviously the conditions $2\lambda_{1}+\lambda
_{2}\overset{!}{>}0\wedge\lambda_{1}+\lambda_{2}$ $\overset{!}{>}0$ have to be
satisfied. In other words:%
\begin{align}
\lambda_{1}\overset{!}{>}-\frac{\lambda_{2}}{2}\text{ for }\lambda_{2}  &
>0\label{l121}\\
\lambda_{1}\overset{!}{>}-\lambda_{2}\text{ for }\lambda_{2}  &  <0\text{.}
\label{l122}%
\end{align}
Additionally,\ we have to ascertain that $\mathcal{V}_{4}(\phi_{N},\phi_{S})$
is a positive-sign contribution to $\mathcal{V}(\phi_{N},\phi_{S})$ in all
directions of the condensates. In order to verify that this is fulfilled, we
set $y_{\sigma}\equiv\eta_{\sigma}x_{\sigma}$ ($\eta_{\sigma}\geq0$) yielding
the following form of $\mathcal{V}_{4}(\phi_{N},\phi_{S})$:%
\begin{equation}
\mathcal{V}_{4}(\phi_{N},\phi_{S})=a_{\sigma}x_{\sigma}^{2}+b_{\sigma}%
\eta_{\sigma}^{2}x_{\sigma}^{2}+c_{\sigma}\eta_{\sigma}x_{\sigma}^{2}\equiv
f_{\sigma}(\eta_{\sigma})x_{\sigma}^{2} \label{V42}%
\end{equation}

with $a_{\sigma}\equiv(2\lambda_{1}+\lambda_{2})/8\overset{!}{>}0$,
$b_{\sigma}\equiv(\lambda_{1}+\lambda_{2})/4\overset{!}{>}0$, $c_{\sigma
}\equiv\lambda_{1}/2$ and $f_{\sigma}(\eta_{\sigma})\equiv b_{\sigma}%
\eta_{\sigma}^{2}+c_{\sigma}\eta_{\sigma}+a_{\sigma}\overset{!}{>}0$. The
latter can be written in the following way:%
\begin{equation}
f_{\sigma}(\eta_{\sigma})\equiv b_{\sigma}\left(  \eta_{\sigma}+\frac
{c_{\sigma}}{2b_{\sigma}}\right)  ^{2}+\left(  a_{\sigma}-\frac{c_{\sigma}%
^{2}}{4b_{\sigma}}\right)  \text{.} \label{fsigma}%
\end{equation}

Thus, additionally to the already stated condition $b_{\sigma}\overset{!}{>}0$
we also need to ascertain that $a_{\sigma}-\frac{c_{\sigma}^{2}}{4b_{\sigma}%
}\overset{!}{>}0$ in order for $f_{\sigma}(\eta_{\sigma})\overset{!}{>}0$ to
be fulfilled. Consequently, we obtain%
\begin{equation}
a_{\sigma}>\frac{c_{\sigma}^{2}}{4b_{\sigma}}\Rightarrow c_{\sigma}%
<2\sqrt{a_{\sigma}b_{\sigma}}%
\end{equation}

or in other words%
\begin{equation}
\frac{\lambda_{1}}{2}<2\sqrt{\frac{(2\lambda_{1}+\lambda_{2})(\lambda
_{1}+\lambda_{2})}{32}}=\frac{1}{2}\sqrt{\left(  \lambda_{1}+\frac{\lambda
_{2}}{2}\right)  (\lambda_{1}+\lambda_{2})}\Leftrightarrow\lambda_{1}%
<\sqrt{\left(  \lambda_{1}+\frac{\lambda_{2}}{2}\right)  (\lambda_{1}%
+\lambda_{2})}\text{.} \label{fsigma1}%
\end{equation}

The square root on the right-hand side of inequality (\ref{fsigma1}) is well
defined due to the already stated conditions (\ref{l121}) and (\ref{l122}).
For $\lambda_{1}<0$, only the condition (\ref{l121}), i.e., $\lambda
_{1}\overset{!}{>}-\lambda_{2}/2$ and $\lambda_{2}>0$ can be fulfilled.
Consequently,%
\begin{equation}
\frac{-\lambda_{2}}{2}<\lambda_{1}<0\text{ and }\lambda_{2}>0\text{.}
\label{fsigma2}%
\end{equation}

For $\lambda_{1}>0$, the square of the inequality (\ref{fsigma1}) yields%
\begin{equation}
\lambda_{1}^{2}<\lambda_{1}^{2}+\frac{3}{2}\lambda_{1}\lambda_{2}%
+\frac{\lambda_{2}^{2}}{2}\Leftrightarrow0<\lambda_{2}(3\lambda_{1}%
+\lambda_{2})\Leftrightarrow\left\{
\begin{tabular}
[c]{l}%
$\lambda_{2}<0\wedge\lambda_{1}<-\lambda_{2}/3$\\
\lbrack contradiction to $\lambda_{1}\overset{!}{>}-\lambda_{2}$\\
from condition (\ref{l122})]\\
$\lambda_{2}>0\wedge\lambda_{1}>-\lambda_{2}/3$\\
(fulfilled per definition because $\lambda_{1}>0$).
\end{tabular}
\ \ \right.  \label{fsigma3}%
\end{equation}

Combining both conditions (\ref{fsigma2}) and (\ref{fsigma3})\ yields%
\begin{equation}
\lambda_{2}>0\text{ and }\lambda_{1}>\frac{-\lambda_{2}}{2}\text{.}
\label{l12}%
\end{equation}

The conditions (\ref{m02}) and (\ref{l12}) will be used in the following
calculation of the decay widths in the scalar meson sector.

\subsection{Scalar Isosinglet Masses} \label{sec.scalarmasses1}

The Lagrangian (\ref{Lagrangian}) yields mixing between the $\sigma_{N}$ and
$\sigma_{S}$ fields with the mixing term given by%
\begin{equation}
\mathcal{L}_{\sigma_{N}\sigma_{S}}=-2\lambda_{1}\phi_{N}\phi_{S}\sigma
_{N}\sigma_{S}\text{.} \label{sigma-sigma}%
\end{equation}

The full $\sigma_{N}$-$\sigma_{S}$\ interaction Lagrangian has the form%
\begin{equation}
\mathcal{L}_{\sigma_{N}\sigma_{S},\,\mathrm{full}}=\frac{1}{2}(\partial_{\mu
}\sigma_{N})^{2}+\frac{1}{2}(\partial_{\mu}\sigma_{S})^{2}-\frac{1}%
{2}m_{\sigma_{N}}^{2}\sigma_{N}{}^{2}-\frac{1}{2}m_{\sigma_{S}}^{2}\sigma
_{S}{}^{2}+z_{\sigma}\sigma_{N}\sigma_{S}\text{,} \label{sigma-sigma_2}
\end{equation}

where $z_{\sigma}$ is the mixing term of the pure states $\sigma_{N}\equiv(\bar
{u}u+\bar{d}d)/\sqrt{2}$ and $\sigma_{S}\equiv\bar{s}s$.

The mixing between the states $\sigma_{N}$ and $\sigma_{S}$\ yields two 
fields, denoted henceforth as $\sigma_{1}$ and $\sigma_{2}$ [analogously to Eq.\ \ref{aaa}]:

\begin{equation}
\left(
\begin{array}
[c]{c}%
\sigma_{1}\\
\sigma_{2}%
\end{array}
\right)  =\left(
\begin{array}
[c]{cc}%
\cos\varphi_{\sigma} & \sin\varphi_{\sigma}\\
-\sin\varphi_{\sigma} & \cos\varphi_{\sigma}%
\end{array}
\right)  \left(
\begin{array}
[c]{c}%
\sigma_{N}\\
\sigma_{S}%
\end{array}
\right)  \text{.}\label{sigma-sigma_1}%
\end{equation}

At this point, it is not possible to assign the fields $\sigma_{1}$ and
$\sigma_{2}$ (considered to be physical just as the resonances $\eta$ and $\eta^{\prime}$ in
Sec.\ \ref{sec.eta-eta}). The reason is that the experimental data suggest a
larger number of physical resonances in the scalar isosinglet channel than can
be accommodated within the model (as discussed in Chapter \ref{sec.scalarexp}). We will therefore
calculate masses and decay widths of the resonances $\sigma_{1}$ and $\sigma_{2}$;
the resonances will then be assigned to physical states depending on the results
regarding the $\sigma_{1,2}$ masses and decay widths.

We can calculate the masses of the mixed sigma states, $m_{\sigma_{1}}$ and
$m_{\sigma_{2}}$, and the $\sigma_{N}$-$\sigma_{S}$ mixing angle
$\varphi_{\sigma}$\ analogously to the calculations concerning $m_{\eta}$,
$m_{\eta^{\prime}}$ and $\varphi_{\eta}$ in Eqs.\ (\ref{m_eta}) - (\ref{phi}).
We obtain%

\begin{align}
m_{\sigma_{1}}^{2}  &  =m_{\sigma_{N}}^{2}\cos^{2}\varphi_{\sigma}%
+m_{\sigma_{S}}^{2}\sin^{2}\varphi_{\sigma}-z_{\sigma}\sin(2\varphi_{\sigma
})\text{,} \label{m_sigma_1}\\
m_{\sigma_{2}}^{2}  &  =m_{\sigma_{N}}^{2}\sin^{2}\varphi_{\sigma}%
+m_{\sigma_{S}}^{2}\cos^{2}\varphi_{\sigma}+z_{\sigma}\sin(2\varphi_{\sigma
}) \label{m_sigma_2}%
\end{align}

with $m_{\sigma_{N}}$ from Eq.\ (\ref{m_sigma_N}), $m_{\sigma_{S}}$ from
Eq.\ (\ref{m_sigma_S}) and\ the mixing term%

\begin{equation}
z_{\sigma}\overset{!}{=}(m_{\sigma_{S}}^{2}-m_{\sigma_{N}}^{2})\tan
(2\varphi_{\sigma})/2\text{.} \label{zsigma}%
\end{equation}

Consequently, from Eqs.\ (\ref{sigma-sigma}) and (\ref{zsigma}) we obtain%

\begin{equation}
(m_{\sigma_{S}}^{2}-m_{\sigma_{N}}^{2})\tan(2\varphi_{\sigma})=-4\lambda
_{1}\phi_{N}\phi_{S} \label{phisigma}%
\end{equation}

or, in other words,%

\begin{align}
\varphi_{\sigma}  &  =-\frac{1}{2}\arctan\left(  \frac{4\lambda_{1}\phi
_{N}\phi_{S}}{m_{\sigma_{S}}^{2}-m_{\sigma_{N}}^{2}}\right) \nonumber\\
&  \overset{\text{Eqs.\ (\ref{m_sigma_N}), (\ref{m_sigma_S})}}{=}\frac{1}%
{2}\arctan\left[  \frac{8\lambda_{1}\phi_{N}\phi_{S}}{(4\lambda_{1}%
+3\lambda_{2})\phi_{N}^{2}-(4\lambda_{1}+6\lambda_{2})\phi_{S}^{2}}\right]
\text{,} \label{phisigma1}%
\end{align}

with $\lambda_{1}$ constrained via $m_{0}^{2}+\lambda_{1}(\phi_{N}^{2}%
+\phi_{S}^{2})=-463425$ MeV$^{2}$.

\begin{figure}
[h]
\begin{center}
\includegraphics[
height=2.0582in,
width=3.9666in
]%
{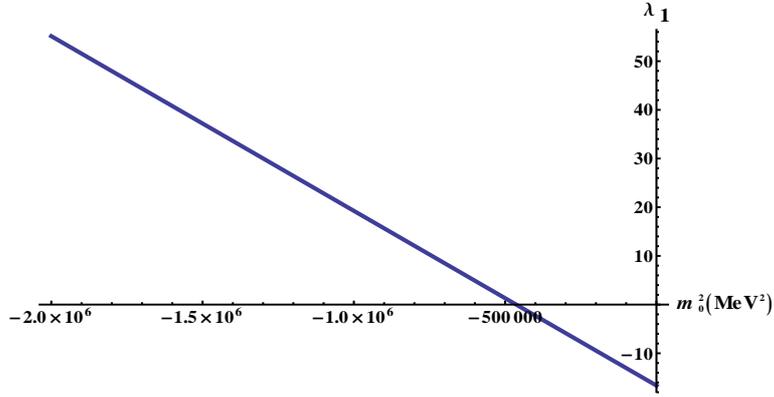}%
\caption{Dependence of parameter $\lambda_{1}$ on $m_{0}^{2}$ from Fit I. The condition
(\ref{l12}), i.e., $\lambda_{1}>-\lambda_{2}/2$, is apparently fulfilled for
all values of $m_{0}^{2}<0$.}%
\label{lambda1}%
\end{center}
\end{figure}

Using the parameter combination $m_{0}^{2}+\lambda_{1}(\phi_{N}^{2}+\phi
_{S}^{2})$\ allows us to remove $\lambda_{1}$ from the mixing term
(\ref{sigma-sigma})\ as well as from the mass terms (\ref{m_sigma_N}) and
(\ref{m_sigma_S}). The parameter $\lambda_{1}$ then fulfills the condition
(\ref{l12}), as is evident from Fig.\ \ref{lambda1}.

\begin{figure}[h]
  \begin{center}
    \begin{tabular}{cc}
      \resizebox{99mm}{!}{\includegraphics{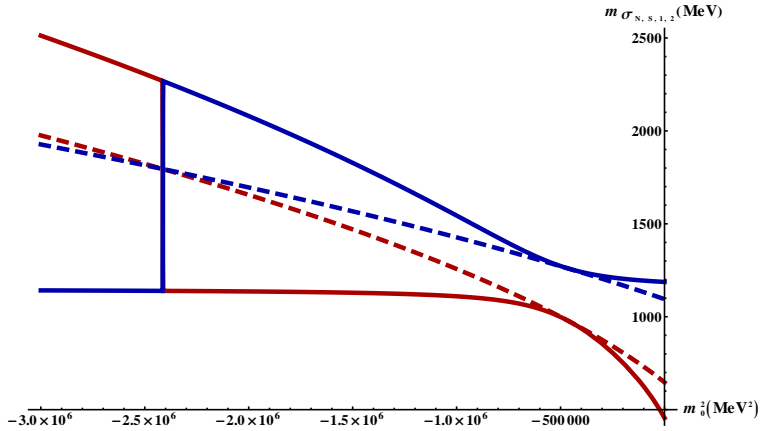}}  
    \end{tabular}
    \caption{Dependence of $m_{\sigma_{1}}$ (full lower curve), $m_{\sigma_{2}}$
(full upper curve), $m_{\sigma_{N}}$ (dashed lower curve) and $m_{\sigma_{S}}$
(dashed upper curve) on $m_{0}^{2}$ under the condition $m_{0}^{2}<0$.}
    \label{Sigmamassen1}
  \end{center}
\end{figure}

This leads to the dependence of\ $m_{\sigma_{1}}$ and
$m_{\sigma_{2}}$, Eqs.\ (\ref{m_sigma_1}) and (\ref{m_sigma_2}),\ on
$m_{0}^{2}$ only. The dependence is depicted in Fig.\ \ref{Sigmamassen1}, with
$m_{0}^{2}\leq0$ in accordance with Eq.\ (\ref{m02}).%

\begin{figure}
[h]
\begin{center}
\includegraphics[
height=2.0582in,
width=3.9666in
]%
{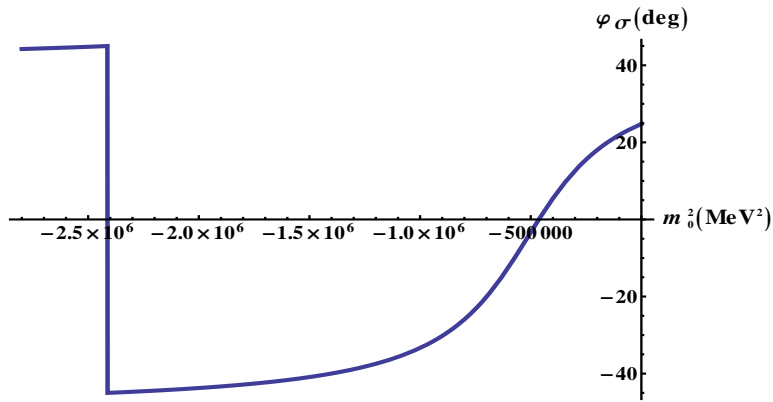}%
\caption{Dependence of the $\sigma_{N}$-$\sigma_{S}$ mixing angle
$\varphi_{\sigma}$ on $m_{0}^{2}$, Eq.\ (\ref{phisigma1}).}%
\label{phi1}%
\end{center}
\end{figure}

We conclude immediately from Fig.\ \ref{Sigmamassen1} that the values of
$m_{\sigma_{1}}$ and $m_{\sigma_{2}}$ vary over wide intervals, respectively,
and that it is therefore not possible to assign the mixed states $\sigma_{1}$
and $\sigma_{2}$ to physical states using only the masses of the mixed states.
Note also that, at $m_{0}^{2}\simeq-2.413\cdot10^{6}$ MeV$^{2}$,
$m_{\sigma_{N}}$ becomes larger than $m_{\sigma_{S}}$, $\varphi_{\sigma}=45%
{{}^\circ}%
$ (see Fig.\ \ref{phi1}) and thus $\sigma_{1}$ and $\sigma_{2}$
interchange places. Therefore, $m_{0}^{2}=-2.413\cdot10^{6}$ MeV$^{2}$
represents the lower limit for $m_{0}^{2}$ and thus, together with
Eq.\ (\ref{m02}), we obtain%

\begin{equation}
-2.413\cdot10^{6}\text{ MeV}^{2}\leq m_{0}^{2}\leq0\text{.} \label{m02b1}%
\end{equation}
From the previous inequality we obtain the following boundaries for
$m_{\sigma_{1,2}}$:%

\begin{align}
456\text{ MeV} &  \leq m_{\sigma_{1}}\leq 1139\text{ MeV}\text{,} \label{ms1}\\
1187\text{ MeV} &  \leq m_{\sigma_{2}}\leq 2268\text{ MeV.}\label{ms2}%
\end{align}
Considering the mass values, $\sigma_{1}$ may correspond either to $f_{0}(600)$ or
$f_{0}(980)$ and $\sigma_{2}$ may correspond to $f_{0}(1370)$, $f_{0}(1500)$
or $f_{0}(1710)$. [We do not consider the as yet unconfirmed states
$f_{0}(2020)$, $f_{0}(2100)$ and $f_{0}(2200)$ although they could also come
within the $m_{\sigma_{2}}$ range. Note also our comments in Sec.\ \ref{sec.f0(1790)} regarding
the $f_{0}(1790)$ resonance that decays predominantly into pions and appears
to be a radial excitation of $f_{0}(1370)$ -- therefore it cannot correspond
to our state $\sigma_{2}$ that is predominantly strange, as we will see
in the following.] Therefore, a mere calculation of scalar
masses does not allow us to assign the scalar states $\sigma_{1}$ and
$\sigma_{2}$ to physical resonances.\ To resolve this ambiguity, we will
calculate various decay widths of the states $\sigma_{1}$ and $\sigma_{2}$;
comparison of the decay widths with experimental data \cite{PDG} will allow
for a definitive statement regarding the assignment of our theoretical states
to the physical ones.%

\begin{figure}[h]
  \begin{center}
    \begin{tabular}{cc}
      \resizebox{78mm}{!}{\includegraphics{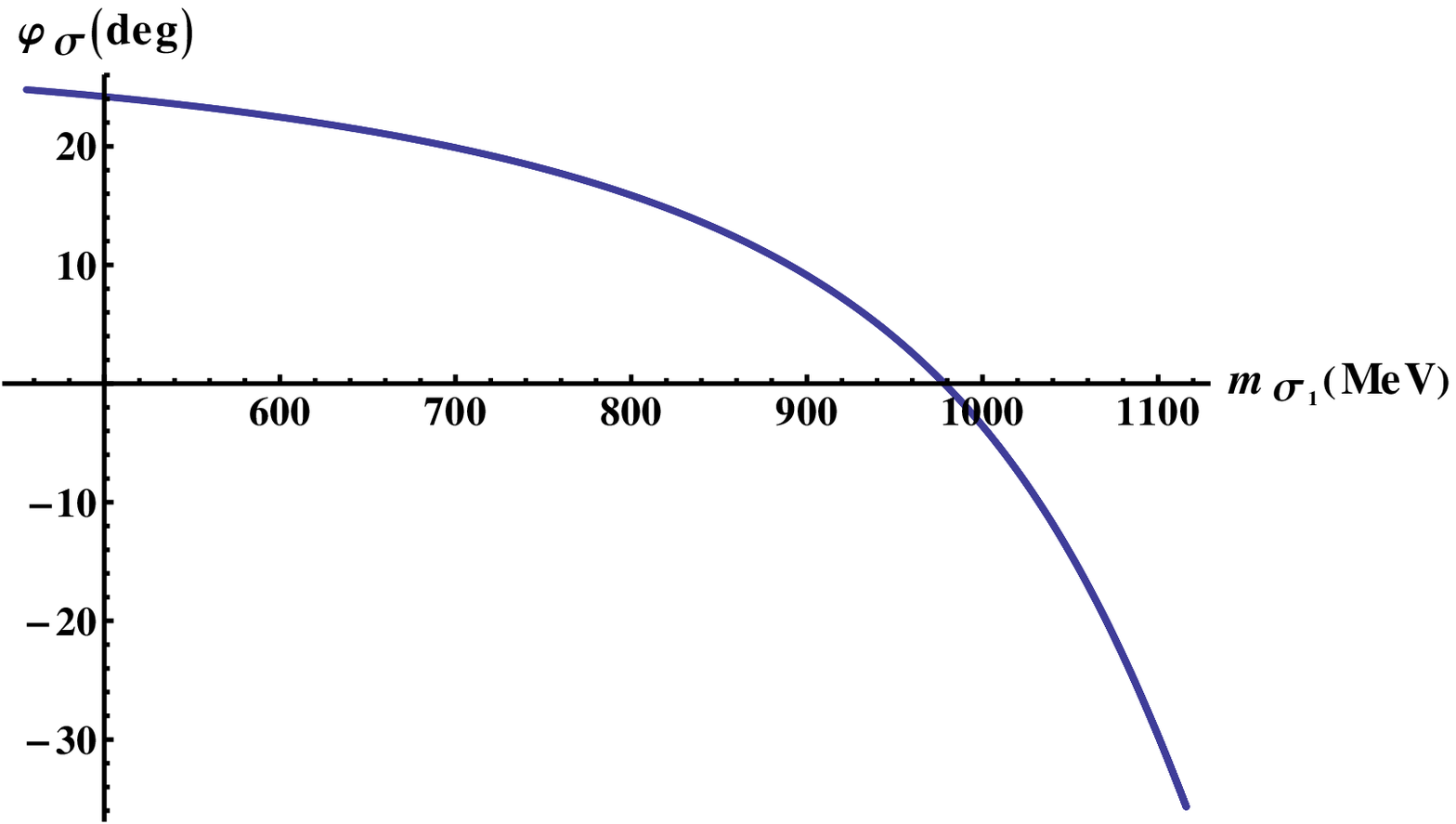}} &
      \resizebox{78mm}{!}{\includegraphics{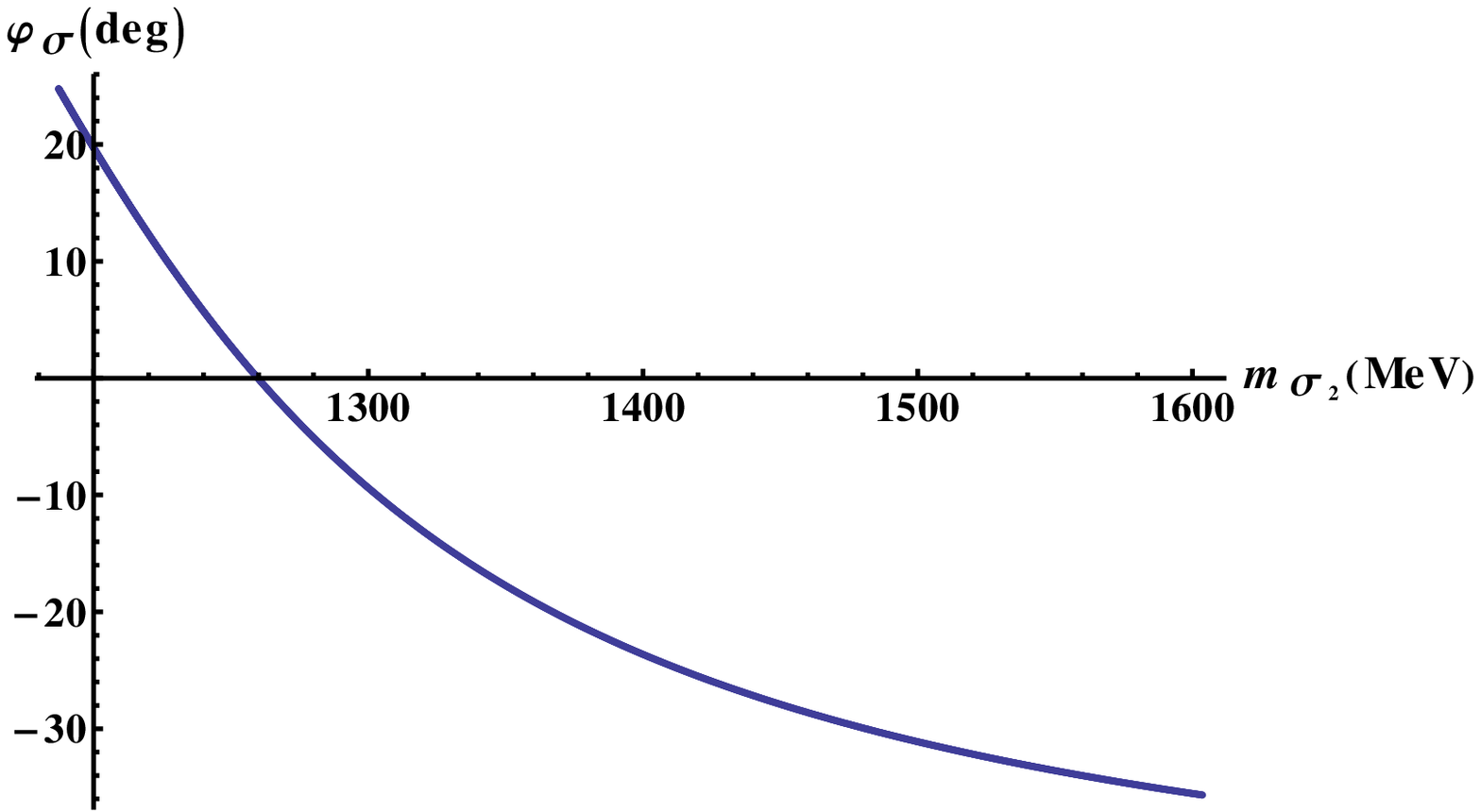}} 
    \end{tabular}
    \caption{The $\sigma_{N}$-$\sigma_{S}$ mixing angle $\varphi_{\sigma}$ as
function of $m_{\sigma_{1,2}}$.}
    \label{phi11}
  \end{center}
\end{figure}

Nonetheless, from the variation of the $\sigma_{N}$ - $\sigma_{S}$ mixing
angle $\varphi_{\sigma}$ we can conclude that the $\sigma_{1}$ field is
predominantly non-strange and the $\sigma_{2}$ field is predominantly composed
of strange quarks, see Fig.\ \ref{phi11}. Note that the two diagrams on
Fig.\ \ref{phi11} were obtained from two simultaneous, implicit plots of
$\varphi_{\sigma}(\lambda_{1})$, Eq.\ (\ref{phisigma1}), and $m_{\sigma_{1,2}%
}[\varphi_{\sigma}(\lambda_{1})]$, Eqs.\ (\ref{m_sigma_1}) and
(\ref{m_sigma_2}), with $m_{0}^{2}+\lambda_{1}(\phi_{N}^{2}+\phi_{S}%
^{2})=-463425$ MeV$^{2}$ and $m_{0}^{2}$ from inequality (\ref{m02b1}).%

\begin{figure}[h]
  \begin{center}
    \begin{tabular}{cc}
      \resizebox{78mm}{!}{\includegraphics{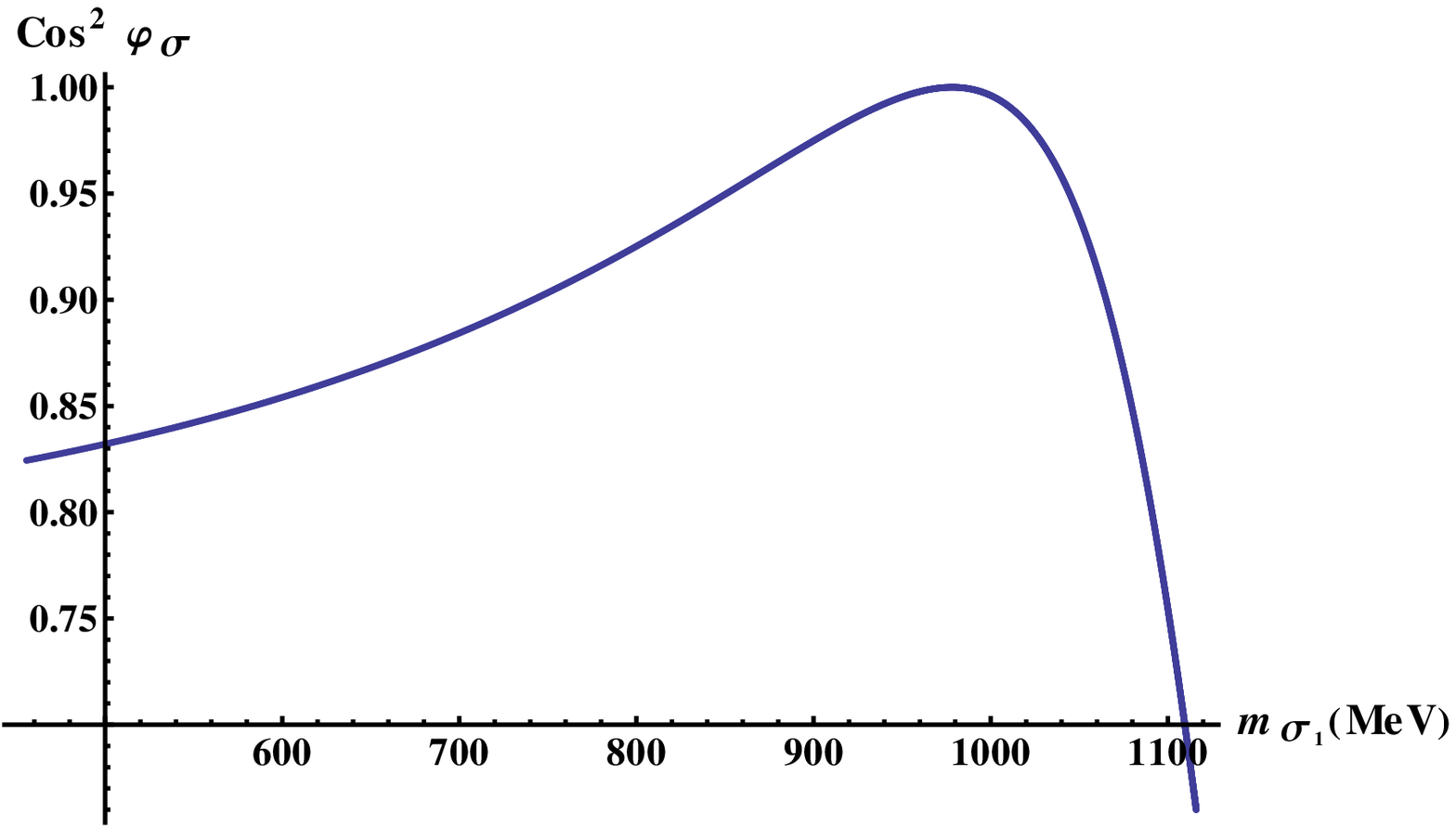}} &
      \resizebox{78mm}{!}{\includegraphics{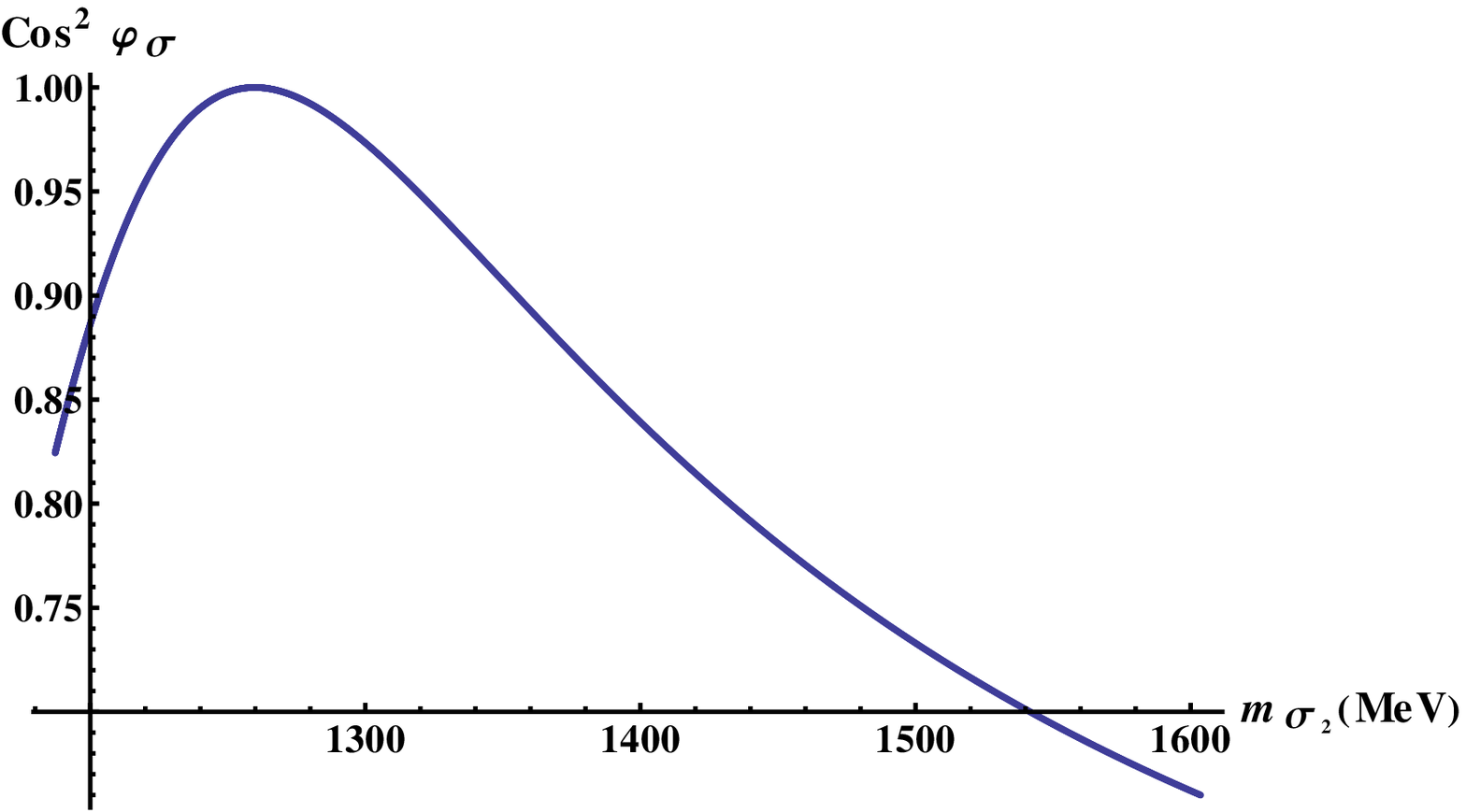}} 
    \end{tabular}
    \caption{Contribution of the pure non-strange field $\sigma_{N}$ to
$\sigma_{1}$ (left panel) and of the pure strange field $\sigma_{S}$ to
$\sigma_{2}$ (right panel), respectively in dependence on $m_{\sigma_{1}}$ and
$m_{\sigma_{2}}$.}
    \label{phi12}
  \end{center}
\end{figure}

We illustrate the contribution of $m_{\sigma_{N}}$ to $m_{\sigma_{1}}$ and of
$m_{\sigma_{S}}$ to $m_{\sigma_{2}}$ in Fig.\ \ref{phi12}. The contributions expectedly decrease with $m_{\sigma_{1,2}}$
because the mixing angle approaches $-45�$ (see Fig.\ \ref{phi11}) where $\sigma_{1}$ and $\sigma_{2}$ interchange places. 

\subsection{Decay Width \boldmath $\sigma_{1,2}\rightarrow\pi\pi$} \label{sec.sigmapionpion1}

The Lagrangian (\ref{Lagrangian}) contains the pure states $\sigma_{N}$ and
$\sigma_{S}$; the interaction Lagrangian of these states with the pions reads:%
\begin{align}
\mathcal{L}_{\sigma\pi\pi}  &  =A_{\sigma_{N}\pi\pi}\sigma_{N}[(\pi^{0}%
)^{2}+2\pi^{+}\pi^{-}]+B_{\sigma_{N}\pi\pi}\sigma_{N}[(\partial_{\mu}\pi
^{0})^{2}+2\partial_{\mu}\pi^{+}\partial^{\mu}\pi^{-}]\nonumber\\
&  +C_{\sigma_{N}\pi\pi}\sigma_{N}(\pi^{0}\square\pi^{0}+\pi^{+}\square\pi
^{-}+\pi^{-}\square\pi^{+})\nonumber\\
&  +A_{\sigma_{S}\pi\pi}\sigma_{S}[(\pi^{0})^{2}+2\pi^{+}\pi^{-}%
]+B_{\sigma_{S}\pi\pi}\sigma_{S}[(\partial_{\mu}\pi^{0})^{2}+2\partial_{\mu
}\pi^{+}\partial^{\mu}\pi^{-}] \label{sigmapionpion}%
\end{align}
with%

\begin{align}
A_{\sigma_{N}\pi\pi}  &  =-\left(  \lambda_{1}+\frac{\lambda_{2}}{2}\right)
Z_{\pi}^{2}\phi_{N}\text{,} \label{ANspp}\\
B_{\sigma_{N}\pi\pi}  &  =-2g_{1}Z_{\pi}^{2}w_{a_{1}}+\left(  g_{1}^{2}%
+\frac{h_{1}+h_{2}-h_{3}}{2}\right)  Z_{\pi}^{2}w_{a_{1}}^{2}\phi
_{N}\text{,} \label{BNspp}\\
C_{\sigma_{N}\pi\pi}  &  =-g_{1}Z_{\pi}^{2}w_{a_{1}}\text{,} \label{CNspp}\\
A_{\sigma_{S}\pi\pi}  &  =-\lambda_{1}Z_{\pi}^{2}\phi_{S}\text{,} \label{ASspp}\\
B_{\sigma_{S}\pi\pi}  &  =\frac{h_{1}}{2}Z_{\pi}^{2}w_{a_{1}}^{2}\phi
_{S}\text{.} \label{BSspp}%
\end{align}

Note that the term $B_{\sigma_{N}\pi\pi}$, Eq.\ (\ref{BNspp}), can be further transformed as follows:
\begin{align}
&  B_{\sigma_{N}\pi\pi}\overset{\text{Eq.\ (\ref{wa1})}}{=}Z_{\pi}^{2}%
\frac{g_{1}^{2}\phi_{N}}{m_{a_{1}}^{2}}\left(  -2+\frac{g_{1}^{2}\phi_{N}^{2}%
}{m_{a_{1}}^{2}}+\frac{h_{1}+h_{2}-h_{3}}{2}\frac{\phi_{N}^{2}}{m_{a_{1}}^{2}%
}\right) \nonumber\\
&  =Z_{\pi}^{2}\frac{g_{1}^{2}\phi_{N}}{m_{a_{1}}^{4}}\left(  -2m_{a_{1}}%
^{2}+g_{1}^{2}\phi_{N}^{2}+\frac{h_{1}+h_{2}-h_{3}}{2}\phi_{N}^{2}\right)
\nonumber\\
&  \overset{\text{Eq.\ (\ref{m_a_1})}}{=}Z_{\pi}^{2}\frac{g_{1}^{2}\phi_{N}%
}{m_{a_{1}}^{4}}\left(  -2m_{a_{1}}^{2}+m_{a_{1}}^{2}-m_{1}^{2}-\frac{h_{1}%
}{2}\phi_{S}^{2}-2\delta_{N}\right)  \equiv-Z_{\pi}^{2}\frac{g_{1}^{2}\phi
_{N}}{m_{a_{1}}^{4}}\left(  m_{a_{1}}^{2}+m_{1}^{2}\right)  \text{,}
\label{BNspp1}%
\end{align}
as $h_{1}=0=\delta_{N}$. Note also that the decay of the pure strange state
$\sigma_{S}$ into pions is driven by the large-$N_{c}$ suppressed couplings
$\lambda_{1}$ and $h_{1}$, see Eq.\ (\ref{largen}).

At this point it is necessary to disentangle the pure states $\sigma_{N}$ and
$\sigma_{S}$ that do not represent asymptotic states in the $\sigma\pi\pi$
Lagrangian (\ref{sigmapionpion}). To obtain decay widths of the physical,
mixed states $\sigma_{1}$ and $\sigma_{2}$, we have to consider the full
Lagrangian containing the $\sigma$ fields [$\mathcal{L}_{\sigma_{N}\sigma
_{S},\,\mathrm{full}}$ from Eq.\ (\ref{sigma-sigma_2})]:
\begin{align}
\mathcal{L}_{\sigma\pi\pi\text{, full}}  &  =\mathcal{L}_{\sigma_{N}\sigma
_{S},\,\mathrm{full}}+\mathcal{L}_{\sigma\pi\pi}\nonumber\\
&  =\frac{1}{2}(\partial_{\mu}\sigma_{N})^{2}+\frac{1}{2}(\partial_{\mu}%
\sigma_{S})^{2}-\frac{1}{2}m_{\sigma_{N}}^{2}-\frac{1}{2}m_{\sigma_{S}}%
^{2}+z_{\sigma}\sigma_{N}\sigma_{S}\nonumber\\
&  +A_{\sigma_{N}\pi\pi}\sigma_{N}[(\pi^{0})^{2}+2\pi^{+}\pi^{-}%
]+B_{\sigma_{N}\pi\pi}\sigma_{N}[(\partial_{\mu}\pi^{0})^{2}+2\partial_{\mu
}\pi^{+}\partial^{\mu}\pi^{-}]\nonumber\\
&  +C_{\sigma_{N}\pi\pi}\sigma_{N}(\pi^{0}\square\pi^{0}+\pi^{+}\square\pi
^{-}+\pi^{-}\square\pi^{+})\nonumber\\
&  +A_{\sigma_{S}\pi\pi}\sigma_{S}[(\pi^{0})^{2}+2\pi^{+}\pi^{-}%
]+B_{\sigma_{S}\pi\pi}\sigma_{S}[(\partial_{\mu}\pi^{0})^{2}+2\partial_{\mu
}\pi^{+}\partial^{\mu}\pi^{-}]\text{.} \label{sigmapionpion2}
\end{align}

Let us now insert the inverted Eq.\ (\ref{sigma-sigma_1}) into Eq.\ (\ref{sigmapionpion2});
analogously to Eq.\ (\ref{eta-eta-1}) we obtain:
\begin{align}
\mathcal{L}_{\sigma\pi\pi\text{, full}}  &  =\frac{1}{2}[(\partial_{\mu}%
\sigma_{1})^{2}(\cos\varphi_{\sigma})^{2}+(\partial_{\mu}\sigma
_{2})^{2}(\sin\varphi_{\sigma})^{2}]\nonumber\\
&  +\frac{1}{2}[(\partial_{\mu}\sigma_{1})^{2}(\sin\varphi_{\sigma})^{2}%
+(\partial_{\mu}\sigma_{2})^{2}(\cos\varphi_{\sigma})^{2}%
]\nonumber\\
&  -\frac{1}{2}m_{\sigma_{N}}^{2}[\sigma_{1}^{2}(\cos\varphi_{\sigma}%
)^{2}+\sigma_{2}^{2}(\sin\varphi_{\eta})^{2}-\sin(2\varphi_{\sigma})\sigma
_{1}\sigma_{2}]\nonumber\\
&  -\frac{1}{2}m_{\sigma_{S}}^{2}[\sigma_{1}^{2}(\sin\varphi_{\sigma}%
)^{2}+\sigma_{2}^{2}(\cos\varphi_{\sigma})^{2}+\sin(2\varphi_{\sigma}%
)\sigma_{1}\sigma_{2}]\nonumber\\
&  +z_{\sigma}[(\sigma_{1}^{2}-\sigma_{2}^{2})\sin\varphi_{\sigma}\cos
\varphi_{\sigma}+\cos(2\varphi_{\sigma})\sigma_{1}\sigma_{2}]\nonumber\\
&  +(A_{\sigma_{N}\pi\pi}\cos\varphi_{\sigma}+A_{\sigma_{S}\pi\pi}\sin
\varphi_{\sigma})\sigma_{1}[(\pi^{0})^{2}+2\pi^{+}\pi^{-}]\nonumber
\end{align}
\begin{align}
&  +(B_{\sigma_{N}\pi\pi}\cos\varphi_{\sigma}+B_{\sigma_{S}\pi\pi}\sin
\varphi_{\sigma})\sigma_{1}[(\partial_{\mu}\pi^{0})^{2}+2\partial_{\mu}\pi
^{+}\partial^{\mu}\pi^{-}]\nonumber\\
&  +C_{\sigma_{N}\pi\pi}\cos\varphi_{\sigma}\sigma_{1}(\pi^{0}\square\pi
^{0}+\pi^{+}\square\pi^{-}+\pi^{-}\square\pi^{+})\nonumber\\
&  +(A_{\sigma_{S}\pi\pi}\cos\varphi_{\sigma}-A_{\sigma_{N}\pi\pi}\sin
\varphi_{\sigma})\sigma_{2}[(\pi^{0})^{2}+2\pi^{+}\pi^{-}]\nonumber\\
&  +(B_{\sigma_{S}\pi\pi}\cos\varphi_{\sigma}-B_{\sigma_{N}\pi\pi}\sin
\varphi_{\sigma})\sigma_{2}[(\partial_{\mu}\pi^{0})^{2}+2\partial_{\mu}\pi
^{+}\partial^{\mu}\pi^{-}]\nonumber\\
&  -C_{\sigma_{N}\pi\pi}\sin\varphi_{\sigma}\sigma_{2}(\pi^{0}\square\pi
^{0}+\pi^{+}\square\pi^{-}+\pi^{-}\square\pi^{+})\nonumber\\
&  =\frac{1}{2}(\partial_{\mu}\sigma_{1})^{2}+\frac{1}{2}(\partial_{\mu}%
\sigma_{2})^{2}-\frac{1}{2}[m_{\sigma_{N}}^{2}(\cos\varphi_{\sigma}%
)^{2}+m_{\sigma_{S}}^{2}(\sin\varphi_{\sigma})^{2}-z_{\sigma}\sin
(2\varphi_{\sigma})]\sigma_{1}^{2}\nonumber\\
&  -\frac{1}{2}[m_{\sigma_{N}}^{2}(\sin\varphi_{\sigma})^{2}+m_{\sigma_{S}%
}^{2}(\cos\varphi_{\sigma})^{2}+z_{\sigma}\sin(2\varphi_{\sigma})]\sigma
_{2}^{2}\nonumber\\
&  -\frac{1}{2}[(m_{\sigma_{S}}^{2}-m_{\sigma_{N}}^{2})\sin(2\varphi_{\sigma
})-2z_{\sigma}\cos(2\varphi_{\sigma})]\sigma_{1}\sigma_{2}\nonumber\\
&  +(A_{\sigma_{N}\pi\pi}\cos\varphi_{\sigma}+A_{\sigma_{S}\pi\pi}\sin
\varphi_{\sigma})\sigma_{1}[(\pi^{0})^{2}+2\pi^{+}\pi^{-}]\nonumber\\
&  +(B_{\sigma_{N}\pi\pi}\cos\varphi_{\sigma}+B_{\sigma_{S}\pi\pi}\sin
\varphi_{\sigma})\sigma_{1}[(\partial_{\mu}\pi^{0})^{2}+2\partial_{\mu}\pi
^{+}\partial^{\mu}\pi^{-}]\nonumber\\
&  +C_{\sigma_{N}\pi\pi}\cos\varphi_{\sigma}\sigma_{1}(\pi^{0}\square\pi
^{0}+\pi^{+}\square\pi^{-}+\pi^{-}\square\pi^{+})\nonumber\\
&  +(A_{\sigma_{S}\pi\pi}\cos\varphi_{\sigma}-A_{\sigma_{N}\pi\pi}\sin
\varphi_{\sigma})\sigma_{2}[(\pi^{0})^{2}+2\pi^{+}\pi^{-}]\nonumber\\
&  +(B_{\sigma_{S}\pi\pi}\cos\varphi_{\sigma}-B_{\sigma_{N}\pi\pi}\sin
\varphi_{\sigma})\sigma_{2}[(\partial_{\mu}\pi^{0})^{2}+2\partial_{\mu}\pi
^{+}\partial^{\mu}\pi^{-}]\nonumber\\
&  -C_{\sigma_{N}\pi\pi}\sin\varphi_{\sigma}\sigma_{2}(\pi^{0}\square\pi
^{0}+\pi^{+}\square\pi^{-}+\pi^{-}\square\pi^{+})\text{.}
\label{sigmapionpion3}%
\end{align}

From Eq.\ (\ref{sigmapionpion3}) we then retrieve the already known
Eqs.\ (\ref{m_sigma_1}) and (\ref{m_sigma_2}) for $m_{\sigma_{1,2}}^{2}$ as
well as the condition (\ref{zsigma}) for $z_{\sigma}$\ ascertaining that there
is no mixing between the physical states $\sigma_{1}$ and $\sigma_{2}$. Let
us now write the Lagrangian (\ref{sigmapionpion3}) in the following form:%

\begin{align}
\mathcal{L}_{\sigma\pi\pi\text{, full}}  &  =\frac{1}{2}(\partial_{\mu}%
\sigma_{1})^{2}-\frac{1}{2}m_{\sigma_{1}}^{2}\sigma_{1}^{2}\nonumber\\
&  +(A_{\sigma_{N}\pi\pi}\cos\varphi_{\sigma}+A_{\sigma_{S}\pi\pi}\sin
\varphi_{\sigma})\sigma_{1}[(\pi^{0})^{2}+2\pi^{+}\pi^{-}]\nonumber\\
&  +(B_{\sigma_{N}\pi\pi}\cos\varphi_{\sigma}+B_{\sigma_{S}\pi\pi}\sin
\varphi_{\sigma})\sigma_{1}[(\partial_{\mu}\pi^{0})^{2}+2\partial_{\mu}\pi
^{+}\partial^{\mu}\pi^{-}]\nonumber\\
&  +C_{\sigma_{N}\pi\pi}\cos\varphi_{\sigma}\sigma_{1}(\pi^{0}\square\pi
^{0}+\pi^{+}\square\pi^{-}+\pi^{-}\square\pi^{+})\nonumber\\
&  +\frac{1}{2}(\partial_{\mu}\sigma_{2})^{2}-\frac{1}{2}m_{\sigma_{2}}%
^{2}\sigma_{2}^{2}\nonumber\\
&  +(A_{\sigma_{S}\pi\pi}\cos\varphi_{\sigma}-A_{\sigma_{N}\pi\pi}\sin
\varphi_{\sigma})\sigma_{2}[(\pi^{0})^{2}+2\pi^{+}\pi^{-}]\nonumber\\
&  +(B_{\sigma_{S}\pi\pi}\cos\varphi_{\sigma}-B_{\sigma_{N}\pi\pi}\sin
\varphi_{\sigma})\sigma_{2}[(\partial_{\mu}\pi^{0})^{2}+2\partial_{\mu}\pi
^{+}\partial^{\mu}\pi^{-}]\nonumber\\
&  -C_{\sigma_{N}\pi\pi}\sin\varphi_{\sigma}\sigma_{2}(\pi^{0}\square\pi
^{0}+\pi^{+}\square\pi^{-}+\pi^{-}\square\pi^{+})\text{.}
\label{sigmapionpion4}%
\end{align}

\begin{figure}[h]
 \begin{align*}
\qquad \qquad \qquad  \qquad \qquad \qquad  \quad \;  \parbox{180mm}{ \begin{fmfgraph*}(180,80)
                  \fmfleftn{i}{1}\fmfrightn{o}{2}
                  \fmf{vanilla,label=\text{\small\(\sigma_{1,,2}(P)\)},label.dist=-18}{i1,v1}\fmf{dashes,label=\text{\small\(\pi(P_2)\)},label.dist=-28}{v1,o1}\fmf{dashes,label=\text{\small\(\pi(P_1)\)},label.dist=-28}{v1,o2}\fmfdot{v1}
                 \end{fmfgraph*}}
 \end{align*}
\caption{Decay process $\sigma_{1,2}\rightarrow\pi\pi$.}
\end{figure}
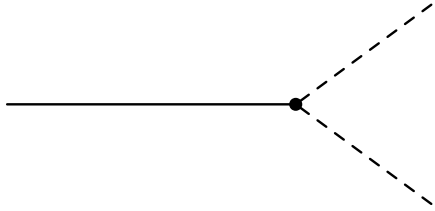
$\,$\\
The decay amplitudes of the mixed states read

\begin{align}
-i\mathcal{M}_{\sigma_{1}\rightarrow\pi\pi}(m_{\sigma_{1}})  &  =i\left\{
\cos\varphi_{\sigma}\left[  A_{\sigma_{N}\pi\pi}-B_{\sigma_{N}\pi\pi}%
\frac{m_{\sigma_{1}}^{2}-2m_{\pi}^{2}}{2}-C_{\sigma_{N}\pi\pi}m_{\pi}%
^{2}\right] \right.  \nonumber\\
& \left. + \sin\varphi_{\sigma}\left[  A_{\sigma_{S}\pi\pi}-B_{\sigma
_{S}\pi\pi}\frac{m_{\sigma_{1}}^{2}-2m_{\pi}^{2}}{2}\right]  \right\}
\nonumber\\
&  =i\left\{  \cos\varphi_{\sigma}\left[  A_{\sigma_{N}\pi\pi}-\frac
{B_{\sigma_{N}\pi\pi}}{2}m_{\sigma_{1}}^{2}+(B_{\sigma_{N}\pi\pi}%
-C_{\sigma_{N}\pi\pi})m_{\pi}^{2}\right] \right.  \nonumber \\
& \left. +\sin\varphi_{\sigma}\left[
A_{\sigma_{S}\pi\pi}-B_{\sigma_{S}\pi\pi}\frac{m_{\sigma_{1}}^{2}-2m_{\pi}%
^{2}}{2}\right]  \right\} \label{Ms1pp}
\end{align}
and
\begin{align}
-i\mathcal{M}_{\sigma_{2}\rightarrow\pi\pi}(m_{\sigma_{2}})  &  =i \left\{
\cos\varphi_{\sigma}\left[  A_{\sigma_{S}\pi\pi}-B_{\sigma_{S}\pi\pi}%
\frac{m_{\sigma_{2}}^{2}-2m_{\pi}^{2}}{2}\right] \right.  \nonumber\\
& \left. -\sin\varphi_{\sigma}\left[
A_{\sigma_{N}\pi\pi}-\frac{B_{\sigma_{N}\pi\pi}}{2}m_{\sigma_{2}}%
^{2}+(B_{\sigma_{N}\pi\pi}-C_{\sigma_{N}\pi\pi})m_{\pi}^{2}\right]  \right\}
\text{.} \label{Ms2pp}%
\end{align}
%
%
%
%
%
Summing over all decay channels $\sigma_{1,2}\rightarrow
\pi^{0}\pi^{0},\pi^{\pm}\pi^{\mp}$ we obtain the following formulas for the
decay widths $\Gamma_{\sigma_{1,2}\rightarrow\pi\pi}$:%

\begin{eqnarray}
\Gamma_{\sigma_{1}\rightarrow\pi\pi}  &  =\frac{3k(m_{\sigma_{1}},m_{\pi
},m_{\pi})}{4\pi m_{\sigma_{1}}^{2}}|-i\mathcal{M}_{\sigma_{1}\rightarrow
\pi\pi}(m_{\sigma_{1}})|^{2}\text{,} \label{Gs1pp}\\
\Gamma_{\sigma_{2}\rightarrow\pi\pi}  &  =\frac{3k(m_{\sigma_{2}},m_{\pi
},m_{\pi})}{4\pi m_{\sigma_{2}}^{2}}|-i\mathcal{M}_{\sigma_{2}\rightarrow
\pi\pi}(m_{\sigma_{2}})|^{2}\text{.} \label{Gs2pp}%
\end{eqnarray}

We have considered an isospin factor of 6 in the above Eqs.\ (\ref{Gs1pp}) and
(\ref{Gs2pp}). The Lagrangian (\ref{sigmapionpion4}) provides us with an
additional factor of $2^{2}=4$ in $\Gamma_{\sigma_{1,2}\rightarrow\pi\pi}$
respectively from the charged ($\pi^{\pm}\pi^{\mp}$) and neutral ($\pi^{0}%
\pi^{0}$)\ modes, i.e., in total with a factor of $8$. However, there is a
symmetrisation factor of $1/\sqrt{2}$\ that also has to be considered for the
neutral modes; therefore, their contribution to $\Gamma_{\sigma_{1,2}%
\rightarrow\pi\pi}$ is actually not $2^{2}$ but rather $(2/\sqrt{2})^{2}=2$
that together with the charged-mode contribution $2^{2}=4$ yields a total
isospin factor of $6$.\\

\begin{figure}[h]
  \begin{center}
    \begin{tabular}{cc}
      \resizebox{78mm}{!}{\includegraphics{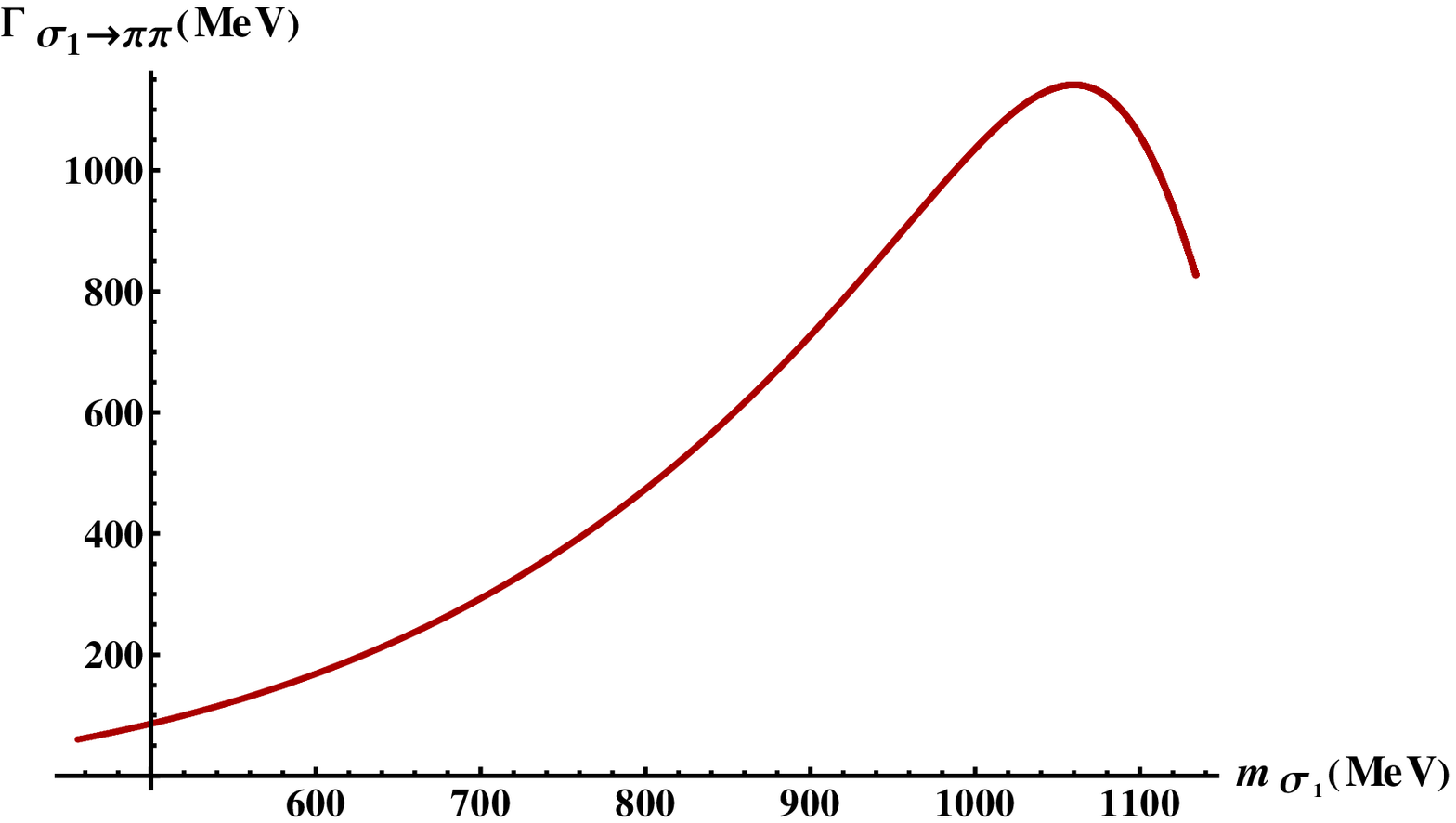}} &
      \resizebox{78mm}{!}{\includegraphics{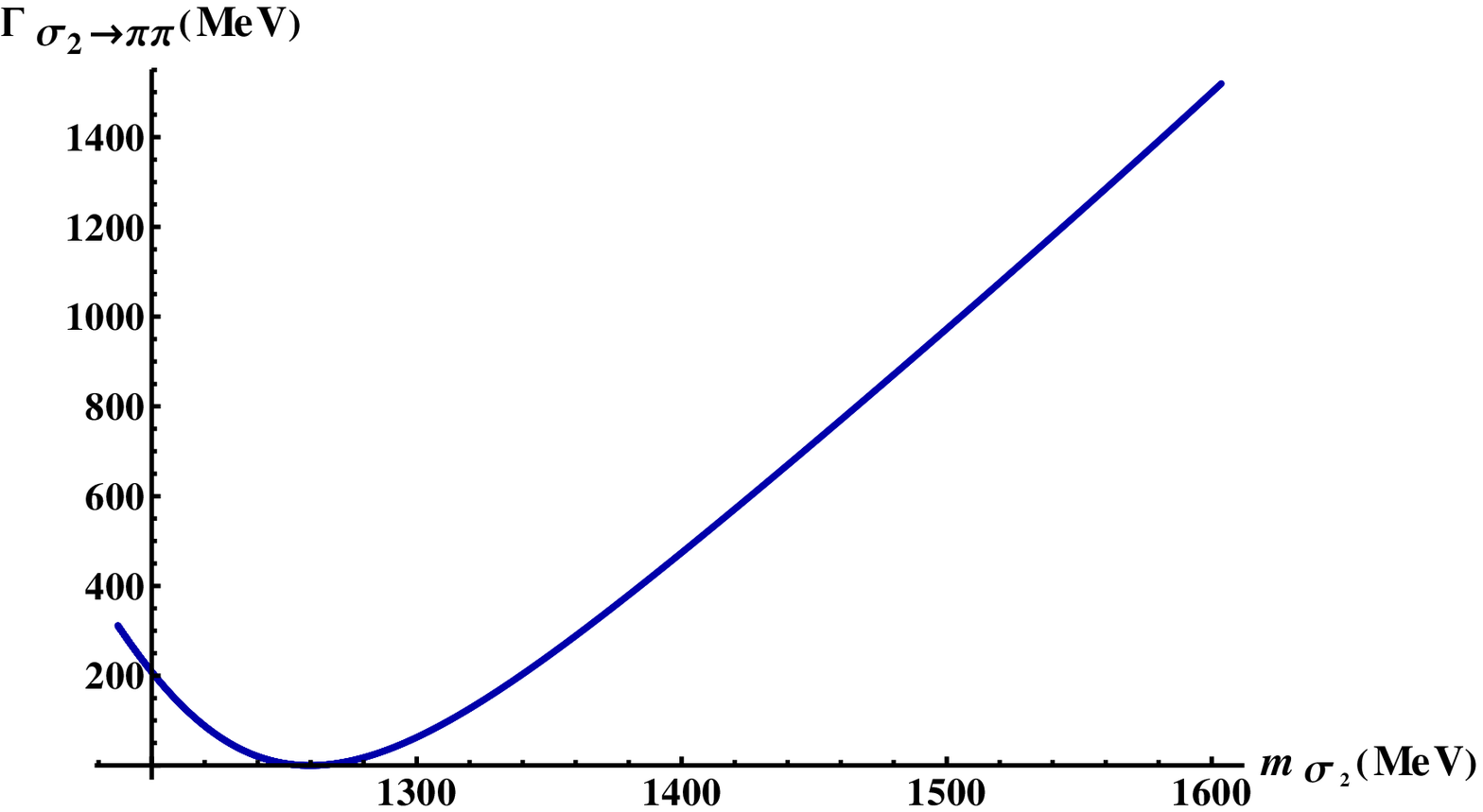}} 
    \end{tabular}
    \caption{$\Gamma_{\sigma_{1}\rightarrow\pi\pi}$ and $\Gamma_{\sigma
_{2}\rightarrow\pi\pi}$ as functions of $m_{\sigma_{1}}$ and $m_{\sigma_{2}}$,
respectively.}
    \label{Spp1}
  \end{center}
\end{figure}

A plot of the two decay widths is presented in Fig.\ \ref{Spp1}. We conclude from the left panel of Fig.\ \ref{Spp1} that our state $\sigma
_{1}$ possesses the best correspondence with the $f_{0}(600)$ resonance. For
example, setting $m_{\sigma_{1}}=800$ MeV yields $\Gamma_{\sigma
_{1}\rightarrow\pi\pi}=473.5$ MeV. From the right panel of Fig.\ \ref{Spp1} we
note that $\Gamma_{\sigma_{2}\rightarrow\pi\pi}$ increases very rapidly with
$m_{\sigma_{2}}$ and therefore the best values are obtained for $m_{\sigma
_{2}}\simeq1300$ MeV. If we consider data from Ref.\ \cite{buggf0}, then our
results are fairly close to the results from this review: Ref.\ \cite{buggf0}
cites the value of 325 MeV at $m_{f_{0}(1370)}=(1309\pm1\pm15)$ MeV from an
$f_{0}(1370)$\ Breit-Wigner fit and we obtain $\Gamma_{\sigma_{2}%
\rightarrow\pi\pi}=325$ MeV at $m_{\sigma_{2}}=1368$ MeV; Ref.\ \cite{buggf0}
cites the value of $207$ MeV for the full width at one-half maximum (FWHM)
with the peak in the decay channel $f_{0}(1370)\rightarrow\pi\pi$
at$\ m_{f_{0}(1370)}=1282$ MeV -- we obtain $\Gamma_{\sigma_{2}\rightarrow
\pi\pi}=207$ MeV at $m_{\sigma_{2}}=1341$ MeV and at $m_{\sigma_{2}}=1200$
MeV. Let us, however, emphasise that these results suggest $f_{0}(1370)$ to be
predominantly a $\bar{s}s$ state as we can see from Fig.\ \ref{phi12}.
Concretely, results suggest that $f_{0}(1370)$ is 88\% a $\bar{s}s$ state at
$m_{\sigma_{2}}=1368$ MeV, 92\% a $\bar{s}s$ state at $m_{\sigma_{2}}=1341$
MeV and 89\% a $\bar{s}s$ state at $m_{\sigma_{2}}=1200$ MeV. Note that we do
not assign error values to our masses because the errors in our model are
determined by errors of experimental data used for our calculations and no
errors were assigned to $\Gamma_{f_{0}(1370)\rightarrow\pi\pi}$ in
Ref.\ \cite{buggf0}.\\

As evident from Fig.\ \ref{Spp1}, $\Gamma_{\sigma_{2}\rightarrow\pi\pi}=0$
for $m_{\sigma_{2}}=1260$ MeV, corresponding to $m_{0}^{2}=-463425$ MeV$^{2}$
and thus $m_{\sigma_{1}}=978$ MeV\ (see Fig.\ \ref{Sigmamassen1}). The reason
is that, due to the constraint $m_{0}^{2}+\lambda_{1}(\phi_{N}^{2}+\phi
_{S}^{2})=-463425$ MeV$^{2}$ (see Table \ref{Fit1-4}), we obtain $\lambda
_{1}=0$ for $m_{0}^{2}=-463425$ MeV$^{2}$; consequently, according to
Eq.\ (\ref{phisigma1}), one also obtains that the $\sigma_{N}$ - $\sigma_{S}$
mixing angle $\varphi_{\sigma}=0$. Thus $\sigma_{N}$ and $\sigma_{S}$ decouple
at this point. Usually, this would merely imply that the $2\pi$\ decay
amplitude $\mathcal{M}_{\sigma_{2}\rightarrow\pi\pi}(m_{\sigma_{2}})$,
Eq.\ (\ref{Ms2pp}), of the (now pure-strange) state $\sigma_{2}\equiv
\sigma_{S}$ would become suppressed but it would not necessarily vanish. It
could still be non-zero by large-$N_{c}$ suppressed parameters, in our
case $\lambda_{1}$ and $h_{1}$ that appear in $A_{\sigma_{S}\pi\pi}$ and
$B_{\sigma_{S}\pi\pi}$ [Eqs.\ (\ref{ASspp}) and (\ref{BSspp}), respectively].
However, we have set $h_{1}\equiv0$ throughout our calculations (see Table
\ref{Fit1-4}) and, as we have just discussed, $\lambda_{1}$ also vanishes at
this point. For this reason, $A_{\sigma_{S}\pi\pi}=0=B_{\sigma_{S}\pi\pi}$ and
consequently also $\mathcal{M}_{\sigma_{2}\rightarrow\pi\pi}=0$. Therefore,
$\Gamma_{\sigma_{2}\rightarrow\pi\pi}=0$. Note that setting $h_{1}\neq0$ would
not alter the fact that $\Gamma_{\sigma_{2}\rightarrow\pi\pi}$ vanishes for a
certain value of $m_{\sigma_{2}}$. The reason is the relative minus sign of
the two terms in $\mathcal{M}_{\sigma_{2}\rightarrow\pi\pi}$ [see
Eq.\ (\ref{Ms2pp})], allowing for a value of $\varphi_{\sigma}$ to be found
where they exactly cancel out.
%

\subsection{Decay Width \boldmath $\sigma_{1,2}\rightarrow K K$} \label{sec.sigmakaonkaon}

The interaction Lagrangian of the pure states $\sigma_{N}$ and $\sigma_{S}$
with the kaons, Eq.\ (\ref{Lagrangian}), reads:%

\begin{align}
\mathcal{L}_{\sigma KK} &  =A_{\sigma_{N}KK}\sigma_{N}(K^{0}\bar{K}^{0}%
+K^{-}K^{+})+B_{\sigma_{N}KK}\sigma_{N}(\partial_{\mu}K^{0}\partial^{\mu}%
\bar{K}^{0}+\partial_{\mu}K^{-}\partial^{\mu}K^{+})\nonumber\\
&  +C_{\sigma_{N}KK}\partial_{\mu}\sigma_{N}(K^{0}\partial^{\mu}\bar{K}%
^{0}+\bar{K}^{0}\partial^{\mu}K^{0}+K^{-}\partial^{\mu}K^{+}+K^{+}%
\partial^{\mu}K^{-})\nonumber\\
&  +A_{\sigma_{S}KK}\sigma_{S}(K^{0}\bar{K}^{0}+K^{-}K^{+})+B_{\sigma_{S}%
KK}\sigma_{S}(\partial_{\mu}K^{0}\partial^{\mu}\bar{K}^{0}+\partial_{\mu}%
K^{-}\partial^{\mu}K^{+})\nonumber\\
&  +C_{\sigma_{S}KK}\partial_{\mu}\sigma_{S}(K^{0}\partial^{\mu}\bar{K}%
^{0}+\bar{K}^{0}\partial^{\mu}K^{0}+K^{-}\partial^{\mu}K^{+}+K^{+}%
\partial^{\mu}K^{-})\nonumber\\
&  =A_{\sigma_{N}KK}\sigma_{N}(K^{0}\bar{K}^{0}+K^{-}K^{+})+(B_{\sigma_{N}%
KK}-2C_{\sigma_{N}KK})\sigma_{N}(\partial_{\mu}K^{0}\partial^{\mu}\bar{K}%
^{0}+\partial_{\mu}K^{-}\partial^{\mu}K^{+})\nonumber\\
&  -C_{\sigma_{N}KK}\sigma_{N}(K^{0}\square\bar{K}^{0}+\bar{K}^{0}\square
K^{0}+K^{-}\square K^{+}+K^{+}\square K^{-})\nonumber\\
&  +A_{\sigma_{S}KK}\sigma_{S}(K^{0}\bar{K}^{0}+K^{-}K^{+})+(B_{\sigma_{S}%
KK}-2C_{\sigma_{S}KK})\sigma_{S}(\partial_{\mu}K^{0}\partial^{\mu}\bar{K}%
^{0}+\partial_{\mu}K^{-}\partial^{\mu}K^{+})\nonumber\\
&  -C_{\sigma_{S}KK}\sigma_{S}(K^{0}\square\bar{K}^{0}+\bar{K}^{0}\square
K^{0}+K^{-}\square K^{+}+K^{+}\square K^{-})\label{sigmakaonkaon}%
\end{align}

with

\begin{align}
A_{\sigma_{N}KK}  &  =\frac{Z_{K}^{2}}{\sqrt{2}}[\lambda_{2}(\phi_{S}-\sqrt
{2}\phi_{N})-2\sqrt{2}\lambda_{1}\phi_{N}]\text{,} \label{ANsKK}\\
B_{\sigma_{N}KK}  &  =\frac{g_{1}}{2}Z_{K}^{2}w_{K_{1}}[-2+g_{1}w_{K_{1}}%
(\phi_{N}+\sqrt{2}\phi_{S})]+\frac{Z_{K}^{2}}{2}w_{K_{1}}^{2}[(2h_{1}%
+h_{2})\phi_{N}-\sqrt{2}h_{3}\phi_{S}]\text{,} \label{BNsKK}\\
C_{\sigma_{N}KK}  &  =\frac{g_{1}}{2}Z_{K}^{2}w_{K_{1}}\text{,} \label{CNsKK}\\
A_{\sigma_{S}KK}  &  =\frac{Z_{K}^{2}}{\sqrt{2}}[\lambda_{2}(\phi_{N}%
-2\sqrt{2}\phi_{S})-2\sqrt{2}\lambda_{1}\phi_{S}]\text{,} \label{ASsKK}
\end{align}
\begin{align}
B_{\sigma_{S}KK}  &  =\frac{\sqrt{2}}{2}Z_{K}^{2}g_{1}w_{K_{1}}[-2+g_{1}%
w_{K_{1}}(\phi_{N}+\sqrt{2}\phi_{S})]+\frac{Z_{K}^{2}}{\sqrt{2}}w_{K_{1}}%
^{2}[\sqrt{2}(h_{1}+h_{2})\phi_{S}-h_{3}\phi_{N}]\text{,} \label{BSsKK}\\
C_{\sigma_{S}KK}  &  =\frac{\sqrt{2}}{2}Z_{K}^{2}g_{1}w_{K_{1}}\equiv\sqrt
{2}C_{\sigma_{N}KK}\text{.} \label{CSsKK}%
\end{align}

Let us consider only the $\sigma_{1,2}\rightarrow K^{0}\bar{K}^{0}$ decay
channel ($\sigma_{1,2}\rightarrow K^{+}K^{-}$ will give the same contribution
to the full decay width due to the isospin symmetry). As in
Eq.\ (\ref{sigmapionpion2}) we obtain from Eqs.\ (\ref{sigma-sigma_2}) and
(\ref{sigmakaonkaon})%

\begin{align}
\mathcal{L}_{\sigma KK\text{, full}}  &  =\mathcal{L}_{\sigma_{N}\sigma
_{S},\,\mathrm{full}}+\mathcal{L}_{\sigma KK}\nonumber\\
&  =\frac{1}{2}(\partial_{\mu}\sigma_{N})^{2}+\frac{1}{2}(\partial_{\mu}%
\sigma_{S})^{2}-\frac{1}{2}m_{\sigma_{N}}^{2}-\frac{1}{2}m_{\sigma_{S}}%
^{2}+z_{\sigma}\sigma_{N}\sigma_{S}\nonumber\\
&  +A_{\sigma_{N}KK}\sigma_{N}K^{0}\bar{K}^{0}+(B_{\sigma_{N}KK}%
-2C_{\sigma_{N}KK})\sigma_{N}\partial_{\mu}K^{0}\partial^{\mu}\bar{K}%
^{0}\nonumber\\
&  -C_{\sigma_{N}KK}\sigma_{N}(K^{0}\square\bar{K}^{0}+\bar{K}^{0}\square
K^{0})\nonumber\\
&  +A_{\sigma_{S}KK}\sigma_{S}K^{0}\bar{K}^{0}+(B_{\sigma_{S}KK}%
-2C_{\sigma_{S}KK})\sigma_{S}\partial_{\mu}K^{0}\partial^{\mu}\bar{K}%
^{0}\nonumber\\
&  -C_{\sigma_{S}KK}\sigma_{S}(K^{0}\square\bar{K}^{0}+\bar{K}^{0}\square
K^{0})\text{.} \label{sigmakaonkaon1}%
\end{align}

Inserting the inverted Eq.\ (\ref{sigma-sigma_1}) into
Eq.\ (\ref{sigmakaonkaon1}), identifying $m_{\sigma_{1,2}}^{2}$ from
Eqs.\ (\ref{m_sigma_1}) and (\ref{m_sigma_2}) and $z_{\sigma}$ from
Eq.\ (\ref{zsigma}) and rearranging parameters as in
Eq.\ (\ref{sigmapionpion3}) leads to

\begin{align}
\mathcal{L}_{\sigma K K \text{, full}}  &  =\frac{1}{2}(\partial_{\mu}%
\sigma_{1})^{2}-\frac{1}{2}m_{\sigma_{1}}^{2}\sigma_{1}^{2}\nonumber\\
&  +(A_{\sigma_{N}KK}\cos\varphi_{\sigma}+A_{\sigma_{S}KK}\sin\varphi_{\sigma
})\sigma_{1}K^{0}\bar{K}^{0}\nonumber\\
&  +[(B_{\sigma_{N}KK}-2C_{\sigma_{N}KK})\cos\varphi_{\sigma}+(B_{\sigma
_{S}KK}-2C_{\sigma_{S}KK})\sin\varphi_{\sigma}]\sigma_{1}\partial_{\mu}%
K^{0}\partial^{\mu}\bar{K}^{0}\nonumber\\
&  -(C_{\sigma_{N}KK}\cos\varphi_{\sigma}+C_{\sigma_{S}KK}\sin\varphi_{\sigma
})\sigma_{1}(K^{0}\square\bar{K}^{0}+\bar{K}^{0}\square K^{0})\nonumber\\
&  +\frac{1}{2}(\partial_{\mu}\sigma_{2})^{2}-\frac{1}{2}m_{\sigma_{2}}%
^{2}\sigma_{2}^{2}\nonumber\\
&  +(A_{\sigma_{S}KK}\cos\varphi_{\sigma}-A_{\sigma_{N}KK}\sin\varphi_{\sigma
})\sigma_{2}K^{0}\bar{K}^{0}\nonumber\\
&  +[(B_{\sigma_{S}KK}-2C_{\sigma_{S}KK})\cos\varphi_{\sigma}-(B_{\sigma
_{N}KK}-2C_{\sigma_{N}KK})\sin\varphi_{\sigma}]\sigma_{2}\partial_{\mu}%
K^{0}\partial^{\mu}\bar{K}^{0}\nonumber\\
&  -(C_{\sigma_{S}KK}\cos\varphi_{\sigma}-C_{\sigma_{N}KK}\sin\varphi_{\sigma
})\sigma_{2}(K^{0}\square\bar{K}^{0}+\bar{K}^{0}\square K^{0})\text{.} \label{sigmakaonkaon2}%
\end{align}

Let us denote the momenta of the two kaons as $P_{1}$ and $P_{2}$. Energy
conservation on the vertex implies $P=P_{1}+P_{2}$, where $P$ denotes the momentum of
$\sigma_{1}$ or $\sigma_{2}$; given that our particles are on-shell, we obtain

\begin{equation}
P_{1}\cdot P_{2}=\frac{P^{2}-P_{1}^{2}-P_{2}^{2}}{2}\equiv\frac{m_{\sigma_{1}
}^{2}-2m_{K}^{2}}{2}\text{.}
\end{equation}

Then the decay amplitudes of the mixed states $\sigma_{1,2}$ read%

\begin{align}
-i\mathcal{M}_{\sigma_{1}\rightarrow K^{0}\bar{K}^{0}}(m_{\sigma_{1}})  &
=i\left\{  \cos\varphi_{\sigma}\left[  A_{\sigma_{N}KK}-(B_{\sigma_{N}%
KK}-2C_{\sigma_{N}KK})P_{1}\cdot P_{2}+2C_{\sigma_{N}KK}m_{K}^{2}\right]
\right. \nonumber\\
&  \left.  +\sin\varphi_{\sigma}\left[  A_{\sigma_{S}KK}-(B_{\sigma_{S}%
KK}-2C_{\sigma_{S}KK})\frac{m_{\sigma_{1}}^{2}-2m_{K}^{2}}{2}+2C_{\sigma
_{S}KK}m_{K}^{2}\right]  \right\} \nonumber\\
&  =i\left\{  \cos\varphi_{\sigma}\left[  A_{\sigma_{N}KK}-(B_{\sigma_{N}%
KK}-2C_{\sigma_{N}KK})\frac{m_{\sigma_{1}}^{2}-2m_{K}^{2}}{2}+2C_{\sigma
_{N}KK}m_{K}^{2}\right]  \right. \nonumber\\
&  \left.  +\sin\varphi_{\sigma}\left[  A_{\sigma_{S}KK}-(B_{\sigma_{S}%
KK}-2C_{\sigma_{S}KK})\frac{m_{\sigma_{1}}^{2}-2m_{K}^{2}}{2}+2C_{\sigma
_{S}KK}m_{K}^{2}\right]  \right\} \label{Ms1KK}
\end{align}
and
\begin{align}
-i\mathcal{M}_{\sigma_{2}\rightarrow K^{0}\bar{K}^{0}}(m_{\sigma_{2}})  &  =i\left\{
\cos\varphi_{\sigma}\left[  A_{\sigma_{S}KK}-(B_{\sigma_{S}KK}-2C_{\sigma
_{S}KK})\frac{m_{\sigma_{2}}^{2}-2m_{K}^{2}}{2}+2C_{\sigma_{S}KK}m_{K}%
^{2}\right]  \right. \nonumber\\
&  \left.  -\sin\varphi_{\sigma}\left[  A_{\sigma_{N}KK}-(B_{\sigma_{N}%
KK}-2C_{\sigma_{N}KK})\frac{m_{\sigma_{2}}^{2}-2m_{K}^{2}}{2}+2C_{\sigma
_{N}KK}m_{K}^{2}\right]  \right\}  \text{.} \label{Ms2KK}%
\end{align}

%

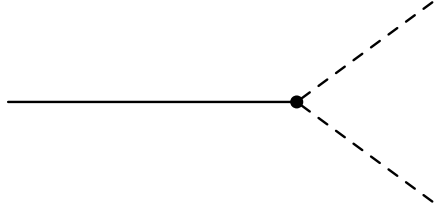
\begin{figure}[h]
 \begin{align*}
\qquad \qquad \qquad  \qquad \qquad \qquad  \quad \;  \parbox{180mm}{ \begin{fmfgraph*}(180,80)
                  \fmfleftn{i}{1}\fmfrightn{o}{2}
                  \fmf{vanilla,label=\text{\small\(\sigma_{1,,2}(P)\)},label.dist=-18}{i1,v1}
\fmf{dashes,label=\text{\small\(\bar{K}^0(P_2)\)},label.dist=-28}{v1,o1}
\fmf{dashes,label=\text{\small\(K^0(P_1)\)},label.dist=-28}{v1,o2}\fmfdot{v1}
                 \end{fmfgraph*}}
 \end{align*}
\caption{Decay process $\sigma_{1,2}\rightarrow K^{0}\bar{K}^{0}$.}
\end{figure}

Finally, taking into account all contributions to the decay widths of the
mixed states $\sigma_{1,2}$, i.e., $\sigma_{1,2}\rightarrow K^{0}\bar{K}%
^{0}+K^{-}K^{+}$, we obtain%

\begin{align}
\Gamma_{\sigma_{1}\rightarrow KK}  &  =\frac{k(m_{\sigma_{1}},m_{K},m_{K}%
)}{4\pi m_{\sigma_{1}}^{2}}|-i\mathcal{M}_{\sigma_{1}\rightarrow K^{0}\bar
{K}^{0}}(m_{\sigma_{1}})|^{2}\text{,} \label{Gs1KK}\\
\Gamma_{\sigma_{2}\rightarrow KK}  &  =\frac{k(m_{\sigma_{2}},m_{K},m_{K}%
)}{4\pi m_{\sigma_{2}}^{2}}|-i\mathcal{M}_{\sigma_{2}\rightarrow K^{0}\bar
{K}^{0}}(m_{\sigma_{2}})|^{2}\text{.} \label{Gs2KK}%
\end{align}

\begin{figure}[h]
  \begin{center}
    \begin{tabular}{cc}
      \resizebox{78mm}{!}{\includegraphics{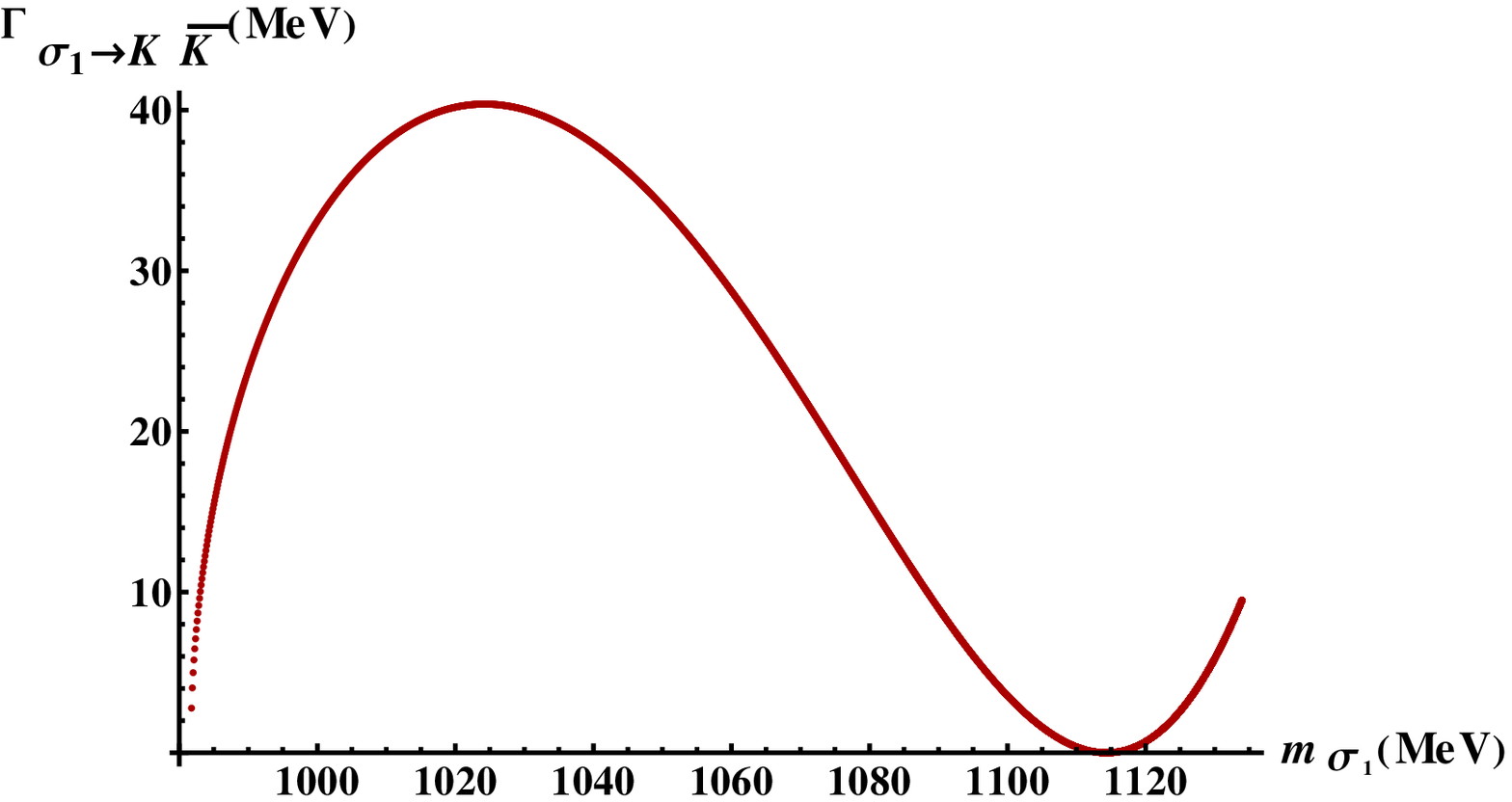}} &
      \resizebox{78mm}{!}{\includegraphics{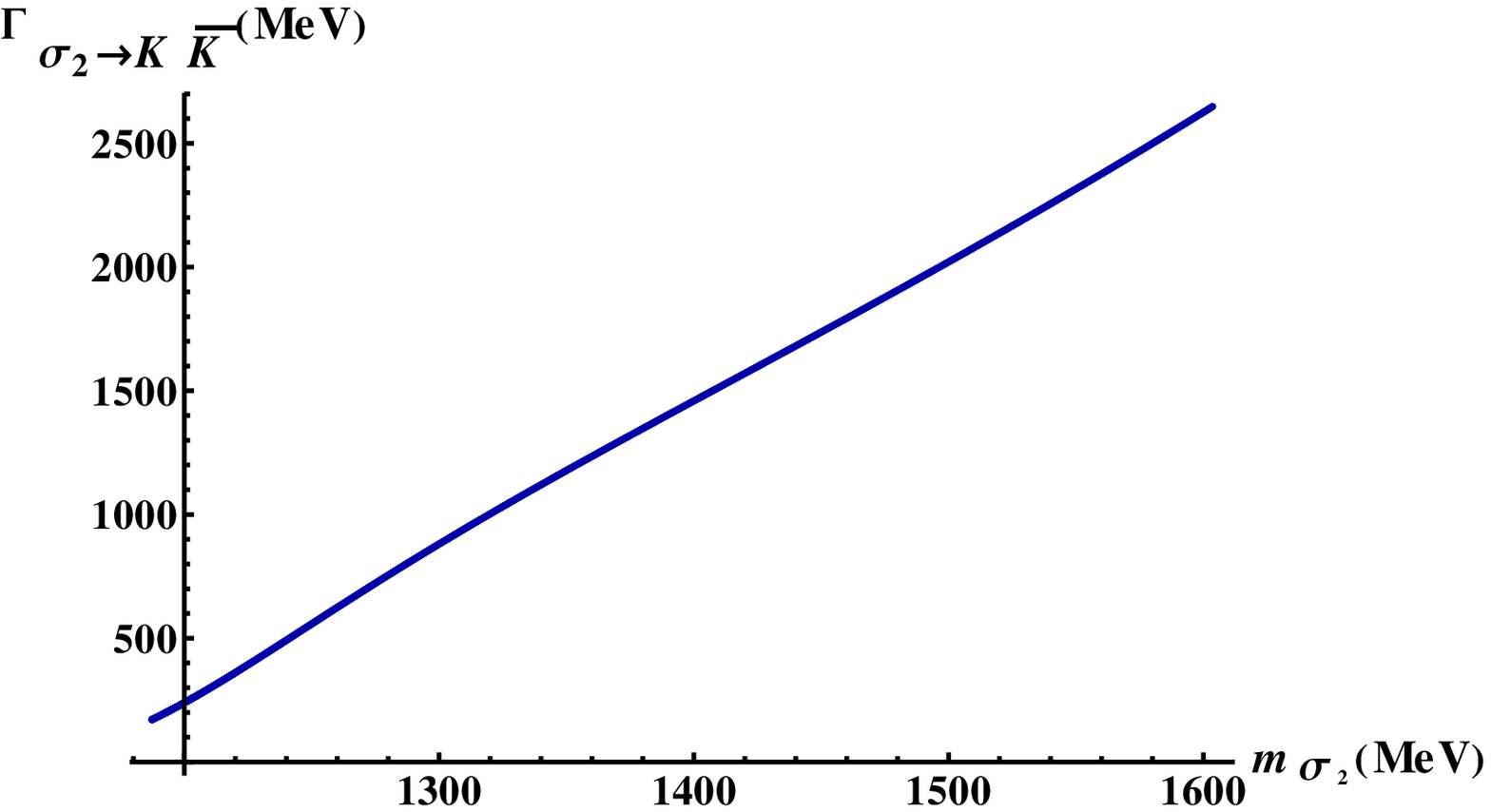}} 
    \end{tabular}
    \caption{$\Gamma_{\sigma_{1}\rightarrow KK}$ and $\Gamma_{\sigma
_{2}\rightarrow KK}$ as functions of $m_{\sigma_{1}}$ and $m_{\sigma_{2}}$,
respectively.}
    \label{SKK1}
  \end{center}
\end{figure}
$\,$\\
The decay widths are depicted in Fig.\ \ref{SKK1}. The kaon-kaon threshold opens at $981.7$ MeV, see Table \ref{Fit1-5}.
Therefore, the decay $\sigma_{1}\rightarrow KK$ is phase-space suppressed.
Nonetheless, these results suggest that $\sigma_{1}\equiv f_{0}(600)$ also
decays into kaons above the threshold, which is in principle possible but has
not been observed \cite{PDG}. The results regarding $\sigma_{1}$\ are actually
more consistent with the decay $f_{0}(980)\rightarrow KK$ but an
interpretation of $\sigma_{1}$\ as $f_{0}(980)$ is not possible due to the
results in the two-pion channel, see previous subsection
\ref{sec.sigmapionpion1} and Fig.\ \ref{Spp1}. We can therefore conclude that
a proper determination of $\Gamma_{\sigma_{1}\rightarrow KK}$ is not possible
if one only considers $\sigma_{1}$ -- we have to consider results regarding
$\sigma_{2}\rightarrow KK$ as well.

The decay width $\Gamma_{\sigma_{2}\rightarrow KK}$ rises rapidly with
$m_{\sigma_{2}}$. The lowest value is $\Gamma_{\sigma_{2}\rightarrow KK}=171$
MeV at $m_{\sigma_{2}}=1187$ MeV, a value consistent with experiment
\cite{Etkin:1981,f01370KK2,Tikhomirov:2003,Polychronakos:1978,f01370KK1,f01370KK3}. Our analysis of the $\sigma_{2}%
\rightarrow\pi\pi$ decay yielded three values of the $m_{\sigma_{2}}$ where
the correspondence with the decay width $f_{0}(1370)\rightarrow\pi\pi$ was
particularly good: $m_{\sigma_{2}}=1200$ MeV, $m_{\sigma_{2}}=1341$ MeV and
$m_{\sigma_{2}}=1368$ MeV. In the $\sigma_{2}\rightarrow KK$ channel we obtain
$\Gamma_{\sigma_{2}\rightarrow KK}=240$ MeV for $m_{\sigma_{2}}=1200$ MeV,
$\Gamma_{\sigma_{2}\rightarrow KK}=1125$ MeV for $m_{\sigma_{2}}=1341$ MeV and
$\Gamma_{\sigma_{2}\rightarrow KK}=1281$ MeV for $m_{\sigma_{2}}=1368$ MeV.
(The width rises to $\Gamma_{\sigma_{2}\rightarrow KK}=2021$ MeV for
$m_{\sigma_{2}}=1500$ MeV.) Thus, there is some discrepancy between the
results in the $\sigma_{2}\rightarrow\pi\pi$ and $\sigma_{2}\rightarrow KK$
channels, unless one works with%

\begin{equation}
m_{\sigma_{2}}=1200\text{ MeV.} \label{ms21}%
\end{equation}

The latter mass implies $m_{0}^{2}=-160233$ MeV from Eq.\ (\ref{m_sigma_2})
leading to%

\begin{equation}
m_{\sigma_{1}}=705\text{ MeV} \label{ms11}%
\end{equation}

via Eq.\ (\ref{m_sigma_1}). Consequently, $\Gamma_{\sigma_{1}\equiv
f_{0}(600)\rightarrow KK}=0$ as $m_{\sigma_{1}}$ is below the kaon-kaon decay threshold,
a result in accordance with experimental data and also consistent with the
assignment $\sigma_{1}\equiv f_{0}(600)$. Note that $m_{\sigma_{1}}=705$ MeV
also yields $\Gamma_{\sigma_{1}\rightarrow\pi\pi}=305$ MeV via
Eq.\ (\ref{Gs1pp}).

\begin{figure}
[h]
\begin{center}
\includegraphics[
height=2.3582in,
width=3.7666in
]%
{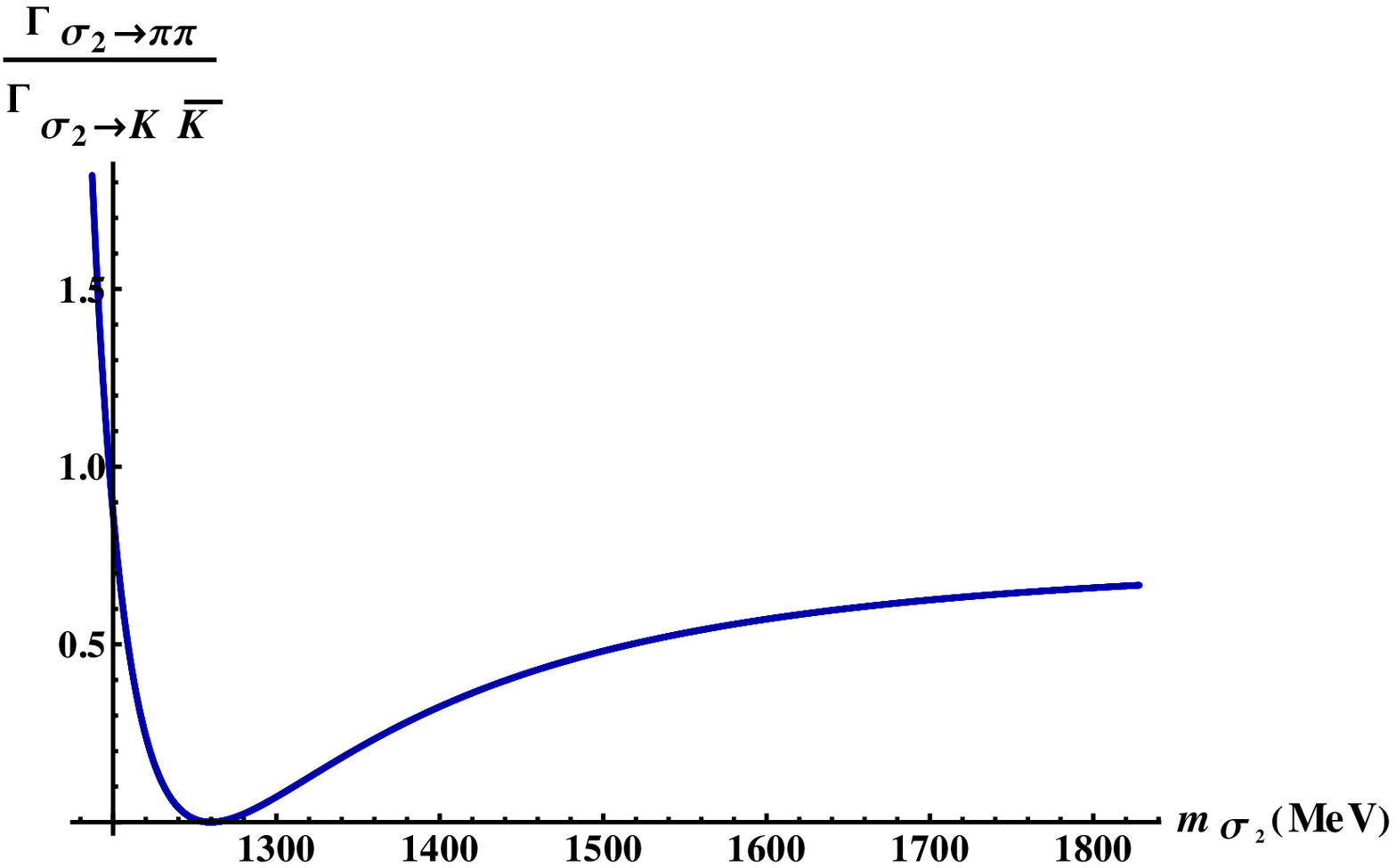}%
\caption{Ratio $\Gamma_{\sigma_{2}\rightarrow\pi\pi}/\Gamma_{\sigma
_{2}\rightarrow KK}$ as function of $m_{\sigma_{2}}$.}%
\label{SPPKK1}%
\end{center}
\end{figure}

We can also look into results regarding the ratio $\Gamma_{\sigma
_{2}\rightarrow KK}/\Gamma_{\sigma_{2}\rightarrow\pi\pi}$, see
Fig.\ \ref{SPPKK1}. Experimental data about this ratio are by far inconclusive
\cite{Barberis:1999,Ablikim:2004,Anisovich:2001,Bargiotti:2003}; we observe $\Gamma_{\sigma_{2}\rightarrow
KK}/\Gamma_{\sigma_{2}\rightarrow\pi\pi}=1.15$ at $m_{\sigma_{2}}=1200$ MeV,
larger than any set of experimental data reported so far. The reason is the
relatively large decay width $\sigma_{2}\rightarrow KK$; in fact, we observe
from Fig.\ \ref{SPPKK1} that $\Gamma_{\sigma_{2}\rightarrow\pi\pi}%
<\Gamma_{\sigma_{2}\rightarrow KK}$ at all values of $m_{\sigma_{2}}$ except
for the lowest ones ($\lesssim1200$ MeV). This would imply that $f_{0}(1370)$
decays predominantly into kaons, as one would expect from a $\bar{s}s$ state,
but it is clearly at odds with the data \cite{PDG}.%

\subsection{Decay Width \boldmath $\sigma_{1,2}\rightarrow\eta\eta$} \label{sec.setaeta1}

The Lagrangian (\ref{Lagrangian}) contains only the pure fields $\sigma_{N,S}$ and
$\eta_{N,S}$; the corresponding interaction Lagrangian reads:%

\begin{align}
\mathcal{L}_{\sigma\eta_{N}\eta_{S}}  &  =-Z_{\pi}^{2}\phi_{N}\left(
\lambda_{1}+\frac{\lambda_{2}}{2}+c_{1}\phi_{S}^{2}\right)  \sigma_{N}\eta
_{N}^{2}-Z_{\eta_{S}}^{2}\phi_{N}\left(  \lambda_{1}+\frac{c_{1}}{2}\phi
_{N}^{2}\right)  \sigma_{N}\eta_{S}^{2}\nonumber\\
&  -\frac{3}{2}c_{1}Z_{\pi}Z_{\eta_{S}}\phi_{N}^{2}\phi_{S}\sigma_{N}\eta
_{N}\eta_{S}\nonumber\\
&  +Z_{\pi}^{2}w_{a_{1}}\left[  g_{1}(g_{1}w_{a_{1}}\phi_{N}-1)+\frac{\phi
_{N}}{2}w_{a_{1}}(h_{1}+h_{2}-h_{3})\right]  \sigma_{N}(\partial_{\mu}\eta
_{N})^{2}\nonumber\\
&  +\frac{h_{1}}{2}Z_{\eta_{S}}^{2}w_{f_{1S}}^{2}\phi_{N}\sigma_{N}%
(\partial_{\mu}\eta_{S})^{2}+g_{1}w_{a_{1}}Z_{\pi}^{2}\partial_{\mu}\sigma
_{N}\partial^{\mu}\eta_{N}\eta_{N}\nonumber\\
&  -(\lambda_{1}+\lambda_{2})Z_{\eta_{S}}^{2}\phi_{S}\sigma_{S}\eta_{S}%
^{2}-Z_{\pi}^{2}\phi_{S}(\lambda_{1}+c_{1}\phi_{N}^{2})\sigma_{S}\eta_{N}%
^{2}-\frac{1}{2}c_{1}Z_{\pi}Z_{\eta_{S}}\phi_{N}^{3}\sigma_{S}\eta_{N}\eta
_{S}\nonumber\\
&  +Z_{\eta_{S}}^{2}w_{f_{1S}}\left[  \sqrt{2}g_{1}(\sqrt{2}g_{1}w_{f_{1S}%
}\phi_{S}-1)+w_{f_{1S}}\phi_{S}\left(  \frac{h_{1}}{2}+h_{2}-h_{3}\right)
\right]  \sigma_{S}(\partial_{\mu}\eta_{S})^{2}\nonumber\\
&  +\frac{h_{1}}{2}Z_{\pi}^{2}w_{a_{1}}^{2}\phi_{S}\sigma_{S}(\partial_{\mu
}\eta_{N})^{2}+\sqrt{2}g_{1}Z_{\eta_{S}}^{2}w_{f_{1S}}\partial_{\mu}\sigma
_{S}\partial^{\mu}\eta_{S}\eta_{S}\text{.} \label{sigmaetaNetaS}%
\end{align}

As in the case of $\mathcal{L}_{\sigma\pi\pi}$, Eq.\ (\ref{sigmapionpion}),
decays of the pure non-strange state $\sigma_{N}\rightarrow\eta_{S}\eta_{S}$
and of the pure strange state $\sigma_{S}\rightarrow\eta_{N}\eta_{N}$\ are
driven by the large-$N_{c}$ suppressed couplings $\lambda_{1}$ and $h_{1}$, see
Eq.\ (\ref{largen}).

Note that the coupling of $\sigma_{N}$ to $(\partial_{\mu}\eta_{N})^{2}$ in
Eq.\ (\ref{sigmaetaNetaS}) can be transformed in the following way:%

\begin{align}
&  Z_{\pi}^{2}w_{a_{1}}\left[  g_{1}(g_{1}w_{a_{1}}\phi_{N}-1)+\frac{\phi_{N}%
}{2}w_{a_{1}}(h_{1}+h_{2}-h_{3})\right] \nonumber\\
&  =Z_{\pi}^{2}w_{a_{1}}\left\{  w_{a_{1}}\left[  g_{1}^{2}\phi_{N}+\frac
{\phi_{N}}{2}(h_{1}+h_{2}-h_{3})\right]  -g_{1}\right\} \nonumber
\end{align}
\begin{align}
&  \overset{\text{Eqs.\ (\ref{m_a_1})}}{=}Z_{\pi}^{2}w_{a_{1}}\left(
w_{a_{1}}\frac{m_{a_{1}}^{2}-m_{1}^{2}-\frac{h_{1}}{2}\phi_{S}^{2}-2\delta
_{N}}{\phi_{N}}-g_{1}\right) \nonumber\\
&  \overset{\text{Eqs.\ (\ref{wa1})}}{=}-Z_{\pi}^{2}\frac{w_{a_{1}}^{2}}%
{\phi_{N}}\left(  m_{1}^{2}+\frac{h_{1}}{2}\phi_{S}^{2}+2\delta_{N}\right)
\text{,} \label{p}%
\end{align}
$\,$\\
where we have used $w_{a_{1}}m_{a_{1}}^{2}/\phi_{N}=g_{1}$, and that the
coupling of $\sigma_{S}$ to $(\partial_{\mu}\eta_{S})^{2}$ in
Eq.\ (\ref{sigmaetaNetaS}) can be similarly transformed in the following way:%

\begin{align}
&  Z_{\eta_{S}}^{2}w_{f_{1S}}\left[  \sqrt{2}g_{1}(\sqrt{2}g_{1}w_{f_{1S}}%
\phi_{S}-1)+w_{f_{1S}}\phi_{S}\left(  \frac{h_{1}}{2}+h_{2}-h_{3}\right)
\right] \nonumber\\
&  =Z_{\eta_{S}}^{2}w_{f_{1S}}\left\{  w_{f_{1S}}\left[  2g_{1}^{2}\phi
_{S}+\phi_{S}\left(  \frac{h_{1}}{2}+h_{2}-h_{3}\right)  \right]  -\sqrt
{2}g_{1}\right\} \nonumber\\
&  \overset{\text{Eqs.\ (\ref{m_f1_S})}}{=}Z_{\eta_{S}}^{2}w_{f_{1S}}\left(
w_{f_{1S}}\frac{m_{f_{1S}}^{2}-m_{1}^{2}-\frac{h_{1}}{2}\phi_{N}^{2}%
-2\delta_{S}}{\phi_{S}}-\sqrt{2}g_{1}\right) \nonumber\\
&  \overset{\text{Eqs.\ (\ref{wf1S})}}{=}-Z_{\eta_{S}}^{2}\frac{w_{f_{1S}}%
^{2}}{\phi_{S}}\left(  m_{1}^{2}+\frac{h_{1}}{2}\phi_{N}^{2}+2\delta
_{S}\right)\text{,}  \label{a}%
\end{align}

where we have used $w_{f_{1S}}m_{f_{1S}}^{2}/(\sqrt{2}\phi_{S})=g_{1}$.
Substituting Eqs.\ (\ref{p}) and (\ref{a}) into Eq.\ (\ref{sigmaetaNetaS}) and
additionally substituting $\eta_{N}$ and $\eta_{S}$ by $\eta$ and
$\eta^{\prime}$ according to Eqs.\ (\ref{etaN}) and (\ref{etaS}), we obtain
the following form of the interaction Lagrangian:
\begin{align}
\mathcal{L}_{\sigma\eta\eta}  &  =A_{\sigma_{N}\eta\eta}\sigma_{N}\eta
^{2}+B_{\sigma_{N}\eta\eta}\sigma_{N}(\partial_{\mu}\eta)^{2}+C_{\sigma
_{N}\eta\eta}\partial_{\mu}\sigma_{N}\partial^{\mu}\eta\eta\nonumber\\
&  +A_{\sigma_{S}\eta\eta}\sigma_{S}\eta^{2}+B_{\sigma_{S}\eta\eta}\sigma
_{S}(\partial_{\mu}\eta)^{2}+C_{\sigma_{S}\eta\eta}\partial_{\mu}\sigma
_{S}\partial^{\mu}\eta\eta\label{sigmaetaeta}%
\end{align}

with%

\begin{align}
A_{\sigma_{N}\eta\eta}  &  =-Z_{\pi}^{2}\phi_{N}\left(  \lambda_{1}%
+\frac{\lambda_{2}}{2}+c_{1}\phi_{S}^{2}\right)  \cos^{2}\varphi_{\eta
}-Z_{\eta_{S}}^{2}\phi_{N}\left(  \lambda_{1}+\frac{c_{1}}{2}\phi_{N}%
^{2}\right)  \sin^{2}\varphi_{\eta}\nonumber\\
&  -\frac{3}{4}c_{1}Z_{\pi}Z_{\eta_{S}}\phi_{N}^{2}\phi_{S}\sin(2\varphi
_{\eta})\nonumber\\
&  =-\phi_{N}\left\{  \lambda_{1}(Z_{\pi}^{2}\cos^{2}\varphi_{\eta}%
+Z_{\eta_{S}}^{2}\sin^{2}\varphi_{\eta})+\frac{\lambda_{2}}{2}Z_{\pi}^{2}%
\cos^{2}\varphi_{\eta}\right. \nonumber\\
&  \left.  +c_{1}\left[  \frac{Z_{\eta_{S}}^{2}}{2}\phi_{N}^{2}\sin^{2}%
\varphi_{\eta}+Z_{\pi}^{2}\phi_{S}^{2}\cos^{2}\varphi_{\eta}+\frac{3}{4}%
Z_{\pi}Z_{\eta_{S}}\phi_{N}\phi_{S}\sin(2\varphi_{\eta})\right]  \right\}\text{,}
\label{ANsetaeta}\\
B_{\sigma_{N}\eta\eta}  &  =-Z_{\pi}^{2}\frac{w_{a_{1}}^{2}}{\phi_{N}}\left(
m_{1}^{2}+\frac{h_{1}}{2}\phi_{S}^{2}+2\delta_{N}\right)  \cos^{2}%
\varphi_{\eta}+\frac{h_{1}}{2}Z_{\eta_{S}}^{2}w_{f_{1S}}^{2}\phi_{N}\sin
^{2}\varphi_{\eta}\text{,} \label{BNsetaeta}\\
C_{\sigma_{N}\eta\eta}  &  =g_{1}w_{a_{1}}Z_{\pi}^{2}\cos^{2}\varphi_{\eta
}\text{,} \label{CNsetaeta}\\
A_{\sigma_{S}\eta\eta}  &  =-(\lambda_{1}+\lambda_{2})Z_{\eta_{S}}^{2}\phi
_{S}\sin^{2}\varphi_{\eta}-Z_{\pi}^{2}\phi_{S}(\lambda_{1}+c_{1}\phi_{N}%
^{2})\cos^{2}\varphi_{\eta}-\frac{1}{4}c_{1}Z_{\pi}Z_{\eta_{S}}\phi_{N}%
^{3}\sin(2\varphi_{\eta})\nonumber\\
&  =-\lambda_{1}\phi_{S}(Z_{\eta_{S}}^{2}\sin^{2}\varphi_{\eta}+Z_{\pi}%
^{2}\cos^{2}\varphi_{\eta})-Z_{\eta_{S}}^{2}\lambda_{2}\phi_{S}\sin^{2}%
\varphi_{\eta}\nonumber\\
&  -Z_{\pi}c_{1}\phi_{N}\left[  Z_{\pi}\phi_{N}\phi_{S}\cos^{2}\varphi_{\eta
}+\frac{1}{4}Z_{\eta_{S}}\phi_{N}^{2}\sin(2\varphi_{\eta})\right]\text{,}
\label{ASsetaeta}\\
B_{\sigma_{S}\eta\eta}  &  =-Z_{\eta_{S}}^{2}\frac{w_{f_{1S}}^{2}}{\phi_{S}%
}\left(  m_{1}^{2}+\frac{h_{1}}{2}\phi_{N}^{2}+2\delta_{S}\right)  \sin
^{2}\varphi_{\eta}+\frac{h_{1}}{2}Z_{\pi}^{2}w_{a_{1}}^{2}\phi_{S}\cos
^{2}\varphi_{\eta}\text{,} \label{BSsetaeta}
\end{align}
\begin{align}
C_{\sigma_{S}\eta\eta}  &  =\sqrt{2}Z_{\eta_{S}}^{2}g_{1}w_{f_{1S}}\sin
^{2}\varphi_{\eta}\text{.} \label{CSsetaeta}%
\end{align}
$\,$\\
As in Eq.\ (\ref{sigmapionpion2}) we obtain from Eqs.\ (\ref{sigma-sigma_2})
and (\ref{sigmaetaeta})

\begin{align}
\mathcal{L}_{\sigma\eta\eta\text{, full}}  &  =\mathcal{L}_{\sigma_{N}%
\sigma_{S},\,\mathrm{full}}+\mathcal{L}_{\sigma\eta\eta}\nonumber\\
&  =\frac{1}{2}(\partial_{\mu}\sigma_{N})^{2}+\frac{1}{2}(\partial_{\mu}%
\sigma_{S})^{2}-\frac{1}{2}m_{\sigma_{N}}^{2}-\frac{1}{2}m_{\sigma_{S}}%
^{2}+z_{\sigma}\sigma_{N}\sigma_{S}\nonumber\\
&  +A_{\sigma_{N}\eta\eta}\sigma_{N}\eta^{2}+B_{\sigma_{N}\eta\eta}\sigma
_{N}(\partial_{\mu}\eta)^{2}+C_{\sigma_{N}\eta\eta}\partial_{\mu}\sigma
_{N}\partial^{\mu}\eta\eta\nonumber\\
&  +A_{\sigma_{S}\eta\eta}\sigma_{S}\eta^{2}+B_{\sigma_{S}\eta\eta}\sigma
_{S}(\partial_{\mu}\eta)^{2}+C_{\sigma_{S}\eta\eta}\partial_{\mu}\sigma
_{S}\partial^{\mu}\eta\eta\text{.} \label{sigmaetaeta1}
\end{align}
$\,$\\
Substituting $\sigma_{N,S}$ by $\sigma_{1,2}$ we obtain

\begin{align}
\mathcal{L}_{\sigma\eta\eta\text{, full}} &  =\frac{1}{2}(\partial_{\mu}
\sigma_{1})^{2}-\frac{1}{2}m_{\sigma_{1}}^{2}\sigma_{1}^{2}\nonumber\\
&  +(A_{\sigma_{N}\eta\eta}\cos\varphi_{\sigma}+A_{\sigma_{S}\eta\eta}
\sin\varphi_{\sigma})\sigma_{1}\eta^{2}\nonumber\\
&  +(B_{\sigma_{N}\eta\eta}\cos\varphi_{\sigma}+B_{\sigma_{S}\eta\eta}
\sin\varphi_{\sigma})\sigma_{1}(\partial_{\mu}\eta)^{2}\nonumber\\
&  +(C_{\sigma_{N}\eta\eta}\cos\varphi_{\sigma}+C_{\sigma_{S}\eta\eta}
\sin\varphi_{\sigma})\partial_{\mu}\sigma_{1}\partial^{\mu}\eta\eta\nonumber\\
&  +\frac{1}{2}(\partial_{\mu}\sigma_{2})^{2}-\frac{1}{2}m_{\sigma_{2}}
^{2}\sigma_{2}^{2}\nonumber\\
&  +(A_{\sigma_{S}\eta\eta}\cos\varphi_{\sigma}-A_{\sigma_{N}\eta\eta}
\sin\varphi_{\sigma})\sigma_{2}\eta^{2}\nonumber\\
&  +(B_{\sigma_{S}\eta\eta}\cos\varphi_{\sigma}-B_{\sigma_{N}\eta\eta}
\sin\varphi_{\sigma})\sigma_{2}(\partial_{\mu}\eta)^{2}\nonumber\\
&  +(C_{\sigma_{S}\eta\eta}\cos\varphi_{\sigma}-C_{\sigma_{N}\eta\eta}
\sin\varphi_{\sigma})\partial_{\mu}\sigma_{2}\partial^{\mu}\eta\eta\text{.}
\end{align}

\begin{figure}[h]
 \begin{align*}
\qquad \qquad \qquad  \qquad \qquad \qquad  \quad \; \parbox{180mm}{ \begin{fmfgraph*}(180,80)
                  \fmfleftn{i}{1}\fmfrightn{o}{2}
                  \fmf{vanilla,label=\text{\small\(\sigma_{1,,2}(P)\)},label.dist=-18}{i1,v1}\fmf{dashes,label=\text{\small\(\eta(P_2)\)},label.dist=-28}{v1,o1}\fmf{dashes,label=\text{\small\(\eta(P_1)\)},label.dist=-28}{v1,o2}\fmfdot{v1}
                 \end{fmfgraph*}}
 \end{align*}
\caption{Decay process $\sigma_{1,2}\rightarrow \eta \eta$.}
\end{figure}
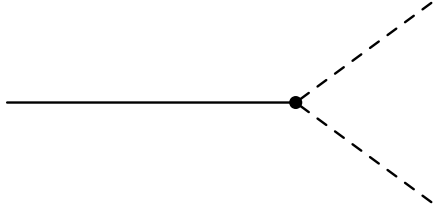
$\,$\\

Let us set $P$ as momentum of $\sigma_{1}$ or $\sigma_{2}$ (depending on the
decaying particle) and $P_{1}$ and $P_{2}$ as the momenta of the $\eta$ fields.
Upon substituting $\partial^{\mu}\rightarrow-iP^{\mu}$\ for the decaying
particles and $\partial^{\mu}\rightarrow iP_{1,2}^{\mu}$ for the decay products,
the decay amplitudes of the mixed states $\sigma_{1,2}$ read

\begin{align}
-i\mathcal{M}_{\sigma_{1}\rightarrow\eta\eta}(m_{\sigma_{1}}) &  =i\left\{
\cos\varphi_{\sigma}(A_{\sigma_{N}\eta\eta}-B_{\sigma_{N}\eta\eta}P_{1}\cdot
P_{2}+C_{\sigma_{N}\eta\eta}P\cdot P_{1})\right.  \nonumber\\
&  \left.  +\sin\varphi_{\sigma}\left[  A_{\sigma_{S}\eta\eta}-B_{\sigma
_{S}\eta\eta}\frac{m_{\sigma_{1}}^{2}-2m_{\eta}^{2}}{2}+C_{\sigma_{S}\eta\eta
}\frac{m_{\sigma_{1}}^{2}}{2}\right]  \right\}  \nonumber
\end{align}
\begin{align}
&  =i\left\{  \cos\varphi_{\sigma}\left[  A_{\sigma_{N}\eta\eta}-B_{\sigma
_{N}\eta\eta}\frac{m_{\sigma_{1}}^{2}-2m_{\eta}^{2}}{2}+C_{\sigma_{N}\eta\eta
}\frac{m_{\sigma_{1}}^{2}}{2}\right]  \right.  \nonumber\\
&  \left.  +\sin\varphi_{\sigma}\left[  A_{\sigma_{S}\eta\eta}-B_{\sigma
_{S}\eta\eta}\frac{m_{\sigma_{1}}^{2}-2m_{\eta}^{2}}{2}+C_{\sigma_{S}\eta\eta
}\frac{m_{\sigma_{1}}^{2}}{2}\right]  \right\}\text{,} \label{Ms1etaeta}\\
-i\mathcal{M}_{\sigma_{2}\rightarrow\eta\eta}(m_{\sigma_{2}}) &  =i\left\{
\cos\varphi_{\sigma}\left[  A_{\sigma_{S}\eta\eta}-B_{\sigma_{S}\eta\eta}%
\frac{m_{\sigma_{2}}^{2}-2m_{\eta}^{2}}{2}+C_{\sigma_{S}\eta\eta}%
\frac{m_{\sigma_{2}}^{2}}{2}\right]  \right.  \nonumber\\
&  \left.  -\sin\varphi_{\sigma}\left[  A_{\sigma_{N}\eta\eta}-B_{\sigma
_{N}\eta\eta}\frac{m_{\sigma_{2}}^{2}-2m_{\eta}^{2}}{2}+C_{\sigma_{N}\eta\eta
}\frac{m_{\sigma_{2}}^{2}}{2}\right]  \right\}  \text{.}\label{Ms2etaeta}%
\end{align}
%

Finally, we obtain the following formulas for the decay widths:

\begin{align}
\Gamma_{\sigma_{1}\rightarrow \eta \eta}  &  =\frac{k(m_{\sigma_{1}},m_{\eta},m_{\eta}
)}{8\pi m_{\sigma_{1}}^{2}}|-i\mathcal{M}_{\sigma_{1}\rightarrow\eta\eta
}(m_{\sigma_{1}})|^{2}\text{,} \label{Gs1etaeta}\\
\Gamma_{\sigma_{2}\rightarrow \eta \eta}  &  =\frac{k(m_{\sigma_{2}},m_{\eta},m_{\eta}
)}{8\pi m_{\sigma_{2}}^{2}}|-i\mathcal{M}_{\sigma_{2}\rightarrow\eta\eta
}(m_{\sigma_{2}})|^{2}\text{.} \label{Gs2etaeta}
\end{align}

\begin{figure}[t]
  \begin{center}
    \begin{tabular}{cc}
      \resizebox{78mm}{!}{\includegraphics{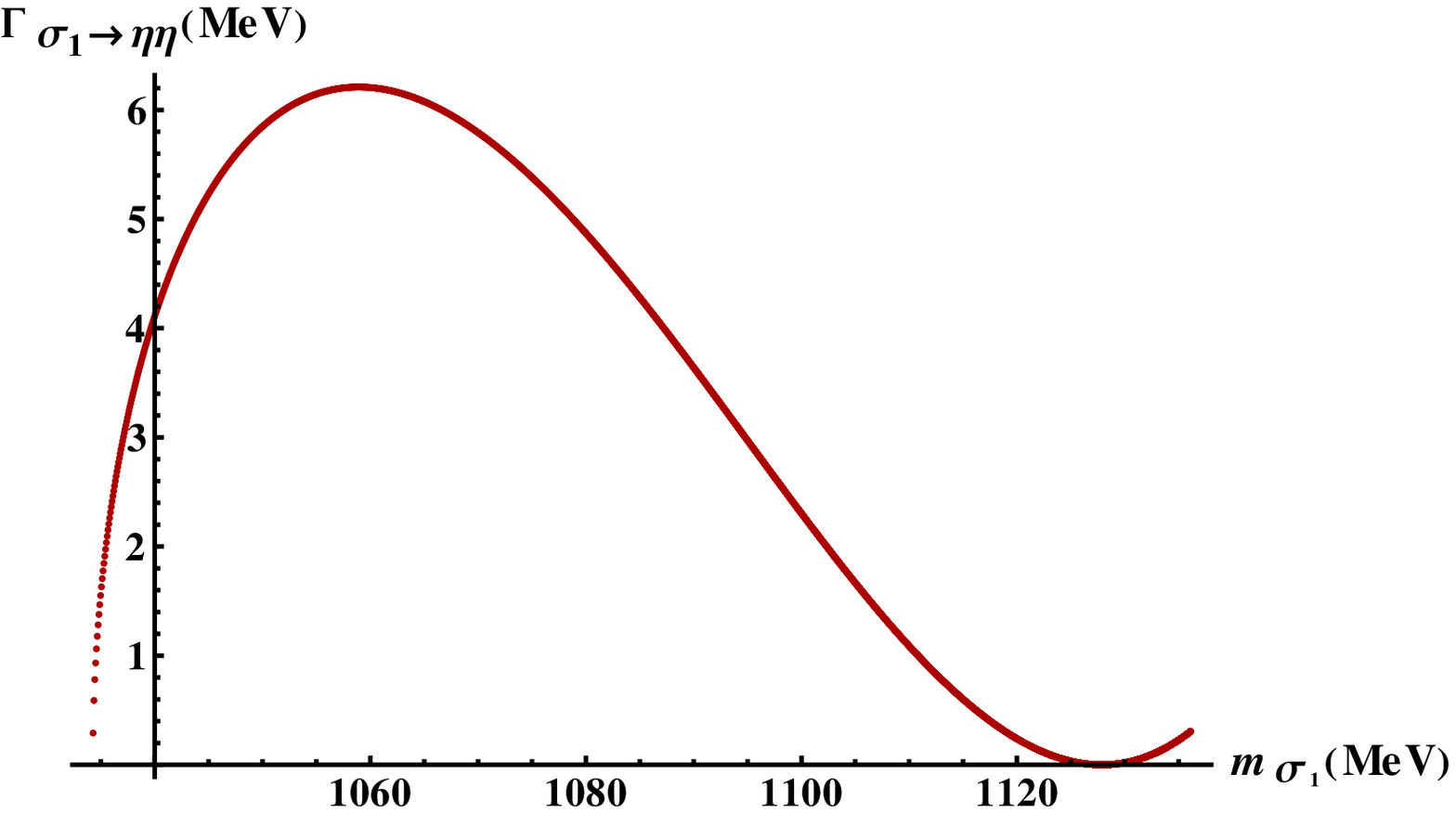}} &
      \resizebox{78mm}{!}{\includegraphics{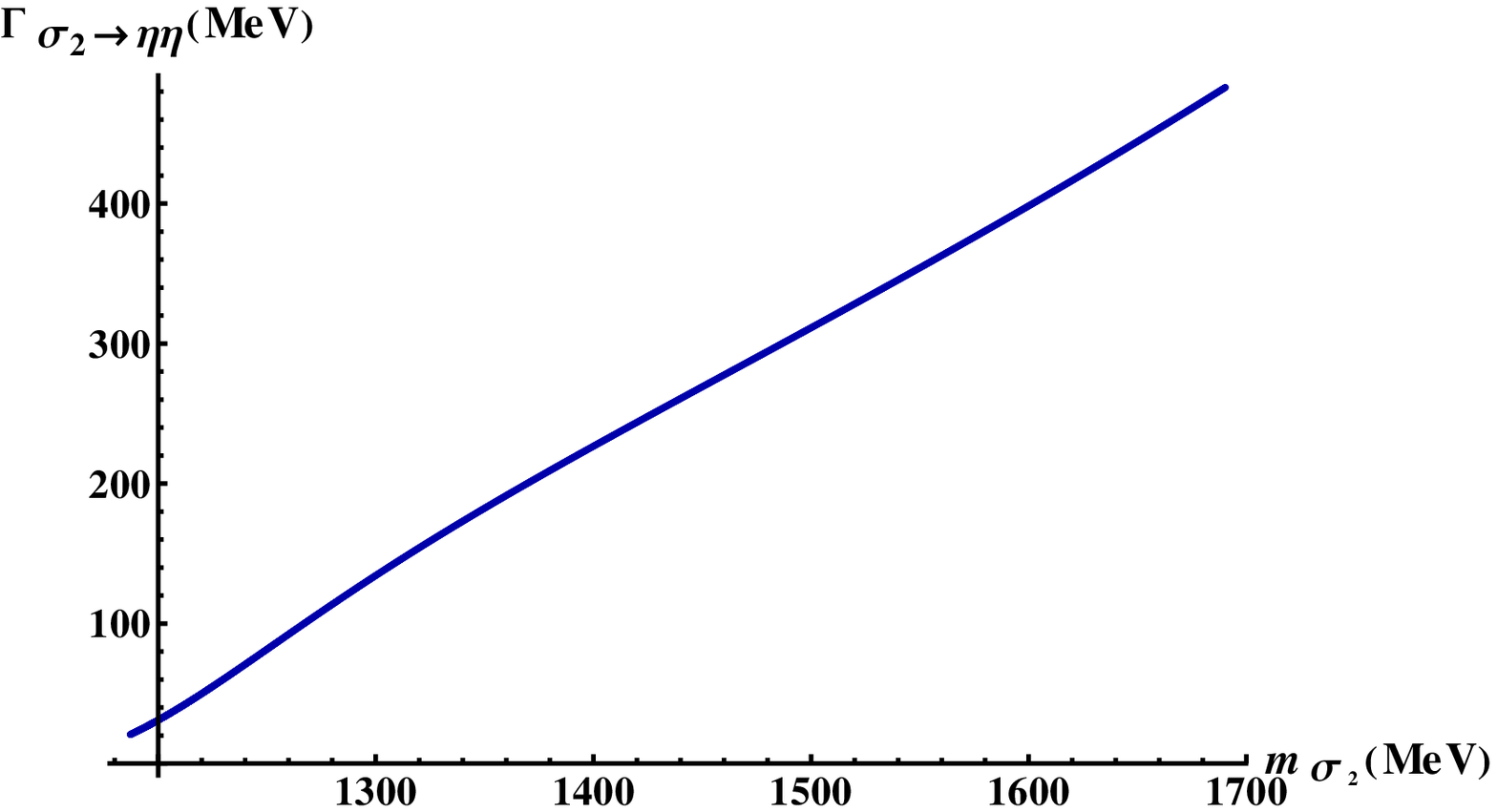}} 
    \end{tabular}
    \caption{$\Gamma_{\sigma_{1}\rightarrow\eta\eta}$ and $\Gamma_{\sigma
_{2}\rightarrow\eta\eta}$ as functions of $m_{\sigma_{1}}$ and $m_{\sigma_{2}%
}$, respectively.}
    \label{Setaeta1}
  \end{center}
\end{figure}

\begin{figure}
[!h]
\begin{center}
\includegraphics[
height=2.3582in,
width=3.7666in
]%
{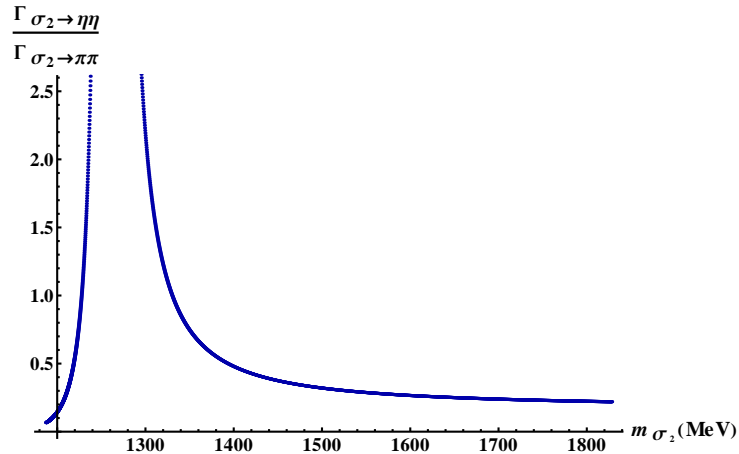}%
\caption{Ratio $\Gamma_{\sigma_{2}\rightarrow\eta\eta}/\Gamma_{\sigma
_{2}\rightarrow\pi\pi}$ as function of $m_{\sigma_{2}}$.}%
\label{Setaetapp1}%
\end{center}
\end{figure}

The decay widths are shown diagrammatically in Fig.\ \ref{Setaeta1}. As expected, $\Gamma_{\sigma_{1}\rightarrow\eta\eta}$ is suppressed due to a
limited phase space for $\eta\eta$. We observe a strong increase of
$\Gamma_{\sigma_{2}\rightarrow\eta\eta}$ over the $f_{0}(1370)$ mass interval
(see the right panel in Fig.\ \ref{Setaeta1}). Our results regarding the decay
channels $\sigma_{2}\rightarrow\pi\pi$ and $\sigma_{2}\rightarrow KK$ favour
$m_{\sigma_{2}}=1200$ MeV for which we obtain $\Gamma_{\sigma_{2}%
\rightarrow\eta\eta}=31$ MeV.\\

\begin{figure}
[!b]
\begin{center}
\includegraphics[
height=2.3582in,
width=3.7666in
]%
{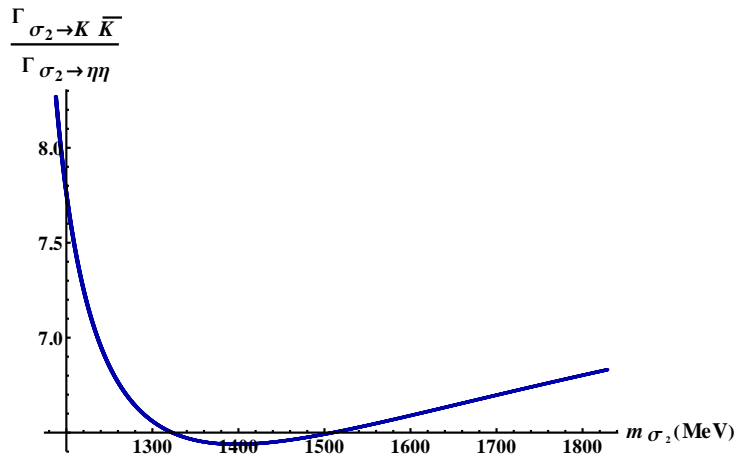}%
\caption{Ratio $\Gamma_{\sigma_{2}\rightarrow KK}/\Gamma_{\sigma
_{2}\rightarrow\eta\eta}$ as function of $m_{\sigma_{2}}$.}%
\label{SKKetaeta1}%
\end{center}
\end{figure}

A plot of $\Gamma_{\sigma_{2}\rightarrow\eta
\eta}/\Gamma_{\sigma_{2}\rightarrow\pi\pi}$ is shown in Fig.\ \ref{Setaetapp1}. There is a discontinuity in the ratio $\Gamma_{\sigma_{2}\rightarrow\eta\eta}%
/\Gamma_{\sigma_{2}\rightarrow\pi\pi}$ at the point $m_{\sigma_{2}}=1260$ MeV;
for $m_{\sigma_{2}}=1200$ MeV we obtain $\Gamma_{\sigma_{2}\rightarrow\eta
\eta}/\Gamma_{\sigma_{2}\rightarrow\pi\pi}=0.15$, in accordance with the
result $\Gamma_{f_{0}(1370)\rightarrow\eta\eta}/\Gamma_{f_{0}(1370)\rightarrow
\pi\pi}=0.19\pm0.07$ from Ref.\ \cite{buggf0}. We also observe that the
$\sigma_{2}\rightarrow KK$ channel is dominant in comparison with the
$\sigma_{2}\rightarrow$ $\eta\eta$ channel, see Fig.\ \ref{SKKetaeta1},
reaffirming the conclusion reached from the comparison of the decays $\sigma
_{2}\rightarrow KK$ and $\sigma_{2}\rightarrow\eta\eta$ (see
Fig.\ \ref{SPPKK1}).%

\section{Decay Amplitude \boldmath $a_{0}(980)\rightarrow\eta\pi$} \label{sec.a0etapion}

We have seen in Sec.\ \ref{sec.fitI} that a calculation of the parameter
$h_{2}$ via the decay width $\Gamma_{f_{1N}\rightarrow a_{0}(980)\pi}$,
Eq.\ (\ref{fit114}), yields two sets of values, a relatively lower and a
relatively higher one. In the $U(2)\times U(2)$ version of the
model, Sec.\ \ref{sec.f1NQ}, the higher set of $h_{2}$ values was found to be preferred.
Conversely, Fit I in Sec.\ \ref{sec.fitI} prefers lower values of $h_{2}$; in
this section we discuss whether, in that case, it is still possible to obtain
a correct value of the $a_{0}(980)\rightarrow\eta\pi$ decay amplitude.\\

The $a_{0}^{0}\eta\pi^{0}$ interaction Lagrangian reads%

\begin{equation}
\mathcal{L}_{a_{0}\eta\pi}=A_{a_{0}\eta_{N}\pi}a_{0}^{0}\eta_{N}\pi
^{0}+B_{a_{0}\eta_{N}\pi}a_{0}^{0}\partial_{\mu}\eta_{N}\partial^{\mu}\pi
^{0}+C_{a_{0}\eta_{N}\pi}\partial_{\mu}a_{0}^{0}(\pi^{0}\partial^{\mu}\eta
_{N}+\eta_{N}\partial^{\mu}\pi^{0})+A_{a_{0}\eta_{S}\pi}a_{0}^{0}\eta_{S}%
\pi^{0} \label{a0etaNetaSpion}%
\end{equation}

with

\begin{align}
A_{a_{0}\eta_{N}\pi}  &  =(-\lambda_{2}+c_{1}\phi_{S}^{2})Z_{\pi}^{2}\phi
_{N}\text{,} \\
B_{a_{0}\eta_{N}\pi}  &  =-2\frac{g_{1}^{2}\phi_{N}}{m_{a_{1}}^{2}}\left[
1-\frac{1}{2}\frac{Z_{\pi}^{4}f_{\pi}^{2}}{m_{a_{1}}^{2}}(h_{2}-h_{3})\right]\text{,}
\end{align}
\begin{align}
C_{a_{0}\eta_{N}\pi}  &  =g_{1}w_{a_{1}}Z_{\pi}^{2}\text{,}\\
A_{a_{0}\eta_{S}\pi}  &  =\frac{1}{2}c_{1}Z_{\pi}Z_{\eta_{S}}\phi_{N}^{2}%
\phi_{S}\text{.}%
\end{align}

Substituting $\eta_{N}$ by $\eta \cos\varphi_{\eta}$ and $\eta_{S}$ by
$\eta \sin\varphi_{\eta}$ [see Eqs.\ (\ref{etaN}) and (\ref{etaS})] we obtain
from the Lagrangian (\ref{a0etaNetaSpion})

\begin{align}
\mathcal{L}_{a_{0}\eta\pi} &  =A_{a_{0}\eta_{N}\pi}\cos\varphi_{\eta}a_{0}
^{0}\eta\pi^{0}+B_{a_{0}\eta_{N}\pi}\cos\varphi_{\eta}a_{0}^{0}\partial_{\mu
}\eta\partial^{\mu}\pi^{0}\nonumber\\
&  +C_{a_{0}\eta_{N}\pi}\cos\varphi_{\eta}\partial_{\mu}a_{0}^{0}(\pi
^{0}\partial^{\mu}\eta+\eta\partial^{\mu}\pi^{0})+A_{a_{0}\eta_{S}\pi}
\sin\varphi_{\eta}a_{0}^{0}\eta\pi^{0}\text{.}\label{a0etaNetaSpion1}
\end{align}

Consequently, the decay amplitude $\mathcal{M}_{a_{0}^{0}\rightarrow\eta
\pi^{0}}$ reads

\begin{align}
-i\mathcal{M}_{a_{0}^{0}\rightarrow\eta\pi^{0}}(m_{\eta}) & = i\left\{
\cos\varphi_{\eta} \left [  A_{a_{0}\eta_{N}\pi}-B_{a_{0}\eta_{N}\pi}%
\frac{m_{a_{0}}^{2}-m_{\eta}^{2}-m_{\pi}^{2}}{2}+C_{a_{0}\eta_{N}\pi
}m_{a_{0}(980)}^{2}\right] \right. \nonumber \\
& \left.  +\sin\varphi_{\eta}A_{a_{0}\eta_{S}\pi}\right\}
\text{.}\label{Ma0etapion}
\end{align}

Note that we can write $\mathcal{M}_{a_{0}^{0}\rightarrow\eta\pi^{0}}(m_{\eta
})$ also as

\begin{equation}
-i\mathcal{M}_{a_{0}^{0}\rightarrow\eta\pi^{0}}(m_{\eta})=-i[\cos\varphi
_{\eta}\mathcal{M}_{a_{0}^{0}\rightarrow\eta_{N}\pi^{0}}(m_{\eta})+\sin
\varphi_{\eta}\mathcal{M}_{a_{0}^{0}\rightarrow\eta_{S}\pi^{0}}(m_{\eta
})]\text{,} \label{Ma0etapion1}%
\end{equation}

where

\begin{equation}
-i\mathcal{M}_{a_{0}^{0}\rightarrow\eta_{N}\pi^{0}}(m_{\eta})=i\left[
A_{a_{0}\eta_{N}\pi}-B_{a_{0}\eta_{N}\pi}\frac{m_{a_{0}}^{2}-m_{\eta}
^{2}-m_{\pi}^{2}}{2}+C_{a_{0}\eta_{N}\pi}m_{a_{0}(980)}^{2}\right]
\label{Ma0etaNpion}
\end{equation}

and

\begin{equation}
-i\mathcal{M}_{a_{0}^{0}\rightarrow\eta_{S}\pi^{0}}(m_{\eta})=iA_{a_{0}%
\eta_{S}\pi}\text{.}\label{Ma0etaSpion}
\end{equation}

In Eqs.\ (\ref{Ma0etaNpion}) and (\ref{Ma0etaSpion}), $\mathcal{M}_{a_{0}%
^{0}\rightarrow\eta_{N}\pi^{0}}$ is obtained only from terms containing
$\eta_{N}$ in Eq.\ (\ref{a0etaNetaSpion}) whereas $\mathcal{M}_{a_{0}%
^{0}\rightarrow\eta_{S}\pi^{0}}$ is obtained from the $a_{0}^{0}\eta_{S}%
\pi^{0}$ coupling in Eq.\ (\ref{a0etaNetaSpion}). Note also that
$\mathcal{M}_{a_{0}^{0}\rightarrow\eta_{N}\pi^{0}}$ possesses an analogous
form to Eq.\ (\ref{Ma0etaNpionQ}).\\ 

All the parameters as well as $m_{a_{0}(980)}$, $m_{\eta}$ and $m_{\pi}$ are
determined uniquely from the fit and can be found in Tables \ref{Fit1-4} and
\ref{Fit1-5}. Consequently, we obtain from Eq.\ (\ref{Ma0etapion})%

\begin{equation}
\mid\mathcal{M}_{a_{0}^{0}(980)\rightarrow\eta\pi^{0}}(m_{\eta})\mid
=3155\text{ MeV.}
\end{equation}

The value is within experimental data stating $\mathcal{M}_{a_{0}\eta\pi
}(m_{\eta})=(3330\pm150)$ MeV \cite{Bugg:1994}. Note that we have used
$\varphi_{\eta}=-42%
{{}^\circ}%
$, in accordance with results derived in \textit{Step 4} of
Sec.\ \ref{sec.fitI}.

\section{Decay Width \boldmath $K^{\star}_{0}(800)\rightarrow K \pi$} \label{sec.KstarKp}

In this section we turn to the phenomenology of the scalar kaon $K_{S}$,
assigned to $K_{0}^{\star}(800)$, or $\kappa$, in Fit I. The scalar kaon is
known to decay into $K\pi$ \cite{PDG}. The corresponding interaction
Lagrangian from Eq.\ (\ref{Lagrangian}) reads (we consider only the neutral component; the
other ones possess analogous forms)

\begin{align}
\mathcal{L}_{K_{S}K\pi}  &  =A_{K_{S}K\pi}K_{S}^{0}(\bar{K}^{0}\pi^{0}%
-\sqrt{2}K^{-}\pi^{+})+B_{K_{S}K\pi}K_{S}^{0}(\partial_{\mu}\bar{K}%
^{0}\partial^{\mu}\pi^{0}-\sqrt{2}\partial_{\mu}K^{-}\partial^{\mu}\pi
^{+})\nonumber\\
&  +C_{K_{S}K\pi}\partial_{\mu}K_{S}^{0}(\pi^{0}\partial^{\mu}\bar{K}%
^{0}-\sqrt{2}\pi^{+}\partial_{\mu}K^{-})+D_{K_{S}K\pi}\partial_{\mu}K_{S}%
^{0}(\bar{K}^{0}\partial^{\mu}\pi^{0}-\sqrt{2}K^{-}\partial^{\mu}\pi^{+})
\label{KSKpion}%
\end{align}

with the following coefficients:

\begin{align}
A_{K_{S}K\pi}  &  =\frac{Z_{\pi}Z_{K}Z_{K_{S}}}{\sqrt{2}}\lambda_{2}\phi
_{S}\text{,} \label{AKSKp}\\
B_{K_{S}K\pi}  &  =-\frac{Z_{\pi}Z_{K}Z_{K_{S}}}{4}w_{a_{1}}w_{K_{1}}\left[
g_{1}^{2}(3\phi_{N}+\sqrt{2}\phi_{S})-2g_{1}\frac{w_{a_{1}}+w_{K_{1}}%
}{w_{a_{1}}w_{K_{1}}}+h_{2}(\phi_{N}+\sqrt{2}\phi_{S})-2h_{3}\phi_{N}\right]\text{,}
\label{BKSKp}\\
C_{K_{S}K\pi}  &  =\frac{Z_{\pi}Z_{K}Z_{K_{S}}}{4}[2g_{1}(\sqrt{2}%
ig_{1}w_{K^{\star}}w_{K_{1}}\phi_{S}-iw_{K^{\star}}-w_{K_{1}})-2\sqrt{2}%
ih_{3}w_{K^{\star}}w_{K_{1}}\phi_{S}]\text{,} \label{CKSKp}\\
D_{K_{S}K\pi}  &  =-\frac{Z_{\pi}Z_{K}Z_{K_{S}}}{4}\{g_{1}[2w_{a_{1}%
}-2iw_{K^{\star}}+ig_{1}w_{a_{1}}w_{K^{\star}}(3\phi_{N}-\sqrt{2}\phi
_{S})]\nonumber\\
&  +i(h_{2}-2h_{3})w_{K^{\star}}w_{a_{1}}\phi_{N}-\sqrt{2}ih_{2}w_{K^{\star}%
}w_{a_{1}}\phi_{S}\}\text{.} \label{DKSKp}%
\end{align}

Note that the coefficients containing the imaginary unit are nonetheless real
because $w_{K^{\star}}$, Eq.\ (\ref{wKstar}), is imaginary.\\

Let us focus on the decay $K_{S}^{0}\rightarrow K^{0}\pi^{0}$\ in the
following. The contribution of the charged modes to the decay width is twice the
contribution of the neutral modes, as apparent from Eq.\ (\ref{KSKpion}).
(We are changing the charge of the decay products in comparison to the one
present in the interaction Lagrangian. The Lagrangian itself has to be
charge-neutral and therefore contains particles and antiparticles
simultaneously; however, decay products are charge-conjugated in the
scattering matrix, see Sec.\ \ref{sec.calcdw}).\ It is then straightforward to calculate the
decay amplitude ($P$, $P_{1}$ and $P_{2}$ denote momenta of $K_{S}$, kaon and
pion, respectively, and we substitute $ \partial^{\mu}\rightarrow - i P^{\mu}$\ for
the decaying particles and $\partial^{\mu}\rightarrow i P_{1,2}^{\mu}$ for the decay products):

\begin{equation}
-i\mathcal{M}_{K_{S}^{0}\rightarrow K^{0}\pi^{0}}=i(A_{K_{S}K\pi}-B_{K_{S}%
K\pi}P_{1}\cdot P_{2}+C_{K_{S}K\pi}P\cdot P_{1}+D_{K_{S}K\pi}P\cdot
P_{2})\text{.}%
\end{equation}

Due to energy conservation on the vertex, $P=P_{1}+P_{2}$; thus we obtain%

\begin{equation}
-i\mathcal{M}_{K_{S}^{0}\rightarrow K^{0}\pi^{0}}=i[A_{K_{S}K\pi}-B_{K_{S}%
K\pi}P_{1}\cdot P_{2}+C_{K_{S}K\pi}(P_{1}^{2}+P_{1}\cdot P_{2})+D_{K_{S}K\pi
}(P_{2}^{2}+P_{1}\cdot P_{2})]\text{.}%
\end{equation}

Kaons and pions in the decay process are on-shell particles; therefore
$P_{1}^{2}=m_{K}^{2}$ and $P_{2}^{2}=m_{\pi}^{2}$. Additionally, $P_{1}\cdot
P_{2}=(P^{2}-P_{1}^{2}-P_{2}^{2})/2\equiv(m_{K_{S}}^{2}-m_{K}^{2}-m_{\pi}%
^{2})/2$. Therefore,%
\begin{align}
-i\mathcal{M}_{K_{S}^{0}\rightarrow K^{0}\pi^{0}} & =i\left[  A_{K_{S}K\pi
}+(C_{K_{S}K\pi}+D_{K_{S}K\pi}-B_{K_{S}K\pi})\frac{m_{K_{S}}^{2}-m_{K}%
^{2}-m_{\pi}^{2}}{2} \right. \nonumber \\
& + \left. C_{K_{S}K\pi}m_{K}^{2}+D_{K_{S}K\pi}m_{\pi}^{2}\right]
\text{.}\label{MKSKp}%
\end{align}

The decay width $\Gamma_{K_{S}^{0}\rightarrow K\pi}$ then reads%

\begin{equation}
\Gamma_{K_{S}^{0}\rightarrow K\pi}=3\frac{k(m_{K_{S}},m_{K},m_{\pi})}{8\pi
m_{K_{S}}^{2}}|-i\mathcal{M}_{K_{S}^{0}\rightarrow K^{0}\pi^{0}}|^{2}%
\text{.}\label{GKSKp}%
\end{equation}

Note that all parameters entering Eqs.\ (\ref{MKSKp}) and (\ref{GKSKp}) are
known from Table \ref{Fit1-4}; $Z_{K_{S}}$, $w_{a_{1}}$, $w_{K_{1}}$ and
$w_{K^{\star}}$ are determined respectively from Eqs.\ (\ref{Z_K_S}),
(\ref{wa1}), (\ref{wK1}) and (\ref{wKstar}). Then the value of $\Gamma
_{K_{S}^{0}\rightarrow K\pi}$ is determined uniquely:%

\begin{equation}
\Gamma_{K_{S}^{0}\rightarrow K\pi}=490\text{ MeV.} \label{GKSKp1}%
\end{equation}

The result is close to the value quoted by the PDG: $\Gamma_{\kappa
}^{\exp}=(548\pm24)$ MeV \cite{PDG}. The $\kappa$ resonance is experimentally
known to be broad and this finding is reproduced in our model. (Note, however,
that our $m_{K_{S}}$ is approximately by a factor of two larger than
$m_{\kappa}^{\exp}=676$ MeV, see Table \ref{Fit1-5} and Sec.\ \ref{sec.fitI}.)

Let us also point out the remarkable influence of the diagonalisation shift,
Eqs.\ (\ref{shift22}) and (\ref{shift24}) - (\ref{shift28}), on this decay
width: omitting the shift ($w_{a_{1}}=w_{K^{\star}}=w_{K_{1}}=0$), i.e.,
ignoring mixing terms from Eq.\ (\ref{mixingterms}), would yield
$\Gamma_{K_{0}^{\star}(800)\rightarrow K\pi}\simeq3$ GeV. Consequently,
coefficients arising from the shift [Eqs.\ (\ref{BKSKp}) - (\ref{DKSKp})]
induce a destructive interference in the Lagrangian (\ref{KSKpion}) decreasing
the decay width by approximately a factor of $6$.

\section{Phenomenology of the Vector and Axial-Vector Mesons in Fit I} \label{sec.VA1}

An important test of our Fit I derived in Sec.\ \ref{sec.fitI} is the
phenomenology of the vector and axial-vector states. In the vector channel,
the exact value of $\Gamma_{\rho\rightarrow\pi\pi}=149.1$ MeV has already been
implemented to determine the parameter $g_{2}$ (see Table \ref{Fit1-4}). In
principle our model also allows for the discussion of the phenomenology for the
isosinglet vector state $\omega_{S}\equiv\varphi(1020)$. This state decays
into kaons; our Fit I yields $m_{\omega_{S}}=870.35$ MeV thus implying that no
tree-level calculation of the decay width can be performed as $m_{\omega_{S}}$ is
below the two-kaon threshold. Therefore, this state is not well described within
Fit I. [Note that decays of the non-strange vector isosinglet $\omega_{N}$
cannot be calculated within the model because there are no corresponding
vertices: $\omega_{N}$ always appears quadratically in the Lagrangian
(\ref{Lagrangian}).] Therefore in this section we only need to consider
the phenomenology of the $K^{\star}$\ meson to complete the vector phenomenology (see
subsection \ref{sec.kstar1}).

Fit I has also yielded $m_{f_{1S}}=1643.4$ MeV, see Table \ref{Fit1-5}. As
discussed in Sec.\ \ref{sec.fitI}, $m_{f_{1S}}$ is too large when compared to
the experimental result $m_{f_{1(1420)}}^{\exp}=(1426.4\pm0.9)$ MeV.\ The
$f_{1S}\equiv f_{1}(1420)$ resonance decays predominantly into $K^{\star}K$
\cite{PDG}. The corresponding decay width can be calculated within our model
and it will represent an important test of Fit I because $f_{1}(1420)$ is a
sharp resonance with $\Gamma_{f_{1}(1420)}^{\exp}=(54.9\pm2.6)$ MeV, see
subsection \ref{sec.f1S1}. The $K_{1}$ phenomenology is discussed in
Sec.\ \ref{sec.K11}. We will, however, begin with the phenomenology of
$a_{1}(1260)$. This state possesses a large decay width [$(250-600)$ MeV
\cite{PDG}] with a dominant $\rho\pi$ decay channel. From our model,
$a_{1}(1260)$ is expected to be the chiral partner of the $\rho$ meson and to become
degenerate with this state upon the chiral transition. However, one first needs to
ascertain whether the $a_{1}(1260)$ phenomenology in vacuum can be described
correctly from Fit I. This is discussed in the following subsections
\ref{sec.a1rhopion1} and \ref{sec.a1sigmapion1}. The phenomenology of the
axial-vector isosinglet $f_{1N}\equiv f_{1}(1285)$ is discussed subsequently
in Sec.\ \ref{sec.f1N1}; as in the case of $f_{1}(1420)$, only the $K^{\star}K$
decay channel can be considered in our model.

\subsection{Decay Width \boldmath $a_{1}(1260)\rightarrow\rho\pi$ in Fit I} \label{sec.a1rhopion1}

\begin{figure}
[h]
\begin{center}
\includegraphics[
height=2.1582in,
width=3.8666in]
{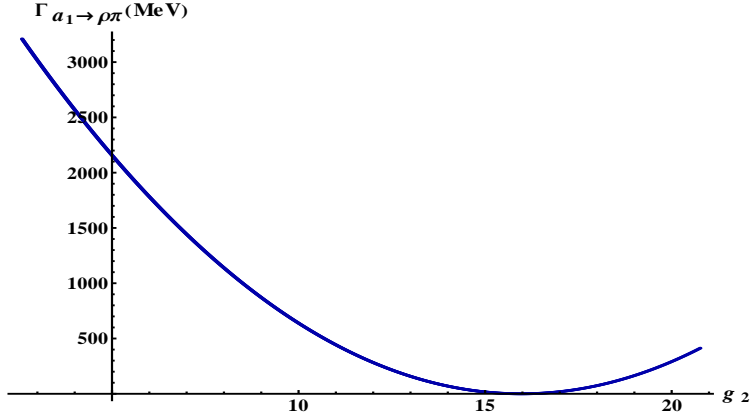}
\caption{$\Gamma_{a_{1}(1260)\rightarrow\rho\pi}$ as function of $g_{2}$ in
Fit I.}
\label{g221}
\end{center}
\end{figure}

The interaction Lagrangian for the decay $a_{1}(1260)\rightarrow\rho\pi$
has the same form as in the $U(2)\times U(2)$ version of the model,
Sec.\ \ref{sec.a1rpQ}. We can therefore make use of the same formula for
$\Gamma_{a_{1}(1260)\rightarrow\rho\pi}$ as in Eq.\ (\ref{a1rhopionQ}). The
decay width depends (among other parameters) on $g_{2}$; this parameter is fixed via the decay width
$\Gamma_{\rho\rightarrow\pi\pi}$ [Eq.\ (\ref{g2Z})] and given that our Fit I
yields a relatively large value of $m_{a_{1}}=1395.5$ MeV, see Table
\ref{Fit1-5}, then we obtain a value of $g_{2}=-11.2$ for $\Gamma
_{\rho\rightarrow\pi\pi}=149.1$ MeV \cite{PDG}\ (see Table \ref{Fit1-4}). The
large magnitude of this parameter influences $\Gamma_{a_{1}(1260)\rightarrow
\rho\pi}$ in a very strong way: we obtain $\Gamma_{a_{1}(1260)\rightarrow
\rho\pi}\simeq13$ GeV for the stated value of $g_{2}$ and other parameters
listed in Table \ref{Fit1-4}. In fact, we would require $g_{2}\gtrsim10$ for
$\Gamma_{a_{1}(1260)\rightarrow\rho\pi}$ to have values within the PDG
interval $(250-600)$ MeV \cite{PDG}, see Fig.\ \ref{g221}. Note that
integrating over the $\rho$ spectral function, just as in Sec.\ \ref{sec.AVP},
yields the decay width of $\sim11$ GeV -- again unphysically large.

\begin{figure}
[h]
\begin{center}
\includegraphics[
height=2.1582in,
width=4.1666in
]%
{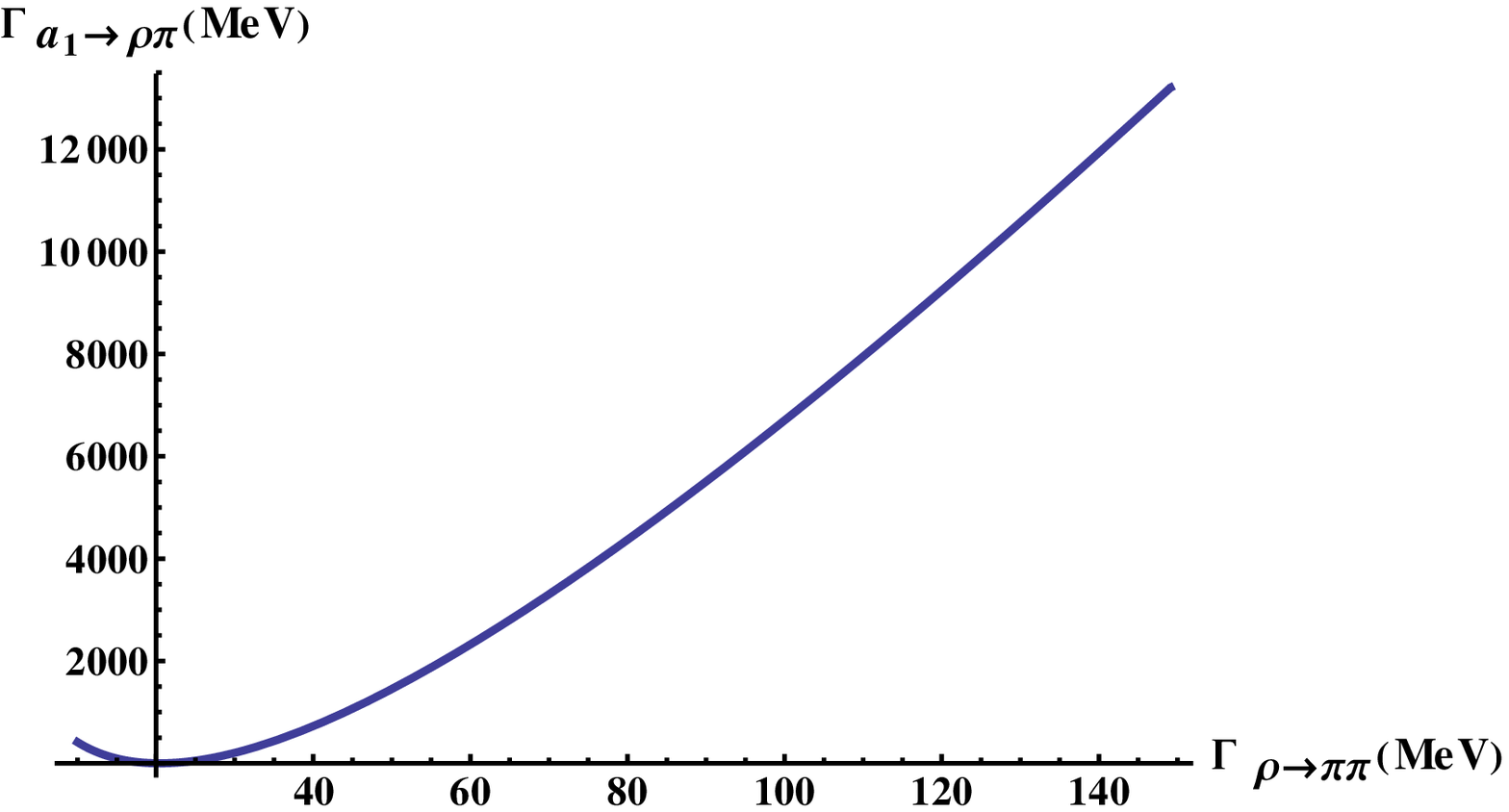}%
\caption{$\Gamma_{a_{1}(1260)\rightarrow\rho\pi}$ as function of $\Gamma
_{\rho\rightarrow\pi\pi}$.}%
\label{a1rhopi1}%
\end{center}
\end{figure}

Our fit determines all parameter values uniquely and therefore we do not have
a possibility to fine-tune $\Gamma_{a_{1}(1260)\rightarrow\rho\pi}$; the only
exception arises by changing $\Gamma_{\rho\rightarrow\pi\pi}$ to increase
$g_{2}$ and consequently decrease $\Gamma_{a_{1}(1260)\rightarrow\rho\pi}$
(note, however, that the experimental uncertainty regarding $\Gamma
_{\rho\rightarrow\pi\pi}$ is actually very small: $\pm0.8$ MeV \cite{PDG}).
Consequent changes in $\Gamma_{a_{1}(1260)\rightarrow\rho\pi}$ are depicted in
Fig.\ \ref{a1rhopi1}.%

We can see from Fig.\ \ref{a1rhopi1} that $\Gamma_{a_{1}(1260)\rightarrow
\rho\pi}<600$ MeV only if $\Gamma_{\rho\rightarrow\pi\pi}<38$ MeV, more than
$100$ MeV smaller than the physical value of $\Gamma_{\rho\rightarrow\pi\pi
}=149.1$ MeV \cite{PDG}. Alternatively, increasing $\Gamma_{\rho\rightarrow
\pi\pi}$ also leads to a very strong increase of $\Gamma_{a_{1}%
(1260)\rightarrow\rho\pi}$ with $\Gamma_{a_{1}(1260)\rightarrow\rho\pi}>1$ GeV
already at $\Gamma_{\rho\rightarrow\pi\pi}\simeq44$ MeV. Therefore, we observe
a strong tension between the decays in the non-strange vector and axial-vector
channels -- it is not possible to obtain correct decay width values in both
channels at the same time as either the decay $\rho\rightarrow\pi\pi$ is
subdominant [and $\Gamma_{a_{1}(1260)\rightarrow\rho\pi}$ within the physical
range] or the decay $\rho\rightarrow\pi\pi$ is correctly described but the
channel $a_{1}(1260)\rightarrow\rho\pi$ is virtually dissolved in the continuum.

\subsection{Decay Width \boldmath $a_{1}(1260)\rightarrow f_{0}(600) \pi$ in Fit I} \label{sec.a1sigmapion1}

Unlike the case of the $a_{1}(1260)\rightarrow\rho\pi$ decay, the interaction
Lagrangian for the process $a_{1}(1260)\rightarrow f_{0}(600)\pi$ with
$f_{0}(600)\equiv\sigma_{1}$ is slightly different than in the $U(2)\times
U(2)$ version of the model, Sec.\ \ref{sec.a1spQ}:

\begin{equation}
\mathcal{L}_{a_{1}\sigma\pi}=A_{a_{1}\sigma_{N}\pi}a_{1}^{\mu0}\sigma
_{N}\partial_{\mu}\pi^{0}+B_{a_{1}\sigma_{N}\pi}a_{1}^{\mu0}\pi^{0}
\partial_{\mu}\sigma_{N}+A_{a_{1}\sigma_{S}\pi}a_{1}^{\mu0}\sigma_{S}
\partial_{\mu}\pi^{0} \label{a1sp}
\end{equation}

with the following coefficients:

\begin{align}
A_{a_{1}\sigma_{N}\pi}  &  =Z_{\pi}\left[  g_{1}(-1+2g_{1}w_{a_{1}}\phi
_{N})+(h_{1}+h_{2}-h_{3})w_{a_{1}}\phi_{N}\right]\text{,} \label{Aa1sNp}\\
B_{a_{1}\sigma_{N}\pi}  &  =g_{1}Z_{\pi}\text{,} \label{Ba1sNp}\\
A_{a_{1}\sigma_{S}\pi}  &  =h_{1}Z_{\pi}w_{a_{1}}\phi_{S}\text{.}
\label{Aa1sSp}
\end{align}

\begin{figure}
[h]
\begin{center}
\includegraphics[
height=2.2582in,
width=4.1666in
]%
{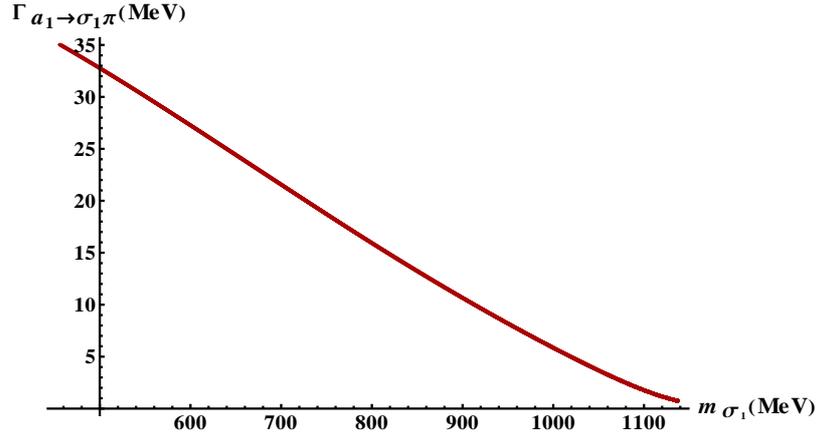}
\caption{$\Gamma_{a_{1}(1260)\rightarrow\sigma_{1}\pi}$ as function of
$m_{\sigma_{1}}$.}
\label{a1s1pi1}
\end{center}
\end{figure}

It is necessary to substitute the pure states $\sigma_{N}$ and $\sigma_{S}$ by
the mixed states $\sigma_{1}$ and $\sigma_{2}$ in Eq.\ (\ref{a1sp}); we are
considering only the decay $a_{1}(1260)\rightarrow\sigma_{1}\pi$ and thus it
suffices to perform the substitutions $\sigma_{N}\rightarrow\cos\varphi_{\sigma
}\sigma_{1}$ and $\sigma_{S}\rightarrow\sin\varphi_{\sigma}\sigma_{2}$:

\begin{equation}
\mathcal{L}_{a_{1}\sigma\pi}=(A_{a_{1}\sigma_{N}\pi}\cos\varphi_{\sigma
}+A_{a_{1}\sigma_{S}\pi}\sin\varphi_{\sigma})a_{1}^{\mu0}\sigma_{1}
\partial_{\mu}\pi^{0}+B_{a_{1}\sigma_{N}\pi}\cos\varphi_{\sigma}a_{1}^{\mu
0}\pi^{0}\partial_{\mu}\sigma_{1}\text{,} \label{a1sp1}
\end{equation}

where $\varphi_{\sigma}$ is the $\sigma_{N}$-$\sigma_{S}$ mixing angle,
Eq.\ (\ref{phisigma1}). The interaction Lagrangian from Eq.\ (\ref{a1sp1})
possesses an analogous form to the one presented in Eq.\ (\ref{ASP}), with the
latter describing a generic decay of an axial-vector state $A$ into a scalar $S$
and a pseudoscalar ${\tilde{P}}$. Thus we can use the generic formula for the
decay width from Eq.\ (\ref{GASP}) upon substituting $A\leftrightarrow a_{1}%
$, $S{\leftrightarrow}\sigma_{1}$, ${\tilde{P}\leftrightarrow}\pi$,
$A_{AS{\tilde{P}}}\leftrightarrow A_{a_{1}\sigma_{N}\pi}\cos\varphi_{\sigma
}+A_{a_{1}\sigma_{S}\pi}\sin\varphi_{\sigma}$ and $A_{AS{\tilde{P}}%
}\leftrightarrow B_{a_{1}\sigma_{N}\pi}\cos\varphi_{\sigma}$. We consequently
obtain $\Gamma_{a_{1}(1260)\rightarrow\sigma_{1}\pi}$ as shown in
Fig.\ \ref{a1s1pi1}.

We observe from Fig.\ \ref{a1s1pi1} that $\Gamma_{a_{1}(1260)\rightarrow
\sigma_{1}\pi}$ rapidly decreases with the available phase space. The exact
value of $\Gamma_{a_{1}(1260)\rightarrow\sigma_{1}\pi}$ is therefore strongly
dependent on $m_{\sigma_{1}}$; e.g., we obtain the result $\Gamma_{a_{1}(1260)\rightarrow
\sigma_{1}\pi} = 21$ MeV for $m_{\sigma_{1}}=705$ MeV [our best value of
$m_{\sigma_{1}}$, see Eq.\ (\ref{ms11})]. Nonetheless, these results show
$\Gamma_{a_{1}(1260)\rightarrow\sigma_{1}\pi}$ to be suppressed in comparison
with $\Gamma_{a_{1}(1260)\rightarrow\rho\pi}$\ and qualitatively similar to
the values in Scenario I of the $U(2)\times U(2)$ version of the model.

\subsection{Decay Width \boldmath $a_{1}(1260)\rightarrow K^{\star} K
\rightarrow K K \pi$ in Fit I} \label{sec.a1KstarK1}

The corresponding interaction Lagrangian is a feature of the $U(3)\times U(3)$
version of the model. The Lagrangian reads

\begin{align}
\mathcal{L}_{a_{1}K^{\star}K} &  =A_{a_{1}K^{\star}K}a_{1}^{\mu0}(K_{\mu
}^{\star0}\bar{K}^{0}+K_{\mu}^{\star-}K^{+})\nonumber\\
&  +B_{a_{1}K^{\star}K}^{\mu}a_{1}^{\mu0}[(\partial_{\nu}K_{\mu}^{\star
0}-\partial_{\mu}K_{\nu}^{\star0})\partial^{\nu}\bar{K}^{0}+(\partial_{\nu
}K_{\mu}^{\star-}-\partial_{\mu}K_{\nu}^{\star-})\partial^{\nu}K^{+}%
]\nonumber\\
&  +\partial^{\nu}a_{1}^{\mu0}(K_{\nu}^{\star0}\partial_{\mu}\bar{K}%
^{0}-K_{\mu}^{\star0}\partial_{\nu}\bar{K}^{0}+K_{\nu}^{\star-}\partial_{\mu
}K^{+}-K_{\mu}^{\star-}\partial_{\nu}K^{+})]+\text{h.c.}\label{a1KstarK}%
\end{align}

with the following coefficients:

\begin{align}
A_{a_{1}K^{\star}K} &  =-\frac{i}{4}Z_{K}\left[  g_{1}^{2}(3\phi_{N}-\sqrt
{2}\phi_{S})+h_{2}(\phi_{N}-\sqrt{2}\phi_{S})-2h_{3}\phi_{N}\right]\text{,}
\label{Aa1KstarK}\\
B_{a_{1}K^{\star}K} &  =-\frac{i}{2}Z_{K}g_{2}w_{K_{1}}\text{,} \label{Ba1KstarK}\\
C_{a_{1}K^{\star}K} &  =-\frac{i}{2}Z_{K}g_{2}w_{K_{1}}\text{.}%
\label{Ca1KstarK}
\end{align}

We can now consider results from Sec.\ \ref{sec.AVP} where a generic decay of
an axial-vector into a vector and pseudoscalar was considered. The decay
$a_{1}\rightarrow\bar{K}^{\star}K$\ is tree-level forbidden because $a_{1}$ is
below the $K^{\star}K$\ threshold. However, if an off-shell $K^{\star}$\ state
is considered then the ensuing decay $a_{1}\rightarrow\bar{K}^{\star
}K\rightarrow\bar{K}K\pi$\ can be studied. We can therefore use
Eq.\ (\ref{GAVP1}) as formula for the decay width (the isospin factor is
$I=4$) and integrate over the $K^{\star}$ spectral function in Eq.\ (\ref{dV}).
The value of the $K^{\star}$ decay width used in the spectral function is given further below, in
Eq.\ (\ref{GKstarKp1}). We obtain

\begin{equation}
\Gamma_{a_{1}\rightarrow\bar{K}^{\star}K\rightarrow\bar{K}K\pi}= 1.97 \text{
GeV.}\label{Ga1KstarK}
\end{equation}

The decay is strongly enhanced for the same reasons as in the previousy
discussed $a_{1}(1260)$ channels.

\subsection{Decay Width \boldmath $f_{1}(1285) \rightarrow K^{\star} K $ in Fit I} \label{sec.f1N1}

The $f_{1N}\equiv f_{1}(1285)$ meson is the non-strange axial-vector
isosinglet state, i.e., the isospin-zero partner of the $a_{1}(1260)$
resonance. These two resonances are degenerate in our model [see
Eq.\ (\ref{m_a_1})] given that the model implements the isospin symmetry exactly.

There are two decays of the $f_{1}(1285)$ resonance that can be
calculated from the $U(3)\times U(3)$ version of our model: a decay involving
non-strange states, $f_{1}(1285)\rightarrow a_{0}(980)\pi$, and a decay into
kaons, $f_{1}(1285)\rightarrow\bar{K}^{\star}K$. The former decay width has
already been utilised to calculate the parameter $h_{2}$ in Fit I, see
Sec.\ \ref{fitstructure} and Eq.\ (\ref{fit114}); therefore, this decay width
corresponds exactly to the experimental value $\Gamma_{f_{1}(1285)\rightarrow
a_{0}(980)\pi}=8.748$ MeV (see Table \ref{Fit1-5}). The latter decay width is
discussed in this section. The PDG actually lists the $f_{1}(1285)\rightarrow
\bar{K}^{\star}K$ process as "not seen" although the three-body decay
$f_{1}(1285)\rightarrow\bar{K}K\pi$ possesses a branching ratio of
$(9.0\pm0.4)\%$; the full decay width of the resonance is $\Gamma
_{f_{1}(1285)}=(24.3\pm1.1)$\ MeV \cite{PDG}. The stated three-body decay can,
within our model, arise from the sequential decay $f_{1}(1285)\rightarrow
\bar{K}^{\star}K\rightarrow\bar{K}K\pi$. Therefore, in this section, we
discuss implications of the interaction Lagrangian for the $f_{1}%
(1285)\rightarrow\bar{K}^{\star}K$ decay.

The $f_{1N}K^{\star}K$ interaction Lagrangian from Eq.\ (\ref{Lagrangian}) reads

\begin{align}
\mathcal{L}_{f_{1N}K^{\star}K}  &  =A_{f_{1N}K^{\star}K}f_{1N}^{\mu}(K_{\mu
}^{\star0}\bar{K}^{0}+K_{\mu}^{\star+}K^{-}-\bar{K}_{\mu}^{\star0}K^{0}%
-K_{\mu}^{\star-}K^{+})\nonumber\\
&  +B_{f_{1N}K^{\star}K} f_{1N}^{\mu}[(\partial_{\nu}K_{\mu}^{\star
0}-\partial_{\mu}K_{\nu}^{\star0})\partial^{\nu}\bar{K}^{0}+(\partial_{\nu
}K_{\mu}^{\star+}-\partial_{\mu}K_{\nu}^{\star+})\partial^{\nu}K^{-}%
\nonumber\\
&  -(\partial_{\nu}\bar{K}_{\mu}^{\star0}-\partial_{\mu}\bar{K}_{\nu}^{\star
0})\partial^{\nu}K^{0}-(\partial_{\nu}K_{\mu}^{\star-}-\partial_{\mu}K_{\nu
}^{\star-})\partial^{\nu}K^{+}]\nonumber\\
&  +C_{f_{1N}K^{\star}K}\partial^{\nu}f_{1N}^{\mu}(K_{\mu}^{\star0}%
\partial_{\nu}\bar{K}^{0}-K_{\nu}^{\star0}\partial_{\mu}\bar{K}^{0}+K_{\mu
}^{\star+}\partial_{\nu}K^{-}-K_{\nu}^{\star+}\partial_{\mu}K^{-}\nonumber\\
&  -\bar{K}_{\mu}^{\star0}\partial_{\nu}K^{0}+\bar{K}_{\nu}^{\star0}%
\partial_{\mu}K^{0}-K_{\mu}^{\star-}\partial_{\nu}K^{+}+K_{\nu}^{\star
-}\partial_{\mu}K^{+}) \label{f1NKstarK}
\end{align}

with the following coefficients:

\begin{align}
A_{f_{1N}K^{\star}K}  &  =\frac{i}{4}Z_{K}\left[  g_{1}^{2}(3\phi_{N}-\sqrt
{2}\phi_{S})+h_{2}(\phi_{N}-\sqrt{2}\phi_{S})-2h_{3}\phi_{N}\right]\text{,}
\label{Af1NKstarK}\\
B_{f_{1N}K^{\star}K}  &  =\frac{i}{2}Z_{K}g_{2}w_{K_{1}}\text{,} \label{Bf1NKstarK}\\
C_{f_{1N}K^{\star}K}  &  =-\frac{i}{2}Z_{K}g_{2}w_{K_{1}}\text{.}
\label{Cf1NKstarK}
\end{align}

Let us now turn to the decay process $f_{1S}\rightarrow\bar{K}^{\star0}K^{0}$
from Eq.\ (\ref{f1NKstarK}); other decay processes from Eq.\ (\ref{f1NKstarK})
will be considered by an appropriate isospin factor. Let us denote the momenta of
$f_{1N}$, $\bar{K}_{\mu}^{\star0}$ and $K^{0}$ as $P$, $P_{1}$ and $P_{2}$.
The decay process involves two vector states: $f_{1N}$\ and$\ K^{\star}$. We
therefore have to consider the corresponding polarisation vectors labelled as
$\varepsilon_{\mu}^{(\alpha)}(P)$ for $f_{1N}$\ and $\varepsilon_{\nu}%
^{(\beta)}(P_{1})$ for $K^{\star}$. Then, upon substituting $\partial^{\mu
}\rightarrow-iP^{\mu}$\ for the decaying particle and $\partial^{\mu}\rightarrow
iP_{1,2}^{\mu}$ for the decay products, we obtain the following Lorentz-invariant
$f_{1N}\bar{K}^{\star0}K^{0}$ scattering amplitude $-i\mathcal{M}%
_{f_{1N}\rightarrow\bar{K}^{\star0}K^{0}}^{(\alpha,\beta)}$:

\begin{align}
-i\mathcal{M}_{f_{1N}\rightarrow\bar{K}^{\star0}K^{0}}^{(\alpha,\beta)}  &
=\varepsilon_{\mu}^{(\alpha)}(P)\varepsilon_{\nu}^{(\beta)}(P_{1}%
)h_{f_{1N}\bar{K}^{\star0}K^{0}}^{\mu\nu}=i\varepsilon_{\mu}^{(\alpha
)}(P)\varepsilon_{\nu}^{(\beta)}(P_{1})\nonumber\\
&  \times\left\{  A_{f_{1N}K^{\star}K}g^{\mu\nu}+[B_{f_{1N}K^{\star}K}%
(P_{1}^{\mu}P_{2}^{\nu}-(P_{1}\cdot P_{2})g^{\mu\nu}]\right. \nonumber\\
&  \left.  +C_{f_{1N}K^{\star}K}[(P\cdot P_{2})g^{\mu\nu}-P_{2}^{\mu}P^{\nu
}]\right\}  \label{Mf1NKstarK}
\end{align}

with
\begin{align}
h_{f_{1N}\bar{K}^{\star0}K^{0}}^{\mu\nu} & =i\left\{  A_{f_{1N}K^{\star}K}%
g^{\mu\nu}+[B_{f_{1N}K^{\star}K}(P_{1}^{\mu}P_{2}^{\nu}-(P_{1}\cdot
P_{2})g^{\mu\nu}] \right. \nonumber \\
& \left. + C_{f_{1N}K^{\star}K}[(P\cdot P_{2})g^{\mu\nu}-P_{2}^{\mu
}P^{\nu}]\right\}\text{,} \label{hf1NKstarK}
\end{align}

where $h_{f_{1N}\bar{K}^{\star0}K^{0}}^{\mu \nu}$ denotes the $f_{1N}\bar
{K}^{\star0}K^{0}$ vertex. We observe that the form of the vertex in
Eq.\ (\ref{hf1NKstarK}) is analogous to that of the $a_{1}\rho\pi$ vertex of
Eq.\ (\ref{ha1rpQ}). Therefore we can use the formulas for the $a_{1}(1260)\rightarrow
\rho\pi$ decay amplitude and decay width to calculate $\Gamma_{f_{1S}%
\rightarrow\bar{K}^{\star}K}$ (naturally, upon substitution of corresponding
coefficients: $A_{a_{1}\rho\pi}\rightarrow A_{f_{1N}K^{\star}K}$,
$B_{a_{1}\rho\pi}\rightarrow B_{f_{1N}K^{\star}K}$, $C_{a_{1}\rho\pi
}\rightarrow-C_{f_{1N}K^{\star}K}$); an isospin factor of four has
to be considered to account for the decays $f_{1N}\rightarrow\bar{K}^{\star0}%
K^{0}$, $\bar{K}^{0}K^{\star0}$, $K^{\star+}K^{-}$ and $K^{\star-}K^{+}$. Note
that all parameters entering the coefficients $A_{f_{1N}K^{\star}K}$,
$B_{f_{1N}K^{\star}K}$ and $C_{f_{1N}K^{\star}K}$ in Eqs.\ (\ref{Af1NKstarK})
- (\ref{Cf1NKstarK}) are known from Table \ref{Fit1-4}; mass values can be
found in Table \ref{Fit1-5}.

The decay $f_{1N}\rightarrow\bar{K}^{\star}K$ is actually tree-level forbidden
if one considers the physical masses of the three resonances concerned:
$m_{f_{1}(1285)}^{\exp}=(1281.8\pm0.6)$ MeV $<m_{K^{\star}}^{\exp}+$
$m_{K}^{\exp}$ because $m_{K^{\star}}^{\exp}=(891.66\pm0.26)$ MeV and
$m_{K}^{\exp}=(493.677\pm0.016)$ MeV.\ However, our Fit I yields $m_{a_{1}%
}=1396$\ MeV and given that (due to the isospin invariance of our model)
$m_{f_{1N}}=m_{a_{1}}$, then the tree-level decay $f_{1N}\rightarrow\bar
{K}^{\star}K$ is nonetheless kinematically allowed. The problem is that the
value of the parameter $g_{2}=-11.2$ possesses a rather large modulus that with
all other parameter values leads to

\begin{equation}
\Gamma_{f_{1N}\rightarrow\bar{K}^{\star}K}=2.15\text{ GeV.} \label{Gf1NKstarK}
\end{equation}

A similar problem was present in Sec.\ \ref{sec.a1rhopion1}; let us again try
to remedy the issue by varying $\Gamma_{\rho\rightarrow\pi\pi}$ to decrease
$g_{2}$.%

\begin{figure}
[h]
\begin{center}
\includegraphics[
height=2.2582in,
width=4.1666in
]%
{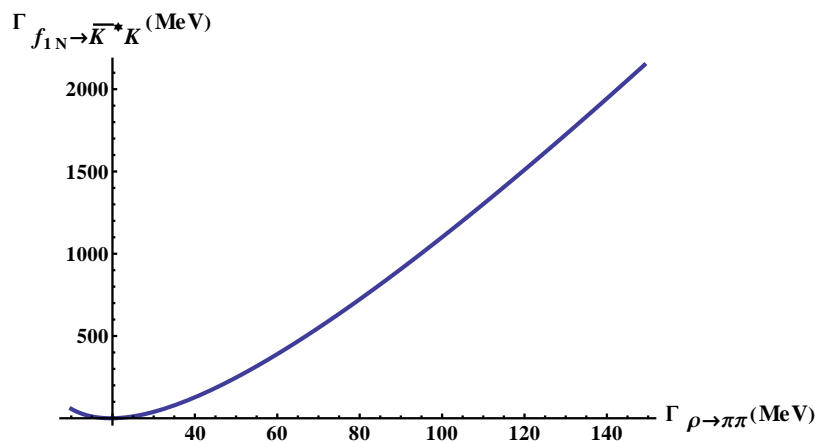}%
\caption{$\Gamma_{f_{1N}\equiv f_{1}(1285)\rightarrow\bar{K}^{\star}K}$ as
function of $\Gamma_{\rho\rightarrow\pi\pi}$.}%
\label{f1NKstarK1}%
\end{center}
\end{figure}

As apparent from Fig.\ \ref{f1NKstarK1}, obtaining reasonable values of
$\Gamma_{f_{1N}\rightarrow\bar{K}^{\star}K}$ (expected to be $\leq (2.2 \pm 0.1)$ MeV
from the PDG branching ratio for $f_{1}(1285)\rightarrow\bar{K}K\pi$ stated at
the beginning of this section) would require $\Gamma_{\rho\rightarrow\pi\pi
}\simeq20$ MeV, clearly at odds with experiment \cite{PDG}. Note that the same
holds if one integrates over the spectral function of the $K^{\star}$ meson
(as in Sec.\ \ref{sec.AVP}): we
obtain $\Gamma_{f_{1N}\rightarrow\bar{K}^{\star}K\rightarrow\bar{K}K\pi}=1.98$
GeV and, again, $\Gamma_{\rho\rightarrow\pi\pi}\overset{!}{\sim}20$ MeV for
$\Gamma_{f_{1N}\rightarrow\bar{K}^{\star}K\rightarrow\bar{K}K\pi}<2$ MeV to be
true. Thus Fit I yields kaon decay widths of the $f_{1}(1285)$ resonance that
are three orders of magnitude larger than suggested by experimental data.

\subsection{Decay Width \boldmath $K^{\star} \rightarrow K \pi$ in Fit I} \label{sec.kstar1}

In this section we describe the phenomenology of the vector kaon $K^{\star}$,
the strange counterpart of the $\vec{\rho}$ state present in our model. Our
$K^{\star}$ state is assigned to $K^{\star}(892)$. This resonance decays to
$\simeq100\%$ into $K\pi$ \cite{PDG}. \\
The $K^{\star0}K\pi$ interaction
Lagrangian from Eq.\ (\ref{Lagrangian}) reads
\begin{align}
\mathcal{L}_{K^{\star}K\pi}  &  =A_{K^{\star}K\pi}K_{\mu}^{\star0}(\pi
^{0}\partial^{\mu}\bar{K}^{0}-\sqrt{2}\pi^{+}\partial^{\mu}K^{-})+B_{K^{\star
}K\pi}K_{\mu}^{\star0}(\bar{K}^{0}\partial^{\mu}\pi^{0}-\sqrt{2}K^{-}%
\partial^{\mu}\pi^{+})\nonumber\\
&  +C_{K^{\star}K\pi}\partial_{\nu}K_{\mu}^{\star0}(\partial^{\mu}\bar{K}%
^{0}\partial^{\nu}\pi^{0}-\sqrt{2}\partial^{\mu}K^{-}\partial^{\nu}\pi
^{+}) \nonumber \\
& + C_{K^{\star}K\pi}^{\ast}\partial_{\nu}K_{\mu}^{\star0}(\partial^{\mu}%
\pi^{0}\partial^{\nu}\bar{K}^{0}-\sqrt{2}\partial^{\mu}\pi^{+}\partial^{\nu
}K^{-})\nonumber\\
&  +A_{K^{\star}K\pi}^{\ast}\bar{K}_{\mu}^{\star0}(\pi^{0}\partial^{\mu}%
K^{0}-\sqrt{2}\pi^{-}\partial^{\mu}K^{+})+B_{K^{\star}K\pi}^{\ast}\bar{K}%
_{\mu}^{\star0}(K^{0}\partial^{\mu}\pi^{0}-\sqrt{2}K^{+}\partial^{\mu}\pi
^{-})\nonumber\\
&  +C_{K^{\star}K\pi}^{\ast}\partial_{\nu}\bar{K}_{\mu}^{\star0}(\partial
^{\mu}K^{0}\partial^{\nu}\pi^{0}-\sqrt{2}\partial^{\mu}K^{+}\partial^{\nu}%
\pi^{-}) \nonumber \\
& + C_{K^{\star}K\pi}\partial_{\nu}\bar{K}_{\mu}^{\star0}(\partial^{\mu
}\pi^{0}\partial^{\nu}K^{0}-\sqrt{2}\partial^{\mu}\pi^{-}\partial^{\nu}K^{+})
\label{KstarKpion}%
\end{align}

with the following coefficients:%

\begin{align}
A_{K^{\star}K\pi}  &  =\frac{i}{2}Z_{\pi}Z_{K}\left[  g_{1}(\sqrt{2}%
g_{1}w_{K_{1}}\phi_{S}-1)-\sqrt{2}h_{3}w_{K_{1}}\phi_{S}\right]\text{,}
\label{AKstarKp}\\
B_{K^{\star}K\pi}  &  =\frac{i}{4}Z_{\pi}Z_{K}\left[  2g_{1}+w_{a_{1}}%
(-3g_{1}^{2}-h_{2}+2h_{3})\phi_{N}+\sqrt{2}w_{a_{1}}(g_{1}^{2}+h_{2})\phi
_{S}\right]\text{,} \label{BKstarKp}\\
C_{K^{\star}K\pi}  &  =\frac{i}{2}Z_{\pi}Z_{K}w_{a_{1}}w_{K_{1}}g_{2}\text{.}
\label{CKstarKp}%
\end{align}

The interaction Lagrangian containing $K^{\star\pm}$ is analogous to the
Lagrangian presented in Eq.\ (\ref{KstarKpion}). Note that the Lagrangian in
Eq.\ (\ref{KstarKpion}) contains not only the parameter combinations $A_{K^{\star
}K\pi}$, $B_{K^{\star}K\pi}$ and $C_{K^{\star}K\pi}$ but also their complex
conjugates. This is necessary to ascertain that the Lagrangian is hermitian;
indeed we obtain $\mathcal{L}_{K^{\star}K\pi}^{\dagger}=\mathcal{L}_{K^{\star
}K\pi}$ upon substituting $A_{K^{\star}K\pi}\rightarrow A_{K^{\star}K\pi
}^{\ast}$, $B_{K^{\star}K\pi}\rightarrow B_{K^{\star}K\pi}^{\ast}$,
$C_{K^{\star}K\pi}\rightarrow C_{K^{\star}K\pi}^{\ast}$, $K_{\mu}^{\star
0}\rightarrow\bar{K}_{\mu}^{\star0}$, $K^{0}\rightarrow\bar{K}^{0}$,
$K^{+}\rightarrow K^{-}$, $\pi^{+}\rightarrow\pi^{-}$. In the following we
will focus only on the decay $K^{\star0}\rightarrow K\pi$; the corresponding
decay of $\bar{K}^{\star0}$ yields the same result due to isospin symmetry (as
do the corresponding $K^{\star\pm}$ decays).

The calculation of $\Gamma_{K^{\star0}\rightarrow K\pi}$ requires knowledge of
decay widths in two distinct channels: $K^{\star0}\rightarrow K^{0}\pi^{0}$
and $K^{\star0}\rightarrow K^{+}\pi^{-}$. (Note the changed charges for the
decay products, as in Sec.\ \ref{sec.KstarKp}). As apparent from
Eq.\ (\ref{KstarKpion}), these differ by a factor of two: $\Gamma_{K^{\star
0}\rightarrow K^{+}\pi^{-}}=2\Gamma_{K^{\star0}\rightarrow K^{0}\pi^{0}}$.
Then $\Gamma_{K^{\star0}\rightarrow K\pi}=3\Gamma_{K^{\star0}\rightarrow
K^{0}\pi^{0}}$. Let us therefore calculate the decay width for the process
$K^{\star0}\rightarrow K^{0}\pi^{0}$.

We denote the momenta of $K^{\star}$, $\pi$ and $K$ as $P$, $P_{1}$ and
$P_{2}$, respectively. $K_{\mu}^{\star}$ is a vector state for
which we have to consider the polarisation vector $\varepsilon
_{\mu}^{(\alpha)}(P)$. Then, upon substituting $\partial^{\mu}\rightarrow
-iP^{\mu}$\ for the decaying particle and $\partial^{\mu}\rightarrow
iP_{1,2}^{\mu}$ for the decay products, we obtain the following Lorentz-invariant $K^{\star
}K\pi$ scattering amplitude $-i\mathcal{M}_{K^{\star0}\rightarrow K^{0}\pi
^{0}}^{(\alpha)}$ from the Lagrangian (\ref{KstarKpion}):%
\begin{align}
-i\mathcal{M}_{K^{\star0}\rightarrow K^{0}\pi^{0}}^{(\alpha)}=\varepsilon
_{\mu}^{(\alpha)}(P)h_{K^{\star}K\pi}^{\mu} & =-\varepsilon_{\mu}^{(\alpha
)}(P)\left\{  A_{K^{\star}K\pi}P_{2}^{\mu}+B_{K^{\star}K\pi}P_{1}^{\mu
}+C_{K^{\star}K\pi}[P_{2}^{\mu}(P\cdot P_{1}) \right. \nonumber \\
& - \left. P_{1}^{\mu}(P\cdot
P_{2})]\right\}  \label{hKstarKp}%
\end{align}
with
\begin{equation}
h_{K^{\star}K\pi}^{\mu}=-\left\{  A_{K^{\star}K\pi}P_{2}^{\mu}+B_{K^{\star
}K\pi}P_{1}^{\mu}+C_{K^{\star}K\pi}[P_{2}^{\mu}(P\cdot P_{1})-P_{1}^{\mu
}(P\cdot P_{2})]\right\}\text{,}  \label{hKstarKp0}%
\end{equation}

where $h_{K^{\star}K\pi}^{\mu}$ denotes the $K^{\star}K\pi$ vertex.\newline

It will be necessary to determine the square of the scattering amplitude in
order to calculate the decay width. Given that the scattering amplitude in
Eq.\ (\ref{hKstarKp}) depends on the polarisation vector $\varepsilon_{\mu
}^{(\alpha)}(P)$, it is necessary to calculate the average of the amplitude
for all values of $\varepsilon_{\mu}^{(\alpha)}(P)$. This has already been
performed in Sec.\ \ref{sec.rhopipi} for the decay $\rho\rightarrow\pi\pi$; we
can calculate $|-i\mathcal{\bar{M}}_{K^{\star0}\rightarrow K^{0}\pi^{0}}|^{2}$
in accordance with Eq.\ (\ref{iM2}). We first calculate the squared vertex
$(h_{K^{\star}K\pi}^{\mu})^{2}$ using Eq.\ (\ref{hKstarKp}):

\begin{align}
(h_{K^{\star}K\pi}^{\mu})^{2}  &  =A_{K^{\star}K\pi}^{2}m_{K}^{2}+B_{K^{\star
}K\pi}^{2}m_{\pi}^{2}+C_{K^{\star}K\pi}^{2}[P_{2}^{\mu}(P\cdot P_{1}%
)-P_{1}^{\mu}(P\cdot P_{2})]^{2}\nonumber\\
&  +2A_{K^{\star}K\pi}B_{K^{\star}K\pi}P_{1}\cdot P_{2}+2A_{K^{\star}K\pi
}C_{K^{\star}K\pi}[P\cdot P_{1}m_{K}^{2}-(P_{1}\cdot P_{2})(P\cdot
P_{2})]\nonumber\\
&  +2B_{K^{\star}K\pi}C_{K^{\star}K\pi}[(P_{1}\cdot P_{2})(P\cdot
P_{1})-P\cdot P_{2}m_{\pi}^{2}]\text{.} \label{hKstarKp2}%
\end{align}

Additionally, again from Eq.\ (\ref{hKstarKp}):%

\begin{align}
(h_{K^{\star}K\pi}^{0})^{2}  &  =A_{K^{\star}K\pi}^{2}E_{K}^{2}+B_{K^{\star
}K\pi}^{2}E_{\pi}^{2}+C_{K^{\star}K\pi}^{2}[E_{K}(P\cdot P_{1})-E_{\pi}(P\cdot
P_{2})]^{2}\nonumber\\
&  +2A_{K^{\star}K\pi}B_{K^{\star}K\pi}E_{\pi}E_{K}+2A_{K^{\star}K\pi
}C_{K^{\star}K\pi}[(P\cdot P_{1})E_{K}^{2}-E_{\pi}E_{K}(P\cdot P_{2}%
)]\nonumber\\
&  +2B_{K^{\star}K\pi}C_{K^{\star}K\pi}[E_{\pi}E_{K}(P\cdot P_{1})-(P\cdot
P_{2})E_{\pi}^{2}]\text{.} \label{hKstarKp3}%
\end{align}

From Eqs.\ (\ref{iM2}), (\ref{hKstarKp2}) and (\ref{hKstarKp3}) we obtain%

\begin{align}
|-i\mathcal{\bar{M}}_{K^{\star0}\rightarrow K^{0}\pi^{0}}|^{2}  &  =\frac
{1}{3}\{(A_{K^{\star}K\pi}^{2}+B_{K^{\star}K\pi}^{2})k^{2}(m_{K^{\star}}%
,m_{K},m_{\pi})\nonumber\\
&  +C_{K^{\star}K\pi}^{2}\{k^{2}(m_{K^{\star}},m_{K},m_{\pi})[(P\cdot
P_{1})^{2}+(P\cdot P_{2})^{2}]\nonumber\\
&  -2(P\cdot P_{1})(P\cdot P_{2})(E_{\pi}E_{K}-P_{1}\cdot P_{2})\}\nonumber\\
&  +2A_{K^{\star}K\pi}B_{K^{\star}K\pi}(E_{\pi}E_{K}-P_{1}\cdot P_{2}%
)\nonumber\\
&  +2A_{K^{\star}K\pi}C_{K^{\star}K\pi}[k^{2}(m_{K^{\star}},m_{K},m_{\pi
})P\cdot P_{1}-(P\cdot P_{2})(E_{\pi}E_{K}-P_{1}\cdot P_{2})]\nonumber\\
&  +2B_{K^{\star}K\pi}C_{K^{\star}K\pi}[(P\cdot P_{1})(E_{\pi}E_{K}-P_{1}\cdot
P_{2})-k^{2}(m_{K^{\star}},m_{K},m_{\pi})(P\cdot P_{2})]\}\nonumber\\
&  =\frac{1}{3}\left\{  \{A_{K^{\star}K\pi}^{2}+B_{K^{\star}K\pi}%
^{2}+C_{K^{\star}K\pi}^{2}[(P\cdot P_{1})^{2}+(P\cdot P_{2})^{2}]\right.
\nonumber\\
&  \left.  +2C_{K^{\star}K\pi}[A_{K^{\star}K\pi}(P\cdot P_{1})-B_{K^{\star
}K\pi}(P\cdot P_{2})]\}k^{2}(m_{K^{\star}},m_{K},m_{\pi})\right. \nonumber\\
&  \left.  +2\{A_{K^{\star}K\pi}B_{K^{\star}K\pi}-C_{K^{\star}K\pi}^{2}(P\cdot
P_{1})(P\cdot P_{2})+C_{K^{\star}K\pi}(B_{K^{\star}K\pi}P\cdot P_{1}\right.
\nonumber\\
&  \left.  -A_{K^{\star}K\pi}P\cdot P_{2})\}(E_{\pi}E_{K}-P_{1}\cdot
P_{2})\right\}\text{.}  \label{MKstarKp}%
\end{align}

Note that Eq.\ (\ref{MKstarKp}) can also be written in a slightly different,
but equivalent, manner. To this end, note that the vertex $h_{K^{\star}K\pi
}^{\mu}$ from Eq.\ (\ref{hKstarKp0}) can be transformed as%

\begin{align}
h_{K^{\star}K\pi}^{\mu}  &  =-[A_{K^{\star}K\pi}P_{2}^{\mu}+B_{K^{\star}K\pi
}P_{1}^{\mu}+C_{K^{\star}K\pi}(m_{K^{\star}}E_{\pi}P_{2}^{\mu}-m_{K^{\star}%
}E_{K}P_{1}^{\mu})]\nonumber\\
&  =-(B_{K^{\star}K\pi}-C_{K^{\star}K\pi}m_{K^{\star}}E_{K})P_{1}^{\mu
}-(A_{K^{\star}K\pi}+C_{K^{\star}K\pi}m_{K^{\star}}E_{\pi})P_{2}^{\mu}\text{.}
\label{hKstarKp4}%
\end{align}

Inserting Eq.\ (\ref{hKstarKp4}) into Eq.\ (\ref{iM2}) yields%

\begin{align}
|-i\mathcal{\bar{M}}_{K^{\star0}\rightarrow K^{0}\pi^{0}}|^{2}  &  =\frac{1}{3}\{-[(B_{K^{\star
}K\pi}-C_{K^{\star}K\pi}m_{K^{\star}}E_{K})P_{1}^{\mu}+(A_{K^{\star}K\pi
}+C_{K^{\star}K\pi}m_{K^{\star}}E_{\pi})P_{2}^{\mu}]^{2}\nonumber\\
&  +\frac{1}{m_{K^{\star}}^{2}}[(B_{K^{\star}K\pi}-C_{K^{\star}K\pi
}m_{K^{\star}}E_{K})P_{1\mu}P^{\mu} \nonumber \\
& +(A_{K^{\star}K\pi}+C_{K^{\star}K\pi
}m_{K^{\star}}E_{\pi})P_{2\mu}P^{\mu}]^{2}\}\text{.} \label{hKstarKp5}%
\end{align}

Using $P_{1}\cdot P=m_{K^{\star}}E_{\pi}$ and $P_{2}\cdot P=m_{K^{\star}}%
E_{K}$ we obtain from Eq.\ (\ref{hKstarKp5})

\begin{align}
|-i\mathcal{\bar{M}}_{K^{\star0}\rightarrow K^{0}\pi^{0}}|^{2}  &  =\frac{1}{3}[(B_{K^{\star}%
K\pi}-C_{K^{\star}K\pi}m_{K^{\star}}E_{K})^{2}+(A_{K^{\star}K\pi}+C_{K^{\star
}K\pi}m_{K^{\star}}E_{\pi})^{2}\nonumber\\
&  -2(A_{K^{\star}K\pi
}+C_{K^{\star}K\pi}m_{K^{\star}}E_{\pi}) (B_{K^{\star}K\pi}-C_{K^{\star}K\pi}m_{K^{\star}}E_{K})]k^{2}(m_{K^{\star}},m_{K},m_{\pi
})\nonumber\\
&  =\frac{1}{3}[A_{K^{\star}K\pi}-B_{K^{\star}K\pi}+C_{K^{\star}K\pi
}m_{K^{\star}}(E_{\pi}+E_{K})]^{2}k^{2}(m_{K^{\star}},m_{K},m_{\pi
}) \nonumber\\
&  =\frac{1}{3}(A_{K^{\star}K\pi}-B_{K^{\star}K\pi}+C_{K^{\star}K\pi
}m_{K^{\star}}^{2})^{2}k^{2}(m_{K^{\star}},m_{K},m_{\pi})\text{.} \label{MKstarKp1}
\end{align}

Using Eq.\ (\ref{MKstarKp1}) -- or, equivalently, Eq.\ (\ref{MKstarKp}) -- we obtain the following formula for $\Gamma_{K^{\star0}\rightarrow K^{0}\pi^{0}}$:
\begin{equation}
\Gamma_{K^{\star0}\rightarrow K \pi}=3\frac{k(m_{K^{\star}},m_{K},m_{\pi})}{8\pi
m_{K^{\star}}^{2}}|-i\mathcal{\bar{M}}_{K^{\star0}\rightarrow K^{0}\pi^{0}}|^{2}\text{,}
\label{GKstarKp}%
\end{equation} 

where we have used the already discussed equality $\Gamma_{K^{\star0}\rightarrow K\pi}=3\Gamma_{K^{\star0}\rightarrow
K^{0}\pi^{0}}$.\\

Note that all parameters entering Eq.\ (\ref{GKstarKp}), i.e., Eqs.\ (\ref{AKstarKp}) - (\ref{CKstarKp}) and (\ref{MKstarKp1}), 
have been determined uniquely from our Fit I, see Table \ref{Fit1-4}. Therefore we can calculate the value of the decay width immediately
and obtain

\begin{equation}
\Gamma_{K^{\star0}\rightarrow K \pi} = 32.8 \text{ MeV.} \label{GKstarKp1}
\end{equation} 

The experimental value reads $\Gamma_{K^{\star0}\rightarrow K \pi}^{\exp} = 46.2 $ MeV \cite{PDG}.
Therefore, the value obtained within Fit I is by approximately 13 MeV (or 30\%) too small.

\subsection{Decay Width \boldmath $f_{1}(1420)\rightarrow K^{\star} K$ in Fit I}  \label{sec.f1S1}

The $f_{1}(1420)\equiv f_{1S}$ resonance represents a sharp peak in the
$K^{\star}K$ channel with a mass of $m_{f_{1}(1420)}^{\exp}=(1426.4\pm0.9)$
MeV and width $\Gamma_{f_{1}(1420)}^{\exp}=(54.9\pm2.6)$ MeV \cite{PDG}.
(There are also other decay channels for this resonance but they are
subdominant.) As discussed in Sec.\ \ref{sec.fitI}, Fit I yields a rather
large value of $m_{f_{1}(1420)}=1643.4$ MeV, see Table \ref{Fit1-5}. In this
section we address the question whether a value of $\Gamma_{f_{1}(1420)}$
close to the experimental value $\Gamma_{f_{1}(1420)}^{\exp}$ can be obtained,
thus improving the $f_{1}(1420)$ phenomenology in Fit I.\newline The
$f_{1S}K^{\star}K$ interaction Lagrangian from Eq.\ (\ref{Lagrangian}) reads

\begin{align}
\mathcal{L}_{f_{1S}K^{\star}K} &  =A_{f_{1S}K^{\star}K}f_{1S}^{\mu}(K_{\mu
}^{\star0}\bar{K}^{0}+K_{\mu}^{\star+}K^{-}-\bar{K}_{\mu}^{\star0}K^{0}%
-K_{\mu}^{\star-}K^{+})\nonumber\\
&  +B_{f_{1S}K^{\star}K}f_{1S}^{\mu}[(\partial_{\nu}K_{\mu}^{\star0}%
-\partial_{\mu}K_{\nu}^{\star0})\partial^{\nu}\bar{K}^{0}+(\partial_{\nu
}K_{\mu}^{\star+}-\partial_{\mu}K_{\nu}^{\star+})\partial^{\nu}K^{-}%
\nonumber\\
&  -(\partial_{\nu}\bar{K}_{\mu}^{\star0}-\partial_{\mu}\bar{K}_{\nu}^{\star
0})\partial^{\nu}K^{0}-(\partial_{\nu}K_{\mu}^{\star-}-\partial_{\mu}K_{\nu
}^{\star-})\partial^{\nu}K^{+}]\nonumber\\
&  +C_{f_{1S}K^{\star}K}\partial^{\nu}f_{1S}^{\mu}(K_{\mu}^{\star0}%
\partial_{\nu}\bar{K}^{0}-K_{\nu}^{\star0}\partial_{\mu}\bar{K}^{0}+K_{\mu
}^{\star+}\partial_{\nu}K^{-}-K_{\nu}^{\star+}\partial_{\mu}K^{-}\nonumber\\
&  -\bar{K}_{\mu}^{\star0}\partial_{\nu}K^{0}+\bar{K}_{\nu}^{\star0}%
\partial_{\mu}K^{0}-K_{\mu}^{\star-}\partial_{\nu}K^{+}+K_{\nu}^{\star
-}\partial_{\mu}K^{+})\label{f1SKstarKaon}
\end{align}

with

\begin{align}
A_{f_{1S}K^{\star}K} &  =\frac{i}{4}Z_{K}\left[  g_{1}^{2}(\sqrt{2}\phi
_{N}-6\phi_{S})+\sqrt{2}h_{2}(\phi_{N}-\sqrt{2}\phi_{S})+4h_{3}\phi
_{S}\right]\text{,} \label{Af1SKstarK}\\
B_{f_{1S}K^{\star}K} &  =-\frac{i}{\sqrt{2}}Z_{K}g_{2}w_{K_{1}}\text{,}
\label{Bf1SKstarK}\\
C_{f_{1S}K^{\star}K} &  =\frac{i}{\sqrt{2}}Z_{K}g_{2}w_{K_{1}}\text{.}%
\label{Cf1SKstarK}
\end{align}

The Lagrangian (\ref{f1SKstarKaon})\ allows us to calculate the decay width for
the process $f_{1S}\rightarrow\bar{K}^{\star0}K^{0}$. Let us denote the momenta of
$f_{1S}$, $\bar{K}_{\mu}^{\star0}$ and $K^{0}$ as $P$, $P_{1}$ and $P_{2}$.
Two vector states are involved in the decay process: $f_{1S}$\ and$\ K_{\mu
}^{\star}$. As in Sec.\ \ref{sec.f1N1}, we consider the corresponding
polarisation vectors labelled as $\varepsilon_{\mu}^{(\alpha)}(P)$ for
$f_{1S}$\ and $\varepsilon_{\nu}^{(\beta)}(P_{1})$ for $K_{\mu}^{\star}$. Then
substituting $\partial^{\mu}\rightarrow-iP^{\mu}$\ the for decaying particle and
$\partial^{\mu}\rightarrow iP_{1,2}^{\mu}$ for the decay products, we obtain the
following $f_{1S}\bar{K}^{\star0}K^{0}$ scattering amplitude $-i\mathcal{M}%
_{f_{1S}\rightarrow\bar{K}^{\star0}K^{0}}^{(\alpha,\beta)}$ from the
Lagrangian (\ref{f1SKstarKaon}):

\begin{align}
-i\mathcal{M}_{f_{1S}\rightarrow\bar{K}^{\star0}K^{0}}^{(\alpha,\beta)}  &
=\varepsilon_{\mu}^{(\alpha)}(P)\varepsilon_{\nu}^{(\beta)}(P_{1}%
)h_{f_{1S}\bar{K}^{\star0}K^{0}}^{\mu\nu}=i\varepsilon_{\mu}^{(\alpha
)}(P)\varepsilon_{\nu}^{(\beta)}(P_{1})\nonumber\\
&  \times\left\{  A_{f_{1S}K^{\star}K}g^{\mu\nu}+[B_{f_{1S}K^{\star}K}%
(P_{1}^{\mu}P_{2}^{\nu}-(P_{1}\cdot P_{2})g^{\mu\nu}] \right. \nonumber\\
&  \left.  + C_{f_{1S}K^{\star}K}[(P\cdot P_{2})g^{\mu\nu}-P_{2}^{\mu}P^{\nu
}]\right\}  \label{Mf1SKstarK}
\end{align}

with
\begin{align}
h_{f_{1S}\bar{K}^{\star0}K^{0}}^{\mu\nu} & =i\left\{  A_{f_{1S}K^{\star}K}%
g^{\mu\nu}+[B_{f_{1S}K^{\star}K}(P_{1}^{\mu}P_{2}^{\nu}-(P_{1}\cdot
P_{2})g^{\mu\nu}] \right. \nonumber \\
& \left. +C_{f_{1S}K^{\star}K}[(P\cdot P_{2})g^{\mu\nu}-P_{2}^{\mu
}P^{\nu}]\right\}\text{,}  \label{hf1SKstarK}
\end{align}

where $h_{f_{1S}\bar{K}^{\star0}K^{0}}^{\mu}$ denotes the $f_{1S}\bar
{K}^{\star0}K^{0}$ vertex. The vertex in Eq.\ (\ref{hf1SKstarK}) is analogous
to the $a_{1}\rho\pi$ vertex of Eq.\ (\ref{ha1rpQ}). Therefore we can use the formulas
for the $a_{1}(1260)\rightarrow\rho\pi$ decay amplitude and decay width to
calculate $\Gamma_{f_{1S}\rightarrow\bar{K}^{\star}K}$. The corresponding
coefficients in the two vertices have to be substituted: $A_{a_{1}\rho\pi
}\rightarrow A_{f_{1S}K^{\star}K}$, $B_{a_{1}\rho\pi}\rightarrow
B_{f_{1S}K^{\star}K}$, $C_{a_{1}\rho\pi}\rightarrow-C_{f_{1S}K^{\star}K}$
and an isospin factor of four has to be considered to account
for the decays $f_{1S}\rightarrow\bar{K}^{\star0}K^{0}$, $\bar{K}^{0}K^{\star0}$,
$K^{\star+}K^{-}$ and $K^{\star-}K^{+}$. However, as in Sec.\ \ref{sec.f1N1},
the large modulus of the parameter $g_{2}=-11.2$ (see Table \ref{Fit1-4})
leads to

\begin{equation}
\Gamma_{f_{1S}\rightarrow\bar{K}^{\star}K}=17.6\text{ GeV.}
\label{Gf1SKstarK1}%
\end{equation}

This value is in stark contrast to the one reported by the PDG: $\Gamma
_{f_{1}(1420)}^{\exp}=(54.9\pm2.6)$ MeV. Therefore Fit I, where the scalar meson
states are assumed to be under $1$ GeV, yields a very poor phenomenology of the
strange axial-vector isosinglet: $m_{f_{1}(1420)}=1643.4$ MeV is by
approximately $200$ MeV too large and $\Gamma_{f_{1S}\rightarrow\bar{K}%
^{\star}K}=17.6$ GeV is unphysical (as it is two orders of magnitude too
large). Note that we have obtained similarly large values of $\Gamma
_{a_{1}(1260)\rightarrow\rho\pi}\approx13$ GeV in Sec.\ \ref{sec.a1rhopion1}
and of $\Gamma_{f_{1N}\rightarrow\bar{K}^{\star}K}\approx2$ GeV in
Sec.\ \ref{sec.f1N1}, again due to the large value of $g_{2}$.

Analogously to considerations in the mentioned sections, let us vary
$\Gamma_{\rho\rightarrow\pi\pi}$ to examine the corresponding change of
$\Gamma_{f_{1S}\rightarrow\bar{K}^{\star}K}$, as $\Gamma_{\rho\rightarrow
\pi\pi}$ determines $g_{2}$ uniquely.

\begin{figure}
[h]
\begin{center}
\includegraphics[
height=2.2582in,
width=4.1666in
]%
{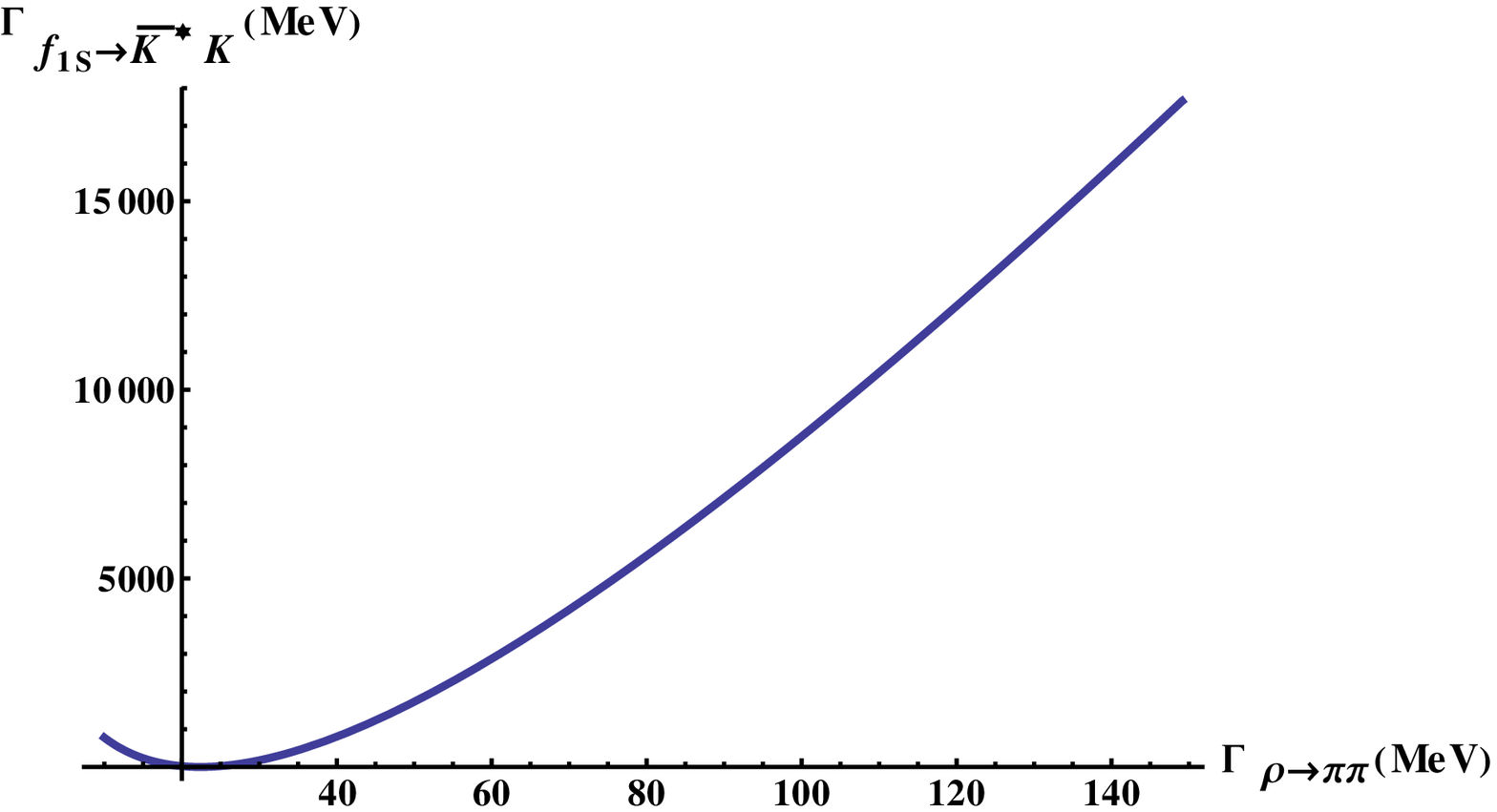}%
\caption{$\Gamma_{f_{1S}\equiv f_{1}(1420) \rightarrow \bar{K}^{\star} K}$ as
function of $\Gamma_{\rho \rightarrow \pi \pi}$.}%
\label{f1SKstarK1}%
\end{center}
\end{figure}

We observe from Fig.\ \ref{f1SKstarK1} that $\Gamma_{f_{1S}\rightarrow\bar
{K}^{\star}K}$ corresponds to $\Gamma_{f_{1}(1420)}^{\exp}=(54.9\pm2.6)$ MeV
only if $\Gamma_{\rho\rightarrow\pi\pi}\sim30$ MeV. Thus we would require
$\Gamma_{\rho\rightarrow\pi\pi}$ that is approximately $120$ MeV smaller than
$\Gamma_{\rho\rightarrow\pi\pi}^{\exp}$. Consequently, there is tension
between $\Gamma_{f_{1S}\rightarrow\bar{K}^{\star}K}$ and $\Gamma
_{\rho\rightarrow\pi\pi}$ as it is not possible to obtain physical values
simultaneously for both decay widths. This problem is analogous to the one
described in Sec.\ \ref{sec.a1rhopion1} for $\Gamma_{a_{1}(1260)\rightarrow
\rho\pi}$ and in Sec.\ \ref{sec.f1N1} for $\Gamma_{f_{1N}\rightarrow\bar
{K}^{\star}K}$ and represents an additional difficulty for Fit I.

\subsection{\boldmath $K_{1}$ Decays in Fit I} \label{sec.K11}

We have seen at the end of Sec.\ \ref{sec.fitI} that Fit I yields $m_{K_{1}%
}=1520$ MeV, a value that is significantly larger than the mass of the
resonance $K_{1}(1270)$ to which our $K_{1}$ state was assigned in
Sec.\ \ref{sec.assignment}. For this reason, we have reassigned our $K_{1}$
state to $K_{1}(1400)$ because the mass of this resonance [$m_{K_{1}%
(1400)}=(1403\pm7)$ MeV] corresponds better to the value of $m_{K_{1}}$
obtained from our Fit I. In this section we discuss whether it is possible to
obtain a correct value for the decay width of the $K_{1}$ field [the experimental
result reads $\Gamma_{K_{1}(1400)}=(174\pm13)$ MeV]. To this end, we will
consider all hadronic decays of $K_{1}(1400)$ that can be calculated within
our model: $K_{1}(1400)\rightarrow K^{\star}\pi$, $\rho K$ and $\omega K$.

We present the relevant interaction Lagrangians in a single equation:

\begin{align}
\mathcal{L}_{K_{1}} &  =A_{K_{1}K^{\star}\pi}K_{1}^{\mu0}\left(  \bar{K}_{\mu
}^{\star0}\pi^{0}-\sqrt{2}K_{\mu}^{\star-}\pi^{+}\right)  \nonumber\\
&  +B_{K_{1}K^{\star}\pi}\left\{  K_{1}^{\mu0}\left[  \left(  \partial_{\nu
}\bar{K}_{\mu}^{\star0}-\partial_{\mu}\bar{K}_{\nu}^{\star0}\right)
\partial^{\nu}\pi^{0}-\sqrt{2}\left(  \partial_{\nu}K_{\mu}^{\star-}%
-\partial_{\mu}K_{\nu}^{\star-}\right)  \partial^{\nu}\pi^{+}\right]  \right.
\nonumber\\
&  \left.  +\partial^{\nu}K_{1}^{\mu0}\left[  \left(  \bar{K}_{\nu}^{\star
0}\partial_{\mu}\pi^{0}-\bar{K}_{\mu}^{\star0}\partial_{\nu}\pi^{0}\right)
-\sqrt{2}\left(  K_{\nu}^{\star-}\partial_{\mu}\pi^{+}-K_{\mu}^{\star
-}\partial_{\nu}\pi^{+}\right)  \right]  \right\}  \nonumber \\
&  +A_{K_{1}\rho K}K_{1}^{\mu0}\left(  \rho_{\mu}^{0}\bar{K}^{0}-\sqrt{2}%
\rho_{\mu}^{+}K^{-}\right)  \nonumber \\
&  +B_{K_{1}\rho K}\left\{  K_{1}^{\mu0}\left[  \left(  \partial_{\nu}%
\rho_{\mu}^{0}-\partial_{\mu}\rho_{\nu}^{0}\right)  \partial^{\nu}\bar{K}%
^{0}-\sqrt{2}\left(  \partial_{\nu}\rho_{\mu}^{+}-\partial_{\mu}\rho_{\nu}%
^{+}\right)  \partial^{\nu}K^{-}\right]  \right.  \nonumber\\
&  \left.  +\partial^{\nu}K_{1}^{\mu0}\left[  \left(  \rho_{\nu}^{0}%
\partial_{\mu}\bar{K}^{0}-\rho_{\mu}^{0}\partial_{\nu}\bar{K}^{0}\right)
-\sqrt{2}\left(  \rho_{\nu}^{+}\partial_{\mu}K^{-}-\rho_{\mu}^{+}\partial
_{\nu}K^{-}\right)  \right]  \right\}  \nonumber \\
&  +A_{K_{1}\omega_{N}K}K_{1}^{\mu0}\omega_{N\mu}\bar{K}^{0}+B_{K_{1}%
\omega_{N}K}\left[  K_{1}^{\mu0}\left(  \partial_{\nu}\omega_{N\mu}%
-\partial_{\mu}\omega_{N\nu}\right)  \partial^{\nu}\bar{K}^{0} \right. \nonumber \\
& \left. +\partial^{\nu
}K_{1}^{\mu0}\left(  \omega_{N\nu}\partial_{\mu}\bar{K}^{0}-\omega_{N\mu
}\partial_{\nu}\bar{K}^{0}\right)  \right]  \label{K1}
\end{align}

with

\begin{align}
A_{K_{1}K^{\star}\pi} &  =\frac{i}{\sqrt{2}}Z_{\pi}(h_{3}-g_{1}^{2})\phi
_{S}\text{,} \label{AK1KstarK}\\
B_{K_{1}K^{\star}\pi} &  =-\frac{i}{2}Z_{\pi}g_{2}w_{a_{1}}\text{,} \label{BK1KstarK}\\
A_{K_{1}\rho K} &  =\frac{i}{4}Z_{K}\left[  g_{1}^{2}(\phi_{N}+\sqrt{2} 
\phi_{S})-h_{2}(\phi_{N}-\sqrt{2}\phi_{S})-2h_{3}\phi_{N}\right]\text{,}
\label{AK1rK}\\
B_{K_{1}\rho K} &  =\frac{i}{2}Z_{K}g_{2}w_{K_{1}}\text{,} \label{BK1rK}\\
A_{K_{1}\omega_{N}K} &  =-\frac{i}{4}Z_{K}[g_{1}^{2}(\phi_{N}+\sqrt{2}\phi
_{S})-h_{2}(\phi_{N}-\sqrt{2}\phi_{S})-2h_{3}\phi_{N}]\text{,} \label{AK1oK}\\
B_{K_{1}\omega_{N}K} &  =-\frac{i}{2}Z_{K}g_{2}w_{K_{1}}\text{.} \label{BK1oK}
\end{align}

We observe from Eq.\ (\ref{K1}) that the interaction Lagrangians for the decay
processes $K_{1}^{0}\rightarrow K^{\star0}\pi^{0}$, $K_{1}^{0}\rightarrow
\rho^{0}K^{0}$ and $K_{1}^{0}\rightarrow\omega_{N}K^{0}$ possess the same form:%

\begin{align}
\mathcal{L}_{K_{1}}  &  =A_{K_{1}}K_{1}^{\mu0}V_{\mu}^{0}{\bar{P}}%
^{0}\nonumber\\
&  +B_{K_{1}}\left[  K_{1}^{\mu0}\left(  \partial_{\nu}V_{\mu}^{0}%
-\partial_{\mu}V_{\nu}^{0}\right)  \partial^{\nu}{\bar{P}}^{0}+\partial^{\nu
}K_{1}^{\mu0}\left(  V_{\nu}^{0}\partial_{\mu}{\bar{P}}^{0}-V_{\mu}
^{0}\partial_{\nu}{\bar{P}}^{0}\right)  \right]\text{,}  \label{K12}
\end{align}

where $A_{K_{1}}=\{A_{K_{1}K^{\star}\pi},A_{K_{1}\rho K},A_{K_{1}\omega_{N}%
K}\}$, $B_{K_{1}}=\{B_{K_{1}K^{\star}\pi},B_{K_{1}\rho K},B_{K_{1}\omega_{N}%
K}\}$,$\ V_{\mu}=\{\bar{K}_{\mu}^{\star},\rho_{\mu},$ $\omega_{N\mu}\}$ and
${\bar{P}}=\{\pi,\bar{K}\}$. Let us therefore consider a generic decay process
of the form $K_{1}\rightarrow$ $V^{0}{\bar{P}}^{0}$ [if applicable,
the contribution of the decays into charged modes to the full decay width will be
larger by a factor of two than the contribution of the neutral modes, see
Eq.\ (\ref{K1})].

To this end, we denote the momenta of $K_{1}$, $V$ and ${\bar{P}}$\ as $P$,
$P_{1}$ and $P_{2}$, respectively. The stated decay process involves two
vector states: $K_{1}$\ and$\ {V}$. As in Sec.\ \ref{sec.AVP}, we have to
consider the corresponding polarisation vectors; let us denote them as
$\varepsilon_{\mu}^{(\alpha)}(P)$ for $K_{1}$\ and $\varepsilon_{\nu}%
^{(\beta)}(P_{1})$ for $V$. Consequently, upon substituting $\partial^{\mu
}\rightarrow-iP^{\mu}$\ for the decaying particle and $\partial^{\mu
}\rightarrow iP_{1,2}^{\mu}$ for the decay products, we obtain the following
Lorentz-invariant $K_{1}V{\bar{P}}$ scattering amplitude $-i\mathcal{M}%
_{K_{1}\rightarrow V^{0}{\bar{P}}^{0}}^{(\alpha,\beta)}$:

\begin{align}
-i\mathcal{M}_{K_{1}\rightarrow V^{0}{\bar{P}}^{0}}^{(\alpha,\beta)}  &
=\varepsilon_{\mu}^{(\alpha)}(P)\varepsilon_{\nu}^{(\beta)}(P_{1}
)h_{K_{1}V{\bar{P}}}^{\mu\nu}=i\varepsilon_{\mu}^{(\alpha)}(P)\varepsilon
_{\nu}^{(\beta)}(P_{1})\nonumber\\
&  \times A_{K_{1}}g^{\mu\nu}+B_{K_{1}}\left[  P_{1}^{\mu}P_{2}^{\nu}
+P_{2}^{\mu}P^{\nu}-(P_{1}\cdot P_{2})g^{\mu\nu}-(P\cdot P_{2})g^{\mu\nu
}\right]  \label{iMK1VP}
\end{align}

with

\begin{equation}
h_{K_{1}V{\bar{P}}}^{\mu\nu}=i\left\{  A_{K_{1}}g^{\mu\nu}+B_{K_{1}}
[P_{1}^{\mu}P_{2}^{\nu}+P_{2}^{\mu}P^{\nu}-(P_{1}\cdot P_{2})g^{\mu\nu
}-(P\cdot P_{2})g^{\mu\nu}]\right\}\text{,}  \label{hK1VP}
\end{equation}

where $h_{K_{1}V{\bar{P}}}^{\mu\nu}$ denotes the $K_{1}V{\bar{P}}$ vertex.

The vertex of Eq.\ (\ref{hK1VP}) corresponds to the vertex of Eq.\ (\ref{hAVP}).
We can therefore utilise the decay width formula derived in
Sec.\ \ref{sec.AVP} for a generic decay of an axial-vector state into a vector
and a pseudoscalar state. Setting $A_{AV{\tilde{P}}}\equiv A_{K_{1}}$ and
$B_{AV{\tilde{P}}}\equiv B_{K_{1}}$ in the Lagrangian (\ref{AVP}) we obtain
from Eq.\ (\ref{GAVP})

\begin{equation}
\Gamma_{K_{1}\rightarrow V{\bar{P}}}=I\frac{k(m_{K_{1}},m_{V},m_{{\bar{P}}}
)}{8\pi m_{K_{1}}^{2}}|-i\mathcal{\bar{M}}_{K_{1}\rightarrow V^{0}{\bar{P}
}^{0}}|^{2} \label{GK1VP}
\end{equation}

with the isospin factor $I=3$ for $K_{1}\rightarrow K^{\star}\pi$ and
$K_{1}\rightarrow\rho K$ and $I=1$ for $K_{1}\rightarrow\omega K$ -- as
apparent from Eq.\ (\ref{K1}), the decay width into charged modes (if
applicable) will be larger by a factor of two than the decay width into
neutral modes.

We turn now to the discussion of results for the three decays in Eq.\ (\ref{K1}).

\subsubsection{Decay Width \boldmath$K_{1}\rightarrow K^{\star}\pi$}

In this case we set $V^{0}\equiv K^{\star}$ and ${\bar{P}\equiv}\pi$ in
Eq.\ (\ref{GK1VP}). Given that all parameters entering Eqs.\ (\ref{AK1KstarK})
and (\ref{BK1KstarK}) are known from Table \ref{Fit1-4}, we consequently
obtain the following value of the decay width

\begin{equation}
\Gamma_{K_{1}\rightarrow K^{\star}\pi}=6.73\text{ GeV.} \label{GK1Kstarp}
\end{equation}

\begin{figure}[!t]
\begin{center}
\includegraphics[
height=2.2582in,
width=4.1666in
]%
{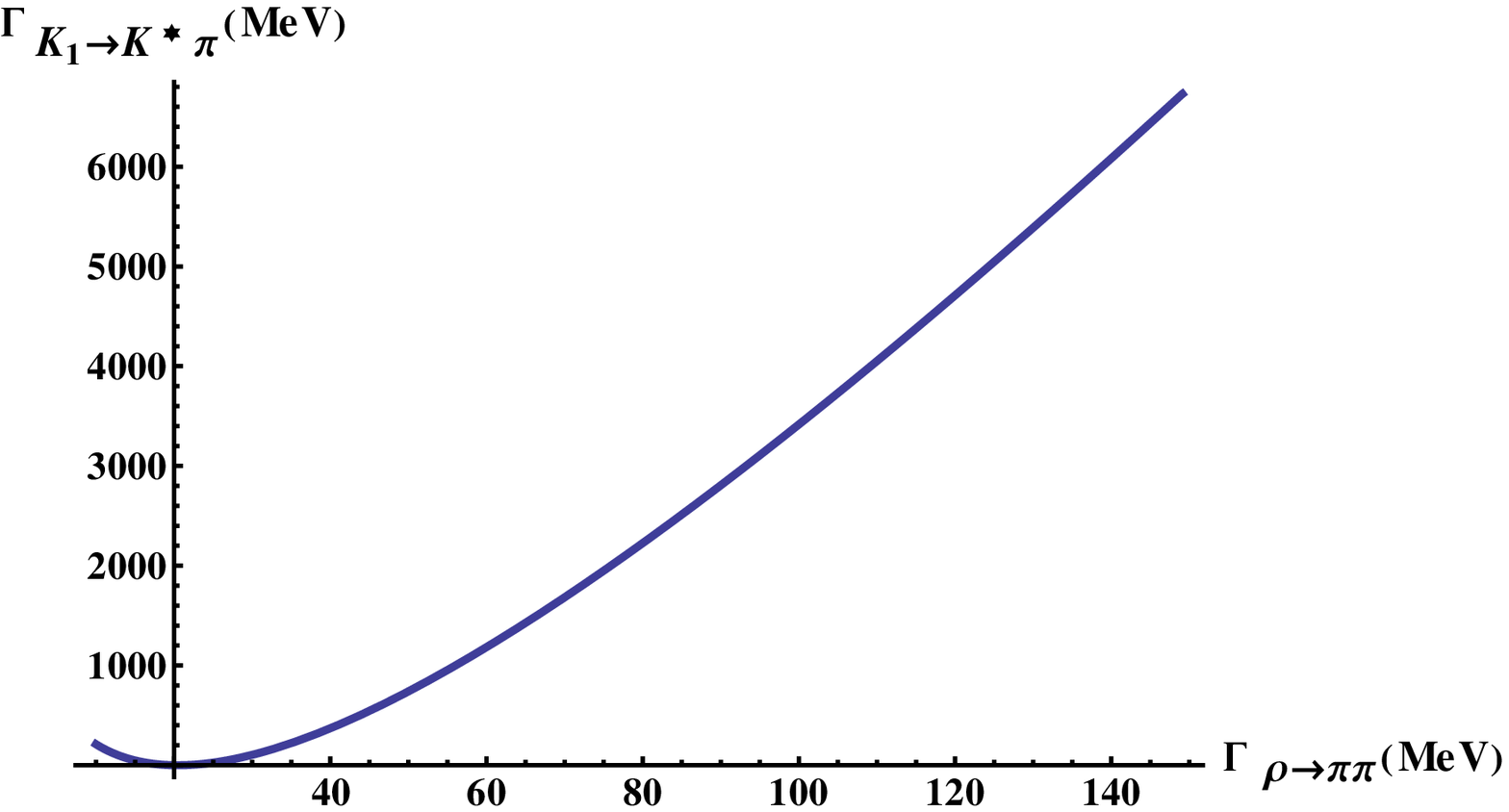}
\caption{$\Gamma_{K_{1}\rightarrow K^{\star}\pi}$ as function of $\Gamma
_{\rho\rightarrow\pi\pi}$.}
\label{K1Kstarp1}%
\end{center}
\end{figure}

This decay width suffers from the same issues as the decay widths in Sections
\ref{sec.a1rhopion1}, \ref{sec.f1N1} and \ref{sec.f1S1}: if we vary
$\Gamma_{\rho\rightarrow\pi\pi}$ to ascertain whether $\Gamma_{K_{1}%
\rightarrow K^{\star}\pi}$ can be sufficiently decreased, then the dependence
in Fig.\ \ref{K1Kstarp1} is obtained. It is apparent from Fig.\ \ref{K1Kstarp1} that $\Gamma_{\rho\rightarrow\pi\pi
}$ would have to be decreased by approximately $120$ MeV for $\Gamma
_{K_{1}(1400)\rightarrow K^{\star}\pi}=(164\pm16)$ MeV \cite{PDG} to be
obtained. The value in Eq.\ (\ref{GK1Kstarp}) is thus by an order of magnitude
too large.

\subsubsection{Decay Width \boldmath $K_{1}\rightarrow\rho K$}

As in the case of $\Gamma_{K_{1}\rightarrow K^{\star}\pi}$, we use the parameter
values from Table \ref{Fit1-4} to calculate the coefficients in Eqs.\ (\ref{AK1rK}) and (\ref{BK1rK}).
Again, there is no freedom to adjust parameters as they
are uniquely determined from the fit. We obtain from Eq.\ (\ref{GK1VP})
\begin{equation}
\Gamma_{K_{1}\rightarrow\rho K}=4.77\text{ GeV.} \label{GK1rK}
\end{equation}

\begin{figure}[!h]
\begin{center}
\includegraphics[
height=2.2582in,
width=4.1666in
]
{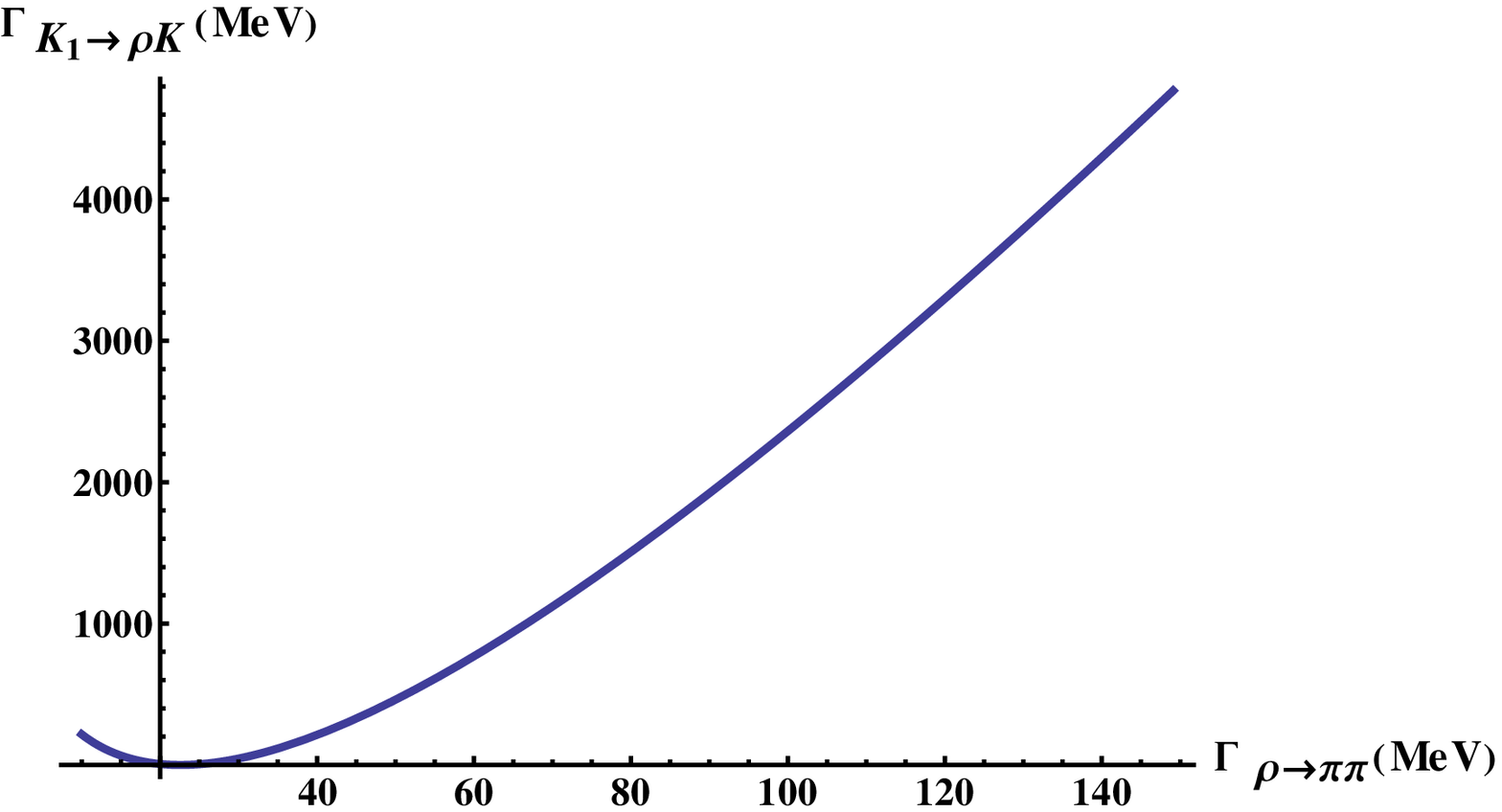}
\caption{$\Gamma_{K_{1}\rightarrow\rho K}$ as function of $\Gamma
_{\rho\rightarrow\pi\pi}$.}
\label{K1rK1}
\end{center}
\end{figure}

This value is of the same order of magnitude as the one in
Eq.\ (\ref{GK1Kstarp}), and equally unphysical. Additionally, the value in
Eq.\ (\ref{GK1rK}) cannot be sufficiently decreased to $\Gamma_{K_{1}%
(1400)\rightarrow\rho K}=(2.1\pm1.1)$ MeV \cite{PDG} unless $\Gamma
_{\rho\rightarrow\pi\pi}\simeq25$ MeV, see Fig.\ \ref{K1rK1}. The value of Eq.\ (\ref{GK1rK}) is thus by three orders of magnitude too large.

\subsubsection{Decay Width \boldmath $K_{1}\rightarrow \omega_{N}K$}

Similarly to the previous two $K_{1}$ decays, we obtain from
Eqs.\ (\ref{AK1oK}), (\ref{BK1oK}) and (\ref{GK1VP}) and Table \ref{Fit1-4}:

\begin{equation}
\Gamma_{K_{1}\rightarrow\omega_{N}K}=1.59\text{ GeV.}\label{GK1oK}
\end{equation}

\begin{figure}
[h]
\begin{center}
\includegraphics[
height=2.2582in,
width=4.1666in
]
{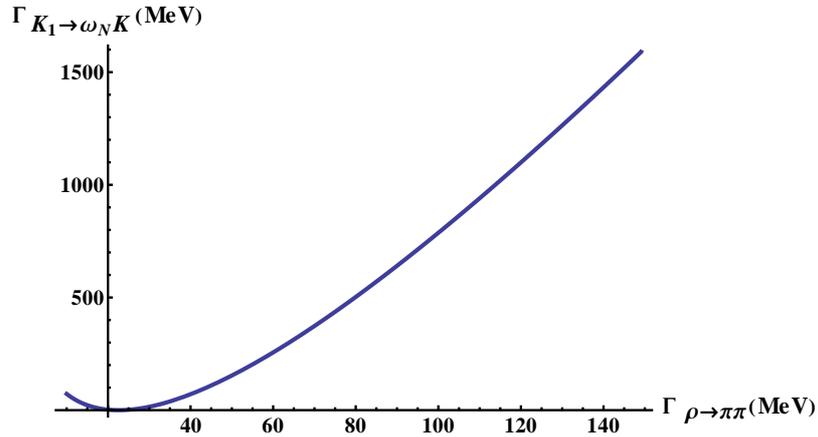}
\caption{$\Gamma_{K_{1}\rightarrow\omega_{N}K}$ as function of $\Gamma
_{\rho\rightarrow\pi\pi}$.}
\label{K1oK1}
\end{center}
\end{figure}

This value is also dramatically larger than the physical value $\Gamma
_{K_{1}(1400)\rightarrow\omega K}=(1.7\pm1.7)$ MeV \cite{PDG}. The large value
of $\Gamma_{K_{1}\rightarrow\omega_{N}K}$ is decreased to the physical value
of the $K_{1}(1400)$ decay width in this channel only if $\Gamma
_{\rho\rightarrow\pi\pi}\simeq25$ MeV is considered, see Fig.\ \ref{K1oK1}.
The value of Eq.\ (\ref{GK1oK}) is thus by three orders of magnitude too large.

The $K_{1}$ phenomenology is therefore very poorly described in Fit I.
Combined results of Eqs.\ (\ref{GK1Kstarp}), (\ref{GK1rK}) and (\ref{GK1oK})
suggest that the full decay width of the $K_{1}(1400)$ resonance should be
$\sim10$ GeV, two orders of magnitude larger than the experimental value
$\Gamma_{K_{1}(1400)}=(174\pm13)$ MeV \cite{PDG}. Such a resonance would then
not be observable in the physical spectrum. These results are consequently
another indication that the fit with the scalar states below $1$ GeV is
not favoured.\\

\begin{figure}
[h]
\begin{center}
\includegraphics[
height=2.2582in,
width=4.1666in
]
{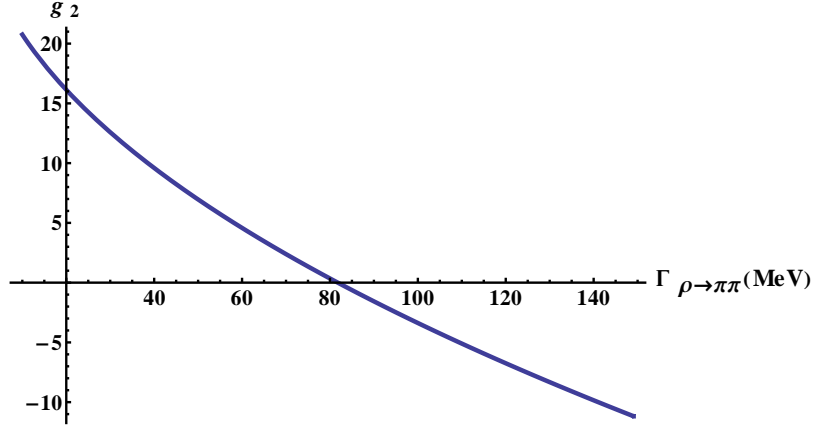}
\caption{Parameter $g_{2}$ as function of $\Gamma_{\rho\rightarrow\pi\pi}$.}
\label{fg21}
\end{center}
\end{figure}

Let us also note that the stated results for the decay widths $\Gamma
_{K_{1}\rightarrow K^{\star}\pi}$, $\Gamma_{K_{1}\rightarrow\rho K}$ and
$\Gamma_{K_{1}\rightarrow\omega K}$ would all require similar values of
$\Gamma_{\rho\rightarrow\pi\pi}\sim(25-30)$ MeV. This in turn implies that
$g_{2}\sim14$ would be needed for the $K_{1}(1400)$ decays to be described
properly, see Fig.\ \ref{fg21}. On the other hand, $g_{2}=-11.2$ used
throughout this chapter is obtained under the condition that $\Gamma
_{\rho\rightarrow\pi\pi}=149.1$ MeV $=\Gamma_{\rho\rightarrow\pi\pi}^{\exp}$.
We thus have $g_{2}$ with the needed modulus, but the sign of $g_{2}$
leads to the mentioned bad results regarding the $K_{1}(1400)$ decay width.

Let us nonetheless consider $\pi\pi$ scattering lengths as well, just as
in\ Scenario I of the $U(2)\times U(2)$ version of the model.

\section{Pion-Pion Scattering Lengths} \label{SL}

In this section we calculate the $\pi\pi$ scattering lengths at threshold,
analogously to the calculation already performed in the two-flavour version of
the model (see Sec.\ \ref{sec.SLQ}). The main difference to the calculation in the
two-flavour case lies in the fact that now the inclusion of an
additional (pure strange) scalar isosinglet $\sigma_{S}$ generates an
additional mixed (and predominantly strange) scalar isosinglet field
($\sigma_{2}$) that in principle also influences the $\pi\pi$ scattering. Note
that an explicit calculation of the $\pi\pi$ scattering terms yields no
further contributing terms other than those already mentioned here and in
Sec.\ \ref{sec.SLQ} -- i.e., "pure" $\pi\pi$ scattering (contact scattering) and
scattering via virtual $\sigma$ and $\rho$ mesons. The former are represented
only by $\sigma_{N}$ in the two-flavour version of the model and by
$\sigma_{1,2}$ in the current version of the model. Therefore, we have to
modify the $\pi\pi$ scattering amplitude to include the contribution from the
additional scalar field. This is implemented by considering the $\sigma\pi\pi$
Lagrangian $\mathcal{L}_{\sigma\pi\pi}$, Eq.\ (\ref{sigmapionpion}), and
substituting the pure fields $\sigma_{N,S}$ by the mixed fields $\sigma_{1,2}%
$. To this end, let us rewrite $\mathcal{L}_{\sigma\pi\pi}$ in the following way:
\begin{align}
\mathcal{L}_{\sigma\pi\pi}  &  =(A_{\sigma_{N}\pi\pi}\cos\varphi_{\sigma
}+A_{\sigma_{S}\pi\pi}\sin\varphi_{\sigma})\sigma_{1}\vec{\pi}^{2}%
+B_{\sigma_{N}\pi\pi}\cos\varphi_{\sigma}\sigma_{1}(\partial_{\mu}\vec{\pi
})^{2}+C_{\sigma_{N}\pi\pi}\cos\varphi_{\sigma}\sigma_{1}\vec{\pi}\cdot
\square\vec{\pi}\nonumber\\
&  +(-A_{\sigma_{N}\pi\pi}\sin\varphi_{\sigma}+A_{\sigma_{S}\pi\pi}\cos
\varphi_{\sigma})\sigma_{2}\vec{\pi}^{2}-B_{\sigma_{N}\pi\pi}\sin
\varphi_{\sigma}\sigma_{2}(\partial_{\mu}\vec{\pi})^{2}-C_{\sigma_{N}\pi\pi
}\sin\varphi_{\sigma}\sigma_{2}\vec{\pi}\cdot\square\vec{\pi} \nonumber \\
\label{sigmapionpion1}%
\end{align}

with $A_{\sigma_{N}\pi\pi}$, $B_{\sigma_{N}\pi\pi}$, $C_{\sigma_{N}\pi\pi}$
and $A_{\sigma_{S}\pi\pi}$ from Eqs.\ (\ref{ANspp}) - (\ref{ASspp}); for
simplicity, we have also made use of $B_{\sigma_{S}\pi\pi}\sim h_{1}=0$,
Eq.\ (\ref{BSspp}).

Then we obtain the following contribution from the the $\pi\pi\sigma$ vertex
to the scattering amplitude $\mathcal{M}_{\pi\pi}$ (the calculation is
analogous to the one described in Ref.\ \cite{DA}):
\begin{align}
\mathcal{M}_{\pi\pi}(s,t,u)  &  \sim-i\,\delta^{ab}\delta^{cd}[-2m_{\pi}%
^{2}C_{\sigma_{N}\pi\pi}\cos\varphi_{\sigma}+B_{\sigma_{N}\pi\pi}\cos
\varphi_{\sigma}(2m_{\pi}^{2}-s)+2(A_{\sigma_{N}\pi\pi}\cos\varphi_{\sigma
}\nonumber\\
&  +A_{\sigma_{S}\pi\pi}\sin\varphi_{\sigma})]^{2}\frac{1}{s-m_{\sigma_{1}%
}^{2}}\nonumber\\
&  -\,i\,\delta^{ac}\delta^{bd}\,[-2m_{\pi}^{2}C_{\sigma_{N}\pi\pi}\cos
\varphi_{\sigma}+B_{\sigma_{N}\pi\pi}\cos\varphi_{\sigma}(2m_{\pi}%
^{2}-t)+2(A_{\sigma_{N}\pi\pi}\cos\varphi_{\sigma}\nonumber\\
&  +A_{\sigma_{S}\pi\pi}\sin\varphi_{\sigma})]^{2}\frac{1}{t-m_{\sigma_{1}%
}^{2}}\nonumber\\
&  -i\,\delta^{ad}\delta^{bc}[-2m_{\pi}^{2}C_{\sigma_{N}\pi\pi}\cos
\varphi_{\sigma}+B_{\sigma_{N}\pi\pi}\cos\varphi_{\sigma}(2m_{\pi}%
^{2}-u)+2(A_{\sigma_{N}\pi\pi}\cos\varphi_{\sigma}\nonumber\\
&  +A_{\sigma_{S}\pi\pi}\sin\varphi_{\sigma})]^{2}\frac{1}{u-m_{\sigma_{1}%
}^{2}}\nonumber\\
&  -i\,\delta^{ab}\delta^{cd}[2m_{\pi}^{2}C_{\sigma_{N}\pi\pi}\sin
\varphi_{\sigma}-B_{\sigma_{N}\pi\pi}\sin\varphi_{\sigma}(2m_{\pi}%
^{2}-s)+2(A_{\sigma_{S}\pi\pi}\cos\varphi_{\sigma}\nonumber\\
&  -A_{\sigma_{N}\pi\pi}\sin\varphi_{\sigma})]^{2}\frac{1}{s-m_{\sigma_{2}%
}^{2}}\nonumber\\
&  -\,i\,\delta^{ac}\delta^{bd}\,[2m_{\pi}^{2}C_{\sigma_{N}\pi\pi}\sin
\varphi_{\sigma}-B_{\sigma_{N}\pi\pi}\sin\varphi_{\sigma}(2m_{\pi}%
^{2}-t)+2(A_{\sigma_{S}\pi\pi}\cos\varphi_{\sigma}\nonumber\\
&  -A_{\sigma_{N}\pi\pi}\sin\varphi_{\sigma})]^{2}\frac{1}{t-m_{\sigma_{2}%
}^{2}}\nonumber\\
&  -i\,\delta^{ad}\delta^{bc}[2m_{\pi}^{2}C_{\sigma_{N}\pi\pi}\sin
\varphi_{\sigma}-B_{\sigma_{N}\pi\pi}\sin\varphi_{\sigma}(2m_{\pi}%
^{2}-u)+2(A_{\sigma_{S}\pi\pi}\cos\varphi_{\sigma}\nonumber\\
&  -A_{\sigma_{N}\pi\pi}\sin\varphi_{\sigma})]^{2}\frac{1}{u-m_{\sigma_{2}%
}^{2}}\text{.} \label{M1}%
\end{align}

Let us now rewrite the $\pi\pi$ scattering amplitude of Eqs.\ (\ref{MSLQ}) in
the following way [we substitute $A_{\rho\pi\pi}$ and $B_{\rho\pi\pi}$ present
in Eqs.\ (\ref{ASL1}) - (\ref{ASL3}) by terms in Eqs.\ (\ref{Arhopipi}) and
(\ref{Brhopipi})]:

\begin{align}
\mathcal{M}_{\pi\pi}(s,t,u) &  =i\delta^{ab}\delta^{cd}\left\{  (g_{1}%
^{2}-h_{3})Z_{\pi}^{4}w_{a_{1}}^{2}s-2\left(  \lambda_{1}+\frac{\lambda_{2}%
}{2}\right)  Z_{\pi}^{4}-(h_{1}+h_{2}+h_{3})Z_{\pi}^{4}w_{a_{1}}^{2}%
(s-2m_{\pi}^{2})\right.  \nonumber\\
&  \left.  -[-2m_{\pi}^{2}C_{\sigma_{N}\pi\pi}\cos\varphi_{\sigma}%
+B_{\sigma_{N}\pi\pi}\cos\varphi_{\sigma}(2m_{\pi}^{2}-s)+2(A_{\sigma_{N}%
\pi\pi}\cos\varphi_{\sigma}\right.  \nonumber\\
&  \left.  +A_{\sigma_{S}\pi\pi}\sin\varphi_{\sigma})]^{2}\frac{1}%
{s-m_{\sigma_{1}}^{2}}\right.  \nonumber\\
&  -[2m_{\pi}^{2}C_{\sigma_{N}\pi\pi}\sin\varphi_{\sigma}-B_{\sigma_{N}\pi\pi
}\sin\varphi_{\sigma}(2m_{\pi}^{2}-s)+2(A_{\sigma_{S}\pi\pi}\cos
\varphi_{\sigma}\nonumber\\
&  \left.  -A_{\sigma_{N}\pi\pi}\sin\varphi_{\sigma})]^{2}\frac{1}%
{s-m_{\sigma_{2}}^{2}}\right.  \nonumber\\
&  \left.  +Z_{\pi}^{4}\left[  g_{1}(1-g_{1}w_{a_{1}}\phi_{N})+h_{3}w_{a_{1}%
}\phi_{N}-g_{2}w_{a_{1}}^{2}\frac{t}{2}\right]  ^{2}\,\frac{u-s}{t-m_{\rho
}^{2}}\right.  \nonumber\\
&  \left.  +Z_{\pi}^{4}\left[  g_{1}(1-g_{1}w_{a_{1}}\phi_{N})+h_{3}w_{a_{1}%
}\phi_{N}-g_{2}w_{a_{1}}^{2}\frac{u}{2}\right]  ^{2}\,\frac{t-s}{u-m_{\rho
}^{2}}\right\}  \nonumber \\
&  +i\delta^{ac}\delta^{bd}\left\{  (g_{1}^{2}-h_{3})Z_{\pi}^{4}w_{a_{1}}%
^{2}t-2\left(  \lambda_{1}+\frac{\lambda_{2}}{2}\right)  Z_{\pi}^{4}%
-(h_{1}+h_{2}+h_{3})Z_{\pi}^{4}w_{a_{1}}^{2}(t-2m_{\pi}^{2})\right.
\nonumber
\end{align}
\begin{align}
&  \left.  -[-2m_{\pi}^{2}C_{\sigma_{N}\pi\pi}\cos\varphi_{\sigma}%
+B_{\sigma_{N}\pi\pi}\cos\varphi_{\sigma}(2m_{\pi}^{2}-t)+2(A_{\sigma_{N}%
\pi\pi}\cos\varphi_{\sigma}\right.  \nonumber\\
&  \left.  +A_{\sigma_{S}\pi\pi}\sin\varphi_{\sigma})]^{2}\frac{1}%
{t-m_{\sigma_{1}}^{2}}\right.  \nonumber \\
&  -[2m_{\pi}^{2}C_{\sigma_{N}\pi\pi}\sin\varphi_{\sigma}-B_{\sigma_{N}\pi\pi
}\sin\varphi_{\sigma}(2m_{\pi}^{2}-t)+2(A_{\sigma_{S}\pi\pi}\cos
\varphi_{\sigma}\nonumber \\
&  \left.  -A_{\sigma_{N}\pi\pi}\sin\varphi_{\sigma})]^{2}\frac{1}%
{t-m_{\sigma_{2}}^{2}}\right.  \nonumber\\
&  \left.  +Z_{\pi}^{4}\left[  g_{1}(1-g_{1}w_{a_{1}}\phi_{N})+h_{3}w_{a_{1}%
}\phi_{N}-g_{2}w_{a_{1}}^{2}\frac{s}{2}\right]  ^{2}\,\frac{u-t}{s-m_{\rho
}^{2}}\right.  \nonumber\\
&  \left.  +Z_{\pi}^{4}\left[  g_{1}(1-g_{1}w_{a_{1}}\phi_{N})+h_{3}w_{a_{1}%
}\phi_{N}-g_{2}w_{a_{1}}^{2}\frac{u}{2}\right]  ^{2}\,\frac{s-t}{u-m_{\rho
}^{2}}\right\}  \nonumber\\
&  +\,i\,\delta^{ad}\delta^{bc}\,\left\{  (g_{1}^{2}-h_{3})Z_{\pi}^{4}%
w_{a_{1}}^{2}u-2\left(  \lambda_{1}+\frac{\lambda_{2}}{2}\right)  Z_{\pi}%
^{4}-(h_{1}+h_{2}+h_{3})Z_{\pi}^{4}w_{a_{1}}^{2}(u-2m_{\pi}^{2})\right.
\nonumber\\
&  \left.  -[-2m_{\pi}^{2}C_{\sigma_{N}\pi\pi}\cos\varphi_{\sigma}%
+B_{\sigma_{N}\pi\pi}\cos\varphi_{\sigma}(2m_{\pi}^{2}-u)+2(A_{\sigma_{N}%
\pi\pi}\cos\varphi_{\sigma}\right.  \nonumber\\
&  \left.  +A_{\sigma_{S}\pi\pi}\sin\varphi_{\sigma})]^{2}\frac{1}%
{u-m_{\sigma_{1}}^{2}}\right.  \nonumber\\
&  -[2m_{\pi}^{2}C_{\sigma_{N}\pi\pi}\sin\varphi_{\sigma}-B_{\sigma_{N}\pi\pi
}\sin\varphi_{\sigma}(2m_{\pi}^{2}-u)+2(A_{\sigma_{S}\pi\pi}\cos
\varphi_{\sigma}\nonumber\\
&  \left.  -A_{\sigma_{N}\pi\pi}\sin\varphi_{\sigma})]^{2}\frac{1}%
{u-m_{\sigma_{2}}^{2}}\right.  \nonumber\\
&  \left.  +Z_{\pi}^{4}\left[  g_{1}(1-g_{1}w_{a_{1}}\phi_{N})+h_{3}w_{a_{1}%
}\phi_{N}-g_{2}w_{a_{1}}^{2}\frac{s}{2}\right]  ^{2}\,\frac{t-u}{s-m_{\rho
}^{2}}\right.  \nonumber\\
&  \left.  +Z_{\pi}^{4}\left[  g_{1}(1-g_{1}w_{a_{1}}\phi_{N})+h_{3}w_{a_{1}%
}\phi_{N}-g_{2}w_{a_{1}}^{2}\frac{t}{2}\right]  ^{2}\,\frac{s-u}{t-m_{\rho
}^{2}}\right\}  \nonumber\\
&  \equiv i\delta^{ab}\delta^{cd}A(s,t,u)+i\delta^{ac}\delta^{bd}%
A(t,u,s)+\,i\,\delta^{ad}\delta^{bc}A(u,s,t)\text{.}\label{M2}%
\end{align}

We can now consider the three components of the scattering amplitude at threshold.

\begin{align}
A(s,t,u)|_{s=4m_{\pi}^{2}}  &  =4(g_{1}^{2}-h_{3})Z_{\pi}^{4}w_{a_{1}}%
^{2}m_{\pi}^{2}-2\left(  \lambda_{1}+\frac{\lambda_{2}}{2}\right)  Z_{\pi}%
^{4}-2(h_{1}+h_{2}+h_{3})Z_{\pi}^{4}w_{a_{1}}^{2}m_{\pi}^{2}\nonumber\\
&  -4[(B_{\sigma_{N}\pi\pi}+C_{\sigma_{N}\pi\pi})m_{\pi}^{2}\cos
\varphi_{\sigma}-(A_{\sigma_{N}\pi\pi}\cos\varphi_{\sigma}+A_{\sigma_{S}\pi
\pi}\sin\varphi_{\sigma})]^{2}\frac{1}{4m_{\pi}^{2}-m_{\sigma_{1}}^{2}%
}\nonumber\\
&  -4[(B_{\sigma_{N}\pi\pi}+C_{\sigma_{N}\pi\pi})m_{\pi}^{2}\sin
\varphi_{\sigma}+(A_{\sigma_{S}\pi\pi}\cos\varphi_{\sigma}-A_{\sigma_{N}\pi
\pi}\sin\varphi_{\sigma})]^{2}\frac{1}{4m_{\pi}^{2}-m_{\sigma_{2}}^{2}%
}\nonumber\\
&  +8[g_{1}Z_{\pi}^{2}(1-g_{1}w_{a_{1}}\phi_{N})+h_{3}Z_{\pi}^{2}w_{a_{1}}%
\phi_{N}]^{2}\frac{m_{\pi}^{2}}{m_{\rho}^{2}}\text{.} \label{A11}%
\end{align}

Using Eq.\ (\ref{wa1}) we can transform the last line of Eq.\ (\ref{A11}) in
the following way:%
\begin{align}
1-g_{1}w_{a_{1}}\phi_{N}  &  =1-\frac{g_{1}^{2}\phi_{N}^{2}}{m_{a_{1}}^{2}%
}=\frac{m_{a_{1}}^{2}-g_{1}^{2}\phi_{N}^{2}}{m_{a_{1}}^{2}}\overset
{\text{Eq.\ (\ref{Z_pi})}}{=}\frac{1}{Z_{\pi}^{2}}\nonumber\\
&  \Rightarrow g_{1}Z_{\pi}^{2}(1-g_{1}w_{a_{1}}\phi_{N})+h_{3}Z_{\pi}%
^{2}w_{a_{1}}\phi_{N}=g_{1}+h_{3}Z_{\pi}^{2}\frac{g_{1}\phi_{N}^{2}}{m_{a_{1}%
}^{2}}\nonumber\\
&  \overset{\text{Eq.\ (\ref{h3})}}{=}g_{1}\left[  1+\frac{1}{m_{a_{1}}^{2}%
}\left(  m_{\rho}^{2}-\frac{m_{a_{1}}^{2}}{Z_{\pi}^{2}}\right)  Z_{\pi}%
^{2}\right]  =g_{1}Z_{\pi}^{2}\frac{m_{\rho}^{2}}{m_{a_{1}}^{2}}%
\end{align}

and thus we obtain%
\begin{align}
A(s,t,u)|_{s=4m_{\pi}^{2}}  &  =4(g_{1}^{2}-h_{3})Z_{\pi}^{4}w_{a_{1}}%
^{2}m_{\pi}^{2}-2\left(  \lambda_{1}+\frac{\lambda_{2}}{2}\right)  Z_{\pi}%
^{4}-2(h_{1}+h_{2}+h_{3})Z_{\pi}^{4}w_{a_{1}}^{2}m_{\pi}^{2}\nonumber\\
&  -4[(B_{\sigma_{N}\pi\pi}+C_{\sigma_{N}\pi\pi})m_{\pi}^{2}\cos
\varphi_{\sigma}-(A_{\sigma_{N}\pi\pi}\cos\varphi_{\sigma}+A_{\sigma_{S}\pi
\pi}\sin\varphi_{\sigma})]^{2}\frac{1}{4m_{\pi}^{2}-m_{\sigma_{1}}^{2}%
}\nonumber\\
&  -4[(B_{\sigma_{N}\pi\pi}+C_{\sigma_{N}\pi\pi})m_{\pi}^{2}\sin
\varphi_{\sigma}+(A_{\sigma_{S}\pi\pi}\cos\varphi_{\sigma}-A_{\sigma_{N}\pi
\pi}\sin\varphi_{\sigma})]^{2}\frac{1}{4m_{\pi}^{2}-m_{\sigma_{2}}^{2}%
} \nonumber \\
&  +8g_{1}^{2}Z_{\pi}^{4}\frac{m_{\pi}^{2}m_{\rho}^{2}}{m_{a_{1}}^{4}}\text{.} \label{A12}
\end{align}

We also obtain%
\begin{align}
A(t,u,s)|_{s=4m_{\pi}^{2}}  &  =-2\left(  \lambda_{1}+\frac{\lambda_{2}}%
{2}\right)  Z_{\pi}^{4}+2(h_{1}+h_{2}+h_{3})Z_{\pi}^{4}w_{a_{1}}^{2}m_{\pi
}^{2}-4g_{1}^{2}Z_{\pi}^{4}\frac{m_{\pi}^{2}m_{\rho}^{2}}{m_{a_{1}}^{4}%
}\nonumber\\
&  +4[(B_{\sigma_{N}\pi\pi}-C_{\sigma_{N}\pi\pi})m_{\pi}^{2}\cos
\varphi_{\sigma}+(A_{\sigma_{N}\pi\pi}\cos\varphi_{\sigma}+A_{\sigma_{S}\pi
\pi}\sin\varphi_{\sigma})]^{2}\frac{1}{m_{\sigma_{1}}^{2}}\nonumber\\
&  +4[(C_{\sigma_{N}\pi\pi}-B_{\sigma_{N}\pi\pi})m_{\pi}^{2}\sin
\varphi_{\sigma}+(A_{\sigma_{S}\pi\pi}\cos\varphi_{\sigma}-A_{\sigma_{N}\pi
\pi}\sin\varphi_{\sigma})]^{2}\frac{1}{m_{\sigma_{2}}^{2}} \label{A21}%
\end{align}

and%
\begin{equation}
A(u,s,t)|_{s=4m_{\pi}^{2}}=A(t,u,s)|_{s=4m_{\pi}^{2}}\text{.} \label{A31}%
\end{equation}
\\
We can now calculate the scattering lengths. We already know from
Sec.\ \ref{sec.SLQ} that

\begin{equation}
T^{0}|_{s=4m_{\pi}^{2}}\equiv32\pi a_{0}^{0}|_{s=4m_{\pi}^{2}}
=3A(s,t,u)|_{s=4m_{\pi}^{2}}+A(t,u,s)|_{s=4m_{\pi}^{2}}+A(u,s,t)|_{s=4m_{\pi
}^{2}}\text{.}
\end{equation}

We then obtain%
\begin{align}
32\pi a_{0}^{0}|_{s=4m_{\pi}^{2}}  &  =12(g_{1}^{2}-h_{3})Z_{\pi}^{4}w_{a_{1}%
}^{2}m_{\pi}^{2}-10\left(  \lambda_{1}+\frac{\lambda_{2}}{2}\right)  Z_{\pi
}^{4}-2(h_{1}+h_{2}+h_{3})Z_{\pi}^{4}w_{a_{1}}^{2}m_{\pi}^{2}\nonumber\\
&  +12[(B_{\sigma_{N}\pi\pi}+C_{\sigma_{N}\pi\pi})m_{\pi}^{2}\cos
\varphi_{\sigma}-(A_{\sigma_{N}\pi\pi}\cos\varphi_{\sigma}+A_{\sigma_{S}\pi
\pi}\sin\varphi_{\sigma})]^{2}\frac{1}{m_{\sigma_{1}}^{2}-4m_{\pi}^{2}%
}\nonumber\\
&  +12[(B_{\sigma_{N}\pi\pi}+C_{\sigma_{N}\pi\pi})m_{\pi}^{2}\sin
\varphi_{\sigma}+(A_{\sigma_{S}\pi\pi}\cos\varphi_{\sigma}-A_{\sigma_{N}\pi
\pi}\sin\varphi_{\sigma})]^{2}\frac{1}{m_{\sigma_{2}}^{2}-4m_{\pi}^{2}%
}\nonumber\\
&  +8[(B_{\sigma_{N}\pi\pi}-C_{\sigma_{N}\pi\pi})m_{\pi}^{2}\cos
\varphi_{\sigma}+(A_{\sigma_{N}\pi\pi}\cos\varphi_{\sigma}+A_{\sigma_{S}\pi
\pi}\sin\varphi_{\sigma})]^{2}\frac{1}{m_{\sigma_{1}}^{2}}\nonumber\\
&  +8[(C_{\sigma_{N}\pi\pi}-B_{\sigma_{N}\pi\pi})m_{\pi}^{2}\sin
\varphi_{\sigma}+(A_{\sigma_{S}\pi\pi}\cos\varphi_{\sigma}-A_{\sigma_{N}\pi
\pi}\sin\varphi_{\sigma})]^{2}\frac{1}{m_{\sigma_{2}}^{2}}\nonumber\\
&  +16g_{1}^{2}Z_{\pi}^{4}\frac{m_{\pi}^{2}m_{\rho}^{2}}{m_{a_{1}}^{4}%
}\text{.} \label{a001}%
\end{align}

Similarly to Eq.\ (\ref{BNspp1}), let us note that the linear combination
$B_{\sigma_{N}\pi\pi}+C_{\sigma_{N}\pi\pi}$ can be transformed in the
following way:%
\begin{align}
&  B_{\sigma_{N}\pi\pi}+C_{\sigma_{N}\pi\pi}\overset{\text{Eq.\ (\ref{wa1})}%
}{=}Z_{\pi}^{2}\frac{g_{1}^{2}\phi_{N}}{m_{a_{1}}^{2}}\left(  -3+\frac
{g_{1}^{2}\phi_{N}^{2}}{m_{a_{1}}^{2}}+\frac{h_{1}+h_{2}-h_{3}}{2}\frac
{\phi_{N}^{2}}{m_{a_{1}}^{2}}\right) \nonumber\\
&  =Z_{\pi}^{2}\frac{g_{1}^{2}\phi_{N}}{m_{a_{1}}^{4}}\left(  -3m_{a_{1}}%
^{2}+g_{1}^{2}\phi_{N}^{2}+\frac{h_{1}+h_{2}-h_{3}}{2}\phi_{N}^{2}\right)
\nonumber\\
&  \overset{\text{Eq.\ (\ref{m_a_1})}}{=}Z_{\pi}^{2}\frac{g_{1}^{2}\phi_{N}%
}{m_{a_{1}}^{4}}\left(  -3m_{a_{1}}^{2}+m_{a_{1}}^{2}-m_{1}^{2}-\frac{h_{1}%
}{2}\phi_{S}^{2}-2\delta_{N}\right)  \equiv-Z_{\pi}^{2}\frac{g_{1}^{2}\phi
_{N}}{m_{a_{1}}^{4}}\left(  2m_{a_{1}}^{2}+m_{1}^{2}\right)  \text{,}
\label{BN+CNspp}%
\end{align}

where we have used $h_{1}=0=\delta_{N}$, and also that the linear
combination $B_{\sigma_{N}\pi\pi}-C_{\sigma_{N}\pi\pi}$ can be written in this way:%
\begin{align}
&  B_{\sigma_{N}\pi\pi}-C_{\sigma_{N}\pi\pi}\overset{\text{Eq.\ (\ref{wa1})}%
}{=}Z_{\pi}^{2}\frac{g_{1}^{2}\phi_{N}}{m_{a_{1}}^{2}}\left(  -1+\frac
{g_{1}^{2}\phi_{N}^{2}}{m_{a_{1}}^{2}}+\frac{h_{1}+h_{2}-h_{3}}{2}\frac
{\phi_{N}^{2}}{m_{a_{1}}^{2}}\right) \nonumber\\
&  =Z_{\pi}^{2}\frac{g_{1}^{2}\phi_{N}}{m_{a_{1}}^{4}}\left(  -m_{a_{1}}%
^{2}+g_{1}^{2}\phi_{N}^{2}+\frac{h_{1}+h_{2}-h_{3}}{2}\phi_{N}^{2}\right)
\nonumber\\
&  \overset{\text{Eq.\ (\ref{m_a_1})}}{=}Z_{\pi}^{2}\frac{g_{1}^{2}\phi_{N}%
}{m_{a_{1}}^{4}}\left(  -m_{1}^{2}-\frac{h_{1}}{2}\phi_{S}^{2}-2\delta
_{N}\right)  \equiv-Z_{\pi}^{2}\frac{g_{1}^{2}\phi_{N}}{m_{a_{1}}^{4}}%
m_{1}^{2}\text{.} \label{BN-CNspp}%
\end{align}

Then, using Eqs.\ (\ref{ANspp}), (\ref{ASspp}), (\ref{BN+CNspp}) and
(\ref{BN-CNspp}), we obtain from Eq.\ (\ref{a001}):%

\begin{align}
32\pi a_{0}^{0}|_{s=4m_{\pi}^{2}}  &  =[12g_{1}^{2}-2(h_{1}+h_{2}%
)-14h_{3}]Z_{\pi}^{4}w_{a_{1}}^{2}m_{\pi}^{2}-10\left(  \lambda_{1}%
+\frac{\lambda_{2}}{2}\right)  Z_{\pi}^{4}+16g_{1}^{2}Z_{\pi}^{4}\frac{m_{\pi
}^{2}m_{\rho}^{2}}{m_{a_{1}}^{4}}\nonumber\\
&  +12Z_{\pi}^{4}\left[  \frac{g_{1}^{2}\phi_{N}}{m_{a_{1}}^{4}}\left(
2m_{a_{1}}^{2}+m_{1}^{2}\right)  m_{\pi}^{2}\cos\varphi_{\sigma}-\lambda
_{1}(\phi_{N}\cos\varphi_{\sigma}+\phi_{S}\sin\varphi_{\sigma})\right.
\nonumber\\
&  \left.  -\frac{\lambda_{2}}{2}\phi_{N}\cos\varphi_{\sigma}\right]
^{2}\frac{1}{m_{\sigma_{1}}^{2}-4m_{\pi}^{2}}\nonumber\\
&  +12Z_{\pi}^{4}\left[  \frac{g_{1}^{2}\phi_{N}}{m_{a_{1}}^{4}}\left(
2m_{a_{1}}^{2}+m_{1}^{2}\right)  m_{\pi}^{2}\sin\varphi_{\sigma}-\lambda
_{1}(\phi_{N}\sin\varphi_{\sigma}-\phi_{S}\cos\varphi_{\sigma})\right.
\nonumber\\
&  \left.  -\frac{\lambda_{2}}{2}\phi_{N}\sin\varphi_{\sigma}\right]
^{2}\frac{1}{m_{\sigma_{2}}^{2}-4m_{\pi}^{2}}\nonumber\\
&  +8Z_{\pi}^{4}\left[  \frac{g_{1}^{2}\phi_{N}}{m_{a_{1}}^{4}}m_{1}^{2}%
m_{\pi}^{2}\cos\varphi_{\sigma}+\lambda_{1}(\phi_{N}\cos\varphi_{\sigma}%
+\phi_{S}\sin\varphi_{\sigma})+\frac{\lambda_{2}}{2}\phi_{N}\cos
\varphi_{\sigma}\right]  ^{2}\frac{1}{m_{\sigma_{1}}^{2}}\nonumber\\
&  +8Z_{\pi}^{4}\left[  \frac{g_{1}^{2}\phi_{N}}{m_{a_{1}}^{4}}m_{1}^{2}%
m_{\pi}^{2}\sin\varphi_{\sigma}+\lambda_{1}(\phi_{N}\sin\varphi_{\sigma}%
-\phi_{S}\cos\varphi_{\sigma})+\frac{\lambda_{2}}{2}\phi_{N}\sin
\varphi_{\sigma}\right]  ^{2}\frac{1}{m_{\sigma_{2}}^{2}} \label{a002}%
\end{align}

and finally the following formula for the $S$-wave, isospin-zero $\pi\pi$
scattering length $a_{0}^{0}$:

\begin{align}
a_{0}^{0}|_{s=4m_{\pi}^{2}}  &  =\frac{Z_{\pi}^{4}}{\pi}\left\{  \left[
\frac{3}{8}g_{1}^{2}-\frac{1}{16}(h_{1}+h_{2})-\frac{7}{16}h_{3}\right]
w_{a_{1}}^{2}m_{\pi}^{2}-\frac{5}{16}\left(  \lambda_{1}+\frac{\lambda_{2}}%
{2}\right)  +\frac{1}{2}g_{1}^{2}Z_{\pi}^{4}\frac{m_{\pi}^{2}m_{\rho}^{2}%
}{m_{a_{1}}^{4}}\right. \nonumber\\
&  \left.  +\frac{3}{8}\left[  \frac{g_{1}^{2}\phi_{N}}{m_{a_{1}}^{4}}\left(
2m_{a_{1}}^{2}+m_{1}^{2}\right)  m_{\pi}^{2}\cos\varphi_{\sigma}-\lambda
_{1}(\phi_{N}\cos\varphi_{\sigma}+\phi_{S}\sin\varphi_{\sigma})\right.
\right. \nonumber\\
&  \left.  \left.  -\frac{\lambda_{2}}{2}\phi_{N}\cos\varphi_{\sigma}\right]
^{2}\frac{1}{m_{\sigma_{1}}^{2}-4m_{\pi}^{2}}\right. \nonumber
\end{align}
\begin{align}
&  \left.  +\frac{3}{8}\left[  \frac{g_{1}^{2}\phi_{N}}{m_{a_{1}}^{4}}\left(
2m_{a_{1}}^{2}+m_{1}^{2}\right)  m_{\pi}^{2}\sin\varphi_{\sigma}-\lambda
_{1}(\phi_{N}\sin\varphi_{\sigma}-\phi_{S}\cos\varphi_{\sigma})\right.
\right. \nonumber\\
&  \left.  \left.  -\frac{\lambda_{2}}{2}\phi_{N}\sin\varphi_{\sigma}\right]
^{2}\frac{1}{m_{\sigma_{2}}^{2}-4m_{\pi}^{2}}\right. \nonumber\\
&  \left.  +\frac{1}{4}\left[  \frac{g_{1}^{2}\phi_{N}}{m_{a_{1}}^{4}}%
m_{1}^{2}m_{\pi}^{2}\cos\varphi_{\sigma}+\lambda_{1}(\phi_{N}\cos
\varphi_{\sigma}+\phi_{S}\sin\varphi_{\sigma})+\frac{\lambda_{2}}{2}\phi
_{N}\cos\varphi_{\sigma}\right]  ^{2}\frac{1}{m_{\sigma_{1}}^{2}}\right.
\nonumber\\
&  \left.  +\frac{1}{4}\left[  \frac{g_{1}^{2}\phi_{N}}{m_{a_{1}}^{4}}%
m_{1}^{2}m_{\pi}^{2}\sin\varphi_{\sigma}+\lambda_{1}(\phi_{N}\sin
\varphi_{\sigma}-\phi_{S}\cos\varphi_{\sigma})+\frac{\lambda_{2}}{2}\phi
_{N}\sin\varphi_{\sigma}\right]  ^{2}\frac{1}{m_{\sigma_{2}}^{2}}\right\}
\text{.} \label{a003}%
\end{align}
$\,$\\
Given that $T^{1}=A(t,u,s)-A(u,s,t)$, we obtain $T^{1}=0$ at threshold because
of $A(t,u,s)|_{s=4m_{\pi}^{2}}=A(u,s,t)|_{s=4m_{\pi}^{2}}$ [see
Eq.\ (\ref{A31})]. Therefore,

\begin{equation}
a_{0}^{1}=0\text{.} \label{a011}%
\end{equation}

We now turn to the calculation of the $S$-wave, isospin-two $\pi\pi$ scattering
length $a_{0}^{2}$. As already known from Sec.\ \ref{sec.SLQ},

\begin{equation}
T^{2}|_{s=4m_{\pi}^{2}}=A(t,u,s)|_{s=4m_{\pi}^{2}}+A(u,s,t)|_{s=4m_{\pi}^{2}}%
\end{equation}

or in other words%

\begin{equation}
T^{2}|_{s=4m_{\pi}^{2}}=2A(t,u,s)|_{s=4m_{\pi}^{2}}%
\end{equation}

because of Eq.\ (\ref{A31}). Given that%
\begin{equation}
32\pi a_{0}^{2}\equiv T^{2}|_{s=4m_{\pi}^{2}}\text{,}
\end{equation}

we consequently obtain%

\begin{equation}
16\pi a_{0}^{2}\equiv A(t,u,s)|_{s=4m_{\pi}^{2}}\text{.}%
\end{equation}

Then substituting Eqs.\ (\ref{ANspp}), (\ref{ASspp}) and (\ref{BN-CNspp}) into
Eq.\ (\ref{A21}) implies%

\begin{align}
16\pi a_{0}^{2}  &  =-2\left(  \lambda_{1}+\frac{\lambda_{2}}{2}\right)
Z_{\pi}^{4}+2(h_{1}+h_{2}+h_{3})Z_{\pi}^{4}w_{a_{1}}^{2}m_{\pi}^{2}-4g_{1}%
^{2}Z_{\pi}^{4}\frac{m_{\pi}^{2}m_{\rho}^{2}}{m_{a_{1}}^{4}}\nonumber\\
&  +4Z_{\pi}^{4}\left[  \frac{g_{1}^{2}\phi_{N}}{m_{a_{1}}^{4}}m_{1}^{2}%
m_{\pi}^{2}\cos\varphi_{\sigma}+\lambda_{1}(\phi_{N}\cos\varphi_{\sigma}%
+\phi_{S}\sin\varphi_{\sigma})+\frac{\lambda_{2}}{2}\phi_{N}\cos
\varphi_{\sigma}\right]  ^{2}\frac{1}{m_{\sigma_{1}}^{2}}\nonumber\\
&  +4Z_{\pi}^{4}\left[  \frac{g_{1}^{2}\phi_{N}}{m_{a_{1}}^{4}}m_{1}^{2}%
m_{\pi}^{2}\sin\varphi_{\sigma}+\lambda_{1}(\phi_{N}\sin\varphi_{\sigma}%
-\phi_{S}\cos\varphi_{\sigma})+\frac{\lambda_{2}}{2}\phi_{N}\sin
\varphi_{\sigma}\right]  ^{2}\frac{1}{m_{\sigma_{2}}^{2}}\text{.} \label{a021}%
\end{align}

Finally, we obtain%
\begin{align}
a_{0}^{2}  &  =\frac{Z_{\pi}^{4}}{\pi}\left\{  \frac{1}{8}(h_{1}+h_{2}%
+h_{3})w_{a_{1}}^{2}m_{\pi}^{2}-\frac{1}{8}\left(  \lambda_{1}+\frac
{\lambda_{2}}{2}\right)  -\frac{1}{4}g_{1}^{2}\frac{m_{\pi}^{2}m_{\rho}^{2}%
}{m_{a_{1}}^{4}}\right. \nonumber\\
&  \left.  +\frac{1}{4}\left[  \frac{g_{1}^{2}\phi_{N}}{m_{a_{1}}^{4}}%
m_{1}^{2}m_{\pi}^{2}\cos\varphi_{\sigma}+\lambda_{1}(\phi_{N}\cos
\varphi_{\sigma}+\phi_{S}\sin\varphi_{\sigma})+\frac{\lambda_{2}}{2}\phi
_{N}\cos\varphi_{\sigma}\right]  ^{2}\frac{1}{m_{\sigma_{1}}^{2}}\right.
\nonumber\\
&  \left.  +\frac{1}{4}\left[  \frac{g_{1}^{2}\phi_{N}}{m_{a_{1}}^{4}}%
m_{1}^{2}m_{\pi}^{2}\sin\varphi_{\sigma}+\lambda_{1}(\phi_{N}\sin
\varphi_{\sigma}-\phi_{S}\cos\varphi_{\sigma})+\frac{\lambda_{2}}{2}\phi
_{N}\sin\varphi_{\sigma}\right]  ^{2}\frac{1}{m_{\sigma_{2}}^{2}}\right\}
\text{.} \label{a022}%
\end{align}

\begin{figure}[h]
  \begin{center}
    \begin{tabular}{cc}
      \resizebox{78mm}{!}{\includegraphics{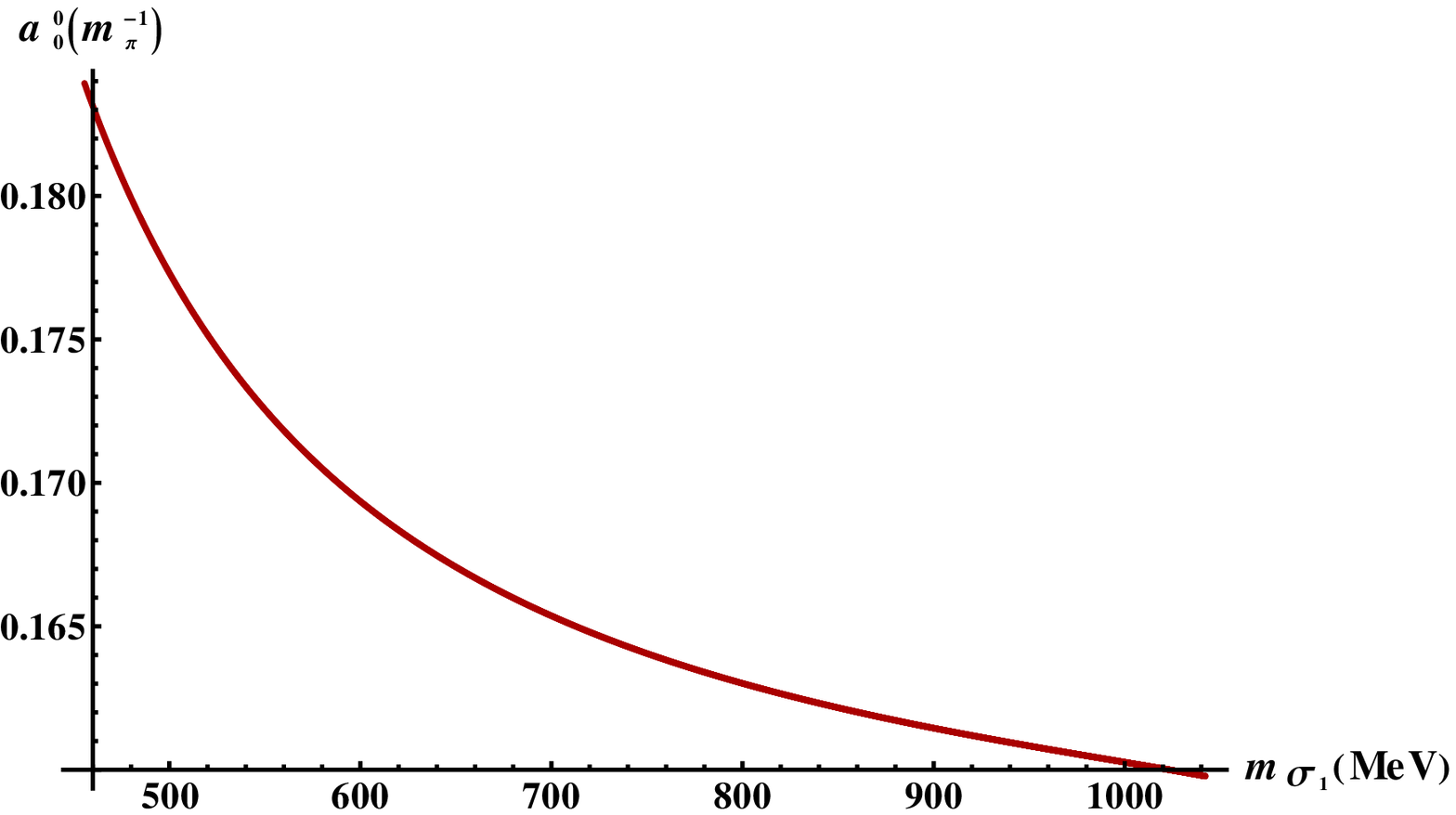}} &
      \resizebox{78mm}{!}{\includegraphics{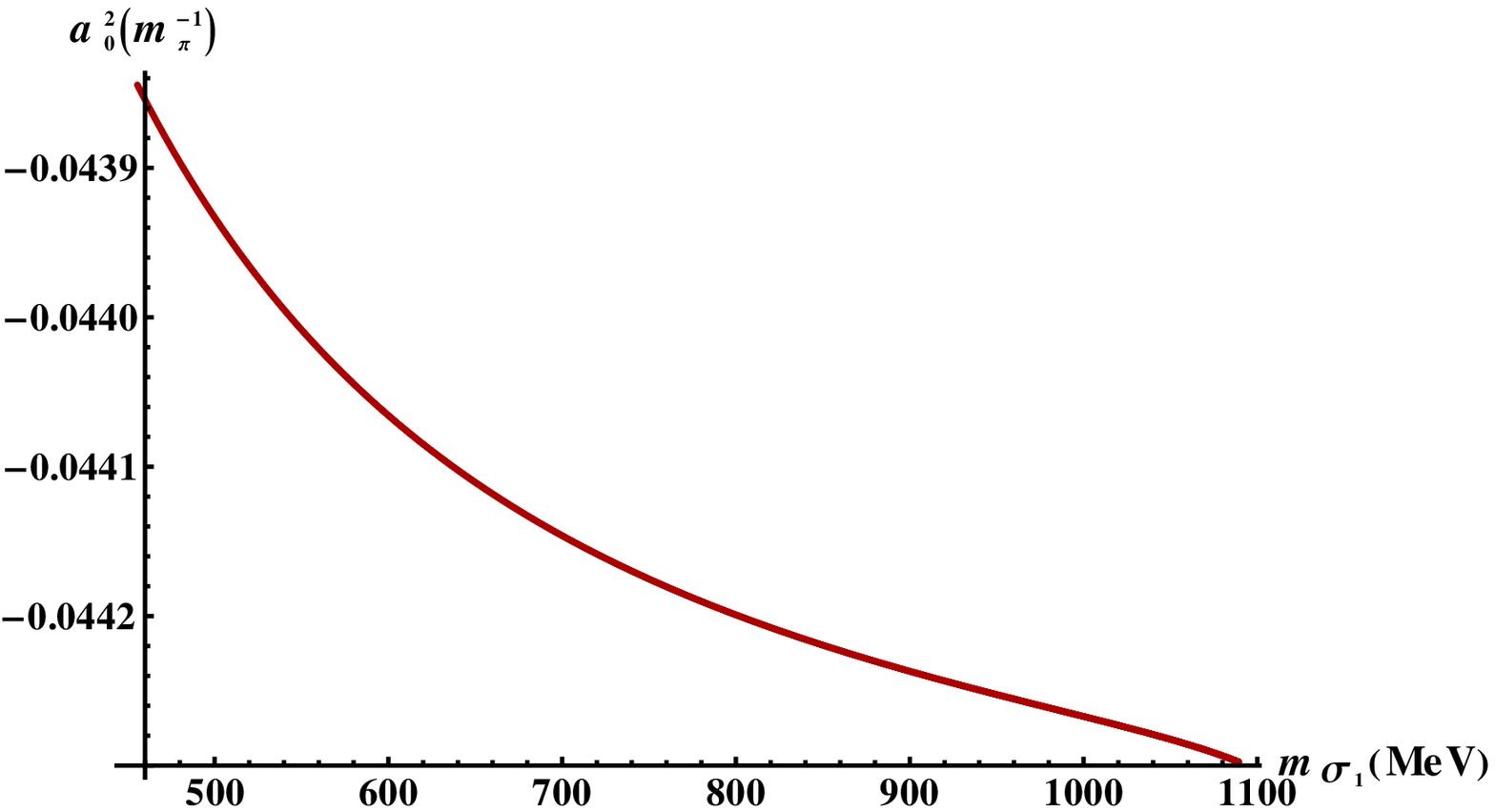}} 
    \end{tabular}
    \caption{Pion-pion scattering lengths from Fit I. We do not indicate the NA48/2 error
bands \cite{Peyaud} because $a_{0}^{0}$ (left panel) is completely outside the
NA48/2 interval $a_{0}^{0}=0.218\pm0.020$ and $a_{0}^{2}$ (left panel) is
completely within the NA48/2 interval $a_{0}^{2}=-0.0457\pm0.0125$. The latter is true because the NA48/2
result possesses large errors and our $a_{0}^{2}$ barely changes with
$m_{\sigma_{1}}$.}
    \label{SL1}
  \end{center}
\end{figure}

We observe from of Eqs.\ (\ref{a003}) and (\ref{a022}) that the $\pi\pi$
scattering lengths now depend on two scalar masses ($m_{\sigma_{1}}$ and
$m_{\sigma_{2}}$) unlike in the two-flavour version of the model where the
dependence was only on one scalar mass ($m_{\sigma_{N}}$). Both scattering
lengths are depicted as functions of $m_{\sigma_{1}}$ in Fig.\ \ref{SL1}.%

We observe that the obtained values of the isospin-zero scattering length
$a_{0}^{0}$ are smaller than those in Scenario I of the two-flavour model as
well as those reported by the NA48/2 Collaboration \cite{Peyaud}. The largest
value of this scattering length is $a_{0}^{0}$ $=0.184$, obtained for
$m_{\sigma_{1}}$ $=456$ MeV [note that this is the smallest value of
$m_{\sigma_{1}}$, determined by the condition $m_{0}^{2}=0$, see
Eq.\ (\ref{ms1})]. The scattering length $a_{0}^{2}$ is within the NA48/2 results.

We can therefore conclude that, as in Scenario I of the two-flavour version of
the model, it is not possible to obtain satisfying results for scattering
lengths as well as scalar decay widths simultaneously: the decay widths
$\Gamma_{\sigma_{1}\rightarrow\pi\pi}$ and $\Gamma_{\sigma_{2}\rightarrow
\pi\pi}$ possess very good values respectively for $m_{\sigma_{1}}=705$ MeV
and $m_{\sigma_{1}}=1200$ MeV [see Eqs.\ (\ref{ms11}) and (\ref{ms21}) and
Fig.\ \ref{Spp1}]; however, the same is not true for $a_{0}^{0}=0.165$ that
is outside the NA48/2 interval reading $a_{0}^{0}=0.218\pm0.020$ although
$a_{0}^{2}=-0.0442$ is within the respective NA48/2 interval (that is,
however, rather broad: $a_{0}^{2}=-0.0457\pm0.0125$ \cite{Peyaud}). Note
that the discrepancy with the NA48/2 result becomes even larger if the
isospin-exact values of $a_{0}^{0\,\mathrm{(I)}}=0.244\pm0.020$ and
$a_{0}^{2\,\mathrm{(I)}}=-0.0385\pm0.0125$ from Sec.\ \ref{sec.ISL} are
considered. Therefore,
our Fit I yields the reverse situation to that of Scenario I in the
$U(2)\times U(2)$ version of the model where we were able to describe the
scattering lengths correctly but the $\sigma_{N}$ decay width was too small.
Nonetheless, it is apparent that the scattering lengths still require the
existence of a light scalar meson as they saturate for large values of
$m_{\sigma_{1}}$.

Note that it is possible to obtain the already-known results for the
scattering lengths within the $N_{f}=2$ model in Scenario I. Setting the
$\sigma_{N}$-$\sigma_{S}$ mixing angle $\varphi_{\sigma}=0$ and considering
the limit $m_{\sigma_{S}}\rightarrow\infty$ leads to the diagrams already
depicted in Scenario I of the two-flavour version of the model (see Fig.\ \ref{a00a02f})
once the parameter values have been adjusted to those from the stated scenario.

\begin{figure}[h]
  \begin{center}
    \begin{tabular}{cc}
      \resizebox{78mm}{!}{\includegraphics{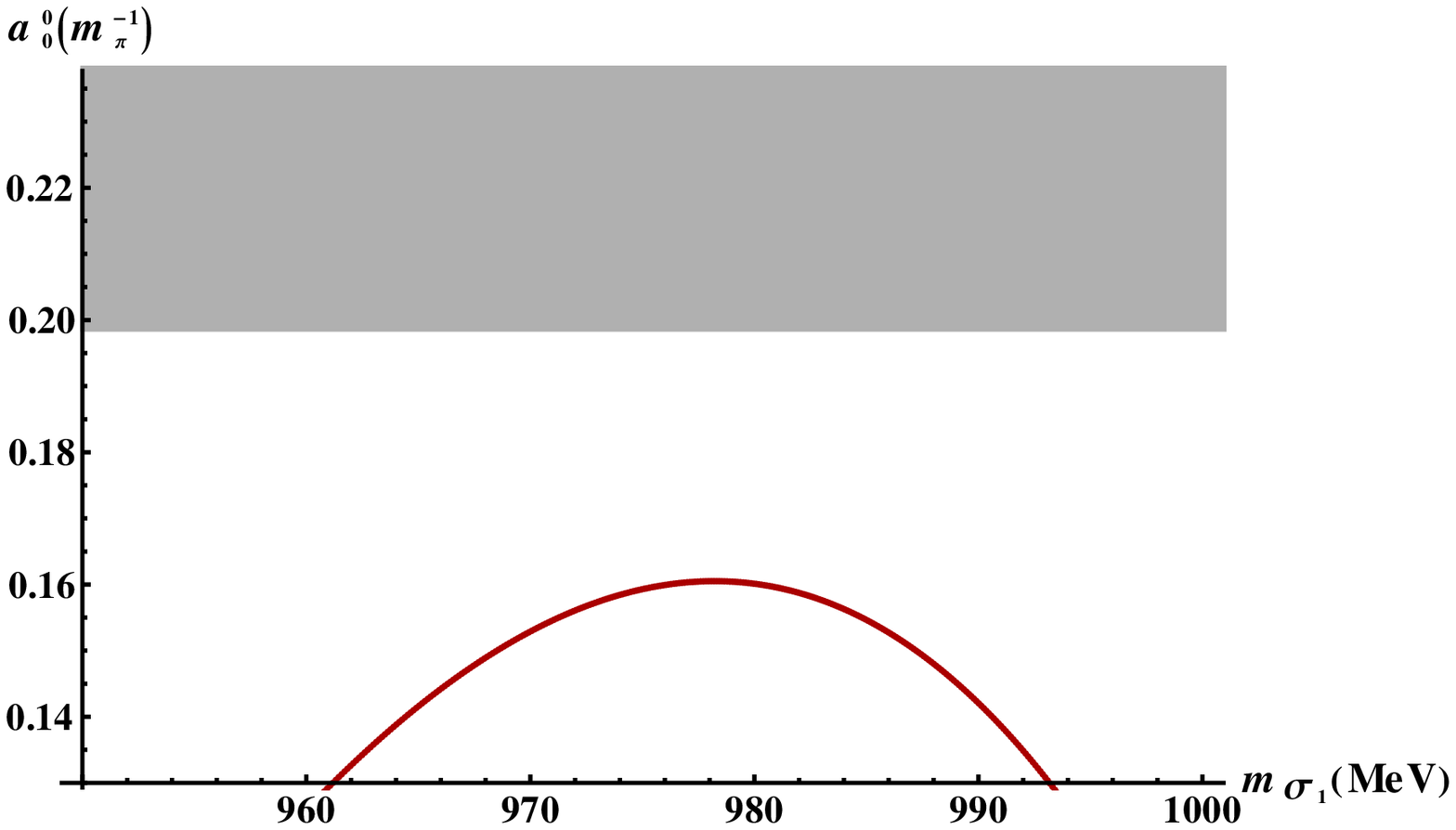}} &
      \resizebox{78mm}{!}{\includegraphics{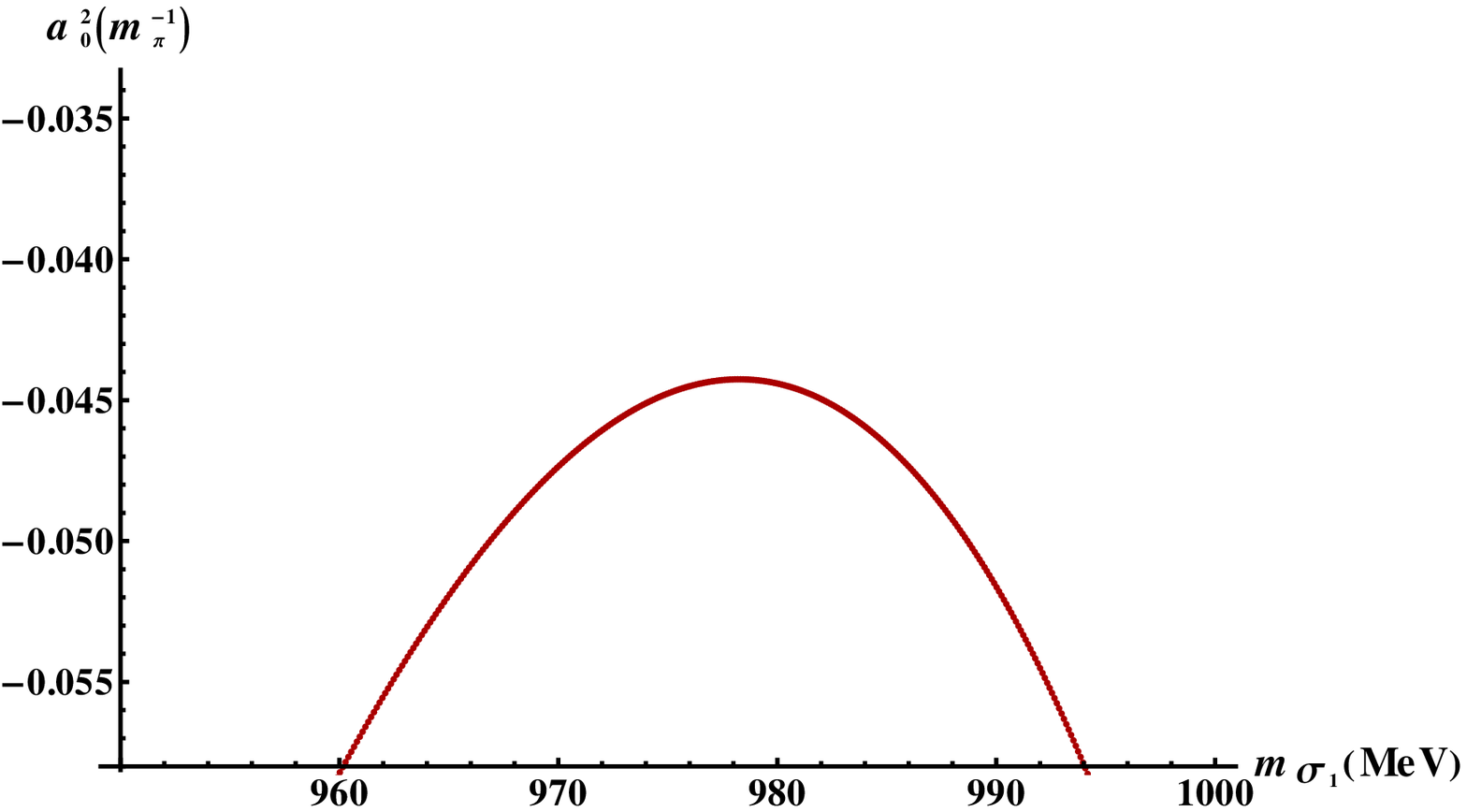}} 
    \end{tabular}
    \caption{Scattering lengths $a_{0}^{0}$ and $a_{0}^{2}$ as functions of
$m_{\sigma_{1}}$ in the limit $m_{\sigma_{2}}\rightarrow\infty$. The shaded
area on the left panel represents the NA48/2 result regarding $a_{0}^{0}$
\cite{Peyaud}; the entire right panel represents the $a_{0}^{2}$ interval from NA48/2.}
    \label{SL2}
  \end{center}
\end{figure}

A different limit is obtained from our Fit I by artificially decoupling
$\sigma_{2}$ (i.e., setting $m_{\sigma_{2}}\rightarrow\infty$) but still
allowing for $m_{\sigma_{1}}$ and $\varphi_{\sigma}$ to change simultaneously
with $m_{0}^{2}$ [see Eqs.\ (\ref{m_sigma_1}) and (\ref{phisigma1})]. In this
limit, $\varphi_{\sigma}$ is not fixed to zero. We observe that the
correspondence of the scattering lengths to data is in this case very much
spoiled. Acceptable values of $a_{0}^{2}$ are obtained only in a small range
of $960$ MeV $\leq m_{\sigma_{1}}\leq994$ MeV while $a_{0}^{0}<0.161$ for all
values of $m_{\sigma_{1}}$, see Fig.\ \ref{SL2}. In the case of $a_{0}^{0}$
with two scalar resonances, Fig.\ \ref{SL1}, the values of $a_{0}^{0}$ were
relatively larger for relatively smaller values of $m_{\sigma_{1}}$ whereas
here, in the one-resonance case, the dependence of scattering lengths on
$m_{\sigma_{1}}$ gains a parabolic form and therefore peaks in a limited
$m_{\sigma_{1}}$ interval. The scattering length $a_{0}^{0}$ then continues to
decrease with decreasing $m_{\sigma_{1}}$ until the contribution of the pole
term $1/(m_{\sigma_{1}}^{2}-4m_{\pi}^{2})$\ becomes sufficiently large and
forces $a_{0}^{0}$\ to rise again (this, however, happens only for
$m_{\sigma_{1}}\simeq300$ MeV, i.e., $m_{0}^{2}>0$, according to
Fig.\ \ref{Sigmamassen1} -- we do not represent this value of $m_{\sigma_{1}}$
in Fig.\ \ref{SL2} and thus do not see an increase of $a_{0}^{0}$ there);
$a_{0}^{2}$ possesses no pole at $\pi\pi$\ threshold and therefore retains a
parabolic form until $m_{\sigma_{1}}=0$.\ %

We can thus conclude that artificially removing $\sigma_{2}$ from the $\pi\pi$
scattering worsens the correspondence with the NA48/2 results considerably although,
given the relatively large values of $m_{\sigma_{2}}$ [see Eq.\ (\ref{ms2})],
one would have expected that the contribution of $\sigma_{2}$ to the
scattering lengths is suppressed. Nonetheless, the scattering lengths depend
decisively on the scalar masses -- as already mentioned, they saturate for large
values of the masses (see Fig.\ \ref{SL1}). Our Fit II will be developed under the
assumption that the scalar $I=1/2$ and $I=1$ states are above $1$ GeV yielding
$m_{\sigma_{1,2}}>1$ GeV as well (see Sec.\ \ref{sec.scalarmasses2}). Our
combined analysis in Sec.\ \ref{sec.scalars2} will subsequently yield
$m_{\sigma_{1}}^{\text{(FIT II)}}=1310$ MeV and $m_{\sigma_{2}}^{\text{(FIT
II)}}=1606$ MeV. This implies that there needs to be no calculation of
scattering lengths in Fit II because the scattering lengths will be in their
respective Weinberg limits \cite{Weinberg:1966}: $a_{0}^{0\text{(FIT II)}}$
$\simeq0.158$, $a_{0}^{2\text{(FIT II)}}$ $\simeq-0.0448$.

\section{Conclusions from Fit with Scalars below 1 GeV} \label{sec.conclusionsfitI}

In the previous sections we have addressed the question whether it is possible
to obtain a reasonable phenomenology of mesons in vacuum under the assumption
that scalar $\bar{q}q$ states possess energies below 1 GeV. To this end, we
have looked for a fit (labelled Fit I) incorporating the masses of $\pi$, $K$,
$\eta$, $\eta^{\prime}$,$\rho$, $K^{\star}$, $\omega_{S}\equiv\varphi(1020)$,
$a_{1}$, $K_{1}$, $f_{1S}\equiv f_{1}(1420)$, decay widths $\Gamma
_{a_{1}\rightarrow\pi\gamma}$ and $\Gamma_{f_{1N}\rightarrow a_{0}(980)\pi}$,
as well as the masses of the scalar states $a_{0}$ and $K_{S}$ assigned to $a_{0}(980)$
and$\ K_{0}^{\star}(800)\equiv\kappa$, respectively. We have not included any
scalar isosinglet masses into the fit in order to let these masses remain a
prediction of the fit.
\\

We summarise the main conclusions of the fit as follows:

\begin{itemize}
\item It is possible to find a fit; masses entering the fit are well described except

\begin{itemize}
\item $m_{\kappa}=1128.7$ MeV, almost by a factor of two larger than the
corresponding PDG value \cite{PDG} (but the $\kappa$ resonance is very broad),

\item $m_{a_{1}}=1395.5$ MeV, approximately $170$ MeV larger than the
corresponding PDG value $m_{a_{1}}^{\exp}=1230$ MeV [but the $a_{1}(1260)$
resonance is also broad and the PDG mass is only an educated guess],

\item $m_{K_{1}}=1520$ MeV, approximately $250$ MeV larger than the mass of
$K_{1}(1270)$; however, assigning our $K_{1}$ field to the (broad) resonance
$K_{1}(1400)$ yields the stated result for $m_{K_{1}}$ acceptable,

\item $m_{\omega_{S}}=870.35$ MeV, approximately $150$ MeV less than
$m_{\varphi(1020)}^{\exp}=1019.46$ MeV and $m_{f_{1S}}=1643.4$ MeV,
approximately $220$ MeV larger than $m_{f_{1}(1420)}^{\exp}=1426.4$ MeV --
this in particular represents a problem because $\varphi(1020)$ and
$f_{1}(1420)$ are rather sharp resonances.
\end{itemize}

Additionally, $\Gamma_{a_{1}\rightarrow\pi\gamma}=0.369$ MeV is outside the
experimental interval $\Gamma_{a_{1}\rightarrow\pi\gamma}^{\exp}%
=0.640\pm0.246$ MeV \cite{PDG}.

\item It is not possible to assign the two mixed scalar isosinglets
$\sigma_{1}$ (predominantly non-strange) and $\sigma_{2}$ (predominantly strange) to measured resonances
if one considers only their masses because 
$m_{\sigma_{1}}$ and
$m_{\sigma_{2}}$ vary in rather large intervals: $456$ MeV $\leq
m_{\sigma_{1}}\leq1139$ MeV and $1187$ MeV $\leq m_{\sigma_{2}}\leq2268$ MeV.
(Interval boundaries are determined by the conditions $m_0^2<0$ and $m_{\sigma_N} < m_{\sigma_S}$.)
Therefore, an analysis of scalar decay widths is called for.

\item We obtain satisfying results in the decay channels $\sigma
_{1,2}\rightarrow\pi\pi$ and $\sigma_{1,2}\rightarrow KK$ if we set
$m_{\sigma_{1}}=705$ MeV and $m_{\sigma_{2}}=1200$ MeV leading to
$\Gamma_{\sigma_{1}\rightarrow\pi\pi}=305$ MeV and $\Gamma_{\sigma
_{2}\rightarrow\pi\pi}=207$ MeV in the former and $\Gamma_{\sigma
_{1}\rightarrow KK}=0$ and $\Gamma_{\sigma_{2}\rightarrow KK}=240$ MeV in the
latter channel. This allows us to assign $\sigma_{1}$ to $f_{0}(600)$ and
$\sigma_{2}$ to $f_{0}(1370)$; $\Gamma_{\sigma_{2}\rightarrow\pi\pi}$ was
chosen such that it corresponds to $\Gamma_{f_{0}(1370)\rightarrow\pi\pi}=207$ MeV
from Ref.\ \cite{buggf0}. Consequently, we interpret $f_{0}(600)$ as a
predominantly non-strange $\bar{q}q$ state while $f_{0}(1370)$ is interpreted
as a predominantly strange quarkonium. The results also suggest, however, that
$f_{0}(1370)$ should predominantly decay into kaons (as $\Gamma_{\sigma
_{2}\rightarrow KK}/\Gamma_{\sigma_{2}\rightarrow\pi\pi}=1.15$) -- not
surprising for a strange quarkonium but clearly at odds with experimental data
\cite{PDG}.

\item Satisfying results are obtained in the $\sigma_{1,2}\rightarrow\eta\eta$
decay channel: $m_{\sigma_{1}}=705$ MeV yields $\Gamma_{\sigma_{1}%
\rightarrow\eta\eta}=0$ (as expected) and $m_{\sigma_{2}}=1200$ MeV yields
$\Gamma_{\sigma_{1}\rightarrow\eta\eta}=31$ MeV (also in line with
expectations). Additionally, one obtains $\Gamma_{\sigma_{2}\rightarrow
\eta\eta}/\Gamma_{\sigma_{2}\rightarrow\pi\pi}=0.15$, in accordance with the
result $\Gamma_{f_{0}(1370)\rightarrow\eta\eta}/\Gamma_{f_{0}(1370)\rightarrow
\pi\pi}=0.19\pm0.07$ from Ref.\ \cite{buggf0}. However, the ratio
$\Gamma_{\sigma_{2}\rightarrow KK}/\Gamma_{\sigma_{2}\rightarrow\eta\eta}\gg1$
again suggests that $f_{0}(1370)$ should decay predominantly into kaons,
problematic from the experimental point of view.

\item We obtain $\Gamma_{K_{0}^{\star}(800)\rightarrow K\pi}=490$ MeV, a
satisfying result predicting a broad scalar kaon resonance in accordance with
the PDG data \cite{PDG}. However, the mass of the resonance is $m_{K_{0}%
^{\star}(800)}=1128.7$ MeV, and thus larger than $m_{K_{0}^{\star}(800)}%
^{\exp}=676$ MeV by a factor of two.

\item The decay amplitude $a_{0}(980)\rightarrow\eta\pi$ is within experimental data.

\item For the scattering lengths, we obtain $a_{0}^{0}<0.184$ for all
values of $m_{\sigma_{1}}$; $a_{0}^{0}$ is thus below NA48/2 results
\cite{Peyaud}. Contrarily, the scattering length $a_{0}^{2}$ is within the
NA48/2 results (that, for this scattering length, possess rather large
uncertainties). We can therefore conclude that Fit I does not allow for
scattering lengths as well as scalar decay widths to be described
simultaneously: although we obtain satisfying values for the decay widths
$\Gamma_{\sigma_{1}\rightarrow\pi\pi}$ and $\Gamma_{\sigma_{2}\rightarrow
\pi\pi}$, the same is not true for $a_{0}^{0}$. Nonetheless, the scattering
lengths still require the existence of a light scalar meson because they saturate
for large values of $m_{\sigma_{1}}$.

\item Additionally, the phenomenology in the vector and axial-vector channels
is not well described.

\begin{itemize}

\item $\Gamma_{K^{\star}\rightarrow K\pi}=32.8$ MeV whereas experimental data
suggest $\Gamma_{K^{\star}\rightarrow K\pi}^{\exp}=46.2$ MeV \cite{PDG}.

\item The decay width $\Gamma_{a_{1}(1260)\rightarrow\rho\pi}$ depends (among others) on
the parameter $g_{2}$, fixed via $\Gamma_{\rho\rightarrow\pi\pi}$. A calculation
of $\Gamma_{a_{1}(1260)\rightarrow\rho\pi}$ then yields values of more than
$10$ GeV if we set $\Gamma_{\rho\rightarrow\pi\pi}=149.1$ MeV (as suggested by
the PDG \cite{PDG}). Alternatively, if one forces $\Gamma_{a_{1}%
(1260)\rightarrow\rho\pi}<600$ MeV to comply with the data, then $\Gamma
_{\rho\rightarrow\pi\pi}<38$ MeV is obtained -- a value that is approximately
$100$ MeV less than the experimental result. We also obtain $\Gamma
_{a_{1}\rightarrow\bar{K}^{\star}K\rightarrow\bar{K}K\pi}=1.97$ GeV.

\item Analogous statements are true for $f_{1}(1285)$ and $f_{1}(1420)$. Fit I
yields $\Gamma_{f_{1}(1285)\rightarrow\bar{K}^{\star}K}\simeq2.15$ GeV for
$\Gamma_{\rho\rightarrow\pi\pi}=149.1$ MeV; the physical value $\Gamma
_{f_{1}(1285)\rightarrow\bar{K}^{\star}K}\lesssim2$ MeV is obtained only for
$\Gamma_{\rho\rightarrow\pi\pi}\sim20$ MeV. The fit also yields $\Gamma
_{f_{1}(1420)\rightarrow\bar{K}^{\star}K}\simeq18$ GeV for $\Gamma
_{\rho\rightarrow\pi\pi}=149.1$ MeV with the physical value $\Gamma
_{f_{1}(1420)\rightarrow\bar{K}^{\star}K}\simeq54.9$ MeV obtained for
$\Gamma_{\rho\rightarrow\pi\pi}\sim27$ MeV.

\item The phenomenology of the $K_{1}(1400)$ meson is described as poorly as the
$a_{1}(1260)$ phenomenology. Combined results in the decay channels
$K_{1}(1400)\rightarrow K^{\star}\pi$, $\rho K$ and $\omega K$ suggest that
the full decay width of the $K_{1}(1400)$ resonance should be $\sim10$ GeV,
two orders of magnitude larger than the experimental value $\Gamma
_{K_{1}(1400)}=(174\pm13)$ MeV \cite{PDG}. The decay widths $\Gamma_{K_{1}%
(1400)\rightarrow\rho K}=4.77$ GeV and $\Gamma_{K_{1}(1400)\rightarrow\omega
K}=1.59$ GeV are three orders of magnitude larger than their respective
experimental values $\Gamma_{K_{1}(1400)\rightarrow\rho K}^{\exp}=(2.1\pm1.1)$
MeV and $\Gamma_{K_{1}(1400)\rightarrow\omega K}^{\exp}=(1.7\pm1.7)$ MeV;
$\Gamma_{K_{1}(1400)\rightarrow K^{\star}\pi}=6.73$ GeV is an order of
magnitude larger than $\Gamma_{K_{1}(1400)\rightarrow K^{\star}\pi}^{\exp
}=(164\pm16)$ MeV. In fact, the only piece of $K_{1}(1400)$ experimental data
correctly described in Fit I is represented by the fact that $K_{1}%
(1400)\rightarrow K^{\star}\pi$ is found to be the dominant decay channel of
this resonance; all other results are not compatible with the data.

\end{itemize}

Thus we cannot accommodate a correct (axial-)vector phenomenology within the fit:
either $a_{1}(1260)$, $f_{1}(1285)$, $f_{1}(1420)$ and $K_{1}(1400)$\ are too
broad [$\sim(1-10)$ GeV] or the $\rho$ meson is too narrow ($\lesssim40$ MeV).
\end{itemize}

Then the fit results, and thus the assumption of scalar $\bar{q}q$ states
below 1 GeV, are extremely problematic.

\chapter{Fit II: Scalars above 1 GeV} \label{sec.fitII}

In this chapter we look for a fit of meson masses (labelled Fit II) assuming
that scalar $\bar{q}q$ states have masses above 1 GeV and discuss the ensuing
phenomenology. We will consequently be able to draw comparative conclusions
with regard to results obtained from Fit I where, conversely, scalar $\bar
{q}q$ states were assigned to resonances below 1 GeV.

The formal structure of Fit II is very similar to that of Fit I. We have
already described in Chapter \ref{sec.remarks} how the initial set of 18
parameters in the Lagrangian (\ref{Lagrangian})\ is reduced to seven unknowns:
$Z_{\pi}$\textit{, }$Z_{K}$\textit{, }$m_{1}^{2}$\textit{, }$h_{2}$\textit{,
}$\delta_{S}$\textit{, }$\lambda_{2}$\textit{ }and\textit{ }$m_{0}^{2}%
+\lambda_{1}(\phi_{N}^{2}+\phi_{S}^{2})$. In order to implement Fit II, we
make use of 16 equations: for $m_{\pi}$, $m_{K}$, $m_{K_{S}}\equiv
m_{K_{0}^{\star}(1430)}$, $m_{a_{0}}\equiv m_{a_{0}(1450)}$, $m_{\eta}$,
$m_{\eta^{\prime}}$ [the latter two via Eqs.\ (\ref{m_eta}) and (\ref{m_eta'})
from $m_{\eta_{N}}$ and $m_{\eta_{S}}$], $m_{\rho}$, $m_{K^{\star}}$,
$m_{\omega_{S}}$, $m_{a_{1}}$, $m_{K_{1}}$, $m_{f_{1S}}$, $\Gamma
_{a_{1}\rightarrow\pi\gamma}$ and $\Gamma_{a_{0}(1450)}$ as well as
Eqs.\ (\ref{Z_K3}) and (\ref{Z_K4}), the latter two for $Z_{K}$. Thus, in this
fit our fields $\vec{a}_{0}$ and $K_{S}$ are assigned respectively to
$a_{0}(1450)$ and $K_{0}^{\star}(1430)$, i.e., to states above 1 GeV.
Conversely, this means that now we are working with the assumption that
$a_{0}(1450)$ and $K_{0}^{\star}(1430)$ are $\bar{q}q$ states. Consequently,
in Fit II there are no states from our model that are assigned to the resonances
$a_{0}(980)$ and $\kappa$; these resonances could be introduced into our model
only as additional fields, such as, for example, tetraquarks \cite{Achim}. Note that there may
also exist mixing in the isotriplet channel between $a_0(980)$ and $a_0(1450)$. The mixing is, however,
small \cite{Giacosa:2006tf} and can be neglected.

As mentioned previously, Fit II will require knowledge of the full $a_{0}(1450)$ decay
width. The corresponding formulas are discussed in the following section.

\section{Full decay width of \boldmath $a_{0}(1450)$}

Experimental data \cite{PDG} suggest that $a_{0}(1450)$ possesses six
decay channels: into $\pi\eta$, $\pi\eta^{\prime}$, $KK$, $\omega\pi\pi$,
$a_{0}(980)\pi\pi$ and $\gamma\gamma$. The latter two are poorly known and
suppressed; we therefore omit these two decay channels from our
considerations. The remaining decay channels can be calculated directly from
our model as follows:

\begin{itemize}
\item The decay width $\Gamma_{a_{0}(1450)\rightarrow\pi\eta}$ is obtained from
the interaction Lagrangian (\ref{a0etaNetaSpion}) as described in
Sec.\ \ref{sec.a0etapion} by assigning our $\vec{a}_{0}$ field to
$a_{0}(1450)$. We use the following formula for the decay width:
\begin{equation}
\Gamma_{a_{0}^{0}(1450)\rightarrow\eta\pi^{0}}=\frac{k(m_{a_{0}(1450)}
,m_{\eta},m_{\pi})}{8\pi m_{a_{0}(1450)}^{2}}|-i\mathcal{M}_{a_{0}
^{0}(1450)\rightarrow\eta\pi^{0}}(m_{\eta})|^{2} \label{Ga0etapion}
\end{equation}

with $\mathcal{M}_{a_{0}^{0}(1450)\rightarrow\eta\pi^{0}}(m_{\eta})$ from
Eq.\ (\ref{Ma0etapion}).

\item The decay width $\Gamma_{a_{0}(1450)\rightarrow\pi\eta^{\prime}}$ is also
obtained from the interaction Lagrangian (\ref{a0etaNetaSpion}). Analogously to
Eq.\ (\ref{Ma0etapion1}) we obtain for the decay amplitude
\begin{equation}
-i\mathcal{M}_{a_{0}^{0}(1450)\rightarrow\eta^{\prime}\pi^{0}}(m_{\eta
^{\prime}})=-i[\cos\varphi_{\eta}\mathcal{M}_{a_{0}^{0}\rightarrow\eta_{S}%
\pi^{0}}(m_{\eta^{\prime}})-\sin\varphi_{\eta}\mathcal{M}_{a_{0}%
^{0}\rightarrow\eta_{N}\pi^{0}}(m_{\eta^{\prime}})] \label{Ma0etappion}
\end{equation}

with $\mathcal{M}_{a_{0}^{0}\rightarrow\eta_{N}\pi^{0}}$\ and $\mathcal{M}
_{a_{0}^{0}\rightarrow\eta_{S}\pi^{0}}$ respectively from
Eqs.\ (\ref{Ma0etaNpion}) and (\ref{Ma0etaSpion}). Then the decay width is
calculated as
\begin{equation}
\Gamma_{a_{0}^{0}(1450)\rightarrow\eta^{\prime}\pi^{0}}=\frac{k(m_{a_{0}
(1450)},m_{\eta^{\prime}},m_{\pi})}{8\pi m_{a_{0}(1450)}^{2}}|-i\mathcal{M}
_{a_{0}^{0}(1450)\rightarrow\eta^{\prime}\pi^{0}}(m_{\eta^{\prime}}
)|^{2}\text{.} \label{Ga0etappion}
\end{equation}

\item The $a_{0}^{0}KK$ interaction Lagrangian obtained from
Eq.\ (\ref{Lagrangian}) has the following form:%
\begin{align}
\mathcal{L}_{a_{0}KK}  &  =A_{a_{0}KK}a_{0}^{0}(K^{0}\bar{K}^{0}-K^{-}%
K^{+})+B_{a_{0}KK}a_{0}^{0}(\partial_{\mu}K^{0}\partial^{\mu}\bar{K}%
^{0}-\partial_{\mu}K^{-}\partial^{\mu}K^{+})\nonumber\\
&  +C_{a_{0}KK}\partial_{\mu}a_{0}^{0}(K^{0}\partial^{\mu}\bar{K}^{0}+\bar
{K}^{0}\partial^{\mu}K^{0}-K^{-}\partial^{\mu}K^{+}-K^{+}\partial^{\mu}K^{-})
\label{a0kaonkaon}%
\end{align}

with
\begin{align}
A_{a_{0}KK}  &  =\frac{\sqrt{2}}{2}\lambda_{2}Z_{K}^{2}(\sqrt{2}\phi_{N}%
-\phi_{S})\text{,} \\
B_{a_{0}KK}  &  =Z_{K}^{2}\left\{  g_{1}w_{K_{1}}\left[  1-\frac{1}{2}%
g_{1}w_{K_{1}}(\phi_{N}+\sqrt{2}\phi_{S})\right]  -\frac{w_{K_{1}}^{2}}%
{2}(h_{2}\phi_{N}-\sqrt{2}h_{3}\phi_{S})\right\}\text{,} \\
C_{a_{0}KK}  &  =-\frac{g_{1}}{2}Z_{K}^{2}w_{K_{1}}\text{.}%
\end{align}

We observe that the Lagrangian in Eq.\ (\ref{a0kaonkaon}) possesses the same
form as $\mathcal{L}_{\sigma KK}$ from Eq.\ (\ref{sigmakaonkaon}). Therefore,
analogously to the calculation performed in Sec.\ \ref{sec.sigmakaonkaon} we
obtain

\begin{equation}
\Gamma_{a_{0}^{0}(1450)\rightarrow K\bar{K}}=\frac{k(m_{a_{0}(1450)}%
,m_{K},m_{K})}{4\pi m_{a_{0}(1450)}^{2}}|-i\mathcal{M}_{a_{0}^{0}%
(1450)\rightarrow K\bar{K}}|^{2}\text{,} \label{Ga0kaonkaon}%
\end{equation}

where we have considered an isospin factor of two for the decays $a_{0}%
^{0}(1450)\rightarrow K^{0}\bar{K}^{0}$ and $a_{0}^{0}(1450)\rightarrow
K^{-}K^{+}$. The decay amplitude $-i\mathcal{M}_{a_{0}^{0}(1450)\rightarrow
K\bar{K}}$ reads

\begin{equation}
-i\mathcal{M}_{a_{0}^{0}(1450)\rightarrow K\bar{K}}=-i\left\{  A_{a_{0}%
KK}-B_{a_{0}KK}\left[  \frac{m_{a_{0}(1450)}^{2}}{2}-m_{K}^{2}\right]
+C_{a_{0}KK}m_{a_{0}(1450)}^{2}\right\}  \text{.}%
\end{equation}

\item The decay width $\Gamma_{a_{0}(1450)\rightarrow\omega\pi\pi}$ is calculated
via the sequential decay $a_{0}(1450)\rightarrow\omega\rho\rightarrow\omega
\pi\pi$. The interaction Lagrangian is already known from Scenario II of the
two-flavour version of our model, Eq.\ (\ref{a0oNr}); the formula for the
decay width $\Gamma_{a_{0}(1450)\rightarrow\omega\rho\rightarrow\omega\pi\pi}$
is stated in Eq.\ (\ref{a0omegapionpion2}).

\item The full decay width of the $a_{0}(1450)$ resonance is obtained from
Eqs.\ (\ref{Ga0etapion}), (\ref{Ga0etappion}), (\ref{Ga0kaonkaon}) and (\ref{a0omegapionpion2}):
\begin{equation}
\Gamma_{a_{0}(1450)}=\Gamma_{a_{0}^{0}(1450)\rightarrow\eta\pi^{0}}
+\Gamma_{a_{0}^{0}(1450)\rightarrow\eta^{\prime}\pi^{0}}+\Gamma_{a_{0}
^{0}(1450)\rightarrow K \bar{K}}+ \Gamma_{a_{0}(1450)\rightarrow\omega\rho\rightarrow\omega\pi\pi}\text{.} \label{Ga01450}
\end{equation}

The experimental value of this decay width is $\Gamma_{a_{0}(1450)} =(265 \pm 13)$ MeV \cite{PDG}.
\end{itemize}

\section{Implementation of Fit II}

Analogously to our calculations in Sec.\ \ref{sec.fitI}, we look for a fit
satisfying the following equations (experimental central values from the PDG
\cite{PDG}; no consideration of experimental uncertainties at this point):

\begin{align}
&  Z_{\pi}^{2}\left[  m_{0}^{2}+\lambda_{1}(\phi_{N}^{2}+\phi_{S}^{2}%
)+\frac{\lambda_{2}}{2}\phi_{N}^{2}\right]  =(139.57\text{ MeV})^{2}\equiv
m_{\pi}^{2}\text{,} \label{fit21}\\
&  Z_{K}^{2}\left[  m_{0}^{2}+\lambda_{1}(\phi_{N}^{2}+\phi_{S}^{2}%
)+\lambda_{2}\left(  \frac{\phi_{N}^{2}}{2}-\frac{\phi_{N}\phi_{S}}{\sqrt{2}%
}+\phi_{S}^{2}\right)  \right]  =(493.677\text{ MeV})^{2}\equiv m_{K}%
^{2}\text{,} \label{fit22}\\
&  Z_{K_{S}}^{2}\left[  m_{0}^{2}+\lambda_{1}(\phi_{N}^{2}+\phi_{S}%
^{2})+\lambda_{2}\left(  \frac{\phi_{N}^{2}}{2}+\frac{\phi_{N}\phi_{S}}%
{\sqrt{2}}+\phi_{S}^{2}\right)  \right]  =(1425\text{ MeV})^{2}\equiv
m_{K_{0}^{\star}(1430)}^{2}\text{,} \label{fit23}\\
&  m_{0}^{2}+\lambda_{1}(\phi_{N}^{2}+\phi_{S}^{2})+\frac{3}{2}\lambda_{2}%
\phi_{N}^{2}=(1474\text{ MeV})^{2}\equiv m_{a_{0}(1450)}^{2}\text{,} \label{fit24}\\
&  Z_{\pi}^{2}\left[  m_{0}^{2}+\lambda_{1}(\phi_{N}^{2}+\phi_{S}^{2}%
)+\frac{\lambda_{2}}{2}\phi_{N}^{2}+c_{1}\phi_{N}^{2}\phi_{S}^{2}\right]
\cos^{2}\varphi_{\eta} \nonumber \\
& + Z_{\eta_{S}}^{2}\left[  m_{0}^{2}+\lambda_{1}(\phi
_{N}^{2}+\phi_{S}^{2})+\lambda_{2}\phi_{S}^{2}+c_{1}\frac{\phi_{N}^{4}}%
{4}\right]  \sin^{2}\varphi_{\eta}\nonumber\\
&  +c_{1}\frac{Z_{\eta_{S}}Z_{\pi}}{2}\phi_{N}^{3}\phi_{S}\sin(2\varphi_{\eta
})=(547.853\text{ MeV})^{2}\equiv m_{\eta}^{2}\text{,} \label{fit25}\\
&  Z_{\pi}^{2}\left[  m_{0}^{2}+\lambda_{1}(\phi_{N}^{2}+\phi_{S}^{2}%
)+\frac{\lambda_{2}}{2}\phi_{N}^{2}+c_{1}\phi_{N}^{2}\phi_{S}^{2}\right]
\sin^{2}\varphi_{\eta} \nonumber \\
& + Z_{\eta_{S}}^{2}\left[  m_{0}^{2}+\lambda_{1}(\phi
_{N}^{2}+\phi_{S}^{2})+\lambda_{2}\phi_{S}^{2}+c_{1}\frac{\phi_{N}^{4}}%
{4}\right]  \cos^{2}\varphi_{\eta}\nonumber\\
&  -c_{1}\frac{Z_{\eta_{S}}Z_{\pi}}{2}\phi_{N}^{3}\phi_{S}\sin(2\varphi_{\eta
})=(957.78\text{ MeV})^{2}\equiv m_{\eta^{\prime}}^{2}\text{,} \label{fit26}\\
&  m_{1}^{2}+(h_{2}+h_{3})\frac{\phi_{N}^{2}}{2}=(775.49\text{ MeV})^{2}\equiv
m_{\rho}^{2}\text{,} \label{fit27}\\
&  m_{1}^{2}+g_{1}^{2}\phi_{N}^{2}+(h_{2}-h_{3})\frac{\phi_{N}^{2}}%
{2}=(1230\text{ MeV})^{2}\equiv m_{a_{1}}^{2}\text{,} \label{fit28}\\
&  m_{1}^{2}+\delta_{S}+\left(  g_{1}^{2}+h_{2}\right)  \frac{\phi_{N}^{2}}%
{4}+\frac{1}{\sqrt{2}}(h_{3}-g_{1}^{2})\phi_{N}\phi_{S}+\left(  g_{1}%
^{2}+h_{2}\right)  \frac{\phi_{S}^{2}}{2}=(891.66\text{ MeV})^{2}\equiv
m_{K^{\star}}^{2}\text{,} \label{fit29}\\
&  m_{1}^{2}+\delta_{S}+\left(  g_{1}^{2}+h_{2}\right)  \frac{\phi_{N}^{2}}%
{4}+\frac{1}{\sqrt{2}}(g_{1}^{2}-h_{3})\phi_{N}\phi_{S}+\left(  g_{1}%
^{2}+h_{2}\right)  \frac{\phi_{S}^{2}}{2}=(1272\text{ MeV})^{2}\equiv
m_{K_{1}}^{2}\text{,} \label{fit210}\\
&  m_{1}^{2}+2\delta_{S}+\left(  h_{2}+h_{3}\right)  \phi_{S}^{2}%
=(1019.455\text{ MeV})^{2}\equiv m_{\omega_{S}}^{2}\text{,} \label{fit211}\\
&  m_{1}^{2}+2\delta_{S}+2g_{1}^{2}\phi_{S}^{2}+\left(  h_{2}-h_{3}\right)
\phi_{S}^{2}=(1426.4\text{ MeV})^{2}\equiv m_{f_{1S}}^{2}\text{,} \label{fit212}\\
&  \frac{e^{2}}{96\pi}(Z_{\pi}^{2}-1)m_{a_{1}}\left[  1-\left(  \frac{m_{\pi}%
}{m_{a_{1}}}\right)  ^{2}\right]  ^{3}=0.640\text{ MeV}\equiv\Gamma
_{a_{1}\rightarrow\pi\gamma}\text{,} \label{fit213}\\
&  \Gamma_{a_{0}^{0}(1450)\rightarrow\eta\pi^{0}}+\Gamma_{a_{0}^{0}%
(1450)\rightarrow\eta^{\prime}\pi^{0}}+\Gamma_{a_{0}^{0}(1450)\rightarrow
K\bar{K}}+\Gamma_{a_{0}(1450)\rightarrow\rho\pi\rightarrow\omega\pi\pi
}=265\text{ MeV}\equiv\Gamma_{a_{0}(1450)}\text{,} \label{fit214}%
\end{align}

as well as the $Z_{K}$ Eqs.\ (\ref{Z_K3}) and (\ref{Z_K4}). Note that also in
this fit we set $h_{1}=0=\delta_{N}$; that $c_{1}=c_{1}(\varphi_{\eta})$ from
Eq.\ (\ref{c1phi}) and that we also use $\phi_{N}=Z_{\pi}f_{\pi}$ ($f_{\pi
}=92.4$ MeV), $\phi_{S}=Z_{K}f_{K}/\sqrt{2}$ ($f_{K}=155.5/\sqrt{2}$ MeV),
$g_{1}$ from Eq.\ (\ref{g1}), $h_{3}$ from Eq.\ (\ref{h3}), $Z_{K_{S}}$ from
Eq.\ (\ref{Z_K_S}) and $Z_{\eta_{S}}$ from Eq.\ (\ref{Z_eta_S}).\\

Now we can make use of the same four-step procedure described in
Sec.\ \ref{sec.fitI} to ascertain whether an acceptable fit can be found.\\

\textit{Step 1.} We first consider the first four equations entering the fit,
i.e., Eqs.\ (\ref{fit21}) - (\ref{fit24}) that depend only on four variables:
$Z_{\pi}$, $Z_{K}$, $\lambda_{2}$ and\textit{ }$m_{0}^{2}+\lambda_{1}(\phi
_{N}^{2}+\phi_{S}^{2})$. As in Fit I, we set $m_{a_{1}}=m_{a_{1}}^{\exp}=1230$
MeV and $m_{K^{\star}}=m_{K^{\star}}^{\exp}=891.66$ MeV \cite{PDG} in order
for the renormalisation coefficient $Z_{K_{S}}$ to be calculated. We again
find that $Z_{K_{S}}$ changes only minutely with $m_{a_{1}}$ and $m_{K^{\star
}}$ and therefore the exact values of these two masses are at this point not
of great importance.

We thus obtain a system of four equations (\ref{fit21}) - (\ref{fit24}) with
four unknowns that can be solved exactly; we obtain the following parameter values:%

\begin{align*}
Z_{\pi}  &  =0.36\\
Z_{K}  &  =0.47\\
\lambda_{2}  &  =1860\\
m_{0}^{2}+\lambda_{1}(\phi_{N}^{2}+\phi_{S}^{2})  &  =-856580\text{ MeV}%
^{2}\text{.}%
\end{align*}

Unfortunately, this set of solutions cannot be used further as it implies
$Z_{\pi}<1$ and $Z_{K}<1$, a condition that by the definitions of these
renormalisation coefficients [Eqs.\ (\ref{Z_pi}) and (\ref{Z_K})] cannot be
fulfilled as otherwise either $g_{1}^{2}<0$ or $\phi_{N}^{2}$ $<0$ [in
Eq.\ (\ref{Z_pi})] and $(\phi_{N}+\sqrt{2}\phi_{S})^{2}<0$ [in Eq.\ (\ref{Z_K}%
)] would have to be true. We do not consider an imaginary scalar-vector coupling
$g_{1}$ or imaginary condensates $\phi_{N,S}$ -- therefore, we have to work
for alternative (approximate) solutions for Eqs.\ (\ref{fit21}) -
(\ref{fit24}). We then obtain the parameter values shown in Table \ref{Fit2-1}.%

\begin{table}[h] \centering
\begin{tabular}
[c]{|c|c|c|c|}\hline
Parameter & Value & Observable & Value [MeV]\\\hline
$Z_{\pi}$ & $1.66$ & $m_{\pi}$ & $138.65$\\\hline
$Z_{K}$ & $1.515$ & $m_{K}$ & $497.96$\\\hline
$\lambda_{2}$ & $89.7$ & $m_{a_{0}(1450)}$ & $1452$\\\hline
$m_{0}^{2}+\lambda_{1}(\phi_{N}^{2}+\phi_{S}^{2})$ & $-1044148$ MeV$^{2}$ &
$m_{K_{0}^{\star}(1430)}$ & $1550$\\\hline
\end{tabular}%
\caption{Best solutions of Eqs.\ (\ref{fit21}) - (\ref{fit24}%
) under the conditions $Z_\pi\overset{!}{>}1$, $Z_K \overset{!}{>}1$.\label
{Fit2-1}}%
\end{table}%

The value of $m_{K_{S}}$ is larger than the PDG value due to the pattern of
explicit symmetry breaking that in our model makes strange mesons
approximately 100 MeV ($\simeq$ strange-quark mass) heavier than their
non-strange counterparts. We also note that the $K_{0}^{\star}(1430)$
resonance is rather broad [$\Gamma_{K_{0}^{\star}(1430)}^{\exp}=(270\pm80)$
MeV] and therefore a deviation of $100$ MeV exhibited by $m_{K_{S}}$ is not
too large.

Additionally, $\Gamma_{a_{1}\rightarrow\pi\gamma}=0.628$ MeV is obtained from
the parameter values in Table \ref{Fit2-1}, within the interval $\Gamma
_{a_{1}\rightarrow\pi\gamma}^{\exp}=(0.640\pm0.246)$ MeV cited by the PDG
\cite{PDG}. This is in contrast to the corresponding result in Fit I where we
obtained $\Gamma_{a_{1}\rightarrow\pi\gamma}=0.322$ MeV (see Table
\ref{Fit1-1}).\\

\textit{Step 2.} We now look for values
of $m_{\rho}$, $m_{a_{1}}$, $m_{K^{\star}}$, $m_{\omega_{S}}$, $m_{K_{1}}$ and
$m_{f_{1S}}$ that lead to the pairwise equality of the three $Z_{K}$ formulas,
Eqs.\ (\ref{Z_K3}) and (\ref{Z_K4}). We use the already known values of
$Z_{\pi}$ and $Z_{K}$ from Table \ref{Fit2-1} and also the PDG values of all
mentioned (axial-)vector masses except $m_{a_{1}}$ [because the value cited by
the PDG is merely an educated guess and also because $a_{1}(1260)$ is a rather
broad resonance]. As in Fit I, it is not possible to equate pairwise the
$Z_{K}$ formulas in Eqs.\ (\ref{Z_K3}) and (\ref{Z_K4}) if we use the PDG values
of the masses. A numerical analysis demonstrates that
Eqs.\ (\ref{Z_K3}) and (\ref{Z_K4}) are fulfilled if the following mass values are used:

\begin{align}
m_{a_{1}}  &  =1219\text{ MeV, }m_{\rho}=775.49\text{ MeV, }m_{K^{\star}%
}=916.52\text{ MeV,}\nonumber\\
m_{\omega_{S}}  &  =1036.90\text{ MeV, }m_{K_{1}}=1343\text{ MeV, }m_{f_{1S}%
}=1457.0\text{ MeV.} \label{s}%
\end{align}

\textit{Steps 3 and 4. }The (axial-)vector fit parameters can now be
determined in such a way that the mass values determined by the three $Z_{K}$
formulas are reproduced. The total decay width of $a_{0}(1450)$,
Eq.\ (\ref{Ga01450}), allow us in principle to determine the value of
parameter $h_{2}$, if all other parameters entering Eq.\ (\ref{Ga01450}) are
known, i.e., if the parameters $Z_{\pi}$, $Z_{K}$, $\lambda_{2}$, $g_{1}$,
$h_{3}$ and $c_{1}$ have been determined. The parameters $Z_{\pi}$, $Z_{K}$ and
$\lambda_{2}$ are known from Table \ref{Fit2-1}; $g_{1}$ and $h_{3}$ can be
calculated from $m_{\rho}$ and $m_{a_{1}}$ via Eqs.\ (\ref{g1}) and
(\ref{h3}). As already discussed in Chapter \ref{sec.remarks}, the parameter
$c_{1}$ influences only the phenomenology of $\eta$ and $\eta^{\prime}$; these
two fields appear in two of the $a_{0}(1450)$ decay channels and for that
reason we first have to determine the value of $c_{1}$ before the value of
$h_{2}$ can be calculated. This is performed using the mass terms for $\eta$ and
$\eta^{\prime}$, Eqs.\ (\ref{fit25}) and (\ref{fit26}). We substitute $c_{1}$
by $\varphi_{\eta}$, Eq.\ (\ref{c1phi}) and use the parameter values from Table
\ref{Fit2-1} as well as the mass values from Eqs.\ (\ref{s}). A subsequent analysis
yields $m_{\eta}=523.20$ MeV, $m_{\eta^{\prime}}=957.78$ MeV and
consequently$\ \varphi_{\eta}=-43.9%
{{}^\circ}%
$. Therefore, as in Fit I, it is possible to exactly obtain the experimental
value of $m_{\eta^{\prime}}$, but not of $m_{\eta}$, due to the condition that
$\varphi_{\eta}<\mid45%
{{}^\circ}%
\mid$ which is necessary to ascertain $m_{\eta_{N}}<m_{\eta_{S}}$. (Enforcing
$m_{\eta}=m_{\eta}^{\exp}$ would require $\varphi_{\eta}>\mid45%
{{}^\circ}%
\mid$ and would spoil the result for $m_{\eta^{\prime}}$.) Then
Eq.\ (\ref{c1phi}) yields $c_{1}=0.00063$ MeV$^{-2}$.

Consequently, all parameters entering the formula for the full decay width of
$a_{0}(1450)$, see Eq.\ (\ref{Ga01450}), are known; we obtain $h_{2}=-0.736$
from the condition $\Gamma_{a_{0}(1450)}=265$ MeV. The value of $h_{2}$ is
considerably smaller than in Fit I that yielded $h_{2}=40.6$. The best values
of the two remaining parameter values ($m_{1}$ and $\delta_{S}$), obtained
from the (axial-)vector mass formulas in Eqs.\ (\ref{m_rho}) - (\ref{m_K_1}) and
mass values in Eqs.\ (\ref{s}), read $m_{1}=762$ MeV and $\delta_{S}=485^{2}$
MeV$^{2}$.

Table \ref{Fit2-4} shows the cumulated results for all parameters from Fit II.%

\begin{table}[h] \centering
\begin{tabular}
[c]{|c|c|c|c|}\hline
Parameter & Value & Parameter & Value\\\hline
$Z_{\pi}$ & $1.66$ & $g_{1}$, Eq.\ (\ref{g1}) & $6.35$\\\hline
$Z_{K}$ & $1.515$ & $g_{2}$, Eq.\ (\ref{g2Z}) & $3.07$\\\hline
$\lambda_{2}$ & $89.7$ & $h_{3}$, Eq.\ (\ref{h3}) & $2.56$\\\hline
$m_{0}^{2}+\lambda_{1}(\phi_{N}^{2}+\phi_{S}^{2})$ & $-1044148$ MeV$^{2}$ &
$h_{0N}$, Eq.\ (\ref{m_pi}) & $1.072\cdot10^{6}$ MeV$^{3}$\\\hline
$m_{1}$ & $762$ MeV & $h_{0S}$, Eq.\ (\ref{m_eta_S}) & $3.388\cdot10^{7}$
MeV$^{3}$\\\hline
$\delta_{S}$ & $485^{2}$ MeV$^{2}$ & $h_{1}$ & $0$\\\hline
$h_{2}$ & $-0.736$ & $\delta_{N}$ & $0$\\\hline
$c_{1}$ & $0.00063$ MeV$^{-2}$ & $g_{3,4,5,6}$ & $0$\\\hline
\end{tabular}%
\caption{Cumulated best values of parameters from Fit II.  \label{Fit2-4}}%
\end{table}%

Table \ref{Fit2-5} shows the cumulated results for all observables from Fit II. We observe that all mass values stemming from Fit II are within $3\%$ of their
respective experimental values, with three exceptions:$\ m_{\eta}$,
$m_{K_{0}^{\star}(1430)}$ and $m_{K_{1}}$.

\begin{table}[h] \centering
\begin{tabular}
[c]{|c|c|c|}\hline
Observable & Our Value [MeV] & Experimental Value [MeV]\\\hline
$m_{\pi}$ & $138.65$ & $139.57$\\\hline
$m_{K}$ & $497.96$ & $493.68$\\\hline
$m_{a_{0}(1450)}$ & $1452$ & $1474$\\\hline
$m_{K_{0}^{\star}(1430)}$ & $1550$ & $1425$\\\hline
$m_{\eta}$ & $523.20$ & $547.85$\\\hline
$m_{\eta^{\prime}}$ & $957.78$ & $957.78$\\\hline
$m_{\rho}$ & $775.49$ & $775.49$\\\hline
$m_{a_{1}}$ & $1219$ & $1230$\\\hline
$\text{ }m_{K^{\star}}$ & $916.52$ & $891.66$\\\hline
$m_{\omega_{S}}\text{ }$ & $1036.90$ & $1019.46$\\\hline
$m_{K_{1}}$ & $1343$ & $1272$\\\hline
$m_{f_{1S}}$ & $1457.0$ & $1426.4$\\\hline
$\Gamma_{a_{1}\rightarrow\pi\gamma}$ & $0.622$ & $0.640$\\\hline
$\Gamma_{a_{0}(1450)}$ & $265$ & $265$\\\hline
\end{tabular}%
\caption
{Cumulated values of observables from Fit II (experimental uncertainties omitted).  \label
{Fit2-5}}%
\end{table}%
$\,$\\
We have already noted that the value of $m_{\eta}$ reproduced by our fit
cannot correspond exactly to the experimental value $m_{\eta}^{\exp}=$
$547.85$ MeV as this would require $\varphi_{\eta}>\mid45%
{{}^\circ}%
\mid$ and thus also $m_{\eta_{N}}>m_{\eta_{S}}$. The values of $m_{\eta}$ and
$m_{\eta^{\prime}}$ present in Table \ref{Fit2-5} imply $\varphi_{\eta}=-43.9%
{{}^\circ}%
$; it is therefore possible to (marginally) decrease $\varphi_{\eta}$ to $-45%
{{}^\circ}%
$ and obtain a slightly larger value of $m_{\eta}$ (but still smaller than
$m_{\eta}^{\exp}$). Then, however, the result for $m_{\eta^{\prime}}$ would be
spoiled. We will therefore work with the values of $m_{\eta}$ and
$m_{\eta^{\prime}}$ as stated in Table \ref{Fit2-5}.\\

We have also already noted that the value of $m_{K_{0}^{\star}(1430)}$ from Table
\ref{Fit2-5} is larger than the corresponding PDG value due to the pattern of
explicit symmetry breaking that in our model made $K_{0}^{\star}(1430)$
approximately 100 MeV ($\simeq$ strange-quark mass) heavier than its
non-strange counterpart, $a_{0}(1450)$. The $K_{0}^{\star}(1430)$ resonance
also possesses a decay width of approximately $270$ MeV and therefore the
stated deviation of $m_{K_{0}^{\star}(1430)}$ from experiment is acceptable.\\

We observe from Table \ref{Fit2-5} that the value of $m_{K_{1}}$ is
approximately $70$ MeV larger than $m_{K_{1}(1270)} $ = $1272$ MeV \cite{PDG}. It
is, however, approximately, $60$ MeV smaller than $m_{K_{1}(1400)}=1403$ MeV
\cite{PDG}. Both mentioned resonances are rather broad [$\Gamma_{K_{1}%
(1270)}=(90\pm20)$ MeV; $\Gamma_{K_{1}(1400)}=(174\pm13)$ MeV]. Therefore, the
field $K_{1}$ from our model can in principle be assigned to either of them.
However, a more plausible explanation is that our $K_{1}$ field is a mixture
of the two physical fields $K_{1}(1270)$ and $K_{1}(1400)$ -- or, in other
words, that the physical resonances $K_{1}(1270)$ and $K_{1}(1400)$ are
mixtures of the field $K_{1}$ from our model and an additional field currently
not present in our model. We discuss this possibility in Sec.\ \ref{2K1}. \\

Finally, let us also note that Fit II yields a large value of $m_{1}=762$ MeV,
just as Fit I. This implies that non-quark contributions are expected to play
a strong role in the mass generation of the $\rho$ meson. However, Fit II also
yields the decay width $\Gamma_{a_{1}\rightarrow\pi\gamma}$ within the
experimental boundaries (unlike Fit I) and we observe additionally that the
correspondence of our mass values to experiment is in Fit II decisively better
than in Fit I (see Tables \ref{Fit1-5} and \ref{Fit2-5}). We can thus conclude
that the results obtained until now give Fit II precedence over Fit I.

\section{Two \boldmath $K_{1}$ Fields} \label{2K1}

We have seen in the previous section that Fit II implies $m_{K_{1}}=1343$ MeV,
a value that is virtually the median of $m_{K_{1}(1270)}$ and $m_{K_{1}(1400)}$.
Thus our previous assignment of the $K_{1}$ field from the model to the
$K_{1}(1270)$ resonance appears to be somewhat in doubt as $m_{K_{1}}$
deviates almost equally from both $m_{K_{1}(1270)}$ and $m_{K_{1}(1400)}$. In
this section we propose an explanation for the value of $m_{K_{1}}$ obtained
from Fit II \cite{Goldman1998,Godfrey,Lipkin,Cheng:2011,Cheng:2003,Close:1997}.

Let us postulate the existence of the following two nonets, labelled
$A_{1}^{\mu}$ and $B_{1}^{\mu}$:%

\begin{equation}
A_{1}^{\mu}=\frac{1}{\sqrt{2}}\left(
\begin{array}
[c]{ccc}%
\frac{f_{1N,A}+a_{1}^{0}}{\sqrt{2}} & a_{1}^{+} & K_{1,A}^{+}\\
a_{1}^{-} & \frac{f_{1N,A}-a_{1}^{0}}{\sqrt{2}} & K_{1,A}^{0}\\
K_{1,A}^{-} & {\bar{K}}_{1,A}^{0} & f_{1S,A}%
\end{array}
\right)  ^{\mu}\text{{\normalsize ,}}\;B_{1}^{\mu}=\frac{1}{\sqrt{2}}\left(
\begin{array}
[c]{ccc}%
\frac{f_{1N,B}+b_{1}^{0}}{\sqrt{2}} & b_{1}^{+} & K_{1,B}^{+}\\
b_{1}^{-} & \frac{f_{1N,B}-b_{1}^{0}}{\sqrt{2}} & K_{1,B}^{0}\\
K_{1,B}^{-} & {\bar{K}}_{1,B}^{0} & f_{1S,B}%
\end{array}
\right)  ^{\mu}\text{.} \label{A1B1}%
\end{equation}

Let us assign the field $a_{1}^{\mu}$ from $A_{1}^{\mu}$ to the $a_{1}(1260)$
resonance and the field $b_{1}^{\mu}$ from $B_{1}^{\mu}$ to the $b_{1}(1235)$
resonance. The $a_{1}(1260)$ meson is a $I(J^{PC})=1(1^{++})$ state whereas
$b_{1}(1235)$ is a $I(J^{PC})=1(1^{+-})$ state. Thus the resonances possess
different charge conjugation $C$.

Due to $P=(-1)^{L+1}$, where $P$ denotes parity and $L$ the orbital angular
momentum, both resonances exhibit $L=1$; however, the difference in $C$
implies $S=1$ for $a_{1}(1260)$ and $S=0$ for $b_{1}(1235)$, with
$C=(-1)^{L+S}$ and $S$ denoting the spin. In the spectroscopic
$^{2S+1}L_{J}$ notation ($J$: total angular momentum), our states are thus
$P$-wave states: $a_{1}^{\mu}$ is a $^{3}P_{1}$ state while $b_{1}^{\mu}$
represents a $^{1}P_{1}$ state. Consequently, all states present in the nonet
$A_{1}^{\mu}$ are $^{3}P_{1}$ states and all states present in the nonet
$B_{1}^{\mu}$ are $^{1}P_{1}$ states. Thus the nonet $A_{1}^{\mu}$ contains
axial-vectors while the nonet $B_{1}^{\mu}$ contains pseudovectors. We then
assign the fields in the two nonets as follows: $f_{1N,A}\equiv f_{1}(1285)$,
$f_{1S,A}\equiv f_{1}(1420)$, $f_{1N,B}\equiv h_{1}(1170)$, $f_{1S,B}\equiv
h_{1}(1380)$. Let us not assign $K_{1,A}$ and $K_{1,B}$ for the moment.\\

It is possible to bring about the mixing of the two nonets using the
explicit symmetry breaking in the axial-vector channel, modelled by the
$\Delta$ matrix in Eq.\ (\ref{Delta}). Indeed a calculation of the following
term containing the commutator of $A_{1}^{\mu}$ and $B_{1}^{\mu}$%

\begin{equation}
\mathrm{Tr}(\Delta [ A_{1\mu},B_{1}^{\mu} ] ) \label{K11}%
\end{equation}

yields
\begin{equation}
\mathrm{Tr}(\Delta [ A_{1\mu},B_{1}^{\mu} ] )=\frac{1}{2}(\delta_{S}%
-\delta_{N})({\bar{K}}_{1\mu,A}^{0}K_{1,B}^{\mu0}-{\bar{K}}_{1\mu,B}%
^{0}K_{1,A}^{\mu0}+K_{1\mu,A}^{-}K_{1,B}^{\mu+}-K_{1\mu,A}^{+}K_{1,B}^{\mu
-})\text{.}%
\end{equation}

Note that the commutator $[A_{1},B_{1}]$ is $CP$ invariant: $P$ invariance is
trivially fulfilled due to $A_{1}\overset{P}{\rightarrow}A_{1}$,
$B_{1}\overset{P}{\rightarrow}B_{1}$ while the $C$ transformation
($A_{1}\overset{C}{\rightarrow}A_{1}^{\text{t}}$, $B_{1}\overset
{C}{\rightarrow}-B_{1}^{\text{t}}$) yields $\mathrm{Tr}(\Delta\lbrack
A_{1\mu},B_{1}^{\mu}])\overset{C}{\rightarrow}\mathrm{Tr}(\Delta(B_{1}^{\mu \text{t}%
}A_{1\mu}^{\text{t}}-A_{1\mu}^{\text{t}}B_{1}^{\mu\text{t}}))=\mathrm{Tr}%
(\Delta(A_{1\mu}B_{1}^\mu-B_{1}^\mu A_{1\mu})^{\text{t}})=\mathrm{Tr}(\Delta\lbrack
A_{1\mu},B_{1}^{\mu}])$.\\

Therefore, the non-vanishing difference of quark masses $m_{s}^{2}-m_{u}%
^{2}\sim\delta_{S}-\delta_{N}$ induces mixing of the axial-vector nonet
$A_{1}^{\mu}$ with the pseudovector nonet $B_{1}^{\mu}$. The term (\ref{K11})
yields mixing of the $K_{1}$ states; in other words, the $K_{1}$ fields from
the two nonets mix due to explicit breaking of the chiral symmetry.
Consequently, we assert that the physical fields $K_{1}(1270)$ and
$K_{1}(1400)$ arise from the mixing of $K_{1,A}$ and $K_{1,B}$. The $K_{1}$
state from our Lagrangian (\ref{Lagrangian}) then corresponds to $K_{1,A}$
whose $^{1}P_{1}$ counterpart is not present in the model. For this reason, it
is not surprising that our model yields $m_{K_{1}}$ different from masses of
both $K_{1}(1270)$ and $K_{1}(1400)$, see Table \ref{Fit2-5}.

Therefore, an extension of our model by a nonet of $^{1}P_{1}$ states may be a
useful tool to further study kaon phenomenology. Such spin-orbit mixing has
been considered, e.g., in Ref.\ \cite{Goldman1998}\ (see also
Ref.\ \cite{Godfrey})\ where, within a non-relativistic constituent quark
model, it was found that two mixing scenarios of the $K_{1,A}$ and $K_{1,B}$
states are possible: (\textit{i}) $K_{1,A}$-$K_{1,B}$ mixing angle
$\varphi_{K_{1}}\simeq37%
{{}^\circ}%
$, $m_{K_{1,A}}=1322$ MeV and $m_{K_{1,B}}=1356$ MeV; (\textit{ii})
$\varphi_{K_{1}}=45%
{{}^\circ}%
$, $m_{K_{1,A}}=m_{K_{1,B}}=1339$ MeV. As shown in Ref.\ \cite{Goldman1998},
possibility (\textit{ii}) would imply $m_{a_{1}}=m_{b_{1}}=1211$ MeV, slightly
at odds with experimental data citing $m_{b_{1}}=(1229.5\pm3.2)$ MeV
\cite{PDG} whereas possibility (\textit{i}) yields $m_{a_{1}}=1191$ MeV and
$m_{b_{1}}=1231$ MeV [and also $m_{K_{1}(1270)}=1273$ MeV, $m_{K_{1}%
(1400)}=1402$ MeV] and thus\ a better correspondence with experiment. Our
model is of course different from that of Ref.\ \cite{Goldman1998}; however, the
qualitative consistency of our (independently obtained) value $m_{K_{1}}=1343$
MeV with the results of Ref.\ \cite{Goldman1998} seems to confirm the notion that
$K_{1}(1270)$ and $K_{1}(1400)$ indeed arise from the mixing of $^{1}P_{1}$ and
$^{3}P_{1}$ nonets.

Note that the inclusion and further study of the term (\ref{K11})\ in our
model would make the mixing of $K_{1,A}$ and $K_{1,B}$ an intrinsic property
of the model; however, there are also alternative mixing mechanisms, not based
on an analysis of mass eigenstates, such as mixing via decay channels as
suggested in Ref.\ \cite{Lipkin}. Additionally, a calculation of
$\varphi_{K_{1}}$ from a QCD-like theory in Ref.\ \cite{Cheng:2011} found
$\varphi_{K_{1}}\simeq35%
{{}^\circ}%
$ to be preferred; for other analyses of $\varphi_{K_{1}}$, see
Ref.\ \cite{Cheng:2003}. It is possible to study mixing of other states from
the two nonets as well, such as $f_{1N,A}$-$f_{1S,A}$ and $f_{1N,B}$%
-$f_{1S,A}$ mixings in Refs.\ \cite{Cheng:2011,Close:1997}.

We will discuss the broader $K_{1}$ phenomenology (decay widths) further on,
in Sec.\ \ref{sec.K12}.

\chapter{Implications of Fit II} \label{ImplicationsFitII}

We now turn to the discussion of meson phenomenology that follows from Fit II. As
in Fit I, we will devote particular attention to hadronic decay widths of
scalar and axial-vector resonances as a matter of comparing results between
Fits I and II but also because these resonances possess the most ambiguities
regarding their structure and decay widths.

\section{Phenomenology in the \boldmath $I(J^{PC})=0(0^{++})$ Channel}

\begin{figure}
[h]
\begin{center}
\includegraphics[
height=2.3582in,
width=3.7666in
]%
{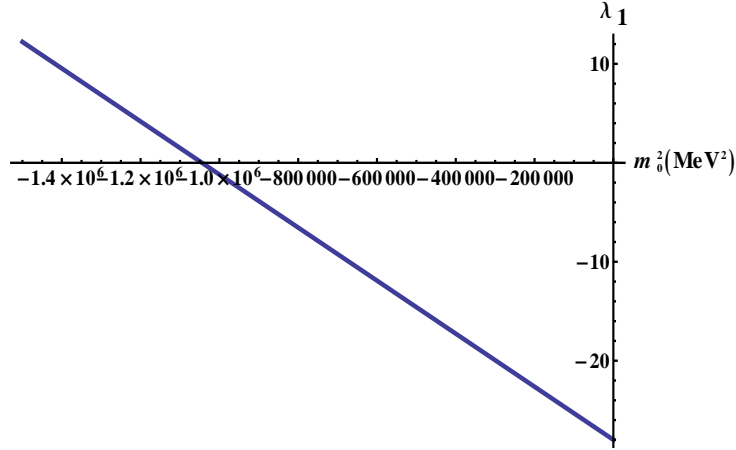}%
\caption{Dependence of parameter $\lambda_{1}$ on $m_{0}^{2}$ from Fit II. The
condition (\ref{l12}), i.e., $\lambda_{1}>-\lambda_{2}/2$, is
fulfilled for all values of $m_{0}^{2}<0$, see Table \ref{Fit2-4}.}%
\label{lambda2}%
\end{center}
\end{figure}

In this section we discuss results regarding the masses and decay widths of
the two scalar states $\sigma_{1}$ and $\sigma_{2}$. These states arise from
mixing of the two pure states $\sigma_{N}$ and $\sigma_{S}$ present in
Lagrangian (\ref{Lagrangian}). The mixing is described at the beginning of
Sec.\ \ref{sec.scalarmasses1}, see Eqs.\ (\ref{sigma-sigma}) - (\ref{phisigma1}%
). We note again that the mass terms $m_{\sigma_{N}}$ and $m_{\sigma_{S}}$
depend on $m_{0}^{2}+3\lambda_{1}\phi_{N}^{2}+\lambda_{1}\phi_{S}^{2}$ and
$m_{0}^{2}+\lambda_{1}\phi_{N}^{2}+3\lambda_{1}\phi_{S}^{2}$, respectively,
and thus cannot be calculated from the knowledge of the linear combination
$m_{0}^{2}+\lambda_{1}(\phi_{N}^{2}+\phi_{S}^{2})$ in Table \ref{Fit2-4}.
Therefore, as in Fit I, we express the parameter $\lambda_{1}$ in terms of the
mass parameter $m_{0}^{2}$ using the mentioned linear combination.
Additionally, the necessary condition for the spontaneous breaking of the
chiral symmetry suggests $m_{0}^{2}<0$ [see inequality (\ref{m02})].\ We note
at this point that, due to the latter condition, the parameter $\lambda_{1}$
obtained from Fit II fulfills the constraint (\ref{l12}), as apparent from
Fig.\ \ref{lambda2}.%

Now we can turn to the calculation of $m_{\sigma_{1,2}}$ and $\sigma_{1,2}$ decay widths.

\subsection{Scalar Isosinglet Masses} \label{sec.scalarmasses2}

\begin{figure}[h]
  \begin{center}
    \begin{tabular}{cc}
      \resizebox{98mm}{!}{\includegraphics{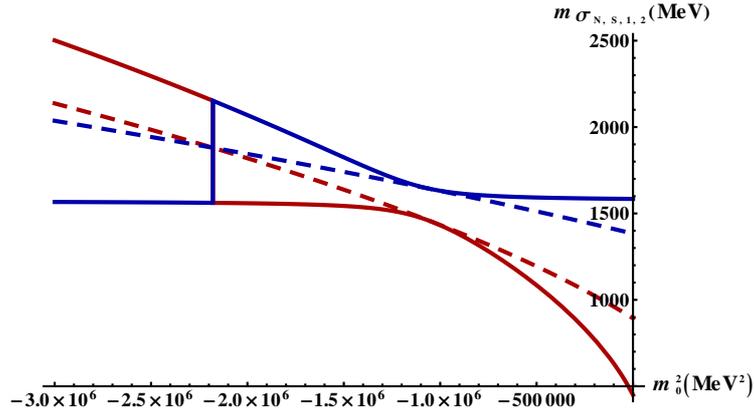}}  
    \end{tabular}
    \caption{Dependence of $m_{\sigma_{1}}$ (full lower curve), $m_{\sigma_{2}}$
(full upper curve), $m_{\sigma_{N}}$ (dashed lower curve) and $m_{\sigma_{S}}$
(dashed upper curve) on $m_{0}^{2}$ under the condition $m_{0}^{2}<0$.}
    \label{Sigmamassen2}
  \end{center}
\end{figure}

As described at the beginning of this section, we substitute $\lambda_{1}$ in
Eqs.\ (\ref{m_sigma_1}) and (\ref{m_sigma_2}) by $m_{0}^{2}$\ [from the linear
combination $m_{0}^{2}+\lambda_{1}(\phi_{N}^{2}+\phi_{S}^{2})$ in Table
\ref{Fit2-4}]. The ensuing dependence of $m_{\sigma_{1}}$ and $m_{\sigma_{2}}%
$\ on $m_{0}^{2}$ is depicted in Fig.\ \ref{Sigmamassen2}, with $m_{0}^{2}%
\leq0$ in accordance with Eq.\ (\ref{m02}).

\begin{figure}
[h]
\begin{center}
\includegraphics[
height=2.2582in,
width=3.7666in
]%
{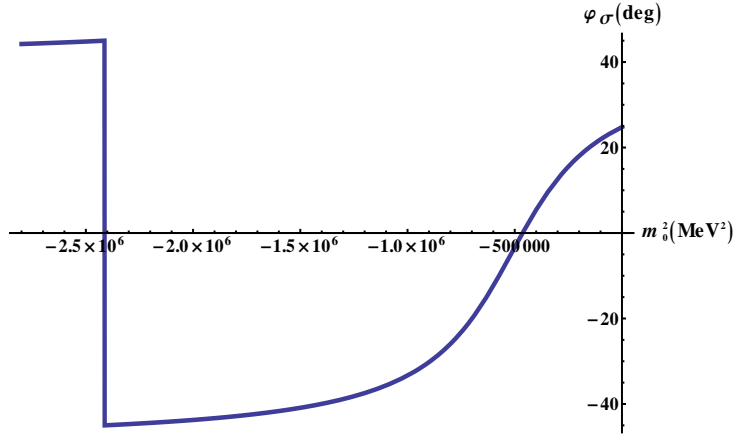}%
\caption{Dependence of the $\sigma_{N}$-$\sigma_{S}$ mixing angle
$\varphi_{\sigma}$ on $m_{0}^{2}$, Eq.\ (\ref{phisigma1}).}%
\label{phi2}%
\end{center}
\end{figure}

As in Fit I, $m_{\sigma_{1}}$ and $m_{\sigma_{2}}$ vary over wide intervals.
We note from Fig.\ \ref{Sigmamassen2} that $m_{\sigma_{N}}$ becomes larger
than $m_{\sigma_{S}}$ at $m_{0}^{2}\simeq-2.179\cdot10^{6}$ MeV$^{2}$ at which
point there is a jump of $\varphi_{\sigma}$ from $-45%
{{}^\circ}%
$ to $45%
{{}^\circ}%
$ (see Fig.\ \ref{phi2}).%

Therefore, $\sigma_{1}$ and $\sigma_{2}$ interchange places for $m_{0}%
^{2}\simeq -2.179\cdot10^{6}$ MeV$^{2}$; we use this value of $m_{0}^{2}$ as an
upper boundary for this parameter. Thus, together with Eq.\ (\ref{m02}), we obtain%

\begin{equation}
-2.179 \cdot 10^{6}\text{ MeV}^{2}\leq m_{0}^{2}\leq0\text{.} \label{m02b2}%
\end{equation}
From the previous inequality we obtain the following boundaries for
$m_{\sigma_{1,2}}$:%
\begin{align}
450\text{ MeV}  &  \leq m_{\sigma_{1}}\leq1561\text{ MeV}\text{,} \label{ms12}\\
1584\text{ MeV}  &  \leq m_{\sigma_{2}}\leq2152\text{ MeV.} \label{ms22}%
\end{align}
The inequalities (\ref{ms12}) and (\ref{ms22}) suggest that the mixed state
$\sigma_{1}$ may correspond to $f_{0}(600)$, $f_{0}(980)$, $f_{0}(1370)$ or
$f_{0}(1500)$ whereas the only confirmed resonance within the range of
$m_{\sigma_{2}}$ is $f_{0}(1710)$. [As in Fit I, we do not consider the states
$f_{0}(1790)$, $f_{0}(2020)$, $f_{0}(2100)$ and $f_{0}(2200)$.] A definitive
assignment of $\sigma_{1}$, and a confirmation whether $\sigma_{2}$
corresponds to $f_{0}(1710)$, require a more detailed analysis of
phenomenology in the scalar channel, performed in the following sections.

\begin{figure}[h]
  \begin{center}
    \begin{tabular}{cc}
      \resizebox{78mm}{!}{\includegraphics{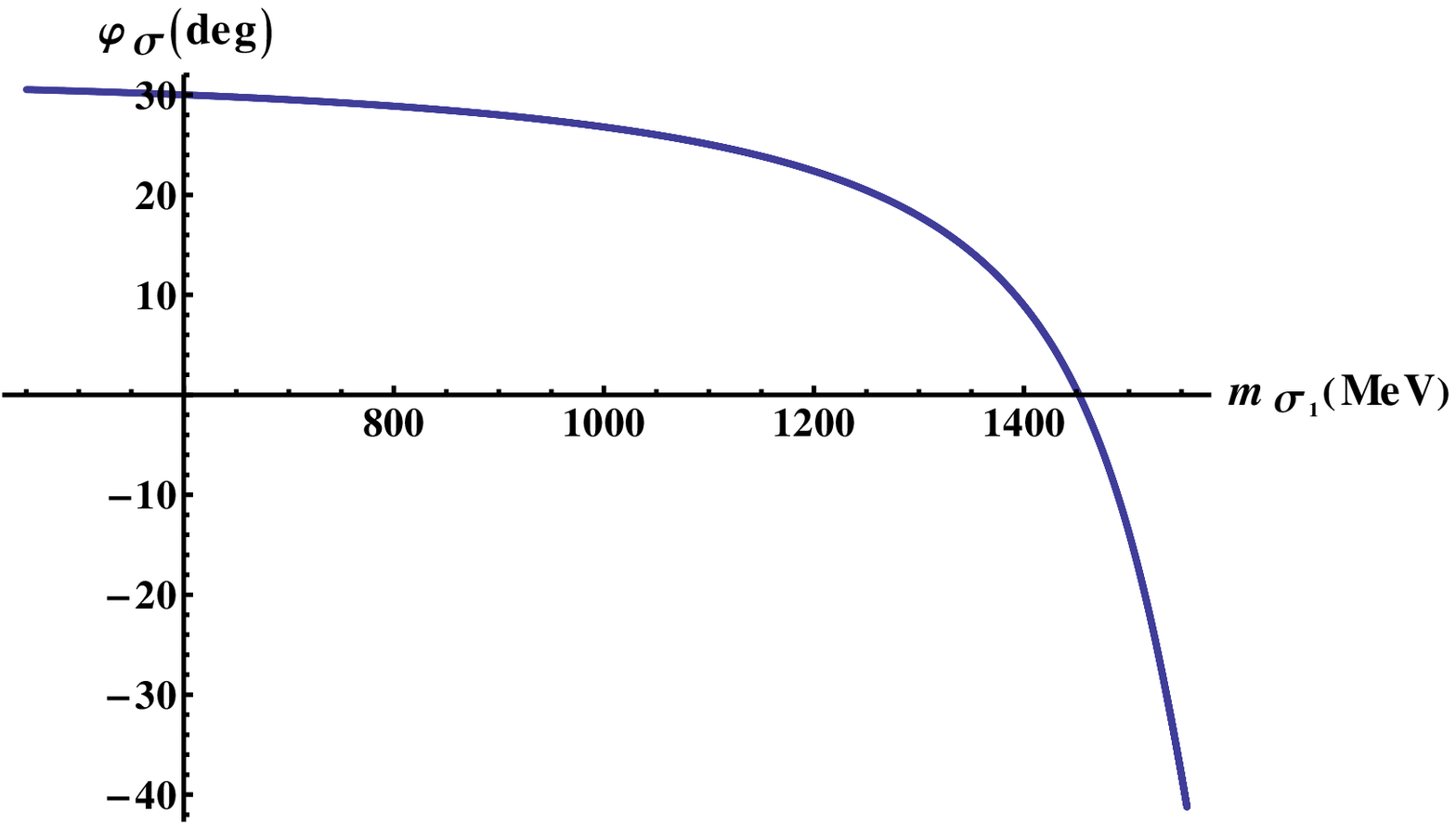}} &
      \resizebox{78mm}{!}{\includegraphics{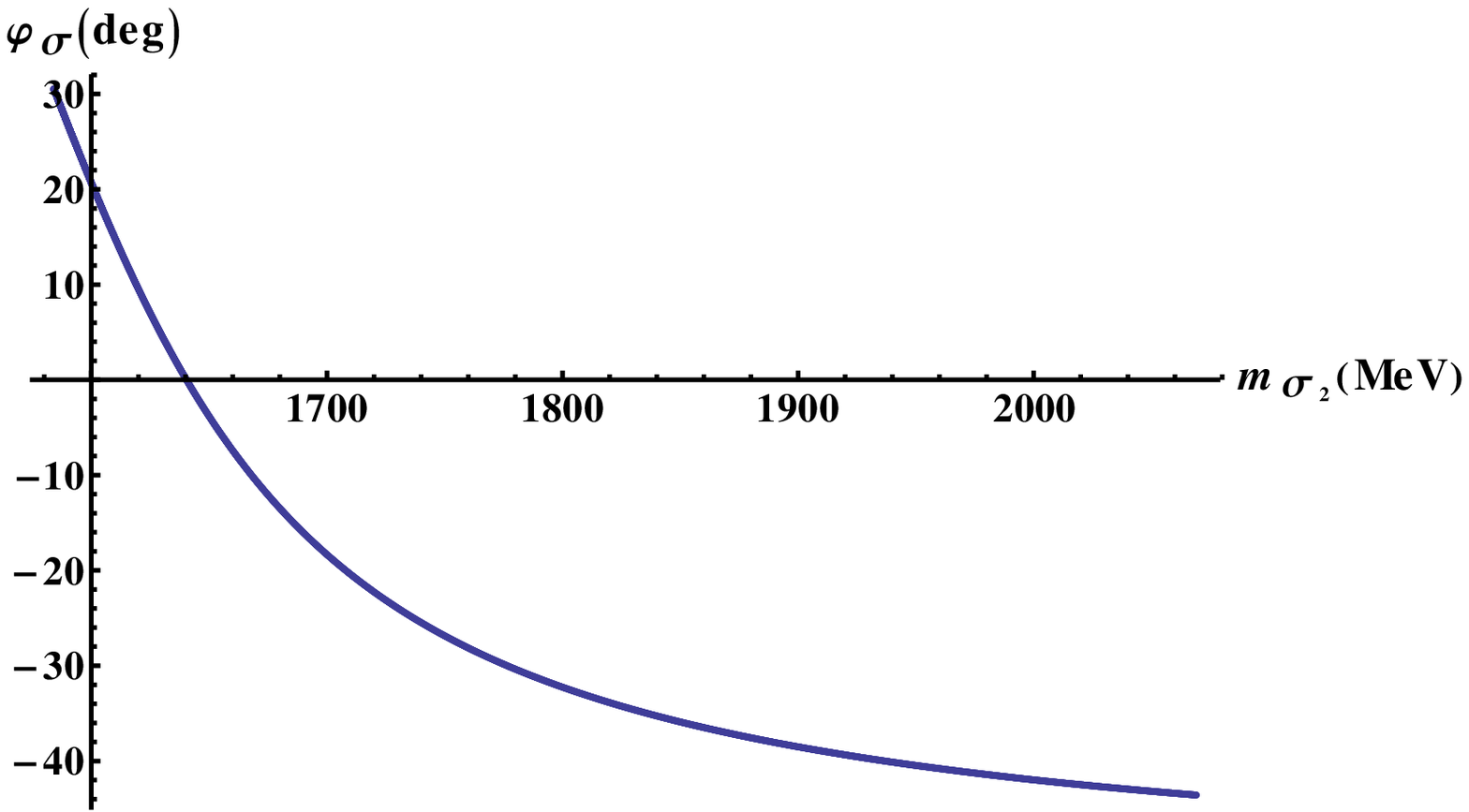}} 
    \end{tabular}
    \caption{The $\sigma_{N}$-$\sigma_{S}$ mixing angle $\varphi_{\sigma}$ as
function of $m_{\sigma_{1,2}}$.}
    \label{phi21}
  \end{center}
\end{figure}

Nonetheless, from the variation of the $\sigma_{N}$ - $\sigma_{S}$ mixing
angle $\varphi_{\sigma}$ we can conclude that $\sigma_{1}$ is predominantly a
$\bar{n}n$ state and the $\sigma_{2}$ field is predominantly composed of
strange quarks, see Fig.\ \ref{phi21}. Note that, as in Fit I, we obtain the
two diagrams in Fig.\ \ref{phi21} from two implicit plots: of $\varphi
_{\sigma}(\lambda_{1})$, Eq.\ (\ref{phisigma1}), and $m_{\sigma_{1,2}}%
[\varphi_{\sigma}(\lambda_{1})]$, Eqs.\ (\ref{m_sigma_1}) and (\ref{m_sigma_2}%
), with $m_{0}^{2}+\lambda_{1}(\phi_{N}^{2}+\phi_{S}^{2})=-1044148$ MeV$^{2}$
from Table \ref{Fit2-4}\ and $m_{0}^{2}$ from the inequality (\ref{m02b1}).%

\begin{figure}[h]
  \begin{center}
    \begin{tabular}{cc}
      \resizebox{78mm}{!}{\includegraphics{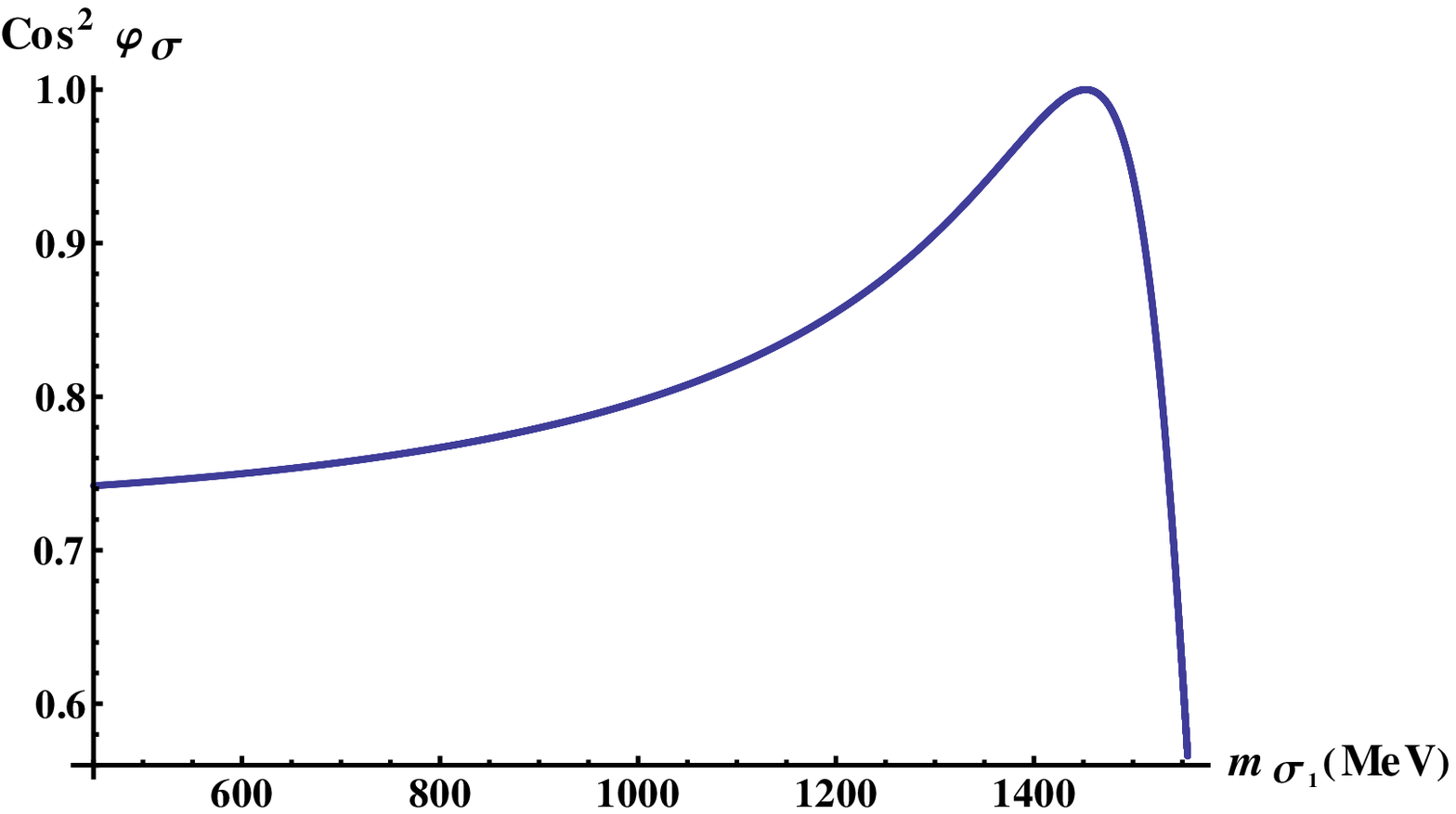}} &
      \resizebox{78mm}{!}{\includegraphics{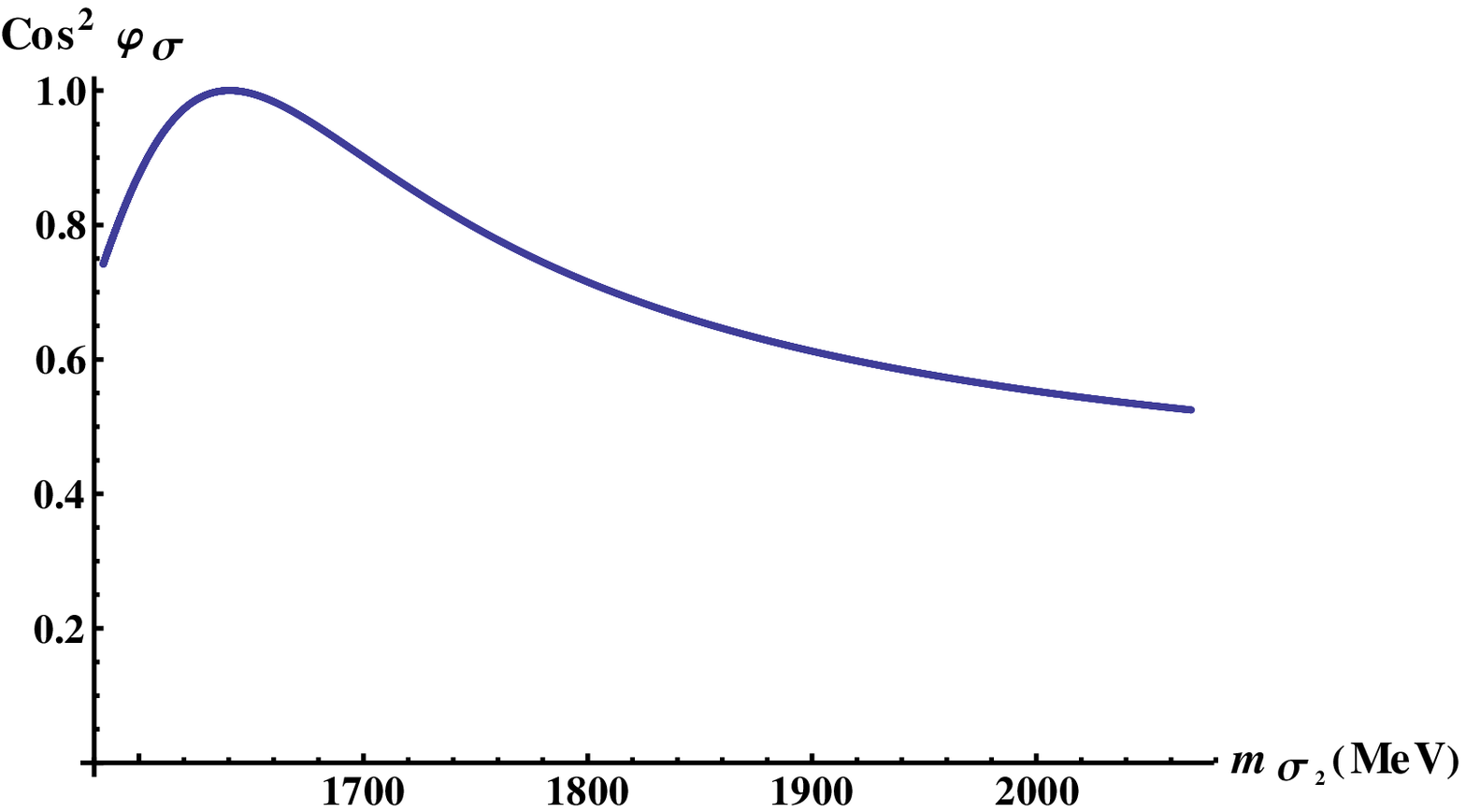}} 
    \end{tabular}
    \caption{Contribution of the pure non-strange field $\sigma_{N}$ to
$\sigma_{1}$ (left panel) and of the pure strange field $\sigma_{S}$ to
$\sigma_{2}$ (right panel), respectively in dependence on $m_{\sigma_{1}}$ and
$m_{\sigma_{2}}$.}
    \label{phi22}
  \end{center}
\end{figure}

Contribution of $m_{\sigma_{N}}$ to $m_{\sigma_{1}}$ and contribution of
$m_{\sigma_{S}}$ to $m_{\sigma_{2}}$ are illustrated in Fig.\ \ref{phi22}.%

Before we continue, let us make an important point: we observe from
Fig.\ \ref{Sigmamassen2} that $m_{\sigma_{1}}$ and $m_{\sigma_{2}}$ are not
independent. Thus, in the following, any determination of either of these
masses (e.g., from a decay width) fixes the other mass to a certain value (and
also determines values of all decay widths depending on this mass). This is true because the two masses are connected via the mass parameter $m_{0}^{2}$ (as
also apparent from Fig.\ \ref{Sigmamassen2}). We will be
making use of this feature in the following sections. 

\subsection{Decay Width \boldmath $\sigma_{1,2}\rightarrow\pi\pi$} \label{sec.sigmapionpion2}

\begin{figure}[h]
  \begin{center}
    \begin{tabular}{cc}
      \resizebox{78mm}{!}{\includegraphics{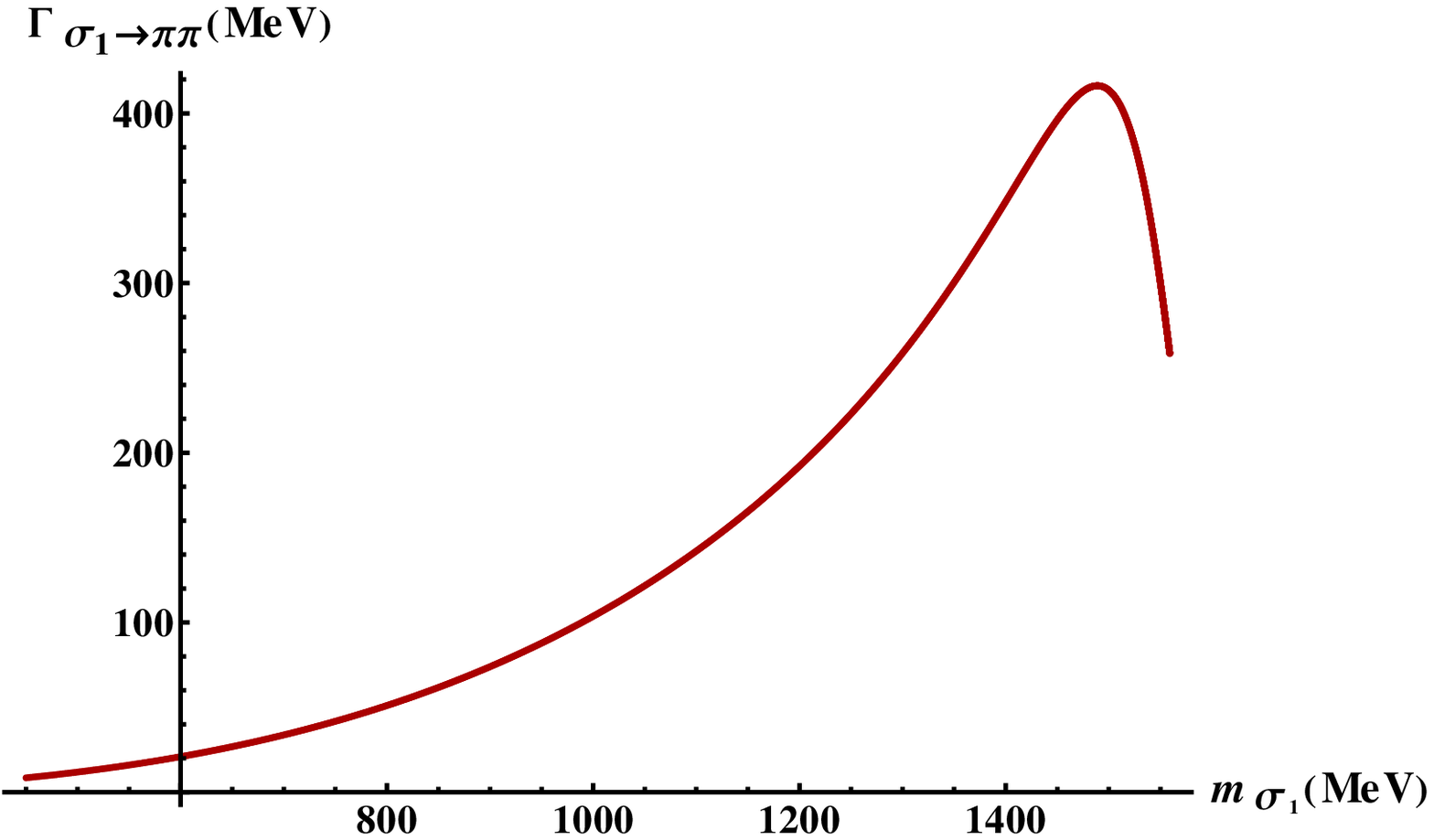}} &
      \resizebox{78mm}{!}{\includegraphics{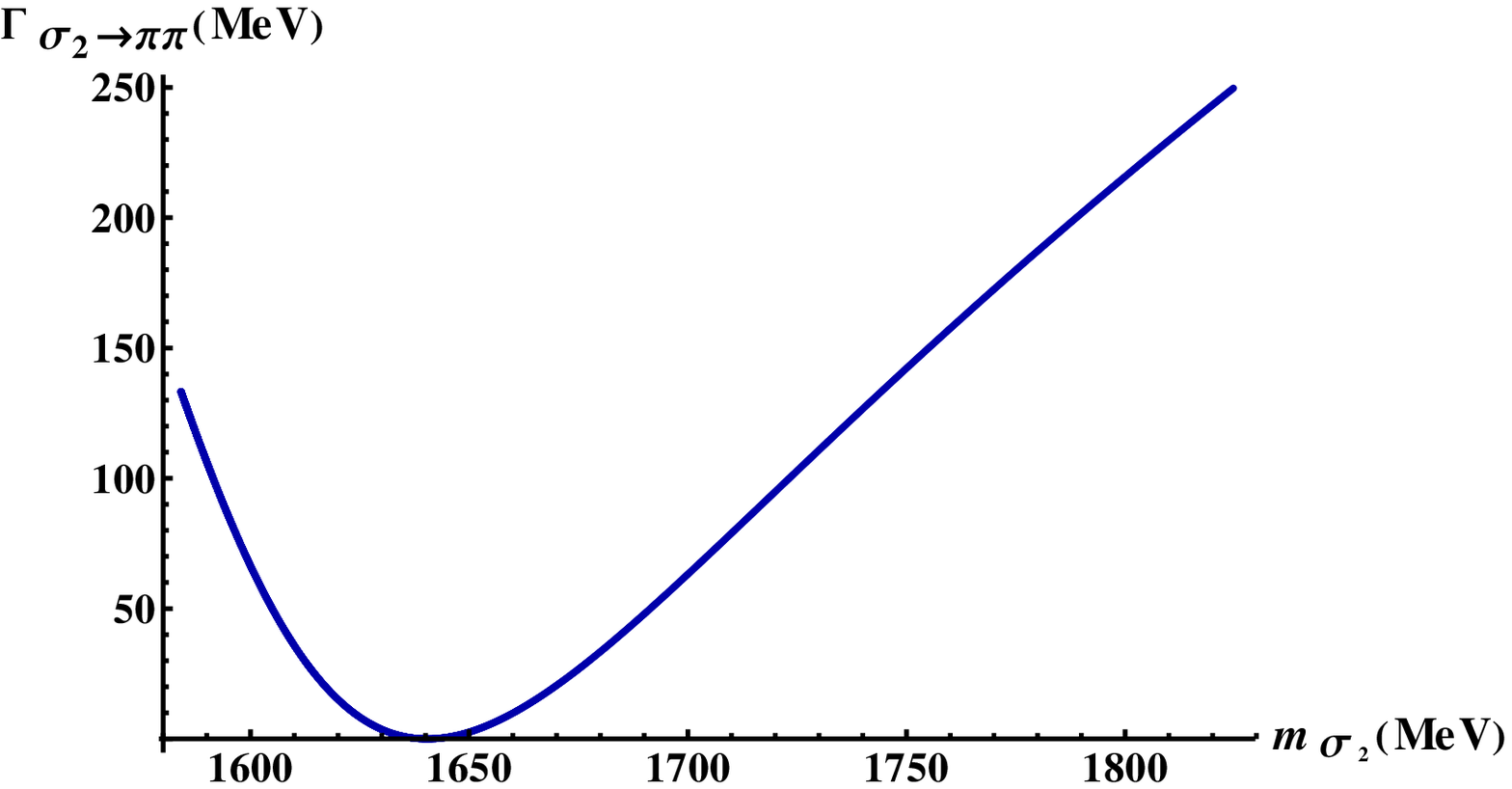}} 
    \end{tabular}
    \caption{$\Gamma_{\sigma_{1}\rightarrow\pi\pi}$ and $\Gamma_{\sigma
_{2}\rightarrow\pi\pi}$ as functions of $m_{\sigma_{1}}$ and $m_{\sigma_{2}}$,
respectively.}
    \label{Spp2}
  \end{center}
\end{figure}

In Sec.\ \ref{sec.sigmapionpion1} we have already performed the calculation of
the decay widths $\Gamma_{\sigma_{1}\rightarrow\pi\pi}$, Eq.\ (\ref{Gs1pp}),
and $\Gamma_{\sigma_{2}\rightarrow\pi\pi}$, Eq.\ (\ref{Gs2pp}), from the
$\sigma\pi\pi$ interaction Lagrangian (\ref{sigmapionpion}). We can therefore
immediately plot the two decay widths, see Fig.\ \ref{Spp2}.%

From the left panel of Fig.\ \ref{Spp2} we can conclude that the state
$\sigma_{1}$ appears to possess the best correspondence with the $f_{0}(1370)$
resonance. Clearly, $\Gamma_{\sigma_{1}\rightarrow\pi\pi}$ is too small in the
mass region of $f_{0}(600)$, i.e., $m_{\sigma_{2}}\lesssim800$ MeV. Therefore,
an assignment of $\sigma_{1}$\ to $f_{0}(600)$ based on the $2\pi$ decay
channel is not possible. Additionally, $\sigma_{1}$ cannot correspond to the
$f_{0}(1500)$ resonance either: we obtain $\Gamma_{\sigma_{1}\rightarrow\pi
\pi}\simeq400$ MeV at $m_{\sigma_{1}}\simeq1500$ MeV, in stark contrast to
experimental data \cite{PDG} reading $\Gamma_{f_{0}(1500)\rightarrow\pi\pi
}\simeq30$ MeV.

Let us now ascertain whether there is indeed a good correspondence of our
predominantly non-strange state $\sigma_{1}$ to $f_{0}(1370)$, and
additionally of $\sigma_{2}$ to $f_{0}(1710)$ as suggested by $m_{\sigma_{2}}%
$, see discussion of Fig.\ \ref{Sigmamassen2}. There are two strategies to
this end: we can first determine $m_{\sigma_{1}}$ necessary to describe
correctly $\Gamma_{f_{0}(1370)\rightarrow\pi\pi}$ from Ref.\ \cite{buggf0} (a
comprehensive fit of several data sets used here because the PDG data
\cite{PDG} are not conclusive), then calculate $m_{\sigma_{2}}$ and
$\Gamma_{\sigma_{2}\rightarrow\pi\pi}$ and compare these results with
$m_{f_{0}(1710)}$ and $\Gamma_{f_{0}(1710)\rightarrow\pi\pi}$. Alternatively,
we can first determine our result for $m_{\sigma_{2}}$ in such a way that
$\Gamma_{\sigma_{2}\rightarrow\pi\pi}$ describes $\Gamma_{f_{0}%
(1710)\rightarrow\pi\pi}$ correctly and then calculate $m_{\sigma_{1}}$ and
$\Gamma_{\sigma_{1}\rightarrow\pi\pi}$ and compare them with results for
$m_{f_{0}(1370)}$ and $\Gamma_{f_{0}(1370)\rightarrow\pi\pi}$ from
Ref.\ \cite{buggf0}.

\begin{itemize}
\item Reference \cite{buggf0} cites the value of $\Gamma_{f_{0}(1370)}=325$ MeV at
$m_{f_{0}(1370)}=(1309\pm1\pm15)$ MeV from an $f_{0}(1370)$\ Breit-Wigner fit
and we obtain $\Gamma_{\sigma_{1}\rightarrow\pi\pi}=325$ MeV at $m_{\sigma
_{1}}=1376$ MeV. Reference \cite{buggf0} also cites the value of $207$ MeV for the
full width at half maximum (FWHM) with the peak in the decay channel
$f_{0}(1370)\rightarrow\pi\pi$ at$\ m_{f_{0}(1370)}=1282$ MeV -- we obtain
$\Gamma_{\sigma_{1}\rightarrow\pi\pi}=207$ MeV at $m_{\sigma_{1}}=1225$ MeV.
Our results are thus qualitatively consistent with results from
Ref.\ \cite{buggf0}. As already noted, assigning a value to $m_{\sigma_{1}}$
implies also a certain value of $m_{\sigma_{2}}$. Consequently, $m_{\sigma
_{1}}=1376$ MeV leads to $m_{\sigma_{2}}=1616$ MeV and to $\Gamma_{\sigma
_{2}\rightarrow\pi\pi}=22.6$ MeV whereas $m_{\sigma_{1}}=1225$ MeV leads to
$m_{\sigma_{2}}=1599$ MeV and to $\Gamma_{\sigma_{2}\rightarrow\pi\pi}=71.2$
MeV (see the right panel of Fig.\ \ref{Spp2}). $\Gamma_{\sigma_{2}%
\rightarrow\pi\pi}=22.6$ MeV is within the PDG-preferred interval of
Eq.\ (\ref{f0(1710)_8}) reading $\Gamma_{f_{0}(1710)\rightarrow\pi\pi
}=29.28_{-7.69}^{+5.42}$ MeV; it is outside the BES II interval $\Gamma
_{f_{0}(1710)\rightarrow\pi\pi}<9.34$ MeV, Eq.\ (\ref{f0(1710)_16}) and also
above the WA102 range $\Gamma_{f_{0}(1710)\rightarrow\pi\pi}=(16.1\pm3.6)$
MeV, see Eq.\ (\ref{f0(1710)_20}). $\Gamma_{\sigma_{2}\rightarrow\pi\pi}=71.2$
MeV is outside all the mentioned intervals.

\item Enforcing $\Gamma_{\sigma_{2}\rightarrow\pi\pi}=29.28_{-7.69}^{+5.42}$
MeV $\equiv$ $\Gamma_{f_{0}(1710)\rightarrow\pi\pi}^{\text{PDG}}$ leads to two
sets of solutions for $m_{\sigma_{2}}$ due to the parabolic form of
$\Gamma_{\sigma_{2}\rightarrow\pi\pi}$, see Fig.\ \ref{Sigmamassen2}. We
obtain (\textit{i}) $m_{\sigma_{2}}=(1613\mp3)$ MeV and (\textit{ii})
$m_{\sigma_{2}}=1677_{-6}^{+4}$ MeV. Both sets of results are below
$m_{f_{0}(1710)}=(1720\pm6)$ MeV \cite{PDG}. From results (\textit{i}) we
obtain $m_{\sigma_{1}}=1360_{+19}^{-15}$ MeV and\ $\Gamma_{\sigma
_{1}\rightarrow\pi\pi}=309_{+19}^{-13}$ MeV. From results (\textit{ii}) we
obtain $m_{\sigma_{1}}=1497_{-5}^{+3}$ MeV and\ $\Gamma_{\sigma_{1}%
\rightarrow\pi\pi}=(415\mp1)$ MeV. The second set of results would imply a
dominant $2\pi$ decay of $f_{0}(1370)$\ at approximately $1.5$ GeV, at odds
with experimental data \cite{buggf0} and therefore we will not consider it.
The first set of results, however, can accommodate $\Gamma_{f_{0}(1370)}=325$
MeV, although the corresponding mass $m_{\sigma_{1}}=1360_{+19}^{-15}$ MeV is
slightly larger than the one cited in Ref.\ \cite{buggf0}. Additionally, the
first set of results describes $\Gamma_{f_{0}(1710)\rightarrow\pi\pi}$
correctly although the obtained mass interval $m_{\sigma_{2}}=(1613\mp3)$ MeV
is approximately $100$ MeV smaller than the PDG result $m_{f_{0}%
(1710)}=(1720\pm6)$ MeV.

Constraining $m_{\sigma_{2}}$ via $\Gamma_{f_{0}(1710)\rightarrow\pi\pi
}^{\text{BES II}}<9.34$ MeV, Eq.\ (\ref{f0(1710)_16}), yields $1624$ MeV $\leq
m_{\sigma_{2}}\leq1659$ MeV, $1411$ MeV $\leq m_{\sigma_{1}}\leq1480$ MeV and
$359$ MeV $\leq\Gamma_{\sigma_{1}\rightarrow\pi\pi}\leq$ $415$ MeV. These
results imply a slightly too large value of $m_{\sigma_{1}}$ where the $2\pi
$\ channel is expected to be dominant for $f_{0}(1370)$ -- Ref.\ \cite{buggf0}
suggests the mass of approximately $1300$ MeV, not $1400$ MeV, where
$f_{0}(1370)$ decays predominantly into $2\pi$ rather than $4\pi$.

We can also utilise $\Gamma_{f_{0}(1710)\rightarrow\pi\pi}^{\text{WA102}%
}=16.1\pm3.6$ MeV, Eq.\ (\ref{f0(1710)_20}), to constrain $m_{\sigma_{2}}$. We
obtain (\textit{i}) $m_{\sigma_{2}}=1619_{+3}^{-2}$ MeV and (\textit{ii})
$m_{\sigma_{2}}=(1666\pm3)$ MeV. From results (\textit{i}) we obtain
$m_{\sigma_{1}}=(1393\mp9)$ MeV and\ $\Gamma_{\sigma_{1}\rightarrow\pi\pi
}=(341\mp9)$ MeV. From results (\textit{ii}) we obtain $m_{\sigma_{1}%
}=(1487\pm3)$ MeV and\ $\Gamma_{\sigma_{1}\rightarrow\pi\pi}=416$ MeV. Results
(\textit{ii}) would suggest a large contribution of the $2\pi$\ channel to
$f_{0}(1370)$ at $\simeq1.49$ MeV and we therefore disregard them; results
(\textit{i}) are then more acceptable but still above the range of
$m_{\sigma_{1}}=1360_{+19}^{-15}$ MeV and\ $\Gamma_{\sigma_{1}\rightarrow
\pi\pi}=309_{+19}^{-13}$ MeV, obtained from $\Gamma_{f_{0}(1710)\rightarrow
\pi\pi}^{\text{PDG}}$. We thus prefer the latter result.
\end{itemize}

We conclude that results regarding the $2\pi$ decay channel allow for a
correct description of the $f_{0}(1370)$ and $f_{0}(1710)$\ decay widths,
although the mass values could be improved. The latter point emphasises the need
to include a glueball state into our model \cite{Stani} because, if it is
found at $\sim1.5$ GeV, this state should induce a level repulsion shifting
$m_{\sigma_{1}}$ downwards and $m_{\sigma_{2}}$ upwards -- i.e., both masses
being shifted in the directions favoured by the experiment.

The best results suggested by comparing $\Gamma_{\sigma_{1}\rightarrow\pi\pi}$
to $\Gamma_{f_{0}(1370)\rightarrow\pi\pi}$ and $\Gamma_{\sigma_{2}%
\rightarrow\pi\pi}$ to $\Gamma_{f_{0}(1710)\rightarrow\pi\pi}$ read
$m_{\sigma_{1}}=1360_{-17}^{+16}$ MeV,\ $\Gamma_{\sigma_{1}\rightarrow\pi\pi
}=309_{-15}^{+16}$ MeV, $m_{\sigma_{2}}=(1613\pm3)$ MeV and $\Gamma
_{\sigma_{2}\rightarrow\pi\pi}=(29.3\pm6.5)$ MeV. These results justify the
assignments $\sigma_{1}\equiv$ $f_{0}(1370)$ and $\sigma_{2}\equiv$
$f_{0}(1710)$; the assignments will also be confirmed in the subsequent
sections (see below). The results also suggest that $f_{0}(1370)$ is
$94.6_{+1.4}^{-1.0}$\% a $\bar{n}n$ state and, conversely, that
$f_{0}(1710)$ is $94.6_{+1.4}^{-1.0}$\% a $\bar{s}s$ state.\\

As apparent from Fig.\ \ref{Spp2}, $\Gamma_{\sigma_{2}\rightarrow\pi\pi}=0$
for $m_{\sigma_{2}}=1640$ MeV, corresponding to $m_{0}^{2}=-1044148$ MeV$^{2}$
and thus $m_{\sigma_{1}}=1452$ MeV\ (see Fig.\ \ref{Sigmamassen2}). As already
noted, the parameter $\lambda_{1}$ in our fit is determined only indirectly, from
the linear combination $m_{0}^{2}+\lambda_{1}(\phi_{N}^{2}+\phi_{S}%
^{2})=-1044148$ MeV$^{2}$ (see Table \ref{Fit2-4}). Therefore, $\lambda_{1}=0$
for $m_{0}^{2}=-1044148$ MeV$^{2}$; consequently, according to
Eq.\ (\ref{phisigma1}), one also obtains that the $\sigma_{N}$ - $\sigma_{S}$
mixing angle $\varphi_{\sigma}=0$. Thus $\sigma_{N}$ and $\sigma_{S}$
decouple. As in Fit I, $\Gamma_{\sigma_{2}\rightarrow\pi\pi}$ then vanishes
identically because $\lambda_{1}=0=h_{1}$ (and only these large-$N_{c}$
suppressed parameters could bring about $\Gamma_{\sigma_{2}\rightarrow\pi\pi
}\neq0$). Setting $h_{1}\neq0$ would not alter $\Gamma
_{\sigma_{2}\rightarrow\pi\pi}=0$ for a certain value of $m_{\sigma_{2}}$
because of the relative minus sign of the two terms in $\mathcal{M}%
_{\sigma_{2}\rightarrow\pi\pi}$, Eq.\ (\ref{Ms2pp}). The relative sign
difference still leads to a cancellation of the two terms in $\mathcal{M}%
_{\sigma_{2}\rightarrow\pi\pi}$ for a certain value of $\varphi_{\sigma}$.\\

Thus our Fit II prefers $f_{0}(1370)$ rather than $f_{0}(600)$ to be the
non-strange quarkonium, just as Scenario II of the $U(2)\times U(2)$ version
of our model. \\

\textit{A note on }$\sigma_{1,2}\rightarrow4\pi$\textit{ decays.} We
have also considered the sequential decay $\sigma_{1,2}\rightarrow\rho
\rho\rightarrow4\pi$ by integrating over the spectral functions of the two
intermediate $\rho$\ mesons, similarly to Sec.\ \ref{sec.sNppQII}. The following Lagrangian
obtained from Eq.\ (\ref{Lagrangian})\ has been utilised:

\begin{align}
\mathcal{L}_{\sigma\rho\rho} &  =\frac{1}{2}(h_{1}+h_{2}+h_{3})\phi_{N}%
\sigma_{N}\left[  (\rho_{\mu}^{0})^{2}+2\rho_{\mu}^{+}\rho_{\mu}^{-}\right]
+\frac{1}{2}h_{1}\phi_{S}\sigma_{S}\left[  (\rho_{\mu}^{0})^{2}+2\rho_{\mu
}^{+}\rho_{\mu}^{-}\right] \nonumber\\
&  =\frac{1}{2}\left[  (h_{1}+h_{2}+h_{3})\phi_{N}\cos\varphi_{\sigma}%
+h_{1}\phi_{S}\sin\varphi_{\sigma}\right]  \sigma_{1}\left[  (\rho_{\mu}%
^{0})^{2}+2\rho_{\mu}^{+}\rho_{\mu}^{-}\right] \nonumber\\
&  +\frac{1}{2}\left[  h_{1}\phi_{S}\cos\varphi_{\sigma}-(h_{1}+h_{2}%
+h_{3})\phi_{N}\sin\varphi_{\sigma}\right]  \sigma_{2}\left[  (\rho_{\mu}%
^{0})^{2}+2\rho_{\mu}^{+}\rho_{\mu}^{-}\right]  \label{srr}
\end{align}

with the substitutions $\sigma_{N}\rightarrow\cos\varphi_{\sigma}\sigma_{1}$ and
$\sigma_{S}\rightarrow\sin\varphi_{\sigma}\sigma_{1}$ that enable us to
calculate decay width of $f_{0}(1370)\equiv\sigma_{1}$ and the substitutions
$\sigma_{N}\rightarrow-\sin\varphi_{\sigma}\sigma_{2}$ and $\sigma
_{S}\rightarrow\cos\varphi_{\sigma}\sigma_{2}$ that enable us to calculate
decay width of $f_{0}(1710)\equiv\sigma_{2}$ [see Eq.\ (\ref{sigma-sigma_1})].
After the substitutions, the Lagrangian in Eq.\ (\ref{srr}) obtains an analogous
form as the one in Eq.\ (\ref{srrQ}). For this reason it is subsequently possible to
perform the mentioned integration over the $\rho$\ spectral functions. 

We\ then observe that results obtained from our $N_{f}=3$ fit are by at least
a factor of ten smaller than those obtained within the realm of Scenario II in
the $U(2)\times U(2)$ version of the model. The reason is the different value
of $h_{2}$: whereas in Scenario II of the two-flavour model this parameter had
the value $\simeq5$, our Fit I in the three-flavour model prefers the value of
$h_{2}\simeq0$ scaling the $4\pi$ decay width of the scalar states downwards.
We expect results in the $4\pi$ channel to improve considerably upon inclusion
of the scalar glueball field into the $U(3)\times U(3)$ version of our model
because we will see in Chapter \ref{chapterglueball} that the glueball-field coupling to the $4\pi$
channel is significantly stronger than the corresponding coupling of the
non-strange quarkonium. The ensuing mixture of the pure glueball and the pure
quarkonium should improve the decay width of the predominantly $\bar{n}n$ state in
the $4\pi$ channel.

\subsubsection{A Putative Assignment of \boldmath $\sigma_{1}$ to
\boldmath $f_{0}(980)$}

Let us briefly discuss our $\sigma_{1}$ state in terms of $f_{0}(980)$,
another resonance\ within the mass range of our $\sigma_{1}$ state.\ We note
that $\Gamma_{\sigma_{1}\rightarrow\pi\pi}=97$ MeV at $m_{\sigma_{1}}=980$ MeV
and that $94$ MeV $\leq\Gamma_{\sigma_{1}\rightarrow\pi\pi}\leq100$ MeV for
$970$ MeV $\leq m_{\sigma_{1}}\leq990$ MeV,\ with the latter mass interval
corresponding to the lower and upper boundaries of $m_{f_{0}(980)}$. Given
that the full decay width $\Gamma_{f_{0}(980)}=(40-100)$ MeV \cite{PDG}, there
would appear to be some parallels between our $\sigma_{1}$ state and the
$f_{0}(980)$ resonance. As noted in Sec.\ \ref{sec.f0(980)}, this resonance is
close to the kaon-kaon threshold; thus an experimental analysis is not always
straightforward with different collaborations and reviews obtaining at times
very different results
\cite{Amsler:1995bf,f0(980),Armstrong:1991,WA102:1999,WA102:1999_1,Barberis:1999,Anisovich-KMatrix,Amsler:1995gf,Bellazzini:1999,Janssen:1994,Bugg:1994,Tornqvist:1995,Anisovich:1997zw,Anisovich:2009zza,Etkin:1981,OBELIX:1997,Tikhomirov:2003}%
. We thus note that there is no universally accepted value of $\Gamma
_{f_{0}(980)}$ that ranges between $\sim14$ MeV \cite{Oller1998} ($T$-matrix
pole)\ and $(201\pm28)$ MeV \cite{Achasov2000}, with the latter result
model-dependent, broad due to inclusion of $KK$-threshold effects and not
considering possible interference with the high-mass tail of $f_{0}(600)$.
Additionally, even if the precise value of $\Gamma_{f_{0}(980)}$ were known,
the branching ratio $\Gamma_{f_{0}(980)\rightarrow\pi\pi}/\Gamma_{f_{0}(980)}$
remains ambiguous. \\

The $f_{0}(980)$ resonance can actually also decay
non-hadronically, into diphotons and dileptons; however, these decays are
known to be suppressed \cite{PDG} and therefore we can set $\Gamma
_{f_{0}(980)}=$ $\Gamma_{f_{0}(980)\rightarrow\pi\pi}+\Gamma_{f_{0}%
(980)\rightarrow KK}$ -- consequently, $\Gamma_{f_{0}(980)\rightarrow\pi\pi
}/\Gamma_{f_{0}(980)}\equiv\Gamma_{f_{0}(980)\rightarrow\pi\pi}/[\Gamma
_{f_{0}(980)\rightarrow\pi\pi}+\Gamma_{f_{0}(980)\rightarrow KK}]$. There are
not many publications discussing both $\Gamma_{f_{0}(980)}$ and $\Gamma
_{f_{0}(980)\rightarrow\pi\pi}$ / $[\Gamma_{f_{0}(980)\rightarrow\pi\pi}%
+\Gamma_{f_{0}(980)\rightarrow KK}]$. Recently, the BABAR Collaboration
\cite{BABAR2006} has published results regarding the $f_{0}(980)$
phenomenology from the $B^{\pm}\rightarrow K^{\pm}K^{\pm}K^{\mp}$ decay
obtaining $\Gamma_{f_{0}(980)\rightarrow\pi\pi}$ / $[\Gamma_{f_{0}(980)\rightarrow
\pi\pi}+\Gamma_{f_{0}(980)\rightarrow KK}]=0.52\pm0.12$. Using $e^{+}e^{-}$
annihilation into kaons and pions and isolating hadronic intermediate
states,\ the same Collaboration also found \cite{BABAR2007} $\Gamma
_{f_{0}(980)}^{(1)}=(65\pm13)$ MeV from the $\varphi(1020)\pi^{+}\pi^{-}$
intermediate state and $\Gamma_{f_{0}(980)}^{(2)}=(81\pm21)$ MeV\ from the
$\varphi(1020)\pi^{0}\pi^{0}$ intermediate state. The mentioned $f_{0}%
(980)\rightarrow\pi\pi$ branching ratio together with $\Gamma_{f_{0}%
(980)}^{(1)}$ suggests $\Gamma_{f_{0}(980)\rightarrow\pi\pi}^{(1)}\simeq
(34\pm15)$ MeV whereas from $\Gamma_{f_{0}(980)}^{(2)}$ we obtain
$\Gamma_{f_{0}(980)\rightarrow\pi\pi}^{(2)}\simeq(42\pm21)$ MeV. Both results
are by approximately a factor of two smaller than our result $\Gamma
_{\sigma_{1}\rightarrow\pi\pi}=97$ MeV. \\
Additionally, a review in
Ref.\ \cite{Oller1997} found $\Gamma_{f_{0}(980)}\sim25$ MeV and
$\Gamma_{f_{0}(980)\rightarrow\pi\pi}/[\Gamma_{f_{0}(980)\rightarrow\pi\pi
}+\Gamma_{f_{0}(980)\rightarrow KK}]=0.68$ from a lowest-order chiral
Lagrangian and unitarity. These results suggest $\Gamma_{f_{0}(980)\rightarrow
\pi\pi}\sim17$ MeV, substantially less than our results for $\Gamma
_{\sigma_{1}\rightarrow\pi\pi}$. Therefore, our analysis does not favour
$f_{0}(980)$ as a predominantly $\bar{q}q$ state. Note also that assigning $m_{\sigma_{1}}$
to the mass range between $970$ MeV and $990$ MeV would imply $m_{\sigma_{2}%
}\simeq1590$ MeV (see Fig.\ \ref{Sigmamassen2}) and thus $\Gamma_{\sigma
_{2}\rightarrow\pi\pi}\simeq100$ MeV (see the right panel of Fig.\ \ref{Spp2}%
). Therefore, $\sigma_{2}\equiv f_{0}(1710)$ would have to saturate in the
$2\pi$ channel. This would clearly be at odds with data \cite{PDG}, and thus
it represents an additional argument against interpreting $f_{0}(980)$ as a predominantly
$\bar{q}q$ state within our model.\\

Nonetheless, it is possible that $f_{0}(980)$ may contain a quarkonium
component \cite{Muenz:1996,Delbourgo:1998,f0(980)asqq}. Alternatively, this state can also be
interpreted as a $\bar{q}^{2}q^{2}$ state, as a glueball, $KK$ bound state or
even as an $\eta\eta$ bound state (see Sec.\ \ref{sec.f0(980)}).

\subsection{Decay Width \boldmath $\sigma_{1,2}\rightarrow K K$} \label{sec.sigmakaonkaon2}

The interaction Lagrangian of the pure states $\sigma_{N,S}$ with the kaons
has already been stated in Eq.\ (\ref{sigmakaonkaon}). The corresponding decay
widths $\Gamma_{\sigma_{1}\rightarrow KK}$ and $\Gamma_{\sigma_{2}\rightarrow
KK}$ are given in Eqs.\ (\ref{Gs1KK}) and (\ref{Gs2KK}), respectively.

\begin{figure}[h]
  \begin{center}
    \begin{tabular}{cc}
      \resizebox{78mm}{!}{\includegraphics{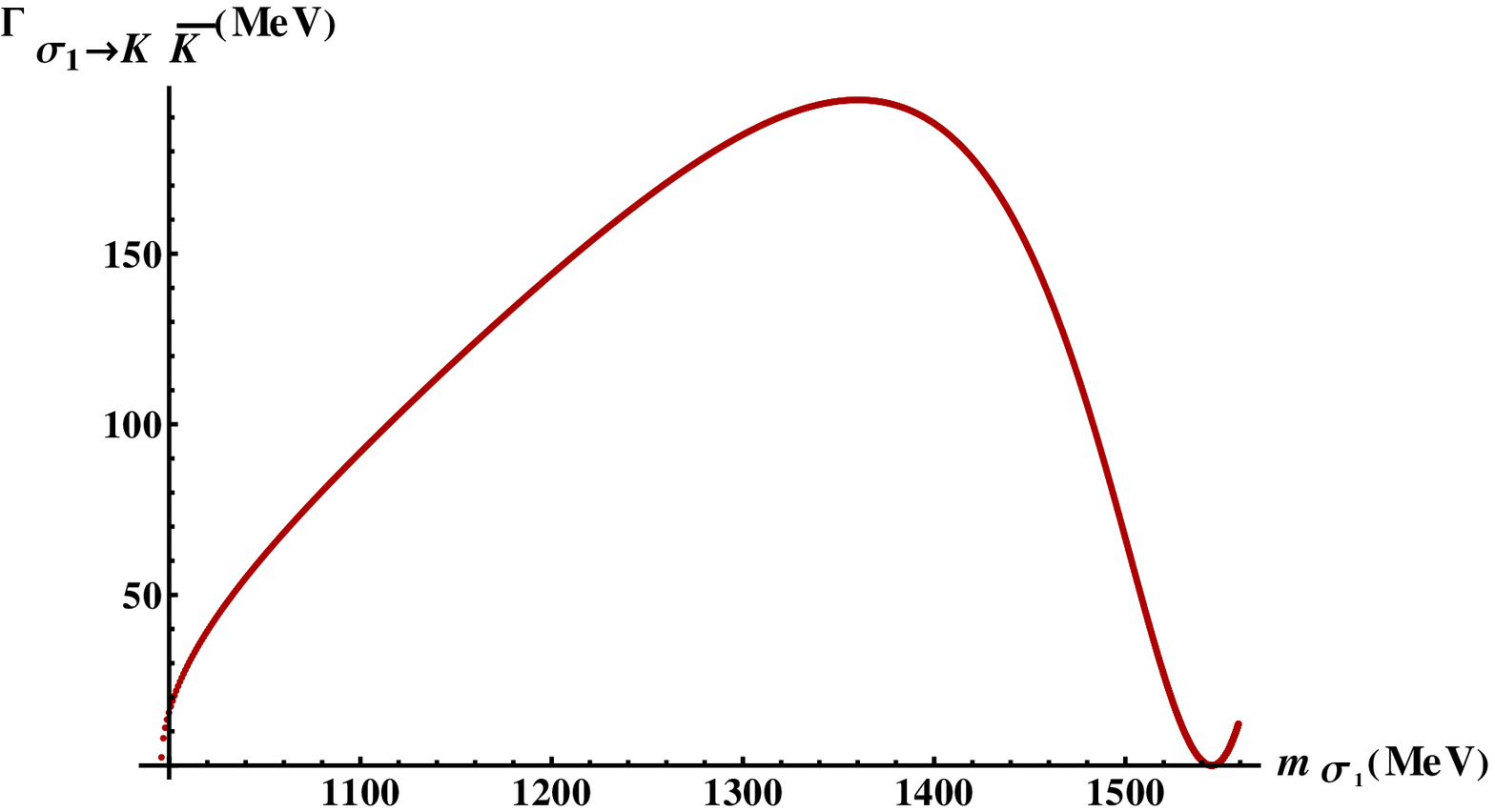}} &
      \resizebox{78mm}{!}{\includegraphics{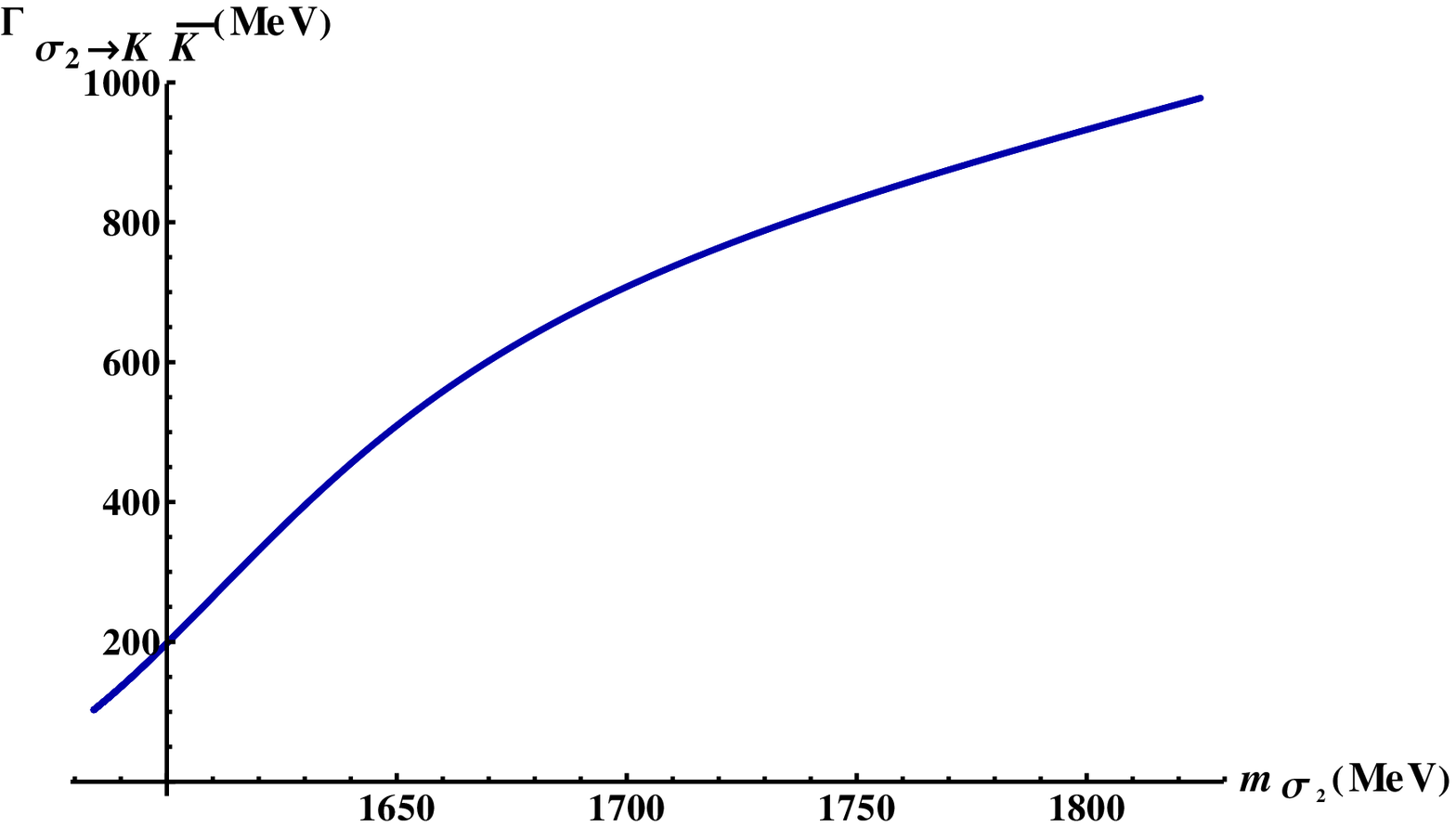}} 
    \end{tabular}
    \caption{$\Gamma_{\sigma_{1}\rightarrow KK}$ and $\Gamma_{\sigma
_{2}\rightarrow KK}$ as functions of $m_{\sigma_{1}}$ and $m_{\sigma_{2}}$,
respectively.}
    \label{SKK2}
  \end{center}
\end{figure}

We can therefore turn directly to a discussion of the decay widths, depicted in
Fig.\ \ref{SKK2}.

From the left panel of Fig.\ \ref{SKK2} we observe that $\Gamma_{\sigma
_{1}\rightarrow KK}$ is within the experimental results of
Refs.\ \cite{Etkin:1981,f01370KK2,Tikhomirov:2003,Polychronakos:1978,f01370KK1,f01370KK3}.
From the right panel of Fig.\ \ref{SKK2} we observe that
$\Gamma_{\sigma_{2}\rightarrow KK}$ rises rapidly with $m_{\sigma_{2}}$. The
PDG data suggest $\Gamma_{f_{0}(1710)\rightarrow KK}^{\text{PDG}%
}=71.44_{-35.02}^{+23.18}$ MeV, Eq.\ (\ref{f0(1710)_9}); note that this is the
dominant decay channel of $f_{0}(1710)$ and thus the reason why, already
from the experimental point of view, this resonance is a $\bar{s}s$ candidate.
Due to the rapid growth of $\Gamma_{\sigma_{2}\rightarrow KK}$, an exact
correspondence of our value with the central value of $\Gamma_{f_{0}%
(1710)\rightarrow KK}^{\text{PDG}}$ would require $m_{\sigma_{2}}=1578$ MeV.
However, $m_{\sigma_{2}}$ would then be outside the interval (\ref{ms22})
determined from the correct implementation of the spontaneous breaking of the
chiral symmetry -- we would require $m_{0}^{2}>0$ in contrast to condition
(\ref{m02}). Due to condition (\ref{ms22}), the lowest value of $m_{\sigma
_{2}}=1584$ MeV, for which we obtain $\Gamma_{\sigma_{2}\rightarrow KK}=102.7$
MeV, is above the interval for $\Gamma_{f_{0}(1710)\rightarrow KK}^{\text{PDG}}$.
As in the $2\pi$ channel, our results again yield $m_{\sigma_{2}}$ that is by
approximately $100$ MeV smaller than $m_{f_{0}(1710)}$. Additionally,
$m_{\sigma_{2}}=1584$ MeV implies $m_{\sigma_{1}}=450$ MeV (see
Fig.\ \ref{Sigmamassen2}), spoiling the correspondence of $\Gamma_{\sigma
_{1}\rightarrow\pi\pi}$ to experiment (see Fig.\ \ref{Spp2}). Note, however,
that our results allow for the WA102 value $\Gamma_{f_{0}(1710)\rightarrow
KK}^{\text{WA102}}=(80.5\pm30.1)$ MeV to be described: considering $1584$ MeV
$\leq m_{\sigma_{2}}\leq1586$ MeV yields $103$ MeV $\leq\Gamma_{\sigma
_{2}\rightarrow KK}\leq110.6$ MeV; the $m_{\sigma_{2}}$ interval is small due
to the steep rise of $\Gamma_{\sigma_{2}\rightarrow KK}$, see Fig.\ \ref{SKK2}%
. The mentioned interval also implies $450$ MeV $\leq$ $m_{\sigma_{1}}\leq688$
MeV, again spoiling the correspondence of $\Gamma_{\sigma_{1}\rightarrow\pi
\pi}$ to experiment as apparent from Fig.\ \ref{Spp2}. Thus using
$\Gamma_{f_{0}(1710)\rightarrow KK}$ does not allow us to constrain
$m_{\sigma_{1}}$ and $m_{\sigma_{2}}$ very well.

\begin{figure}[h]
  \begin{center}
    \begin{tabular}{cc}
      \resizebox{78mm}{!}{\includegraphics{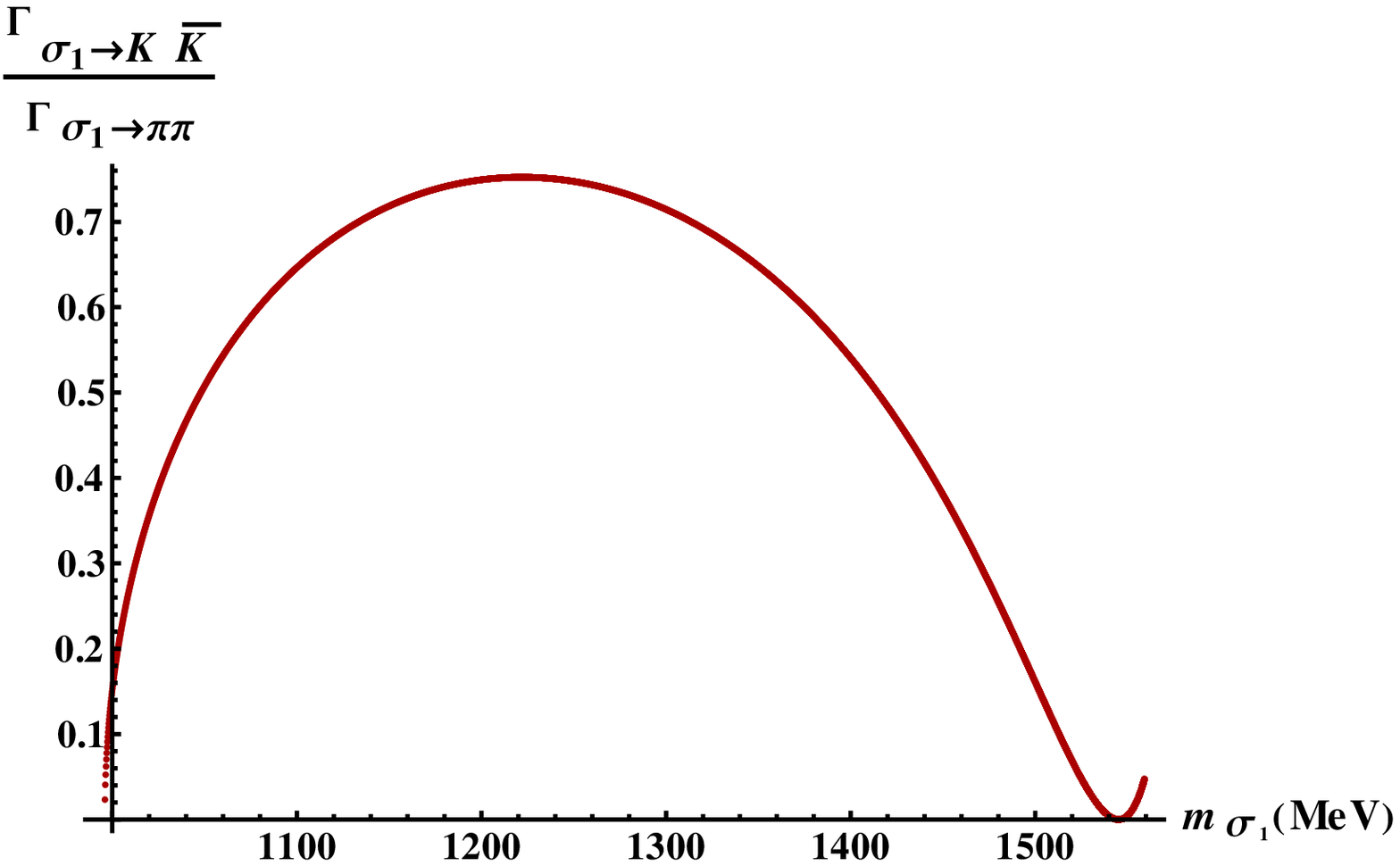}} &
      \resizebox{78mm}{!}{\includegraphics{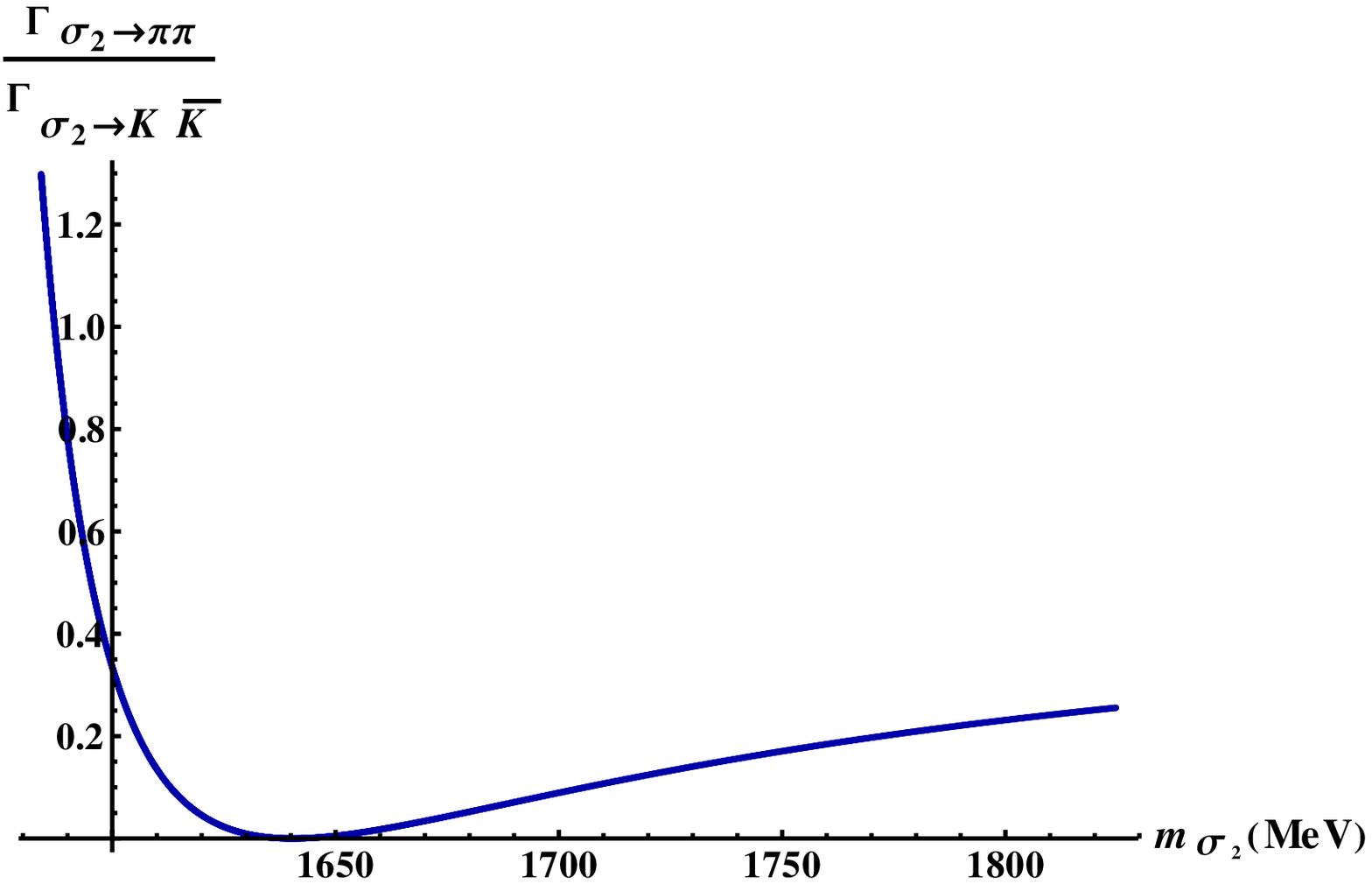}} 
    \end{tabular}
    \caption{Left panel: ratio $\Gamma_{\sigma_{1}\rightarrow KK}/\Gamma
_{\sigma_{1}\rightarrow\pi\pi}$ as function of $m_{\sigma_{1}}$. Right panel:
ratio $\Gamma_{\sigma_{2}\rightarrow\pi\pi}/\Gamma_{\sigma_{2}\rightarrow KK}$
as function of $m_{\sigma_{2}}$.}
    \label{SPPKK2}
  \end{center}
\end{figure}

Let us therefore look into the ratios $\Gamma_{\sigma_{1}\rightarrow
KK}/\Gamma_{\sigma_{1}\rightarrow\pi\pi}$ and $\Gamma_{\sigma_{2}%
\rightarrow\pi\pi}/\Gamma_{\sigma_{2}\rightarrow KK}$, depicted in
Fig.\ \ref{SPPKK2}.%

Let us first discuss results for $\Gamma_{\sigma_{1}\rightarrow KK}%
/\Gamma_{\sigma_{1}\rightarrow\pi\pi}$ (left panel in Fig.\ \ref{SPPKK2}). We
observe that the ratio varies between $0.16$ for $m_{\sigma_{1}} = 1500$ MeV
and $0.75$ for $m_{\sigma_{1}} = 1200$ MeV. Experimental data regarding this
ratio are unfortunately inconclusive \cite{PDG}.

\begin{itemize}
\item In 2005, the BESII Collaboration \cite{Ablikim:2004} noted the ratio
value of $0.08\pm0.08$ from the hadronic decay of the $J/\psi$ meson
($J/\psi\rightarrow\varphi\pi^{+}\pi^{-}$ and $J/\psi\rightarrow\varphi
K^{+}K^{-}$).

\item In 2003, the OBELIX Collaboration \cite{Bargiotti:2003} published a
coupled-channel analysis of $\bar{p}p$ annihilation into light mesons with the
result $\Gamma_{f_{0}(1370)\rightarrow KK}/\Gamma_{f_{0}(1370)\rightarrow
\pi\pi}=$ $0.91\pm0.20$.

\item A combined fit of Crystal Barrel, GAMS and BNL data performed by
Anisovich, \textit{et al.}, from 2002 found $\Gamma_{f_{0}(1370)\rightarrow
KK}/\Gamma_{f_{0}(1370)\rightarrow\pi\pi}=$ $0.12\pm0.06$
\cite{Anisovich:2001}.

\item The WA102 Collaboration found in 1999 the ratio $\Gamma_{f_{0}%
(1370)\rightarrow KK}/\Gamma_{f_{0}(1370)\rightarrow\pi\pi}=$ $0.46\pm
0.15\pm0.11$ \cite{Barberis:1999}.
\end{itemize}

Thus, the data vary over a large range of values. If we assign our $\sigma
_{1}$ state to $f_{0}(1370)$ and vary $m_{\sigma_{1}}$ from $1200$ MeV to
$1500$ MeV, then our results can be accommodated within all data sets. Clearly,
more conclusive data would allow for more conclusive results regarding our
theoretical predictions.\newline

Regarding the ratio $\Gamma_{f_{0}(1710)\rightarrow\pi\pi}%
/\Gamma_{f_{0}(1710)\rightarrow KK}$, experimental results are rather
ambiguous, as discussed in Sections \ref{f0(1710)-PDG-BESII}, \ref{aa} and
\ref{bb}.

\begin{itemize}
\item From the PDG-preferred ratio $\Gamma_{f_{0}(1710)\rightarrow\pi\pi
}^{\text{PDG}}/\Gamma_{f_{0}(1710)\rightarrow KK}^{\text{PDG}}=0.41_{-0.17}%
^{+0.11}$ (see Sec.\ \ref{f0(1710)-PDG-BESII}) we obtain $m_{\sigma_{2}%
}=1598_{-3}^{+6}$ MeV. This result implies $\Gamma_{\sigma_{2}\rightarrow
\pi\pi}=75_{+11}^{-21}$ MeV from Eq.\ (\ref{Gs2pp}), too large when compared
to data, see Eq.\ (\ref{f0(1710)_8}). Additionally, we obtain $m_{\sigma_{1}%
}=1209_{-56}^{+82}$ MeV and $\Gamma_{\sigma_{1}\rightarrow\pi\pi}%
=197_{-30}^{+55}$ MeV from Eq.\ (\ref{Gs1pp}). These results are within
the boundaries cited in Ref.\ \cite{buggf0}.

Thus using $\Gamma_{f_{0}(1710)\rightarrow\pi\pi}^{\text{PDG}}/\Gamma
_{f_{0}(1710)\rightarrow KK}^{\text{PDG}}$\ constrains $m_{\sigma_{2}}$\ in a
way that does not allow us to describe simultaneously $\Gamma_{f_{0}%
(1710)\rightarrow\pi\pi}$ as well as $\Gamma_{f_{0}(1370)\rightarrow\pi\pi}%
$.\newline Note that $\Gamma_{f_{0}(1710)\rightarrow\pi\pi}^{\text{PDG}%
}/\Gamma_{f_{0}(1710)\rightarrow KK}^{\text{PDG}}$ could also be described by
the high-mass tail of $\Gamma_{\sigma_{2}\rightarrow\pi\pi}$ / $\Gamma_{\sigma
_{2}\rightarrow KK}$ in Fig.\ \ref{SPPKK2}; however, this would imply
$m_{\sigma_{2}}\gtrsim1800$ MeV leading to unphysically large values of
$\Gamma_{\sigma_{2}\rightarrow KK}$, see Fig.\ \ref{SKK2}.

\item From the BES II ratio $\Gamma_{f_{0}(1710)\rightarrow\pi\pi}^{\text{BES
II}}/\Gamma_{f_{0}(1710)\rightarrow KK}^{\text{BES II}}<0.11$,
Eq.\ (\ref{f0(1710)_12}), we obtain $1612$ MeV $\leq m_{\sigma_{2}}\leq1712$
MeV. Given the parabolic form of $\Gamma_{\sigma_{2}\rightarrow\pi\pi}%
/\Gamma_{\sigma_{2}\rightarrow KK}$, let us separate the mentioned interval
into two subintervals: (\textit{i}) $1612$ MeV $\leq m_{\sigma_{2}}\leq1640$
MeV and (\textit{ii}) $1640$ MeV $\leq m_{\sigma_{2}}\leq1712$ MeV with
$m_{\sigma_{2}}=1640$ MeV the point where the ratio vanishes (see
Fig.\ \ref{SPPKK2}). Interval (\textit{i}) yields $1356$ MeV $\leq
m_{\sigma_{1}}\leq1452$ MeV and $306$ MeV $\leq\Gamma_{\sigma_{1}%
\rightarrow\pi\pi}\leq398$ MeV. Interval (\textit{ii}) yields $1452$ MeV $\leq
m_{\sigma_{1}}\leq1517$ MeV and $397$ MeV $\leq\Gamma_{\sigma_{1}%
\rightarrow\pi\pi}\leq416$ MeV, see Fig.\ \ref{Spp2}. As noted in
Sec.\ \ref{aa}, it is not possible to calculate $\Gamma_{f_{0}%
(1710)\rightarrow KK}$ from these data.

\item The WA102 ratio $\Gamma_{f_{0}(1710)\rightarrow\pi\pi}^{\text{WA102}%
}/\Gamma_{f_{0}(1710)\rightarrow KK}^{\text{WA102}}=0.2\pm0.06$,
Eq.\ (\ref{f0(1710)_17}), also yields two intervals for $m_{\sigma_{2}}$:
(\textit{i}) $m_{\sigma_{2}}=1606_{+4}^{-3}$ MeV and (\textit{ii})
$m_{\sigma_{2}}=1772_{-42}^{+58}$ MeV. We disregard the interval (\textit{ii})
because it leads to a very large value of $\Gamma_{\sigma_{2}\rightarrow KK}$,
see Fig.\ \ref{SKK2}. From interval (\textit{i}) we obtain $m_{\sigma_{1}%
}=1310_{+30}^{-29}$ MeV and $\Gamma_{\sigma_{1}\rightarrow\pi\pi}%
=267_{+25}^{-50}$ MeV. These results are consistent with the experimental values
of Ref.\ \cite{buggf0}.
\end{itemize}

In summary: it is not possible to constrain $m_{\sigma_{2}}$ via
$\Gamma_{f_{0}(1710)\rightarrow KK}$\ in a way that yields acceptable values
of $\Gamma_{f_{0}(1710)\rightarrow\pi\pi}$ (because our values $\Gamma
_{\sigma_{2}\rightarrow KK}$ increase rapidly with $m_{\sigma_{2}}$). However,
utilising the ratio $\Gamma_{f_{0}(1710)\rightarrow\pi\pi}/\Gamma
_{f_{0}(1710)\rightarrow KK}$ allows us to constrain $m_{\sigma_{2}}$ such that both
$m_{\sigma_{1}}$ and $\Gamma_{\sigma_{1}\rightarrow\pi\pi}$ are within values
published in Ref.\ \cite{buggf0}. This can be accomplished using either
PDG-preferred or WA102 values for the mentioned ratio. Given the issues
regarding $\Gamma_{f_{0}(1710)\rightarrow\pi\pi}^{\text{PDG}}/\Gamma
_{f_{0}(1710)\rightarrow KK}^{\text{PDG}}$ discussed in
Sec.\ \ref{f0(1710)-PDG-BESII}, we prefer results obtained from $\Gamma
_{f_{0}(1710)\rightarrow\pi\pi}^{\text{WA102}}/\Gamma_{f_{0}(1710)\rightarrow
KK}^{\text{WA102}}$, i.e., $m_{\sigma_{1}}=1310_{+30}^{-29}$ MeV,
$m_{\sigma_{2}}=1606_{+4}^{-3}$ MeV, $\Gamma_{\sigma_{1}\rightarrow\pi\pi
}=267_{+25}^{-50}$ MeV. Note that these results yield $\Gamma_{\sigma
_{2}\rightarrow KK}\sim200$ MeV [larger than experimental results
but\ consistent with the notion of a predominant $2K$ decay channel of $f_{0}%
(1710)$] and also $\Gamma_{\sigma_{2}\rightarrow\pi\pi}=47_{-10}^{+9}$ MeV
[larger than the WA102 value of Eq.\ (\ref{f0(1710)_20}) but consistent with
the notion of a subdominant $2\pi$ decay channel of  $f_{0}(1710)$]. These combined
results from the $2\pi$ and $2K$ channels suggest that $f_{0}(1370)$ is
$(95.5\pm1.0)$\% a $\bar{n}n$ state and that, conversely, that $f_{0}(1710)$
is $(95.5\pm1.0)$\% a $\bar{s}s$ state.

\subsection{Decay Width \boldmath $\sigma_{1,2}\rightarrow \eta \eta$} \label{sec.sigmaetaeta2}

We have already discussed the $\sigma\eta\eta$ interaction Lagrangian in
Sec.\ \ref{sec.setaeta1}, formulas for the decay widths $\Gamma_{\sigma
_{1}\rightarrow\eta\eta}$ and $\Gamma_{\sigma_{2}\rightarrow\eta\eta}$ are
stated in Eqs.\ (\ref{Gs1etaeta}) and (\ref{Gs2etaeta}), respectively.

\begin{figure}[h]
  \begin{center}
    \begin{tabular}{cc}
      \resizebox{78mm}{!}{\includegraphics{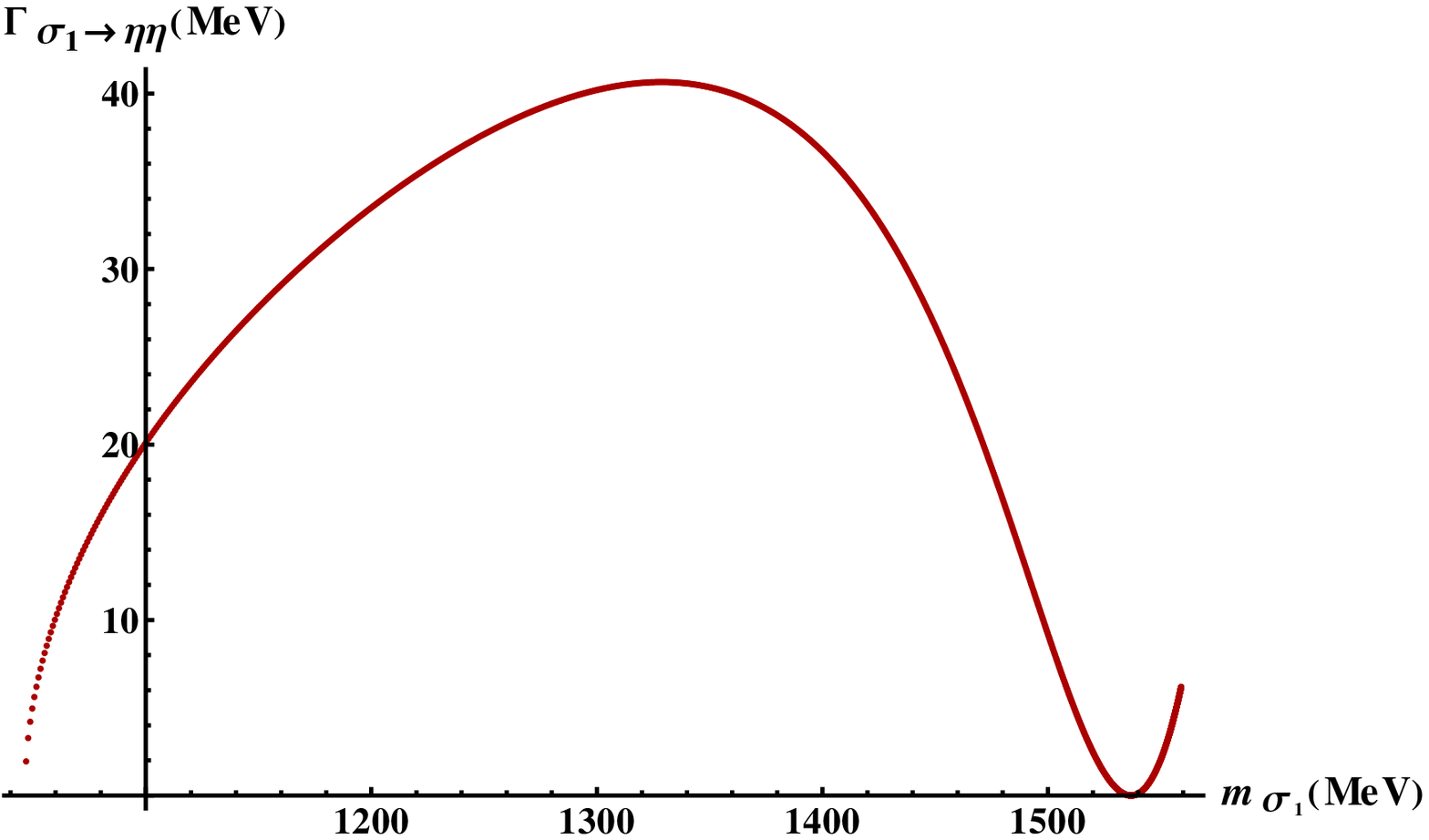}} &
      \resizebox{78mm}{!}{\includegraphics{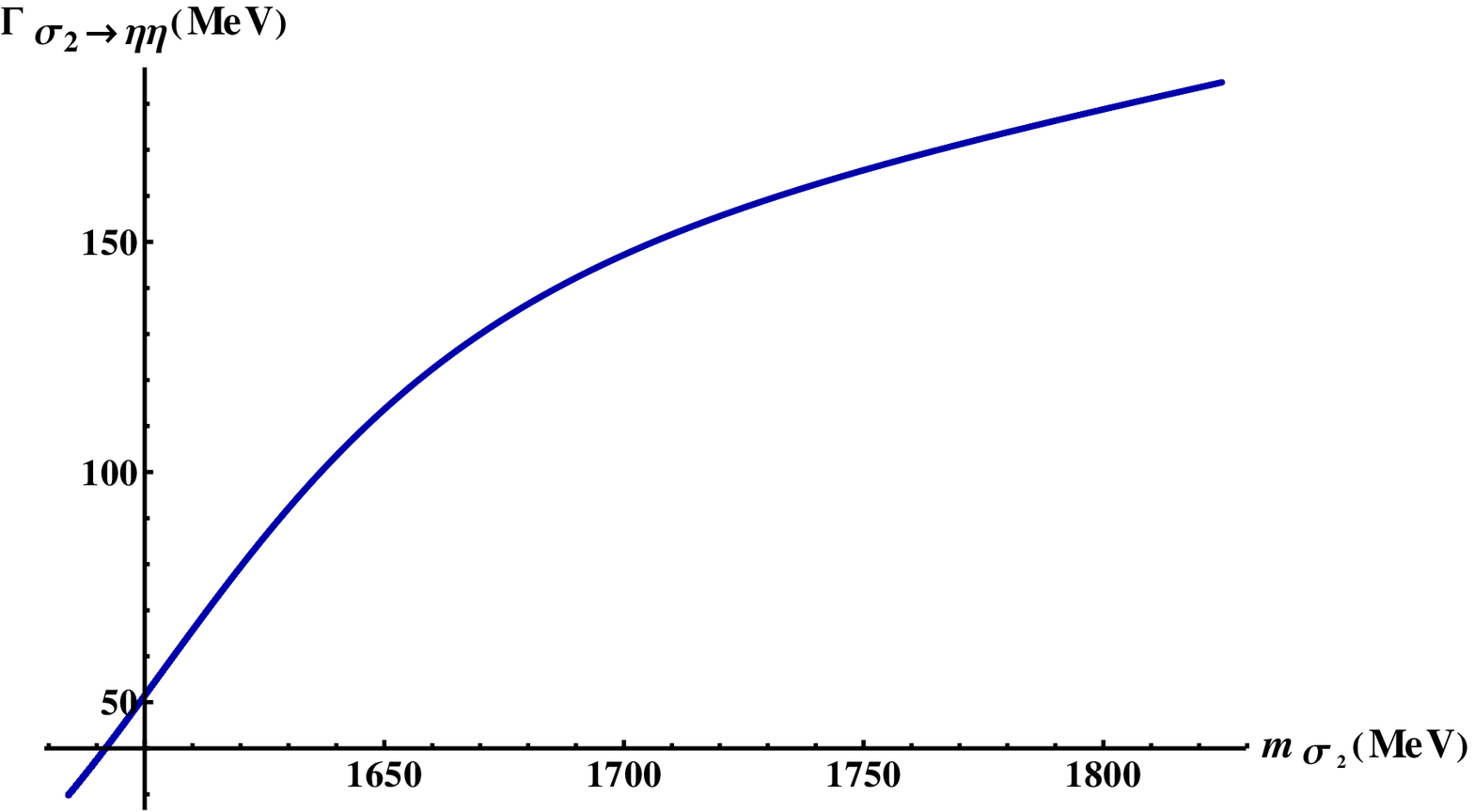}} 
    \end{tabular}
    \caption{$\Gamma_{\sigma_{1}\rightarrow\eta\eta}$ and $\Gamma_{\sigma
_{2}\rightarrow\eta\eta}$ as functions of $m_{\sigma_{1}}$ and $m_{\sigma_{2}%
}$ in Fit II.}
    \label{Setaeta2}
  \end{center}
\end{figure}

The dependence of the decay widths on $m_{\sigma_{1,2}}$ is shown diagramatically
in Fig.\ \ref{Setaeta2}.

We observe from the left panel of Fig.\ \ref{Setaeta2} that $\Gamma
_{\sigma_{1}\rightarrow\eta\eta}$ is suppressed over the entire mass range of
$\sigma_{1}$. Contrarily, $\Gamma_{\sigma_{2}\rightarrow\eta\eta}$ rises
rapidly over the mass range of $\sigma_{2}$.

Experimental results regarding the decay $f_{0}(1370)\rightarrow\eta\eta$ are
ambiguous; there are Crystal Barrel ${\bar{p}p}$ data \cite{f0(1500)-CB-1992}
and GAMS ${\pi}^{-}{p}$ \cite{Alde:1985kp} suggesting a decay width of
$\sim(250-300)$ MeV in\ this decay channel from Breit-Wigner fits. These are
known, however, to be very sensitive to the opening of new channels (such as
$4\pi$, see Sec.\ \ref{sec.f0(1370)}). For this reason, we will consider only
the (more unambiguously determined) values of $\Gamma_{f_{0}(1710)\rightarrow
\eta\eta}$ from Sec.\ \ref{f0(1710)channels}.\\

Our discussion of $\Gamma_{\sigma_{1,2}\rightarrow\eta\eta}$ will be
constrained by the following entries: (\textit{i}) the experimental result for
$\Gamma_{f_{0}(1710)\rightarrow\eta\eta}$; (\textit{ii}) the condition $m_{0}%
^{2}\leq0$ from formula (\ref{m02}), necessary to utilise because the lower
boundaries of $\Gamma_{f_{0}(1710)\rightarrow\eta\eta}$ from
Sec.\ \ref{f0(1710)channels} may imply $m_{\sigma_{2}}<1584$ MeV and thus
$m_{0}^{2}>0$ [see condition (\ref{ms22})]; (\textit{iii}) given that the
$\eta\eta$ channel represents a confirmed decay mode of $f_{0}(1370)$
\cite{PDG} (although, as already mentioned, the corresponding decay width is
by no means unambiguous), we also require that $m_{\sigma_{1}}$ is above the
$\eta\eta$ threshold, i.e., $m_{\sigma_{1}}\geq2m_{\eta}=1046$ MeV with
$m_{\eta}$ from Table \ref{Fit2-5}. [Remember that $m_{\sigma_{1}}$ determines
uniquely the values of $m_{0}^{2}$ and $m_{\sigma_{2}}$ from
Fig.\ \ref{Sigmamassen2} or, equivalently, from Eqs.\ (\ref{m_sigma_1}) -
(\ref{phisigma1}); $m_{\sigma_{2}}$ then allows for a determination of
$\Gamma_{\sigma_{2}\rightarrow\eta\eta}$ from the right panel of
Fig.\ \ref{Setaeta2}, or, equivalently, from Eqs.\ (\ref{Gs1etaeta}) and
(\ref{Gs2etaeta}).] The consequences of the stated three entries are as follows:

\begin{itemize}
\item The PDG-preferred result reads $\Gamma_{f_{0}(1710)\rightarrow\eta\eta
}^{\text{PDG}}=34.26_{-20.0}^{+15.42}$ MeV, see Eq.\ (\ref{f0(1710)_10}). It
is not possible to accommodate the full experimental interval within our model
as utilising the lower boundary of $\Gamma_{f_{0}(1710)\rightarrow\eta\eta
}^{\text{PDG}}$ would violate the above condition (\textit{ii}). Then
combining $\Gamma_{f_{0}(1710)\rightarrow\eta\eta}^{\text{PDG}}=34.26_{-20.0}%
^{+15.42}$ MeV with condition (\textit{ii}) yields $m_{\sigma_{2}}%
=1588_{-4}^{+11}$ MeV (the upper boundary for $m_{\sigma_{2}}$ was determined
from the upper boundary of $\Gamma_{f_{0}(1710)\rightarrow\eta\eta
}^{\text{PDG}}$) and, in turn, $\Gamma_{f_{0}(1710)\rightarrow\eta\eta
}=34.26_{-4.41}^{+15.42}$ MeV. However, condition (\textit{iii}), i.e.,
$m_{\sigma_{1}}\geq1046$ MeV, implies $m_{0}^{2}\leq-457456$ MeV$^{2}$ and
thus $m_{\sigma_{2}}\geq1591$ MeV. Combining the latter inequality with the
interval $m_{\sigma_{2}}=1588_{-4}^{+11}$ MeV yields $1591$ MeV $\leq
m_{\sigma_{2}}\leq1599$ MeV and, consequently, $39.12$ MeV $\leq\Gamma
_{f_{0}(1710)\rightarrow\eta\eta}\leq49.68$ MeV. The latter two sets of
inequalities also imply $1046$ MeV $\leq m_{\sigma_{1}}\leq1227$ MeV (or
$1200$ MeV $\leq m_{\sigma_{1}}\leq1227$ MeV considering the PDG data
\cite{PDG}) and $0\leq\Gamma_{\sigma_{1}\equiv f_{0}(1370)\rightarrow\eta\eta
}\leq35.92$ MeV. The $\eta\eta$ decay of $f_{0}(1370)$ is then suppressed in
comparison with the $2\pi$ decay, see Sec.\ \ref{sec.sigmapionpion2}.

\item There is another set of experimental data discussed in
Sec.\ \ref{f0(1710)channels}: $\Gamma_{f_{0}(1710)\rightarrow\eta\eta
}^{\text{WA102}}=(38.6\pm18.8)$ MeV from Eq.\ (\ref{f0(1710)_22}). As in the
case of the PDG-preferred data, we combine $\Gamma_{f_{0}(1710)\rightarrow
\eta\eta}^{\text{WA102}}=(38.6\pm18.8)$ MeV with the above condition
(\textit{ii}) and obtain $m_{\sigma_{2}}=1591_{-7}^{+13}$ MeV. Note that the
lower boundary of $m_{\sigma_{2}}=1594$ MeV implies $\Gamma_{f_{0}%
(1710)\rightarrow\eta\eta}=29.8$ MeV hence modifying the WA102 result to
$\Gamma_{f_{0}(1710)\rightarrow\eta\eta}=38.6_{-8.8}^{+18.8}$ MeV. As already
mentioned, condition (\textit{iii}) implies $m_{\sigma_{1}}\geq1046$ MeV,
i.e., $m_{0}^{2}\leq-457456$ MeV$^{2}$ and thus also $m_{\sigma_{2}}\geq1591$
MeV. The latter inequality in conjunction with $m_{\sigma_{2}}=1591_{-7}%
^{+13}$ MeV yields $1591$ MeV $\leq m_{\sigma_{2}}\leq1604$ MeV and,
consequently, $38.6$ MeV $\leq\Gamma_{\sigma_{2}\equiv f_{0}(1710)\rightarrow
\eta\eta}\leq56.6$ MeV. The latter two sets of inequalities also imply $1046$
MeV $\leq m_{\sigma_{1}}\leq1289$ MeV (i.e., $1200$ MeV $\leq m_{\sigma_{1}%
}\leq1289$ MeV considering the PDG data \cite{PDG}) and $0\leq\Gamma
_{\sigma_{1}\equiv f_{0}(1370)\rightarrow\eta\eta}\leq39.8$ MeV. Therefore,
$\Gamma_{f_{0}(1370)\rightarrow\eta\eta}$ is in this case slightly larger than
in the case of the PDG-preferred data but still smaller than the $2\pi$ decay
width discussed in Sec.\ \ref{sec.sigmapionpion2}.
\end{itemize}

Given the ambiguities in the BES II data utilised by the PDG (as discussed in
Sec.\ \ref{f0(1710)-PDG-BESII}), we prefer the results obtained from the WA102
data.\newline

\begin{figure}[h]
  \begin{center}
    \begin{tabular}{cc}
      \resizebox{78mm}{!}{\includegraphics{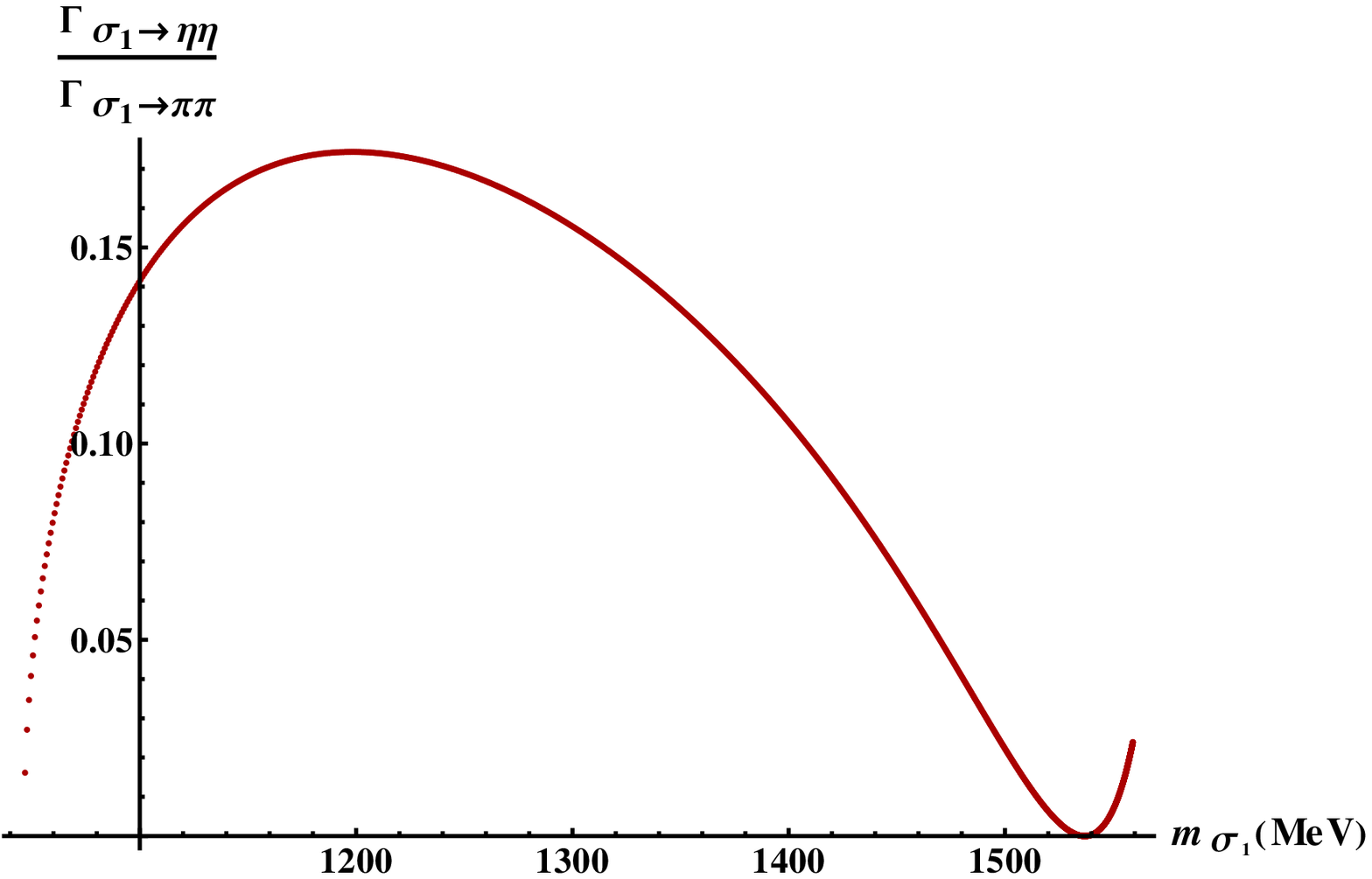}} &
      \resizebox{78mm}{!}{\includegraphics{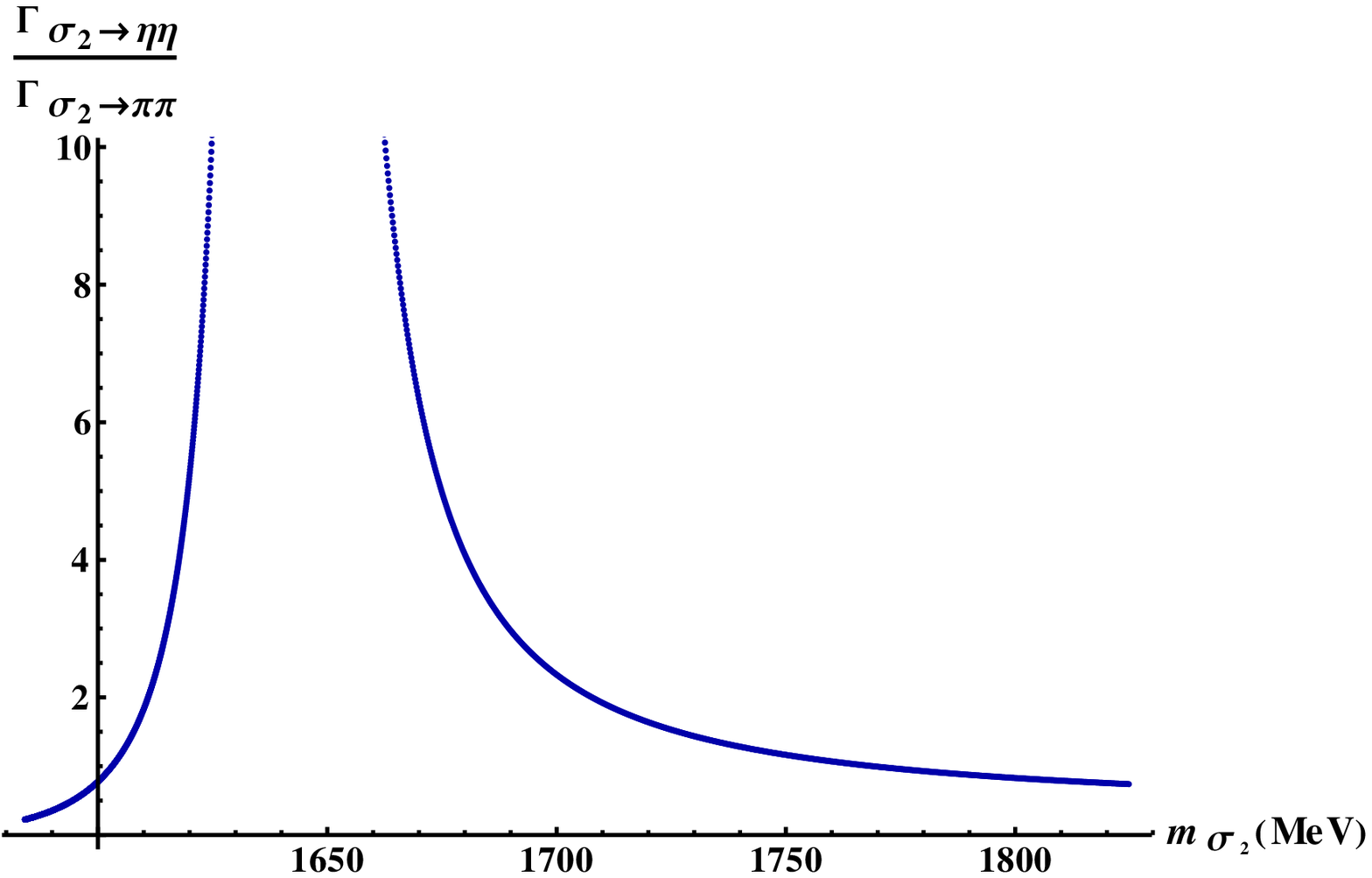}} 
    \end{tabular}
    \caption{Ratios $\Gamma_{\sigma_{1}\rightarrow\eta\eta}/\Gamma_{\sigma
_{1}\rightarrow\pi\pi}$ as function of $m_{\sigma_{1}}$ and $\Gamma
_{\sigma_{2}\rightarrow\eta\eta}/\Gamma_{\sigma_{2}\rightarrow\pi\pi}$ as
function of $m_{\sigma_{2}}$ in Fit II.}
    \label{Setaetapp2}
  \end{center}
\end{figure}

Let us now consider the ratios of the decay widths discussed so far. A plot of
$\Gamma_{\sigma_{1}\rightarrow\eta\eta}/\Gamma_{\sigma_{1}\rightarrow\pi\pi}$
and $\Gamma_{\sigma_{2}\rightarrow\eta\eta}/\Gamma_{\sigma_{2}\rightarrow
\pi\pi}$ is shown in Fig.\ \ref{Setaetapp2}.

Our results for $\Gamma_{\sigma_{1}\rightarrow\eta\eta}/\Gamma_{\sigma
_{1}\rightarrow\pi\pi}$ are within the ratio $\Gamma_{f_{0}(1370)\rightarrow
\eta\eta}/\Gamma_{f_{0}(1370)\rightarrow\pi\pi}=0.19\pm0.07$ \cite{buggf0}%
\ for a rather large mass interval: $1081$ MeV $\leq m_{\sigma_{1}}\leq1377$
MeV. Due to the constraints regarding $m_{f_{0}(1370)}$ \cite{PDG} we obtain
$1200$ MeV $\leq m_{\sigma_{1}}\leq1377$ MeV. Note that the largest value of
the ratio obtained (and shown in Fig.\ \ref{Setaetapp2}) is $0.174$, for
$m_{\sigma_{1}}=1200$ MeV.\newline

Additionally, there are three sets of data regarding the ratio
$\Gamma_{f_{0}(1710)\rightarrow\eta\eta}/\Gamma_{f_{0}(1710)\rightarrow\pi\pi
}$ that need to be considered (see Sec.\ \ref{f0(1710)channels}).

\begin{itemize}
\item Data preferred by the PDG suggest $\Gamma_{f_{0}(1710)\rightarrow
\eta\eta}^{\text{PDG}}/\Gamma_{f_{0}(1710)\rightarrow\pi\pi}^{\text{PDG}}$
$=1.17_{-0.61}^{+0.48}$, Eq.\ (\ref{f0(1710)_5}). As apparent from
Fig.\ \ref{Setaetapp2}, the stated ratio can be accommodated within our model
for two sets of $m_{\sigma_{2}}$ values. The higher of these two sets
($m_{\sigma_{2}}\sim1800$ MeV) is not considered because it would lead to very
large values of the $2\pi$ and $2K$ decay widths for this resonance (see figures
\ref{Spp2} and \ref{SKK2}). For this reason, we consider the lower of the
intervals reading $m_{\sigma_{2}}=1605_{-9}^{+4}$ MeV. This value implies
$m_{\sigma_{1}}=1302_{-102}^{+32}$ MeV by Fig.\ \ref{Sigmamassen2}, with the
lower boundary limited to $m_{\sigma_{1}\equiv f_{0}(1370)}=1200$ MeV
\cite{PDG}, and $0.14\leq\Gamma_{\sigma_{1}\rightarrow\eta\eta}/\Gamma
_{\sigma_{1}\rightarrow\pi\pi}\leq0.17$, see\ Fig.\ \ref{Setaetapp2}.

\item BES II data from condition (\ref{f0(1710)_14}) suggest$\ \Gamma
_{f_{0}(1710)\rightarrow\eta\eta}^{\text{BES II}}/\Gamma_{f_{0}%
(1710)\rightarrow\pi\pi}^{\text{BES II}}>4.36$. This ratio implies
$m_{\sigma_{2}}>1618$ MeV, $m_{\sigma_{1}}>1389$ MeV (see
Fig.\ \ref{Sigmamassen2}) and $\Gamma_{\sigma_{1}\rightarrow\eta\eta}%
/\Gamma_{\sigma_{1}\rightarrow\pi\pi}<0.11$, see Fig.\ \ref{Setaetapp2} (we
again disregard the high-mass tail of $m_{\sigma_{2}}$ that would also fulfill
the stated ratio). The lower boundaries for $m_{\sigma_{1,2}}$ are
incompatible with the best values in the $2\pi$ and $2K$ decay channels of
$\sigma_{1,2}$, as discussed at the end of Sec.\ \ref{sec.sigmakaonkaon2}. The
obtained ratio for $\Gamma_{\sigma_{1}\rightarrow\eta\eta}/\Gamma_{\sigma
_{1}\rightarrow\pi\pi}$ is outside of the interval $\Gamma_{f_{0}%
(1370)\rightarrow\eta\eta}/\Gamma_{f_{0}(1370)\rightarrow\pi\pi}=0.19\pm0.07$
suggested by Ref.\ \cite{buggf0}. For this reason, the BES II result regarding
the ratio of $\Gamma_{f_{0}(1710)\rightarrow\eta\eta}/\Gamma_{f_{0}%
(1710)\rightarrow\pi\pi}$ is not supported by our model.

\item WA102 data from Eq.\ (\ref{f0(1710)_19}) suggest $\Gamma_{f_{0}%
(1710)\rightarrow\eta\eta}^{\text{WA102}}/\Gamma_{f_{0}(1710)\rightarrow\pi
\pi}^{\text{WA102}}=2.4\pm1.04$. As apparent from Fig.\ \ref{Setaetapp2}, this
ratio also implies two possible $m_{\sigma_{2}}$ intervals, a relatively lower
one and a relatively higher one. The latter interval is disregarded because it
would yield $m_{\sigma_{2}}\sim1700$ MeV, a value that -- although close to
the experimental value of $m_{f_{0}(1710)}=1720$ MeV -- nonetheless yields
very large values of $\Gamma_{\sigma_{2}\rightarrow\pi\pi}$ and $\Gamma
_{\sigma_{2}\rightarrow KK}$, see figures \ref{Spp2} and \ref{SKK2}. We
therefore consider only the lower set of $m_{\sigma_{2}}$ values reading
$m_{\sigma_{2}}=1613_{-6}^{+3}$ MeV. This interval implies $m_{\sigma_{1}%
}=1360_{-43}^{+19}$ MeV (see Fig.\ \ref{Sigmamassen2}) and $0.12\leq
\Gamma_{\sigma_{1}\rightarrow\eta\eta}/\Gamma_{\sigma_{1}\rightarrow\pi\pi
}\leq0.15$, see Fig.\ \ref{Setaetapp2}. The latter ratio is within the
interval $\Gamma_{f_{0}(1370)\rightarrow\eta\eta}/\Gamma_{f_{0}%
(1370)\rightarrow\pi\pi}=0.19\pm0.07$ suggested by Ref.\ \cite{buggf0}.
\end{itemize}

\begin{figure}[h]
  \begin{center}
    \begin{tabular}{cc}
      \resizebox{78mm}{!}{\includegraphics{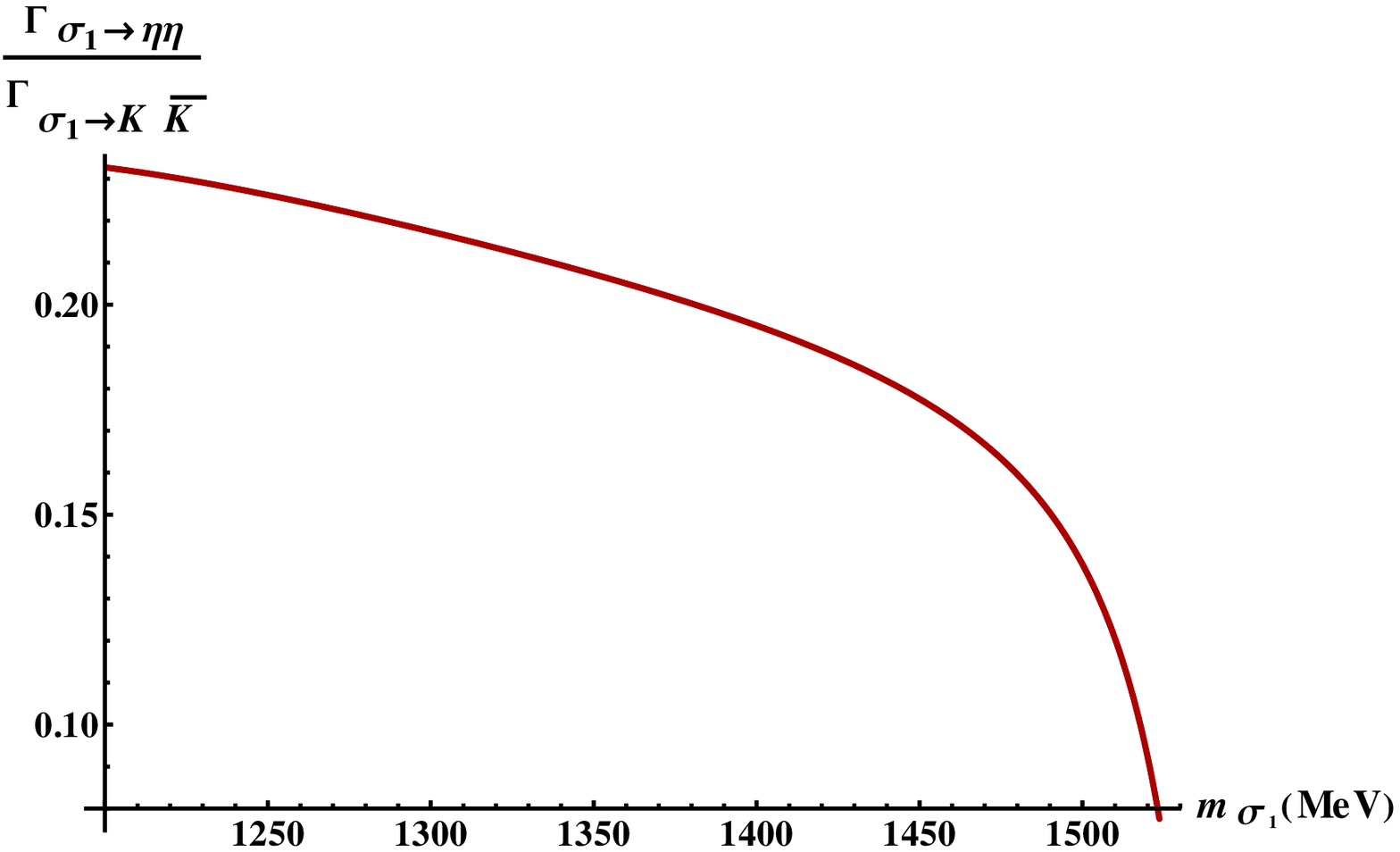}} &
      \resizebox{78mm}{!}{\includegraphics{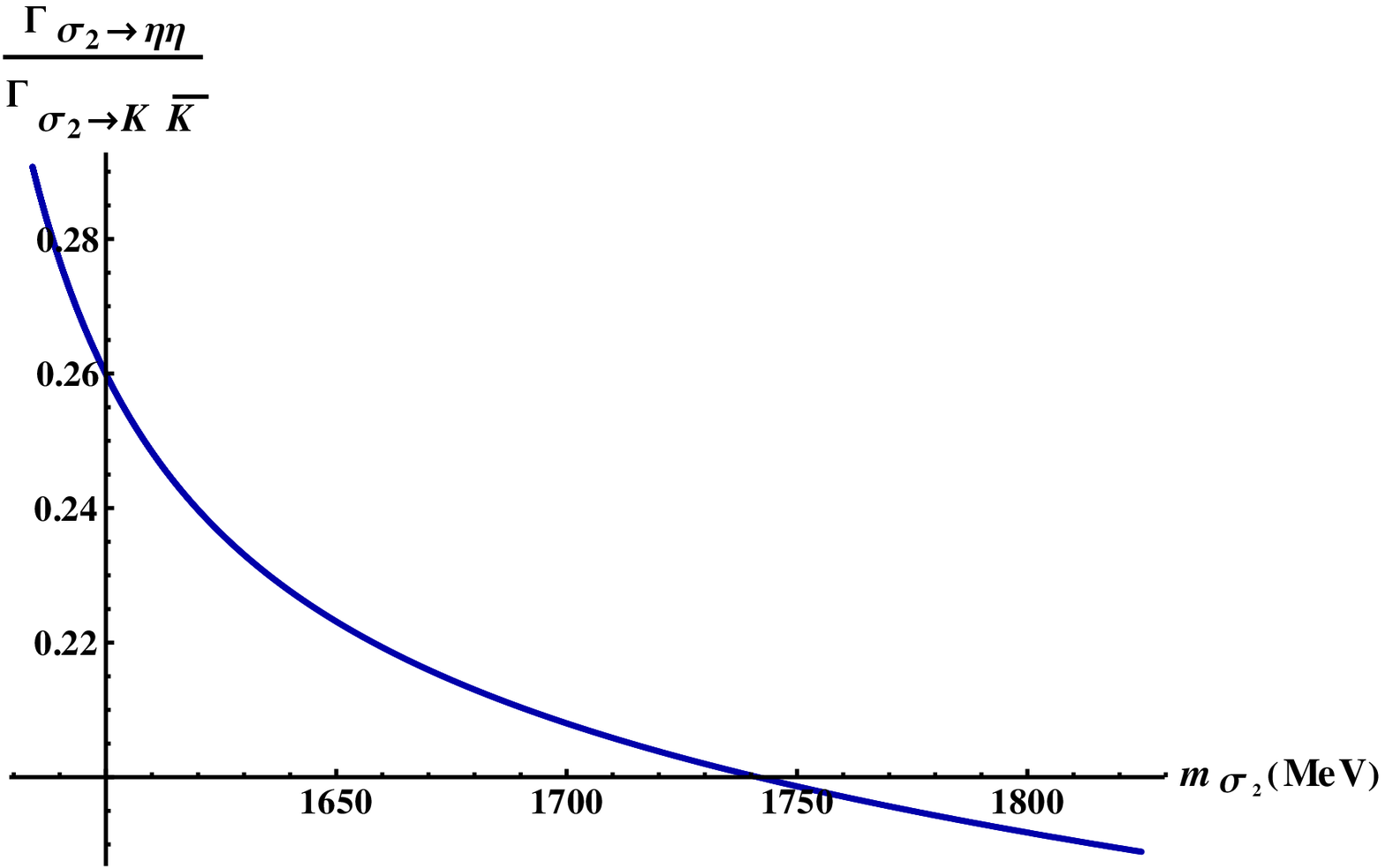}} 
    \end{tabular}
    \caption{Ratios $\Gamma_{\sigma_{1}\rightarrow\eta\eta}/\Gamma_{\sigma
_{1}\rightarrow KK}$ as function of $m_{\sigma_{1}}$ and $\Gamma_{\sigma
_{2}\rightarrow\eta\eta}/\Gamma_{\sigma_{2}\rightarrow KK}$ as function of
$m_{\sigma_{2}}$ in Fit II.}
    \label{SetaetaKK2}
  \end{center}
\end{figure}

Let us now consider the ratio $\Gamma_{\sigma_{1,2}\rightarrow\eta\eta}
/\Gamma_{\sigma_{1,2}\rightarrow KK}$ shown in Fig.\ \ref{SetaetaKK2}.
The corresponding ratio for $f_{0}(1710)$ has been determined by the WA102
Collaboration \cite{Barberis:2000} with data from $pp$ collisions yielding
$\Gamma_{f_{0}(1710)\rightarrow\eta\eta}/\Gamma_{f_{0}(1710)\rightarrow
KK}=0.48\pm0.15$ and in a combined-fit analysis of Ref.\ \cite{Anisovich:2001}
where $\Gamma_{f_{0}(1710)\rightarrow\eta\eta}/\Gamma_{f_{0}(1710)\rightarrow
KK}=0.46_{-0.38}^{+0.70}$ was obtained. The results are obviously mutually
compatible; the PDG cites the WA102 result as the referential one. We observe,
however, that the WA102 interval is outside the ratio on the right panel of
Fig.\ \ref{SetaetaKK2} and that the result from Ref.\ \cite{Anisovich:2001}
cannot be utilised to constrain $m_{\sigma_{2}}$ because the entire interval
on the right panel of Fig.\ \ref{SetaetaKK2} is within the result
$\Gamma_{f_{0}(1710)\rightarrow\eta\eta}/\Gamma_{f_{0}(1710)\rightarrow
KK}=0.46_{-0.38}^{+0.70}$. It is therefore not possible to utilise the ratio
$\Gamma_{f_{0}(1710)\rightarrow\eta\eta}/\Gamma_{f_{0}(1710)\rightarrow KK}$
in order to constrain $m_{\sigma_{2}}$.

\subsubsection{Short Summary of Results}

Let us now summarise results obtained so far. The ratio $\Gamma_{f_{0}%
(1710)\rightarrow\pi\pi}^{\text{WA102}}/\Gamma_{f_{0}(1710)\rightarrow
KK}^{\text{WA102}}=0.2\pm0.06$ allows us to determine $m_{\sigma_{2}}$ and
then observables for $\sigma_{1}$. We obtain $m_{\sigma_{1}}=1310_{+30}^{-29}$
MeV, $m_{\sigma_{2}}=1606_{+4}^{-3}$ MeV and $\Gamma_{\sigma_{1}\rightarrow
\pi\pi}=267_{+25}^{-50}$ MeV. The results for $\sigma_{1}$ are consistent with
interpretation of this state as $f_{0}(1370)$. In particular $m_{\sigma_{1}}$
is consistent with the combined-fit value of $m_{f_{0}(1370)}=(1309\pm1\pm15)$
MeV from Ref.\ \cite{buggf0}; $\Gamma_{\sigma_{1}\rightarrow\pi\pi}$ is
consistent with both the Breit-Wigner decay width and the FWHM value of
Ref.\ \cite{buggf0}. The value of $m_{\sigma_{2}}$ is approximately $100$ MeV
smaller than $m_{f_{0}(1710)}$; however, a pure glueball state [that would
very probably shift $m_{\sigma_{2}}$ in the direction of $m_{f_{0}(1710)}$] is
not present in the $U(3)\times U(3)$ version of our model. Additionally, we
observe that our state $\sigma_{2}$ possesses a strongly enhanced kaon decay,
also consistent with the corresponding feature of $f_{0}(1710)$ although the
absolute value of the decay width in this channel is too large.\ Additionally,
$\Gamma_{\sigma_{2}\rightarrow\pi\pi}=47_{-10}^{+9}$ MeV is larger than the
value expected for the $f_{0}(1710)$ resonance; this may be a consequence of
the missing glueball field that, if included, could modify decay amplitudes in
such a way that $\Gamma_{\sigma_{2}\rightarrow\pi\pi}$ and $\Gamma_{\sigma
_{2}\rightarrow KK}$ obtain values closer to those of $f_{0}(1710)$.\newline

Additionally, the decay channel $\sigma_{1,2}\rightarrow\eta\eta$ is well
accommodated within the model: our results for $\Gamma_{\sigma_{2}\equiv
f_{0}(1710)\rightarrow\eta\eta}$\ are within $\Gamma_{f_{0}(1710)\rightarrow
\eta\eta}^{\text{WA102}}=(38.6\pm18.8)$ MeV if we set $1591$ MeV $\leq
m_{\sigma_{2}}\leq1604$ MeV. Then we obtain simultaneously $1200$ MeV $\leq
m_{\sigma_{1}}\leq1289$ MeV, $0\leq\Gamma_{\sigma_{1}\equiv f_{0}%
(1370)\rightarrow\eta\eta}\leq39.8$ MeV and $38.6$ MeV $\leq\Gamma_{\sigma
_{2}\equiv f_{0}(1710)\rightarrow\eta\eta}\leq56.6$ MeV. [$\Gamma_{\sigma
_{2}\equiv f_{0}(1710)\rightarrow\eta\eta}$ does not correspond exactly to
$\Gamma_{f_{0}(1710)\rightarrow\eta\eta}^{\text{WA102}}$ because we have
required $m_{\sigma_{1}}\geq2m_{\eta}$ hence constraining $m_{0}^{2}$ and
consequently other observables as well.] Note, however, that the obtained
$m_{\sigma_{1}}$ and $m_{\sigma_{2}}$ overlap with $m_{\sigma_{1}}$ and
$m_{\sigma_{2}}$ determined from $\Gamma_{f_{0}(1710)\rightarrow\pi\pi
}^{\text{WA102}}/\Gamma_{f_{0}(1710)\rightarrow KK}^{\text{WA102}}$ within errors.

Finally, it is not possible to constrain $m_{\sigma_{1,2}}$ and other
observables from $\Gamma_{f_{0}(1710)\rightarrow\eta\eta}/\Gamma
_{f_{0}(1710)\rightarrow KK}$; however, the opposite is true for
$\Gamma_{f_{0}(1710)\rightarrow\eta\eta}/\Gamma_{f_{0}(1710)\rightarrow\pi\pi
}$. We prefer the result of the WA102 Collaboration $\Gamma_{f_{0}(1710)\rightarrow\eta\eta
}^{\text{WA102}}/\Gamma_{f_{0}(1710)\rightarrow\pi\pi}^{\text{WA102}}%
=2.4\pm1.04$ because of reliability issues of an alternative, PDG-preferred
ratio value (discussed at the beginning of Sec.\ \ref{f0(1710)-PDG-BESII}).
Utilising the stated WA102 interval we obtain $m_{\sigma_{1}}=1360_{-43}%
^{+19}$ MeV, $m_{\sigma_{2}}=1613_{-6}^{+3}$ MeV [suggesting that
$f_{0}(1370)$ is $94.7_{-3.0}^{+1.4}\%$ a $\bar{n}n$ state and that,
conversely, that $f_{0}(1710)$ is $94.7_{-3.0}^{+1.4}\%$ a $\bar{s}s$ state]
and $0.12\leq\Gamma_{\sigma_{1}\rightarrow\eta\eta}/\Gamma_{\sigma
_{1}\rightarrow\pi\pi}\leq0.15$. It is obvious that these results are also
compatible with the previous two (within errors).

\subsection{Combined Results in the Pion, Kaon and Eta Channels} \label{sec.scalars2}

Until now we have considered experimental information regarding the $\pi\pi$, $KK$
and $\eta\eta$\ channels by exploring the possibility to describe each of
these\ decay channels separately. However, the already noted compatibility of
thus obtained results (within errors)\ prompts us to investigate whether
similarly good results can be obtained considering a single observable. Let
that observable be the ratio $\Gamma_{f_{0}(1710)\rightarrow\pi\pi}%
/\Gamma_{f_{0}(1710)\rightarrow KK}$ due to the importance of pion and kaon
decays in discriminating between predominantly non-strange and predominantly
strange states. Utilising the WA102 result $\Gamma_{f_{0}(1710)\rightarrow
\pi\pi}^{\text{WA102}}/\Gamma_{f_{0}(1710)\rightarrow KK}^{\text{WA102}%
}=0.2\pm0.06$ \cite{Barberis:1999} allows us to exactly determine the only
parameter that we have varied until now: $m_{0}^{2}=-791437_{-46053}^{+42287}$
MeV$^{2}$. (Note that until now the only conditions set upon\ $m_{0}^{2}$ were
$m_{0}^{2}\overset{!}{<}0$ and that the values of this parameter must imply
$m_{\sigma_{N}}<m_{\sigma_{S}}$.) Then we obtain the following results
[remember -- our assignment is $\sigma_{1}\equiv f_{0}(1370)$ and $\sigma
_{2}\equiv f_{0}(1710)$]:

\begin{itemize}
\item \textit{Masses:} we obtain $m_{\sigma_{1}}=1310_{+30}^{-29}$ MeV and
$m_{\sigma_{2}}=1606_{+4}^{-3}$ MeV. The former is virtually the same as the
combined-fit Breit-Wigner mass in Ref.\ \cite{buggf0} where $m_{f_{0}%
(1370)}=(1309\pm1\pm15)$ MeV was obtained (our error results are dictated by
uncertainties in $\Gamma_{f_{0}(1710)\rightarrow\pi\pi}^{\text{WA102}}%
/\Gamma_{f_{0}(1710)\rightarrow KK}^{\text{WA102}}$) and also very close to
the $f_{0}(1370)$ peak mass in the $2\pi$ channel, found to be $1282$ MeV in
Ref.\ \cite{buggf0}. The latter is approximately $100$ MeV smaller than
$m_{f_{0}(1710)}=(1720\pm6)$ MeV because the glueball field has not been
included in the current version of the model. This implies that $f_{0}(1370)$
is $91.2_{+2.0}^{-1.7}\%$ a $\bar{n}n$ state and that, conversely,
$f_{0}(1710)$ is $91.2_{+2.0}^{-1.7}\%$ a $\bar{s}s$ state.

\item \textit{Pion decay channel:} we obtain $\Gamma_{\sigma_{1}\rightarrow
\pi\pi}=267_{+25}^{-50}$ MeV and $\Gamma_{\sigma_{2}\rightarrow\pi\pi
}=47_{-10}^{+9}$ MeV. The former is virtually a median of (and thus consistent
with both) the Breit-Wigner decay width $\Gamma_{f_{0}(1370)\rightarrow\pi\pi
}=325$ MeV and the $f_{0}(1370)$ FWHM in the $2\pi$ channel, the value of
which was determined as $207$ MeV in Ref.\ \cite{buggf0}. The latter is too
large when compared to the WA102 result in Eq.\ (\ref{f0(1710)_20}) but still
demonstrates that the decay $f_{0}(1710)\rightarrow\pi\pi$ is suppressed in
comparison with other decay modes (see below) -- a fact that is in accordance
with the data \cite{PDG}.

\item \textit{Kaon decay channel:} we obtain $\Gamma_{\sigma_{1}\rightarrow
KK}=188_{+6}^{-9}$ MeV and $\Gamma_{\sigma_{2}\rightarrow KK}=237_{+25}^{-20}$
MeV. The two-kaon decay width for $f_{0}(1370)$ has not been determined
unambiguously, but our result is consistent with experimental data in
Refs.\ \cite{Etkin:1981,f01370KK2,Tikhomirov:2003,Polychronakos:1978,f01370KK1,f01370KK3}%
. We find $\Gamma_{f_{0}(1370)\rightarrow KK}<\Gamma_{f_{0}(1370)\rightarrow
\pi\pi}$, consistent with the interpretation of $f_{0}(1370)$ as a predominantly
non-strange $\bar{q}q$ state. $\Gamma_{\sigma_{2}\rightarrow KK}$ is larger
than the WA102 data presented in Eq.\ (\ref{f0(1710)_21}); however, our results
suggest nonetheless that $f_{0}(1710)\rightarrow KK$ is the most dominant
decay channel for this resonance -- in accordance with the data (see
Sec.\ \ref{bb}).

\item \textit{Eta decay channel:} we obtain $\Gamma_{\sigma_{1}\rightarrow
\eta\eta}=(40\mp1)$ MeV and $\Gamma_{\sigma_{2}\rightarrow\eta\eta}%
=60_{+5}^{-4}$ MeV. The former is lower than the values cited in
Refs.\ \cite{f0(1500)-CB-1992,Alde:1985kp} but note that the cited
publications did not consider in their Breit-Wigner fits that new decay
channels may open over the broad $f_{0}(1370)$ decay interval. The latter is
marginally (within errors)\ consistent with the value $\Gamma_{f_{0}%
(1710)\rightarrow\eta\eta}^{\text{WA102}}=(38.6\pm18.8)$ MeV from
Eq.\ (\ref{f0(1710)_21}).

\item \textit{Pion-kaon ratio:} $\Gamma_{\sigma_{1}\rightarrow\pi\pi}%
/\Gamma_{\sigma_{1}\rightarrow KK}=1.42_{+0.09}^{-0.05}$ is consistent with
the WA102 result stating $\Gamma_{f_{0}(1370)\rightarrow\pi\pi}/\Gamma_{f_{0}%
(1370)\rightarrow KK}=$ $2.17\pm1.23$ obtained from Ref.\ \cite{Barberis:1999}
and also qualitatively consistent with the result $\Gamma_{f_{0}%
(1370)\rightarrow\pi\pi}/\Gamma_{f_{0}(1370)\rightarrow KK}=$ $1.10\pm0.24$
obtained from the OBELIX data in Ref.\ \cite{Bargiotti:2003}.

\item \textit{The eta-pion ratios }read $\Gamma_{\sigma_{1}\rightarrow\eta\eta
}/\Gamma_{\sigma_{1}\rightarrow\pi\pi}=0.15\pm0.01$ and $\Gamma_{\sigma
_{2}\rightarrow\eta\eta}/\Gamma_{\sigma_{2}\rightarrow\pi\pi}=1.26_{+0.52}%
^{-0.27}$. The former is within the ratio $\Gamma_{f_{0}(1370)\rightarrow
\eta\eta}/\Gamma_{f_{0}(1370)\rightarrow\pi\pi}=0.19\pm0.07$ of
Ref.\ \cite{buggf0}. The latter is corresponds almost completely to the WA102
ratio $\Gamma_{f_{0}(1710)\rightarrow\eta\eta}^{\text{WA102}}/\Gamma
_{f_{0}(1710)\rightarrow\pi\pi}^{\text{WA102}}=2.4\pm1.04$ from
Eq.\ (\ref{f0(1710)_19}).

\item \textit{The eta-kaon ratios} read $\Gamma_{\sigma_{1}\rightarrow\eta\eta
}/\Gamma_{\sigma_{1}\rightarrow KK}=0.22\pm0.01$ and $\Gamma_{\sigma
_{2}\rightarrow\eta\eta}/\Gamma_{\sigma_{2}\rightarrow KK}=0.25\pm0.004$. To
our knowledge, there are no experimental results for the ratio $\Gamma
_{f_{0}(1370)\rightarrow\eta\eta}$ $/\Gamma_{f_{0}(1370)\rightarrow KK}$. Our
result for $\Gamma_{\sigma_{1}\equiv f_{0}(1370)\rightarrow\eta\eta}%
/\Gamma_{\sigma_{1}\equiv f_{0}(1370)\rightarrow KK}$ is hence a prediction.
Our value of the ratio $\Gamma_{\sigma_{2}\rightarrow\eta\eta}/\Gamma_{\sigma
_{2}\rightarrow KK}$ is completely within the combined-fit result of
Ref.\ \cite{Anisovich:2001} reading $\Gamma_{f_{0}(1710)\rightarrow\eta\eta
}/\Gamma_{f_{0}(1710)\rightarrow KK}=0.46_{-0.38}^{+0.70}$ and within
$2\sigma$\ of the WA102 result where $\Gamma_{f_{0}(1710)\rightarrow\eta\eta
}/\Gamma_{f_{0}(1710)\rightarrow KK}=0.48\pm0.15$ was obtained
\cite{Barberis:2000}.
\end{itemize}

For these reasons, the assumption of scalar $\bar{q}q$ states above $1$ GeV is
strongly preferred over the assumption that the same states are present below
$1$ GeV. Fit II describes non-strange scalars decisively better than Fit I
(see Sec.\ \ref{sec.conclusionsfitI}). Additionally, results obtained in this
section will allow us to explore three more decay channels of our $\sigma
_{1}\equiv f_{0}(1370)$ state: into $\eta\eta^{\prime}$, $a_{1}(1260)\pi$, and
$2\omega(782)$. Experimental information regarding these decays is scarce
\cite{PDG}; thus our results will have strong predictive power.

\subsection{Decay Width \boldmath $\sigma_{1,2}\rightarrow\eta\eta^{\prime}$} \label{sec.sigmaetaetap}

The interaction Lagrangian for this decay has already been presented in
Eq.\ (\ref{sigmaetaNetaS}). The Lagrangian contains the pure states $\sigma_{N,S}$
and $\eta_{N,S}$ and, as in Sec.\ \ref{sec.setaeta1}, we will first introduce
the fields $\eta$ and $\eta^{\prime}$ in accordance with Eqs.\ (\ref{etaN})
and (\ref{etaS}). The Lagrangian in Eq.\ (\ref{sigmaetaNetaS}) then obtains
the following form:

Substituting Eqs.\ (\ref{p}) and (\ref{a}) into Eq.\ (\ref{sigmaetaNetaS}) and
additionally substituting $\eta_{N}$ and $\eta_{S}$ by $\eta$ and
$\eta^{\prime}$ according to Eqs.\ (\ref{etaN}) and (\ref{etaS}), we obtain
the following form of the interaction Lagrangian:

\begin{align}
\mathcal{L}_{\sigma\eta\eta^{\prime}}  &  =A_{\sigma_{N}\eta\eta^{\prime}
}\sigma_{N}\eta\eta^{\prime}+B_{\sigma_{N}\eta\eta^{\prime}}\sigma
_{N}(\partial_{\mu}\eta)(\partial^{\mu}\eta^{\prime})+C_{\sigma_{N}\eta
\eta^{\prime}}\partial_{\mu}\sigma_{N}(\eta\partial^{\mu}\eta^{\prime}
+\eta^{\prime}\partial^{\mu}\eta)\nonumber\\
&  +A_{\sigma_{S}\eta\eta^{\prime}}\sigma_{S}\eta\eta^{\prime}+B_{\sigma
_{S}\eta\eta^{\prime}}\sigma_{S}(\partial_{\mu}\eta)(\partial^{\mu}
\eta^{\prime})+C_{\sigma_{S}\eta\eta^{\prime}}\partial_{\mu}\sigma_{S}
(\eta\partial^{\mu}\eta^{\prime}+\eta^{\prime}\partial^{\mu}\eta)
\label{sigmaetaetap}
\end{align}

with
\begin{align}
A_{\sigma_{N}\eta\eta^{\prime}}  &  =Z_{\pi}^{2}\phi_{N}\left(  \lambda
_{1}+\frac{\lambda_{2}}{2}+c_{1}\phi_{S}^{2}\right)  \sin(2\varphi_{\eta
})-Z_{\eta_{S}}^{2}\phi_{N}\left(  \lambda_{1}+\frac{c_{1}}{2}\phi_{N}
^{2}\right)  \sin(2\varphi_{\eta})\nonumber\\
&  -\frac{3}{2}c_{1}Z_{\pi}Z_{\eta_{S}}\phi_{N}^{2}\phi_{S}\cos(2\varphi
_{\eta})\nonumber\\
&  =\phi_{N}\left\{  \lambda_{1}(Z_{\pi}^{2}-Z_{\eta_{S}}^{2})\sin
(2\varphi_{\eta})+\frac{\lambda_{2}}{2}Z_{\pi}^{2}\sin(2\varphi_{\eta})\right.
\nonumber\\
&  \left.  +c_{1}\left[  \left(  Z_{\pi}^{2}\phi_{S}^{2}-\frac{Z_{\eta_{S}
}^{2}}{2}\phi_{N}^{2}\right)  \sin(2\varphi_{\eta})-\frac{3}{2}Z_{\pi}
Z_{\eta_{S}}\phi_{N}\phi_{S}\cos(2\varphi_{\eta})\right]  \right\}\text{,}
\label{ANsetaetap}\\
B_{\sigma_{N}\eta\eta^{\prime}}  &  =\left[  Z_{\pi}^{2}\frac{w_{a_{1}}^{2}
}{\phi_{N}}\left(  m_{1}^{2}+\frac{h_{1}}{2}\phi_{S}^{2}+2\delta_{N}\right)
+\frac{h_{1}}{2}Z_{\eta_{S}}^{2}w_{f_{1S}}^{2}\phi_{N}\right]  \sin
(2\varphi_{\eta})\text{,} \label{BNsetaetap}
\end{align}
\begin{align}
C_{\sigma_{N}\eta\eta^{\prime}}  &  =-\frac{g_{1}}{2}w_{a_{1}}Z_{\pi}^{2}
\sin(2\varphi_{\eta})\text{,} \label{CNsetaetap}\\
A_{\sigma_{S}\eta\eta^{\prime}}  &  =-(\lambda_{1}+\lambda_{2})Z_{\eta_{S}
}^{2}\phi_{S}\sin(2\varphi_{\eta})+Z_{\pi}^{2}\phi_{S}(\lambda_{1}+c_{1}
\phi_{N}^{2})\sin(2\varphi_{\eta})-\frac{1}{2}c_{1}Z_{\pi}Z_{\eta_{S}}\phi
_{N}^{3}\cos(2\varphi_{\eta})\nonumber\\
&  =\phi_{S}\left\{  \left[  \lambda_{1}(Z_{\pi}^{2}-Z_{\eta_{S}}^{2}
)-\lambda_{2}Z_{\eta_{S}}^{2}\right]  \sin(2\varphi_{\eta})+c_{1}Z_{\pi}
\phi_{N}^{2}\left[  Z_{\pi}\sin(2\varphi_{\eta})-\frac{Z_{\eta_{S}}}{2\phi
_{S}}\phi_{N}\cos(2\varphi_{\eta})\right]  \right\}\text{,} \label{ASsetaetap}\\
B_{\sigma_{S}\eta\eta^{\prime}}  &  =\left[  -Z_{\eta_{S}}^{2}\frac{w_{f_{1S}
}^{2}}{\phi_{S}}\left(  m_{1}^{2}+\frac{h_{1}}{2}\phi_{N}^{2}+2\delta
_{S}\right)  -\frac{h_{1}}{2}Z_{\pi}^{2}w_{a_{1}}^{2}\phi_{S}\right]
\sin(2\varphi_{\eta})\text{,} \label{BSsetaetap}\\
C_{\sigma_{S}\eta\eta^{\prime}}  &  =\frac{\sqrt{2}}{2}Z_{\eta_{S}}^{2}
g_{1}w_{f_{1S}}\sin(2\varphi_{\eta})\text{.} \label{CSsetaetap}
\end{align}

As in Eq.\ (\ref{sigmapionpion2}) we obtain from Eqs.\ (\ref{sigma-sigma_2})
and (\ref{sigmaetaetap})

\begin{align}
\mathcal{L}_{\sigma\eta\eta^{\prime}\text{, full}}  &  =\mathcal{L}
_{\sigma_{N}\sigma_{S},\,\mathrm{full}}+\mathcal{L}_{\sigma\eta\eta^{\prime}
}\nonumber\\
&  =\frac{1}{2}(\partial_{\mu}\sigma_{N})^{2}+\frac{1}{2}(\partial_{\mu}
\sigma_{S})^{2}-\frac{1}{2}m_{\sigma_{N}}^{2}-\frac{1}{2}m_{\sigma_{S}}
^{2}+z_{\sigma}\sigma_{N}\sigma_{S}\nonumber\\
&  +A_{\sigma_{N}\eta\eta^{\prime}}\sigma_{N}\eta\eta^{\prime}+B_{\sigma
_{N}\eta\eta^{\prime}}\sigma_{N}(\partial_{\mu}\eta)(\partial^{\mu}
\eta^{\prime})+C_{\sigma_{N}\eta\eta^{\prime}}\partial_{\mu}\sigma_{N}
(\eta\partial^{\mu}\eta^{\prime}+\eta^{\prime}\partial^{\mu}\eta)\nonumber\\
&  +A_{\sigma_{S}\eta\eta^{\prime}}\sigma_{S}\eta\eta^{\prime}+B_{\sigma
_{S}\eta\eta^{\prime}}\sigma_{S}(\partial_{\mu}\eta)(\partial^{\mu}
\eta^{\prime})+C_{\sigma_{S}\eta\eta^{\prime}}\partial_{\mu}\sigma_{S}
(\eta\partial^{\mu}\eta^{\prime}+\eta^{\prime}\partial^{\mu}\eta)\text{.}
\label{sigmaetaetap1}
\end{align}

$\mathcal{L}_{\sigma\eta\eta^{\prime}\text{, full}}$ can be transformed in the
following way:

\begin{align}
\mathcal{L}_{\sigma\eta\eta\text{, full}}  &  =\frac{1}{2}(\partial_{\mu
}\sigma_{1})^{2}-\frac{1}{2}m_{\sigma_{1}}^{2}\sigma_{1}^{2}\nonumber\\
&  +(A_{\sigma_{N}\eta\eta^{\prime}}\cos\varphi_{\sigma}+A_{\sigma_{S}\eta
\eta^{\prime}}\sin\varphi_{\sigma})\sigma_{1}\eta\eta^{\prime}\nonumber\\
&  +(B_{\sigma_{N}\eta\eta^{\prime}}\cos\varphi_{\sigma}+B_{\sigma_{S}\eta
\eta^{\prime}}\sin\varphi_{\sigma})\sigma_{1}(\partial_{\mu}\eta
)(\partial^{\mu}\eta^{\prime})\nonumber\\
&  +(C_{\sigma_{N}\eta\eta^{\prime}}\cos\varphi_{\sigma}+C_{\sigma_{S}\eta
\eta^{\prime}}\sin\varphi_{\sigma})\partial_{\mu}\sigma_{1}(\eta\partial^{\mu
}\eta^{\prime}+\eta^{\prime}\partial^{\mu}\eta)\nonumber\\
&  +\frac{1}{2}(\partial_{\mu}\sigma_{2})^{2}-\frac{1}{2}m_{\sigma_{2}}
^{2}\sigma_{2}^{2}\nonumber\\
&  +(A_{\sigma_{S}\eta\eta^{\prime}}\cos\varphi_{\sigma}-A_{\sigma_{N}\eta
\eta^{\prime}}\sin\varphi_{\sigma})\sigma_{2}\eta\eta^{\prime}\nonumber\\
&  +(B_{\sigma_{S}\eta\eta^{\prime}}\cos\varphi_{\sigma}-B_{\sigma_{N}\eta
\eta^{\prime}}\sin\varphi_{\sigma})\sigma_{2}(\partial_{\mu}\eta
)(\partial^{\mu}\eta^{\prime})\nonumber\\
&  +(C_{\sigma_{S}\eta\eta^{\prime}}\cos\varphi_{\sigma}-C_{\sigma_{N}\eta
\eta^{\prime}}\sin\varphi_{\sigma})\partial_{\mu}\sigma_{2}(\eta\partial^{\mu
}\eta^{\prime}+\eta^{\prime}\partial^{\mu}\eta)\text{.}
\end{align}

Let us set $P$ as the momentum of $\sigma_{1}$ or $\sigma_{2}$ (depending on the
decaying particle) and $P_{1}$ and $P_{2}$ as the momenta of the $\eta$ and
$\eta^{\prime}$ fields, respectively. Upon substituting $\partial^{\mu
}\rightarrow-iP^{\mu}$\ for the decaying particles and $\partial^{\mu}\rightarrow
iP_{1,2}^{\mu}$ for the decay products, the decay amplitudes of the mixed states
$\sigma_{1,2}$ read

\begin{align}
-i\mathcal{M}_{\sigma_{1}\rightarrow\eta\eta^{\prime}}(m_{\sigma_{1}})  &
=i\left\{  \cos\varphi_{\sigma}(A_{\sigma_{N}\eta\eta^{\prime}}-B_{\sigma
_{N}\eta\eta^{\prime}}P_{1}\cdot P_{2}+C_{\sigma_{N}\eta\eta^{\prime}}
P\cdot(P_{1}+P_{2})\right. \nonumber\\
&  \left.  +\sin\varphi_{\sigma}\left[  A_{\sigma_{S}\eta\eta^{\prime}
}-B_{\sigma_{S}\eta\eta^{\prime}}P_{1}\cdot P_{2}+C_{\sigma_{S}\eta
\eta^{\prime}}P\cdot(P_{1}+P_{2})\right]  \right\} \nonumber\\
&  =i\left\{  \cos\varphi_{\sigma}\left[  A_{\sigma_{N}\eta\eta^{\prime}
}-B_{\sigma_{N}\eta\eta^{\prime}}\frac{m_{\sigma_{1}}^{2}-m_{\eta}^{2}
-m_{\eta^{\prime}}^{2}}{2}+C_{\sigma_{N}\eta\eta^{\prime}}m_{\sigma_{1}}
^{2}\right]  \right. \nonumber\\
&  \left.  +\sin\varphi_{\sigma}\left[  A_{\sigma_{S}\eta\eta^{\prime}
}-B_{\sigma_{S}\eta\eta^{\prime}}\frac{m_{\sigma_{1}}^{2}-m_{\eta}^{2}
-m_{\eta^{\prime}}^{2}}{2}+C_{\sigma_{S}\eta\eta^{\prime}}m_{\sigma_{1}}
^{2}\right]  \right\}\text{,} \label{Ms1etaetap}
\end{align}
\begin{align}
-i\mathcal{M}_{\sigma_{2}\rightarrow\eta\eta^{\prime}}(m_{\sigma_{2}})  &
=i\left\{  \cos\varphi_{\sigma}\left[  A_{\sigma_{S}\eta\eta^{\prime}
}-B_{\sigma_{S}\eta\eta^{\prime}}\frac{m_{\sigma_{2}}^{2}-m_{\eta}^{2}
-m_{\eta^{\prime}}^{2}}{2}+C_{\sigma_{S}\eta\eta^{\prime}}m_{\sigma_{2}}
^{2}\right]  \right. \nonumber\\
&  \left.  -\sin\varphi_{\sigma}\left[  A_{\sigma_{N}\eta\eta^{\prime}
}-B_{\sigma_{N}\eta\eta^{\prime}}\frac{m_{\sigma_{2}}^{2}-m_{\eta}^{2}
-m_{\eta^{\prime}}^{2}}{2}+C_{\sigma_{N}\eta\eta^{\prime}}m_{\sigma_{2}}
^{2}\right]  \right\}  \text{.} \label{Ms2etaetap}
\end{align}

Note that we have used the identity $P^{2}=(P_{1}+P_{2})^{2}\Leftrightarrow
P_{1}\cdot P_{2}=(P^{2}-P_{1}^{2}-P_{2}^{2})/2=(m_{\sigma_{1}}^{2}-m_{\eta
}^{2}-m_{\eta^{\prime}}^{2})/2$ in Eqs.\ (\ref{Ms1etaetap}) and
(\ref{Ms2etaetap}).

Finally, we obtain the following decay widths formulas:

\begin{align}
\Gamma_{\sigma_{1}\rightarrow\eta\eta^{\prime}}  &  =\frac{k(m_{\sigma_{1}
},m_{\eta},m_{\eta^{\prime}})}{8\pi m_{\sigma_{1}}^{2}}|-i\mathcal{M}
_{\sigma_{1}\rightarrow\eta\eta^{\prime}}(m_{\sigma_{1}})|^{2}\text{,}
\label{Gs1etaetap}\\
\Gamma_{\sigma_{2}\rightarrow\eta\eta^{\prime}}  &  =\frac{k(m_{\sigma_{2}
},m_{\eta},m_{\eta^{\prime}})}{8\pi m_{\sigma_{2}}^{2}}|-i\mathcal{M}
_{\sigma_{2}\rightarrow\eta\eta^{\prime}}(m_{\sigma_{2}})|^{2}\text{.}
\label{Gs2etaetap}
\end{align}

Note that the $\eta\eta^{\prime}$ threshold in our model lies at $1481$ MeV
according to Table \ref{Fit2-5}. For this reason, a non-vanishing value
$\Gamma_{\sigma_{1}\rightarrow\eta\eta^{\prime}}$ could only be obtained for
correspondingly large $m_{\sigma_{1}}$ (that can actually be smaller than the
threshold value if the state is sufficiently broad). Our model yields
$m_{\sigma_{1}}=1310_{+30}^{-29}$ MeV, see Sec.\ \ref{sec.scalars2}, and thus
a value that does not allow for a tree-level $\sigma_{1}\rightarrow\eta
\eta^{\prime}$ decay and renders an off-shell-$\sigma_{1}$ decay extremely suppressed.

The situation is quite different for $\sigma_{2}$. Constraining $\Gamma
_{\sigma_{2}\rightarrow\eta\eta^{\prime}}$ in Eq.\ (\ref{Gs2etaetap}) via
$m_{\sigma_{2}}=1606_{+4}^{-3}$ MeV (determined in Sec.\ \ref{sec.scalars2})
and using the parameters in Table \ref{Fit2-4} and masses in Table \ref{Fit2-5},
we obtain

\begin{equation}
\Gamma_{\sigma_{2}\rightarrow\eta\eta^{\prime}}=41_{-5}^{+4}\text{ MeV.}
\label{Gs2etaetap1}
\end{equation}

This result is a prediction because the PDG does not report an
$\eta\eta^{\prime}$ channel for $f_{0}(1710)\equiv\sigma_{2}$. Note, however,
that results for the absolute values of the partial $f_{0}(1710)$ decay widths
tend to be larger than experimental data (as discussed in
Sec.\ \ref{sec.scalars2}); nonetheless, they also have correct relative
magnitudes and for this reason we conclude that a non-vanishing value of
$\Gamma_{f_{0}(1710)\rightarrow\eta\eta^{\prime}}$ is expected, suppressed
when compared to $f_{0}(1710)\rightarrow KK$ but of approximately equal
magnitude as $f_{0}(1710)\rightarrow\pi\pi$ and $f_{0}(1710)\rightarrow
\eta\eta$. Indeed using Eqs.\ (\ref{Gs2etaetap}) and (\ref{Gs1KK}) as well as
$m_{\sigma_{2}}=1606_{+4}^{-3}$ MeV we obtain

\begin{equation}
\Gamma_{\sigma_{2}\rightarrow\eta\eta^{\prime}}/\Gamma_{\sigma_{2}\rightarrow
KK}=0.17_{-0.03}^{+0.04}\text{,} \label{Gs2etaetap2}
\end{equation}

using Eqs.\ (\ref{Gs2etaetap}) and (\ref{Gs2pp}) we obtain

\begin{equation}
\Gamma_{\sigma_{2}\rightarrow\eta\eta^{\prime}}/\Gamma_{\sigma_{2}
\rightarrow\pi\pi}=0.86_{+0.11}^{-0.06}\text{,} \label{Gs2etaetap3}
\end{equation}

using Eqs.\ (\ref{Gs2etaetap}) and (\ref{Gs2etaeta}) we obtain

\begin{equation}
\Gamma_{\sigma_{2}\rightarrow\eta\eta^{\prime}}/\Gamma_{\sigma_{2}
\rightarrow\eta\eta}=0.68\pm0.13\text{.} \label{Gs2etaetap4}
\end{equation}

There are no experimental results for these ratios -- the results are pure
predictions. We know from Sec.\ \ref{sec.scalars2} that experimental
ratios of scalar decay widths are better described in our model than
absolute values of decay widths. For this reason, experimental measurements
regarding $f_{0}(1710)\rightarrow\eta\eta^{\prime}$ would be strongly
appreciated and would represent a valuable test for our results in
Eqs.\ (\ref{Gs2etaetap2}) - (\ref{Gs2etaetap4}).

\subsection{Decay Width \boldmath $\sigma_{1,2}\rightarrow a_{1}(1260)\pi
\rightarrow\rho\pi\pi$} \label{sec.sigmaa1pion}

This decay width can be calculated from the same Lagrangian as the one stated
in Eq.\ (\ref{a1sp}). However, the calculation of the decay width is in this case
slightly different than the one presented in Sec.\ \ref{sec.ASP} because the
$a_{1}(1260)$ spectral function has to be considered. Additionally, $a_{1}$ is
no longer at rest (unlike the decaying axial-vector in Sec.\ \ref{sec.ASP}).
This has to be considered while the decay amplitude is calculated (see below).

The decay width is determined as follows. The pure states $\sigma_{N,S}$ in
Eq.\ (\ref{a1sp}) need to be replaced by the physical states $\sigma_{1,2}%
$\ according to the inverted Eq.\ (\ref{sigma-sigma_1}):

\begin{align}
\mathcal{L}_{\sigma a_{1}\pi}  &  =A_{a_{1}\sigma_{N}\pi}a_{1}^{\mu0}
(\cos\varphi_{\sigma}\sigma_{1}-\sin\varphi_{\sigma}\sigma_{2})\partial_{\mu
}\pi^{0}+B_{a_{1}\sigma_{N}\pi}a_{1}^{\mu0}\pi^{0}\partial_{\mu}(\cos
\varphi_{\sigma}\sigma_{1}-\sin\varphi_{\sigma}\sigma_{2})\nonumber\\
&  +A_{a_{1}\sigma_{S}\pi}a_{1}^{\mu0}(\sin\varphi_{\sigma}\sigma_{1}
+\cos\varphi_{\sigma}\sigma_{2})\partial_{\mu}\pi^{0}\nonumber\\
&  =(A_{a_{1}\sigma_{N}\pi}\cos\varphi_{\sigma}+A_{a_{1}\sigma_{S}\pi}
\sin\varphi_{\sigma})\sigma_{1}a_{1}^{\mu0}\partial_{\mu}\pi^{0}
+B_{a_{1}\sigma_{N}\pi}\cos\varphi_{\sigma}a_{1}^{\mu0}\pi^{0}\partial_{\mu
}\sigma_{1}\nonumber\\
&  +(A_{a_{1}\sigma_{S}\pi}\cos\varphi_{\sigma}-A_{a_{1}\sigma_{N}\pi}
\sin\varphi_{\sigma})\sigma_{2}a_{1}^{\mu0}\partial_{\mu}\pi^{0}
-B_{a_{1}\sigma_{N}\pi}\sin\varphi_{\sigma}a_{1}^{\mu0}\pi^{0}\partial_{\mu
}\sigma_{2}\nonumber\\
&  =A_{a_{1}\sigma_{1}\pi}\sigma_{1}a_{1}^{\mu0}\partial_{\mu}\pi^{0}
+B_{a_{1}\sigma_{1}\pi}a_{1}^{\mu0}\pi^{0}\partial_{\mu}\sigma_{1}\nonumber\\
&  +A_{a_{1}\sigma_{2}\pi}\sigma_{2}a_{1}^{\mu0}\partial_{\mu}\pi^{0}
+B_{a_{1}\sigma_{2}\pi}a_{1}^{\mu0}\pi^{0}\partial_{\mu}\sigma_{2}
\label{sa1p}
\end{align}

with $A_{a_{1}\sigma_{N}\pi}$, $B_{a_{1}\sigma_{N}\pi}$ and $A_{a_{1}
\sigma_{S}\pi}$ from Eqs.\ (\ref{Aa1sNp}) - (\ref{Aa1sSp}), $A_{a_{1}
\sigma_{1}\pi}=A_{a_{1}\sigma_{N}\pi}\cos\varphi_{\sigma}+A_{a_{1}\sigma
_{S}\pi}\sin\varphi_{\sigma}$, $B_{a_{1}\sigma_{1}\pi}=B_{a_{1}\sigma_{N}\pi
}\cos\varphi_{\sigma}$, $A_{a_{1}\sigma_{2}\pi}=A_{a_{1}\sigma_{S}\pi}
\cos\varphi_{\sigma}-A_{a_{1}\sigma_{N}\pi}\sin\varphi_{\sigma}$ and
$B_{a_{1}\sigma_{2}\pi}=-B_{a_{1}\sigma_{N}\pi}\sin\varphi_{\sigma}$.

Let us consider only the decay $\sigma_{1}\rightarrow a_{1}\pi$ in the
following; the calculation of $\Gamma_{\sigma_{2}\rightarrow a_{1}\pi}$ is
analogous. We denote the momenta of $\sigma_{1}$, $a_{1}$ and $\pi$ as $P$,
$P_{1}$ and $P_{2}$, respectively. Then, upon substituting $\partial^{\mu
}\rightarrow-iP^{\mu}$\ for the decaying particle and $\partial^{\mu
}\rightarrow iP_{1,2}^{\mu}$ for the decay products, we obtain the following
Lorentz-invariant $\sigma_{1}a_{1}\pi$ scattering amplitude $-i\mathcal{M}
_{\sigma_{1}\rightarrow a_{1}\pi}^{(\alpha)}$:

\begin{equation}
-i\mathcal{M}_{\sigma_{1}\rightarrow a_{1}\pi}^{(\alpha)}=\varepsilon_{\mu
}^{(\alpha)}(P_{1})h_{\sigma_{1}a_{1}\pi}^{\mu}=-\varepsilon_{\mu}^{(\alpha
)}(P_{1})\left(  A_{\sigma_{1}a_{1}\pi}P_{2}^{\mu}-B_{\sigma_{1}a_{1}\pi
}P^{\mu}\right)\text{,}
\end{equation}

where $\varepsilon_{\mu}^{(\alpha)}(P_{1})$ denotes the polarisation tensor of
$a_{1}$\ and

\begin{equation}
h_{\sigma_{1}a_{1}\pi}^{\mu}=-\left(  A_{\sigma_{1}a_{1}\pi}P_{2}^{\mu
}-B_{\sigma_{1}a_{1}\pi}P^{\mu}\right)  \label{hsa1p}
\end{equation}

denotes the $\sigma_{1}a_{1}\pi$ vertex.\newline

It is now necessary to calculate the square of the averaged decay amplitude.
This is performed analogously to Sec.\ \ref{sec.ASP}:

\begin{align}
-i\mathcal{M}_{\sigma_{1}\rightarrow a_{1}\pi}^{(\alpha)}  &  =\varepsilon
_{\mu}^{(\alpha)}(P_{1})h_{\sigma_{1}a_{1}\pi}^{\mu}\Rightarrow\left\vert
-i\mathcal{\bar{M}}_{\sigma_{1}\rightarrow a_{1}\pi}\right\vert ^{2}=\frac
{1}{3}\sum\limits_{\alpha=1}^{3}\left\vert -i\mathcal{M}_{\sigma
_{1}\rightarrow a_{1}\pi}^{(\alpha)}\right\vert ^{2}\nonumber\\
&  =\frac{1}{3}\sum\limits_{\alpha,\beta=1}^{3}\varepsilon_{\mu}^{(\alpha
)}(P_{1})h_{\sigma_{1}a_{1}\pi}^{\mu}\varepsilon_{\nu}^{(\alpha)}
(P_{1})h_{\sigma_{1}a_{1}\pi}^{\ast\nu}\nonumber\\
&  \overset{\text{Eq.\ (\ref{iMASP1})}}{=}\frac{1}{3}\left[  -\left\vert
h_{\sigma_{1}a_{1}\pi}^{\mu}\right\vert ^{2}+\frac{\left\vert h_{\sigma
_{1}a_{1}\pi}^{\mu}P_{1\mu}\right\vert ^{2}}{m_{a_{1}}^{2}}\right]  \text{.}
\label{iMsa1p}
\end{align}

Let us determine the two contributions to $\left\vert -i\mathcal{\bar{M}
}_{\sigma_{1}\rightarrow a_{1}\pi}\right\vert ^{2}$ in Eq.\ (\ref{iMsa1p}).
The square of the vertex reads

\begin{align}
\left\vert h_{\sigma_{1}a_{1}\pi}^{\mu}\right\vert ^{2}  &  =A_{\sigma
_{1}a_{1}\pi}^{2}m_{\pi}^{2}+B_{\sigma_{1}a_{1}\pi}^{2}m_{\sigma_{1}}
^{2}-2A_{\sigma_{1}a_{1}\pi}B_{\sigma_{1}a_{1}\pi}P\cdot P_{2}\nonumber\\
&  =A_{\sigma_{1}a_{1}\pi}^{2}m_{\pi}^{2}+B_{\sigma_{1}a_{1}\pi}^{2}
m_{\sigma_{1}}^{2}-2A_{\sigma_{1}a_{1}\pi}B_{\sigma_{1}a_{1}\pi}m_{\sigma_{1}
}E_{2}(x_{a_{1}})\text{.} \label{iMsa1p1}
\end{align}

In the second line of Eq.\ (\ref{iMsa1p1}) we have used $P\cdot P_{2}
=m_{\sigma_{1}}E_{2}(x_{a_{1}})$ with the pion energy $E_{2}(x_{a_{1}})=\sqrt{k^{2}
(m_{\sigma_{1}},x_{a_{1}},m_{{\pi}})+m_{{\pi}}^{2}}$, $k(m_{\sigma_{1}
},x_{a_{1}},m_{{\pi}})$ from Eq.\ (\ref{kabc}) and $x_{a_{1}}$ the running
mass of the $a_{1}$ state.

The second term from Eq.\ (\ref{iMsa1p}) is calculated from

\begin{equation}
\left\vert h_{\sigma_{1}a_{1}\pi}^{\mu}P_{1\mu}\right\vert ^{2}=(A_{\sigma
_{1}a_{1}\pi}P_{1}\cdot P_{2}-B_{\sigma_{1}a_{1}\pi}P\cdot P_{1})^{2}
\label{iMsa1p2}
\end{equation}

and the equalities $P_{1}\cdot P_{2}=(m_{\sigma_{1}}^{2}-x_{a_{1}}^{2}-m_{{\pi}
}^{2})/2$ and $P\cdot P_{1}=m_{\sigma_{1}}E_{1}(x_{a_{1}})$ with the $a_1$ energy
$E_{1}(x_{a_{1}})=\sqrt{k^{2}(m_{\sigma_{1}},x_{a_{1}},m_{{\pi}})+x_{a_{1}
}^{2}}$. Then we obtain

\begin{align}
\left\vert h_{\sigma_{1}a_{1}\pi}^{\mu}P_{1\mu}\right\vert ^{2} & =A_{\sigma
_{1}a_{1}\pi}^{2}\frac{(m_{\sigma_{1}}^{2}-x_{a_{1}}^{2}-m_{{\pi}}^{2})^{2}
}{4}+B_{\sigma_{1}a_{1}\pi}^{2}m_{\sigma_{1}}^{2}E_{1}^{2}(x_{a_{1}
}) \nonumber \\
& - A_{\sigma_{1}a_{1}\pi}B_{\sigma_{1}a_{1}\pi}m_{\sigma_{1}}E_{1}(x_{a_{1}
})(m_{\sigma_{1}}^{2}-x_{a_{1}}^{2}-m_{{\pi}}^{2})\text{.} \label{iMsa1p3}
\end{align}

Inserting Eqs.\ (\ref{iMsa1p1}) and (\ref{iMsa1p3}) into Eqs.\ (\ref{iMsa1p})
we obtain

\begin{align}
\left\vert -i\mathcal{\bar{M}}_{\sigma_{1}\rightarrow a_{1}\pi}(x_{a_{1}
})\right\vert ^{2}  &  =\frac{1}{3}\left\{  A_{\sigma_{1}a_{1}\pi}^{2}
\frac{(m_{\sigma_{1}}^{2}-x_{a_{1}}^{2}-m_{{\pi}}^{2})^{2}-4m_{a_{1}}
^{2}m_{{\pi}}^{2}}{4m_{a_{1}}^{2}}+B_{\sigma_{1}a_{1}\pi}^{2}m_{\sigma_{1}
}^{2}\left[  \frac{E_{1}^{2}(x_{a_{1}})}{m_{a_{1}}^{2}}-1\right]  \right.
\nonumber\\
&  \left.  -A_{\sigma_{1}a_{1}\pi}B_{\sigma_{1}a_{1}\pi}m_{\sigma_{1}}\left[
\frac{E_{1}(x_{a_{1}})(m_{\sigma_{1}}^{2}-x_{a_{1}}^{2}-m_{{\pi}}^{2}
)}{m_{a_{1}}^{2}}-2E_{2}(x_{a_{1}})\right]  \right\}  \text{.}
\label{iMs1a1p4}
\end{align}

The decay width needs to consider three possible decay channels: $\sigma
_{1}\rightarrow a_{1}^{0}\pi^{0}$ and $\sigma_{1}\rightarrow a_{1}^{\pm}
\pi^{\mp}$. Then we obtain

\begin{equation}
\Gamma_{\sigma_{1}\rightarrow a_{1}\pi}(x_{a_{1}})=\frac{3k(m_{\sigma_{1}
},x_{a_{1}},m_{{\pi}})}{8\pi m_{\sigma_{1}}^{2}}\left\vert -i\mathcal{\bar{M}
}_{\sigma_{1}\rightarrow a_{1}\pi}(x_{a_{1}})\right\vert ^{2}\text{.}
\label{Gs1a1p1}%
\end{equation}

Additionally, we introduce the $a_{1}$ spectral function $d_{a_{1}}(x_{a_{1}
})$ as in Sec.\ \ref{sec.AVP} assuming the decay width of $\Gamma_{a_{1}\rightarrow\rho
\pi}^{\exp}=425$ MeV for $a_{1}(1260)$ -- this is the mean value of the
corresponding PDG interval reading $(250-600)$ MeV. Then we can determine the
decay width $\Gamma_{\sigma_{1}\rightarrow a_{1}\pi\rightarrow\rho\pi\pi}$:

\begin{equation}
\Gamma_{\sigma_{1}\rightarrow a_{1}\pi\rightarrow\rho\pi\pi}=
{\displaystyle\int\limits_{0}^{\infty}}
\text{d}x_{a_{1}}\Gamma_{\sigma_{1}\rightarrow a_{1}\pi}(x_{a_{1}})d_{a_{1}
}(x_{a_{1}})\text{,} \label{Gs1a1p2}
\end{equation}

where the spectral function reads

\begin{equation}
d_{a_{1}}(x_{a_{1}})=N_{a_{1}}\,\frac{x_{a_{1}}^{2}\Gamma_{a_{1}
\rightarrow\rho\pi}^{\exp}}{(x_{a_{1}}^{2}-m_{a_{1}}^{2})^{2}+\left(
x_{a_{1}}\Gamma_{a_{1}\rightarrow\rho\pi}^{\exp}\right)  ^{2}}\,\theta
(x_{a_{1}}-m_{\rho}-m_{\pi}) \label{d_a_1}
\end{equation}

with the constant $N_{a_{1}}$ determined such that $\int_{0}^{\infty
}\mathrm{d}x_{a_{1}}\,d_{a_{1}}(x_{a_{1}})=1$. [We are using $\Gamma
_{a_{1}\rightarrow\rho\pi}^{\exp}$ in $d_{a_{1}}(x_{a_{1}})$ as a first
approximation although in principle the fully parametrised $\Gamma
_{a_{1}\rightarrow\rho\pi}$ from our model should be used. The ensuing results
are thus more of qualitative nature.]

Analogously, we obtain
\begin{equation}
\Gamma_{\sigma_{2}\rightarrow a_{1}\pi}(x_{a_{1}})=\frac{3k(m_{\sigma_{2}
},x_{a_{1}},m_{{\pi}})}{8\pi m_{\sigma_{2}}^{2}}\left\vert -i\mathcal{\bar{M}
}_{\sigma_{2}\rightarrow a_{1}\pi}(x_{a_{1}})\right\vert ^{2}\text{.}
\label{Gs2a1p1}
\end{equation}

and

\begin{equation}
\Gamma_{\sigma_{2}\rightarrow a_{1}\pi\rightarrow\rho\pi\pi}=
{\displaystyle\int\limits_{0}^{\infty}}
\text{d}x_{a_{1}}\Gamma_{\sigma_{2}\rightarrow a_{1}\pi}(x_{a_{1}})d_{a_{1}
}(x_{a_{1}})\text{.} \label{Gs2a1p2}
\end{equation}

We use the parameter values stated in Table \ref{Fit2-4} to determine the coefficients
in Eq.\ (\ref{sa1p}). Mass values can be found in Table \ref{Fit2-4}, except
for $m_{\sigma_{1}}=1310_{+30}^{-29}$ MeV and $m_{\sigma_{2}}=1606_{+4}^{-3}$
MeV determined in Sec.\ \ref{sec.scalars2}. Then Eq.\ (\ref{Gs1a1p2}) yields

\begin{equation}
\Gamma_{\sigma_{1}\rightarrow a_{1}\pi\rightarrow\rho\pi\pi}=12.7_{-4.2}
^{+5.8}\text{ MeV.}
\end{equation}

Consequently, the decay channel $f_{0}(1370)\rightarrow a_{1}(1260)\pi
\rightarrow\rho\pi\pi$ is strongly suppressed in our model. The PDG does not
state a value for this decay width but rather notes the Crystal Barrel ratio
$\Gamma_{f_{0}(1370)\rightarrow a_{1}(1260)\pi}/\Gamma_{f_{0}(1370)\rightarrow
4\pi}=0.06\pm0.02$ \cite{Abele:2001pv}. Our results do not reproduce the
stated ratio because the absence of the glueball field in our $U(3)\times
U(3)$\ model implies a very small $\Gamma_{f_{0}(1370)\rightarrow4\pi}$ (see
the note on $\sigma_{1,2}\rightarrow4\pi$ decays in
Sec.\ \ref{sec.sigmapionpion2}). Nonetheless, the results of
Ref.\ \cite{Abele:2001pv} imply a suppressed decay of $f_{0}(1370)$ into
$a_{1}(1260)$ and this is consistent with our finding.

Additionally, from Eq.\ (\ref{Gs2a1p2}) we obtain

\begin{equation}
\Gamma_{\sigma_{2}\rightarrow a_{1}\pi\rightarrow\rho\pi\pi}=15.2_{-3.1}
^{+2.6}\text{ MeV.}
\end{equation}

Current PDG data do not suggest the existence of the decay channel $f_{0}
(1710)\rightarrow a_{1}(1260)\pi\rightarrow\rho\pi\pi$ \cite{PDG}; indeed we
find it to be strongly suppressed in comparison to pion, kaon and eta decays
of $f_{0}(1710)$. Nonetheless, our results imply that a small but definite
signal should be observed for this resonance as well.

\subsection{Decay Width \boldmath $\sigma_{2} \rightarrow\omega\omega$} \label{sec.sigmaomegaomega}

The $\sigma\omega_{N}\omega_{N}$ interaction Lagrangian reads

\begin{align}
\mathcal{L}_{\sigma\omega\omega}  &  =\frac{1}{2}(h_{1}+h_{2}+h_{3})\phi
_{N}\sigma_{N}(\omega_{N}^{\mu})^{2}+\frac{1}{2}h_{1}\phi_{S}\sigma_{S}
(\omega_{N}^{\mu})^{2}\nonumber\\
&  =\frac{1}{2}\left[  (h_{1}+h_{2}+h_{3})\phi_{N}\cos\varphi_{\sigma}
+h_{1}\phi_{S}\sin\varphi_{\sigma}\right]  \sigma_{1}(\omega_{N}^{\mu}
)^{2}\nonumber\\
&  +\frac{1}{2}\left[  h_{1}\phi_{S}\sigma_{S}\cos\varphi_{\sigma}
-(h_{1}+h_{2}+h_{3})\phi_{N}\sin\varphi_{\sigma}\right]  \sigma_{2}(\omega
_{N}^{\mu})^{2} \label{sigmaomegaomega}
\end{align}

with the substitutions $\sigma_{N}\rightarrow\cos\varphi_{\sigma}\sigma_{1}%
-\sin\varphi_{\sigma}\sigma_{2}$ and $\sigma_{S}\rightarrow\sin
\varphi_{\sigma}\sigma_{1}+\cos\varphi_{\sigma}\sigma_{2}$. Let us remind
ourselves of the assignment of the relevant fields: $\sigma_{1}\equiv
f_{0}(1370)$, $\sigma_{2}\equiv f_{0}(1710)$, $\omega_{N}\equiv\omega
(782)=\omega$. The state $\sigma_{1}$ is below the $\omega\omega$ threshold
and, for that reason, we do not consider the corresponding decay. Conversely,
the state $\sigma_{2}$ is above, but not far away from, the $\omega\omega$
threshold: $m_{\sigma_{2}}=1606_{+4}^{-3}$ MeV (see Sec.\ \ref{sec.scalars2}).
Nonetheless, the decay is kinematically possible. We have already considered
the decay of a scalar state into two vectors in Sec.\ \ref{sec.SVV}. We can modify the formula
for the decay width obtained there for the purposes of this section:

\begin{align}
\Gamma_{\sigma_{2}\rightarrow\omega_{N}\omega_{N}}(x_{\omega_{N}}) & =\frac{k(m_{\sigma_{2}},x_{\omega_{N}},x_{\omega_{N}})}{16\pi m_{\sigma_{2}
}^{2}}\left[  (h_{1}+h_{2}+h_{3})\phi_{N}\sin\varphi_{\sigma}-h_{1}\phi
_{S}\cos\varphi_{\sigma}\right]  ^{2} \nonumber \\
& \times \left[  1+\frac{(m_{\sigma_{2}}
^{2}-2x_{\omega_{N}}^{2})^{2}}{8m_{\omega_{N}}^{4}}\right]  \label{Gsoo}
\end{align}

with $x_{\omega_{N}}$ denoting the running $\omega$ mass and $k(m_{\sigma_{1}
},x_{a_{1}},m_{{\pi}})$ from Eq.\ (\ref{kabc}). There are two ways to proceed
with Eq.\ (\ref{Gsoo}). It is possible to evaluate $\Gamma_{\sigma
_{2}\rightarrow\omega\omega}$ at the point $x_{\omega_{N}}=m_{\omega_{N}
}=775.49$ MeV [$=m_{\rho}$ according to Eq.\ (\ref{m_rho})]. The mass value
stems from Table \ref{Fit2-5}. Then using the parameter values from Table
\ref{Fit2-4} we obtain the following result from Eq.\ (\ref{Gsoo}):

\begin{equation}
\Gamma_{\sigma_{2}\rightarrow\omega_{N}\omega_{N}}(m_{\omega_{N}}
)\simeq0.02\text{ MeV.} \label{Gsoo2}
\end{equation}

Therefore the value is extremely small. The experimental situation regarding the
decay $\sigma_{2}\equiv$ $f_{0}(1710)\rightarrow\omega\omega$ is uncertain:
the existence of a weak signal was claimed only by the BES II Collaboration in
$J/\psi\rightarrow\gamma\omega\omega$ decays but no values were cited for the
partial decay width \cite{Ablikim:2006}. Our results are in qualitative
agreement with the BES II result, and our model does not suggest a strong
enhancement of $f_{0}(1710)$ in the $\omega\omega$ channel.

Equation (\ref{Gsoo}) can also be used to consider the sequential decay
$\sigma_{2}\equiv$ $f_{0}(1710)\rightarrow\omega\omega\rightarrow6\pi$:

\begin{equation}
\Gamma_{\sigma_{2}\rightarrow\omega\omega\rightarrow6\pi}=
{\displaystyle\int\limits_{0}^{\infty}}
\text{d}x_{\omega_{N}}\Gamma_{\sigma_{2}\rightarrow\omega_{N}\omega_{N}
}(x_{\omega_{N}})d_{\omega_{N}}(x_{\omega_{N}})\text{,} \label{Gsoo1}
\end{equation}

where $d_{\omega_{N}}(x_{\omega_{N}})$ denotes the spectral function of the
$\omega_{N}$ state:

\begin{equation}
d_{\omega_{N}}(x_{\omega_{N}})=N_{\omega_{N}}\,\frac{x_{\omega_{N}}^{2}
\Gamma_{\omega_{N}\rightarrow3\pi}^{\exp}}{(x_{\omega_{N}}^{2}-m_{\omega_{N}
}^{2})^{2}+\left(  x_{\omega_{N}}\Gamma_{\omega_{N}\rightarrow3\pi}^{\exp
}\right)  ^{2}}\,\theta(x_{\omega_{N}}-3m_{\pi})
\end{equation}

with $\Gamma_{\omega_{N}\rightarrow3\pi}^{\exp}=8.49$ MeV assuming, in
excellent approximation, that $\omega(782)$ decays only into $3\pi
$\ \cite{PDG} and $N_{\omega_{N}}$ determined such that $\int_{0}^{\infty
}\mathrm{d}x_{\omega_{N}}\,d_{\omega_{N}}(x_{\omega_{N}})=1$. The
$\omega(782)$ resonance is very narrow and therefore utilisation of
$\Gamma_{\omega_{N}\rightarrow3\pi}^{\exp}$ (rather than a formula
parametrised in our model) in $d_{\omega_{N}}(x_{\omega_{N}})$ is fully
justified. Nonetheless, the result obtained is the same as the one in
Eq.\ (\ref{Gsoo2})
\begin{equation}
\Gamma_{\sigma_{2}\rightarrow\omega\omega\rightarrow6\pi}\simeq0.02\text{
MeV.} \label{Gsoo3}
\end{equation}

The reason is the narrowness of the $\omega(782)$ resonance. This result
indicates that the $6\pi$ decay channel of $f_{0}(1710)$ is strongly
suppressed if virtual $\omega$ states are considered. Note, however, that
this decay channel might still arise from the more prominent $f_{0}(1710)$
decays into $KK$ and $\eta\eta$.

\section{Decay Width \boldmath$K_{0}^{\star}(1430)\rightarrow K\pi$} \label{sec.K0starKp}

The $K_{S}K\pi$ interaction Lagrangian, Eq.\ (\ref{KSKpion}), has already been
discussed in Sec.\ \ref{sec.KstarKp}. The scalar kaon field $K_{S}$ is
reassigned to $K_{0}^{\star}(1430)$ in Fit II but the decay width formula for
the process $K_{S}^{0}\rightarrow K\pi$, presented in Eq.\ (\ref{GKSKp}), is
of course valid nonetheless. There are no free parameters -- utilising
parameter values from Table \ref{Fit2-4} and mass values from Table
\ref{Fit2-5}, $\Gamma_{K_{S}^{0}\rightarrow K\pi}$ is determined uniquely as

\begin{equation}
\Gamma_{K_{S}^{0}\rightarrow K\pi}=263\text{ MeV.} \label{GKSKp2}
\end{equation}

The result is within the PDG value $\Gamma_{K_{0}^{\star}(1430)}^{\exp
}=(270\pm80)$ MeV \cite{PDG}. The stated PDG value actually depicts the full
decay width of $K_{0}^{\star}(1430)$ but the resonance is known to decay
almost exclusively to $K\pi$ \cite{Aston:1987}. Our result is therefore in
excellent correspondence with experimental data and it justifies the
assignment of $K_{S}$ to $K_{0}^{\star}(1430)$.

Let us, as in Sec.\ \ref{sec.KstarKp}, point out the influence of the
diagonalisation shift, Eqs.\ (\ref{shift22}) and (\ref{shift24}) -
(\ref{shift28}), on this decay width: omitting the shift ($w_{a_{1}
}=w_{K^{\star}}=w_{K_{1}}=0$) would yield $\Gamma_{K_{0}^{\star}
(1430)\rightarrow K\pi}\simeq12$ GeV. We thus conclude again that the
coefficients arising from the shift [Eqs.\ (\ref{BKSKp}) - (\ref{DKSKp})]
induce a destructive interference in the Lagrangian (\ref{KSKpion}) decreasing
the decay width by two orders of magnitude. The necessity to perform the shift
arises from the inclusion of vectors and axial-vectors into our model.
Therefore $\Gamma_{K_{S}\rightarrow K\pi}$ demonstrates that a reasonable
description of scalars requires that the (axial-)vectors be included into the
model. We can note, of course, that decoupling of (axial-)vectors from our
model would spoil the result for the decay width of the low-lying scalar kaon
$\kappa$ as well (as described in Sec.\ \ref{sec.KstarKp}) whereas, e.g.,
nonstrange-scalar decay widths would still be fine in this case. Nonetheless,
it is clear that our scalar kaon (how ever it may be assigned)\ is only
described properly if the (axial-)vectors are present in the model as well. We
will discuss the phenomenology of the vectors and axial-vectors in the subsequent sections.

\section{Phenomenology of Vector and Axial-Vector Mesons in Fit II}

Vector and axial-vector states are extremely important for our model. They are
known to decay into scalar and pseudoscalar states discussed so far \cite{PDG}
and their mixing with scalar and pseudoscalar degrees of freedom observed in
Eq.\ (\ref{mixingterms}) yields the diagonalisation shift of Eqs.\ (\ref{shift22})
and (\ref{shift24}) - (\ref{shift28}) originating in new terms in our Lagrangian
(\ref{Lagrangian}) that in turn influence the phenomenology of other states (but
also of vectors and axial-vectors themselves).\newline

Fit I showed considerable tension between the decay widths of $a_{1}(1260)$,
$f_{1}(1285)$, $f_{1}(1420)$ and $K_{1}(1400)$ on the one side and
$\Gamma_{\rho\rightarrow\pi\pi}$ on the other: either the axial-vectors were
too broad [$\sim(1-10)$ GeV] or the $\rho$ meson was too narrow ($\lesssim40$
MeV), see Sec.\ \ref{sec.conclusionsfitI}. Therefore, a major task in the
following sections will be to ascertain whether this state of affairs is
changed in Fit II where a fundamentally different assumption is implemented --
that the scalar quarkonia are above $1$ GeV.

In the vector channel, the exact value of $\Gamma_{\rho\rightarrow\pi\pi
}=149.1$ MeV has already been implemented to determine the parameter $g_{2}$ (see
Table \ref{Fit2-4}). Our model also allows for a calculation of the $2K$ decay width
of the strange isosinglet vector state $\omega_{S}\equiv\varphi(1020)$. It was
not possible to calculate $\Gamma_{\omega_{S}\rightarrow KK}$ because our Fit
I implied $m_{\omega_{S}}^{\text{FIT I}}=870.35$ MeV -- a value below the $2K$
threshold. Therefore, $\varphi(1020)$ was not well described within Fit I.
This is not the case in Fit II that yields $m_{\omega_{S}}=1036.90$ MeV
$>2m_{K}$, see Table \ref{Fit2-5}. We will calculate $\Gamma_{\omega
_{S}\rightarrow KK}$ in Sec.\ \ref{sec.oSKK2}. We will also consider
the phenomenology of the $K^{\star}$\ meson in Sec.\ \ref{sec.kstar2}.

In the axial-vector channel, we will consider the phenomenology of both
non-strange and strange isosinglets, $f_{1N}\equiv f_{1}(1285)$ in
Sec.\ \ref{sec.f1N2} and $f_{1S}\equiv f_{1}(1420)$\ in Sec.\ \ref{sec.f1S2}
(only $K^{\star}K$ decay channel can be considered in our model for both
resonances). Important considerations will regard the $a_{1}(1260)$ resonance,
the putative chiral partner of the $\rho$ meson,
in\ Sec.\ \ref{sec.a1rhopion2}. Note that Fit II does not allow for
$\Gamma_{a_{1}(1260)\rightarrow f_{0}(600)\pi}$ to be calculated because our
$\sigma_{1}$ field has now been reassigned to $f_{0}(1370)$ whereas we
determine the width for the sequential decay $a_{1}\rightarrow\bar{K}^{\star
}K\rightarrow\bar{K}K\pi$ in Sec.\ \ref{sec.a1KstarK2}.\ Note also that our
$K_{1}$ field (the phenomenology of which is discussed in Sec.\ \ref{sec.K12})
no longer corresponds to $K_{1}(1400)$ as in Fit I because, in Fit II, it is
found to be a member of an axial-vector nonet that in principle
requires consideration of the mixing with the corresponding pseudovector nonet
before phenomenology statements can be made [see Sec.\ \ref{2K1} and
Eq.\ (\ref{A1B1})]. For this reason, our calculations in Sec.\ \ref{sec.K12}
will be more of informative nature.

\subsection{Decay Width \boldmath$a_{1}(1260)\rightarrow\rho\pi$ in Fit II} \label{sec.a1rhopion2}

The interaction Lagrangian for the decay $a_{1}(1260)\rightarrow\rho\pi$ has
the same form as in the $U(2)\times U(2)$ version of the model. We have
considered the decay of an axial-vector state into a vector and a pseudoscalar
in Sec.\ \ref{sec.AVP}; utilising parameters in Table \ref{Fit2-4} allows us
to calculate $\Gamma_{a_{1}\rightarrow\rho\pi}$ from Eq.\ (\ref{GAVP}) with
$I=2$:

\begin{equation}
\Gamma_{a_{1}\rightarrow\rho\pi}=861\text{ MeV.} \label{Ga1rp2}
\end{equation}

\begin{figure}
[h]
\begin{center}
\includegraphics[
height=2.3582in,
width=3.9666in
]%
{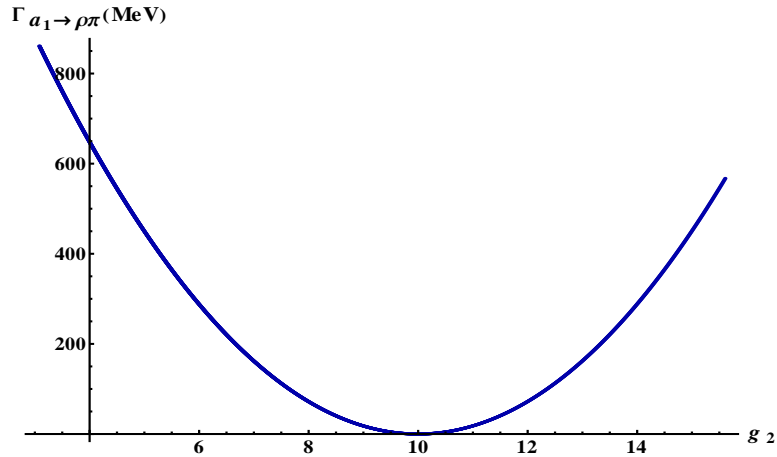}%
\caption{$\Gamma_{a_{1}\rightarrow\rho\pi}$ as a function of the parameter $g_{2}$
in Fit II.}%
\label{g222}%
\end{center}
\end{figure}

The result is outside the PDG value for the full width $\Gamma_{a_{1}%
(1260)}=(200-600)$ MeV but it is by more than an order of magnitude smaller
than the corresponding result $\Gamma_{a_{1}(1260)\rightarrow\rho\pi
}^{\text{FIT I}}\simeq13$ GeV obtained from Fit I in
Sec.\ \ref{sec.a1rhopion1}. The decisive difference is the value of $g_{2}$:
Fit II yields $g_{2}=3.07$ (Table \ref{Fit2-4}) whereas Fit I implied
$g_{2}=-11.2$ (Table \ref{Fit1-4}). $\Gamma_{a_{1}\rightarrow\rho\pi}<600$ MeV
would actually require $g_{2}\gtrsim4$ in Fit II, see Fig.\ \ref{g222}, but
the difference to $g_{2}=3.07$ is obviously not as large as in the case of Fit
I where $g_{2}\gtrsim10$ was necessary but $g_{2}=-11.2$ was obtained.

\begin{figure}
[b]
\begin{center}
\includegraphics[
height=1.8582in,
width=3.9666in
]%
{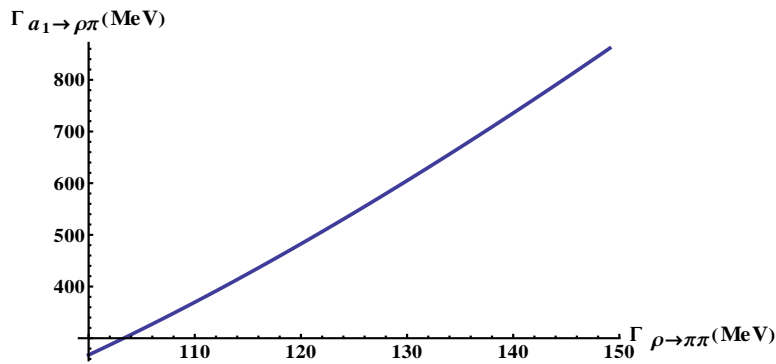}%
\caption{$\Gamma_{a_{1}(1260)\rightarrow\rho\pi}$ as function of $\Gamma
_{\rho\rightarrow\pi\pi}$ in Fit II.}%
\label{a1rhopi2}%
\end{center}
\end{figure}

Let us also consider, as in Sec.\ \ref{sec.a1rhopion1}, how far $\Gamma
_{\rho\rightarrow\pi\pi}$ (the decay width that determines $g_{2}$) would have
to be changed to enable us to obtain reasonable values of $\Gamma
_{a_{1}\rightarrow\rho\pi}$. The result is shown in Fig.\ \ref{a1rhopi2}.

We observe $a_{1}(1260)$ as a very broad resonance. We can see from
Fig.\ \ref{a1rhopi2} that values of $\Gamma_{a_{1}\rightarrow\rho\pi}$ within
the PDG range are obtained if we set $\Gamma_{\rho\rightarrow\pi\pi}$
approximately $20$ MeV lower than the PDG value. Smaller values of
$\Gamma_{a_{1}\rightarrow\rho\pi}$ follow if $\Gamma_{\rho\rightarrow\pi\pi}$
is decreased further. We thus require $\Gamma_{\rho\rightarrow\pi\pi}%
\lesssim130$ MeV for $\Gamma_{a_{1}\rightarrow\rho\pi}<600$ MeV. The value of
$\Gamma_{\rho\rightarrow\pi\pi}$ is somewhat smaller than $\Gamma
_{\rho\rightarrow\pi\pi}^{\text{PDG}}=149.1$ MeV \cite{PDG}; nonetheless, this
result is a strong improvement in comparison with $\Gamma_{\rho
\rightarrow\pi\pi}\lesssim38$ MeV in Fit I.

However, decreasing $\Gamma_{\rho\rightarrow\pi\pi}$ may not be the only
possibility to decrease $\Gamma_{a_{1}\rightarrow\rho\pi}$. We could, for
example, take into account the spectral function of the $\rho$ meson in order
to obtain the decay width $a_{1}\rightarrow\rho\pi\rightarrow3\pi$. Note that,
in this case, $g_{2}$ is not varied but rather fixed via on-shell values of
$\Gamma_{\rho\rightarrow\pi\pi}$, $m_{a_{1}}$ and $m_{\rho}$ (i.e.,
$g_{2}=3.07$ as in Table \ref{Fit2-4}). Integration over the $\rho$ spectral
function, analogously to Sec.\ \ref{sec.AVP}, yields

\begin{equation}
\Gamma_{a_{1}\rightarrow\rho\pi\rightarrow3\pi}=706\text{ MeV.}
\label{Ga1rp21}
\end{equation}

$\Gamma_{a_{1}\rightarrow\rho\pi}$ is thus decreased by approximately $160$
MeV once an off-shell $\rho$ meson is considered. Note that decreasing
$\Gamma_{\rho\rightarrow\pi\pi}$ by only $10$ MeV and re-integrating over the
$\rho$ spectral function with the thus obtained $g_{2}=3.65$ leads to
$\Gamma_{a_{1}\rightarrow\rho\pi}=600$ MeV, corresponding to the upper
boundary of the $a_{1}(1260)$ decay width as stated by the PDG.

Another possibility to decrease $\Gamma_{a_{1}\rightarrow\rho\pi}$ in
Eq.\ (\ref{Ga1rp2}) would be to introduce a form factor such as $\exp
[-k^{2}(x_{a_{1}},m_{\rho},m_{\pi})/\Lambda^{2}]$ to account for the finite
range of strong interactions; $x_{a_{1}}$ denotes the off-shell mass of
$a_{1}(1260)$ and $\Lambda$ is a cut-off with values between, e.g., $0.5$ GeV
and $1$ GeV. However, introduction of form factors would need to be performed
consistently throughout the model rather than \textit{ad hoc} for a single
decay width. For this reason, we will not utilise a form factor here, although
we note that the trial value of $\Lambda\sim0.5$ GeV allows for $\Gamma
_{a_{1}\rightarrow\rho\pi}\sim400$ MeV to be obtained once the decay width of
Eq.\ (\ref{GAVP}) is modulated by the stated exponential and the $a_{1}(1260)$
spectral function. Modulating Eq.\ (\ref{GAVP}) with the spectral functions of
both $\rho$ and $a_{1}(1260)$ as well as the stated form factor decreases the
decay width $\Gamma_{a_{1}\rightarrow\rho\pi\rightarrow3\pi}$ even further, to
$\sim300$ MeV, for $\Lambda\sim0.5$ GeV -- to less than one half of the value
presented in Eq.\ (\ref{Ga1rp21}).

For these reasons, the value of the decay width in Eq.\ (\ref{Ga1rp2}) is not
too problematic not only because it is close the PDG decay width interval but
also because, as we have seen, there exist means of decreasing it to the
values preferred by the PDG.

\subsubsection{A Remark on the Decay Width \boldmath $a_{1}(1260)\rightarrow
f_{0}(1370)\pi$}

It is not possible to calculate the decay width for the process $a_{1}%
(1260)\rightarrow f_{0}(600)\pi$\ within Fit II. This was performed in
Sec.\ \ref{sec.a1sigmapion1}, i.e., within Fit I, where our predominantly
non-strange scalar $\sigma_{1}$ was assigned to the $f_{0}(600)$ resonance.
This assignment is different in Fit II: $\sigma_{1}$ is identified with
$f_{0}(1370)$. This allows us in principle to calculate $\Gamma_{a_{1}%
(1260)\rightarrow f_{0}(1370)\pi}$. However, this decay is forbidden for
on-shell masses: utilising $m_{\sigma_{1}}=1310_{+30}^{-29}$ MeV from
Sec.\ \ref{sec.scalars2} and $m_{\pi}=138.65$ MeV requires $m_{a_{1}}>1419.65$
MeV. We obtained $m_{a_{1}}=1219$ MeV in Table \ref{Fit2-5}. Therefore, a
non-vanishing value of $\Gamma_{a_{1}\rightarrow\sigma_{1}\pi}$ could only be
obtained if a large $a_{1}(1260)$ decay width were considered in this decay
channel such that the high-mass tail of $a_{1}$ would allow for the decay
$a_{1}(1260)\rightarrow f_{0}(1370)\pi$\ to occur. Performing a calculation
analogous to the one in Sec.\ \ref{sec.ASP} to obtain the tree-level width for
$a_{1}(1260)\rightarrow f_{0}(1370)\pi$ and integrating over the $a_{1}$
spectral function in Eq.\ (\ref{d_a_1}) yields $\Gamma_{a_{1}(1260)\rightarrow
f_{0}(1370)\pi}\simeq0$. For this reason, we cannot confirm the existence of the
$a_{1}(1260)\rightarrow f_{0}(1370)\pi$ decay mode. Indeed the only piece of
experimental data asserting the existence of this decay channel
\cite{Asner:1999} assumed the $f_{0}(1370)$ mass and width of $1186$ and $350$
MeV respectively. The assumed mass value is too small according to the PDG and
also according to the review in Ref. \cite{buggf0}. Consequently, even
experimental data regarding this decay seem to be uncertain. We thus conclude
that the contribution of this decay channel to the $f_0(1370)$ width is negligible (if not zero).

\subsection{Decay Width \boldmath $a_{1}(1260)\rightarrow K^{\star} K
\rightarrow K K \pi$ in Fit II} \label{sec.a1KstarK2}

The corresponding interaction Lagrangian has already been stated in
Eq.\ (\ref{a1KstarK}). As in Sec.\ \ref{sec.a1KstarK1}, the decay
$a_{1}\rightarrow\bar{K}^{\star}K$\ is tree-level forbidden because $a_{1}$ is
below the $K^{\star}K$\ threshold (see Table \ref{Fit2-5}). However, if we
consider an off-shell $K^{\star}$\ state (just as in Sec.\ \ref{sec.AVP}) then
the ensuing decay $a_{1}\rightarrow\bar{K}^{\star}K\rightarrow\bar{K}K\pi
$\ can be studied. We use Eq.\ (\ref{GAVP1}) with an isospin factor $I=4$ and
integrate over the $K^{\star}$ spectral function in Eq.\ (\ref{dV}). The value
of the $K^{\star}$ decay width is given further below, in
Eq.\ (\ref{GKstarKp2}). Equation (\ref{GAVP1}) yields

\begin{equation}
\Gamma_{a_{1}\rightarrow\bar{K}^{\star}K\rightarrow\bar{K}K\pi}=0.55\text{
MeV.} \label{Ga1KstarK1}
\end{equation}

The above result is four orders of magnitude smaller than the one in
Eq.\ (\ref{Ga1KstarK}) because of the different value of the parameter $g_{2}$ (as
discussed in the previous sections) and also because $K^{\star}$ is a rather
narrow resonance. We thus find that the kaon decay of the $a_{1}(1260)$
resonance is strongly suppressed. The value is below the branching ratio
$\Gamma_{a_{1}(1260)\rightarrow\bar{K}^{\star}K}/\Gamma_{a_{1}(1260)}%
\lesssim0.04$ (our estimate from Refs.\ \cite{Barate:1999hj,Asner:1999,Coan:2004ep}) and
also below the result $\Gamma_{a_{1}(1260)\rightarrow\bar{K}^{\star}K}%
/\Gamma_{a_{1}(1260)}\lesssim(0.08-0.15)$ \cite{Drutskoy:2002ib}. The reason
is that we have actually considered a sequential decay ($a_{1}\rightarrow
\bar{K}^{\star}K\rightarrow\bar{K}K\pi$) rather than merely the tree-level
decay $a_{1}\rightarrow\bar{K}^{\star}K$ (although this decay will inevitably
lead to $\bar{K}K\pi$ upon $\bar{K}^{\star}$ decay). Additionally, the full
decay width of the $a_{1}(1260)$ state is ambiguous, $\Gamma_{a_{1}
(1260)}=(250-600)$ MeV \cite{PDG}, and therefore an exact experimental value
for a decay width such as $\Gamma_{a_{1}\rightarrow\bar{K}^{\star}
K\rightarrow\bar{K}K\pi}$ cannot be trivially determined.

\subsection{Decay Width \boldmath $\varphi(1020)\rightarrow K^{+}K^{-}$ in Fit II} \label{sec.oSKK2}

The $KK$ decay width of the $\omega_{S}\equiv\varphi(1020)$ state is specific
within our model because it can only be calculated from Fit II. It was not
possible to determine $\Gamma_{\omega_{S}\rightarrow KK}$ in the case of Fit I
because, as apparent from Table \ref{Fit1-5}, this fit yielded $m_{\omega_{S}
}^{\text{FIT I}}=870.35$ MeV -- a value below the $2K$ threshold. Contrarily,
Fit II yields $m_{\omega_{S}}=1036.90$ MeV $>2m_{K}$, see Table \ref{Fit2-5}.
The tree-level decay is therefore possible and the $\omega_{S}KK$ interaction
Lagrangian reads

\begin{align}
\mathcal{L}_{\omega_{S}KK}  &  =A_{\omega_{S}KK}\omega_{S}^{\mu}(\bar{K}
^{0}\partial_{\mu}K^{0}-K^{0}\partial_{\mu}\bar{K}^{0}+K^{-}\partial_{\mu
}K^{+}-K^{+}\partial_{\mu}K^{-})\nonumber\\
&  +B_{\omega_{S}KK}\partial^{\nu}\omega_{S}^{\mu}(\partial_{\nu}\bar{K}
^{0}\partial_{\mu}K^{0}-\partial_{\nu}K^{0}\partial_{\mu}\bar{K}^{0}
+\partial_{\nu}K^{-}\partial_{\mu}K^{+}-\partial_{\nu}K^{+}\partial_{\mu}
K^{-}) \label{oSKK2}
\end{align}

with

\begin{align}
A_{\omega_{S}KK}  &  =\frac{i}{2\sqrt{2}}Z_{K}^{2}\left\{  -2g_{1}+w_{K_{1}
}\left[  g_{1}^{2}(\phi_{N}+\sqrt{2}\phi_{S})+h_{2}(\phi_{N}-\sqrt{2}\phi
_{S})-2\sqrt{2}h_{3}\phi_{S}\right]  \right\}\text{,} \label{AoSKK2}\\
B_{\omega_{S}KK}  &  =\frac{i}{\sqrt{2}}Z_{K}^{2}g_{2}w_{K_{1}}^{2}\text{.}
\label{BoSKK2}
\end{align}

Note that the PDG data \cite{PDG} cite the $\varphi(1020)$ decay width into
charged as well as neutral kaon modes. They are not the same due to isospin
violation [whereas our model is isospin-symmetric, as apparent, e.g., from
Eq.\ (\ref{oSKK2})]. Our Fit II implemented $m_{K^{\pm}}$ [see
Eq.\ (\ref{fit22})] and therefore we will in the following, for consistency,
focus on the decay $\varphi(1020)\rightarrow K^{+}K^{-}$.

The calculation of $\Gamma_{\varphi(1020)\rightarrow K^{+}K^{-}}$ is analogous to
the generic calculation described in Sec.\ \ref{sec.ASP}. Equation (\ref{oSKK2})
yields the following decay amplitude upon substituting $\partial^{\mu
}\rightarrow-iP^{\mu}$ for the decaying particle and $\partial^{\mu
}\rightarrow iP_{1,2}^{\mu}$ for the decay products:

\begin{equation}
-i\mathcal{M}_{\omega_{S}\rightarrow K^{+}K^{-}}^{(\alpha)}=\varepsilon_{\mu
}^{(\alpha)}(P)h_{\omega_{S}KK}^{\mu}=-\varepsilon_{\mu}^{(\alpha)}(P)\left[
(A_{\omega_{S}KK}+B_{\omega_{S}KK}P\cdot P_{1})(P_{1}^{\mu}-P_{2}^{\mu
})\right]\text{,}  \label{iMoKK}
\end{equation}

where the momenta of $\omega_{S}$, $K^{+}$ and $K^{-}$ are denoted as $P$,
$P_{1}$ and $P_{2}$, respectively; $\varepsilon_{\mu}^{(\alpha)}(P)$
represents the polarisation vector of $\omega_{S}$\ and the vertex
$h_{\omega_{S}KK}^{\mu}$ reads

\begin{equation}
h_{\omega_{S}KK}^{\mu}=-(A_{\omega_{S}KK}+B_{\omega_{S}KK}P\cdot P_{1}
)(P_{1}^{\mu}-P_{2}^{\mu})\text{.} \label{hoKK}%
\end{equation}

According to Eq.\ (\ref{iMASP1}), there are two contributions to the averaged
squared amplitude $|-i\mathcal{\bar{M}}_{\omega_{S}\rightarrow K^{+}K^{-}
}|^{2}$ that involves a vector state: the first one is the squared vertex
$\left\vert h_{\omega_{S}KK}^{\mu}\right\vert ^{2}$ and the second one,
$\left\vert h_{\omega_{S}KK}^{\mu}P_{\mu}\right\vert ^{2}$, contains the
vertex $h_{\omega_{S}KK}^{\mu}$ contracted with the vector-state momentum
$P_{\mu}=(m_{A},\vec{0})$. Consequently, $h_{\omega_{S}KK}^{\mu}P_{\mu}\equiv
h_{\omega_{S}KK}^{0}P_{0}=0$ because $h_{\omega_{S}KK}^{0}=0$, see
Eq.\ (\ref{hoKK}). Therefore, only $\left\vert h_{\omega_{S}KK}^{\mu
}\right\vert ^{2}$ contributes to $|-i\mathcal{\bar{M}}_{\omega_{S}\rightarrow
K^{+}K^{-}}|^{2}$:

\begin{align}
\left|  -i\mathcal{\bar{M}}_{\omega_{S}\rightarrow K^{+}K^{-}} \right|  ^{2}
&  =-\frac{1}{3}\left\vert h_{\omega_{S}KK}^{\mu}\right\vert ^{2}=-\frac{1}
{3}(A_{\omega_{S}KK}+B_{\omega_{S}KK}P\cdot P_{1})^{2}(P_{1}^{\mu}-P_{2}^{\mu
})^{2}\nonumber\\
&  =-\frac{2}{3} \left(  A_{\omega_{S}KK}+B_{\omega_{S}KK}\frac{m_{\omega_{S}
}^{2}}{2} \right)  ^{2}(m_{K}^{2}-P_{1}\cdot P_{2}) \nonumber\\
& = \frac{1}{3} \left(
A_{\omega_{S}KK}+B_{\omega_{S}KK}\frac{m_{\omega_{S}}^{2}}{2} \right)
^{2}(m_{\omega_{S}}^{2}-4m_{K}^{2})\text{.} \label{iMoKK1}
\end{align}

Then the decay width reads

\begin{align}
\Gamma_{\omega_{S}\rightarrow K^{+}K^{-}} & =\frac{k(m_{\omega_{S}},m_{K},m_{{K}
})}{8\pi m_{\omega_{S}}^{2}}|-i\mathcal{\bar{M}}_{\omega_{S}\rightarrow
K^{+}K^{-}}|^{2} \nonumber \\
& =\frac{k(m_{\omega_{S}},m_{K},m_{{K}})}{24\pi m_{\omega_{S}
}^{2}} \left(  A_{\omega_{S}KK}+B_{\omega_{S}KK}\frac{m_{\omega_{S}}^{2}}
{2}\right)  ^{2}(m_{\omega_{S}}^{2}-4m_{K}^{2}) \label{GoKK}
\end{align}

with $k(m_{\omega_{S}},m_{K},m_{{K}})$ from Eq.\ (\ref{kabc}). Using the parameter
values from Table \ref{Fit2-4} to determine the coefficients in
Eqs.\ (\ref{AoSKK2}) and (\ref{BoSKK2})\ and the mass values from Table
\ref{Fit2-5}, Eq.\ (\ref{GoKK}) yields

\begin{equation}
\Gamma_{\omega_{S}\rightarrow K^{+}K^{-}}=2.33\text{ MeV.} \label{GoKK1}
\end{equation}

This value is slightly larger than the one suggested by the PDG data reading
$\Gamma_{\varphi(1020)\rightarrow K^{+}K^{-}}^{\exp}=(2.08\pm0.04)$ MeV. The
reason is that our $m_{\omega_{S}}=1036.90$ MeV is by approximately $20$ MeV
larger than $m_{\varphi(1020)}=1019.46$ MeV \cite{PDG}. Nonetheless, our
chiral-model result is remarkably close to the experimental value; it
represents an additional statement in favour of Fit II when compared to Fit I
(with the latter not permitting for this decay width to be calculated at all,
see introductory remarks in Sec.\ \ref{sec.VA1}).

\subsection{Decay Width \boldmath $K^{\star} \rightarrow K \pi$ in Fit II} \label{sec.kstar2}

Phenomenology of the vector kaon $K^{\star}\equiv K^{\star}(892)$ has already
been discussed within Fit I in Sec.\ \ref{sec.kstar1}. The value
$\Gamma_{K^{\star0}\rightarrow K\pi}^{\text{FIT I}}=32.8$ MeV was obtained,
see Eq.\ (\ref{GKstarKp1}). The experimental value reads $\Gamma_{K^{\star
}\rightarrow K\pi}^{\exp}=46.2$ MeV and the resonance decays to $\simeq100\%$
into $K\pi$ \cite{PDG}. Thus the value obtained in Fit I was by approximately
13 MeV (or 30\%) smaller than the experimental result.

The $K^{\star0}K\pi$ interaction Lagrangian has been presented in
Eq.\ (\ref{KstarKpion}). Repeating the calculation performed in
Sec.\ \ref{sec.kstar1} with the set of parameters from Table \ref{Fit2-4} and
masses in Table \ref{Fit2-5} we obtain the following value in Fit II:

\begin{equation}
\Gamma_{K^{\star0}\rightarrow K\pi}=44.2\text{ MeV.} \label{GKstarKp2}
\end{equation}

This result is only $2$ MeV smaller than the stated experimental result.
Correspondence with experiment is hence excellent: the assumption of scalar
$\bar{q}q$ states above $1$ GeV and the ensuing Fit II shift the value
obtained in Fit I in the correct direction and allow us to describe the
vector-kaon decay width almost exactly. This is a strong indication in favour
of Fit II.

\subsection{Decay Width \boldmath $f_{1}(1285)\rightarrow K^{\star} K $ in Fit II} \label{sec.f1N2}

As stated in Sec.\ \ref{sec.f1N1}, there are two decay widths of the
$f_{1N}\equiv f_{1}(1285)$ state that can be calculated from our model:
$f_{1N}\rightarrow a_{0}\pi$ and $f_{1N}\rightarrow\bar{K}^{\star}K$. The
former can only be considered within Fit I where the scalar states were
assumed to be below $1$ GeV\ [$a_{0}\equiv\vec{a}_{0}(980)$]. In Fit II, the
correspondence $a_{0}\equiv\vec{a}_{0}(1450)$ holds and thus the decay width
for the process $f_{1N}\rightarrow a_{0}\pi$ cannot be calculated (as it is
kinematically forbidden). Note also that the decay $a_{0}(1450)\rightarrow
f_{1}(1285)\pi$ has not been observed \cite{PDG}.

We then only need to consider the decay $f_{1}(1285)\rightarrow\bar{K}^{\star
}K$, analogously to the calculations performed in Sec.\ \ref{sec.f1N1}. Let us
again note that the PDG lists the $f_{1}(1285)\rightarrow\bar{K}^{\star}K$
process as "not seen" although the three-body decay $f_{1}(1285)\rightarrow
\bar{K}K\pi$ possesses a branching ratio of $(9.0\pm0.4)\%$ whereas the full
decay width of the resonance is $\Gamma_{f_{1}(1285)}=(24.3\pm1.1)$\ MeV
\cite{PDG}. The $f_{1}(1285)$\ decay into $\bar{K}^{\star}$ and $K$ is
forbidden for the on-shell masses of the three particles considered, as
apparent from experimental data and also from our mass values in Table
\ref{Fit2-5}. However, the three-body $\bar{K}K\pi$ decay can, within our
model, arise from the sequential decay $f_{1}(1285)\rightarrow\bar{K}^{\star
}K\rightarrow\bar{K}K\pi$. The latter decay was discussed in
Sec.\ \ref{sec.f1N1} within Fit I and the value $\Gamma_{f_{1N}\rightarrow
\bar{K}^{\star}K\rightarrow\bar{K}K\pi}=1.98$ GeV was obtained -- three order
of magnitude larger than the experimental limit $\Gamma_{f_{1N}\rightarrow
\bar{K}^{\star}K\rightarrow\bar{K}K\pi}\leq(2.2\pm0.1)$ MeV\ expected from the
mentioned branching ratio for $f_{1}(1285)\rightarrow\bar{K}K\pi$ and the full
$f_{1}(1285)$\ decay width. In this section we discuss whether $\Gamma
_{f_{1N}\rightarrow\bar{K}^{\star}K\rightarrow\bar{K}K\pi}$ is improved in Fit II.

To this end we need to repeat the calculations from Sec.\ \ref{sec.f1N1} utilising
the parameter set stated in Table \ref{Fit2-4}. The $f_{1N}K^{\star}K$
interaction Lagrangian is stated in Eq.\ (\ref{f1NKstarK}). We need to
consider the $K^{\star}$ spectral function (because the decay is enabled by an
off-shell $K^{\star}$ state), as described in Sec.\ \ref{sec.f1N1}.
Consequently, we obtain

\begin{equation}
\Gamma_{f_{1N}\rightarrow\bar{K}^{\star}K\rightarrow\bar{K}K\pi}=0.9\text{
MeV.} \label{Gf1NKstarK2}
\end{equation}

The value in Eq.\ (\ref{Gf1NKstarK2}) represents an improvement by three
orders of magnitude compared to $\Gamma_{f_{1N}\rightarrow\bar{K}^{\star
}K\rightarrow\bar{K}K\pi}=1.98$ GeV obtained in Fit I. It is smaller than the
PDG value $\Gamma_{f_{1}(1285)\rightarrow\bar{K}K\pi}=(2.2\pm0.1)$ MeV. Thus,
at this point, we do not expect $\Gamma_{f_{1N}\rightarrow\bar{K}^{\star
}K\rightarrow\bar{K}K\pi}$ to be the only contribution to $\Gamma
_{f_{1N}\rightarrow\bar{K}K\pi}$ -- for example, a direct three-body decay
into $\bar{K}K\pi$ might also contribute to the total decay width in this
channel. Nonetheless, from results presented until now we conclude that
approximately $40\%$ of the decay $f_{1}(1285)\rightarrow\bar{K}K\pi$ is
generated via the sequential process $f_{1}(1285)\rightarrow\bar{K}^{\star
}K\rightarrow\bar{K}K\pi$. This is contrary to the PDG conclusion stating that
no such contribution exists. Note that the PDG conclusion is based on
${\bar{p}p}$ annihilation data from Ref.\ \cite{Nacasch:1977}; there are,
however, newer data (but with limited statistics) from the L3 Collaboration
\cite{Achard:2007} that suggest a non-vanishing contribution of $f_{1}%
(1285)\rightarrow\bar{K}^{\star}K\rightarrow\bar{K}K\pi$ to $f_{1}%
(1285)\rightarrow\bar{K}K\pi$. Our results seem to corroborate those of the L3
Collaboration.\newline

\begin{figure}
[h]
\begin{center}
\includegraphics[
height=2.1082in,
width=3.9666in
]%
{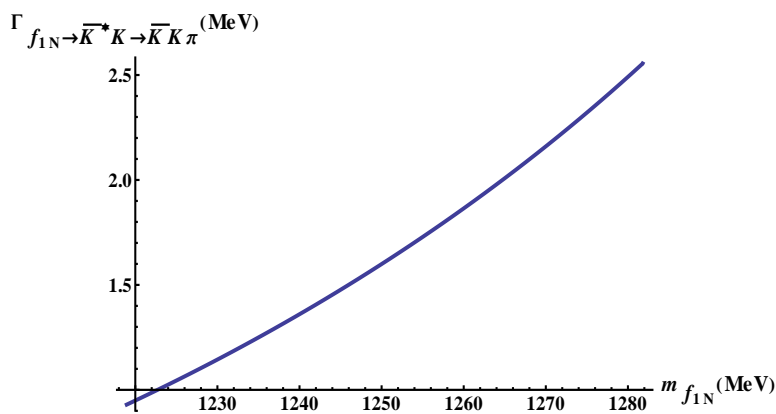}%
\caption{$\Gamma_{f_{1N}\rightarrow\bar{K}^{\star}K\rightarrow\bar{K}K\pi}$ as
function of $m_{f_{1N}}$.}%
\label{f1NKstarK2}
\end{center}
\end{figure}

$\Gamma_{f_{1N}\rightarrow\bar{K}^{\star}K\rightarrow\bar{K}K\pi}$ in
Eq.\ (\ref{Gf1NKstarK2}) was obtained assuming $m_{f_{1N}}=1219$ MeV $\equiv
m_{a_{1}}$, see Eq.\ (\ref{m_a_1}). Considering finite-width effects for the
rather broad $a_{1}(1260)$\ resonance (analogously to calculations in
Ref.\ \cite{Giacosa:2007bn}) might, however, induce a mass splitting of $f_{1N}$
and $a_{1}$. Then $m_{f_{1N}}=m_{a_{1}}$ would no longer hold and
Fig.\ \ref{f1NKstarK2} demonstrates the change of $\Gamma_{f_{1N}
\rightarrow\bar{K}^{\star}K\rightarrow\bar{K}K\pi}$ if $m_{f_{1N}}$ is
increased, e.g., to the experimental value $m_{f_{1}(1285)}=1281.8$ MeV.

We observe that $\Gamma_{f_{1N}\rightarrow\bar{K}^{\star}K\rightarrow\bar
{K}K\pi}$ is strongly dependent on $m_{f_{1N}}$ and increases by almost three
times in the mass region of interest: from $0.9$ MeV for $m_{f_{1N}}=1219$
MeV\ to $2.6$ MeV for $m_{f_{1N}}=m_{f_{1}(1285)}=1281.8$ MeV. Therefore,
varying $m_{f_{1N}}$ implies that the decay $f_{1}(1285)\rightarrow\bar{K}%
K\pi$ is completely saturated by the sequential decay$\ f_{1}(1285)\rightarrow
\bar{K}^{\star}K\rightarrow\bar{K}K\pi$. It is actually possible to determine
$m_{f_{1N}}$ such that $\Gamma_{f_{1N}\rightarrow\bar{K}K\pi}$ corresponds
exactly to $\Gamma_{f_{1}(1285)\rightarrow\bar{K}K\pi}=(2.2\pm0.1)$ MeV. As
apparent from Fig.\ \ref{f1NKstarK2}, this is realised for $m_{f_{1N}%
}=1271_{-7}^{+6}$ MeV, in a very good agreement with $m_{f_{1}(1285)}%
=(1281.8\pm0.6)$ MeV.

We thus conclude that the $f_{1}(1285)$ phenomenology is decisively better
described in Fit II than in Fit I.

\subsection{Decay Width \boldmath $f_{1}(1420)\rightarrow K^{\star} K$ in Fit II} \label{sec.f1S2}

Let us remind ourselves that the $f_{1}(1420)$ decays predominantly into
$K^{\star}K$; the resonance possesses a mass of $m_{f_{1}(1420)}^{\exp
}=(1426.4\pm0.9)$ MeV and width $\Gamma_{f_{1}(1420)}^{\exp}=(54.9\pm2.6)$ MeV
\cite{PDG}. We have seen in Sec.\ \ref{sec.f1S1} that Fit I implies the
$f_{1S}\equiv f_{1}(1420)$ resonance to be $17.6$ GeV broad. Clearly, this
result cannot be regarded as physical and in this section we discuss whether
the stated large value of the decay width is decreased in Fit II.
Additionally, we obtained $m_{f_{1}(1420)}=1643.4$ MeV in Fit I (see Table
\ref{Fit1-5}). As apparent from\ Table \ref{Fit2-5}, this value is decreased
substantially to $1457$ MeV within Fit II -- the correspondence with the mentioned
experimental value $m_{f_{1}(1420)}^{\exp}$ is therefore much better than in
Fit I. In this section we discuss the phenomenology of $f_{1}(1420)$ within
Fit II. Given the predominance of the decay into $K^{\star}K$, it suffices to
discuss this decay channel only.

The calculation of the decay width is performed analogously to the one in
Sec.\ \ref{sec.AVP}. The $f_{1S}K^{\star0}\bar{K}^{0}$ interaction Lagrangian
presented in Eq.\ (\ref{f1SKstarKaon}) is analogous to the one presented in
Eq.\ (\ref{AVP}); the same is true for the vertices in Eqs.\ (\ref{hAVP}) and
(\ref{hf1SKstarK}). Consequently, we can utilise the generic formula for an
axial-vector decay width presented in Eq.\ (\ref{GAVP}). Setting $I=4$ to
consider the decays $f_{1S}\rightarrow\bar{K}^{\star0}K^{0}$, $\bar{K}
^{0}K^{\star0}$, $K^{\star+}K^{-}$ and $K^{\star-}K^{+}$ we obtain

\begin{equation}
\Gamma_{f_{1S}\rightarrow\bar{K}^{\star}K}=274\text{ MeV.} \label{Gf1SKstarK2}
\end{equation}

This value improves the Fit I value $\Gamma_{f_{1S}\rightarrow\bar{K}^{\star
}K}^{\text{FIT I}}=17.6$ GeV by two orders of magnitude. Nonetheless, it is
still larger than the one reported by the PDG: $\Gamma_{f_{1}(1420)}^{\exp
}=(54.9\pm2.6)$ MeV. Therefore Fit II, where scalar meson states are assumed
to be above $1$ GeV, shifts $\Gamma_{f_{1S}\rightarrow\bar{K}^{\star}K}$ in
the correct direction but does not yield the experimental result. We will see
in the next section that the analogous problem persists in the $K_{1}$
phenomenology as well. The reason may be that the current form of our model
does not implement mixing of our $1^{++}$ field [$\equiv f_{1N,A}$ in
Eq.\ (\ref{A1B1})] with the $C$-conjugated $1^{+-}$ partner [$\equiv f_{1N,B}$
in Eq.\ (\ref{A1B1})]. Note that the value in Eq.\ (\ref{Gf1SKstarK2}) is
decreased by approximately $40$ MeV upon integration over the $K^{\star}$
spectral function. Thus the sequential decay $f_{1}(1420)\rightarrow\bar
{K}^{\star}K\rightarrow\bar{K}K\pi$ appears to be dominant in the
$f_{1}(1420)\rightarrow\bar{K}K\pi$ decay channel, just as in the case of the
$f_{1}(1285)$ resonance.

\begin{figure}
[h]
\begin{center}
\includegraphics[
height=2.0582in,
width=3.9666in
]%
{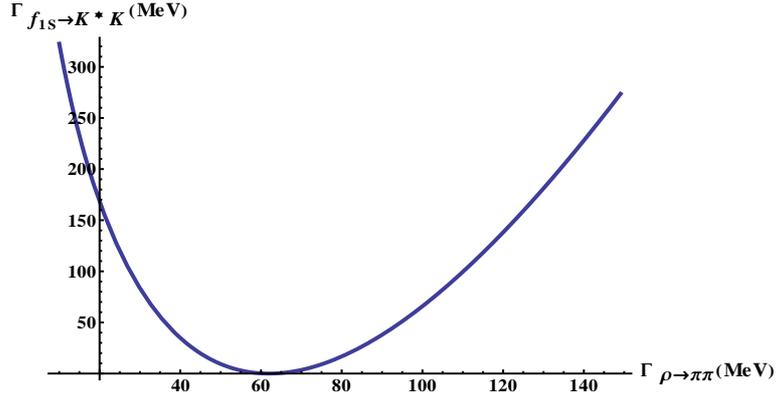}%
\caption{$\Gamma_{f_{1S}\equiv f_{1}(1420)\rightarrow\bar{K}^{\star}K}$ as
function of $\Gamma_{\rho\rightarrow\pi\pi}$ in Fit II.}%
\label{f1SKstarK2}
\end{center}
\end{figure}

Let us also note that the correct value of $\Gamma_{f_{1S}\rightarrow\bar
{K}^{\star}K}$ can be obtained by decreasing $\Gamma_{\rho\rightarrow\pi\pi}$
to approximately $96$ MeV (Fig.\ \ref{f1SKstarK2}). However, this statement
must be viewed with caution because, as already mentioned, the current form of
the model lacks $f_{1N,A}$-$f_{1N,B}$ mixing upon which no decreasing of
$\Gamma_{\rho\rightarrow\pi\pi}$ may be needed to obtain the correct value of
$\Gamma_{f_{1S}\rightarrow\bar{K}^{\star}K}$.%

\subsection{\boldmath $K_{1}$ Decays in Fit II} \label{sec.K12}

There are two important remarks regarding the $K_{1}$ phenomenology in Fit II.
Firstly, our $K_{1}$ field can no longer \textit{a priori} be assigned to a
physical resonance because Fit II yields $m_{K_{1}}=1343$ MeV, a value that is
virtually the mass median of $K_{1}(1270)$ with $m_{K_{1}(1270)}=(1272\pm7)$ MeV
and $K_{1}(1400)$ with $m_{K_{1}(1400)}=(1403\pm7)$ MeV. Indeed our discussion
in Sec.\ \ref{2K1} has suggested that the stated value of $m_{K_{1}}$ is an
indication that a $1^{+-}$ nonet needs to be considered in our model together
with the (already present) $1^{++}$ nonet. The second remark is about results
obtained in Fit I, Sec.\ \ref{sec.K11}. Three decay channels were considered:
$K_{1}(1400)\rightarrow K^{\star}\pi$, $\rho K$ and $\omega K$. Each one of
them was found to be more than $1$ GeV broad; in fact, the sum of all partial
decay widths in Eqs.\ (\ref{GK1Kstarp}), (\ref{GK1rK}) and (\ref{GK1oK})
suggests $\Gamma_{K_{1}(1400)}\sim10$ GeV -- a value that is two orders of
magnitude larger than $\Gamma_{K_{1}(1400)}=(174\pm13)$ MeV \cite{PDG}.

In this section we discuss whether it is possible to amend the unphysically
large values of the decay widths obtained in Sec.\ \ref{sec.K11}. A word of
caution is nonetheless necessary: given the absence of the $1^{+-}$ nonet from
the model and the consequent mass value of the $1^{++}$ field $m_{K_{1}}=1343$
MeV corresponding to neither $K_{1}(1270)$ nor $K_{1}(1400)$, it cannot be
expected that our results in this section will yield exact experimental
values. However, we can still observe whether the results from Fit I are
shifted in the correct direction by Fit II.

The interaction Lagrangian of the $K_{1}$ state with the already mentioned
decay products has been presented in Eq.\ (\ref{K1}). The calculation of the decay
widths proceeds exactly as described in Sec.\ \ref{sec.K11}; in this section
we utilise parameter values from Table \ref{Fit2-4} and mass values from Table
\ref{Fit2-5}.

\begin{equation}
\Gamma_{K_{1}\rightarrow K^{\star}\pi}=307\text{ MeV, }\Gamma_{K_{1}
\rightarrow\rho K}=128\text{ MeV, }\Gamma_{K_{1}\rightarrow\omega_{N}
K}=41\text{ MeV.} \label{GK12}%
\end{equation}

For comparison, Fit I yielded

\begin{equation}
\Gamma_{K_{1}\rightarrow K^{\star}\pi}^{\text{FIT I}}=6.73\text{ GeV, }
\Gamma_{K_{1}\rightarrow\rho K}^{\text{FIT I}}=4.77\text{ GeV, }\Gamma
_{K_{1}\rightarrow\omega_{N}K}^{\text{FIT I}}=1.59\text{ GeV.} \label{GK13}
\end{equation}

The sum of the decay widths in Eq.\ (\ref{GK12}) suggests a full $K_{1}$ decay
width of $\sim480$ MeV, larger than both $\Gamma_{K_{1}(1400)}=(174\pm13)$ MeV
and $\Gamma_{K_{1}(1270)}=(90\pm20)$ MeV but two orders of magnitude less than
the value $\sim10$ GeV obtained in Fit I. We observe that $\Gamma
_{K_{1}\rightarrow K^{\star}\pi}$ and $\Gamma_{K_{1}\rightarrow\rho K}$ have
been improved by an order of magnitude in comparison with Fit I;
$\Gamma_{K_{1}\rightarrow\omega_{N}K}$ has been improved by two orders of
magnitude. Thus we conclude that the values of the stated $K_{1}$ decay
widths, while still not satisfactory, are nonetheless strongly improved in
comparison with Fit I.

Note that all the mentioned decay widths could be improved if $\Gamma
_{\rho\rightarrow\pi\pi}$ were decreased by $\sim100$ MeV thus implying
\ $g_{2}\sim10$. However, it is not necessary to include the corresponding
diagrams into this work because, as already stated, it is not possible to
assign our $K_{1}$ field to a physical resonance. Nonetheless, we find Fit II
to be favoured over Fit I.

\subsubsection{A Note on Decay \boldmath $K_{1}\rightarrow K f_{0}(1370)$}

The decay into $K$ and $f_{0}(1370)$ has been observed for both $K_{1}(1270)$
and $K_{1}(1400)$; it can, in principle, be calculated from our model as the
$K_{1}K\sigma_{N,S}$ interaction Lagrangian obtained from
Eq.\ (\ref{Lagrangian}) reads

\begin{align}
\mathcal{L}_{K_{1}K\sigma}  &  =A_{K_{1}K\sigma_{N}}K_{1}^{\mu0}\sigma
_{N}\partial_{\mu}\bar{K}^{0}+B_{K_{1}K\sigma_{N}}K_{1}^{\mu0}\partial_{\mu
}\sigma_{N}\bar{K}^{0}\nonumber\\
&  +A_{K_{1}K\sigma_{S}}K_{1}^{\mu0}\sigma_{S}\partial_{\mu}\bar{K}
^{0}+B_{K_{1}K\sigma_{S}}K_{1}^{\mu0}\partial_{\mu}\sigma_{S}\bar{K}^{0}
\label{K1Ks}
\end{align}

with

\begin{align}
A_{K_{1}K\sigma_{N}}  &  =\frac{Z_{K}}{2}\left\{  g_{1}\left(  -1+g_{1}
w_{K_{1}}\phi_{N}+\sqrt{2}g_{1}w_{K_{1}}\phi_{S}\right)  +w_{K_{1}}\left[
\left(  2h_{1}+h_{2}\right)  \phi_{N}-\sqrt{2}h_{3}\phi_{S}\right]  \right\}
\text{,} \\
B_{K_{1}K\sigma_{N}}  &  =\frac{g_{1}}{2}Z_{K}\text{,} \\
A_{K_{1}K\sigma_{S}}  &  =\frac{Z_{K}}{\sqrt{2}}\left\{  g_{1}\left(
-1+g_{1}w_{K_{1}}\phi_{N}+\sqrt{2}g_{1}w_{K_{1}}\phi_{S}\right)  +w_{K_{1}
}\left[  \sqrt{2}\left(  h_{1}+h_{2}\right)  \phi_{S}-h_{3}\phi_{N}\right]
\right\}\text{,} \\
B_{K_{1}K\sigma_{S}}  &  =\frac{g_{1}}{\sqrt{2}}Z_{K}\text{.}
\end{align}

Inserting the inverted Eq.\ (\ref{sigma-sigma_1}) into Eq.\ (\ref{K1Ks}) would
allow us to determine the interaction Lagrangians for the processes $K_{1}\rightarrow
K\sigma_{1}$ and also $K_{1}\rightarrow K\sigma_{2}$. However, the decay
$K_{1}\rightarrow K\sigma_{2}$ is kinematically forbidden due to the
assignment of the fields $\sigma_{1}$ and $\sigma_{2}$ to $f_{0}(1370)$ and
$f_{0}(1710)$, respectively. For this reason we only consider the part of the
Lagrangian in Eq.\ (\ref{K1Ks}) containing $\sigma_{1}$:

\begin{align}
\mathcal{L}_{K_{1}K\sigma_{1}}  &  =(A_{K_{1}K\sigma_{N}}\cos\varphi_{\sigma
}+A_{K_{1}K\sigma_{S}}\sin\varphi_{\sigma})K_{1}^{\mu0}\sigma_{1}\partial
_{\mu}\bar{K}^{0} \nonumber \\
& + (B_{K_{1}K\sigma_{N}}\cos\varphi_{\sigma}+B_{K_{1}
K\sigma_{S}}\sin\varphi_{\sigma})K_{1}^{\mu0}\bar{K}^{0}\partial_{\mu}
\sigma_{1}\nonumber\\
&  \equiv A_{K_{1}K\sigma_{1}}K_{1}^{\mu0}\sigma_{1}\partial_{\mu}\bar{K}
^{0}+B_{K_{1}K\sigma_{1}}K_{1}^{\mu0}\bar{K}^{0}\partial_{\mu}\sigma_{1}
\label{K1Ks1}
\end{align}

with $A_{K_{1}K\sigma_{1}}=A_{K_{1}K\sigma_{N}}\cos\varphi_{\sigma}%
+A_{K_{1}K\sigma_{S}}\sin\varphi_{\sigma}$ and $B_{K_{1}K\sigma_{1}}%
=B_{K_{1}K\sigma_{N}}\cos\varphi_{\sigma}+B_{K_{1}K\sigma_{S}}\sin
\varphi_{\sigma}$. We note that the Lagrangian in Eq.\ (\ref{K1Ks1}) possesses
the same form as the generic Lagrangian presented Sec.\ \ref{sec.ASP},
Eq.\ (\ref{ASP}). This allows us in principle to calculate the decay width
$\Gamma_{K_{1}\rightarrow K\sigma_{1}}$ from Eq.\ (\ref{GASP}). However, such
a calculation would only be possible for an off-shell $K_{1}$ state, i.e.,
an integration over the $K_{1}$ spectral function would be required. This is not
feasible in the current form of the model because the absence of the $1^{+-}$
nonet still yields too large values of the $K_{1}$ decay width [see
Eq.\ (\ref{GK12})]. Nonetheless, a discussion of the decay $K_{1}\rightarrow
K\sigma_{1}$ upon inclusion of the pseudovector nonet into our model would
represent a valuable extension of this work.

\section{Conclusions from Fit with Scalars above 1 GeV} \label{sec.conclusionsfitII}

The previous sections have addressed the question whether it is possible to obtain
a reasonable phenomenology of mesons in vacuum under the assumption that
scalar $\bar{q}q$ states possess energies above 1 GeV. This is the cardinal
difference between the consequent fit (labelled as Fit II) and Fit I from
Chapters \ref{sec.fitI} and \ref{ImplicationsFitI} where, conversely, scalar
$\bar{q}q$ states were assumed to be below 1 GeV.

The parameters in Fit II were determined from the masses of $\pi$, $K$, $\eta$,
$\eta^{\prime}$,$\rho$, $K^{\star}$, $\omega_{S}\equiv\varphi(1020)$, $a_{1}$,
$K_{1}$, $f_{1S}\equiv f_{1}(1420)$, decay widths $\Gamma_{a_{1}\rightarrow
\pi\gamma}$ and $\Gamma_{a_{0}(1450)}$ as well as the masses of the scalar states $a_{0}$
and $K_{S}$ assigned to $a_{0}(1450)$ and$\ K_{0}^{\star}(1430)$,
respectively. We have not included any scalar isosinglet masses into the fit
in order to let these masses remain a prediction of the fit. \newline

We summarise the main conclusions of Fit II as follows:

\begin{itemize}

\item It is possible to find a fit; unlike Fit I, all masses obtained from Fit
II are within 3\%\ of the respective experimental values except

\begin{itemize}

\item $m_{\eta}=523.20$ MeV, approximately 25 MeV ($\simeq4.5\%$) smaller than
$m_{\eta}^{\exp}=547.85$ MeV due to the condition $m_{\eta_{N}}\overset{!}{<}m_{\eta_{S}}$ ($\eta_{N,S}$ are,
respectively, pure-nonstrange and
pure-strange contributions to the $\eta$ wave function),

\item $m_{K_{0}^{\star}(1430)}=1550$ MeV, a value that is $125$ MeV
($\simeq8.8\%$) larger than the corresponding PDG value $m_{K_{0}^{\star
}(1430)}^{\exp}=1425$ MeV due to the pattern of explicit symmetry breaking in
our model that always renders strange states approximately 100 MeV
($\simeq$ strange-quark mass) heavier than their corresponding non-strange
counterparts [note, for example, that Fit II also yields $m_{a_{0}%
(1450)}=1452$ MeV, approximately 100 MeV less than $m_{K_{0}^{\star}%
(1430)}=1550$ MeV].
\end{itemize}

In particular the narrow resonances $\varphi(1020)$\ and $f_{1}(1420)$ are
decisively better described in Fit II than in Fit I. They exhibited strong
deviations from the experimental results in Fit I [$(150-200)$ MeV mass
difference]. However, in Fit II, their masses differ by only $\simeq2\%$\ from
the experimental values. Additionally, $\Gamma_{a_{1}\rightarrow\pi\gamma
}=0.622$ MeV is within the experimental interval $\Gamma_{a_{1}\rightarrow
\pi\gamma}^{\exp}=0.640\pm0.246$ MeV \cite{PDG} and $\Gamma_{a_{0}(1450)}=265$
MeV corresponds exactly to the experimental result.

\item Fit II yields a somewhat unexpected result for the mass of the
axial-vector kaon: $m_{K_{1}}=1343$ MeV, a value representing virtually a
mass-median of the two experimentally established axial-vector kaons,
$K_{1}(1270)$ and $K_{1}(1400)$. Consequently, Fit II does not allow for the
$K_{1}$ state in our model to be assigned to either of the two physical
fields. This statement is compatible with our explanation of the stated value
of $m_{K_{1}}$: our $K_{1}$ field is a member of a $1^{++}$ nonet that first
mixes with a $1^{+-}$ (pseudovector) nonet and then leads to the physical
fields $K_{1}(1270)$ and $K_{1}(1400)$. The $1^{+-}$ nonet is absent from the
model presented in this work; however, inclusion of the nonet into the model
is clearly demanded by our results, also because the full decay width
$\Gamma_{K_{1}}$ corresponds to neither $\Gamma_{K_{1}(1270)}$ nor
$\Gamma_{K_{1}(1400)}$. (See Sec.\ \ref{2K1} for more details.)

\item It is not possible to assign the two mixed isoscalar singlets
$\sigma_{1,2}$ if one varies $m_{0}^{2}<0$ and requires $m_{\sigma_{N}
}<m_{\sigma_{S}}$ because the ensuing intervals are rather large: $450$ MeV
$\leq m_{\sigma_{1}}\leq1561$ MeV and $1584$ MeV $\leq m_{\sigma_{2}}\leq2152$
MeV. Note, however, that $f_{0}(1710)$ is the only resonance confirmed by the
PDG in the mass range of the predominantly strange field $\sigma_{2}$.
Nonetheless, $m_{\sigma_{2}}$ varies too strongly for a definitive assignment
of $\sigma_{2}$ to $f_{0}(1710)$ to be performed. Consequently, the assignment
of the predominantly non-strange field $\sigma_{1}$ is, at this point, also uncertain.

\item As indicated above, the ambiguity in the determination of $m_{\sigma_{1,2}}$
is caused by uncertainty in the determination of $m_{0}^{2}<0$. The $m_{\sigma
_{2}}$ interval seems to suggest the correspondence $\sigma_{2}\equiv f_{0}
(1710)$. We therefore determine $m_{0}^{2}$ (with corresponding errors) from
the experimentally known ratio $\Gamma_{f_{0}(1710)\rightarrow\pi\pi}
(m_{0}^{2})/\Gamma_{f_{0}(1710)\rightarrow KK}(m_{0}^{2})=0.2\pm0.06$ and test
the ensuing phenomenology. [Note that the stated ratio did not enter Fit II
because otherwise our results would have been inclined to a certain assignment
of our scalar states. Note also that the stated ratio was obtained by the
WA102 Collaboration and that it does not correspond to the one preferred by
the PDG because the latter suffers from a large background and possible
interference of $f_{0}(1710)$ with $f_{0}(1790)$, see Sections
\ref{sec.f0(1790)} and \ref{f0(1710)-PDG-BESII}.] A large number of
scalar-meson observables can consequently be determined using \emph{only} the experimental ratio $\Gamma_{f_{0}(1710)\rightarrow\pi\pi} 
/\Gamma_{f_{0}(1710)\rightarrow KK} =0.2\pm0.06$ as input:

\begin{itemize}

\item \textit{Masses:} we obtain $m_{\sigma_{1}}=1310_{+30}^{-29}$ MeV and
$m_{\sigma_{2}}=1606_{+4}^{-3}$ MeV. The central value of $m_{\sigma_{1}}$
corresponds almost exactly to the combined-fit Breit-Wigner mass of
Ref.\ \cite{buggf0} where $m_{f_{0}(1370)}=(1309\pm1\pm15)$ MeV was obtained
and also to the $f_{0}(1370)$ peak mass in the $2\pi$ channel, found to be
$1282$ MeV in Ref.\ \cite{buggf0}. Additionally, we observe that
$m_{\sigma_{2}}$\ is $\sim100$ MeV smaller than $m_{f_{0}(1710)}=(1720\pm6)$
MeV because the glueball field has not been included in the current version of
the model. The mass values imply that $f_{0}(1370)$ is $91.2_{+2.0}^{-1.7}\%$
a $\bar{n}n$ state and that, conversely, $f_{0}(1710)$ is $91.2_{+2.0}%
^{-1.7}\%$ a $\bar{s}s$ state.

\item \textit{Two-pion decays:} we obtain $\Gamma_{\sigma_{1}\rightarrow\pi
\pi}=267_{+25}^{-50}$ MeV and $\Gamma_{\sigma_{2}\rightarrow\pi\pi}%
=47_{-10}^{+9}$ MeV. The former is consistent with the Breit-Wigner decay
width $\Gamma_{f_{0}(1370)\rightarrow\pi\pi}=325$ MeV and the $f_{0}(1370)$
full width at half-maximum in the $2\pi$ channel, the value of which was
determined as $207$ MeV in Ref.\ \cite{buggf0}. The latter is too large (see
Sec.\ \ref{bb}) but nonetheless demonstrates that $\Gamma_{f_{0}%
(1710)\rightarrow\pi\pi}$ is suppressed in comparison with $\Gamma
_{f_{0}(1710)\rightarrow KK}$, in accordance with the PDG data \cite{PDG}.

\item \textit{Two-kaon decays:} we obtain $\Gamma_{\sigma_{1}\rightarrow
KK}=188_{+6}^{-9}$ MeV and $\Gamma_{\sigma_{2}\rightarrow KK}=237_{+25}^{-20}$
MeV. The former is consistent with results in
Refs.\ \cite{Etkin:1981,f01370KK2,Tikhomirov:2003,Polychronakos:1978,f01370KK1,f01370KK3}
and implies $\Gamma_{f_{0}(1370)\rightarrow KK}<\Gamma_{f_{0}(1370)\rightarrow
\pi\pi}$, consistent with interpretation of $f_{0}(1370)$ as a predominantly
$\bar{n}n$ state. $\Gamma_{\sigma_{2}\rightarrow KK}$ is larger than the
corresponding experimental result $\sim80$ MeV (see Sec.\ \ref{bb}); however,
it is also dominant in comparison with decay widths in other channels --
consistent with the data \cite{PDG}.

\item \textit{Two-eta decays:} we obtain $\Gamma_{\sigma_{1}\rightarrow
\eta\eta}=(40\mp1)$ MeV and $\Gamma_{\sigma_{2}\rightarrow\eta\eta}%
=60_{+5}^{-4}$ MeV. The former is lower than the values from
Refs.\ \cite{f0(1500)-CB-1992,Alde:1985kp} that, however, did not consider
the opening of new channels over the broad $f_{0}(1370)$ decay interval. The
latter is marginally (within errors)\ consistent with the experimental result
presented in Sec.\ \ref{bb}. We also obtain $\Gamma_{\sigma_{2}\rightarrow
\eta\eta^{\prime}}=41_{-5}^{+4}$ MeV; this result is a prediction.

\item \textit{Decays with }$a_{1}$\textit{: }we predict $\Gamma_{f_{0}%
(1370)\rightarrow a_{1}(1260)\pi\rightarrow\rho\pi\pi}=12.7_{-4.2}^{+5.8}$ MeV
and additionally $\Gamma_{f_{0}(1710)\rightarrow a_{1}(1260)\pi\rightarrow\rho\pi\pi
}=15.2_{-3.1}^{+2.6}$ MeV (strongly suppressed in comparison with other decay
channels of the two resonances).

\item \textit{The pion-kaon ratio} $\Gamma_{\sigma_{1}\rightarrow\pi\pi}%
/\Gamma_{\sigma_{1}\rightarrow KK}=1.42_{+0.09}^{-0.05}$ is consistent with
the WA102 result $\Gamma_{f_{0}(1370)\rightarrow\pi\pi}/\Gamma_{f_{0}%
(1370)\rightarrow KK}=$ $2.17\pm1.23$ \cite{Barberis:1999}.

\item \textit{The eta-pion ratios }read $\Gamma_{\sigma_{1}\rightarrow\eta\eta
}/\Gamma_{\sigma_{1}\rightarrow\pi\pi}=0.15\pm0.01$ and $\Gamma_{\sigma
_{2}\rightarrow\eta\eta}/\Gamma_{\sigma_{2}\rightarrow\pi\pi}=1.26_{+0.52}%
^{-0.27}$. The former is within the ratio $\Gamma_{f_{0}(1370)\rightarrow
\eta\eta}/\Gamma_{f_{0}(1370)\rightarrow\pi\pi}=0.19\pm0.07$ of
Ref.\ \cite{buggf0}. The latter corresponds almost completely to the WA102
ratio $\Gamma_{f_{0}(1710)\rightarrow\eta\eta}^{\text{WA102}}/\Gamma
_{f_{0}(1710)\rightarrow\pi\pi}^{\text{WA102}}=2.4\pm1.04$, see Sec.\ \ref{bb}.

\item \textit{The eta-kaon ratios} read $\Gamma_{\sigma_{1}\rightarrow\eta\eta
}/\Gamma_{\sigma_{1}\rightarrow KK}=0.22\pm0.01$ and $\Gamma_{\sigma
_{2}\rightarrow\eta\eta}/\Gamma_{\sigma_{2}\rightarrow KK}=0.25\pm0.004$. The
former is purely a prediction. The latter is completely within the combined-fit
result $\Gamma_{f_{0}(1710)\rightarrow\eta\eta}/\Gamma_{f_{0}(1710)\rightarrow
KK}=0.46_{-0.38}^{+0.70}$ \cite{Anisovich:2001} and within $2\sigma$\ of the
WA102 result $\Gamma_{f_{0}(1710)\rightarrow\eta\eta}/\Gamma_{f_{0}%
(1710)\rightarrow KK}=0.48\pm0.15$ \cite{Barberis:2000}.

\item \textit{Ratios with }$\eta^{\prime}$\textit{:} our model predicts the values $\Gamma
_{f_{0}(1710)\rightarrow\eta\eta^{\prime}}/\Gamma_{f_{0}(1710)\rightarrow
KK}=0.17_{-0.03}^{+0.04}$, $\Gamma_{f_{0}(1710)\rightarrow\eta\eta^{\prime}%
}/\Gamma_{f_{0}(1710)\rightarrow\pi\pi}=0.86_{+0.11}^{-0.06}$ and
$\Gamma_{f_{0}(1710)\rightarrow\eta\eta^{\prime}}/\Gamma_{f_{0}%
(1710)\rightarrow\eta\eta}=0.68\pm0.13$.

\item \textit{Decays with }$\omega(782)$\textit{:} we predict $\Gamma
_{f_{0}(1710)\rightarrow\omega\omega}\simeq0.02$ MeV and\textit{\ }%
$\Gamma_{f_{0}(1710)\rightarrow\omega\omega\rightarrow6\pi}\simeq0.02$ MeV,
thus virtually no $f_{0}(1710)$ decay in the $\omega\omega$ channel.
\end{itemize}

\item We obtain $\Gamma_{K_{0}^{\star}(1430)\rightarrow K\pi}=263$ MeV, within
the PDG interval $\Gamma_{K_{0}^{\star}(1430)}^{\exp}=(270\pm80)$ MeV
\cite{PDG}.

\item The scattering lengths $a_{0}^{0,2}$ are saturated to their
Weinberg limits $a_{0}^{0}$ $\simeq0.158$, $a_{0}^{2}$ $\simeq-0.0448$ (see
the end of Sec.\ \ref{SL}) in Fit II. This implies the necessity to include the scalars
below 1 GeV into our model -- but they cannot be of $\bar{q}q$ structure.

\item Additionally, the phenomenology in the vector and axial-vector channels is
extremely improved in comparison with Fit I.

\begin{itemize}

\item We obtain $\Gamma_{K^{\star}\rightarrow K\pi}=44.2$ MeV, only $2$ MeV
less than the PDG result $\Gamma_{K^{\star}\rightarrow K\pi}^{\exp}=46.2$ MeV
\cite{PDG}. Note that Fit I implied $\Gamma_{K^{\star}\rightarrow K\pi
}^{\text{FIT I}}=32.8$ MeV. Note that the PDG mass and decay width of the
non-strange vector state $\rho(770)$ are implemented exactly in our model.

\item We obtain $\Gamma_{\varphi(1020)\rightarrow K^{+}K^{-}}=2.33$ MeV
whereas the PDG suggests $\Gamma_{\varphi(1020)\rightarrow K^{+}%
K^{-}}^{\exp}=(2.08\pm0.04)$ MeV \cite{PDG}. Our result is slightly larger
because $m_{\varphi(1020)}$ from our model is $\sim20$ MeV heavier than
the experimental value inducing an increase in phase space. It was not
possible to calculate this decay width from Fit I because $\varphi(1020)$ was
well below the $KK$ threshold.

\item The decay width $\Gamma_{a_{1}(1260)\rightarrow\rho\pi}$ is improved by
two orders of magnitude and now has the value 861 MeV whereas Fit I yielded
$ \simeq 13$ GeV. The decay width is decreased once the $\rho$ meson is
considered an off-shell state: $\Gamma_{a_{1}\rightarrow\rho\pi\rightarrow
3\pi}=706$ MeV. Nonetheless, it is still somewhat above the PDG interval
$\Gamma_{a_{1}(1260)\rightarrow\rho\pi}=(250-600)$ MeV \cite{PDG}. This may
imply (\textit{i}) that one needs to consider the finiteness of the strong
interaction using a suitable form factor or (\textit{ii}) that $a_{1}(1260)$
is not predominantly a quarkonium (although the overlap with the $\bar{q}q$
wave function is large, as suggested by our results). Alternatively,
decreasing $\Gamma_{\rho\rightarrow\pi\pi}$ (the decay width determining the
parameter $g_{2}$ that in turn decisively influences $\Gamma_{a_{1}%
(1260)\rightarrow\rho\pi}$) by $\sim20$ MeV yields $\Gamma_{a_{1}%
(1260)\rightarrow\rho\pi}<600$ MeV. Note that Fit I required decreasing
$\Gamma_{\rho\rightarrow\pi\pi}$ by $\sim100$ MeV for $\Gamma_{a_{1}%
(1260)\rightarrow\rho\pi}<600$ MeV to be obtained. We also obtain
$\Gamma_{a_{1}(1260)\rightarrow\bar{K}^{\star}K\rightarrow\bar{K}K\pi}=0.55$
MeV and find $\Gamma_{a_{1}(1260)\rightarrow f_{0}(1370)\pi}\simeq0$. Fit I
implied $\Gamma_{a_{1}(1260)\rightarrow\bar{K}^{\star}K\rightarrow\bar{K}K\pi
}^{\text{FIT I}}=1.97$ GeV.

\item Only one partial decay width of the $f_{1}(1285)$ resonance can be calculated within
our model: $f_{1}(1285)\rightarrow\bar{K}^{\star}K\rightarrow\bar{K}K\pi$. The
PDG does not cite a value for the decay width of the sequential decay but
rather $\Gamma_{f_{1}(1285)\rightarrow\bar{K}K\pi}=(2.2\pm0.1)$ MeV, stating
that no there is no contribution to this decay width from the stated
sequential decay. Contrarily, we find $\Gamma_{f_{1}(1285)\rightarrow\bar
{K}^{\star}K\rightarrow\bar{K}K\pi}=0.9$ MeV, implying a 40\% contribution to
$\Gamma_{f_{1}(1285)\rightarrow\bar{K}K\pi}$. The result is obtained for
$m_{f_{1}(1285)\equiv f_{1N}}=m_{a_{1}}=1219$ MeV; increasing $m_{f_{1}%
(1285)\equiv f_{1N}}$ to the PDG value of $1281.8$ MeV yields $\Gamma
_{f_{1}(1285)\rightarrow\bar{K}^{\star}K\rightarrow\bar{K}K\pi}=$
$\Gamma_{f_{1}(1285)\rightarrow\bar{K}K\pi}$. Note that Fit I yielded
$\Gamma_{f_{1}(1285)\rightarrow\bar{K}^{\star}K}^{\text{FIT I}}\simeq2.15$ GeV
and thus Fit II represents a strong improvement of the results from Fit I.

\item We also obtain $\Gamma_{f_{1}(1420)\rightarrow\bar{K}^{\star}K}=274$
MeV. This result is two orders of magnitude smaller than the (unphysically
large) value $\Gamma_{f_{1S}\rightarrow\bar{K}^{\star}K}^{\text{FIT I}}=17.6$
GeV. Nonetheless, it is larger than the one reported by the PDG:
$\Gamma_{f_{1}(1420)}^{\exp}=(54.9\pm2.6)$ MeV. The reason is assumed to be
the absence of the $1^{+-}$ nonet from our model expected to mix with the
$1^{++}$ nonet already present in the model [and containing $f_{1}(1420)$].

\item The full $K_{1}$ decay width is still larger than those of the two
physical states: we obtain $\Gamma_{K_{1}}\sim480$ MeV whereas the data
suggest $\Gamma_{K_{1}(1400)}=(174\pm13)$ MeV and $\Gamma_{K_{1}(1270)}%
=(90\pm20)$ MeV. The reason has already been discussed: mixing of our $1^{++}$
nonet with the partner $1^{+-}$ nonet has to be implemented in the model.
Nonetheless, Fit II improves not only the full decay widths but also the
partial ones: we obtain $\Gamma_{K_{1}\rightarrow K^{\star}\pi}=307$ MeV,
$\Gamma_{K_{1}\rightarrow\rho K}=128$ MeV, $\Gamma_{K_{1}\rightarrow\omega
_{N}K}=41$ MeV whereas Fit I yielded $\Gamma_{K_{1}\rightarrow K^{\star}\pi
}^{\text{FIT I}}=6.73$ GeV, $\Gamma_{K_{1}\rightarrow\rho K}^{\text{FIT I}%
}=4.77$ GeV, $\Gamma_{K_{1}\rightarrow\omega_{N}K}^{\text{FIT I}}=1.59$ GeV.
Thus the full and partial $K_{1}$ decay widths, although still too large, have
strongly improved in Fit II.
\end{itemize}

Thus Fit II accommodates the correct (axial-)vector phenomenology into the model
[$\rho$, $K^{\star}$, $\varphi(1020)$, $f_{1}(1285)$], yields qualitative
consistence [$a_{1}(1260)$] or suggests the necessity to include further states
into the model [$f_{1}(1420)$, $K_{1}$].
\end{itemize}

Fit II yields a decisively better description of the overall phenomenology:
meson masses and decay widths are either described correctly or stand closer
to the data than in Fit I. For these reasons, the assumption of scalar
$\bar{q}q$ states above $1$ GeV is strongly preferred over the assumption that
the same states are present below $1$ GeV.\\

A comparison of results from the two fits is presented in Table \ref{Comparison}%
; experimental uncertainties are omitted. Table \ref{Comparison} contains all
the masses except $m_{\sigma_{1,2}}$ because of the experimental
uncertainties (the $\sigma_{1,2}$ results are discussed above in this section).%

\begin{table}[!b] \centering
\begin{tabular}
[c]{|c|c|c|c|}\hline
Observable & Fit I [MeV] & Fit II [MeV] & Experiment [MeV]\\\hline
$m_{\pi}$ & $138.04$ & $138.65$ & $139.57$\\\hline
$m_{K}$ & $490.84$ & $497.96$ & $493.68$\\\hline
$m_{\eta}$ & $517.13$ & $523.20$ & $547.85$\\\hline
$m_{K_{S}}$ & $1128.7$ & $1550$ & $676$ (FitI)/$1425$ (FitII) \\\hline
$m_{\eta^{\prime}}$ & $957.78$ & $957.78$ & $957.78$\\\hline
$m_{\rho}$ & $775.49$ & $775.49$ & $775.49$\\\hline
$\text{ }m_{K^{\star}}$ & $832.53$ & $916.52$ & $891.66$\\\hline
%
$\text{ }m_{a_{0}}$ & $978$ & $1452$ & $980$ (Fit I)/$1474$ (FitII)\\\hline
$m_{\varphi(1020)}\text{ }$ & $870.35$ & $1036.90$ & $1019.46$\\\hline
$m_{a_{1}(1260)}$ & $1396$ & $1219$ & $1230$\\\hline
$m_{K_{1}}$ & $1520$ & $1343$ & $1272$ or $1403$\\\hline
$m_{f_{1}(1420)}$ & $1643.4$ & $1457.0$ & $1426.4$\\\hline
$\Gamma_{a_{1}(1260)\rightarrow\pi\gamma}$ & $0.369$ & $0.622$ &
$0.640$\\\hline
$\Gamma_{\rho\rightarrow\pi\pi}$ & $149.1$ & $149.1$ & $149.1$\\\hline
$\Gamma_{K^{\star}\rightarrow K\pi}$ & $32.8$ & $44.2$ & $46.2$\\\hline
$\Gamma_{\varphi(1020)\rightarrow K^{+}K^{-}}$ & $0$ & $2.33$ & $2.08$\\\hline
$\Gamma_{a_{1}(1260)\rightarrow\rho\pi}$ & $\sim13000$ & $861$ &
$<600$\\\hline
$\Gamma_{a_{1}(1260)\rightarrow\rho\pi\rightarrow3\pi}$ & $\sim11000$ & $706$
& $<600$\\\hline
$\Gamma_{a_{1}(1260)\rightarrow\bar{K}^{\star}K\rightarrow\bar{K}K\pi}$ &
$1970$ & $0.55$ & small\\\hline
$\Gamma_{f_{1}(1285)\rightarrow\bar{K}^{\star}K\rightarrow\bar{K}K\pi}$ &
$1980$ & $0.9$ & $<2.2$\\\hline
$\Gamma_{f_{1}(1420)\rightarrow\bar{K}^{\star}K\rightarrow\bar{K}K\pi}$ &
$17600$ & $274$ & $\sim54.9$\\\hline
$\Gamma_{K_{1}}$ & $\sim13000$ & $\sim480$ & $\lesssim170$\\\hline
$a_{0}^{0}$ & $0.165$ MeV$^{-1}$ & $0.158$ MeV$^{-1}$ & $0.218$ MeV$^{-1}%
$\\\hline
$a_{0}^{2}$ & $-0.0442$ MeV$^{-1}$ & $-0.0448$ MeV$^{-1}$ & $-0.0457$
MeV$^{-1}$\\\hline
\end{tabular}%
\caption{Results from Fit I and Fit II compared with experiment.  \label
{Comparison}}%
\end{table}%
\end{fmffile}

%% file: Glueball.tex
\chapter{Incorporating the Glueball into the Model} \label{chapterglueball}

In the previous chapters we have discussed the phenomenology of states
containing an antiquark and a quark. However, gluons -- the gauge bosons of
QCD -- can also build composite states of their own: the so-called
\textit{glueballs}. Thus we expect in particular a scalar glueball to exist;
if it does, then it could mix with the scalar $\bar{q}q$\ states already presented in
this work. The mixing is discussed in this chapter.

\section{Introduction}

Glueballs, bound states of gluons, are naturally expected in QCD
due to the non-Abelian nature of the theory: gluons interact strongly with
themselves and thus they can bind and form colorless states, analogously to
what occurs in the quark sector. The existence of glueballs has been studied
in the framework of the effective bag model for QCD already four decades ago
\cite{bag-glueball} and it has been further investigated in a variety of
approaches \cite{Close:1987,f0(1710)asglueball,Close}. Numerical calculations
of the Yang-Mills sector of QCD also find a full glueball spectrum in which
the scalar glueball is the lightest state \cite{Morningstar}.

Glueballs can mix with quarkonium ($\bar{q}q$) states with the same quantum
numbers. This makes the experimental search for glueballs more complicated,
because physical resonances emerge as mixed states. The scalar sector
$J^{PC}=0^{++}$ has been investigated in many works in the past. The resonance
$f_{0}(1500)$ is relatively narrow when compared to other scalar-isoscalar
states: for this reason it has been considered as a convincing candidate for a
glueball state. Mixing scenarios in which two quark-antiquark isoscalar states
$\bar{n}n$ and $\bar{s}s$ and one scalar glueball $gg$ mix and generate the
physical resonances $f_{0}(1370),$ $f_{0}(1500)$, and $f_{0}(1710)$ have been
discussed in Refs.\ \cite{refs,longglueball}.

In this chapter we discuss how to extend the calculations presented in this
work to include a glueball field. The discussion will regard the $U(2)\times
U(2)$\ version of the model from Chapter \ref{chapterQ} only; a corresponding extension of
the $U(3)\times U(3)$ model is a very interesting project in itself that will
be treated in a separate work \cite{StaniD}.

The first attempt to incorporate a glueball into a linear sigma model was
performed long ago in Ref.\ \cite{schechter}. The novel features of the study
in this chapter are the following: (\textit{i}) The glueball is introduced as
a dilaton field within a theoretical framework where not only scalar and
pseudoscalar mesons, but also vector and axial-vector mesons are present from
the very beginning. This fact allows also for a calculation of decays into
vector mesons. As already indicated, the model is explicitly evaluated for the
case of $N_{f}=2,$ for which only one scalar-isoscalar quarkonium state
exists: $\sigma_{N}\equiv\bar{n}n$ which mixes with the glueball. The two
emerging mixed states are assigned to the resonances $f_{0}(1370)$ which is,
in accordance with Sec.\ \ref{sec.scalars2}, predominantly a $\bar{n}n$ state,
and with $f_{0}(1500)$ which is predominantly a glueball state. (\textit{ii})
We consequently test -- to our knowledge for the first time -- this mixing
scenario above 1 GeV in the framework of a chiral model.

Let us emphasise again that our model is built in accordance with the
symmetries of the QCD Lagrangian. It possesses the known degrees of freedom of
low-energy QCD [(pseudo)scalar and (axial-)vector mesons] as well as the same
global chiral invariance. In this chapter, we model another feature of the QCD
Lagrangian: the scale (or dilatation) invariance $x^{\mu}\rightarrow
\lambda^{-1}x^{\mu}$ (where $x^{\mu}$ is a Minkowski-space coordinate and
$\lambda$ the scale parameter of the conformal group), see Eq.\ (\ref{Dt1})
and the discussion thereafter. The scale invariance is realised at the
classical level but broken at the quantum level due to the loop corrections in
the Yang-Mills sector (scale anomaly). In this chapter the breaking of scale
invariance is implemented at tree-level by means of a dilaton field
(representing a glueball) with the usual logarithmic dilaton potential
\cite{schechter}. However, all the other interaction terms (with the exception
of the chiral anomaly) are dilatation-invariant in the chiral limit.

Having constructed the Lagrangian of the effective model, we calculate the
masses of the pure $\bar{n}n$ and glueball states in the $J^{PC}=0^{++}$
channel, study their mixing and calculate the decay widths of the mixed
states. Although we work with $N_{f}=2$ in this chapter, the use of flavour
symmetry enables us to calculate the decay widths of the scalar resonances
into kaons and into both the $\eta$ and $\eta^{\prime}$ mesons which contain
the $s$-quark in their flavour wave functions. After the study of the already
mentioned scenario where $f_{0}(1370)$ and $f_{0}(1500)$ are predominantly
quarkonium and glueball, respectively, we also test the alternative scenario
in which the resonance $f_{0}(1710)$ is predominantly glueball and scenarios
in which $f_{0}(600)$ is predominantly quarkonium. They, however, lead to
inconsistencies when compared to the present data and are therefore regarded
as less favourable. Additionally, our results discussed in
Sec.\ \ref{sec.conclusionsfitII} also favour $f_{0}(1710)$ to be a
predominantly $\bar{s}s$ state rather than a glueball.

\section{The Model}

The Yang-Mills (YM) sector of QCD (QCD without quarks) is classically
invariant under dilatations [see Eqs.\ (\ref{Dt1}) -- (\ref{Tm})]. This
symmetry is, however, broken at the quantum level. The divergence of the
corresponding current is the trace of the energy-momentum tensor
$T_{\text{YM}}^{\mu\nu}$ of the YM Lagrangian
\begin{equation}
\left(  T_{\text{YM}}\right)  _{\mu}^{\mu}=\frac{\beta(g)}{4g}\,G_{\mu\nu}%
^{a}G^{a,\mu\nu}\neq0\text{,} \label{ta}%
\end{equation}
where $G_{\mu\nu}^{a}$ is the field-strength tensor of the gluon fields,
$g=g(\mu)$ is the renormalised coupling constant at the scale $\mu$, and the
$\beta$-function is given by $\beta(g)=\partial g/\partial\ln\mu.$ At the
one-loop level $\beta(g)=-bg^{3}$ with $b=11N_{c}/(48\pi^{2})$. This implies
$g^{2}(\mu)=\left[  2b\ln(\mu/\Lambda_{\text{YM}})\right]  ^{-1}$, where
$\Lambda_{\text{YM}}\simeq200$ MeV is the Yang-Mills scale. A finite energy
scale thus emerges in a theory which is classically invariant under dilatation
(dimensional transmutation). The expectation value of the trace anomaly does
not vanish and represents the so-called gluon condensate:
\begin{equation}
\left\langle T_{\text{YM},\mu}^{\mu}\right\rangle =-\frac{11N_{c}}%
{48}\left\langle \frac{\alpha_{s}}{\pi}\,G_{\mu\nu}^{a}G^{a,\mu\nu
}\right\rangle =-\frac{11N_{c}}{48}C^{4}\text{,} \label{gc}%
\end{equation}
where \cite{Sumrules,Lattice}
\begin{equation}
C^{4}\simeq(300-600\text{ MeV})^{4}\text{.} \label{gc1}%
\end{equation}

At the composite level one can build an effective theory of the YM sector of
QCD by introducing a scalar dilaton field $G$ which describes the trace
anomaly. The dilaton Lagrangian reads \cite{schechter}%
\begin{equation}
\mathcal{L}_{dil}=\frac{1}{2}(\partial_{\mu}G)^{2}-\frac{1}{4}\frac{m_{G}^{2}%
}{\Lambda^{2}}\left(  G^{4}\ln\left\vert \frac{G}{\Lambda}\right\vert
-\frac{G^{4}}{4}\right)  \text{.} \label{ldil}%
\end{equation}
The minimum $G_{0}$ of the dilaton potential is realised for $G_{0}=\Lambda$.
Upon shifting $G\rightarrow G_{0}+G,$ a particle with mass $m_{G}$ emerges,
which is interpreted as the scalar glueball. The numerical value has been
determined in Lattice QCD and reads $m_{G}\sim1.5$ GeV \cite{Morningstar}. The
logarithmic term of the potential explicitly breaks the invariance under a
dilatation transformation. The divergence of the corresponding current reads
$\partial_{\mu}J_{dil}^{\mu}=T_{dil,\,\mu}^{\;\mu}=-\frac{1}{4}m_{G}%
^{2}\Lambda^{2}$. This can be compared with the analogous quantity in
Eq.\ (\ref{gc}) which implies $\Lambda=\sqrt{11}\,C^{2}/(2m_{G})$.

As demonstrated in Sec.\ \ref{sec.othersymmetries}, QCD\ with quarks is
also classically invariant under dilatation transformations in the limit of
zero quark masses (chiral limit). The scale of all hadronic phenomena is given
by the previously introduced energy scale $\Lambda_{\text{YM}}$. This fact
holds true also when the small but nonzero values of the quark masses are
considered. In order to describe these properties in a hadronic model we now
extend the linear sigma model of the previous chapters by including the
dilaton. To this end, the following criteria are applied \cite{dynrec}:
(\textit{i}) With the exception of the chiral anomaly, the parameter $\Lambda$
from Eq.\ (\ref{ldil}), which comes from the Yang-Mills sector of the theory
in accordance with QCD, is the only dimensionful parameter of the Lagrangian
in the chiral limit. (\textit{ii}) The Lagrangian is required to be finite for
every finite value of the gluon condensate $G_{0}$. This, in turn, also
assures that no singular terms arise in the limit $G_{0}\rightarrow0$. In
accordance with the requirements (\textit{i}) and (\textit{ii}) only terms
with dimension exactly equal to 4 are allowed in the chiral limit.

The hadronic Lagrangian obeying these requirements reads
\begin{align}%
\mathcal{L}%
&  =\mathcal{L}_{dil}+\text{\textrm{Tr}}\left[  (D^{\mu}\Phi)^{\dag}(D_{\mu
}\Phi)-m_{0}^{2}\left(  \frac{G}{G_{0}}\right)  ^{2}\Phi^{\dag}\Phi
-\lambda_{2}(\Phi^{\dag}\Phi)^{2}\right]  -\lambda_{1}(\text{\textrm{Tr}%
}\left[  \Phi^{\dag}\Phi\right]  )^{2}\nonumber\\
&  +c[\text{\textrm{det}}(\Phi^{\dag})+\text{\textrm{det}}(\Phi
)]+\text{\textrm{Tr}}\left[  H\left(  \Phi^{\dag}+\Phi\right)  \right]
-\frac{1}{4}\text{\textrm{Tr}}\left[  (L^{\mu\nu})^{2}+(R^{\mu\nu})^{2}\right]
\nonumber\\
&  +\mathrm{Tr}\left\{  \left[  \frac{m_{1}^{2}}{2}\left(  \frac{G}{G_{0}%
}\right)  ^{2}+\Delta\right]  (L_{\mu}^{2}+R_{\mu}^{2})\right\}  +\frac{h_{1}%
}{2}\text{\textrm{Tr}}[\Phi^{\dag}\Phi]\text{\textrm{Tr}}[L_{\mu}L^{\mu
}+R_{\mu}R^{\mu}]\nonumber\\
&  +h_{2}\text{\textrm{Tr}}[\Phi^{\dag}L_{\mu}L^{\mu}\Phi+\Phi R_{\mu}R^{\mu
}\Phi^{\dag}]+2h_{3}\text{\textrm{Tr}}[\Phi R_{\mu}\Phi^{\dag}L^{\mu}%
]+\ldots\text{ ,} \label{LagrangianG}%
\end{align}
where $\Phi$ denotes the $N_{f}\times N_{f}$ (pseudo)scalar multiplet and
$L^{\mu}$ and $R^{\mu}$ the left- and right-handed vector multiplets,
respectively. The dots represent further terms which do not affect the
processes studied in this work.

The Lagrangian presented in Eq.\ (\ref{LagrangianG}) possesses the generic
form for any number of flavours $N_{f}$. It is a generalisation of the
Lagrangian (\ref{LagrangianGe}) constructed in Chapter \ref{chapterC}. In this
chapter, the Lagrangian is evaluated for two flavours only. Consequently, as
in Chapter \ref{chapterQ}, we define $\Phi=(\sigma_{N}+i\eta_{N})\,t^{0}%
+(\vec{a}_{0}+i\vec{\pi})\cdot\vec{t}$ ($\eta_{N}$ contains only non-strange
degrees of freedom), $L^{\mu}=(\omega^{\mu}+f_{1}^{\mu})\,t^{0}+(\vec{\rho
}^{\mu}+\vec{a}_{1}^{\mu})\cdot\vec{t}$ and $R^{\mu}=(\omega_{N}^{\mu}%
-f_{1N}^{\mu})\,t^{0}+(\vec{\rho}^{\mu}-\vec{a}_{1}^{\mu})\cdot\vec{t}$ ;
$t^{0}$, $\vec{t}$ are the generators of $U(2)$. Moreover, $D^{\mu}%
\Phi=\partial^{\mu}\Phi-ig_{1}(L^{\mu}\Phi-\Phi R^{\mu})$, $L^{\mu\nu
}=\partial^{\mu}L^{\nu}-\partial^{\nu}L^{\mu}$, $R^{\mu\nu}=\partial^{\mu
}R^{\nu}-\partial^{\nu}R^{\mu}$.

The explicit breaking of the global chiral symmetry is described by the term
\textrm{Tr}$[H(\Phi+\Phi^{\dagger})]\equiv h\sigma$ $(h=const$. $\sim
m_{u,d}^{2})$, which allows us to take into account the non-vanishing value
$m_{u.d}$ of the non-strange quark mass. This term contains the dimensionful
parameter $h$ with $[h]$ $=[\mathrm{energy}^{3}]$ and also explicitly breaks
the dilatation invariance, just as the quark masses do in the underlying QCD
Lagrangian. Finally, the chiral anomaly is described by the term $c\,(\det
\Phi+\det\Phi^{\dagger})$. This term corresponds to the one utilised in the
model containing quarkonia only, see Chapter \ref{chapterQ} (and also
Sec.\ \ref{sec.anomaly}). For $N_{f}=2$ the parameter $c$ carries the
dimension $[\mathrm{energy}^{2}]$ and represents a further breaking of
dilatation invariance. This term arises from instantons which are also a
property of the Yang-Mills sector of QCD. 

The identification of the fields of the model with the resonances listed by
the PDG \cite{PDG} is the same as in Chapter \ref{chapterQ}. We assign the
fields $\vec{\pi}$ and $\eta_{N}$ to the pion and the $SU(2)$ counterpart of
the $\eta$ meson, respectively, $\eta_{N}\equiv(\bar{u}u+\bar
{d}d)/\sqrt{2}$, with a mass of about $700$ MeV. This value can be obtained by
`unmixing' the physical $\eta$ and $\eta^{\prime}$ mesons which also contain
$\bar{s}s$ contributions. The fields $\omega_{N}^{\mu}$ and $\vec{\rho
}^{\;\mu}$ represent the $\omega(782)$ and $\rho(770)$ vector mesons,
respectively, while the fields $f_{1N}^{\mu}$ and $\vec{a}_{1}^{\;\mu}$
represent the $f_{1}(1285)$ and $a_{1}(1260)$ axial-vector mesons,
respectively. [In principle, the physical $\omega(782)$ and $f_{1}(1285)$
states also contain $\bar{s}s$ contributions but their admixture is
small.] As shown in Sec.\ \ref{sec.sigmapionpionQ}, the $\sigma_{N}$ field
should be interpreted as a $\bar{n}n$ state because its decay width decreases
as $1/N_{c}$ in the limit of a large number of colors. The $\sigma_{N}$ and
$G$ fields mix: the physical fields $\sigma^{\prime}$ and $G^{\prime}$ are
obtained through an $SO(2)$ rotation, as we shall show in the following. Then
the first and most natural assignment is $\{\sigma^{\prime},G^{\prime
}\}=\{f_{0}(1370),f_{0}(1500)\},$ see Sec.\ \ref{A}. Note that the $\vec
{a}_{0}$ state is assigned to the physical $a_{0}(1450)$ resonance in
accordance with results of Sec.\ \ref{sec.a0(1450)Q}, confirmed by results
from the $U(3)\times U(3)$ version of our model in Chapter \ref{sec.fitII}.
Other assignments for $\{\sigma^{\prime},G^{\prime}\}$ will be also tested in
Sections \ref{B} and \ref{C} and turn out to be less favourable.

In order to study the non-vanishing vacuum expectation values (vev's) of the
two $J^{PC}=0^{++}$ scalar-isoscalar fields of the model $\sigma_{N}$ and $G$,
we set all the other fields in Eq.\ (\ref{LagrangianG}) to zero and obtain:
\begin{equation}
\mathcal{L}%
_{\sigma G}=\mathcal{L}_{dil}+\frac{1}{2}(\partial^{\mu}\sigma_{N})^{2}
-\frac{1}{2}\left[  m_{0}^{2}\left(  \frac{G}{G_{0}}\right)  ^{2}-c\right]
\sigma_{N}^{2}-\frac{1}{4}\left(  \lambda_{1}+\frac{\lambda_{2}}{2}\right)
\sigma_{N}^{4}+h\sigma_{N}\text{.} \label{Lagrangian1}
\end{equation}
Upon shifting the fields by their vacuum expectation values, $\sigma
_{N}\rightarrow\sigma_{N}+\phi_{N}$ and $G\rightarrow G+G_{0}$, we obtain the
masses of the states $\sigma_{N}=(\bar{u}u+\bar{d}d)/\sqrt{2}$ and $G=gg$,
\begin{equation}
M_{\sigma_{N}}^{2}=m_{0}^{2}-c+3\left(  \lambda_{1}+\frac{\lambda_{2}}
{2}\right)  \phi_{N}^{2}\;,\;\;\;\;\;M_{G}^{2}=m_{0}^{2}\,\frac{\phi_{N}^{2}
}{G_{0}^{2}}+m_{G}^{2}\,\frac{G_{0}^{2}}{\Lambda^{2}}\,\left(  1+3\ln
\left\vert \frac{G_{0}}{\Lambda}\right\vert \right)  \text{.} \label{M_G}
\end{equation}
Note that the pure glueball mass $M_{G}$ depends also on the quark condensate
$\phi_{N}$, but correctly reduces to $m_{G}$ in the limit $m_{0}^{2}=0$
(decoupling of quarkonia and glueball). In the presence of quarkonia,
$m_{0}^{2}\neq0$, the vev $G_{0}$ is given by the equation
\begin{equation}
-\frac{m_{0}^{2}\phi_{N}^{2}\Lambda^{2}}{m_{G}^{2}}=G_{0}^{4}\ln\left\vert
\frac{G_{0}}{\Lambda}\right\vert \text{.} \label{G0}
\end{equation}

The shift of the fields by their vev's introduces a bilinear mixing term
$\sim\sigma_{N}G$ in the Lagrangian (\ref{Lagrangian1}). The physical fields
$\sigma^{\prime}$ and $G^{\prime}$ can be obtained through an $SO(2)$
rotation,
\begin{equation}
\left(
\begin{array}
[c]{c}
\sigma^{\prime}\\
G^{\prime}
\end{array}
\right)  =\left(
\begin{array}
[c]{cc}%
\cos\varphi_{G} & \sin\varphi_{G}\\
-\sin\varphi_{G} & \cos\varphi_{G}%
\end{array}
\right)  \left(
\begin{array}
[c]{c}
\sigma_{N}\\
G
\end{array}
\right)  \text{,}
\end{equation}
with%
\begin{align}
M_{\sigma^{\prime}}^{2}  &  =M_{\sigma_{N}}^{2}\cos^{2}\varphi_{G}+M_{G}
^{2}\sin^{2}\varphi_{G}+2\,m_{0}^{2}\,\frac{\phi_{N}}{G_{0}}\sin(2\varphi
_{G})\text{,}\label{m_sigma_r}\\
M_{G^{\prime}}^{2}  &  =M_{G}^{2}\cos^{2}\varphi_{G}+M_{\sigma_{N}}^{2}
\sin^{2}\varphi_{G}-2\,m_{0}^{2}\,\frac{\phi_{N}}{G_{0}}\sin(2\varphi
_{G})\text{,} \label{m_G_r}
\end{align}
where the mixing angle $\theta_{G}$ reads
\begin{equation}
\varphi_{G}=\frac{1}{2}\arctan\left[  -4\,\frac{\phi_{N}}{G_{0}}\,\frac
{m_{0}^{2}}{M_{G}^{2}-M_{\sigma_{N}}^{2}}\right]  \text{ .} \label{theta}
\end{equation}
The quantity $m_{0}^{2}$ can be calculated from the masses of the pion,
$\eta_{N}$, and the bare $\sigma_{N}$ mass [Eqs.\ (\ref{sigma}), (\ref{eta1}) and (\ref{pion})]:
\begin{equation}
m_{0}^{2}=\left(  \frac{m_{\pi}}{Z_{\pi}}\right)  ^{2}+\frac{1}{2}\left[
\left(  \frac{m_{\eta_{N}}}{Z_{\pi}}\right)  ^{2}-M_{\sigma_{N}}^{2}\right]
\text{.} \label{m_0}
\end{equation}
If $m_{0}^{2}-c<0$, spontaneous breaking of chiral symmetry is realised.

\section{Results and Discussion}

The Lagrangian (\ref{LagrangianG}) contains the following twelve free
parameters: $m_{0}$, $\lambda_{1}$, $\lambda_{2}$, $m_{1}$, $g_{1}$, $c$, $h$,
$h_{1}$, $h_{2}$, $h_{3}$, $m_{G}$, $\Lambda=\sqrt{11}\,C^{2}/(2m_{G})$. The
processes that we shall consider depend only on the combination $h_{1}+h_{2}+h_{3}$,
thus reducing the number of parameters to ten. We replace the
set of ten parameters by the following equivalent set: $m_{\pi}$, $m_{\eta
_{N}}$, $m_{\rho}$, $m_{a_{1}}$, $\phi_{N}$, $Z_{\pi}$, $M_{\sigma_{N}}$,
$m_{G}$, $m_{1}$, $C$. The masses $m_{\pi}$ ($=139.57$ MeV) and $m_{\rho}$
($=775.49$ MeV) are fixed to their PDG values \cite{PDG}.

As outlined in Sec.\ \ref{sec.eta-eta}, the mass of the $\eta_{N}$ meson can
be calculated using the mixing of strange and non-strange contributions in the
physical fields $\eta$ and $\eta^{\prime}$:
\begin{align}
\eta &  =\eta_{N}\cos\varphi_{\eta}+\eta_{S}\sin\varphi_{\eta}\text{,} \nonumber\\
\text{ }\eta^{\prime}  &  =-\eta_{N}\sin\varphi_{\eta}+\eta_{S}\cos
\varphi_{\eta}\text{,} \label{phiG}%
\end{align}
where $\eta_{S}$ denotes a pure $\bar{s}s$ state and $\varphi_{\eta}%
\simeq-36^{\circ}$ \cite{Giacosa:2007up}. In this way, we obtain the value
$m_{\eta_{N}}=716$ MeV. [Given the well-known uncertainty of the value of the
angle $\varphi_{\eta}$, one could also consider other values, e.g., our result
$\varphi_{\eta}=-43.9%
{{}^\circ}%
$ from Chapter \ref{sec.fitII} (see discussion of Table \ref{Fit2-5}), which
corresponds to $m_{\eta_{N}}=764$ MeV, or the value $\varphi_{\eta
}=-41.4^{\circ}$ from the KLOE Collaboration \cite{KLOE}, which
corresponds to $m_{\eta_{N}}=755$ MeV. Variations of the pseudoscalar mixing
angle affect the results presented in this chapter only slightly.]

The value of $m_{a_{1}}$ is fixed to $1050$ MeV according to the study of
Ref.\ \cite{UBW}. (We note that taking the value 1219 MeV from
Sec.\ \ref{sec.fitII} or the present PDG estimate of 1230 MeV does not change
the conclusions of this chapter.) The chiral condensate is fixed as $\phi
_{N}=Z_{\pi}f_{\pi}$ and the renormalization constant $Z_{\pi}$ is determined
by the study of the process $a_{1}\rightarrow\pi\gamma$: $Z_{\pi}=1.67\pm0.2$
in Sec.\ \ref{sec.a1pg}.

\subsection{Assigning \boldmath $\sigma^{\prime}$ and \boldmath $G^{\prime}$ to \boldmath $f_{0}(1370)$ and
\boldmath $f_{0}(1500)$} \label{AG}

The $\sigma^{\prime}$ field denotes an isoscalar $J^{PC}=0^{++}$ state and its
assignment to a physical state is a long-debated problem of low-energy QCD
\cite{Jaffeq2q2,f0(980)q2q2-f0(600)qq-f0(1370)q2q2,Close:1987,f0(1710)asglueball,refs,fariborz,SchechterQ,Giacosa:2006tf,Close,Morningstar}%
. The two major candidates are the $f_{0}(600)$ and
$f_{0}(1370)$ resonances, see Sections \ref{sec.conclusionsfitI} and
\ref{sec.conclusionsfitII}. We have concluded in
Sec.\ \ref{sec.conclusionsfitII} that $f_{0}(1370)$ is favoured to be
predominantly a $\bar{n}n$ state. As already stated, the resonance $f_{0}(1500)$
is a convincing glueball candidate. For these reasons we first test the
scenario in which $\{\sigma^{\prime},G^{\prime}\}=\{f_{0}(1370),f_{0}%
(1500)\}$, which turns out to be phenomenologically successful, see below.

We are left with the following four free parameters: $C$, $M_{\sigma_{N}}$,
$m_{G}$, $m_{1}$. They can be obtained by a fit to the five experimental
quantities of Table \ref{Table1G}: the masses of the resonances $f_{0}(1500)$
[$M_{G^{\prime}}\equiv M_{f_{0}(1500)}=1505$ MeV \cite{PDG}] and $f_{0}%
(1370)$, for which we use the mean value $M_{\sigma_{N}^{\prime}}^{\exp
}=(1350\pm150)$ MeV taking into account the PDG mass range between 1200 MeV
and 1500 MeV \cite{PDG}), and the three well-known decay widths of the
well-measured resonance $f_{0}(1500)$: $f_{0}(1500)\rightarrow\pi\pi$,
$f_{0}(1500)\rightarrow\eta\eta$, and $f_{0}(1500)\rightarrow K\bar{K}$.

\begin{table}[h] \centering
\begin{tabular}
[c]{|c|c|c|}\hline
Quantity & Our Value [MeV] & Experiment [MeV]\\\hline
$M_{\sigma^{\prime}}$ & $1191\pm26$ & $1200$-$1500$\\\hline
$M_{G^{\prime}}$ & $1505\pm6$ & $1505\pm6$\\\hline
$G^{\prime}\rightarrow\pi\pi$ & $38\pm5$ & $38.04\pm4.95$\\\hline
$G^{\prime}\rightarrow\eta\eta$ & $5.3\pm1.3$ & $5.56\pm1.34$\\\hline
$G^{\prime}\rightarrow K\bar{K}$ & $9.3\pm1.7$ & $9.37\pm1.69$\\\hline
\end{tabular}%
\caption{Fit in the scenario $\{\sigma'$, $G'\}$ = $\{f_0(1370)$, $f_0(1500)\}$. Note that the $f_0(1370)$ mass ranges between 1200 MeV and
1500 MeV \cite{PDG}
and therefore, as an estimate, we are using the value $m_{\sigma'}%
=(1350 \pm150)$ MeV in the fit.\label{Table1G}}
\end{table}%

Using the Lagrangian (\ref{LagrangianG}), these observables can be expressed
as functions of the parameters listed above. Note that, although our framework
is based on $N_{f}=2$, we can calculate the amplitudes for the decays into
mesons containing strange quarks by making use of the flavour symmetry
$SU(N_{f}=3)$ \cite{longglueball}. It is then possible to calculate the
following $f_{0}(1500)$ decay widths into pseudoscalar mesons containing
$s$ quarks: $f_{0}(1500)\rightarrow K\bar{K}$, $f_{0}(1500)\rightarrow\eta
\eta$, and $f_{0}(1500)\rightarrow\eta\eta^{\prime}$.

The $\chi^{2}$ method yields $\chi^{2}/$d.o.f. $=0.29$ (thus very small),
$C=(699\pm40)$ MeV, $M_{\sigma_{N}}=(1275\pm30)$ MeV, $m_{G}=(1369\pm26)$ MeV
and $m_{1}=(809\pm18)$ MeV. We have also examined the uniqueness of our fit.
To this end, we have considered $\chi^{2}$ fixing three of four parameters
entering the fit at their best values and varying the remaining fourth
parameter. In each of the four cases we observe only one minimum of the
$\chi^{2}$ function; each minimum leads exactly to the parameter values stated
in Table \ref{Table1G}. We also observe no changes of the results for the
errors of the parameters. These findings give us confidence that the obtained
minimum corresponds to the absolute minimum of the $\chi^{2}$ function.

The consequences of this fit are the following:

(i) The quarkonium-glueball mixing angle reads $\theta_{G}=\left(
29.7\pm3.6\right)
{{}^\circ}%
$. This, in turn, implies that the resonance $f_{0}(1500)$ consists to $76\%$
of a glueball and to the remaining $24\%$ of a quark-antiquark state. An
inverted situation holds for $f_{0}(1370)$. Given our results discussed in
Sec.\ \ref{sec.conclusionsfitII}, we conclude that $f_{0}(1370)$ possesses
admixtures from both $\bar{n}n$ and glueball; a detailed discussion will be
presented in Ref.\ \cite{StaniD}.

(ii) Our fit allows us to determine the gluon condensate: $C=(699\pm40)$ MeV.
This result implies that the upper value in Eq.\ (\ref{gc}) is favoured by our
analysis. It is remarkable that insights into this basic quantity of QCD can
be obtained from the PDG data on mesons.

(iii) Further results for the $f_{0}(1500)$ meson are reported in the first
two entries of Table \ref{Table2G}. The decay into $4\pi$ is calculated as a
product of an intermediate $\rho\rho$ decay. To this end the usual integration
over the $\rho$ spectral function is performed. Our result yields 30 MeV in
the $4\pi$ decay channel and is about half of the experimental value
$\Gamma_{f_{0}(1500)\rightarrow4\pi}$ $=(54.0\pm7.1)$ MeV. However, it should
be noted that an intermediate state consisting of two $f_{0}(600)$ mesons
(which is also expected to contribute in this decay channel) is not included
in the present model. The decay into the $\eta\eta^{\prime}$ channel is also
evaluated; this channel is subtle because it is exactly on the threshold of
the $f_{0}(1500)$ mass. Therefore, an integration over the spectral function
of the decaying meson $f_{0}(1500)$ is necessary. The result is in a
qualitative agreement with the experiment. Note also that the enhanced value
of the $4\pi$ decay width is a consequence of the inclusion of the glueball
field into the model [identified predominantly with $f_{0}(1500)$] as
otherwise the $4\pi$ decay channel is known to be suppressed [see the note on
$\sigma_{1,2}\rightarrow4\pi$ decays in Sec.\ \ref{sec.sigmapionpion2} for the
case of our non-strange and strange quarkonia].

(iv) The results for the $f_{0}(1370)$ meson are reported in the last four
rows of Table \ref{Table2G}. They are in agreement with the experimental data
regarding the full width: $\Gamma_{f_{0}(1370)}=(200$ -- $500)$\ MeV
\cite{PDG}. Unfortunately, the experimental results in the different channels
are not yet conclusive. Our theoretical results point towards a dominant
direct $\pi\pi$ and a non-negligible $\eta\eta$ contribution; these results
correspond well to the experimental analysis of Ref.\ \cite{buggf0} where
$\Gamma_{f_{0}(1370)\rightarrow\pi\pi}=325$ MeV and $\Gamma_{f_{0}%
(1370)\rightarrow\eta\eta}/\Gamma_{f_{0}(1370)\rightarrow\pi\pi}=$
$0.19\pm0.07$ are obtained. [Note that Ref.\ \cite{buggf0} also cites the
Breit-Wigner mass of $1309$ MeV whereas our result $M_{\sigma^{\prime}\equiv
f_{0}(1370)}=(1191\pm26)$ MeV is $\sim100$ MeV smaller.] We find that the
four-pion decay of $f_{0}(1370)\rightarrow\rho\rho\rightarrow4\pi$ is strongly
suppressed (as was also determined in Sec.\ \ref{sec.sigmapionpion2}). As
stated in Sec.\ \ref{sec.f0(1370)}, the values of the $f_{0}(1370)$ decay
widths are strongly mass-dependent with Ref.\ \cite{buggf0} citing the value
of $\Gamma_{f_{0}(1370)\rightarrow\pi\pi}\sim50$ MeV for $m_{f_{0}(1370)}%
\sim1309$ MeV. The $4\pi$ phase space decreases rapidly with a decreasing
resonance mass and is virtually negligible for our result $M_{\sigma^{\prime
}\equiv f_{0}(1370)}=(1191\pm26)$ MeV. For this reason, our results are
qualitatively consistent with statements in Ref.\ \cite{buggf0}. Additionally,
it should be noted that due to interference effects our result for this decay
channel varies strongly when the parameters are even slightly modified.

(v) The mass of the $\rho$ meson can be expressed as $m_{\rho}^{2}=m_{1}%
^{2}+\phi^{2}\left(  h_{1}+h_{2}+h_{3}\right)  /2$, see Eq.\ (\ref{m_rho}). In
order that the contribution of the chiral condensate is not negative, the
condition $m_{1}\leq m_{\rho}$ should hold. In the framework of our fit this
condition is fulfilled at the two-sigma level. This result points towards a
dominant $m_{1}$ contribution to the $\rho$ mass. This property, in turn,
means that the $\rho$ mass is predominantly generated from the gluon
condensate and not from the chiral condensate, as confirmed by our result
$m_{1}=762$ MeV in the $U(3)\times U(3)$ version of the model, see Table
\ref{Fit2-5}.\ It is therefore expected that the $\rho$ mass in the medium
scales as the gluon condensate rather than as the chiral condensate. In view
of the fact that $m_{1}$ is slightly larger than $m_{\rho}$ we have also
repeated the fit by fixing $m_{1}=m_{\rho}$: the minimum has a $\chi^{2}%
/$d.o.f. $\simeq1$ and the results are very similar to the previous case. The
corresponding discussion about the phenomenology is unchanged.

(vi) As already stressed in Refs.\ \cite{Paper1,Zakopane}, the inclusion of
(axial-)vector mesons plays a central role to obtain the present results. The
artificial decoupling of (axial-)vector states would generate a by far too
wide $f_{0}(1370)$ state. For this reason the glueball-quarkonium mixing
scenario above 1 GeV has been previously studied only in phenomenological
models with flavour symmetry
\cite{Close:1987,f0(1710)asglueball,longglueball,Close} but not in the context
of chirally invariant models.

\begin{table}[h] \centering
\begin{tabular}
[c]{|c|c|c|}\hline
Quantity & Our Value [MeV] & Experiment [MeV]\\\hline
$G^{\prime}\rightarrow\rho\rho\rightarrow4\pi$ & $30$ & $54.0\pm7.1$\\\hline
$G^{\prime}\rightarrow\eta\eta^{\prime}$ & $0.6$ & $2.1\pm1.0$\\\hline
$\sigma_{N}^{\prime}\rightarrow\pi\pi$ & $284\pm43$ & -\\\hline
$\sigma_{N}^{\prime}\rightarrow\eta\eta$ & $72\pm6$ & -\\\hline
$\sigma_{N}^{\prime}\rightarrow K\bar{K}$ & $4.6\pm2.1$ & -\\\hline
$\sigma_{N}^{\prime}\rightarrow\rho\rho\rightarrow4\pi$ & $0.09$ & -\\\hline
\end{tabular}%
\caption{Further results regarding the $\sigma' \equiv
f_0(1370)$ and $G' \equiv f_0(1500)$ decays.\label{Table2G}}%
\end{table}%

Given that the resonance $f_{0}(1370)$ has a large mass uncertainty, we have
also examined the behaviour of the fit at different points of the PDG mass
interval. Considering the minimal value $m_{f_{0}(1370)}^{\min}=(1220\pm20)$
MeV we obtain $\chi^{2}=0.2/$d.o.f. The resulting value of the mixing angle
$\theta_{G}=(30.3\pm3.4)%
{{}^\circ}%
$ is practically the same as the value $\theta_{G}=(29.7\pm3.6)%
{{}^\circ}%
$ obtained in the case where $m_{f_{0}(1370)}=(1350\pm150)$ MeV was
considered. Other results are also qualitatively similar to the case of
$m_{f_{0}(1370)}=(1350\pm150)$ MeV.

For the upper boundary of the $f_{0}(1370)$ mass, the error interval of
$\pm20$ MeV turns out to be too restrictive as it leads to unacceptably large
$\chi^{2}$ values. Consequently, increasing the error interval decreases the
$\chi^{2}$ values -- we observe that $m_{f_{0}(1370)}^{\max}=(1480\pm120)$ MeV
leads to an acceptable $\chi^{2}$ value of $1.14/$d.o.f. Then we obtain
$\theta_{G}=(30.0\pm3.5)%
{{}^\circ}%
$, practically unchanged in comparison with the value $\theta_{G}=(29.7\pm3.6)%
{{}^\circ}%
$ in the case where $m_{f_{0}(1370)}=(1350\pm150)$ MeV. Also other quantities
remain basically the same as in the case of $m_{f_{0}(1370)}=(1350\pm150)$ MeV.

We have also considered the fit at several points between the lower and upper
boundaries of the $m_{f_{0}(1370)}$ mass range. We have chosen points of 50
MeV difference starting at $m_{f_{0}(1370)}=1250$ MeV (i.e., we have
considered $m_{f_{0}(1370)}\in\{1250,1300,1350,1400,1450\}$ MeV) with errors
chosen such that the $\chi^{2}/$d.o.f.\ becomes minimal (error values are
between $\pm30$ MeV for $m_{f_{0}(1370)}=1250$ MeV and $\pm100$ MeV for
$m_{f_{0}(1370)}=1450$ MeV). We observe that the previous results presented in
this section do not change significantly; most notably, the mixing angle
$\theta_{G}$ attains values between $30.2%
{{}^\circ}%
$ and $30.7%
{{}^\circ}%
$, with an average error value of $\pm3.4%
{{}^\circ}%
$.

We therefore conclude that considering different values of $m_{f_{0}(1370)}$
within the $(1200-1500)$ MeV interval does not change the results
significantly. In particular, the quarkonium-glueball mixing angle $\theta
_{G}$ changes only slightly (by approximately $1%
{{}^\circ}%
$) and thus we confirm our conclusion that $f_{0}(1370)$ is predominantly a
quarkonium and $f_{0}(1500)$ is predominantly a glueball.

\subsection{Assigning \boldmath $\sigma_{N}^{\prime}$ and \boldmath $G^{\prime}$ to \boldmath $f_{0}(1370)$
and \boldmath $f_{0}(1710)$} \label{B}

Although the resonance $f_{0}(1710)$ has also been regarded as a glueball
candidate in a variety of works \cite{refs2}, its enhanced decay into kaons
and its rather small decay width make it compatible with a dominant ${\bar{s}%
}s$ contribution in its wave function. This was also confirmed by our results
from the $U(3)\times U(3)$ version of the model, see
Sec.\ \ref{sec.conclusionsfitII}. Nonetheless, we have also tested the
assumption that the pure quarkonium and glueball states mix to produce the
resonances $f_{0}(1370)$ and $f_{0}(1710)$.

Some experimental results regarding the resonance $f_{0}(1710)$ suffer
from uncertainties stemming from the overlap with the nearby state
$f_{0}(1790)$, see Sections \ref{sec.f0(1790)} and \ref{f0(1710)channels}.
Decays of $f_{0}(1710)$ into $\pi\pi,$ $\bar{K}K$, and $\eta\eta$ have been
seen while no decays into $\eta\eta^{\prime}$ and into $4\pi$ have been
detected; partial decay widths of this resonance based on PDG-preferred
results as well as those of the WA102 Collaboration have already been
presented in Sections \ref{f0(1710)-PDG-BESII} and \ref{bb}. The values of
decay widths into $\pi\pi$, $\bar{K}K$, and $\eta\eta$ obtained in
Sec.\ \ref{bb}\ are stated in Table \ref{SzenarioIII}.

A fit analogous to the one in Table \ref{Table1G} yields too large errors for
the decay width $\sigma_{N}^{\prime}\equiv f_{0}(1370)\rightarrow\pi\pi$. For
this reason we repeat our fit by adding the following constraint:
$\Gamma_{\sigma_{N}^{\prime}\rightarrow\pi\pi}=(250\pm150)$ MeV. The large
error assures that this value is in agreement with experimental data on this
decay width. The results of the fit are reported in Table \ref{SzenarioIII}.%

\begin{table}[h] \centering
\begin{tabular}
[c]{|c|c|c|}\hline
Quantity & Our Value [MeV] & Experiment ($G^{\prime}$ Decays from WA102 data, Sec.\ \ref{bb}) [MeV]\\\hline
$M_{\sigma_{N}^{\prime}}$ & $1450\pm34$ & $1350\pm150$\\\hline
$M_{G^{\prime}}$ & $1720\pm6$ & $1720\pm6$\\\hline
$G^{\prime}\rightarrow\pi\pi$ & $16.0\pm3.6$ & $16.1\pm3.6$\\\hline
$G^{\prime}\rightarrow\eta\eta$ & $4.1\pm1.0$ & $38.6\pm18.8$\\\hline
$G^{\prime}\rightarrow K\bar{K}$ & $5.1\pm2.7$ & $80.5\pm30.1$\\\hline
$\sigma_{N}^{\prime}\rightarrow\pi\pi$ & $313\pm49$ & $250\pm150$\\\hline
\end{tabular}%
\caption{Fit in the scenario $\{\sigma
'$, $G'\} = \{f_0(1370)$, $f_0(1710)\}$. Experimental data for $G'$ decays are from Sec.\ \ref
{bb}, other data from the PDG \cite{PDG}.
\label{SzenarioIII}}%
\end{table}%

We obtain $C=(1070\pm65)$ MeV, $M_{\sigma_{N}}=(1483\pm47)$ MeV,
$m_{G}=(1670\pm20)$ MeV and $m_{1}=(817\pm16)$ MeV; Eq.\ (\ref{theta}) yields
the mixing angle between the pure quarkonium and the pure glueball $\theta
_{G}=(19.6\pm5.8)%
{{}^\circ}%
$. Note that the gluon condensate is in this case much larger than what would
be expected from the QCD sum rules or the lattice, see Eq.\ (\ref{gc1}). The
$\chi^{2}$ is worse than in the previous case: $\chi^{2}/$d.o.f. $=2.5$.
Additionally, $\Gamma_{\sigma_{N}^{\prime}\rightarrow\pi\pi}$ is too large for
the mass value $M_{\sigma_{N}^{\prime}}\simeq1450$ MeV, see
Sec.\ \ref{sec.f0(1370)} -- the data suggest that the decay into $4\pi$ (and
not into $2\pi$) is dominant at such large values of the $f_{0}(1370)$ mass.
We also observe that both $\Gamma_{G^{\prime}\rightarrow\eta\eta}$ and
$\Gamma_{G^{\prime}\rightarrow K\bar{K}}$ are by an order of magnitude smaller
than their respective experimental values. Thus the WA102 data do not seem to
favour a fit where $f_{0}(1370)$ and $f_{0}(1710)$ are, respectively,
predominantly quarkonium and glueball. This statement is confirmed if further
decays are considered: as evident from Table \ref{SzenarioIII1}, $G^{\prime
}\equiv f_{0}(1710)\rightarrow4\pi$ should be the largest contribution to the
full $f_{0}(1710)$ decay width (branching ratio $\sim2/3$) while
experimentally it has not been seen.%

\begin{table}[h] \centering
\begin{tabular}
[c]{|c|c|c|}\hline
Decay Width & Our Value [MeV] &
Experimental value [MeV]\\\hline
$G^{\prime}\rightarrow4\pi$ & $41$ & -\\\hline
$G^{\prime}\rightarrow\eta\eta^{\prime}$ & $4.3$ & -\\\hline
$\sigma^{\prime}\rightarrow\eta\eta$ & $100\pm8$ & -\\\hline
$\sigma^{\prime}\rightarrow K\bar{K}$ & $18.7\pm5.3$ & -\\\hline
\end{tabular}%
\caption{Further results from the fit with $\{ \sigma
'$, $G$'\} = \{$f_0(1370)$, $f_0(1710)$\}.\label{SzenarioIII1}}%
\end{table}%

Note that a virtually unchanged picture emerges if the fit utilises the
PDG-preferred data for the decays of $G^{\prime}=f_{0}(1710)$, see
Sec.\ \ref{f0(1710)-PDG-BESII}.\ We then obtain $\chi^{2}/$d.o.f. $=1.7$,
$C=(764\pm256)$ MeV (now within expectations), $M_{\sigma_{N}}=(1516\pm80)$
MeV, $m_{G}=(1531\pm233)$ MeV and $m_{1}=(827\pm36)$ MeV \cite{Stani}. The
mixing angle calculated from Eq.\ (\ref{theta}) is $\theta_{G}=(37.2\pm21.4)%
{{}^\circ}%
$. The mixing angle is large and could also overshoot the value of $45%
{{}^\circ}%
$, which would imply a somewhat unexpected and unnatural reversed ordering, in
which $f_{0}(1370)$ is predominantly glueball and $f_{0}(1710)$ predominantly
quarkonium. Additionally, we still obtain $\Gamma_{G^{\prime}\rightarrow
\eta\eta}=(6.9\pm5.8)$ MeV, five times smaller than $\Gamma_{f_{0}%
(1710)\rightarrow\eta\eta}^{\text{PDG}}=34.26_{-20.0}^{+15.4}$ MeV in
Eq.\ (\ref{f0(1710)_10}), and also $\Gamma_{G^{\prime}\rightarrow K\bar{K}%
}=(16\pm14)$ MeV, again strongly suppressed in comparison with $\Gamma
_{f_{0}(1710)\rightarrow KK}^{\text{PDG}}=71.44_{-35.02}^{+23.18}$ MeV,
Eq.\ (\ref{f0(1710)_9}). Finally, the $4\pi$ decay of $f_{0}(1710)$ should
again be dominant ($\sim115$ MeV), clearly at odds with the data.

Therefore, we conclude that this scenario is not favoured. Moreover, in this
scenario the remaining resonance $f_{0}(1500)$ should then be interpreted as a
predominantly $\bar{s}s$ state, contrary to what its experimentally dominant
$\pi\pi$ decay pattern suggests. Consequently, $f_{0}(1710)$ is unlikely to be
predominantly a glueball state; this is also in accordance with the results
from the ZEUS Collaboration \cite{ZEUS:2008}.

\subsection{Scenarios with \boldmath $\sigma^{\prime}\equiv f_{0}(600)$} \label{C}

The scenarios $\{\sigma^{\prime},G^{\prime}\}=$ $\{f_{0}(600),$ $f_{0}%
(1500)\}$ and $\{\sigma^{\prime},G^{\prime}\}=$ $\{f_{0}(600),$ $f_{0}%
(1710)\}$ have also been tested. In both cases the mixing angle turns out to
be small ($\lesssim15^{\circ}$), thus the state $f_{0}(600)$ is predominantly
quarkonium. Then, in these cases the analysis of Chapter \ref{chapterQ} applies: a
simultaneous description of the $\pi\pi$ scattering lengths and the
$\sigma^{\prime}\rightarrow\pi\pi$ decay width cannot be achieved. For these
reasons the mixing scenarios with the resonance $f_{0}(600)$ as a quarkonium
state are not favoured.

\section{Summary of the Results with the Dilaton Field} \label{sec.summaryglueball}

Once a dilaton field is included into the chirally invariant linear sigma
model with (axial-)vectors, a favoured scenario emerges: the resonance
$f_{0}(1500)$ is predominantly a glueball with a subdominant $\bar{n}n$
component and, conversely, $f_{0}(1370)$ is predominantly a quark-antiquark
$(\bar{u}u+\bar{d}d)/\sqrt{2}$ state with a subdominant glueball contribution.
It is interesting to observe that the success of the phenomenological
description of these scalar resonances is due to the inclusion of the
(axial-)vector mesons in the model. The gluon condensate is also an outcome of
our study and turns out to be in agreement with lattice QCD results. Different
scenarios in which $f_{0}(1710)$ is predominantly glueball and/or $f_{0}(600)$
is predominantly quarkonium do not seem to be in agreement with the present
experimental data.

%% file: Summary.tex
\chapter{Conclusions} \label{sec.summary}

In this thesis we have presented an effective model of Quantum
Chromodynamics (QCD), the theory of strong interactions. The model was
utilised to study all experimentally observed two-body decays of mesons for
which there exist vertices in the model under the assumption that the said
experimental states are of $\bar q q$ nature. In addition, three-body and four-body decay
widths have also been calculated utilising sequential decays; $\pi\pi$
scattering lengths have been calculated as well. Particular attention was
devoted to the question whether scalar ${\bar{q}q}$ states are located below
or above 1 GeV in the physical spectrum. Our results clearly favour the scalar $\bar q q$ states to be above 1 GeV.\newline

A realistic model of QCD with $N_{f}$\ quark flavours should possess at least
two features. Firstly, the model has to implement the symmetries present in
QCD and described in Chapter \ref{sec.QCD}, most notably the local $SU(3)_{c}$
colour symmetry, the discrete $CPT$ symmetry, the global $U(N_{f})_{L}\times
U(N_{f})_{R}$ chiral symmetry and the breaking mechanisms of the latter
symmetry: spontaneous (due to the chiral condensate), explicit (due to
non-vanishing quark masses) as well as at the quantum level [the $U(1)_{A}$
anomaly]. Secondly, the model has to incorporate as many degrees of freedom as
possible, within the energy interval of interest (typically determined by the
mass of the highest resonance in the model, in our case $\sim1.8$ GeV).

This also implies that the resonances should not be considered independently
of each other -- they may mix (if they possess the same quantum numbers) or
stand connected via decay modes. For this reason, in the concrete case of our
meson model, we have considered not only scalar and pseudoscalar but also
vector and axial-vector mesons as well.\\

The model has implemented the linear realisation of the chiral symmetry of QCD
\cite{gellmanlevy} with two ($u$, $d$) and three flavours ($u$, $d$, $s$) --
linear sigma model. The symmetry-breaking mechanisms have also been
considered: the explicit symmetry breaking was modelled with terms
proportional to (non-degenerate) quark masses, the chiral anomaly by a
determinant term and the spontaneous symmetry breaking by means of
condensation of the scalar isosinglet states: $\sigma_{N}\equiv(\bar{u}%
u+\bar{d}d)/\sqrt{2}$ in the $N_{f}=2$ case and $\sigma_{N}$ as well as
$\sigma_{S}\equiv\bar{s}s$ in the $N_{f}=3$ case. Thus combining our meson
states from constituent quarks and antiquarks we are able to construct two
scalar isosinglet states $\sigma_{N,S}$. (Note that all the states present in
our model are of ${\bar{q}q}$ structure, as we demonstrate in
Sec.\ \ref{sec.largen}.)

However, as we have discussed in Chapter \ref{sec.scalarexp}, current
experimental data suggest that there are actually six non-strange scalar
isosinglets: $f_{0}(600)$ or $\sigma$, $f_{0}(980)$, $f_{0}(1370)$,
$f_{0}(1500)$, $f_{0}(1710)$ and $f_{0}(1790)$. At most two of them can be
${\bar{q}q}$ states -- and the thesis has addressed the question \textit{which
two}.
To this end, we have constructed three versions of the sigma
model: in two flavours (Chapter \ref{chapterQ}), three flavours (Chapters
\ref{sec.remarks} -- \ref{ImplicationsFitII}) and two flavours + a scalar
glueball state in Chapter \ref{chapterglueball}.\\

In Chapter \ref{chapterQ}, the two-flavour version of the linear sigma model
with vector and axial-vector mesons was discussed in two scenarios. In
Scenario I, Sec.\ \ref{sec.scenarioI}, we assigned our $\sigma_{N}$ state to
the $f_{0}(600)$ resonance [or, in other words, $f_{0}(600)$ was assumed to be
a ${\bar{q}q}$ state]. However, the ensuing $f_{0}(600)$ decay width was
several times smaller than the result suggested by the data
\cite{PDG,Leutwyler,Pelaez1}. For this reason we have considered an
alternative scenario where $f_{0}(1370)$ was assumed to be of ${\bar{q}q}$
structure obtaining $\Gamma_{f_{0}(1370)\rightarrow\pi\pi}\simeq (300$-$500)$ MeV
for $m_{f_{0}(1370)}=(1200$-$1400)$ MeV. Thus, already in the two-flavour model,
the scenario in which the scalar states above 1 GeV, $f_{0}(1370)$ and
$a_{0}(1450)$, are considered to be (predominantly) $\bar{q}q$ states appears
to be favoured over the assignment in which $f_{0}(600)$ and $a_{0}(980)$ are
considered (predominantly) $\bar{q}q$ states. It is important to stress that
the role of the (axial-)vector states was crucial in obtaining these results
because, in the (unrealistic) limit where the (axial-)vectors are removed from
the model, we observed that the $f_{0}(600)$ decay width was within the data.
Note also that we have calculated a range of other decay widths, in particular
$\Gamma_{a_{1}(1260)\rightarrow\rho\pi}$, found to be consistent with
experiment if $m_{a_{1}(1260)}\simeq1130$ MeV.\\

In Chapters \ref{sec.remarks} -- \ref{ImplicationsFitII} we have addressed the
question whether the conclusions from the two-flavour model remain the same
once the model is generalised by inclusion of strange mesons (kaons). Thus,
upon inclusion of the strange degrees of freedom we have again considered two
possibilities: (\textit{i}) that the scalar ${\bar{q}q}$ states are below 1
GeV and (\textit{ii}) that the scalar ${\bar{q}q}$ states are above 1 GeV. We
refer to these two possibilities as Fits I and II, respectively.

Although the phenomenology of scalar states is found to be acceptable, Fit I
is nonetheless found to be strongly disfavoured for two reasons (see
Sec.\ \ref{sec.conclusionsfitI}). Firstly, the obtained mass values deviate by
up to $\sim200$ MeV (for the $\kappa$ meson: $\sim600$ MeV) from the
experimental results (see Table \ref{Fit1-5}). This is in particular
problematic for the very narrow resonances $\varphi(1020)$ and $f_{1}(1420)$.
Secondly, the axial-vector states are found to be extremely broad:
$a_{1}(1260)$, $f_{1}(1285)$, $f_{1}(1420)$ and $K_{1}(1400)$ possess
decay widths $\sim(1-10)$ GeV. These values are unphysically large. The only
possibility to remedy these large decay widths would be to work with the $\rho$
meson that has a decay width $\lesssim40$ MeV. However, then the $\rho$ meson would be too narrow.\\

We thus consider possibility (\textit{ii}): scalar ${\bar{q}q}$ states above 1
GeV. All masses obtained from Fit II are within 3\%\ of their respective
experimental values with the exception of $m_{\eta}$ ($\simeq4.5\%$ too small)
and $m_{K_{0}^{\star}(1430)}$, found to be $\simeq8.8\%$ too large because the
pattern of explicit symmetry breaking in our model sets masses of strange
states approximately 100 MeV ($\simeq$ strange-quark mass) heavier than their
corresponding non-strange counterparts. Nonetheless, the phenomenology is
massively improved in comparison with Fit I (see
Sec.\ \ref{sec.conclusionsfitII} and in particular Table \ref{Comparison}).
For example, our results for the (axial-)vector states are either within the data
[$\rho$, $K^{\star}$, $\varphi(1020)$, $f_{1}(1285)$] or qualitatively
consistent with the data [$a_{1}(1260)$]. The mixing of the pure-nonstrange
and the pure-strange scalar isosinglet states allows us to determine
$f_{0}(1370)$ as $91.2_{+2.0}^{-1.7}\%$ a $\bar{n}n$ state and $f_{0}(1710)$
as $91.2_{+2.0}^{-1.7}\%$ a $\bar{s}s$ state. Utilising only the ratio
$\Gamma_{f_{0}(1710)\rightarrow\pi\pi}/\Gamma_{f_{0}(1710)\rightarrow
KK}=0.2\pm0.06$ \cite{Barberis:1999} enables us to determine a large range of
other observables with \textit{no free parameters}. We calculate
$\Gamma_{f_{0}(1370)\rightarrow\pi\pi}$, $\Gamma_{f_{0}(1370)\rightarrow KK}$,
$\Gamma_{f_{0}(1710)\rightarrow\eta\eta}$, $\Gamma_{f_{0}(1370)\rightarrow
\pi\pi}/\Gamma_{f_{0}(1370)\rightarrow KK}$, $\Gamma_{f_{0}(1370)\rightarrow
\eta\eta}/\Gamma_{f_{0}(1370)\rightarrow\pi\pi}$, $\Gamma_{f_{0}%
(1710)\rightarrow\eta\eta}/\Gamma_{f_{0}(1710)\rightarrow\pi\pi}$, $\Gamma_{f_{0}%
(1710)\rightarrow\eta\eta}/\Gamma_{f_{0}(1710)\rightarrow KK}$ and
$\Gamma_{K_{0}^{\star}(1430)\rightarrow K\pi}$ and their values are all within the
data. We can even predict $\Gamma_{f_{0}(1370)\rightarrow\eta\eta}%
/\Gamma_{f_{0}(1370)\rightarrow KK}=0.22\pm0.01$, $\Gamma_{f_{0}%
(1710)\rightarrow\eta\eta^{\prime}}/\Gamma_{f_{0}(1710)\rightarrow
KK}=0.17_{-0.03}^{+0.04}$, $\Gamma_{f_{0}(1710)\rightarrow\eta\eta^{\prime}}$
$/\Gamma_{f_{0}(1710)\rightarrow\pi\pi}=0.86_{+0.11}^{-0.06}$, $\Gamma
_{f_{0}(1710)\rightarrow\eta\eta^{\prime}}/\Gamma_{f_{0}(1710)\rightarrow
\eta\eta}=0.68\pm0.13,$ $\Gamma_{f_{0}(1710)\rightarrow\eta\eta^{\prime}%
}=41_{-5}^{+4}$ MeV, $\Gamma_{f_{0}(1370)\rightarrow a_{1}(1260)\pi
\rightarrow\rho\pi\pi}=12.7_{-4.2}^{+5.8}$ MeV, $\Gamma_{f_{0}%
(1710)\rightarrow a_{1}(1260)\pi\rightarrow\rho\pi\pi}=15.2_{-3.1}^{+2.6}$
MeV, $\Gamma_{f_{0}(1710)\rightarrow\omega\omega}\simeq0.02$ MeV as well as
$\Gamma_{f_{0}(1710)\rightarrow\omega\omega\rightarrow6\pi} \simeq0.02$ MeV
(the latter four strongly suppressed). Note, however, that the model also
obtains too large absolute values of the $f_{0}(1710)$ decay widths (although,
as already mentioned, the ratios of the decay widths are correct).

A reason for this may be the missing glueball field as all the calculations
described so far have only been performed with $\bar{q}q$ states. There are
three nearby isoscalar singlet states above 1 GeV: $f_{0}(1370)$,
$f_{0}(1500)$ and $f_{0}(1710)$. The $f_{0}(1500)$ state has long been
discussed as a possible glueball candidate. In Chapter \ref{chapterglueball}
we discuss the sigma model in two flavours + glueball to test this hypothesis.
Indeed we find $f_{0}(1500)$\ to be predominantly a glueball and $f_{0}(1370)$
to be predominantly a non-strange $\bar{q}q$ state. The study in Chapter
\ref{chapterglueball} has been performed in the light-quark sector only and
thus an extension of the study to $N_{f}=3$ + glueball would represent a
valuable continuation of the work presented in this thesis. Nonetheless, the
scenario where $f_{0}(1710)$ is predominantly a glueball was also tested by a
corresponding redefinition of our scalar states; the scenario was found to
be not favoured.\newline

Therefore, the main conclusion of this thesis is that the scalar
$\bar{q}q$ states are strongly favoured to be above 1 GeV because then the
description of the scalar but also of the vector and axial-vector
phenomenology is decisively better than under the assumption that the scalar
quarkonia are below 1 GeV. \newline

We note that the work can be extended in many directions.

\begin{itemize}
\item An obvious point is to extend the $U(3)\times U(3)$ model of Chapter
\ref{sec.fitII} to include the pure glueball field and implement the mixing of
this field with the pure $\bar{n}n$ and $\bar{s}s$ to study the
quarkonium/glueball content of $f_{0}(1370)$, $f_{0}(1500)$ and $f_{0}(1710)$.

\item This work finds the scalars above 1 GeV to be predominantly quarkonia.
This implies that the model can make no statement regarding the nature of the
scalar states below 1 GeV [$f_{0}(600)$, $a_{0}(980)$, $\kappa$]. They may be
interpreted as tetraquark states \cite{Jaffeq2q2,Achim}. Thus a further
extension of the model would entail scalar $\bar{n}n$, $\bar{s}s$, glueball
and tetraquark states (the latter with and without the $s$ quark) -- six
scalar states the mixing of which would be extremely interesting to study
within a chiral model that contains vectors and axial-vectors as well.

\item We have seen in Sec.\ \ref{2K1} that the mass of the $K_{1}$ state
obtained from our model corresponds neither to the mass of $K_{1}(1270)$ nor
to that of $K_{1}(1400)$. The $K_{1}$\ phenomenology is also not well
described (see Sec.\ \ref{sec.conclusionsfitII}). The reason is that our model
currently contains only an axial-vector nonet of states that is, however,
expected to mix with a pseudovector nonet yielding the physical $K_{1}(1270)$
and $K_{1}(1400)$ states \cite{Goldman1998}. Building on this point, one can
study the mixing of the pseudovector and axial-vector nonets within an
extended version of the model in this thesis to determine the features of the
$K_{1}(1270)$ and $K_{1}(1400)$ resonances.

\item The model can be extended to the charm mesons \cite{Walaa}.

\item The hadronic decays of the $\tau$ lepton can also be studied within a
version of the model incorporating the weak interaction (building on work in
Ref.\ \cite{Anja}).

\item Further studies of the nucleon and its chiral partner (as well as, e.g.,
hyperons) can be performed on the line of Ref.\ \cite{Susanna}.\\
An important remark is in order about nucleon-nucleon scattering in the context of results presented in this work.
Baryon-baryon interaction is usually mediated by the exchange of scalar [$f_0(600)$] and vector [$\omega(782)$, $\rho(770)$] mesons
\cite{Schaffner:1995th}. Usually the $f_0(600)$ state is considered to be of $\bar q q$ structure. However,
our results suggest the opposite: that the scalar $\bar q q$ state is actually in the region above 1 GeV.
For this reason, nucleon-nucleon scattering does not appear to be performed by exchange of a quark and an antiquark;
indeed if the states below 1 GeV are interpreted as tetraquarks then, consequently, exchange
of a tetraquark state would occur \cite{Gallas00}.

\item Finally, the issue of restoration of chiral symmetry at nonzero
temperature and density is one of the fundamental questions of modern hadron
and nuclear physics. Linear sigma models
constitute an effective approach to study chiral symmetry restoration because
they contain from the onset not only pseudoscalar and vector mesons, but also
their chiral partners with which they become degenerate once the chiral
symmetry has been restored. Given that the vacuum phenomenology is reasonably
well reproduced within our model, then the model can also be applied to
studies of chiral symmetry restoration at nonzero temperatures (similarly to Refs.\ \cite{RS,Achim}) and densities
(similarly to Ref.\ \cite{Papazoglou:1998vr}).
\end{itemize}

And let us end this thesis along the line of
Ref.\ \cite{vanBeveren:2004bz}: "It took mankind only about one century to
resolve the mystery of the spectral lines in visible light reported by Joseph
Fraunhofer in 1814 \cite{Fraunhofer}. The collection of sufficient data lasted
several decades, during which some progress was made by the discovery of
striking patterns in the spectra. An important step that provided the key to
the analysis of spectra was the classification of hydrogen lines made by
Johann Balmer in 1885 \cite{Balmer}. This allowed Niels Bohr \cite{Bohr} later
on to account for those lines, resulting in a spectacular advance in our
understanding of Nature." Nowadays the mysteries are related to far more
miniature objects but they are nonetheless a large inspiration for anyone
interested in understanding the way how nature functions.

%% file: Zusammenfassung.tex
\chapter{Zusammenfassung}

Die vorliegende Dissertation behandelt eine der grundlegenden Fragen der
menschlichen Existenz: den Zustand der Materie im Universum kurz nach dem
Urknall. Damals (vor ungef\"{a}hr 13 Milliarden Jahren) war die Materie in
ihre mikroskopischen Bauteile zerlegt: beispielsweise waren die Elektronen
nicht an Atomkerne gebunden -- es existierten keine Atome, sondern die
Elektronenn stellten freie Teilchen dar. Die Elektronen waren indes nicht
die einzigen freien Teilchen -- auch andere so genannte \textit{Leptonen}
(mit den Elektronen verwandte Teilchen) bildeten keinerlei gebundene Zust%
\"{a}nde.

Der Materieaufbau im Universum in der jetzigen Zeit ist anders:
beispielsweise sind Elektronen in einem Atom an den Atomkern gebunden (die
entprechende elektrische Wechselwirkung wird als Coulomb-Kraft bezeichnet,
nach dem franz\"{o}sischen Physiker Charles Coulomb, der im 18. Jahrhundert
lebte). Der Atomkern ist aber keine kompakte Einheit - er besitzt selbst
eine innere Struktur, da er aus Protonen (positive eletrische Ladung) und
Neutronen (keine elektrische Ladung) aufgebaut ist. Die Anzahl der Protonen
im Atomkern ist f\"{u}r die Klassifikation der Atome von grundlegender
Bedeutung: jedes Atom eines Naturelements besitzt eine genau festgelegte
Anzahl von Protonen in seinem Kern (Wasserstoff: 1, Helium: 2, Lithium: 3,
..., Ununoctium: 118). Da die Protonen, wie erw\"{a}hnt, elektrisch positiv
geladen sind, m\"{u}ssen sie sich auch im Atomkern absto\ss en; der Atomkern
m\"{u}sste folglich instabil sein, wodurch Atome (und Molek\"{u}le)
ebenfalls instabil sein m\"{u}ssten. \textit{Dies ist nat\"{u}rlich nicht
der Fall }-- stabile Materie ist auf der Erde (und, nach unserem Verst\"{a}%
ndnis, auch im Universum) in der Tat vorhanden. Folglich ist also zu
diskutieren, warum sich die Protonen in der Summe aller Kr\"{a}fte doch
anziehen (und stabile Atomkerne bilden k\"{o}nnen), obwohl sie sich
elektrisch absto\ss en. \newline

Die Antwort liegt in der Betrachtung einer neuen Wechselwirkung: der so
genannten \textit{starken} Kraft. Diese ist nur auf den Atomkern beschr\"{a}%
nkt (also extrem kurzreichweitig), aber innerhalb des Kerns ist sie
dominanter als die elektrische Absto\ss ung der Protonen. In der Summe
ziehen sich also die Protonen in Atomkernen an und Atomkerne und Atome sind
folglich stabil. \\

Die Protonen sind aber nicht die einzigen Teilchen, die der starken
Wechselwirkung unterliegen. Schon die Neutronen, die anderen in Atomkernen pr%
\"{a}senten Teilchen, sind ebenfalls stark wechselwirkend; dies ist auch der
Fall f\"{u}r Hyperonen, Pionen, Kaonen und mehrere Hundert anderer Teilchen.
Daher stellt es einen nat\"{u}rlichen Schritt dar, nach einem
Klassifikationsschema f\"{u}r all diese Teilchen zu suchen. Dieses
Klassifikationsschema erfordert die Annahme, dass die Protonen, Neutronen,
Pionen, Kaonen, ..., eine innere Struktur besitzen - und aus noch
elementareren Teilchen, den so genannteb Quarks, aufgebaut sind.
Unterschiedliche Quark-Kombinationen ergeben dann unterschiedliche Teilchen,
so wie unterschiedliche Quantit\"{a}ten von Protonen unterschiedliche
Atomkerne (und Atome) ergeben. \\

Die aus Quarks aufgebauten Teilchen werden als \textit{Hadronen} bezeichnet.
Die Hadronen unterteilen sich in zwei gro\ss e Gruppen in Abh\"{a}ngigkeit
von ihrem Spin: jene mitganzzahligem Spin (0, 1, 2, ...) werden als Mesonen
bezeichnet (Pionen, Kaonen, ...), w\"{a}hrend die Hadronen mit halbzahligem
Spin (1/2, 3/2, ...) als Baryonen bezeichnet werden (Protonen, Neutronen,
...). Die Quarks kommen in der Natur nicht als freie Teilchen vor -- sie
sind immer in den Hadronen eigenschlossen. Diese experimentelle Beobachtung
wird als Quark-Confinement bezeichnet; die Quarks k\"{o}nnen nur in
hochenergetischen Protonen- oder Schwerionen-St\"{o}\ss en (wie gegenw\"{a}%
rtig bei dem Large Hadron Collider am CERN in Genf oder bald bei der
Facility for Antiproton and Ion Research bei der Gesellschaft f\"{u}r
Schwerionenforschung in Darmstadt) erforscht werden. \newline

Die Quarks waren nicht immer in den komplexeren Teilchen eingeschlossen:
kurz nach dem Urknall waren die Quarks freie Teilchen, genau wie die
Leptonen (wie schon erw\"{a}hnt). Die Expansion des fr\"{u}hen Universums f%
\"{u}hrte zu seiner Abk\"{u}hlung; so konnte die gegenw\"{a}rtig bekannte
Materie nach ungef\"{a}hr 10$^{-10}$ Sekunden anfangen zu kondensieren. Mit
anderen Worten: es entstanden Teilchen, die aus Quarks aufgebaut sind. Es
ist klar, dass das einfachste aus Quarks aufgebaute Teilchen zwei Quarks
besitzen musste - dies ist nach der obigen Definition ein Meson, und daher
ist die Erforschung der Mesonen f\"{u}r die Erforschung des fr\"{u}hen
Universums von au\ss erordentlicher Bedeutung: sie erm\"{o}glicht uns,
Kenntnisse \"{u}ber das Universum kurz nach dem Urknall zu erlangen. \newline

Lassen Sie uns eine kurze Anmerkung einf\"{u}gen. Mesonische Teilchen
bestehen eigentlich nicht aus zwei Quarks, sondern aus einem Quark und einem
Antiquark. Der Grund hierf\"{u}r besteht in der Tatsache, dass die Quarks
neben der elektrischen auch eine zus\"{a}tzliche Ladungsform tragen: die
Farbladung. (Dies ist nicht die Farbe im herk\"{o}mmlichen Sinne, sondern
eine Quanteneigenschaft der Quarks; die Farben werden trotzdem als rot, gr%
\"{u}n und blau bezeichnet und die Experimentaldaten deuten darauf hin, dass
genau drei Quarkfarben existieren.) Die Quarks sind die einzigen bekannten
Teilchen in der Natur, welche diese Farbladung besitzen; alle anderen
Teilchen sind farbneutral und folglich ordnen sich die Quarks so an, dass
das entstehende komposite Teilchen farbneutral ist. Konkret impliziert dies,
dass ein Meson (wie zum Beispiel das Pion) aus einem Quark (mit Farbe) und
einem Antiquark (mit Antifarbe) bestehen muss, damit sich die Farbe und die
Antifarbe aufheben und das Meson, wie vom Experiment verlangt, keine
Farbladung tr\"{a}gt. (Es kann im Rahmen der Gruppentheorie gezeigt werden,
dass beispielsweise Protonen und Neutronen drei Quarks besitzen m\"{u}ssen,
um farbneutral zu sein.) \newline

Die Spins des Quarks und des Antiquarks in einem Meson k\"{o}nnen auf
unterschiedliche Arten kombiniert werden. Die Quarks selbst sind
Spin-1/2-Teilchen. Im Prinzip k\"{o}nnen sie also zu einem Spin-1-Teilchen
(ein so genanntes Vektor-Meson) und zu einem Spin-0-Teilchen (skalares
Meson) kombiniert werden. Die genaue Anzahl von so entstehenden Teilchen h%
\"{a}ngt von der Anzahl der Quarks ab, die in Betracht gezogen wurden. Gegenw%
\"{a}rtige Experimentaldaten deuten darauf hin, dass es sechs Quarks in der
Natur gibt: Up ($u$), Down ($d$), Strange ($s$), Charm ($c$), Bottom ($b$)
und Top ($t$). Das u-Quark besitzt die kleinste Masse, w\"{a}hrend die Masse
des schwersten Top-Quarks etwa 57000 Mal gr\"{o}\ss er ist. Die Massen der
Up- und Down-Quarks sind fast gleich (diese Quarks entarten also) und daher
kann man sie als gleiche Teilchen betrachten. Das Strange-Quark
unterscheidet sich in der Masse vom Up-Down-Paar um etwa Faktor 30. Die Up-
und Down-Quarks werden oftmals als \textit{nichtseltsame} Quarks bezeichnet
(und die Mesonen, welche die Up- und Down-Quarks enthalten, als
nichtseltsame Mesonen). Da die $c$-, $b$- und $t$-Quarks um eine bis drei Gr%
\"{o}\ss enordnungen schwerer als das $s$-Quark sind, kann man diese als
praktisch entkoppelt von den $u$-, $d$- und $s$-Quarks betrachten.
Betrachten wir also die nichtseltsamen $u$- und $d$-Quarks sowie das
seltsame $s$-Quark.

Wegen der erw\"{a}hnten Massenentartung bei den nichtseltsamen Quarks werden
nichtseltsame Mesonen immer sowohl aus $u$- als auch aus $d$-Quars gebildet.
F\"{u}r den konkreten Fall der skalaren Mesonen wird die Wellenfunktion wie
folgt konstruiert:

\[
\sigma _{N}\equiv (\bar{u}u+\bar{d}d)/\sqrt{2}\text{,} 
\]

wo $\sigma _{N}$ das nichtseltsame skalare Meson und $\bar{u}$ und $\bar{d}$
respektive das Anti-Up- und das Anti-Down-Quarks\ kennzeichnen, und f\"{u}r
ein seltsames

skalares Meson $\sigma _{S}$:

\[
\sigma _{S}\equiv \bar{s}s\text{,} 
\]

wo $\bar{s}$ das seltsame Antiquark kennzeichnet.

Also w\"{u}rden wir nach einem Vergleich der oben genannten beiden
Wellenfunktionen mit dem Experiment erwarten, dass die Experimentaldaten
genau zwei nichtseltsame skalare Mesonen aufweisen. \textit{Tats\"{a}chlich
sind es sechs.} \textbf{-- Und die Suche nach den Antiquark-Quark-Teilchen
unter diesen sechs ist einer der Hauptarbeitspunkte der vorliegenden
Dissertation.} \newline

Die allgemein anerkannte physikalische Theorie, welche die Quarks und die
aus den Quarks gebildeten Teilchen beschreibt, hei\ss t

\[
\text{\textbf{Quantenchromodynamik (QCD).} } 
\]

Die Quantenchromodynamik legt eine grundlegende Gleichung fest, den so
genannten QCD-Lagrangian [siehe Gl.\ (\ref{lqcd})]. Der QCD-Lagrangian zeigt
gewisse Eigenschaften auf, die nicht nur eine elegante mathematische
Konstruktion darstellen, sondern auch die tats\"{a}chlichen EIgenschaften
physikalischer (aus Quarks gebildeter) Zust\"{a}nde widerspiegeln. Dies
wurde durch viele Experimente best\"{a}tigt \cite{PDG}. \\

Falls man aber beabsichtigt, diese Zust\"{a}nde der Natur theoretisch n\"{a}%
her zu behandeln, so bedient man sich der so genannten durch die QCD
erlaubten Modelle. Diese Modelle m\"{u}ssen die erw\"{a}hnten Eigenschaften
(die \textit{Symmetrien der QCD},\ siehe Kapitel \ref{sec.QCD}) erf\"{u}%
llen; alle Modelle der QCD erf\"{u}llen die QCD-Symmetrien, aber auf
unterschiedliche Arten -- dies stellt den Hauptunterschied zwischen ihnen
dar.

Das in dieser Doktorarbeit vorgestellte Modell wird als das Lineare
Sigma-Modell bezeichnet und es beinhaltet die in der Natur beobachteten
mesonischen Teilchen. Wir beschreiben in Kapitel \ref{chapterC} die
Konstruktion eines solchen Sigma-Modells. Die Implikationen des Modells
werden in den Kapiteln \ref{chapterQ} -- \ref{ImplicationsFitII} diskutiert.
Insbesondere wird die Frage erforscht, wo sich die skalaren
Antiquark-Quark-Teilchen $\sigma _{N}$ und $\sigma _{S}$ sich im
physikalischen Spektrum befinden. Diese Frage ist aus mindestens zwei Gr\"{u}%
nden interessant:

\begin{itemize}
\item Da die experimentellen Messungen (wie erw\"{a}hnt) mehr skalare
Teilchen nachgewiesen haben als von der theoretischen Seite erwartet, stellt
sich die Frage der Klassifikation solcher Teilchen, oder in anderen Worten
derer Struktur: da h\"{o}chstens zwei von diesen Teilchen von
Antiquark-Quark-Struktur (${\bar{q}q}$) sein k\"{o}nnen, stellt sich die
Frage, welche von den gemessenen Teilchen tats\"{a}chlich die ${\bar{q}q}$%
-Teilchen sind und welche Struktur die \"{u}brig gebliebenen Teilchen
besitzen.

\item Das Pion ist ein wohlbekanntes ${\bar{q}q}$-Teilchen (dies ist seit
langer Zeit sowohl theoretisch als auch experimentell best\"{a}tigt); die
QCD sagt vorher, dass das Pion unter gewissen Bedingungen (sehr hohe
Temperaturen von ungef\"{a}hr einer Billion Grad Celsius) dieselbe Masse wie 
$\sigma _{N}$ besitzen muss -- wir k\"{o}nnen aber zwischen \textit{sechs}
skalaren Teilchen w\"{a}hlen, die allesamt unserem $\sigma _{N}$-Teilchen
entsprechen k\"{o}nnen. Die Frage ist also: \textit{Welches von den skalaren
Teilchen ist es?}
\end{itemize}

Allerdings w\"{a}re eine theoretische Betrachtung von nur Pionen und
skalaren Teilchen nicht gerechtfertigt, da die experimentellen Daten
eindeutig die Existenz anderer Teilchen nachweisen. Zum Beispiel ist
experimentell wohlbekannt, dass auch Teilchen mit Spin 1 existieren (die so
genannten \textit{Vektoren}), welche mit den Pionen und den skalaren
Teilchen wechselwirken.\ Aus diesem Grunde beinhaltet das in dieser
Dissertation disktierte Modell sowohl die Skalaren als auch die Vektoren;
ein Modell mit all diesen Teilchen muss mathematisch konsistent konstruiert
werden, was im Kapitel \ref{chapterC} beschrieben wird. \newline

Die skalaren Mesonen werden in zwei Gruppen geteilt: auf jene mit
Ruheenergie unterhalb 1 GeV und auf jene mit Ruheenergie oberhalb 1 GeV (die
Bezeichnung \textit{GeV} bedeutet \textit{Gigaelektronvolt}, also eine
Milliarde Elektronvolt, wobei ein Elektronvolt der Energie eines Elektrons
im elektrischen Feld von einem Volt St\"{a}rke entspricht). Im Kapitel \ref%
{chapterQ} wird mittels Vergleich der theoretischen Ergebnissen mit
Experimentaldaten diskutiert, ob sich unser skalares ${\bar{q}q}$-Teilchen
unterhalb oder oberhalb 1 GeV befindet -- und es scheint die Ruheenergie
mehr als 1 GeV zu \ besitzen.

Dies ist eigentlich etwas \"{u}berraschend. \"{U}blich ist die Erwartung,
dass ein Teilchen mit blo\ss\ einem Quark und einem Antiquark eher eine
relativ kleine Ruheenergie besitzt (in unserem Fall also weniger als 1 GeV).
Der Grund hierf\"{u}r ist, dass alle anderen skalaren Teilchen, die \textit{%
keine} Antiquark-Quark-Struktur besitzen, aus mehr als zwei Quarks bestehen
und deren Ruheenergie folglich relativ gr\"{o}\ss er ist. Die Ergebnisse des
Kapitels \ref{chapterQ} (und letztendlich dieser Dissertation) deuten auf
ein umgekehrtes Bild hin. \newline

Die Ergebnisse im Kapitel \ref{chapterQ} sind aber nur unter Betrachtung der
Mesonen zustande gekommen, die nur das Up- und das Down-Quark besitzen. Es
ist folglich eine wohldefinierte Frage, ob sich die Ergebnisse wom\"{o}glich 
\"{a}ndern, wenn auch Teilchen mit seltsamen Quarks (die so genannten 
\textit{Kaonen}) in das Modell hunzugef\"{u}gt werden.

Aus diesem Grunde wird in den Kapiteln \ref{sec.remarks} -- \ref%
{ImplicationsFitII} eine ausf\"{u}hrliche Diskussion des Linearen
Sigma-Modells mit skalaren und vektoriellen Mesonen sowohl im nichtseltsamen
als auch im seltsamen Sektor durchgef\"{u}hrt. Die Er\"{o}rterungen \"{u}ber
die skalaren Mesonen sind hierbei nicht die einzigen, welche behandelt
werden -- in den genannten Kapiteln werde \textit{alle} hadronischen Zerf%
\"{a}lle der Mesonen betrachtet, die aus dem Modell ausgerechnet werden k%
\"{o}nnen. Auf diese Weise entsteht eine breite ph\"{a}nomenologische
Abhandlung der experimentell bekannten mesonischen Teilchen, die uns eine
Klassifikation der Teilchen nach ihrer Quark-Struktur (ob ${\bar{q}q}$ oder
nicht) durchzuf\"{u}hren, aber auch Einblicke in das Verhalten der Teilchen
bei sehr hohen Temperaturen erm\"{o}glicht. \newline

Die in den Kapiteln \ref{sec.remarks} -- \ref{ImplicationsFitII} durchgef%
\"{u}hrten\ Berechnungen best\"{a}tigen das (wie erw\"{a}hnt) \"{u}%
berraschende Ergebnis aus Kapitel \ref{chapterQ}: dass die skalaren ${\bar{q}%
q}$-Teilchen $\sigma _{N}$ und $\sigma _{S}$ eine Ruheenergie von mehr als 1
GeV besitzen. Diese Aussage hat mindestens zwei Implikationen:

\begin{itemize}
\item Das skalare Teilchen, welches bei sehr hohen Temperaturen ($\sim
10^{12}$ Grad Kelvin) die gleiche Masse wie das Pion besitzt, hat eine viel
gr\"{o}\ss ere Ruheenergie als das Pion. Dies hat Konsequenzen f\"{u}r
andere Signaturen des so genannten \textit{Quark-Gluon-Plasmas}, eine
Materieform, deren Entstehung bei den erw\"{a}hnten sehr hohen Tenperaturen
erwartet wird und die aus Quarks, aber auch Gluonen besteht -- dabei sind
die Gluonen Teilchen, welche die Wechselwirkung zwischen den Quarks \"{u}%
bertragen (die \textit{Botenteilchen}).

\item Falls (nur) die skalaren Teilchen \"{u}ber 1 GeV die ${\bar{q}q}$%
-Struktur besitzen, dann bleibt die Frage offen, welche Struktur die
(ebenfalls bekannten) Teilchen unter 1 GeV haben k\"{o}nnten. Dazu ist immer
noch keine definitive Antwort vorhanden (auch nicht im Rahmen anderer
Studien), es wird aber schon seit Langem dar\"{u}ber diskutiert, ob die
Teilchen unterhalb 1 GeV aus zwei Quarks und zwei Antiquarks (statt wie
bisher diskutiert aus einem Quark und einem Antiquark) bestehen k\"{o}nnten.
\end{itemize}

In der vorliegenden Dissertation wird aber noch eine zus\"{a}tzliche
Mesonenart diskutiert: die Glueb\"{a}lle. Diese Mesonen bestehen nicht aus
Quarks, sondern ausschlie\ss lich aus Gluonen, den (schon erw\"{a}hnten) 
\textit{Botenteilchen}, \"{u}ber welche die Quarks ihre Wechselwirkungen ausf%
\"{u}hren. In Kapitel \ref{chapterglueball} wird das skalare (also spinlose)
Glueball-Teilchen in das Modell eingef\"{u}hrt und dessen Wechselwirkungen
mit dem Antiquark-Quark-Teilchen $\sigma _{N}$ diskutiert. Es wird wiederum
die Aussage best\"{a}tigt, das die Ruheenergie von $\sigma _{N}$ \"{u}ber 1
GeV liegt und zus\"{a}tzlich die Folgerung diskutiert, dass die Ruheenergie
des Glueball-Teilchens ebenfalls \"{u}ber 1 GeV ist. \newline

Die Hauptaussage dieser Dissertation ist aber, dass die Spin-0-Teilchen aus
einem Antiquark und einem Quark (die skalaren Mesonen) eine h\"{o}here
Ruheenergie besitzen als gew\"{o}hnlich angenommen. Dies hat viele
Implikationen f\"{u}r die weitere Mesonen- und, allgemeiner,
Hadronenerforschung: die Frage nach der Struktur der skalaren Mesonen im
Energiebereich unter 1 GeV bleibt offen genau so wie die Frage nach dem
Materiezustand und -verhalten bei sehr hohen Temperaturen (also jenen wie
kurz nach dem Urknall). Die Erforschung der Materie bei sehr hohen Temperaturen
ist seit langer Zeit das haupts\"{a}chliche Thema vieler Projekte sowohl in
der theoretischen als auch in der experimentellen Hadronenphysik, aber eine definitive
Aussage \"{u}ber das Materieverhalten unter den extremen Bedingungen und bei einer
gro\ss en Anzahl der mikroskopischen Teilchen (und folglich einer fast
unendlichen Anzahl m\"{o}glicher Wechselwirkungswege der betreffenden Teilchen
und der Zerfallswege der instabilen Teilchen) kann noch nicht erfolgen. Aus diesem
Grunde ist die theoretische und die experimentelle Erforschung der Elementarteilchenphysik
bei sehr hohen Energien ein sehr spannendes Feld der Physik -- mit vielf\"{a}ltigen Anwendungsm\"{o}glichkeiten
der hier präsentierten Dissertation. Die vorliegende Dissertation bildet daher durch ihre
Untersuchungen der Antiquark-Quark-Zust\"{a}nde genau die notwendige
Basis f\"{u}r weitere Projekte bez\"{u}glich des Materiezustands bei sehr hohen Temperaturen und
folglich f\"{u}r die Erforschung der Materie kurz nach dem Urknall und dem
Anfang der noch unvollst\"{a}ndig erforschten Ph\"{a}nomene der modernen
Wissenschaft.


%% file: Danksagung.tex
\chapter*{\vspace{-3.3cm} \Large{Acknowledgments}}

\thispagestyle{empty}

\small Many people have contributed to this work. I owe a great debt of gratitude to my mentor and PhD advisor Prof.\ Dirk Rischke
who has been a steady companion, a helping hand and a wonderful teacher ever since I started my diploma studies. I thank him for all the help he 
has provided me on more than one (and any) occasion.\\

Furthermore I thank Dr.\ Francesco Giacosa for answering literally all of my questions any time and with due profoundness and paedagogical cunning.
My understanding of hadronic physics has improved massively in collaboration with Francesco, and for this I am very grateful.\\

I thank Prof.\ J\"{u}rgen Schaffner-Bielich for carefully reading my manuscript. I am very grateful to Dr.\ Gy\"{o}rgy Wolf and Dr.\ P\'{e}ter Kov\'{a}cs
for the calculations in the three-flavour model and for very useful discussions about this thesis.\\

I also thank Prof.\ David Bugg for directing my attention to the issues regarding the $f_0(1790)$ resonance and
for reading the chapter about experimental data on scalar states. The disentanglement of $f_0(1710)$ and $f_0(1790)$ was enormously important
for this thesis and I am very grateful for the discussion about scalars and related issues.\\

Thanks go to Prof.\ Jos\'{e} Pel\'{a}ez and Dr.\ Brigitte Bloch-Devaux for discussions about pion scattering lengths and to Dr.\ Veljko Dmitra\v{s}inovi\'{c} 
for pointing my attention to earlier work about modelling of the chiral anomaly. Thanks also go to Prof.\ Joe Schechter for useful discussions
about sigma models and to Prof.\ Pedro Bicudo for discussions of QCD at non-zero temperatures.\\

I thank Dr.\ Tom\'{a}\v{s} Brauner for extremely useful discussions about physics; for the same reason,
I also thank PD Dr.\ Dennis D.~Dietrich who has given me the first introduction into physics beyond the standard model as well.
I am grateful to Prof.\ Francesco Sannino for useful discussions about the Standard Model, and beyond, and also to
Stefano Di Chiara, Oleg Antipin and Eugenio Del Nobile for providing me with information about technical tools to study Technicolour.\\

I am grateful to Anja Habersetzer for doing a wonderful job creating Feynman diagrams, to Stanislaus Janowski for calculations
with the dilaton field and to Achim Heinz, Martin Grahl and Susanna Gallas for discussions about physics and to all of them for years of friendship.\\

Herzlichen Dank an Peter H\"{a}hnel, ohne den meine Ankunft in Deutschland nicht m\"{o}glich gewesen w\"{a}re. Meine besondere Dankbarkeit
geb\"{u}hrt der Stiftung Polytechnische Gesellschaft, die mein Doktorandenstudium per Stipendium und durch viele interessante Seminare unterst\"{u}tzt hat.
Ich danke insbesondere Herrn Dr.\ Roland Kaehlbrandt, Herrn Dr.\ Wolfgang Eimer und Herrn Tobias Ullrich f\"{u}r sehr gute Zusammenarbeit im Rahmen
des MainCampus-Projektes der Stiftung Polytechnische Gesellschaft.\\

Veliko hvala profesoru Ilijasu Hod\v{z}i\'{c}u iz zeni\v{c}ke Prve gimnazije koji me svojim predavanjima potaknuo da upi\v{s}em fiziku.\\
Porodici Kahriman se zahvaljujem za svu podr\v{s}ku tokom ovih godina u Frankfurtu.\ Aleksandru Mastilovi\'{c}u se zahvaljujem za razgovore o fizici,
a Dervi\v{s}u Buri\'{c}u za razgovore o svemu osim o fizici.\\

Mojoj Enili koja me obogatila na toliko mnogo na\v{c}ina.\\

Mojoj sestri Dajani za podr\v{s}ku tokom doktorskog studija.\\

I mojoj majci Ljiljani koja mi je stalna i ogromna podr\v{s}ka u \v{z}ivotu.

\thispagestyle{empty}

%% file: cv.tex
\chapter*{\vspace{-1.8cm} \Large{Lebenslauf}}

\thispagestyle{empty}

Name: Denis Parganlija\\
Geburtsort: Zenica, Bosnien-Herzegowina\\

{\it 2002} \\
Abitur\\
Allgemeines Gymnasium, Zenica, Bosnien-Herzegowina\\

Aufnahme des Physik-Studiums\\ 
Goethe-Universit\"at, Frankfurt am Main\\

{\it 2004} \\
Stipendium\\
Friedrich-Ebert-Stiftung (bis 2007)\\

{\it 2007} \\
Diplom in Physik\\
Institut f\"ur Theoretische Physik\\
Goethe-Universit\"at, Frankfurt am Main\\
Betreuer: Prof.\ Dr.\ Dirk H.\ Rischke\\

{\it 2007} \\
Aufnahme des Promotionsstudiums in Physik\\
Institut f\"ur Theoretische Physik\\
Goethe-Universit\"at, Frankfurt am Main\\
Betreuer:  Prof.\ Dr.\ Dirk H.\ Rischke\\

{\it 2007} \\
Wissenschaftliche Mitarbeit\\ 
Institut f\"ur Theoretische Physik der Goethe-Universit\"at Frankfurt am Main\\
Betreuung von \"Ubungsgruppen in der Theoretischen Physik\\
Organisation der Ferienkurse in der Theoretischen Physik (Mechanik I, Mechanik II, Elektrodynamik, Quantenmechanik I)\\

{\it 2008} \\
Stipendium\\
Stiftung Polytechnische Gesellschaft, Frankfurt am Main (bis 2012)\\

{\it 2011-2012} \\
Stellvertreter in den Vorlesungen \textit{Elektrodynamik}, \textit{Quantenmechanik I} und \textit{Statistische Mechanik} \\
Stellvertretend f\"ur: Prof.\ Dr.\ Dirk H.\ Rischke